# Process Based Unification for Multi-Model Software Process Improvement

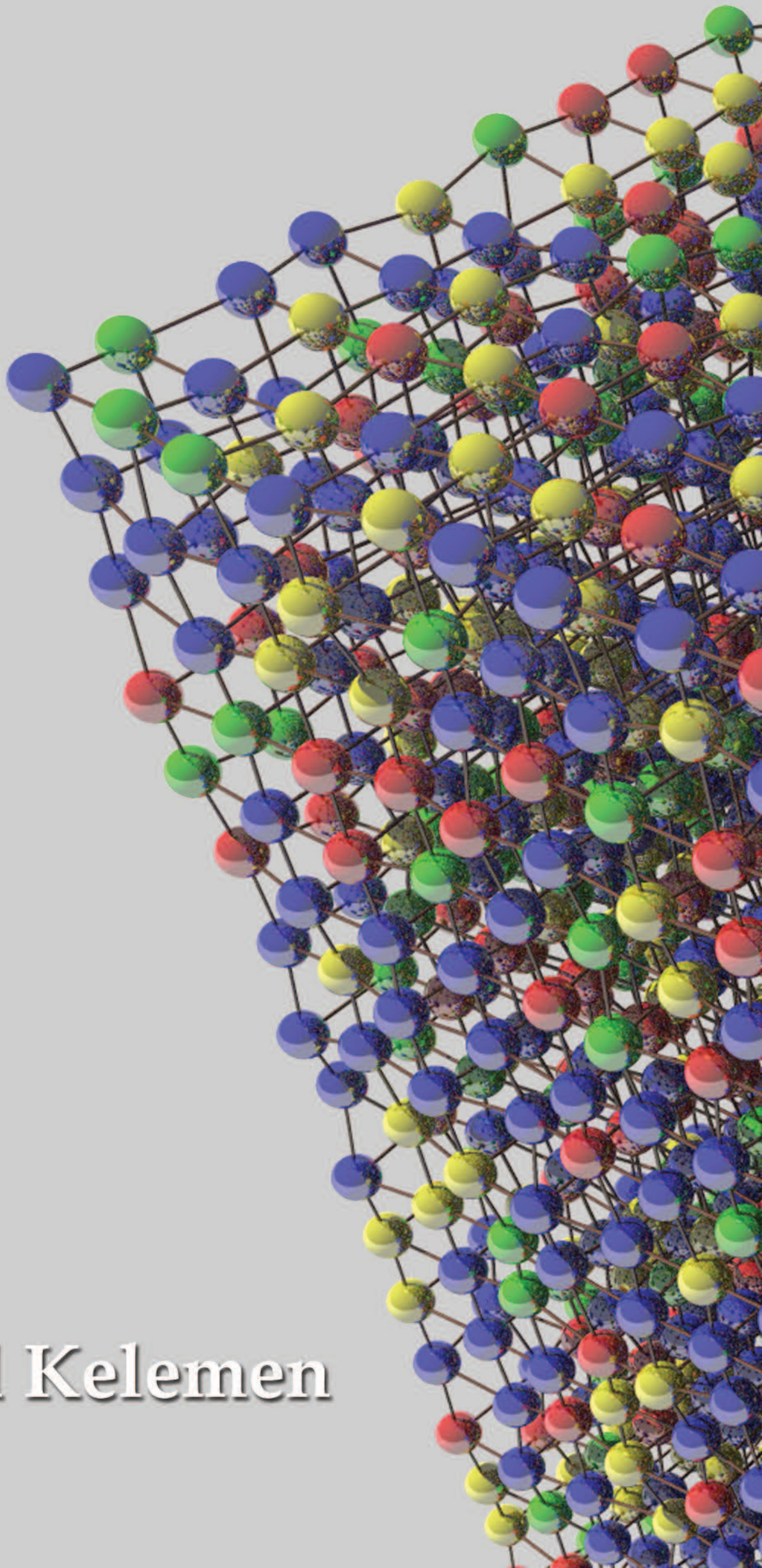

Zádor Dániel Kelemen

# Process Based Unification for Multi-model Software Process Improvement

PROEFSCHRIFT

ter verkrijging van de graad van doctor aan de
Technische Universiteit Eindhoven, op gezag van de
rector magnificus, prof.dr.ir. C.J. van Duijn, voor een
commissie aangewezen door het College voor
Promoties in het openbaar te verdedigen
op woensdag 30 januari 2013 om 16.00 uur

door

Zádor Dániel Kelemen

geboren te Odorheiu Secuiesc (Székelyudvarhely), Roemenië

Dit proefschrift is goedgekeurd door de promotoren:

prof.dr. R.J. Kusters
en
prof.dr.ir. P.W.P.J. Grefen

Copromotor:
dr.ir. J.J.M. Trienekens

# ACKNOWLEDGEMENTS

This thesis summarizes the research that I have conducted between 2007 and 2012.

Firstly, I would like to express my deepest gratitude to God, who continuously provided everything I needed all the way long in completing this research. I want to thank everybody who supported me and contributed to this work.

Special thanks go to prof. dr. Rob Kusters, dr. ir. Jos Trienekens and prof. dr. Katalin Balla who led me all the way through my PhD research. I cannot count how much they have helped me and how many things I have learned from them.

I also thank their encouragement and support in critical episodes of this research. I visited Eindhoven several times for monthly intervals and I am very grateful to Katalin for all the support and for letting me out from the Budapest office and to Jos and Rob for welcoming me so warmly at TU/e. I want to thank prof. dr. ir. Paul Grefen for being my second promoter and providing useful insights into methodology. I would like to thank for committee members of my dissertation prof. dr. ir. Aarnout Brombacher, prof. dr. Sjoerd Romme and prof. dr. ir. Rini van Solingen.

The case study presented in this thesis was conducted at Polygon Ltd. within the frame of the Q-works project. I thank the whole team for the perfect cooperation, especially to dr. György Gál (Gyuri), Melinda, Zsuzsa, Alexandra, József and László. I also thank my colleague Tibor and ex-colleague Gábor for their feedback on my work.

I would like to thank Pieter for reviewing important BPMN figures, Imteaz for English corrections of the text and Daan for translating the English summary to Dutch.

I thank my prior and current students at Budapest University of Technology and Economics who contributed to the progress of my work. I especially thank Vince for working on a topic which if evolves enough, could be a good addition to this research.

I must thank the secretaries Annemarie and Ada who always helped me to get accommodation, bike, room and everything I needed to stay in Eindhoven. I am really impressed how ef-ficient they are!

I thank my friends and colleagues from Eindhoven who made my visits pleasant and cheerful: Ariel, Botond, Davide, Deepak, Egon, Gwenda, Heidi, Jana, Joel, Júlia, Marco, Maryam, Ricardo, Rob, Ronny, Samuil, Vassil, Zhao and Zhiqiang among many others. Thank you for spending your precious time with me, you all made me feel home.

I also thank my friends who influenced my work primarily by their work attitude or by sharing their thoughts: András, Anna, Barna, Gyula, Gyurka, Imti, István, Korn, Linda, Péter, Rajmond and Zoli.

I thank all my family members for their support!

Special thanks to Szilvi, who is the owner of the biggest heart I know. She recognized that this is important to me, so she kept supporting me in all situations. Thank you and to your family!

Huge thanks to my mother, who always believed in me and biffed me with good humor when I was eager to receive some serious back-kicking in my dissatisfied moments. Thanks to my grandparents for showing me the best example of hard work and perseverance. Special thank goes to Irma néni, sister of my grandmother, who helped me to settle in Hungary and to begin my work!

# TABLE OF CONTENTS

# 1. INTRODUCTION

This dissertation is an account of a research project aimed at understanding the problem of *multi-model software process improvement* and providing a *Process Based Unification* (abbreviated as: PBU) framework for simultaneously using multiple software quality approaches. This chapter provides a generic introduction to the research field where the research problem is identified. First we review evolution of software process improvement in chapter 1.1, then terms used in this thesis are discussed in chapter 1.2, multi-model software process improvement is introduced in chapter 1.3 and related issues are presented in chapter 1.4.

## 1.1. Process improvement

A definitive goal in software quality assurance is to ensure and improve the quality of a software product. In developing products, resources implement processes, therefore quality of resources and processes have effect on the quality of the product (Tsui & Karam, 2007).

The process management premise in CMMI-DEV v1.3 says: *"the quality of a system or product is highly influenced by the quality of the process used to develop and maintain it"* and continues with: *"The belief in this premise is seen worldwide in quality movements..."* (CMMI Product Team, 2010a, p. 5).

In practice, due to the diversity and the complexity of software products, it is more difficult to measure and compare their quality than to measure/improve the quality of processes producing them. Moreover, the software product can only be measured at the end of the development, which is late for making changes, while the quality of the process can be measured and improved during the development. Despite that the product quality improvement through processes may be indirect, a mapping between process and product oriented quality approaches can be developed in many cases (García-Mireles, Moraga, García, & Piattini, 2012) and a process improvement can be product focused (Bekkers, Weerd, Spruit, & Brinkkemper, 2010;

Solingen, 2000).

For further discussion on different views on software quality such as product, process and resource oriented approaches see (Balla, 2001; Fenton & Pfleeger, 1997).

Similarly to CMMI, Balla suggests the following fundamental objects of software quality: product, resource and process quality (Balla, 2001). In this thesis from the three objects defined by Balla we focus on the *process*, however without neglecting the associated product and resource of software quality.

Table 1 – Evolution of software processes – a summary of Cugola and Ghezzi's view

| Approach | Example | Strength | Weakness |
| --- | --- | --- | --- |
| Lifecycle models | Waterfall model | Well structured, clear documentation | Idealised processes |
| Methodologies | JSD (Jackson System Development), JSP | Based on experiences from previous projects | Informal notation, increased paperwork |
| Formal development | Program development by stepwise refinement | Transforms specification to correct implementation | Not scalable, applicable only for small programs |
| Automation | SDEs (Software Development Environment) | Automation of some areas of software production | Requirements specification, design decisions cannot be automated |
| Management (Process Orientation) | ISO 9001, CMMI, TSP, PSP | Indirect assurance of quality products | Increased bureaucracy |

In "Software Processes: a Retrospective and a Path to the Future"(Cugola & Ghezzi, 1998) Cugola and Ghezzi give a good summary of the main steps of software process evolution starting from the early 60's. In Table 1 strengths and weaknesses of lifecycle models, methodologies, formal development, automation, management (software processes) are shown.

According to (Cugola & Ghezzi, 1998; Fuggetta, 2000), after these approaches a new era came: process modeling and process programming. There are several process modeling initiatives such as the development of PMLs (Process Modeling Languages), Little JIL (Osterweil, 1997, 2007), Oz (Oz and Oz Web – the first "decentralized" Process-centred Software Engineering Environment was developed at Columbia University), Endeavors, BPM or enterprise modeling (Wortmann & Kusters, 2007) in this field.

In 2009 the SPI manifesto was defined by software process improvement (SPI) researchers and practitioners stating core values and principles of the field (Korsaa et al., 2010). After more than a decade of Cugola and Ghezzi's article on software process evolution, software process improvement is used in practice at software companies and is subject of research from multiple angles and views. Some of the emerging areas of software process improvement are:

- Agile software process improvement (Chow & Cao, 2008; Dybå & Dingsøyr, 2008; Linders, 2011),

- Software process improvement in global environment, (Gray & Smith, 1998; Lanubile, Ebert, Prikladnicki, & Vizcaino, 2010; Portillo-Rodriguez, Vizcaino, Ebert, & Piattini, 2010),
- Software process improvement at (very) small (Garcia, Graettinger, & Kost, 2005; Habra, Alexandre, Desharnais, Laporte, & Renault, 2008; Richardson, 2006) and medium enterprises (Pino, García, & Piattini, 2007),
- Software process improvement at start-ups (Blank, 2006; Cooper, Vlaskovits, & Blank, 2010; Ries, 2011) and
- Multi-model software process improvement involving various approaches such as ISO/IEC/IEEE standards, various improvement frameworks, models, LEAN or agile methods (Heston & Phifer, 2010; SEI, 2008).

More and more organizations have to deal with several of these areas. In this thesis we focus on multi-model process improvement – the synergic usage of multiple process improvement standards, models, technologies and methods. In 1.2 we introduce fundamental terms used in this research, in 1.3 we introduce multi-model software process improvement and in 1.4 issues of the field will be introduced.

## 1.2. Fundamental terms

In this chapter we clarify fundamental terms used in this research. This is needed because there is no consistent terminology used in the field of multi-model software process improvement. The lack of consistent terminology in multi-model software process improvement was also recognized by Pardo et al. and ontological discussions were published in (C. Pardo, Pino, García, Piattini, & Baldassarre, 2011).

When we use the term *standard* we refer to the materials officially standardized and published by standardization organizations. Such international standardization organizations are e.g. the International Organization for Standardization (ISO, 2010), Institute of Electrical and Electronics Engineers (IEEE, 2010) or International Electrotechnical Commission (IEC, 2010).

We use the term *model* when we refer to a published material which calls itself a model, e.g. CMM, CMMI or TMM. The models or standards are not necessarily standardised although some of them are (e.g. SPICE is standardised as ISO 15504).

The terms *methodology*, *method*, *technique* and *notation* are defined by (Blokdijk & Blokdijk, 1987) as follows:

*“Methodology means the science of method: a treatise or dissertation on method. A method is a systematic procedure, technique or mode of inquiry, employed by or proper to a particular discipline; or: a body of skills or techniques. A technique is a procedure or body of technical actions. … A notation is a system of characters, symbols or abbreviated expressions used to express technical facts of quantities.”*

The Merriam-Webster Online Dictionary (Merriam-Webster Inc., 1996) defines *methodology* in a very similar way:

- *“a body of methods, rules, and postulates employed by a discipline: a particular procedure or set of procedures*

- *the analysis of the principles or procedures of inquiry in a particular field"*

The dictionary defines the *method* as:

- *"a procedure or process for attaining an object: as a (1): a systematic procedure, technique, or mode of inquiry employed by or proper to a particular discipline or art (2): a systematic plan followed in presenting material for instruction b (1): a way, technique, or process of or for doing something (2): a body of skills or techniques*
- *a discipline that deals with the principles and techniques of scientific inquiry*
- *orderly arrangement, development, or classification : plan*
- *the habitual practice of orderliness and regularity*
- *a dramatic technique by which an actor seeks to gain complete identification with the inner personality of the character being portrayed."*

We use the term *method* as it is defined by (Blokdijk & Blokdijk, 1987) or for a published material which is recognized as a method (e.g. Agile methods).

In on-going research in the area of software process improvement, different terms are used for software quality approaches. Examples are: quality standard, quality assurance method, improvement framework (M. C. Paulk, 2008), software process improvement (SPI) framework (Halvorsen & Conradi, 2001), quality model, improvement technology (Siviy, Kirwan, Marino, & Morley, 2008a) and process improvement model (A. Ferreira & Machado, 2009) among others. In order to emphasize that each standard (e.g. ISO 9001 or ISO 12207), method and (improvement) technology framework (e.g. CMMI, SPICE) is a specific approach to software quality; we call each of them an *approach*.

An approach which is connected to quality is called *quality approach.* We also emphasize that we mainly focus only on the approaches which can be used to create/improve/maintain software specific processes. A quality approach which can be used in software industry is called *software quality approach*.

A (software) quality approach which mainly focuses on a process or more processes is called *process-oriented* (software) quality approach (abbreviated as po(s)qa). In this way we exclude quality approaches which are not primarily process-oriented. Such approaches could be product (e.g. ISO 9126 (ISO/IEC, 2001, 2003a, 2003b, 2004a) or IEEE 829-2008(IEEE, 2008a)) or resource-oriented (e.g. People-CMM(Curtis, Hefley, & Miller, 2001)).

We call *quality approach element (class)* – an element of a quality approach (e.g. activity, artifact, role, chapter or requirement) and *quality approach element instance* the instance of a quality approach element. E.g. "prepare for review" can be an instance of "activity" element or a "project plan" can be an instance of the "artifact" element. We use the terms "quality approach element" and "quality approach element class" as synonyms.

*Multi-approach process improvement* or *multi-model process improvement* (abbreviated as MSPI) mean process improvement based on multiple software quality approaches. According to our terminology, the first term would be more logical, but we use the latter one because it is getting emphasized in the field of process improvement (Apithanataveepa, 2008; A. L Ferreira, Machado, & Paulk, 2010; Malzahn, 2008; C. J. Pardo, Pino, García, & Piattini, 2009; Salviano, 2009a; Siviy, Kirwan, Marino, & Morley, 2008e). These two terms have the same meaning.

In a research at SEI the term *improvement technology* is used with a very similar meaning to our quality approach, defined as follows:
"...we use the terms *improvement technologies*, technologies, or models somewhat interchangeably as shorthand when we are referring in general to the long list of reference models, standards, best practices, regulatory policies, and other types of practice-based improvement technologies that an organization may use simultaneously." (Siviy, Kirwan, Marino, & Morley, 2008b).
In this research we call the *multi-model problem* the problem of the simultaneous usage of multiple quality approaches. We call *multi-model initiative* all the initiatives which are aimed at solving the multi-model problem. An initiative which solves the multi-model problem is called a *(multi-model) solution*. The output (or result) of a multi-model initiative is called a *multi-model result*.
For further terms and abbreviations used in this document see Terms and definitions and Acronyms.

## 1.3. Multi-model software process improvement

Many different quality approaches are available in the software industry. Discovery of approximately 315 quality approaches of 46 different organizations has been reported by Moore (Moore, 1999). A picture of interrelations among 39 different quality approaches was published in (Sheard, 2001). For a view of quality approaches discovered by Sheard see Figure 1.

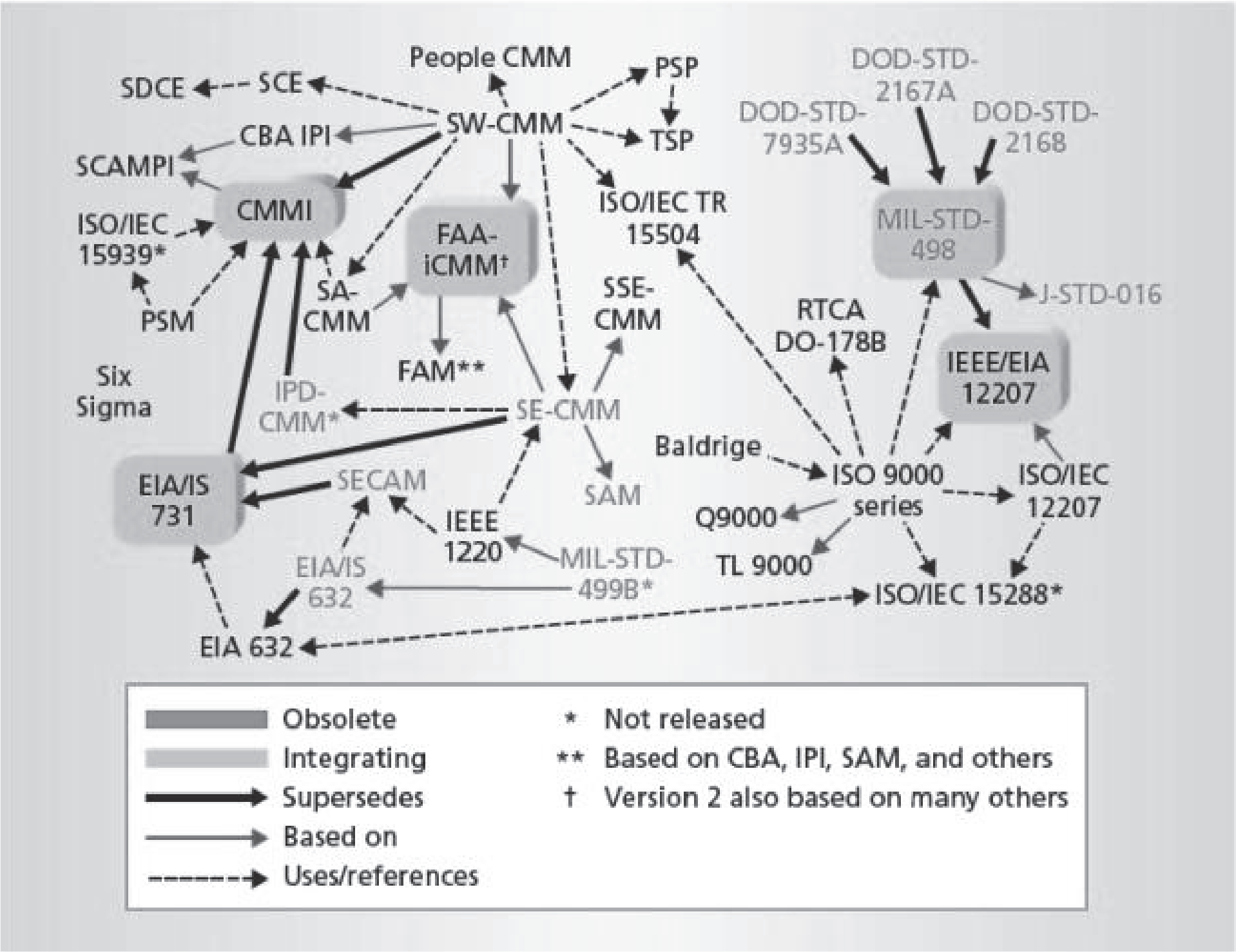

Figure 1 – A "framework quagmire" from 2001, Source: (Sheard, 2001)

After more than a decade of Sheard's article, the situation is becoming even more complicated with the appearance of new (and new versions of prior) quality approaches: see e.g. the review of 52 Software Process Capability/Maturity Models in (von Wangenheim, Hauck, Salviano, & von Wangenheim, 2010) or a review of developing maturity models in (García-Mireles, Ángeles Moraga, & García, 2012).

A number of differences among quality approaches exist. Some of the approaches, such as ISO 9001 (ISO, 2008) are not software specific, i.e. they define general requirements for an organization and they can be used at any company. Others such as Automotive SPICE (Automotive Special Interest Group (SIG), 2008) have been derived from a software specific approach (ISO/IEC, 2004b), and can be used for improving specific (in this case automotive) processes. Some are created to improve development processes (CMMI Product Team, 2010a; FAA, 2001; Ibrahim, 2010; ISO/IEC/IEEE, 2008), others focus on services (CMMI Product Team, 2010b), and again others are related to particular processes such as software testing (TMMi Foundation, 2009) or resource management (Curtis et al., 2001).

Differences in structure and granularity also exist between quality approaches, e.g. both CMMI for Services and ITIL focus on services. However, while CMMI defines process areas, goals and practices, it does not define the concrete steps of the processes. ITIL contains very detailed descriptions and provides process flowcharts for guiding service implementations.

If we look for one process we can find its best practices and requirements in many quality approaches. Taking as an example the peer review process we can see that the whole idea of applying peer reviews in software development comes from Fagan (Fagan, 1976, 1986). Later, the concept of peer review has been widely applied by different parties, e.g. in CMM requirements for a peer review process are represented as a key process area (M. Paulk, Weber, Curtis, & Chrissis, 1995), in CMMI for Development the peer review appears in specific goal levels (CMMI Product Team, 2010a), while in CMMI for Services the peer review is represented as a specific practice (CMMI Product Team, 2010b). The peer review is also applied in many other quality approaches, in many different ways, such as SPICE (ISO/IEC, 2004b), ISO 12207 (ISO/IEC/IEEE, 2008) or IEEE 1028 (IEEE, 2008b). Software testing books also highlight the importance of peer reviews, describing it as a preliminary testing technique (Graham, Van Veenendaal, Evans, & Black, 2007; Hambling, Morgan, Samaroo, Thompson, & Williams, 2007; Morgan, 2010). Wiegers described how to "humanize" peer reviews, giving recommendations and templates for process implementation (K. E. Wiegers, 2002a).

The problem of creating processes conforming to multiple quality approaches can also be recognized in other areas, for example in configuration management, requirements management, requirements engineering or software project management among many others.

There can be various situations in which the usage of multiple quality approaches is needed (Mirna, Jezreel, Giner, A., & Tom´s, 2011; Siviy et al., 2008b), e.g. to strengthen a particular process with various aspects of multiple quality approaches, or to reach certification of the compliance to a number of standards.

Some of the typical situations as we identified them based on literature and practice are:

- One typical case of using multi-model solutions is when the quality approach used by the organization does not contain full description of a selected process (e.g. the peer review process in CMMI Verification Specific Goal 2). CMMI contains the "what" part of the process and not the “how”. Therefore, there is a need for using further quality approaches which contain the missing information for implementing/improving the process (Siviy et al., 2008b).

- Another situation is when certifying multiple quality approaches is required (e.g. required by external parties). In this case one effective solution is the full mapping of quality approaches (and processes) focusing on the requirements of quality approaches. If these approaches (and the mapping of these approaches) are not covering fully the selected process, it might be useful to include specific parts of further approaches (Siviy et al., 2008b).
- A green field case, when the organization decides freely about the quality approaches to use. Since there is no certification pressure on the company, the company can choose a (primary) quality approach to implement. The primary approach will be the quality approach describing the selected process in the best way (fits best of company’s business goals and needs) and the missing parts could be completed from other approaches (Apithanataveepa, 2008; Siviy et al., 2008b).
- There are particular critical processes e.g. in the military, aviation systems or the closing of the water defence in the Netherlands which require as good as possible support for which multi-model SPI can be used (DTIC, 2008; NDIA, 2009).
- Continuous improvement of software development processes would entail monitoring developments in quality approaches and picking out those which can serve as a basis for interesting improvements (CMMI Product Team, 2010a; Hammer, 2002).

Given the existence of so many approaches which focus on processes, an organization has to take decisions. First of all it has to be decided which approaches have potential for the organization. Consequently, the organization may need to use more approaches and the decision has to be made how the chosen approaches have to be used simultaneously.

## 1.4. Issues and solutions in multi-model software process improvement

Despite that numerous software companies use more quality approaches simultaneously, they often struggle with interpreting them due to differences in structure, terminology, content and many other characteristics (Kugler, 2008; Siviy et al., 2008b). Tailoring multiple quality approaches to the company’s needs is a time-consuming process and needs special expertise (Kasser, 2005; Kugler, 2008; Siviy et al., 2008b).

In multi-model (software) process improvement two major issues can be faced, which are usually two consecutive steps:

(1) selecting from multiple approaches (Balla, Bemelmans, Kusters, & Trienekens, 2001; Halvorsen & Conradi, 2001; M. C. Paulk, 2008) and

(2) the simultaneous usage of chosen approaches (A. Ferreira & Machado, 2009; Siviy et al., 2008a, 2008e).

Since there are initiatives which focus on solving the first issue, we review them in this chapter and later we focus on the second issue – which we call the multi-model problem, and which will be addressed in this thesis.

**Initiatives for the first issue: selecting from multiple approaches**

Quality approaches can be classified on the basis of particular characteristics (e.g. based on orientation, on level of detail, on a specialization, on their results, on the authority of a process group using them etc.). Such classifications can help companies to choose among quality approaches on the basis of their specific needs and wishes.

**QMIM** – An initiative for the classification of quality approaches is the QMIM framework (Quality through Managed Improvement and Measurement) (Balla, 2004; Balla et al., 2001). This framework shows how quality standards and quality models are connected to three different types of quality objects, i.e. process, product and resource, and their specifications, i.e. definitions, quality attributes and metrics. The QMIM framework supports companies in selecting suitable quality approaches, e.g. standards and/or models, for the software quality problems they are confronted with.

**Taxonomy based classification** – In the same period that the QMIM framework has been developed a very similar idea has started to evolve, namely: the definition of taxonomies for the identification of the main characteristics of quality approaches, enabling comparison between quality approaches in a structured and consistent way. An example is the taxonomy by Halvorsen and Conradi. They proposed 25 characteristics of quality approaches (so called "SPI Frameworks") grouped in 5 categories (Halvorsen & Conradi, 1999, 2001). Halvorsen and Conradi's taxonomy has been discussed and elaborated further, first by Paulk (M. C. Paulk, 2008) and subsequently by Ferreira et al. (A. L Ferreira et al., 2010). Another taxonomy was developed by Rahman et al. for comparing process improvement frameworks (Rahman, Sahibuddin, & Ibrahim, 2011).

**PrIME** – Process Improvement in Multi-model Environments (PrIME) is a research project on multi-model process improvement launched in 2008 by the Carnegie Mellon Software Engineering Institute (SEI, 2008). The project seems to be finished and results are published in 6 whitepapers (Siviy et al., 2008a, 2008b; Siviy, Kirwan, Marino, & Morley, 2008c, 2008d; Siviy et al., 2008e; Siviy, Kirwan, Marino, & Morley, 2008f). Among other outcomes, it supports the idea of taxonomies; it defines a strategic classification taxonomy for quality approaches. The classification helps companies in choosing from different quality approaches.

We call all these initiatives classifications of quality approaches. A *classification of quality approaches* supports companies in selecting quality approaches and in deciding on directions for improvement in accordance with the specific requirements of the company. Classifications describe important characteristics of quality approaches (e.g. with respect to their purpose, their application domain or their level of detail). Although comparison and selection is supported by these classifications, the second issue – the simultaneous usage of multiple quality approaches – is not fully supported as it will be discussed in chapter 3. In the remainder of this thesis we focus on filling the gap identified: the simultaneous usage of multiple quality approaches (which we refer to as the multi-model problem).

# 2. RESEARCH APPROACH

As we have discussed previously there are several initiatives for the first multi-model issue: selecting quality approaches for multi-model software process improvement. However, there is no widely accepted solution for the second issue: the simultaneous usage of the selected quality approaches. Current multi-model initiatives (as will be shown in chapter 3) suffer from one or more problems and do not provide a complete solution to the simultaneous usage of multiple quality approaches. This is mainly caused because quality approaches have differences in structure, terminology, content and many other characteristics which make their simultaneous usage a complex task. The lack of solution causes problems, e.g. in case of incomplete quality approaches or in satisfying certification requirements (see for more details in chapter 1.3 and chapter 3).

In this chapter we will propose an approach to deal with this problem. The basic hypothesis is that by developing a reference process model we can deal with this problem. The existence of such a reference process model allows us to map individual quality approaches to this single reference process model. We will call this resulting model a unified process and a process that leads to this model a Process Based Unification (PBU) process. This leads to the following terminology:

**The PBU concept** – Our hypothesis is that mapping quality approaches to a process can provide a multi-model solution. The task can be divided to decomposing quality approaches and mapping them to a single unified process. The concept of mapping quality approaches to process will be called the concept of Process Based Unification (PBU) or PBU concept. From now if we mention PBU it means the PBU concept.

**A PBU process** – In order to guide the practical implementation of the PBU concept, we need to provide a process. This process will be called PBU process. The PBU process relies on the PBU concept. The goal of the proposed PBU process is to provide practical guidance for the implementation of the PBU concept.

**The PBU result** – The PBU result consists of all the outputs of a PBU process.

**The unified process** – The key PBU result is a single, unified process, to which quality approaches are mapped. In order to ensure that the unified process conforms to multiple quality approaches, elements of quality approaches will be mapped to process elements. For this mapping, decomposition of quality approaches into their elements is required.

**The PBU framework** – This consists of the PBU concept, the PBU process and the resulting unified process. In order to emphasize the coherence between the elements defined we will call the set of these elements a framework, since we do not just need to understand elements singly, but also their mutual relations.

A PBU process can facilitate the usage of multiple quality approaches, having a positive impact on the work of process experts and on the whole organization by providing a usable and sufficient tool to create or enhance processes on a multi-model basis, also keeping the traceability and maintainability of processes to the source quality approaches. Companies using multiple software quality approaches can follow the steps defined in the PBU process and tailor their own software processes from different parts of different quality approaches. The results achieved by applying the PBU framework to multiple quality approaches can be used in situations discussed in chapter 1.3.

The unified process can provide an organization a solution for the multi-model problem by unifying structure, elements, content and terminology among others to one single process. It can enhance the usage of multiple process oriented quality approaches and thus can accelerate multi-model software process improvement. Users of the unified process do not need to understand each quality approach, handle differences in their characteristics (e.g. granularity, structure, terminology) but they use a single approach which is a unified process. There can be three main levels of process descriptions: theoretical, organizational or project level (Bóka, Balla, Kusters, & Trienekens, 2006). Such a unified process could be used at all these three levels.

In chapter 2.1 the research approach (objective and methodology) and in chapter 2.2 the thesis structure are introduced.

## 2.1. Research objective and methodology

As stated above, the basic hypothesis is that mapping quality approaches to a unified process can provide a multi-model solution. On the basis of this we can identify the research objective of the thesis.

**Research objective:** investigation whether the PBU concept can lead to an acceptable solution for the simultaneous usage of multiple quality approaches, by designing a PBU process.

This means that the final objective can be formulated as a Y/N question, but the research has a strong design orientation, since we need to design a PBU process to validate the PBU concept.

**On this basis we can formulate the following research question:**

*Does the PBU framework provide a solution of sufficient quality for current problems of simultaneous usage of multiple quality approaches?*

The main research question can be broken down into several high level questions (H1-H5).

H1 How can we recognize a solution of sufficient quality for current problems of simultaneous usage of multiple quality approaches?
H2 How can we design a PBU process that allows mapping of quality approaches to a unified process?
H3 How can we provide a proof of concept for the PBU framework?
H4 How can we validate the PBU framework?
H5 What improvements can be made on the proposed PBU framework?

High level question (H1-H5) can be broken down into operational research questions (Q1-Q11).

To answer question H1, in research question 1 (called Q1) criteria are identified that represent a solution of sufficient quality. These are called *MSPI criteria*. Our assertion was that current approaches do not provide sufficient solutions. Given that we now have identified the characteristics of a sufficient solution we can now check whether this assertion is correct. This is done in research question Q2.

Designing a PBU process (H2) requires a number of preliminary steps. First we must understand options and limitations of the basic concept: we need to discover both elements of processes (Q3) and elements of quality approaches (Q4) and analyse if a mapping between them is indeed possible (Q5). On this basis we can design a PBU process (Q6).

After having designed a PBU process, in order to show that the PBU concept is feasible (H3) we will perform a case study in which we will apply the PBU process resulting in a unified process (Q7). Deviations in execution from the designed PBU process will be recorded. These will serve as an input for refinements on the PBU process (Q10).

Having provided a proof of concept we will validate the results (H4). Consequently, we need to show that the PBU framework satisfies the MSPI criteria (Q8). If it satisfies the MSPI criteria then it is a solution of sufficient quality. We also need to ensure and discuss whether the whole research was performed in a valid and reliable way (Q9). Q8 and Q9 are based on the PBU process as actually executed in the case study, including the deviations from the originally designed PBU process as identified in Q10.

Finally, we will identify lessons learned (H5) which will be discussed based on deviations from the PBU framework as they happened when the case study was executed (Q10) and required extensions based on a comparison of case study results to MSPI criteria (Q11). The extensions identified by Q11 were identified after execution of the case study. As they are after the fact extensions they are not part of the analysis performed in Q8 and Q9.

Table 2 includes operational research questions developed based on the discussion above. For each of research questions we identify steps and for each step we define deliverables produced by the step. Deliverables are used to answer the research question.

Table 2 – High level, operational research questions, research steps and deliverables

| High level question | Research question | Research step | Deliverable | Ch. |
|---|---|---|---|---|
| H1 | Q1 What criteria should a multi-model solution satisfy? | RS1 Identify MSPI criteria | MSPI criteria based on literature review | 3 |
| | Q2 Do current MSPI initiatives satisfy the MSPI criteria? | RS2 Review current MSPI initiatives based on MSPI criteria | Characterization of current MSPI initiatives based on MSPI criteria | 3 |
| H2 | Q3 What is a suitable set of process elements to base the unified process on? | RS3 Identify process elements required for mapping | Set of process elements | 4 |
| | Q4 What are the elements of quality approaches? | RS4 Identify elements of quality approaches | Structure and elements of a representative set of quality approaches | 4 |
| | Q5 Which characteristics of (a number of) current quality approaches further and/or hinder mapping of these approaches to a process? | RS5 Identify options and limitations of mapping quality approaches to processes | Options and limitations of mapping quality approaches to processes | 4 |
| | Q6 How can mapping of a set of quality approaches to a unified process take place? | RS6 Design a PBU process | Design of a PBU process | 5 |
| H3 | Q7 How can we provide proof of concept for the PBU framework? | RS7 Perform a case study on the PBU framework | Case study report on the application of the PBU framework | 6-8 |
| H4 | Q8 Is the PBU framework adhering to MSPI criteria? | RS8 Verify the PBU framework against MSPI criteria | Discussion of the PBU framework based on MSPI criteria | 9 |
| | Q9 Which conclusions to validity and reliability of the PBU framework can be drawn from the research results? | RS9 Investigate validity and reliability of the PBU framework | Discussion of validity and reliability aspects of the PBU framework | 9 |
| H5 | Q10 What did we do differently in applying the PBU framework as compared to its original design? | RS10 Modify the PBU framework based on RS7 (based on concrete experiences of the case study) | Discussion of possible refinements of the PBU framework based on concrete experiences of the case study | 10 |
| | Q11 What omissions can we identify when comparing the results of applying the PBU framework to the MSPI criteria? | RS11 Modify the PBU framework based on RS8 | Discussion of possible extensions of the PBU framework based on the comparison with MSPI criteria | 10 |

In the followings research steps (RS1-11) are discussed in more detail:

**RS1:** Identify MSPI criteria.

In order to investigate what will be included in MSPI criteria, various quantitative and qualitative research methods can be applied. One qualitative research method can be conducting structured interviews with domain experts. As the multi-model software process improvement is an emerging field (C. Pardo, Pino, García, Piattini Velthius, & Baldassarre, 2011), finding real experts with considerable experience in solving the multi-model problem would be difficult (Heston & Phifer, 2010). Similar problems could be faced when applying quantitative methods. Another solution is to assess current problems and initiatives based on literature on the problems of MSPI. As an emerging literature is available, we choose the systematic literature review as the research instrument for identifying MSPI criteria.

In the MSPI literature several sources identify fundamental problems in using multiple quality approaches. However, there is no consistent basis on what criteria should an MSPI solution satisfy. In the literature review we will consider these literature sources as a starting point, then a systematic search and assessment of current problems will be performed which result in the MSPI criteria.

The PBU framework will be developed in accordance with the MSPI criteria identified. At the end of this research the resulting PBU framework will be verified if it satisfies the MSPI criteria. Thus the MSPI criteria will play an important role in the validation of the research. The MSPI criteria can also be used by other researchers to refine it, to develop it further or as a basis in developing new MSPI solutions.

**RS2:** Review current MSPI initiatives based on MSPI criteria.

The MSPI criteria resulting from literature review in RS1 gives the opportunity to assess current initiatives. If current initiatives do not satisfy the criteria identified, then the development of an MSPI solution is relevant.

In order to assess problems mentioned in literature we perform a systematic literature review on multi-model software process improvement. The systematic literature review gives a structured result including of current problems and initiatives. Current initiatives will be assessed based on the MSPI problems identified.

**RS3**: Identify process elements required for mapping.

Identifying process elements is important because quality approach elements will be mapped to elements of the unified process. Without having process elements identified, the mapping is not possible.

Looking at literature we found that no uniquely accepted process description format (notation) exists. Processes are usually represented textually and/or graphically, but there are several options available. For textual representation some of the description formats include entry/exit criteria, risks or key success factors while others do not. For graphical notation there are again several choices such as BPMN, Petri Nets, various types of UML diagrams etc. Various process descriptions contain different elements; therefore external literature sources will be considered in defining process elements. Since RS3 is a crucial point of this research, it will be performed in three iterations:

a) Identify process elements based on literature – this iteration provides a preliminary basis for process elements.
b) Refine process elements based on quality approach elements – In this research step we will analyse the structure and elements of process oriented quality approaches. Since these quality approaches are process oriented, they contain elements which describe processes.

The initial list of process elements identified in the first iteration (a) will be refined based on well-defined process related quality approach elements.

c) Refine process elements based on a case study – we perform a case study on applying PBU framework which gives a practical feedback on what process elements can be used in describing a certain process. Experiences of the case study will be used in refining the process elements.

It is important to mention that our main goal is not to define a process modeling language but rather to identify those process elements which can be used in a mapping.

**RS4**: Identify elements of quality approaches.

In order to determine if the mapping of quality approach elements to process elements is possible besides the identification of process elements, the analysis of quality approach elements is also needed. In order to do this we will analyse the structure and elements of a number of relevant quality approaches.

More than 300 quality approaches were identified in (Moore, 1999) and 52 Software Process Capability/Maturity Models in (von Wangenheim et al., 2010). Covering all quality approaches is infeasible within the frame of this research. In order to overcome this problem we will choose 7 different quality approaches which represent a high variety of quality approach characteristics. These will include quality approaches released by important bodies of standardization such as ISO, IEEE, IEC standards, models from SEI CMU, a governmental quality approach as well as a non-standardized quality approach. These approaches will be chosen so that they differ at least in the scope, content, terminology, granularity, structure, size and complexity (the characteristics current multi-model initiatives have problems to handle). In chapter 4 this will be discussed in more detail.

**RS5**: Identify options and limitations of mapping quality approaches to processes.

At this step we will discuss how quality approach elements identified in the previous step can be mapped to process elements. If semantic (meaning of elements) similarities between quality approach elements and process elements are recognized we consider these can be mapped. Results of this research step will be used as a basis for developing the PBU process.

**RS6**: Design a PBU process.

The research is aimed at providing arguments for the basic hypothesis: the relevance of the concept of mapping several quality approaches to a unified process. To support this, a PBU process will be designed and in a case study an example of a unified process will be developed. Design is an iterative process. In this thesis we will perform one full cycle of this process, starting with analysis, developing a first version, executing tests in a case study and leading to assessment and refinement of the original version.

**RS7:** Perform a case study on the PBU framework.

In investigating the feasibility of the PBU framework we can choose from several research instruments such as experiments, expert interviews or a case study. One advantage of the case study is that it gives the opportunity to practically, deeply and rigorously perform a PBU process and to understand issues faced during unification. A drawback of a case study compared to quantitative research methods is, that it is more difficult to ensure validity and reliability of results (Golafshani, 2003). However, for understanding if such a PBU framework is practical-ly applicable, a case study is an appropriate choice since it will be performed within a practical, real-life project.

The case study will answer if the PBU framework is indeed feasible and it will also provide inputs for refinement of the originally designed PBU framework.

**RS8**: Verify the PBU framework against MSPI criteria.

This research starts with the identification of an MSPI criteria based on a systematic literature review (RS1). As a result of the literature review problems and initiatives of MSPI are shown with a conclusion that no MSPI initiative satisfies the MSPI criteria. The PBU framework is intended to serve as an MSPI solution, therefore an assessment is needed on whether it satisfies the MSPI criteria. In order to assess if the PBU framework adheres to MSPI criteria the following steps will be performed for each criterion:

a) **Problem definition** – short reflection of the problem based on the MSPI criteria identified in research step 1-2. In RS1 we identify problems based on literature, here we include possible problem refinements based on experiences of the case study.
b) **Satisfying the MSPI criteria in the PBU framework** – discussion if the PBU framework satisfies the MSPI criteria or with which modifications are required to achieve.
c) **Satisfying the MSPI criteria in the case study** – discussion if the current application of the PBU framework satisfies the MSPI criteria or which modifications are required.
d) **Conclusion –** discussion of applying the PBU framework in other settings.

We will go through each criterion performing on each step defined above. We will always focus on finding information on how the PBU framework and its application in the case study satisfy the MSPI criteria.

If the PBU framework satisfies the MSPI criteria, it means the main research objective is achieved and a solution of sufficient quality is provided for the multi-model problem. A solution of sufficient quality in this case means an initiative which not just supports the usage of multiple quality approaches, but it also satisfies the MSPI criteria, thus handles important issues faced in MSPI. Some of the criteria can be supported on multiple levels. E.g. appraisals can be supported by ensuring traceability, but can also be further supported by tools. We will address levels and options of satisfying criteria where applicable. An MSPI initiative is considered a solution if it satisfies each MSPI criterion. Where multiple criteria levels are present at least the lowest level should be achieved.

**RS9:** Investigate the validity and reliability of the PBU framework.

In order to understand if the PBU framework can be applied in practice we will apply it in a case study at RS7. Since we will use a qualitative research method to test the PBU framework in practice, its validity and reliability should also be discussed as part of the validation. Construct, internal, external and ecological validity and reliability will be discussed based on multiple sources of the validation literature (Fisher, 2010; Gibbert & Ruigrok, 2010; Golafshani, 2003; Trochim, 2006; Yin, 2009).

**RS10:** Modify the PBU framework based on RS7 (concrete experiences of the case study).

We can expect during the case study that deviations from the PBU framework will be deemed necessary or useful. Based on these deviations we will identify modifications of the original PBU framework. This step will be performed after the case study (RS7).

**RS11:** RS11 Modify the PBU framework based on RS8.
During the verification of the application of the PBU framework against the MSPI criteria we will identify omissions (RS8). Based on these omissions, we will define required extensions to the PBU framework.

## 2.2. The structure of this thesis

In this chapter we discuss the structure of the remaining parts of this thesis (starting from chapter 3). Figure 2 gives an overview of the research process including the relation between research steps and thesis chapters. The figure is represented in Business Process Modeling Notation (BPMN), research steps are represented as subprocesses and tasks and chapters are represented as data objects. RS denotes research steps, RSn-m denotes the m. substep of the research step n., while RSo-ip denotes the iteration p. of the research step o. where applicable.
Parts of early versions of chapters 1 and 2 were published in (Kelemen, Balla, Trienekens, & Kusters, 2008a; Kelemen, Kusters, & Trienekens, 2011; Kelemen, Kusters, Trienekens, & Balla, 2009).

**Chapter 3**

In chapter 3 we identify criteria for multi-model solutions (RS1) and analyse current initiatives in multi-model software process improvement (RS2). Chapter 3 discusses first the current problems regarding the use of multiple software quality approaches. Subsequently, multi-model initiatives are categorized into three different groups, respectively: quality approach harmonization, quality approach integration and quality approach mapping. Based on an analysis of the strengths and weaknesses of current multi-model initiatives in these three classes, we derive a set of criteria which can provide a basis for multi-model software process improvement solutions.
A version of chapter 3 was accepted and published online in (Kelemen et al., 2011).

**Chapter 4**

In chapter 4 we identify process elements (RS3) and quality approach elements (RS4). Process elements are first identified based on literature then refined using results of analysing quality approach elements. Quality approach elements of 7 different quality approaches are identified. At the end of this chapter a mapping is provided between quality approach elements and the refined process elements (RS5). Results of this chapter serve as a basis for developing the PBU process. Chapter 4 is described according to research steps 3-5.
A version of chapter 4 was published in (Kelemen et al., 2008a).

**Chapter 5**

The literature shows that the current multi-model initiatives such as quality approach harmonization, quality approach integration and respectively the quality approach mapping do not provide a solution for the multi-model problem and do not satisfy the MSPI criteria (see chapter 3). Chapter 4 shows that a mapping between quality approaches and processes is possible so we can apply the PBU concept.

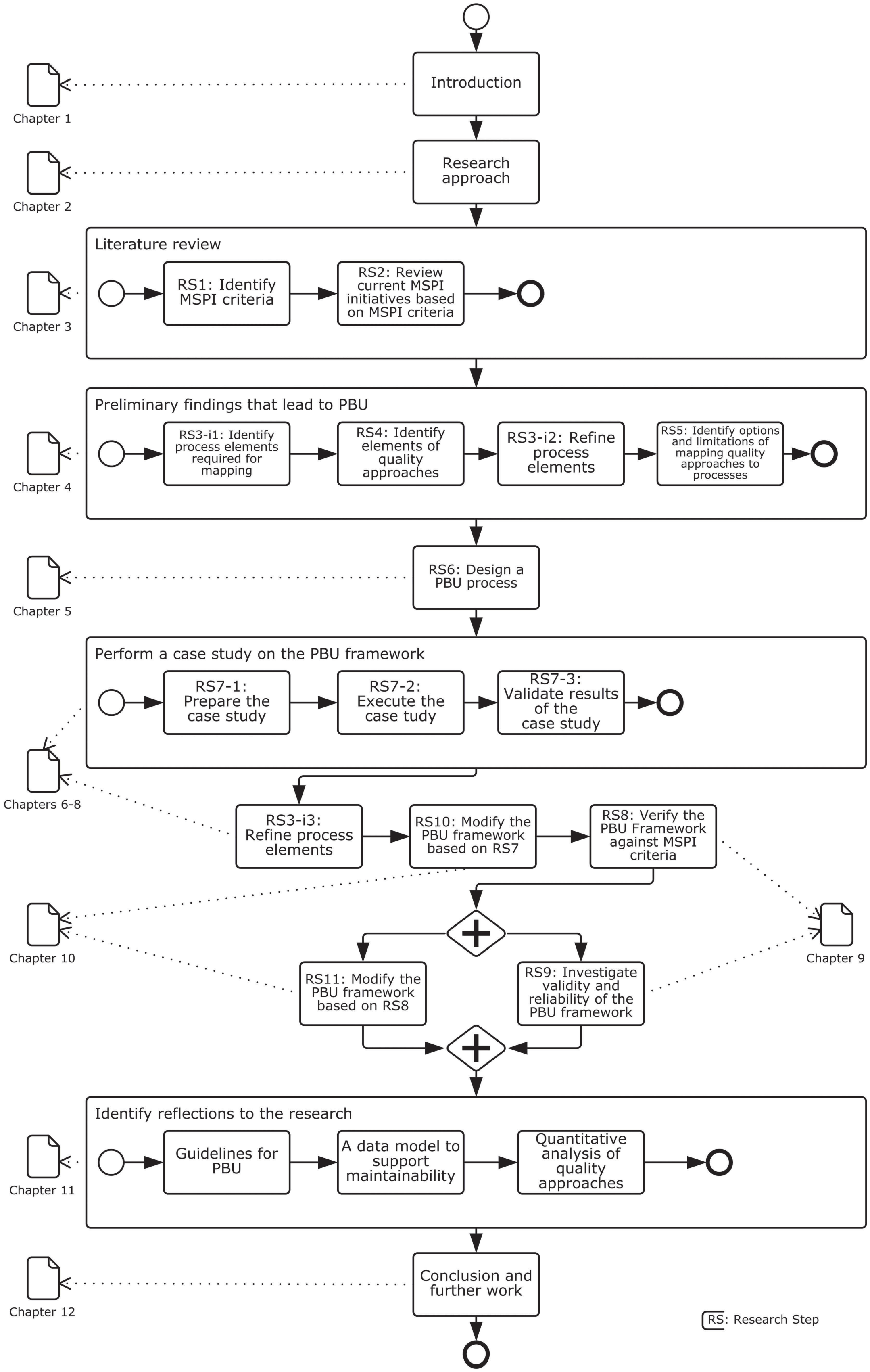

Figure 2 – Research process

In order to make the PBU concept operational, we need a PBU process. In this chapter the focus is on designing the PBU framework and its core component: a PBU process (RS6). In this chapter we describe a PBU process.

A version of PBU concept which led to the development of chapter 5 was published in (Kelemen, 2009; Kelemen et al., 2009).

*Chapters 6-8 are related to RS7 and are structured as follows:*

**Chapter 6**

In order to practically validate the PBU framework, we used it in a real world case study. Chapter 6 presents the preparatory steps of the case study. Chapter 6.1 describes the approach of preparing for the case study. Chapter 6.2 describes the project in which the PBU framework was applied, chapter 6.3 presents the reasons for selecting a process, chapter 6.4 describes the selection of quality approaches to be used for creating a unified process and chapter 6.5 describes the reasons of selecting a process representation format. In chapter 6.6 we discuss deviations from the original PBU framework, in chapter 6.7 we present limitations and in chapter 6.8 we summarise results of preparation steps. These steps are performed according to the first subprocess of the PBU process presented in chapter 5.4.1.

**Chapter 7**

In chapter 7 we continue applying the PBU framework and activities of the second and third subprocesses of PBU process “Analysis of quality approaches” and “Deriving process from quality approaches” described in chapters 5.4.1.3 and 5.4.3 are performed within the case study. This chapter essentially shows how quality approach element instances of multiple quality approaches are identified, how these are mapped to process element instances and how a unified peer review process is developed.

**Chapter 8**

In chapter 8 the validation of the case study is described. This chapter is described according to the last subprocess of the PBU process “Validation” presented in chapter 5.4.4. This chapter mainly focuses on discussing how case study participants accepted the resulting unified process, including a discussion on what kind of validation was chosen, what steps were performed in order to validate the resulting unified peer review process.

Results of the case study (chapters 6-8) were presented on the Q-works workshop organized at IBM Budapest (Kelemen, 2010b) and at a monthly organized Benchmarking Conference in Szeged, Hungary (Kelemen, 2010a).

**Chapter 9**

According to RS8 and RS9, chapter 9 focuses on the validation of the research. In this chapter we verify the PBU framework against MSPI criteria (RS8), furthermore we also assess how valid and reliable the results obtained in the case study are (RS9).

**Chapter 10**

In chapter 10 lessons learned from the case study (RS10) and the comparison of the application of the PBU framework with MSPI criteria (RS11) will be presented. Discussion focus on the modifications needed on the PBU framework based on RS10 and RS11.

**Chapter 11**

The objective of this chapter is to present a number of issues that arose during the research and are relevant, but outside of the scope of the researched thesis. These include (1) guidelines for those are wishing to use the PBU process in their work, (2) a data model for further supporting maintainability of the PBU results and (3) preliminary results of quantitatively analysing a quality approach: the CMMI model.

A version of chapter 11 – analysing CMMI was presented in (Balla & Kelemen, 2011; Kelemen, 2011b).

**Chapter 12**

In chapter 12 (1) results achieved, (2) limitations of the results and (3) further work will be presented.

# 3. IDENTIFYING CRITERIA FOR MULTI-MODEL SOLUTIONS[1]

In this chapter we analyse current initiatives in multi-model software process improvement and identify criteria for multi-model solutions. With multi-model we mean the simultaneous usage of more than one quality approach (e.g. standards, methods, techniques to improve software processes). This chapter discusses first the current problems regarding the use of multiple software quality approaches. Subsequently, multi-model initiatives are categorized into three different groups, respectively: quality approach harmonization, quality approach integration, and quality approach mapping. Based on an analysis of the strengths and weaknesses of current multi-model initiatives in these three classes, we derive a set of criteria, which can provide a basis for multi-model software process improvement solutions.

The structure of this chapter is as follows: in 3.1 we briefly describe our research approach, 3.2 describes details of literature search, 3.3 highlights the main problems in the simultaneous usage of multiple quality approaches and 3.4 gives an overview of MSPI initiatives. In 3.5 the analysis and identification of the criteria for multi-model solutions are presented. 3.7 describes limitation of literature review and 3.8 ends with conclusions.

## 3.1. Research approach

The objective of this chapter is to identify criteria for multi-model solutions. This will be based on answering the following questions:

Q1 What criteria should a multi-model solution satisfy? (We call these criteria MSPI criteria.)

Q2 Do current MSPI initiatives satisfy the MSPI criteria?

In order to answer these questions we follow the research steps 1 (RS1) and 2 (RS2) presented

---

[1] A version of this chapter has been published in (Kelemen, Kusters, & Trienekens, 2011).

in chapter 2.

The work is based on a literature review. This review has been executed using the guidelines from Kitchenham's guide on performing systematic reviews (Kitchenham, 2004).

The resulting criteria can be used for the development of MSPI solutions.

Since the MSPI criteria are crucial for this research, research steps were broken down to research sub-steps:

1. Literature search on MSPI – this step will serve as a basis for identifying current problems and initiatives of MSPI.
2. Identifying problems of MSPI – based on literature search results, MSPI problems will be identified.
3. Analysis of MSPI initiatives based on problems of MSPI – problems of MSPI provide the basis for assessing MSPI initiatives.
4. Identifying MSPI criteria – MSPI criteria will be finalized based on problems and initiatives discussed.
5. Assessing the hypothesis against MSPI criteria – As MSPI criteria are identified, we perform a preliminary assessment if the PBU framework can theoretically meet MSPI criteria.

Table 3 summarizes research questions, research steps, sub-steps and the chapters in which the research steps are discussed.

Table 3 – Research questions, steps and related chapters

| Research question | Research step | Research sub-step (order) | Chapter |
|---|---|---|---|
| Q1 What criteria should a multi-model solution satisfy? | RS1 Identify MSPI criteria | RS1-1 Literature search on MSPI | 3.2 |
| | | RS1-2 Identifying problems of MSPI | 3.3 |
| | | RS1-4 Identifying MSPI criteria | 3.5 |
| | | RS1-5 Assessing the hypothesis against MSPI criteria | 3.6 |
| Q2 Do current MSPI initiatives satisfy MSPI criteria? | RS2 Review current MSPI initiatives based on MSPI criteria | RS2-3 Analysing MSPI initiatives based on problems of MSPI | 3.4 |

# 3.2. Literature search

We performed a literature search with two goals: first to serve as input to RS1 of the research – identifying problems of MSPI, and also to serve as input to RS2 – looking for current initiatives in using multiple quality approaches.

## 3.2.1. Search space

The literature search had the same search space for both goals which included books, journals, conferences, theses, webpages, presentations and technical reports in the area of process improvement, standardization, integrated models and multi-model software process improvement.

The literature search was conducted in a number of public electronic databases including IEEE, ACM, ScienceDirect, Citeseer and Google Scholar. Additional materials were gathered from non-public sources, e.g. from SEIR (CMU SEI Information Repository) which included reports on practical results of SPI. The World-Wide-Web was also searched for relevant webpages and presentations.

### 3.2.2. Conducting the literature search

Choosing the right search terms is essential to the result of the literature search. For finding the most accurate keywords, we followed two steps:

1. Well-known papers in the field were manually analyzed to collect suitable search terms, e.g.: (Balla, 2004; Halvorsen & Conradi, 2001; M. C. Paulk, 2008; Siviy et al., 2008b).
2. Papers found during the literature search were processed using an open-source text mining tool, RapidMiner. All texts were tokenized to words; stopwords were filtered using an English dictionary and a specialized dictionary for filtering useless words (e.g. “page”). Tokens were also filtered by length excluding one letter words and then upper cases were transformed into lower cases. Later, different stemming algorithms were applied.

The application of basic text mining techniques helped us:

1. To find close variants of the keywords found in articles analyzed previously (e.g. granularity – grain);
2. To find all occurrences of different variants of keywords.

**Keyword search** – Our targeted field was the multi-model software process improvement. Search terms were combined by Boolean operators: combining terms of the main field by AND operator (e.g. “multi-model” AND “software process improvement”). The OR operator was also used between terms within the field. Different spellings and synonyms of search terms were applied as discussed previously. For example in the case of the term “multi-model”, the following terms were used: “multi-model”, “multimodel” and “multi model” and for the terms “software process improvement” also “SPI” was used. Identical search terms were applied to all databases mentioned in 3.2.1.

**Search through references and manual search** – In order to discover multi-model-based quality approaches – as an addition to keyword search – several quality approaches were checked (e.g. CMMI, SPICE, iCMM, Enterprise SPICE etc.). As another addition to the keyword-based search, references of relevant sources were checked manually, e.g. all references of (Siviy et al., 2008b).

**Inclusion criteria** – All materials of the final set were checked manually and all sources connected to multi-model software process improvement were included while irrelevant sources were excluded from the library.

**Exclusion criteria** – After checking samples of articles found in keyword search, similarly worded research fields were excluded using exclusion operators (“-“): e.g. publications containing the term “multi-model” in the field of process modeling were manually checked and in case of irrelevancy, sources were excluded by applying exclusion terms ( e.g. one such exclusion term was: -“multi-model view of process modeling“).

Sources addressing different research topics containing useless information (e.g. there were sources containing only references) and duplicates under different names were also excluded.

### 3.2.3. Search results

Articles, technical reports, theses, methodologies, mapping documents, presentations, and quality approaches connected to MSPI were collected using the "Zotero" (Center for History and New Media, 2009) open source reference management tool. After applying exclusion criteria, a final set of literature sources has been developed which contains 58 articles and further sources, 78 in total (see the lists in Appendix A and Appendix B).

For RS1, problems of MSPI have been identified based on the final set of sources which are discussed in more detail in chapter 3.3. For RS2, an identification of MSPI initiatives has been performed (see the discussion in chapter 3.4).

Further details of the literature search are given in (Kelemen, 2011a). This includes the list of literature sources, the pure and stemmed wordlists (using the Porter and Snowball algorithms), wordlist summary, occurrences and variants of important search terms.

## 3.3. Problems of MSPI

In this chapter we address problems of the simultaneous usage of multiple quality approaches as found in the literature search. As described in chapter 3.2.2, a variety of search terms was used to achieve these results. We found 6 problems listed below.

1. **The problem of recognizing and handling differences of source quality approaches (in terms of structure, granularity, terminology, content, size and complexity)**

One of the main problems in multi-model environments is the difference in structure and terminology of the quality approaches and the difficulty of recognizing the similarities between them (Siviy et al., 2008a). As (Thiry, Zoucas, & Tristão, 2010) states: "besides that, the structure of the models is not necessarily similar", it is "difficult to establish relations without the previous and careful assessment of each one, and establishing the granularity to be adopted". Misunderstanding the granularity of quality approaches can erode the benefits of SPI efforts (A. Ferreira & Machado, 2009; Siviy et al., 2008a). Granularity is also an issue when making comparisons and mappings of different quality approaches (Yoo et al., 2006).

Rout and Tuffley state that models "may have differences in structure and content", and "when the structure of the model is significantly different from the reference model, the mapping might be quite complex" (Rout & Tuffley, 2007). In (DMO, 2007) it is stated that the SAFE+ CMMI extension "was developed so that CMMI appraisers and users can become familiar with the structure, style, and informative content provided, to reduce dependence on safety domain expertise". As a consequence of differences in terminology and structure of quality approaches, the CMMI framework was developed in such a way that it includes "a common terminology, common model components, common appraisal methods, and common training materials" (CMMI Product Team, 2010a, 2010b, 2010c). Size and complexity can influence the selection (A. L Ferreira et al., 2010) and simultaneous usage of multiple quality approaches.

2. **The problem of traceability of multi-model results**

Siviy et al. suggest the mapping of quality approach requirements to organizational processes in order to achieve traceability which can be used for audits, assessments and benchmarks (Siviy et al., 2008a). Salviano confirms that traceability can be achieved via a 'mapping' process (Salviano, 2009b). Traceability of process models back to source artifacts is considered important by (Ghose, Koliadis, & Chueng, 2007). Salviano suggests a number of characteristics for his process improvement methodology called PRO2PI. Among other issues he highlights traceability as the necessity to trace back multi-model results to relevant process improvement models (Salviano, 2009b). Rout and Tuffley (Rout & Tuffley, 2007), Halvorsen and Conradi (Halvorsen & Conradi, 2001) suggest how to 'map' different approaches. Also Rout and Tuffley consider that references from the 'mapping' to quality approaches could serve as a solution for using multiple quality approaches.

The need for traceability also appears in models such as CMMI. CMMI for Services OPD SP 1.2 Subpractice 4 states that “adherence to applicable process standards and models is typically demonstrated by developing a mapping from the organization’s set of standard processes to relevant process standards and models. This mapping is a useful input to future appraisals” (CMMI Product Team, 2010a, 2010b, 2010c).

3. **The problem of changeability of multi-model results**

The literature recognizes the need for changeability, however it uses different terms for it, such as flexibility (DMO, 2007; Fonseca & de Almeida Júnior, 2005), adaptability (Fonseca & de Almeida Júnior, 2005) or dynamism (Salviano, 2009b). When changes occur in processes, the impact on the source quality approaches should be checked (Siviy et al., 2008b). In developing a unification of CMMI and CENELEC standards, a proposal was to unify these two frameworks into a single structure, but ‘flexible’ enough to support the necessary ‘adaptations’ (Fonseca & de Almeida Júnior, 2005).

4. **The problem of the completeness of multi-model results**

Reaching a sufficient level of completeness of multi-model results (or outputs) can be an issue of a multi-model solution. In case the multi-model result is not sufficiently complete, it could be difficult to understand and build a process from it. (M. Baldassarre, Caivano, Pino, Piattini, & Visaggio, 2010; Rout, Tuffley, & Cahill, 2001). In the opposite case, Cugola and Ghezzi mention with respect to process modeling languages: “PMLs tend to force process designers to over-specifying the process for completeness” (Cugola & Ghezzi, 1998). Similarly to over-specification of process models, over-specification of multi-model results can occur. In the latter case, if the multi-model result is too detailed, it can be confusing and people can be lost in details.

5. **The problem of supporting multi-model appraisals**

We define appraisal as assessments, audits, reviews, benchmarks or measurements, which are aimed at the discovery of the conformity of organizational processes to a quality approach. Multi-model appraisal support means the conformity of processes can be appraised against multiple quality approaches. This can happen preferably at once or separately. This aspect was considered important in MSPI by different authors (CMMI Product Team, 2010a; DMO, 2007; Ekert, 2009; Mutafelija & Stromberg, 2003, 2009a; Tuffley, Rout, Stone-Tolcher, & Gray, 2004).

6. **The problem of repeatability and documentation of the multi-model solution**

Several multi-model initiatives describe only the multi-model result. For example, many mappings describe only the mapped elements; CMMI gives a list of quality approaches used during its development, but no description is available on how the final multi-model result was achieved. Questions may arise such as “What process/methodology was followed?”, “How repeatable is the process?” or “Is the solution itself documented?”.

As source quality approaches are often described textually, comparisons, mappings or integrations are frequently done in a subjective manner. In order to ensure the repeatability and transparency, a description of the process by which a multi-model result is achieved should be available (Yoo et al., 2006).

Summarizing the problems which influence the quality of a MSPI solution, we can define three problem categories which cover the process of multi-model solution:

- Problems connected to the differences (in terms of structure, granularity, terminology, content, size and complexity) among source quality approaches. These quality approaches are the inputs of MSPI solutions (1).
- Problems connected to process repeatability and documentation (6).
- Problems connected to multi-model results in terms of traceability, changeability, completeness and appraisal support (2, 3, 4, and 5).

These problems are considered by the software process improvement community to be of importance in using multiple quality approaches. We will use them as a basis for a discussion in the next chapter on the strengths and weaknesses of the available multi-model initiatives.

## 3.4. Multi-model initiatives

After collecting multi-model initiatives, we categorized them based on their view on how to solve the multi-model problem. We will distinguish in this chapter three different categories: quality approach harmonization (which we will address in chapter 3.4.1), quality approach integration (to be addressed in chapter 3.4.2), and quality approach mapping (which we will address in chapter 3.4.3). We call a multi-model initiative Harmonization when characteristics of standalone quality approaches are aligned to each other (such characteristics can be e.g. structure or terminology). A multi-model initiative is called Integration when instead of having standalone quality approaches, approaches are put into a single “integrated” one. Finally we call a multi-model initiative a Mapping when specific parts of different quality approaches, such as requirements or terminologies are compared. In the following subchapters we will discuss each of these three categories and examples of multi-model initiatives.

### 3.4.1. Quality approach harmonization

Sheard states: “As process and quality frameworks continue to change, a consensus is emerging on the need for greater compatibility” (Sheard, 2001). In this chapter we summarize initiatives aimed to enhance the compatibility among quality approaches.

**Category definition:** Quality approach harmonization is the process of releasing a modified quality approach or the extension of an existing quality approach in accordance (or in a harmonized way) with one or more quality approach(es). In this sense harmonization results in a modified approach or addition that carries several characteristics of one or more quality approach(es) with which it is harmonized. Such common characteristics can be e.g. the common terminology, the common structure, or the common way of process descriptions. Furthermore, harmonized quality approaches often take into account what the other quality approach states with which they are harmonized, they avoid contradictions and contain references to existing approaches etc.

The International Organization for Standardization (ISO) harmonizes its most widely used standards. For example ISO 9001 and ISO 90003 have been developed with the same structure (i.e. chapters) and terminology in order to help companies in extending their ISO 9001 quality management system with ISO 90003. Moreover, if we look at the reason of releasing the latest version of ISO 9001 (released in 2008), we can find that a goal of this release was to enhance the compatibility with other ISO standards (ISO, 2008). No new requirements were added and all the modifications are focused on a concretization and a refinement of expressing requirements.

A further example of a harmonization of standards is the relation of IEEE 1028:2008 and IEEE 12207:2008. The first can be used to facilitate the achievement of different outcomes of the latter one: namely the chapters 7.2.6 – Software Review Process and 7.2.7 – Software Audit Process (IEEE, 2008b).

Extending a certain approach or tailoring it to other areas in a harmonized way is also quite usual, particularly in the case of CMMI. CMMI started to be a model for software and system development. Now it has got three different constellations: one for development (CMMI Product Team, 2010a), one for services (CMMI Product Team, 2010b) and one for acquisition (CMMI Product Team, 2010c). In addition to the constellations, different extensions exist (e.g. SAFE+ for safety (DMO, 2007) or RAMS in railways industry (Fonseca & de Almeida Júnior, 2005)).

TMMi (TMMi Foundation, 2009) is another example of an extension of a quality approach. TMMi states: “The development of the TMMi Reference Model has used the TMM framework as one of its major sources. In addition to the TMM, the TMMi model development was guided by the work done on the Capability Maturity Model Integration”.

These examples show that both quality standardization and SPI organizations are concerned with the problem of MSPI, especially in increasing the compatibility among quality approaches. ISO, IEEE and SEI make clearly visible steps towards the facilitation of companies to deal with the multi-model problem.

**Strengths**

The main strength of a quality approach harmonization is the increased compatibility among quality approaches. This is implemented in different ways. E.g. in the case of the ISO 9001 standards family, a unified terminology and structure is provided for ISO 9001 and 90003. In the case of IEEE 1028 and ISO/IEEE/IEC 12207 requirements are aligned, which means that the requirements of one standard satisfy the requirements of the other one, see for example

IEEE 1028. In addition to this, IEEE 1028 can be used as a subroutine of other quality approaches which define requirements for reviews and audits, e.g. SPICE or CMMI.

**Weaknesses**

An initiative for solving the multi-model problem is a well-planned and maintained quality approach harmonization. From a MSPI point of view, a weakness of the quality approach harmonization initiatives is that they are applied only to a limited number of standards. Therefore extending a Quality Management System (QMS) with quality approaches with a different structure could be difficult, e.g. the ISO 9001 - 90003 line works only for ISO 9001-like standards. In case that CMMI would be used in combination with ISO 9001-like standards, the multi-model problem would appear again. The process of harmonizing quality approaches is also usually vaguely documented.

Table 4 summarizes characteristics of quality approach harmonization.

Table 4 – Summary of strengths and weaknesses of quality approach harmonization

| Problem | S/W | Explanation |
|---|---|---|
| 1. Handling differences of quality approaches | S | Common structure, granularity, terminology, size and complexity are aligned in many cases. |
| 2. Traceability | S | Traceability is supported; in many cases the harmonized approaches often carry the same characteristics. E.g. in case of ISO 9001 -90003 there is a clear traceability of requirements. |
| 3. Changeability | W | Changes occurring in one standard can be taken into account in development of other standards, but these are not reflected automatically. |
| 4. Completeness | S | In this case quality approaches are standalone. Completeness depends on the authors and independent from harmonization. |
| 5. Appraisal support | – | Multi-model appraisals can be performed easier because of commonalities, but there is no clearly documented solution for this issue. |
| 6. Repeatability and documentation | W | Usually there is no guidance or documentation how the harmonization is developed, how it could be repeated. |

## 3.4.2. Quality approach integration

**Category definition:** An integrated quality approach is a quality approach which has been established on the basis of multiple quality approaches. Quality approach integration is the process of developing an integrated quality approach. A clear difference between this category and the previous harmonization category is that the source quality approaches are not left standalone, but that they are put together into a single, integrated quality approach carrying new shared characteristics, often replacing the source approaches.

Recognizing the need for using multiple quality approaches simultaneously, integrated quality approaches were developed both on a commercial and a non-commercial basis. From many amongst such integrated quality approaches some of the best known are CMMI (CMMI Product Team, 2010a, 2010b, 2010c), iCMM (FAA, 2001) and Enterprise SPICE (Ibrahim, 2010).

A strong point of iCMM is that it clearly defines the source of the terms used and also provides a mapping table between itself and the source quality approaches used (FAA, 2001). In this way the traceability back to the source quality approaches is ensured. Enterprise SPICE is a continuation of iCMM and tries to integrate even more approaches than its ancestor.

A commercial example of an integrated quality approach is CITIL, a combination of CMMI and ITIL. It "supports the improvement of both the development and the operation aspects of IT products and services" (Wibas, 2009).

**Strengths**

The main strengths of integrated quality approaches such as CMMI, SPICE, iCMM or Enterprise SPICE are the unified terminology, structure, content, granularity, size and complexity. Furthermore, iCMM and Enterprise SPICE provide traceability back to the source models via a mapping table.

**Weaknesses**

Integrated quality approaches are the multi-model result of an integration process and do not provide information on the way the integration problem was solved nor on how to integrate new approaches.

Table 5 summarizes characteristics of quality approach integration.

Table 5 – Summary of strengths and weaknesses of quality approach integration

| Problem | S/W | Explanation |
|---|---|---|
| 1. Handling differences of quality approaches | S | Common structure, granularity, terminology, size and complexity are achieved. |
| 2. Traceability | – | In the case of iCMM and Enterprise SPICE traceability is achieved via a mapping table, in case of CMMI (as no mapping table included into the model) traceability is questionable. |
| 3. Changeability | W | There is no information on how to include new quality approaches or new versions of existing ones into an integrated quality approach. |
| 4. Completeness | – | There is no information on how complete the integrated quality approaches are (there is some information in case mapping tables are provided). Since the community accepts these models we consider them complete enough. |
| 5. Appraisal support | S | Since multi-model approaches often replace the source quality approaches, one appraisal is enough to satisfy all the needed requirements. |
| 6. Repeatability and documentation | W | There is no guidance or documentation how the integration is performed and how to perform new integrations. |

### 3.4.3. Quality approach mapping

**Category definition: Quality approach mapping**s focus on the identification of the requirements of two different quality approaches. Subsequently the identified requirements of the first approach are mapped to the requirements of the second approach. Quality approach mappings often include a terminology mapping as well, see e.g. (Halvorsen & Conradi, 2001), (Mutafelija & Stromberg, 2009a) and (Rout & Tuffley, 2007). Many different comparison methods exist for the mapping of different quality approaches, e.g.: (Ekert, 2009; Halvorsen & Conradi, 2001). While Halvorsen and Conradi define four types of such methods, respectively: characteristics comparison, framework mapping, bilateral comparison and needs mapping comparison, other researchers such as Ekert (Ekert, 2009), Mutafelija (Mutafelija & Stromberg, 2003, 2009a) or Rout and Tuffley (Rout & Tuffley, 2007) use their own way of mapping.

A wide literature is available on the comparison of CMMI with other approaches such as ISO 9001, TSP, SPICE, Agile/Lean Development, Six Sigma (Siviy, Penn, & Stoddard, 2007), PMBOK and IEEE software engineering standards (SEI, 2010). A huge library of SPI materials, which include several mappings, is the Software Engineering Information Repository (SEIR) maintained by SEI (SEI, 2010).

In part 3 of the PrIME whitepapers (Siviy et al., 2008a), granularity of quality approaches are discussed in brief, highlighting that not all the mappings provide information on granularity. Relaying only on mappings could lead an organization to satisfy the requirements of an approach, but at the same time may fail on the audit of a finer grained quality approach. Understanding the granularity of quality approaches can thus be crucial. These whitepapers highlight (besides the compliance in assessments) the importance of traceability of processes back to the process requirement sources and the change management of derived processes.

One of the most widely known mappings of CMMI is presented in (Mutafelija & Stromberg, 2003). These authors describe a detailed mapping of CMMI-v1.1 and ISO 9001:2000 requirements, comparing each ISO 9001:2000 requirement to the CMMI specific and generic practices. In CMMI a practice is the description of an activity that is considered important in achieving an associated goal (CMMI Product Team, 2010a). This mapping is very well applicable at companies which already have an ISO 9001 quality management system, and try to move towards CMMI process improvement. In (Mutafelija & Stromberg, 2009a) a mapping of CMMI to ISO 9001, ISO 15288, ISO 12207, and ISO 20000 is described.

Software applications are also available to support mappings of quality approaches, e.g. the Appraisal Assistant. The Beta 3 version of this software application facilitates the appraisal of multiple constellations, versions and extensions of CMMI, People CMM, an extension of iCMM, SPICE and Automotive SPICE (Griffith University, 2007). Another example of such a software application is the Capability Adviser in the automotive industry (Ekert, 2009). The International Software Consulting Network (ISCN) makes efforts to provide software for automotive companies to support an Integrated Automotive & Safety SPICE Assessment Approach. The software application also contains cross-references to ISO 9001 (Ekert, 2009; ISCN, 2010).

**Strengths**

One of the main strengths of quality approach mapping is traceability. Mappings provide concrete, traceable information back to the source approaches. Another strength of mapping is that it can act as a basis for multi-model appraisals (e.g. multi-model appraisal software such as Appraisal Assistant or Capability Adviser are based on mappings) (Ekert, 2009; Griffith University, 2007; ISCN, 2010).

**Weaknesses**

Mappings are specific solutions for (often only) two selected quality approaches, e.g. a mapping of ISO 9001-CMMI or CMMI-Six Sigma. There is no mechanism provided to (automatically) reflect changes of source approaches. Additionally, most of the current mappings pro-vide only the result of mapping without providing general guidance on how to perform the mapping process and how to extend current mappings to further quality approaches.
Since quality approaches are described textually, mappings are usually performed in a subjective manner, e.g. in many cases the level of overlapping can be argued. One disadvantage of quality approach mapping is that the complexity and amount of the work increases extremely with the number of quality approaches included. When a new quality approach is included, new mappings are needed to all quality approaches. Table 6 summarizes characteristics of mappings.

Table 6 – Summary of strengths and weaknesses of quality approach mapping

| Problem | S/W | Explanation |
|---|---|---|
| 1. Handling differences of quality approaches | – | Several mappings take into account the differences of quality approaches. Users of mappings get an overview on how terms and requirements overlap, but they will still need to handle the differences during the simultaneous usage of multiple quality approaches. Therefore we consider this problem only partially solved. |
| 2. Traceability | S | Traceability is well established in mappings. |
| 3. Changeability | W | The most common is to map only two approaches in one or two directions. In case of 3 or more approaches there is an explosion in the number of mappings, thus including new quality approaches into mappings can be extremely difficult. |
| 4. Completeness | S | Completeness is often achieved as all requirements of source approaches are mapped. |
| 5. Appraisal support | S | Appraisal support is a strong point of mappings. Software tools also have been developed in order to support appraisals of mapped quality approaches. |
| 6. Repeatability and documentation | – | In some cases mapping are well documented (e.g. direction of mapping, mapping confidence, mapping process etc.) and thus can be repeated or used as a basis for creating other mappings; in other cases this information is completely missing.<br>Repeatability is limited because of complexity growth with including new quality approaches into the mapping. |

Summarizing this chapter, we can state that each multi-model initiative we identified could be clearly positioned in one of our three categories. The discussions and the positioning made also clear the strengths and weaknesses of the investigated multi-model initiatives.

## 3.5. Analysis

In this chapter we describe findings related to the current initiatives identified (3.5.1). Subsequently we discuss criteria for multi-model solutions (3.5.2).

### 3.5.1. Findings regarding multi-model initiatives

A significant effort has been spent on the multi-model problem by different researchers, organizations and practical process engineers and they have achieved considerable results.
**Here we summarize the findings related to current multi-model initiatives:**

1. Current initiatives can be categorized into three different categories:
- *Quality approach harmonization* is delivered by standardization organizations. Harmonized characteristics of quality approaches help companies to understand and implement the selected standards. Due to the commonalities of these standards, they can be used easier than more different / less harmonized ones.
- *Quality approach integration* has advantages when an organization decides to choose a particular integrated approach, instead of looking for mappings of source approaches or creating an integration of source approaches on its own. This way the company will save resources which would be spent on the integration or mapping. It is e.g. easier to use Enterprise SPICE than to use its separate sources.
- *Quality approach mapping* is useful in case when a company tries to use two different models or standards with different terminologies, contents and structures. Users of mappings can understand how certain terms and requirements are represented in another quality approach which they are also going to use.

In order to clarify the taxonomy, we can say that a mapping is an addition to quality approaches (it does not change the source quality approaches), integration is the replacement of quality approaches (it incorporates source quality approaches), while harmonization is the modification of quality approaches (in order to achieve common characteristics such as structure or terminology).

2. No current multi-model initiative fulfils all the criteria that we identified. Therefore we conclude that there is a serious need for a general multi-model solution.

### 3.5.2. MSPI criteria

We consider problems discovered during the review of multi-model initiatives important and only partially solved, therefore we use exactly these problems as a basis for the MSPI criteria.

**Handling the differences among quality approaches**

Differences among quality approaches exist (in terminology, granularity, structure, content, size and complexity). As we have shown in previous chapters, not all the initiatives handle, nor document the handling of these differences. In order to ensure the "goodness" of results, a multi-model solution should handle these differences in quality approaches.

**Traceability of multi-model results**

Ensuring traceability of the multi-model results back to source quality approaches is crucial when the changeability, adaptability or appraisal support of results need to be guaranteed. It also plays an important role in ensuring the validity of the solution. Though in some of the current solutions traceability is ensured by mapping, many of current initiatives do not provide clear traceability back to source quality approaches. The result of a multi-model solution should clearly be traceable back to source quality approaches.

**Adaptability and expandability of multi-model results**

When a new quality approach is released it may have huge change effects on the multi-model results and a company's processes. This could also happen when a new version of already used quality approach is released. In order to emphasize the importance of the changes occurring in quality approaches we divide changeability into two criteria, respectively *adaptability* (in case that a new version of an existing approach appears) and *expandability* (in case that a new quality approach appears).

**Completeness of multi-model results**

There should be a clear indication on the coverage of source quality approaches. The completeness of multi-model results should be proven and sufficient for users.

**Appraisal support**

Some of the current initiatives provide multiple appraisal support, others do not. In order to ensure the conformance to source quality approaches, results of a multi-model solution should be appraisable based on multiple (source) quality approaches.

**Repeatability and clear documentation**

As we summarized in Table 4, Table 5 and Table 6, one common weakness of current initiatives is the lack of clear documentation and repeatability of multi-model process. In order to ensure that the outputs are valid and repeatable, a transparent and clear documentation of the solution is needed, describing how multi-model results (outputs) are created from inputs.

## 3.6. Assessing the hypothesis against MSPI criteria

In this chapter we assess our hypothesis against the MSPI criteria. We argue how the PBU framework would satisfy the MSPI criteria.

**Hypothesis:** that mapping quality approaches to a unified process can provide a multi-model solution. By MSPI solution we mean an MSPI initiative which satisfies the MSPI criteria.

Below we discuss how the PBU framework would satisfy the criteria.

1. **Handling the differences among quality approaches**

Differences among quality approaches exist. We think that understanding the structure and elements of quality approaches can serve as a basis for decomposing quality approaches and mapping their elements to elements of a process model. If the elements of multiple quality approaches can be mapped to a single process model, this process model can be a new layer hiding differences of quality approaches. People using the unified process model do not have to handle and struggle with various quality approaches but to use a single process model.

In order to understand the crucial points in handling the differences, the structure and main elements of 6 different quality approaches are discussed in (Kelemen et al., 2008a, 2009). In chapter 4.4 we will present 8 further quality approaches having the focus on the structure, content and quality approach elements. After identification of quality approach elements, our proposed solution, the PBU framework will take into consideration further differences (in scope, terminology, granularity, content, size and complexity) among quality approaches. Furthermore, we will tailor a real life process from multiple quality approaches, presenting how to handle the differences among the selected quality approaches.

**Terminology:** With the PBU concept, we map quality approach elements to process elements. All quality approaches are process based, therefore mapping of terms should also be possible. Take the peer review process as an example: terms such as “Inspection checklist” in IEEE 1028 and “peer review checklist” in CMMI-DEV v1.3 (VER SP 2.1 TWP 2) would be mapped to a single term.

**Granularity:** Quality approaches have their descriptions in different levels of granularity (see e.g. the discussion of the peer review process in the introduction). Fortunately, processes carry a similar characteristic; they also can be decomposed, and described at various levels of granularity. This coincidence in characteristics of quality approaches and processes will allow handling granularity differences in quality approaches. In the PBU process for each quality approach element which will be mapped, we will find a process element on a proper, corresponding granularity level. This way granularity differences in quality approaches will be handled by using a single process model.

**Structure and elements:** Knowing the structure and elements of quality approaches will help to determine which elements can and which cannot be mapped to process elements. If (the majority of) quality approach elements can be mapped to process elements, the PBU concept can work; as processes could be build/enhanced using quality approach elements of multiple quality approaches.

In order to handle the differences in structure and elements of quality approaches, their elements will be mapped to process elements of a single process model. This process model will have a single, unified structure, having only process model elements, this way differences in structure and elements of quality approaches will be hidden.

**Content:** the application of the PBU framework will result in a unified process to which the content of quality approaches will be mapped. Content will be handled by using structural and elemental recognitions in quality approaches. For example content present in quality approach elements will be mapped to proportional process elements (e.g. in the most cases, content of CMMI subpractices or ISO 9001 shall statements will be mapped to process activities).

**Size and complexity:** Size and complexity can influence the selection (A. L Ferreira et al., 2010) and simultaneous usage of multiple quality approaches. Complexity can be measured by the coupledness and the number of cross-references inside a quality approach. For example in case of CMMI, it can be observed that there are process areas which are highly interconnected, and there are more separated ones (Balla & Kelemen, 2011; Kelemen, 2011b). These different kinds of process areas may be handled differently. In the PBU process cross-references among quality approaches will be discovered and referred elements will be checked for further mapping. Size and focus of quality approaches can affect the granularity of descriptions; granularity will be handled as discussed previously.

2. **Traceability of PBU results**

Traceability of multi-model results is again crucial to prove the validity of the multi-model solution. In the PBU process both processes and quality approaches will be decomposed, and clear relationships (mapping) among process element instances and quality approach element instances will be identified. These relationships will describe traceability.

3. **Adaptability and expandability of PBU results**

In order to ensure adaptability and expandability of PBU results, the PBU process will provide a solution for including new (versions of) quality approaches into the PBU results.

**Adaptability:** In order to ensure adaptability, changes in the new versions of quality approaches will be discovered and reflected in quality approach element mappings. In case of a new version of a quality approach 3 things can happen regarding quality approach elements:

*deletion, modification* and appearance of *new* quality approach elements. The PBU framework can be supported by a database structure to store the quality approach element – process element mappings and handle adaptability issues related to these three kinds of changes.

**Expandability:** The expandability of the PBU results will be ensured by an iterative process, so that in each iteration, a new quality approach can be added to the PBU result. When a new quality approach is included into the unified process then a new iteration should be performed. If such an iterative process can be defined, then the expandability of the PBU result can be ensured.

4. **Completeness of PBU results**

A multi-model result should indicate the coverage of source quality approaches. Furthermore, the information provided by PBU result should be sufficient for users. A well-documented mapping between the PBU results and source quality approaches should provide the level of coverage (e.g. what percentage of source quality approaches are covered by the PBU result). In order to ensure sufficient level of details of PBU result, the resulting process will be represented in a Process Modeling Language (PML). PMLs (e.g. BPM or EPM) have the abilities to provide appropriate completeness as processes can be decomposed to unlimited levels and each (sub)process can be described in a chosen level of granularity.

5. **Appraisal support**

Processes built/enhanced using the PBU framework should be conformant to multiple source quality approaches, and there should be a possibility to appraise this conformity. Using a well-documented mapping, the relations between the PBU results (process model(s) resulting from the application of the PBU framework) and source quality approaches will be ensured.

6. **Repeatability of the PBU process**

The repeatability and clear documentation of a multi-model solution is crucial in ensuring the validity of the multi-model results. Many of multi-model solutions (quality approach mappings, integrations, harmonizations) are not documented (Kelemen et al., 2011), therefore it is arguable how the multi-model result was achieved. In order to satisfy the repeatability criterion, the PBU framework will provide a clearly documented and repeatable process. The PBU process will be described using a process modeling language (PML), defining its process activities, inputs, outputs (PBU results) and guidelines for applying the PBU process.

## 3.7. Limitations

Here we discuss limitations of the work presented in this chapter.

**Literature Search** – We tried to perform a comprehensive literature search within the field of multi-model software process improvement. However, there might be other sources pointing to this field which include different terminology, or are present in different databases we have not checked.

**MSPI problems** – Besides the problems that we found in implementing a multi-model solution in an organization, other problems might become important as well. For instance we could also refer to implementation-related problems such as: abstraction, understandability, accuracy, predictiveness, inexpensiveness and others mentioned by (Salviano, 2009b), or to people-related problems, such as: integrated teaming or multi-model trainings (Siviy et al., 2008e).

**MSPI initiatives and initiative categories** – As we analyzed the current literature we found that harmonization, integration and mapping are very often used, but not usually clearly defined (see word occurrences in (Kelemen, 2011a)). Therefore we tried to define them and use as a basis for our categories. New categories may arise when different solution will be developed and used in practice (e.g. some formal initiatives already exist, but they are still not widespread, e.g. (Malzahn, 2008)).

**MSPI criteria** – We tried to discover problems mentioned in literature and tailor the criteria based on these problems. The detailed list of criteria can be and probably will be extended / refined as current problems tend to be solved or new problems and initiatives arise.

## 3.8. Conclusion

In this chapter we focused on answering two research questions:

Q1 What criteria should a multi-model solution satisfy?

Q2 Do current MSPI initiatives satisfy the MSPI criteria?

A literature search was performed (RS1-1) where various problems of MSPI were discovered (RS1-2): problems caused by differences in quality approaches which are the inputs of MSPI solutions, problems of documentation and repeatability of the MSPI process, and problems connected to the quality of multi-model results.

(RS2-3) Regarding the current initiatives for solving the problem of using multiple quality approaches, we derived three categories: quality approach harmonization, quality approach integration and quality approach mapping.

Based on literature review and an analysis of the strengths and weaknesses of current initiatives in these three categories, we derived criteria for multi-model software process improvement solutions (RS1-4). These are: criteria for handling the differences of inputs of MSPI solutions (in terminology, structure, granularity, content, size and, complexity), criteria for documentation and repeatability of the MSPI process and finally criteria related to the quality of multi-model results or outputs (such as adaptability, expandability, completeness, traceability and appraisal support).

We could conclude that currently none of the initiatives fulfil all the criteria and therefore no generally applicable solution exists. However, with the identified criteria, a basis is provided for the development of multi-model solutions. In this chapter we have also discussed that if we would develop the PBU framework it would probably satisfy the MSPI criteria (RS1-5).

# 4. OPTIONS AND LIMITATIONS OF MAPPING QUALITY APPROACH ELEMENTS TO PROCESS ELEMENTS[2]

In chapter 3 we analysed current initiatives in multi-model software process improvement and identified the MSPI criteria was identified. We also showed that no initiative satisfies the MSPI criteria. This means we can design a solution for the multi-model problem.

Designing a multi-model solution requires a number of preliminary steps. First we must understand options and limitations of the basic concept: we need to discover both elements of processes and elements of quality approaches and analyse if a mapping between them is indeed possible. On this basis we can design a PBU process.

In this chapter we describe fundamental observations in MSPI which led us to develop the PBU framework, respectively:

- discovering process elements based on literature,
- analysing the of structure and elements of quality approaches,
- refining process elements based on quality approach elements,
- and discussing options and limitations of mapping quality approach elements to process elements.

## 4.1. Research approach

Table 7 summarizes research questions, research steps, sub-steps and the chapters in which the research steps are discussed.

[2] A preliminary version of discussions included in this chapter have been published in (Kelemen, Balla, Trienekens, & Kusters, 2008a).

Table 7 – Research questions, steps, sub-steps, chapters

| Research question | Research step | Research sub-step | Ch. |
|---|---|---|---|
| Q3 What is a suitable set of process elements to base the unified process on? | RS3 Identify process elements required for mapping (will be performed in 3 iterations – see step description and Figure 2) | RS3-i1: Identify process elements required for mapping based on literature | 4.3 |
| | | RS3-i2: Refine process elements | 4.5 |
| Q4 What are the elements of quality approaches? | RS4 Identify elements of quality approaches | | 4.4 |
| Q5 Which characteristics of (a number of) current quality approaches further and/or hinder mapping of these approaches to a process? | RS5 Identify options and limitations of mapping quality approaches to processes | | 4.6 |

In 4.2 we describe a brief introduction to quality approach mapping – which is needed for understanding how quality approaches can be mapped, in 4.3 we describe basic process elements to which quality approaches could be mapped, then in 4.4 we describe the structure of quality approaches and possible elements for mapping. In 4.5 refinements on process elements after identifying quality approach elements will be discussed. In 4.6 options and limitations of mapping quality approach elements to process elements will be discussed. Limitations of and conclusion of the research presented in this chapter are included in 4.7 and 4.8 respectively.

# 4.2. Quality approach mapping

Weidlich et al. define process matching types: "a match refers to a correspondence, which is an element of the powerset of activities of a first model times the powerset of activities of a second model" (Weidlich, Dijkman, & Mendling, 2010). Applying this definition to our scope a *match* refers to a correspondence, which is an element of the powerset of quality approach elements of a first quality approach times the power set of quality approach elements of a second quality approach.

In (Weidlich et al., 2010), a collection of matches are called mapping. Since the process improvement community uses the term mapping for both (M. T. Baldassarre, Piattini, Pino, & Visaggio, 2009; Diaz, Garbajosa, & Calvo-Manzano, 2009; M. De Oliveira, De Oliveira, & Belchior, 2006), we use term mapping when referring to both a single match and a set of matches.

The logic of mapping presented in (Weidlich et al., 2010) can be directly translated to our case without additions needed: the power set of any set S, P(S), is the set of all subsets of S, including the empty set and S itself. A single mapping is denoted by a tuple (A, B) of two sets of quality approach elements. A single mapping (A, B) is called *elementary mapping*, if $|A| = |B| = 1$. A single mapping (A, B) is called *complex* if at least one quality approach element set in the pair contains more than one element, i.e., $|A| > 1$ or $|B| > 1$ (Weidlich et al., 2010).

If we would have only 1:1 mappings, it might be feasible to analyse the whole set of potential mappings, which corresponds to the "Cartesian product of" quality approach elements in A and quality approach elements in B (Weidlich et al., 2010).

For a quality approach element set in A with n elements (e.g. n specific and generic practices from CMMI) which will be mapped to m quality approach elements in B (e.g. “shall” statements in an ISO standard), the sum of 1:x and x:1 mappings “is given by $n * \binom{m}{x} + m * \binom{n}{x}$ ”, where “the binomial coefficient $\binom{n}{m}$ defines the number of x-element subset of an n-element set” (Weidlich et al., 2010).

In mapping quality approaches we can encounter the following mapping types:

- no mapping,
- 1:1 – when one quality approach element of quality approach A is mapped to one quality approach element of quality approach B (elementary mapping),
- 1:N – when one quality approach element from quality approach A is mapped to N quality approach elements from quality approach B (complex mapping),
- M:N – when M quality approach elements from quality approach A is mapped to N quality approach elements from quality approach B (complex mapping).

We do not discuss the mapping types for quality approach instances, since from a mapping point of view there is no difference between quality approach elements and quality approach element instances.

Table 8 – Example mapping of quality approach element instances

| **Mapping type** | **A (Process description from Process Impact)** | **B (IEEE 1028:2008)** |
|---|---|---|
| 1:0, not mapped | “Sign Inspection Summary Report: All participants sign the Inspection Summary Report to indicate their agreement with the inspection outcome.” | – |
| 1:1, simple | “Present Work Product: Describe portions of the work product to the inspection team.” | “The reader shall present the software product to the inspection team.” |
| 1:N, complex where N = 2, | “Record Issues: Capture the information in Table 2 on the Issue Log for each issue raised. State aloud what was recorded to make sure it was recorded accurately.” | “The recorder shall enter each anomaly, location, description, and classification on the anomaly list.” |
| | | “If there is disagreement about an anomaly, the potential anomaly shall be logged and marked for resolution at the end of the meeting.” |
| M:N complex, where M = 3, N = 4 | “Work aid 2) Issue Log” | “Shall level input d) Inspection reporting forms” |
| | “Deliverable 3) Completed Issue Log” | “Shall level input e) Anomalies or issues list” |
| | “Deliverable 5) Counts of defects found and defects corrected” | “Shall level output h) The anomaly list, containing each anomaly location, description, and classification” |
| | | “Should level output m) The inspection anomaly summary listing the number of anomalies identified by each anomaly category” |

For example if quality approach A is a peer review process description of Process Impact (Process Impact, 2010) and quality approach B is the IEEE 1028:2008 standard for Peer Re-views (IEEE, 2008b), both approaches can be divided into small entities and these entities can be mapped. Table 8 shows an example mapping of quality approach element instances, including the mapping type and mapped quality approach element instances of two quality approaches.

Further information about quality approach mapping can be found in (Andre L. Ferreira, Machado, & Paulk, 2010) and on process mapping in (Weidlich et al., 2010).

# 4.3. Identifying process elements based on literature

In this chapter the research step "RS3-i1: Identifying process elements required for mapping based on literature" will be discussed.

According to Curtis et al. "any component of a process is a process element" (Curtis, Kellner, & Over, 1992). They distinguish the following elements: "*process step* – as an atomic action of a process that has no externally visible substructure"; "*agent* – an actor (human or machine) who performs a process element"; "*role*-a coherent set of process elements to be assigned to an agent as a unit of functional responsibility" and "*artefact*-a product created or modified by the enactment of a process element" (Curtis et al., 1992). At the time of writing their article there was no consensus on the constructs that collectively form an essential basis of a process model. It is also mentioned that a task is often synonymous with a process and an activity is synonymous with a process element or step, but they decided to use term "process step".

According to (Kellner, Madachy, & Raffo, 1999) the following are important to be identified in planning and developing a process simulation model: "key *activities* and *tasks*; *primary objects* (e.g., code units, designs, problem reports);vital *resources* (e.g., staff, hardware); *activity dependencies*, *flows of objects* among activities and *sequencing*; iteration *loops*, feedback loops and *decision points*; and other *structural interdependencies* (e.g., the effort to revise a code unit after an inspection, as a function of the number of defects found during the inspection)". They also mention that "the focus should be on those aspects of the process that are especially relevant to the purpose of the model" (Kellner et al., 1999).

Bhuta et al. define process elements as "a group of project activities, and/or other process elements related by logical dependencies which when executed (or enacted) provides value to the project". Process Elements, like software components, have "input and output interfaces, defined by pre-conditions and postconditions" (Bhuta, Boehm, & Meyers, 2006).

Relying on 8 different sources for identifying basic process elements, Prado et al. created an UML model for homogenization of quality approaches with the following element classes: objective, process, process group, process category, indicator, measurement, activity, special activity, product, task, resource, role and tool (C. J. Pardo et al., 2009).

In CMMI basic process elements are the following: inputs, activities, outputs, purpose, entry and exit criteria, roles, measures, and verification steps (CMMI Product Team, 2010a, 2010b, 2010c).

Table 9 summarizes process elements found in (Bhuta et al., 2006; Curtis et al., 1992; Kellner et al., 1999; C. J. Pardo et al., 2009; CMMI Product Team, 2010a, 2010b, 2010c) and a derived list of elements which we will use in the following.

Table 9 – Process elements in literature and our derived terms

| (Curtis et al., 1992) | (Kellner et al., 1999) | (Bhuta et al., 2006) | (C. J. Pardo et al., 2009) | CMMI v1.3 (2010) | Derived Process Element |
|---|---|---|---|---|---|
| process step | key activity and task | project activity | activity, special activity, task | activity | activity |
| agent | resources | – | resource | – | resource |
| role | – | – | role | role | role, responsibility |
| artefact | primary objects | input and output (indirectly) | product | input, output | artifact |
| – | activity dependency, flow of object, sequence, loop, structural interdependency | logical dependency | – | – | element relation |
| – | – | input and output interface, precondition, postcondition | objective, process, process group, process category, indicator, measurement, tool | purpose, measures, verification step, entry and exit criteria | further process elements |

As Table 9 summarizes, most of the publications discuss process step, activity and task with a similar meaning. Role, responsibility, resource, artifact, inputs/outputs, different types of element relations, entry and exit criteria are used also in most of the papers.

Based on the sources discussed we define the following process elements:

- A *(sub)process* can contain any type of process elements and also subprocesses. The only difference between subprocess and process is that the process has no parent process while the subprocess always has a parent process. The parent process contains the subprocess.
- An *activity* is a process element which is needed to be performed by a resource on one or more artifact(s) to create value. The activity is atomic.
- An *artifact* represents an input or output to a (sub)process or activity.
- A *resource* can be a human or machine (e.g. a person or a computer) performing activities, acting in a predefined *role* (e.g. project manager or database server). Roles can have assigned *responsibility* (e.g. planning and monitoring the project, or serving database queries).
- One process element can have *relation* with other elements. Typical relations among sub(processes) and activities can be e.g. *sequence* or *loop*. An artifact can be the *output* of an activity and can serve as an *input* to the activity.

Particular process elements are not taken into account because they are not discussed in most of the articles e.g.: input interface, measurement, tool and many more – we call these further process elements.

In the next chapter we analyse quality approaches, searching especially for quality approach elements which could be mapped to our basic process elements. We will collect further quality approach elements not directly mappable to process elements, which also could be useful in building processes.

Each quality approach description concludes with a table containing the main elements of the quality approach in case, their descriptions and possible mapping to process elements. Further, we analyse in each case whether the conclusion drawn from the structure of a certain quality approach triggers refinements on the process element list identified in 4.3.

## 4.4. Identifying quality approach elements

In this chapter the research step "RS4 Identify elements of quality approaches" will be discussed.

We believe that understanding the structure of quality approaches could help us finding quality approach elements, which could be used in building/enhancing organizational processes. Therefore in this chapter we present the structure and elements of 7 widely used quality approaches carrying differences from a number of perspectives. We include into this analysis quality approaches released by important bodies of standardization such as ISO, IEEE, IEC standards, models from SEI CMU, ITIL which is a governmental quality approach originated from the United Kingdom, and also a non-standardized quality approach: a peer review process description from a commercial organization. These approaches differ in their scope and content (e.g. ISO 9001 is widely acceptable to any organization which wants to build an organizational quality management system, ISO/IEC 90003 and CMMI for development are software specific), terminology (e.g. CMMI has a different terminology than IEEE 1028), granularity (e.g. if we compare CMMI for Development, CMMI for Services and IEEE 1028 these all are representing the peer review process on different levels and different granularity), structure (e.g. ISO standards and CMMI have a completely different structure), size (e.g. each constellation of CMMI has about 500 pages, while the shortest analyzed quality approach is just above 10 pages), and complexity (e.g. thousands of references among CMMI process areas and Process Impact's simple peer review descriptions).

Quality approaches included into the analysis are the following:

1. ISO 9001:2008 Quality management systems – requirements (ISO, 2008),
2. ISO/IEC 90003:2004 Software Engineering – Guidelines for the application of ISO 9001:2000 to computer software (ISO/IEC, 2004c),
3. ISO/IEC/IEEE 12207-2008 – "Information technology - Software life cycle process"(ISO/IEC/IEEE, 2008),
4. CMMI version 1.3 (CMMI Product Team, 2010a, 2010b, 2010c),
5. IT Infrastructure Library (ITIL) v3 (TSO, 2007),
6. IEEE 1028 -2008 IEEE Standard for Software Reviews and Audits (IEEE, 2008b),
7. Structure of a peer review process description from Process Impact (Process Impact, 2010).

Here we mention that we already discussed the structure of previous versions of 6 quality approaches in (Kelemen et al., 2008a; Kelemen, Balla, Trienekens, & Kusters, 2008b) which were:

1. CMMI for Development, Version 1.2,
2. ISO 9001:2000 Quality management systems – requirements,
3. ISO 9004:2000 Quality management systems – Guidelines for performance improvements,
4. ISO/IEC 90003:2000 Software Engineering – Guidelines for the application of ISO9001:2000 to computer software,
5. ISO/IEC 15939-2002 – “Information technology - Software measurement process” and
6. ISO/IEC 12207-95 – “Information technology - Software life cycle process”.

For representing quality approaches we use UML class diagrams, because our focus is on their elements. In order to keep figures simple, we present only the main elements and we use only simple relationships between the elements at describing quality approaches in UML.

### 4.4.1. Structure of ISO 9001:2008 and ISO/IEC 90003:2004

ISO 9001:2008 (ISO, 2008) is an international standard which contains general requirements for quality management systems (QMS). The requirements included in this standard are so general that they can be applied to any company. E.g. one general requirement included into this standard is: “The adoption of a quality management system should be a strategic decision of an organization”.

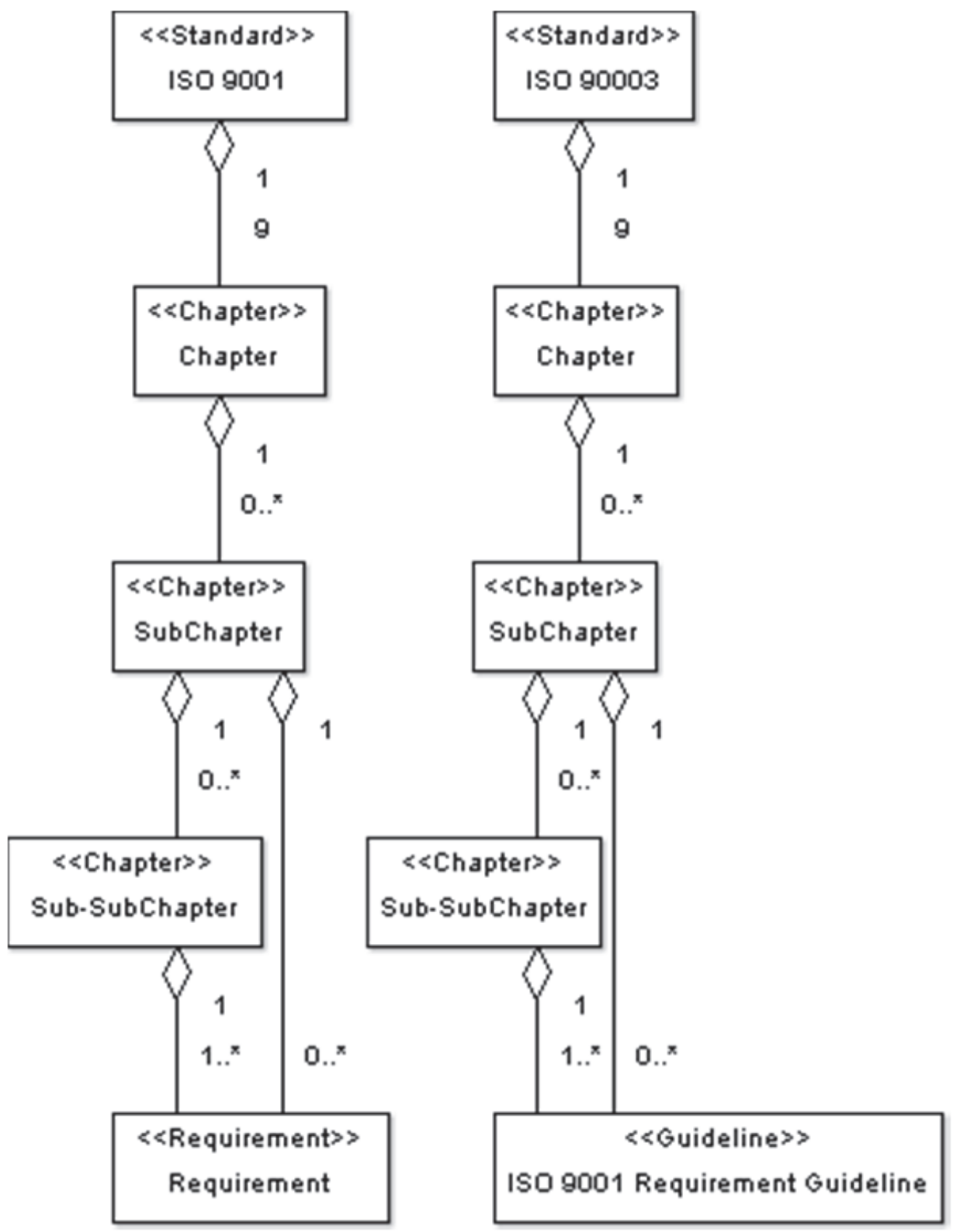

Figure 3 – The structure of ISO 9001:2008 and ISO/IEC 90003:2004

Figure 3 shows the structure of 2 similar quality approaches ISO 9001 and ISO 90003. We call ISO 9001-like standard each quality approach which follows the structure of ISO 9001, e.g. the former version of ISO 9004:2009 (9004:2000) has also the same structure. Looking at the figure, we can see that both quality approaches contain 9 chapters, which could contain subchapters and the subchapters can also contain further subchapters.

Requirements of ISO 9001 can be found at subchapter and sub-subchapter levels in “shall” statements.

The structure of ISO/IEC 90003:2004 “Software Engineering – Guidelines for the application of ISO 9001:2000 to computer software” is identical to the structure of ISO 9001 because it is using ISO 9001 as a basis, containing the same chapters. The only difference between them is that the latter one defines guidelines for implementing ISO 9001 requirements. E.g. chapter 4.2 of ISO 9001 describes the general documentation requirements, while the same chapter of ISO 90003 describes how to implement these requirements in a software organization. For example it includes the following guidance: “documents for the effective planning, operation, and control of processes for software may cover” … “descriptions of life cycle models used”.

**Quality approach elements**

In standards carrying the structure of ISO 9001 there are chapters and subchapters containing shall sentences which form the requirements or guidelines for processes. These sentences include all the information which can be mapped to process elements (e.g. activities, artifacts). Unfortunately, these are not represented in a well-defined format, but described textually.

Despite that there is very little structure in these standards, the meaning of these elements can be analysed further, many of the chapter titles and sentences can be decomposed and processed.

Table 10 – Elements of ISO 9001 and ISO/IEC 90003

| Quality approach element | Description | Process element |
|---|---|---|
| Chapter | The chapter titles can give insight into the process scope. Titles need to be processed manually. | Need to be processed manually |
| Requirements and guidelines in Sentences | Sentences include the necessary information (requirements and guidelines) for developing or refining processes. Various parts of sentences can be mapped to process elements like activity, resource or artifact. (See the discussion in the text above). | No direct mapping is possible. Need to be decomposed and the mapping should be performed manually. |

For example, the chapter titles of ISO 9001 are as follows: Chapter 1: Scope, Chapter 2: Normative Reference, Chapter 3: Terms and definitions (specific to ISO 9001, not specified in ISO 9000), Chapter 4: Quality Management System, Chapter 5: Management Responsibility, Chapter 6: Resource Management, Chapter 7: Product Realization, Chapter 8: Measurement, analysis and improvement and at the end there are tables representing the correspondence between ISO 9001 and other standards. These titles contain many organizational and process aspects, thus information useful for building or refining specific processes can be recognized: e.g. chapter 8 can be used in defining the measurement and analysis process. Furthermore, requirements are described in sentences regarding quality policy, quality manual, internal audits, control of nonconforming product / service and many others. These requirements can be mapped to process elements e.g. quality manual and quality policy to artifact, or internal audit and control of nonconforming product / service to activity or (sub)process.
Table 10 gives a summary of analysis of ISO 9001 and ISO 90003 standards.
**Refinements on the process element list:** at this step there were no quality approach elements discovered on which the list of process elements could be refined.

### 4.4.2. Structure of ISO/IEC/IEEE 12207-2008

ISO/IEC 12207 IEEE Std 12207-2008 – "Information technology - Software life cycle process" (ISO/IEC/IEEE, 2008) describes requirements for processes, activities, tasks, entry and exit conditions, responsibilities and documentation for software lifecycle processes.
Due to its size, Figure 46 representing the elements of this quality approach is moved to the appendix. Figure 46 shows the structure of ISO/IEC/IEEE 12207 representing two major categories of processes: system lifecycle processes and software specific processes. Both types include several different subcategories of processes e.g. agreement, project, technical or support and reuse processes. Process requirements can be found in low level, including descriptions of purpose, outcomes activities and tasks.
**Quality approach elements**
ISO 12207 contains the following elements: chapter, process category, process name, scope, purpose, activity, task and outcome.
The standard defined its process elements as follows:
„Each process of this standard is described in terms of the following attributes:
- The title conveys the scope of the process as a whole
- The purpose describes the goals of performing the process
- The outcomes express the observable results expected from the successful performance of the process
- The activities are a list of actions that are used to achieve the outcomes
- The tasks are requirements, recommendations, or permissible actions intended to support the achievement of the outcomes" (ISO/IEC/IEEE, 2008).

Several of these elements can be mapped to process elements e.g. activity and task can be mapped to the activity process element, or outcome can be mapped to artifact.
Understanding the structure of this quality approach can help handling it in a different way than the elements of ISO 9001, namely certain elements can be mapped directly to process elements, while others (e.g. annexes) might need to be processed manually similarly to ISO 9001 elements.
There are elements in this standard which refer to the standard itself (e.g. overview, scope purpose, limitations of the standard). We focus on process-related elements, therefore we consider these (not processes- related) elements out of scope.
Table 11 gives a summary of the elements of ISO/IEC/IEEE 12207 standard and their mapping to process elements.

Table 11 – Elements of ISO/IEC/IEEE 12207

| Quality approach element | Description | Process element |
|---|---|---|
| Process category | Processes are ordered into categories and subcategories. Main categories are system lifecycle processes and software specific processes. | – |
| Process title | "The title conveys the scope of the process as a whole". | Process name (New) |
| Process purpose | "The purpose describes the goals of performing the process". | Process purpose (New) |
| Activity | "The activities are a list of actions that are used to achieve the outcomes". | Activity |
| Task | "The tasks are requirements, recommendations, or permissible actions intended to support the achievement of the outcomes". | Activity |
| Outcome | The result of the process. | Artifact |
| Terms and definitions | Terms and definitions used through the standard. These can help to understand the content while processing it. | Indirectly mapped, used while understanding the process |
| Non-process related elements | E.g. overview, scope purpose, limitations of the standard, intended usage of the standard, normative references etc. | – |

**Refinements on the process elements list:** up till now our process element list does not contain *process name and process description*. However, after analysing this quality approach it seems to be wise to split the process element into these two elements, because we consider that a description could significantly contribute to understand the processes. The process description can include an introduction, the purpose, objective, goal and any further textual description of the process.

We define new elements as follows:

*Process name*: The process name conveys the scope of the process as a whole.

*Process description:* The process description is a generic introduction to the process which can include elements such as:

- introduction/purpose/goal/objective of the process,
- a list of activities of the process,
- textual description of the process..

The *process category* element could depend on the number and characteristics of processes. In the case of a small number of processes no categorization is needed. In the case of high number of processes various categories might be needed. Therefore, we do not consider process category as a necessary process element (attribute) and we do not include it into our process element list.

*Non-process related elements* are also left out from our process element list, because they are not affecting processes and are relevant to other quality approach elements or to the quality approach itself and not to the process.

### 4.4.3. Structure of CMMI

The current version of CMMI, v1.3 defines 3 constellations: CMMI for Development(CMMI Product Team, 2010a), CMMI for Acquisition(CMMI Product Team, 2010c) and CMMI for Services(CMMI Product Team, 2010b). Figure 4 shows the main elements of CMMI and their interrelations.

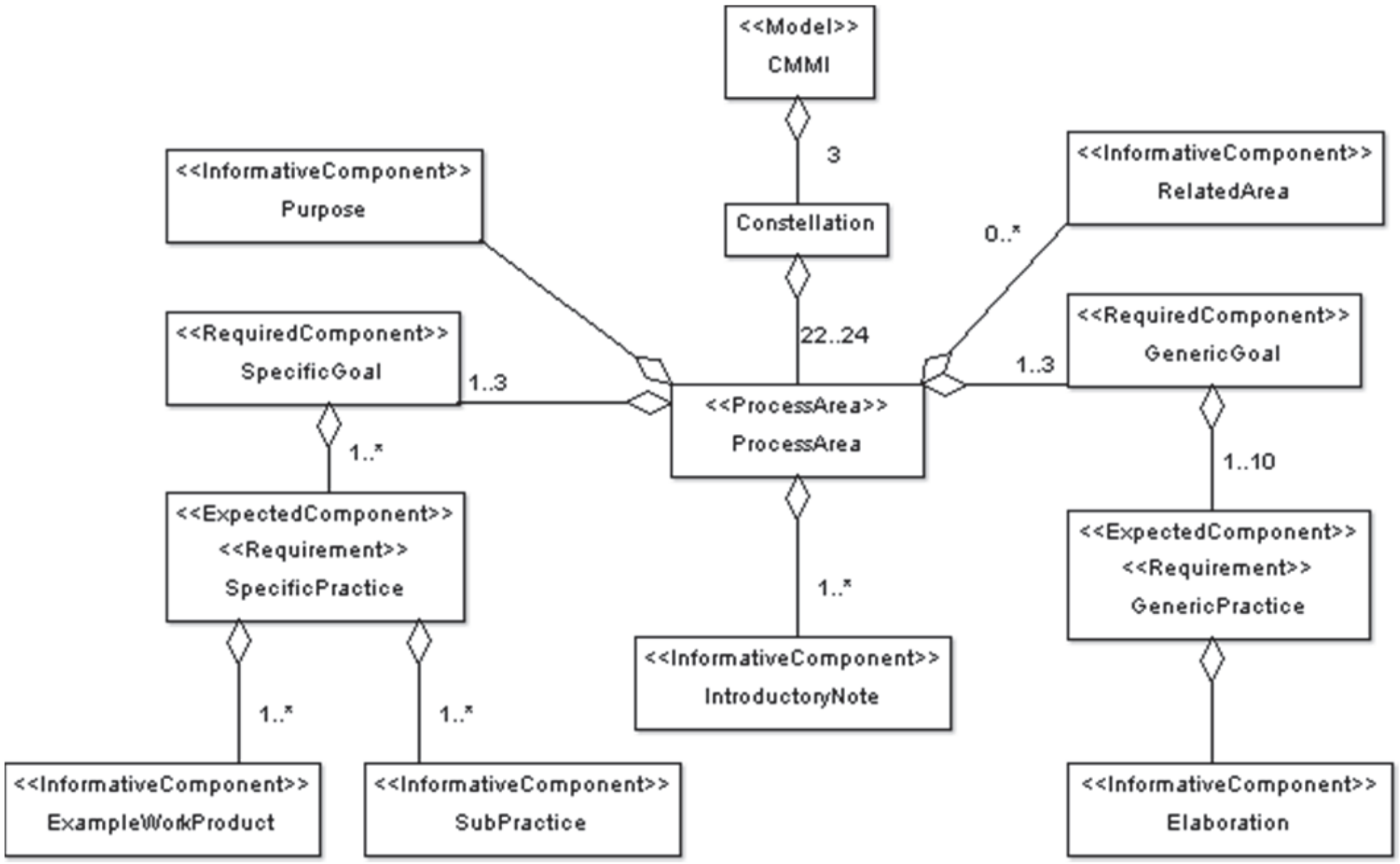

Figure 4 – The structure of CMMI v1.3

CMMI defines its structure as follows:

“All CMMI models are produced from the CMMI Framework. This framework contains all of the goals and practices that are used to produce CMMI models that belong to CMMI constellations.”... “Model components are grouped into three categories - required, expected, and informative - that reflect how to interpret them.” “Required components are CMMI components that are essential to achieving process improvement in a given process area.”… “The required components in CMMI are the specific and generic goals. Goal satisfaction is used in appraisals as the basis for deciding whether a process area has been satisfied.”

“Expected components are CMMI components that describe the activities that are important in achieving a required CMMI component. Expected components guide those who implement improvements or perform appraisals. The expected components in CMMI are the specific and generic practices. Before goals can be considered to be satisfied, either their practices as described, or acceptable alternatives to them, must be present in the planned and implemented processes of the organization.”

“Informative components are CMMI components that help model users understand CMMI required and expected components. These components can be example boxes, detailed explanations, or other helpful information. Subpractices, notes, references, goal titles, practice titles, sources, example work products, and generic practice elaborations are informative model components." (CMMI Product Team, 2010a)

**Quality approach elements**

CMMI is a well-structured quality approach, it clearly defines its elements, and different element types are described using different text styles in the document so that it is easy to recognize which text belongs to which element. Many CMMI elements can be mapped to process elements depending on the context and the process e.g. specific and generic practices, subpractices can be mapped to activities or subprocesses and example work products can be mapped to artifacts. Sometimes roles are also mentioned in the text of CMMI which can be mapped to process roles. Table 12 gives a summary of elements of CMMI-v1.3.

Table 12 – Elements of CMMI-v1.3

| **Quality approach element** | **Description** | **Process element** |
|---|---|---|
| Constellation | "A collection of CMMI components that are used to construct models, training materials, and appraisal related documents for an area of interest (e.g., acquisition, development, services)." (CMMI Product Team, 2010a, 2010b, 2010c) | – |
| Process area | "A cluster of related practices in an area that, when implemented collectively, satisfies a set of goals considered important for making improvement in that area." (CMMI Product Team, 2010a, 2010b, 2010c) | Process name |
| Purpose statement | "A purpose statement describes the purpose of the process area and is an informative component. " (CMMI Product Team, 2010a, 2010b, 2010c) | Process purpose |
| Introductory note | "The introductory notes section of the process area describes the major concepts covered in the process area and is an informative component." (CMMI Product Team, 2010a, 2010b, 2010c) | Process scope |
| Related areas | "The Related Process Areas section lists references to related process areas and reflects the high-level relationships among the process areas. The Related Process Areas section is an informative component." | Processes should be developed with accordance to the related process areas. |
| Specific goal | "A specific goal describes the unique characteristics that must be present to satisfy the process area. A specific goal is a required model component and is used in appraisals to help determine whether a process area is satisfied." | Activity (Needed to be checked manually) |
| Generic goal | "Generic goals are called "generic" because the same goal statement applies to multiple process areas. A generic goal describes the characteristics that must be present to institutionalize processes that implement a process area. A generic goal is a required model component and is used in appraisals to determine whether a process area is satisfied." | Activity (Needed to be checked manually) |
| Specific Goal and Practice Summaries | "The specific goal and practice summary provides a high-level summary of the specific goals and specific practices. The specific goal and practice summary is an informative component." | Activity description |

| Quality approach element | Description | Process element |
|---|---|---|
| Specific practice | "A specific practice is the description of an activity that is considered important in achieving the associated specific goal. The specific practices describe the activities that are expected to result in achievement of the specific goals of a process area. A specific practice is an expected model component." | Activity |
| Example work product | "The example work products section lists sample outputs from a specific practice. An example work product is an informative model component." | Artifact |
| Subpractice | "A subpractice is a detailed description that provides guidance for interpreting and implementing a specific or generic practice. Subpractices can be worded as if prescriptive, but they are actually an informative component meant only to provide ideas that may be useful for process improvement." | Activity |
| Generic practice | "Generic practices are called "generic" because the same practice applies to multiple process areas. The generic practices associated with a generic goal describe the activities that are considered important in achieving the generic goal and contribute to the institutionalization of the processes associated with a process area. A generic practice is an expected model component." | Activity |
| Generic Practice Elaboration | "Generic practice elaborations appear after generic practices to provide guidance on how the generic practices can be applied uniquely to process areas. A generic practice elaboration is an informative model component." | Activity |
| Addition | "Additions are clearly marked model components that contain information of interest to particular users. An addition can be informative material, a specific practice, a specific goal, or an entire process area that extends the scope of a model or emphasizes a particular aspect of its use." | Needed to be checked manually. |
| Note | "A note is text that can accompany nearly any other model component. It may provide detail, background, or rationale. A note is an informative model component." | Needed to be checked manually. |
| Example | "An example is a component comprising text and often a list of items, usually in a box, that can accompany nearly any other component and provides one or more examples to clarify a concept or described activity. An example is an informative model component." | Needed to be checked manually. |
| Reference | "A reference is a pointer to additional or more detailed information in related process areas and can accompany nearly any other model component. A reference is an informative model component." | – |

**Refinements on the process elements list**: at this point it can be seen that similarly to the previous quality approach, this quality approach contains a purpose and a scope related component. After analysing CMMI no further refinements are needed into the process element list.

### 4.4.4. Structure of ITIL v3

The ITIL v3 core set includes 6 books, namely: Official introduction to the ITIL Service Lifecycle, Service Strategy, Service Design, Service Transition, Service Operation and Continual Service Improvement. These books have very similar structure; Figure 5 presents the structure of Service Design book which includes practical guidelines, principles, processes, service design technology-related activities, implementation guides, critical success factors and risks. These terms are used by ITIL as process elements.

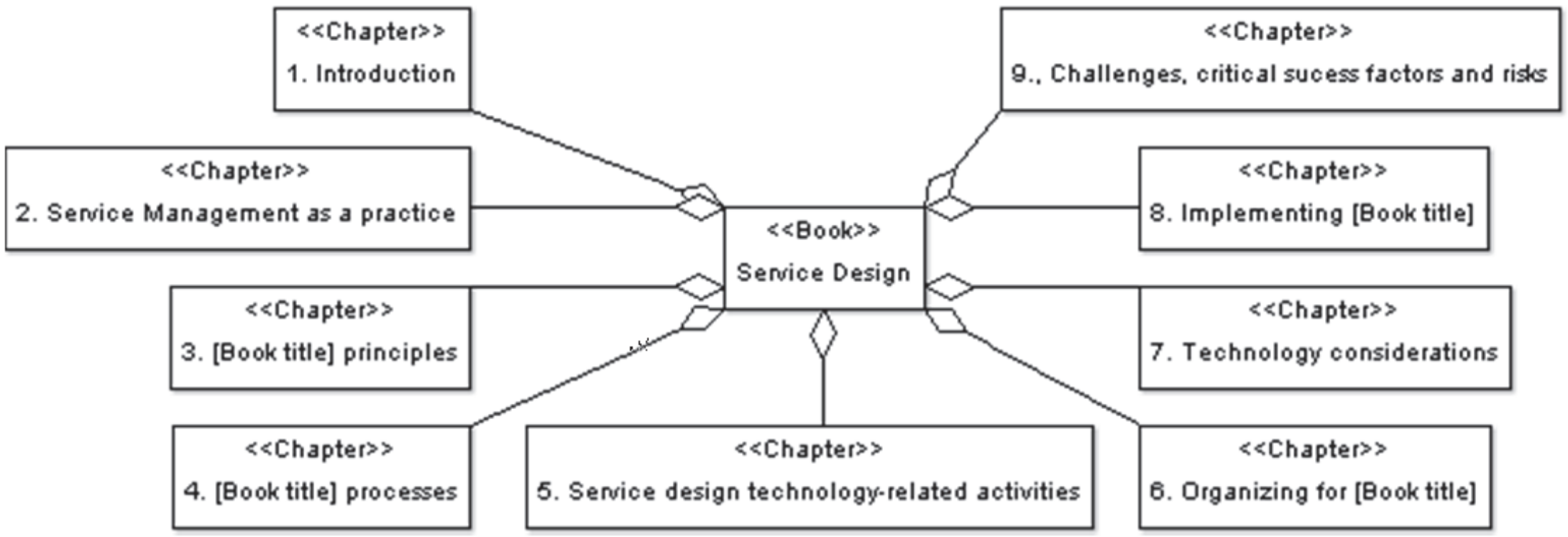

Figure 5 – The structure of ITIL v3 – Service Design book

Book chapters which describe the processes can be further decomposed. Figure 6 shows the composition of a generic ITIL process.

**Quality approach elements**

It can be seen that beside the usual quality approach elements such as activities, inputs and outputs, further information of the process are described i.e. purpose/goal/objective, scope, value to business, policies, information management KPIs and CSFs/risks.

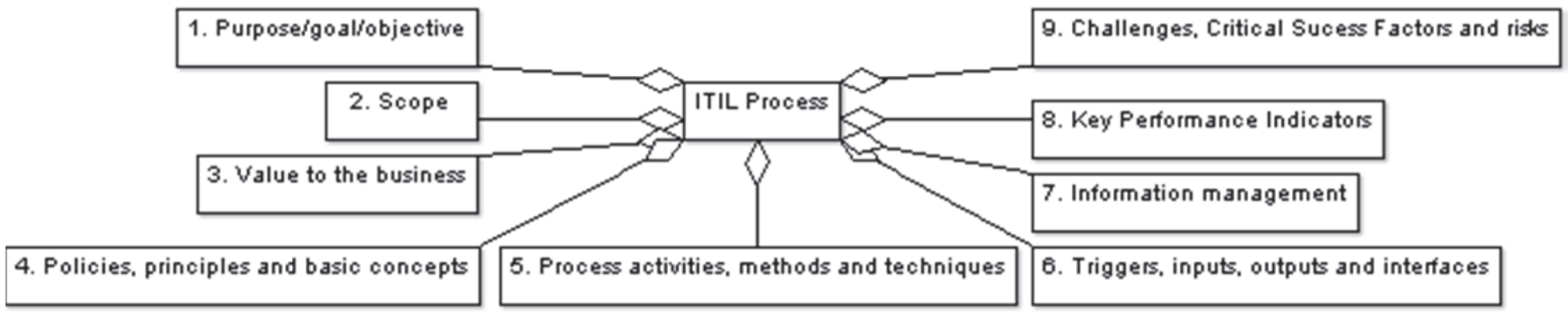

Figure 6 – The structure of an ITIL v3 process

Activity, method and technique can be mapped to activity process element, input and output can be mapped to artifact. Further quality approach elements can also be used to enrich processes e.g. KPI-s are useful to understand the process performance, CSF-s and risks can help understanding and focusing on the important activities of processes.

Table 13 – Elements of an ITIL v3 process description

| Quality approach element | Description | Process element |
| --- | --- | --- |
| Process | „A structured set of Activities designed to accomplish a specific Objective. A Process takes one or more defined inputs and turns them into defined outputs. A Process may include any of the Roles, responsibilities, tools and management Controls required to reliably deliver the outputs. A Process may define Policies, Standards, Guidelines, Activities, and Work Instructions if they are needed.” | Process |
| Purpose/goal/objective | The “Purpose/goal/objective” chapter of ITIL v3 processes give an overview on the purpose, goals and objectives of the processes. | Process purpose |
| Scope | The “scope” chapter of the processes describe the context and scope in which the process should be performed. | Purpose, name |
| Value to business | The “value to business” chapter describes the abilities which can be achieved by performing the process, emphasizing the (business) value of the process. | – |
| Policies/principles/basic concepts | Policy: “Formally documented management expectations and intentions. Policies are used to direct decisions, and to ensure consistent and appropriate development and implementation of Processes, Standards, Roles, Activities, IT Infrastructure, etc.”<br>The “policies/principles/basic concepts” chapter of ITIL v3 include basic information related to certain processes, e.g. in case of incident management the followings are explained in detail: process timescales, incident models and major incidents. | – |
| Process activities, methods and techniques | The “process activities, methods and techniques” chapters of ITIL v3 is probably the most valuable part of the process descriptions since it contains a process flowchart. The process flowchart depicts the complete process flow including process “steps”, sequence of steps or decision points. Later, this chapter also includes the detailed discussion of each step presented in the process flowchart. | Activity, element relations |
| Triggers, input and output/inter-process interfaces | Triggers describe how certain events can cause to start a process. E.g. one possible trigger in case of incident management, an incident can be triggered when a user rings the service desk.<br>Interfaces with a certain process mean relations with other processes. E.g. the whole incident management process is a part of problem management process, so there is an interface with the problem management process. | New notion: Subprocess.<br>New process element: parent |
| Information management | The “information management” chapter describes information related to sources of information, handling and storing the information related to the process. | – |

| Quality approach element | Description | Process element |
|---|---|---|
| Metrics and key performance indicators | The “metrics” and “key performance indicators” chapters describe metrics and KPIs which should be monitored upon to judge the efficiency and effectiveness of a process. It also describes what reports should be produced and what roles should be included to the distribution list. | – |
| Challenges, critical success factors and risks | This chapter describe possible challenges, the factors that are critical and risk to successful process. | – |

**Refinements on the process element list:**
It is usual that in an organization there are a number of processes and these are often related to each other. As it was discussed previously, in CMMI we can find a “related process areas” section in the description of each process area, while in ITIL process relations are also noted in the chapter of “inter-process interface”.
In order to be able to decompose processes we identified the notion of *“subprocess”* in 4.3. Subprocess is a part of a process and it can be decomposed to activities and subprocesses. The difference between process and subprocess is that the subprocess has a parent process. In order to show in the process description that a process is a subprocess of another process we add the *“parent”* element to our process element list. The parent element denotes the possible parent(s) of a subprocess. If the parent element is empty in a process description then the process has no parent, therefore it is not a subprocess, but a process. Similarly to a subprocess, an activity can also have a parent process.

## 4.4.5. Structure of IEEE 1028 -2008

Chapters 4-8 of this standard cover requirements against 5 different types of reviews, namely: management reviews, technical reviews, inspections, walk-throughs and audits.

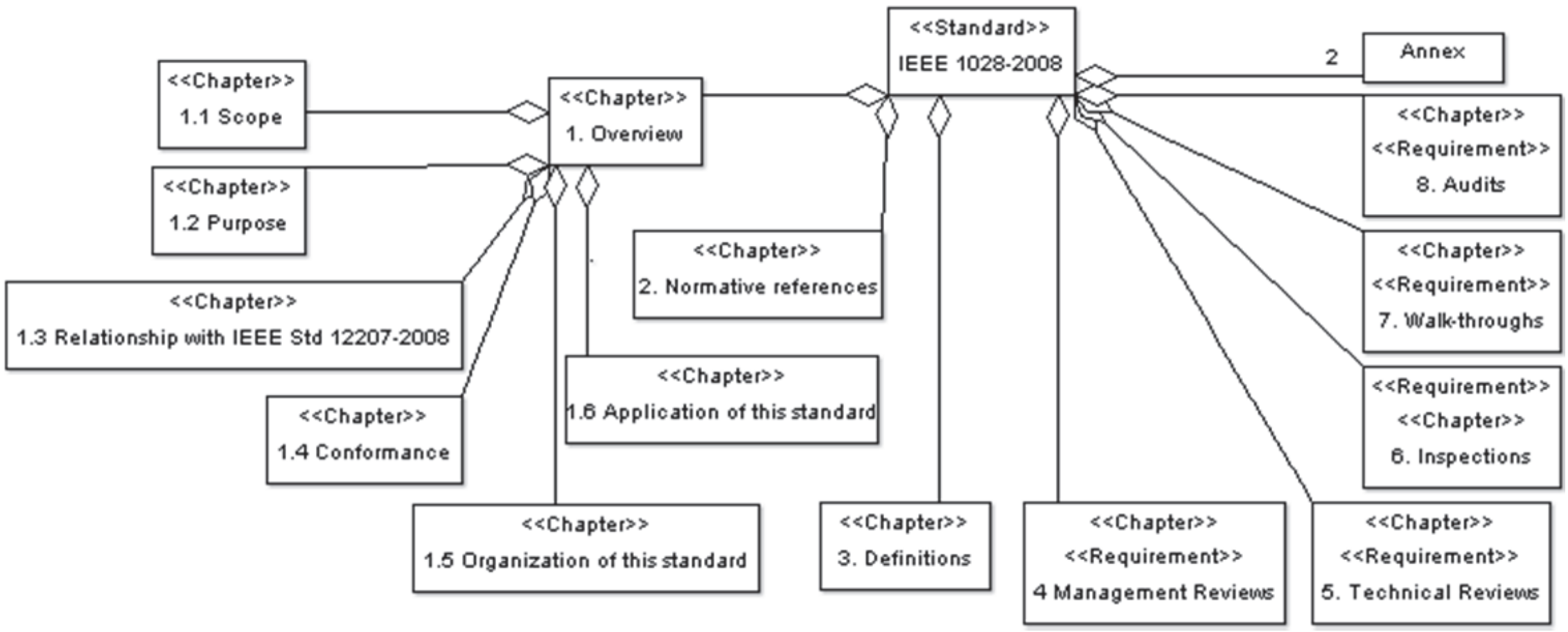

Figure 7 – The structure of IEEE 1028 -2008

IEEE 1028 “applies throughout the scope of any selected software life-cycle model and provides a standard against which software review and audit plans can be prepared and assessed.

Maximum benefit can be derived from this standard by planning for its application early in the project life cycle.” (IEEE, 2008b).

Figure 7 shows how “IEEE 1028 -2008 IEEE Standard for Software Reviews and Audits” can be decomposed based on its chapters.
If we go to a lower level, we can see that chapters 4-8 of IEEE 1028-2008 are quite similar in their structure. Figure 8 represents their structure, which contain typical process elements such as inputs, outputs, entry/exit criteria, procedures, responsibilities, but also a consistent structure starting with an introduction going through a process ending with exit criteria and outputs.

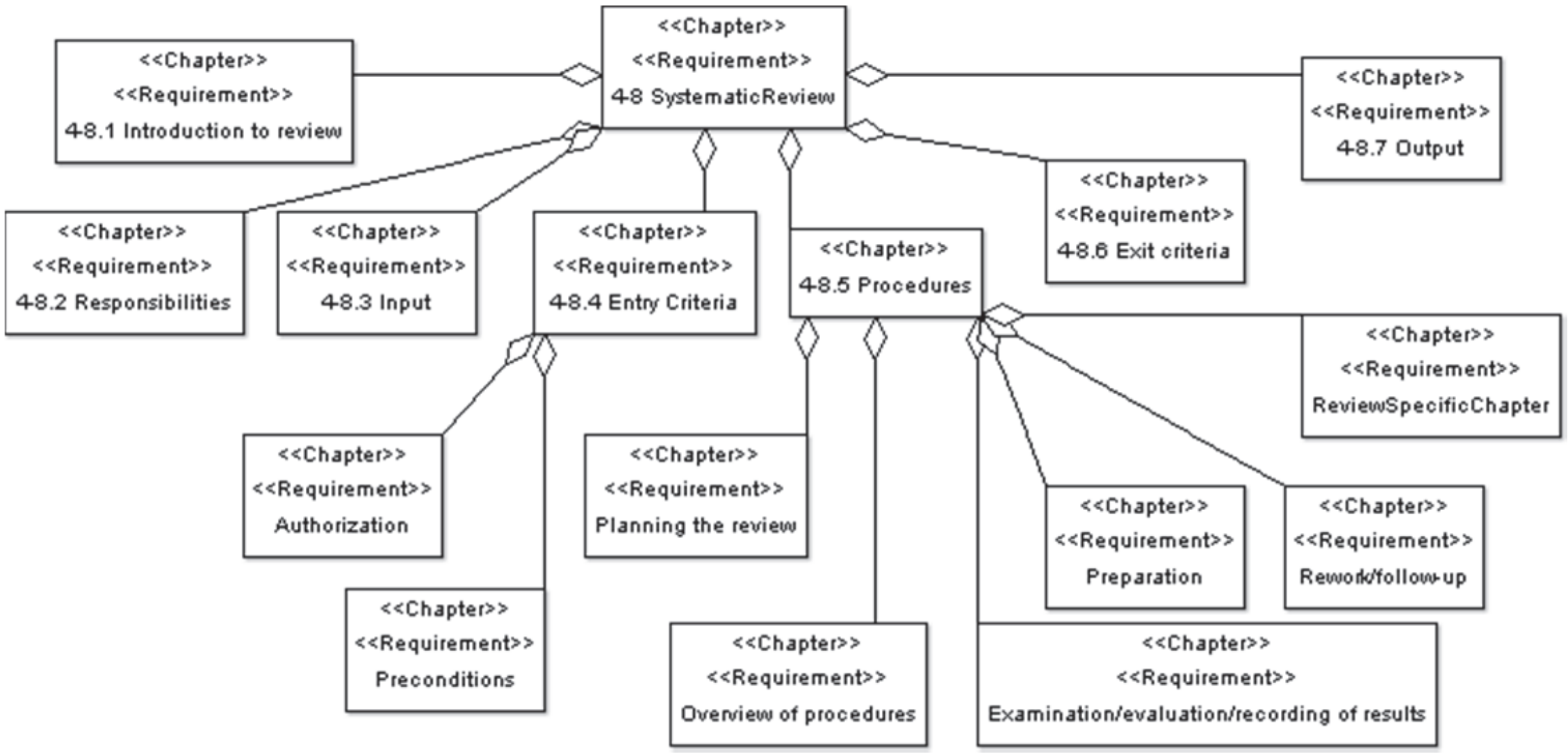

Figure 8 – The structure of chapters 4-8 in IEEE 1028-2008

**Quality approach elements**

In case of IEEE 1028 several quality approaches can be mapped to process elements. Responsibility can be mapped to responsibility process element, procedures can be mapped to (sub)process and procedure subchapter, steps and descriptions included into the procedures can be mapped to activity process element, inputs and outputs can be mapped to artifact process element type.

Table 14 – Elements of an IEEE 1028 process description

| Quality approach element | Description | Process element |
|---|---|---|
| Introduction to review | „Describes the objectives of the systematic review and provides an overview of the systematic review procedures.” | Purpose |
| Responsibilities | „Defines the roles and responsibilities needed for the systematic review.” | Role, Responsibility |
| Input | “Describes the requirements for input needed by the systematic review.” | Artifact |

| Quality approach element | Description | Process element |
|---|---|---|
| Entry criteria | „Describes the criteria to be met before the systematic review can begin, including the following:<br>1) Authorization<br>2) Initiating event" | New: Entry criteria |
| Procedures | „Details the procedures for the systematic review, including the following:<br>1) Planning the review<br>2) Overview of procedures<br>3) Preparation<br>4) Examination/evaluation/recording of results<br>5) Rework/follow-up" | Process, Activity |
| Exit criteria | "Describes the criteria to be met before the systematic review can be considered complete." | New: Exit criteria |
| Output | "Describes the minimum set of deliverables to be produced by the systematic review." | Artifact |

**Refinements on the process element list:**

Entry and exit criteria are mentioned in CMMI as a process description element for "defined" processes and they appear in process descriptions of IEEE 1028 as well. Therefore we add them to our process element list defining them as follows:

*Entry criteria* – Describes the criteria to be met before the process can begin.

*Exit criteria* – Describes the criteria to be met before the process can be considered complete.

### 4.4.6. Structure of peer review process description of Process Impact

Finally we present the structure of a peer review process description by Process Impact (a private company in the field of software process improvement). This quality approach includes an overview of peer review process and work aids which serve as input tools for the process, a guidance on risk assessment including risk criteria, possible participants of a peer review and then three types of peer review procedures: inspection, walkthrough and passaround.

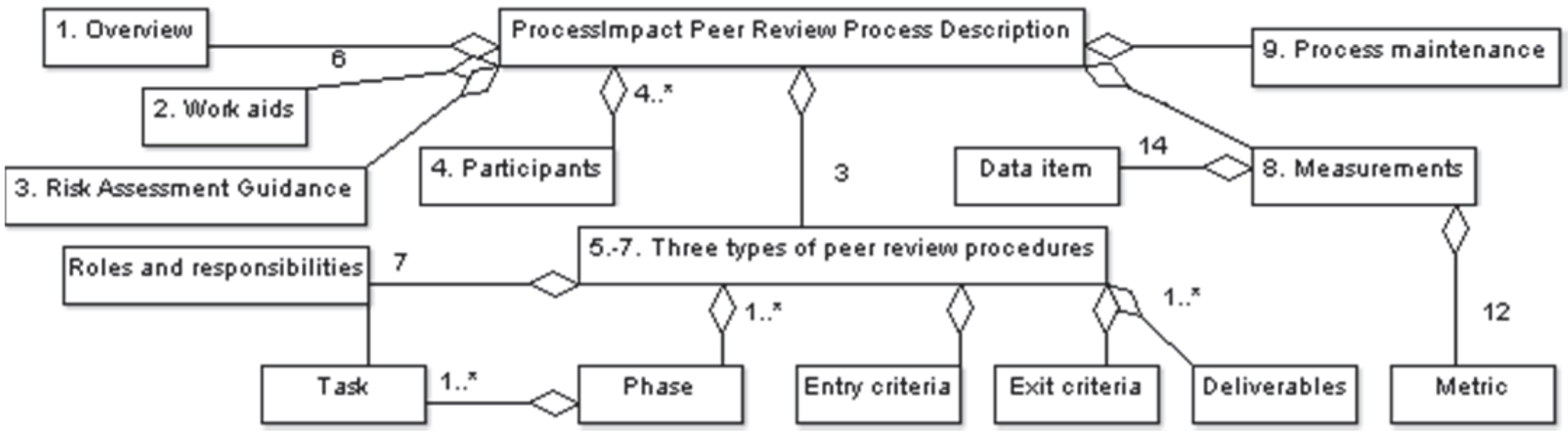

Figure 9 – The structure of Process Impact peer review process description

All these three procedures include the same quality approach elements: phases and tasks, entry and exit criteria and deliverables. Tasks are associated with responsibilities. At the end there is guidance on measurements, providing data items and metrics and also a short suggestion on process maintenance. Figure 9 shows an overview of the structure described.

**Quality approach elements**

Despite that the peer review description from Process Impact is the shortest document from all we discussed, it defines the peer review process in a structured way. Many of its elements can be mapped to basic process elements. E.g. work aid and deliverable can be mapped to artifact, procedure and phase can be mapped to (sub)process, task can be mapped to activity process element, role and responsibility can be mapped to role and responsibility process elements whereas entry and exit criteria are mappable to process entry and exit criteria. It also defines the phases and the order of tasks so that sequences and loops can easily be discovered. Further quality approach elements such as metrics or data items can also be used to enhance the process.

Table 15 – Elements of Process Impact peer review description

| Quality approach element | Description | Process element |
|---|---|---|
| Work Aid | Work aid describes a template document useful in performing the peer review. | Artifact |
| Risk assessment guidance | The risk assessment guidance gives an overview of possible risks of a peer review and provides guidance for typically occurred issues. | – |
| Participants (project role-work product association) | The peer review process can be applied in various context, therefore participants may vary upon the work product reviewed. In the participants section of this document project roles are associated with work products. | – |
| Procedure | In this quality approach, procedures describe 3 different types of peer reviews: inspection, walkthrough and passaround. Procedures contain various quality approach elements which are needed in describing the peer review (participants, role, responsibility, entry-exit criteria, phase, task, sequence of task etc.). | Sub(Process) |
| Role | Role is not explicitly defined in this quality approach, but the usage of this element reflects the definition of Wikipedia: a set of connected behaviours, rights and obligations as conceptualised by actors in a situation and it is attached to a set of responsibilities. | Role |
| Responsibility | Role is not explicitly defined in this quality approach, but the usage of this element reflects the definition of Wikipedia: a duty, obligation or liability for which someone is held accountable. In this quality approach responsibilities are connected to roles. | Responsibility |
| Entry criteria | Entry criteria describe things needed before the process starts. | Entry criteria |
| Phase | A phase contains a set of tasks which should be performed together. A procedure can contain more phases. | (Sub)Process |
| Task | A task describes an activity during a phase. | Activity |
| Sequence of tasks | Tasks are numbered and numbers give the sequence of tasks. | Sequence of activities |
| Task-role association | Tasks are associated to roles, denoting the required person for performing that task. | – |

| Quality approach element | Description | Process element |
|---|---|---|
| Deliverable | A deliverable is an output of a procedure. | Artifact |
| Exit criteria | Entry criteria describe things needed before the process ends. | Exit criteria |
| Verification | Verification describes things should be checked at the end of the procedure. | – |
| Data item | Data items are used to calculate the process metrics. Such items are e.g. planning effort or major defects found. | – |
| Metric | Metrics are calculated from inspection data items and help generating various reports which can give insight into the process health. | – |

**Refinements on the process element list:** at this step there were no quality approach elements discovered on which the list of process elements could be refined.

## 4.5. Refinements on process elements after identifying quality approach elements

After identifying quality approach elements based on literature (chapter 4.3), in the second iteration (chapter 4.4) we have refined the set of process elements based on quality approach elements. According to our research approach (introduced in chapter 2) the third refinement of the process elements will be discussed after using them in practice (chapter 8). Here we summarize process elements after the second iteration (RS3-i2: Refinements on process elements).

**Changes in the process elements:**

Many of quality approaches contain elements which can be mapped to specific process elements, therefore the following process elements were added to the process element list in chapter 4.4: process name, process description, parent, entry and exit criteria.

Since no quality approach elements mappable to "resource" element were identified in chapter 4.4, the "resource" process element was removed.

Table 16 – Description of process elements after the second iteration

| Process Element name | Process Element Description |
|---|---|
| **Process name** | The process name conveys the scope of the process as a whole. |
| **Parent** | The *parent* element denotes the possible parent(s) of an activity or a process. If the parent element is empty or equals to 0 the process has no parent, therefore it is a process, otherwise it is a subprocess. Activities always have parent process. |
| **Process description** | The process description is a generic introduction to the process which can include elements such as:<br>- introduction/purpose/goal/objective of the process,<br>- a list of activities of the process,<br>- textual description of the process. |

| Process Element name | Process Element Description |
|---|---|
| **Activity** | An *activity* is a process element which is needed to be performed by a resource on one or more artifact(s) to create value. The activity is atomic. The activity has a name and description.<br>The activity description is a generic introduction to the process which can include elements such as:<br>- introduction/purpose/goal/objective of the activity,<br>- textual description of the activity. |
| **Role, responsibility** | A *role* (e.g. project manager, database server) can be performed by a human or machine (e.g. one person, a computer) resource.<br>*Roles* can have assigned *responsibilities* (e.g. planning and monitoring the project, or serving database queries). |
| **Data object** | A *data object* (or *artefact*) represents an input or output to a (sub)process or activity. |
| **Element relation** | A process element can have *relation* with other elements. Typical relations among sub(processes) and activities can be e.g. *sequence* or *loop*. An artifact can be the *output* of an activity and can serve as an *input* to the activity. An element relation can have a name and a description. |
| **Entry / exit criteria** | *Entry criteria* – Describes the criteria to be met before the process can begin.<br>*Exit criteria* – Describes the criteria to be met before the process can be considered complete. |

A *(sub)process* can contain any type of process elements and also subprocesses. The only difference between subprocess and process is that the process has no container parent process while the subprocess always has, therefore we only distinguish process and subprocess by parent. Processes can have elements which are summarized in Table 16.

It is important to emphasize that the process element list is not a formal process description or process modeling language; it provides rather a basis for mapping. Identification and development of a process description/modeling language would need more research and it is out of scope of this thesis.

## 4.6. Options and limitations of mapping quality approaches to processes

In this chapter we summarize options and limitations of mapping quality approach elements to process elements (RS5 Identify options and limitations of mapping quality approaches to processes).

In the foregoing we have presented the structure of 7 different quality approaches. In each of them we recognized quality approach elements and we found that many of them can be mapped to different process elements. We also found quality approach elements which cannot be directly mapped to process elements.

We found several coincidences among quality approaches elements to process elements, e.g. shall statements, activities, tasks, practices and other quality approach elements could be mapped to activity process elements, or outcomes and example work products found in quality approaches coincide to process artifacts.

Table 17 – Process elements mapped to quality approach elements

| Process Element / Quality approach elements | ISO 9001:2008, 90003:2004 | ISO/IEC/IEEE 12207-2008 | CMMI v1.3 | ITIL v3 | IEEE 1028:2008 | Process Impact peer review process description |
|---|---|---|---|---|---|---|
| **Process name** | Chapter titles | Process name | Process area name | Process (name), scope | Peer review procedure names | Procedure name |
| **Parent** | Requirements and guidelines in sentences (non-directly mappable) | Process category | Process area, goal, practice, subpractice relations, related areas | Inter-process interfaces, process flowcharts, | Procedure phases | Process, Procedure, phase relations |
| **Process description** | requirements and guidelines in sentences (non directly mappable) | Process purpose | Introductory notes, Purpose statement | Purpose/ goal/objective | Introduction to review | Document overview |
| **Activity** | | Activity, task | goal, practice, subpractice | Process activities, methods and techniques | Procedure description | Task |
| **Role, responsibility** | | – | – | – | Responsibility | Role, responsibility |
| **Data object** | | Outcome | example work product | – | Input, output | Work aid, deliverable |
| **Element relation** | | – | – | flowcharts | | Sequence of tasks |
| **Entry / exit criteria** | | – | – | – | Entry/exit criteria | Entry/exit criteria |

Table 17 gives an overview of process elements identified in chapter 4.3, refined in chapter 4.4 and mapped to quality approach elements identified in chapter 4.4. We make a new refinement at this point: the resource element is an element for which we did not find corresponding quality approach elements (see Table 17) in the quality approach elements analysed. Thus, in our final list of process elements we exclude this element.

## 4.7. Limitations

Here we discuss limitations of the work presented in this chapter.

**Identifying and refinement of process elements** – In identifying a set of process elements five different literature sources were taken into account. Based on these literature sources a set of process elements was identified. A possible limitation of this result could be that further elements could be identified by using more literature sources. In order to overcome this problem, we performed this step in three iterations. In the second iteration we refined process elements based on the elements of a set of well-chosen quality approaches. In the third iteration we will refine process elements based on case study experiences.

**Identifying quality approach elements** – A limitation of this step is that we analysed 7 quality approaches from hundreds available. However, this covered a high variety of quality approaches. The most important types were discussed including the structured CMMI and the unstructured ISO 9001 approach.

**Mapping quality approach elements to process elements** – We showed that elements of 7 different quality approaches can be mapped to process elements. Since we covered a high variety of quality approaches and used multiple literature sources for identifying process elements, furthermore we refined the process elements based on quality approach elements, we expect that other quality approaches are also mappable to processes.

## 4.8. Conclusion

In this chapter we addressed three research questions:

Q3 What is a suitable set of process elements to base the unified process on?

Q4 What are the elements of quality approaches?

Q5 Which characteristics of (a number of) current quality approaches further and/or hinder mapping of these approaches to a process model?

In order to answer Q3-Q5 first we discussed fundamentals of quality approach mapping, identified process elements based on literature (RS3-i1), analysed elements of 7 quality approaches (RS4), refined process elements based on quality approach elements (RS3-i2) and discussed options and limitations of mapping quality approach elements to process elements (RS5).

We showed that quality approaches have similar elements to process elements and these can be mapped to each other. A number of carefully chosen quality approaches were analysed and their elements were mapped to process elements. We showed that in order to map quality approaches, their structure should be understood and elements and element instances need to be identified.

For example KPIs, CSFs and risks contained in ITIL or metrics provided by Process Impact to measure process performance are definitely useful in building processes. These elements could also be used in a mapping.

There are other quality approach elements present in different quality approaches which are not important from our point of view e.g. page numbers, authors, page break, formatting elements etc. Since these quality approach elements are not useful for building processes, their discussion is not included into this study.

As it can be seen, many of quality approach elements found in different quality approaches are relevant and can be used in building/enhancing organizational processes. We also conclude that there is no type level solution and a mapping should always be performed on instance level because it is not trivial to which process element instance to map the quality approach element instances identified. This might be supported by automatic recognition of similarities between the text of quality approach element instances and process element instances, but this still cannot provide full type level mapping, because automatic mapping of texts should be confirmed by users.

In the next chapters we present the Process Based Unification which will be based on notions and results presented in this chapter.

# 5. PROCESS BASED UNIFICATION[3]

Looking at the literature we have seen that the current multi-model initiatives such as quality approach harmonization, quality approach integration and respectively the quality approach mapping do not provide a solution for the multi-model problem and do not fully satisfy the MSPI criteria (see chapter 3).
In chapter 4 we discussed options and limitations of mapping quality approaches to processes. This included the identification of both elements of processes and elements of quality approaches and an analysis of their mapping. We showed that a mapping between quality approaches and processes is possible. On this basis we can design a PBU process.
In 5.1 the research approach and components of PBU framework are presented, in 5.2 the rationale behind the PBU process is presented (RS6-1), 5.3 describes the process representation format used in representing the PBU process (RS6-2), while in 5.4 a possible PBU process is described (RS6-3). Limitations are included in 5.5 and 5.6 concludes the work presented in this chapter.

## 5.1. Research approach

In this chapter we propose a multi-model solution, which we call Process Based Unification (or PBU) framework. The core concept of Process Based Unification is the unification of quality approaches by mapping their elements to elements of one single (unified) process. Later we will show that the PBU framework is not just a multi-model initiative, but it is indeed a multi-model solution which satisfies the MSPI criteria.
Accordingly, in this chapter the main question addressed is the following:
*Q6 How can mapping of a set of quality approaches to a unified process model take place?*

---

[3] Early concepts of Process Based Unification have been published in (Kelemen, Kusters, Trienekens, & Balla, 2009).

In order to answer research question 6 two research steps will be performed: “RS6 Design a PBU process”.

Our proposed multi-model solution, the PBU is framework which includes the following components:

**The PBU concept** – Our hypothesis is that mapping quality approaches to a process can provide a multi-model solution. The task can be divided to decomposing quality approaches to quality approach element instances and then mapping quality approach element instances to process elements. The concept of mapping quality approaches to process is called the concept of Process Based Unification or PBU concept.

**The unified process** – The key PBU result is a single, unified process, to which quality approaches are mapped. This resulting process is called the unified process. In order to ensure that the unified process conforms to multiple quality approaches, quality approaches are decomposed and their element instances are mapped to process elements. A PBU unified process is both usable in practice (e.g. at a software company) and also supports the simultaneous usage of multiple quality approaches. The unified process hides differences in terminology, structure, granularity, content, size and complexity of quality approaches – so that users of PBU unified processes do not need to spend resources on handling differences, they can rather focus on using the unified process.

(In our terminology the unified process is a PBU result, and since the PBU result is a multi-model result, the unified process is also a multi-model result.)

**A PBU process** – In order to guide the practical implementation of the PBU concept, we provide a process. This process is called PBU process. Our PBU process relies on the PBU concept. The goal of the proposed PBU process is to provide practical guidance for the implementation of the PBU concept.

Our proposed PBU process is one possible process of implementing the PBU concept, and it is developed taking into account the logical and functional relations of the needed activities. This means that such a process could be described in various ways: e.g. organizing activities into various subprocesses, performing non-dependent activities in various orders, or describing different activities on different levels of detail. For example the activity of selecting a process modeling language is related to the process creation activity: the selection of a process modeling language can be placed anywhere in the process flow, but should be placed before creating a process model.

If one wants to define a process, process activities need to be placed in the process flow to explicit positions. In the case mentioned, the activity can be placed to various places in the process flow, but the decision on where to put the selection of process modeling language activity cannot be avoided. Further examples of multiple options in creating a process representation are: the detailedness of textual process description or the decomposition of processes to subprocesses and activities. Many process modeling languages offer a theoretically unlimited number of process levels, so there are no technical limitations for the detailedness of a process representation. Thus the detailedness of a process model is up to the author’s decision as well as the length of the textual activity descriptions.

Taking these examples, it is clearly visible that in representing a process, choices should be made due to practical, logical and functional aspects. As a consequence of multiple possibilities in representing a process, the resulting process will be one of probably many options. This is also true in the case of proposed PBU process: our PBU process is one possible process for implementing the PBU concept and obviously not the only and ultimate one. However, we think it is still important to make the concept operational and get a clear understating on the activities needed during the process based unification. This is the reason we defined a PBU process.

We tried to define a PBU process as simple and as logical as we could at this moment. Later we will use this PBU process in a case study and we will refine it based on practical results of the case study.

In order to define a PBU process, first we need to discuss the main rationales behind the proposed PBU process, then a representation format should be chosen and finally the process itself should be described in the chosen format. The resulting process should also be validated. Accordingly, RS6 can be broken down to the following operational steps:

RS6-1: Discussing rationale of the subprocesses of a PBU process;

RS6-2: Identifying textual and graphical process representation formats for PBU process;

RS6-3: Describing a PBU process based on rationale discussed in RS6-1 and process representation identified in step RS6-2;

## 5.2. Rationale of the subprocesses of a PBU process

The most important activities of the PBU process are the identification of quality approach element instances, mapping them to the (unified) process and making refinements on the (unified) process based on the mapping. These activities are in fact the core of process based unification.

In chapter 4 we identified elements of quality approaches on type level. For example in chapter 4 the “role” element was identified in several quality approaches and we argued that this can be mapped to the role process element. We concluded that mapping should be performed on instance level. According to chapter 4, we include instance level identification and mapping of quality approach elements into the PBU process. Instances of the “role” quality approach element will be for example “project manager” or “developer”. Quality approach element instances will be mapped to process element instances.

We organized the PBU process into four subprocesses. We made this distinction first in order to separate it functionally/logically and second to ease its readability. The decomposition of the subprocesses is as follows:

1. selection (of processes, quality approaches and process representation),
2. analysis of quality approaches,
3. deriving process from quality approaches and
4. validation.

In the followings we present the main concepts behind these subprocesses.

### 5.2.1. Selection (of processes, quality approaches and process representation)

**Goal:** the goals of this subprocess are: selection of processes to be improved, selection of quality approaches to be used in the improvement and selection of process representation.

**Rationale of activities of the "Selection (of processes, quality approaches and process representation)" subprocess:**

1. Selection of processes:

In order to start a process improvement project in an organization, processes should be selected for improvement. For selection the actual status of processes (e.g. strengths and weaknesses) need to be known. One common way to get information about processes is process assessment; therefore, the first activity of the „Selection (of processes, quality approaches and process representation)" subprocess is the assessment of the current situation. During the assessment, process strengths and weaknesses are discovered. This assessment provides the basis for the second activity, which is the selection of a process for improvement. The assessment can help in this selection, e.g. in case the assessment reveals that a process is weak, it may be improved. Of course, this decision should be made by the SEPG team, agreed by other relevant stakeholders such as the management and potential process performers and the process owner.

2. Selection of quality approaches:

After selecting a process to be improved, the basis of the improvement should be defined. Since the PBU process tries to improve processes by using multiple quality approaches, more quality approaches need to be selected. Therefore, the second activity of „Selection (of processes, quality approaches and process representation)" is the identification of the quality approach set to be used in improving the selected process.

3. Selection of a process representation format:

As we have shown in chapters 4.3 and 4.5, the elements of processes can vary depending on view the on processes and also depending on the format of the graphical and textual representations. Therefore, before the mapping, identification of graphical and textual process representation should be done. For identifying the textual and graphical process representation formats, results of chapters 4.3, 4.5, 4.6, 5.3, and 6.5.3 can be used.

It also could happen that an organization already uses a textual and graphical representation for processes. In this case, there is no need for a new representation format (this should be decided by the SEPG team).

After having a selected process for improvement and the potential quality approaches for improving that process, as well as a process representation format, a next subprocess can be started: the "Analysis of quality approaches".

### 5.2.2. Analysis of quality approaches

**Goal:** the goal of "Analysis of quality approaches" subprocess is to analyse quality approach elements and to identify quality approach element instances for mapping.

In order to improve processes by using multiple quality approaches, the information included in the quality approaches need to be mapped to the process. This can be done by identifying quality approach elements and by mapping these elements to process elements. At the beginning of this thesis our main hypothesis was that this mapping is possible (chapter 2). Later, in chapter 4.4 we have shown that elements of process oriented quality approaches can be mapped to process elements. We also have shown that the mapping can be done first by identifying quality approach elements and then by mapping these elements to process elements.

Analogically, in order to serve as a basis for the mapping, elements of quality approaches will be identified. Therefore, the goal of the „Analysis of quality approaches" subprocess is to iteratively analyse the selected quality approaches (one by one), and to identify their elements and element instances.

**Rationale of activities of the „Analysis of quality approaches" subprocess:**

The „Analysis of quality approaches" subprocess should include 1. the selection of a quality approach, 2. the analysis of main characteristics of the selected quality approach, and 3. the identification of directly and indirectly mappable quality approach elements and element instances.

After the selection (activity 1), important terms, chapters including explanations on the quality approach (e.g. the quality approach introduction, purpose, objectives, appendices), the structure of the quality approach should be analysed and understood in order to achieve a vision on the possible quality approach elements (activity 2). After selecting and getting an understanding of the quality approach, quality approach elements and element instances should be identified (activity 3). Quality approach elements and element instances will be used later, and will serve the basis for the mapping and process improvement at the "Deriving process from quality approaches" subprocess.

### 5.2.3. Deriving process from quality approaches

**Goal:** The goal of "Deriving process from quality approaches" subprocess is to map quality approach element instances to process and refine the process based on the mapping.

After analysing a quality approach and identifying the quality approach elements and element instances, the quality approach element instances will be used in improving or creating the targeted process.

We have shown that quality approach elements can be mapped to process elements (chapter 4.6) and we assume this can serve as a basis for process refinement/improvement. We also think if the quality approaches are complete enough in describing process-related elements (e.g. requirements or guidelines for a process), even whole processes could be tailored from quality approach element instances.

Since the identification of quality approach elements and quality approach element instances are performed in the previous „Analysis of quality approaches" subprocess, the proposed mapping can be done at this point.

**Rationale of activities of the „Deriving process from quality approaches" subprocess:**

After having analysed the quality approach elements and quality approach element instances were identified, the mapping of quality approach element instances to process element instances can be started. During the mapping various problems can be encountered, some of these problems are mentioned in chapters 4.2 and 4.4, and will be discussed in more detail through a real case in chapter 7.

Next after the mapping, the refinement of the process should also be performed based on the information in the quality approach element instances. Refinements mean that process elements can be modified, e.g. insertion of a new activity, modification of a role, deletion of an artifact, modification of the process flow, simplification or extension of subprocesses etc. All these modification should be done with the goal of improvement, with the involvement of the process owner and should be based on the information contained in quality approach element instance mapped to process element instances.

In case a process was modified based on a mapping, the previous mapping should also be checked for consistency.

In case if the process does not exist, then a new process should be created based on the quality approach elements, using the previously selected representation formats. Of course it can happen that not all the information needed is contained in the selected quality approaches and further sources of information are needed. (E.g. due to the fact that some of the quality approaches do not define the process flow, this information might be missing. If the process flow information is missing, then a process flow could be defined based on discussions and previous experiences, or involving process experts and consultants.) Later, the whole improvement needs to be validated by the relevant process stakeholders.

If there are more quality approaches selected for improving the targeted process, then the „Analysis of quality approaches" and "Deriving process from quality approaches" subprocesses should be repeated iteratively.

### 5.2.4. Validation

**Goal:** The goal of the "Validation" subprocess is to validate the results of this PBU process and achieve commitment to the process.

A process improvement cannot end without stakeholder's (e.g. process owner, process performers) validation. They should be informed and asked to provide feedback on the changes. Moreover, refinements based on their feedback should also be performed if needed.

For validating the result of a process improvement, various techniques can be used, e.g. different types of reviews, questionnaires, interviews, presentations and feedback collecting forms.

**Rationale of activities of the „Validation" subprocess:**

Choosing a validation technique: In order to validate results validation technique should be chosen. Brainstorming, questionnaires and walk-throughs among others can be used for collecting user feedback (validating results).

Applying validation: validation activities should be planned, performed and results should be collected. Results collected in the apply validation activity will be subjected to an analysis.
Analysing validation results: After collecting the feedback, the results of the feedback should be analysed. The analysis should show if refinements are needed (e.g. majority of people cannot accept the process in the current form then refinement is needed). Refinements should be made based on the feedback collected. The analysis and new refinements are performed by SEPG team and process owner.
If there are more processes which need to be refined, then the same PBU process should be repeated starting from the process selection activity in the „Selection (of processes, quality approaches and process representation)" subprocess.

### 5.2.5. Loops in the proposed PBU process

Figure 10 shows an overview of the subprocesses and loops of a possible PBU process. According to the previous discussions we define an iterative PBU process in which two main loops can happen:
L1: In case the subprocess "Deriving process from quality approaches" was performed for a quality approach, then the "Analysis of quality approach" can be performed for the next quality approach.
L2: After finishing the refinement of one process, the next process and the quality approaches needed can be selected. In order to keep the process representations consistent, the process representation needs to be selected only for the first time.

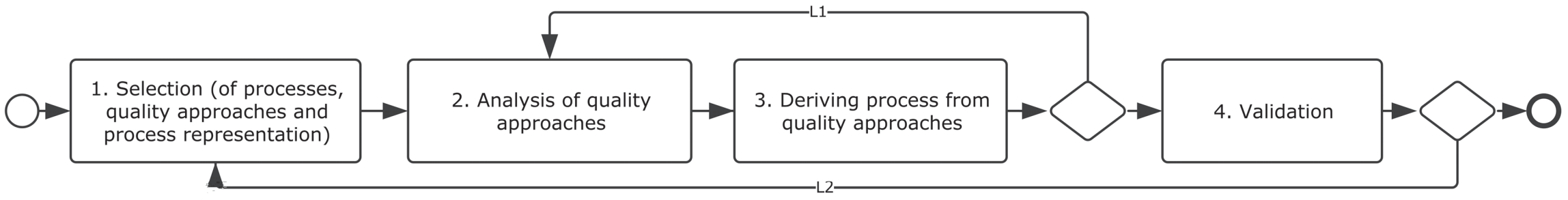

Figure 10 – Overview of subprocesses and loops in a PBU process

## 5.3. The representation format of the PBU process

In order to represent the PBU process in an easily readable and consistent way we decided to represent it both textually and graphically. The graphical representation will help to follow the process flow and to understand the order of tasks and decision points, the textual description will help to get more insight into specific subprocesses, activities, roles and the most important artifacts. Regarding the graphical representation of processes we came to the conclusion that BPMN can be an easily understandable process modeling language (see chapter 6.5.3), we choose it for representing the PBU process as well.
We agree that many different textual representations can represent processes consistently and clearly, but a choice should be made. Therefore, we propose the following format for textual process descriptions using process elements identified in 4.6:
**A table for describing (sub)processes, including the following rows:**
1. "Process ID" – the unique identifier of the process;

2. “Parent ID” –the process to which the subprocess belongs; The parent ID is 0 if the activity belongs to the main process.
3. “Process name” – the name of the process;
4. “Process description” – purpose and goals to achieve by performing the process.
5. “Entry criteria” – criteria need to be satisfied before starting the process.
6. “Inputs” – data objects that serve as input for the process.
7. “Outputs” – data objects that serve as output for the process.
8. “Exit criteria” – criteria needed to be satisfied before exiting the process.

**A table for describing activities, including the following rows:**

1. “Activity ID” – the unique identifier of the activity.
2. “Parent ID” – for identifying the process to which the activity belongs;
3. “Activity name” – the name of the activity;
4. “Activity description” – textual description of the activity;
5. “Inputs” – data objects that serve as input for the process.
6. “Outputs” – data objects that serve as output for the process.
7. “Roles/responsibilities” – roles and responsibilities for performing the activity.

**A table for describing data objects, including the following rows:**

1. “Data object ID” – the unique identifier of the artifact.
2. “Data object name” – the name of the artifact.
3. “Data object description” – textual description of the artifact.

**A table for describing roles and responsibilities, including the following rows:**

1. “Role ID” – the unique identifier of the role.
2. “Role name” – the (name of the) role.
3. “Responsibilities” – textual description of the responsibilities of the role.

The PBU process representation is a combination of the above.

## 5.4. A PBU process

In this chapter the PBU process is presented according to the main concepts presented in chapter 5.2 and the textual and graphical process representation format chosen in 5.3. Figure 11 represents the main PBU process flow using BPMN including the subprocesses: “Selection (of processes, quality approaches and process representation)”, “Analysis of quality approaches”, “Deriving process from quality approaches” and “Validation” and refinement, with their main activities.

These subprocesses are detailed in chapters 5.4.1, 5.4.1.3, 5.4.3 and 5.4.4. In order to make the process clear enough, in certain cases (subprocess 1.3 and 2.3) additional process models (Figure 12 and Figure 13 respectively) are presented. Main loops of the process were previously described in chapter 5.2.5.

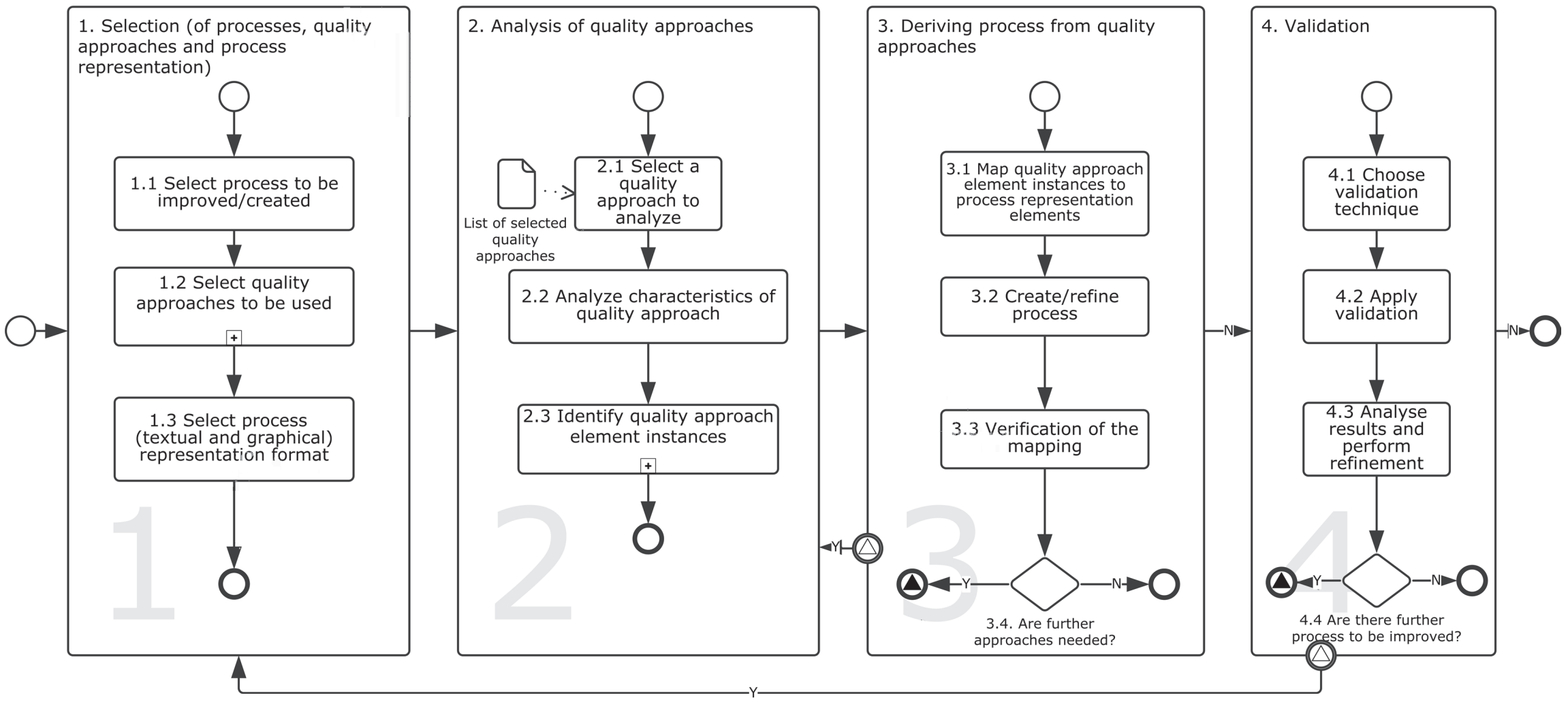

Figure 11 – A PBU process

## 5.4.1. Selection (of processes, quality approaches and process representation)

| | |
|---|---|
| **Process ID** | 1 |
| **Parent ID** | 0 (PBU process) |
| **Process name** | Selection (of processes, quality approaches and process representation) |
| **Process description** | The purpose of the „Selection (of processes, quality approaches and process representation)” subprocess is to:<br>- select processes to be improved/created,<br>- select quality approaches to be used,<br>- select process (textual and graphical) representation format. |
| **Entry criteria** | - The management has decided to support (multi-model) process improvement<br>- Resources are available |
| **Inputs** | - Formal decision of starting a process improvement project,<br>- assessment,<br>- a set of quality approaches. |
| **Outputs** | - List of processes to be improved,<br>- list of quality approaches used in the improvement. |
| **Exit criteria** | Improvement areas and quality approaches have been selected |
| **Roles/ responsibilities** | The leader of the Software Engineering Process Group |

### 5.4.1.1. **Select processes to be improved/created**

| | |
|---|---|
| **Activity ID** | 1.1 |
| **Parent ID** | 1 Selection (of processes, quality approaches and process representation) |
| **Activity name** | Select process to be improved/created |
| **Activity description** | The selection of processes depends on the goals of the organization. The SEPG team assesses the current situation of processes. For performing the assessment, various assessment methods can be used, e.g. the widely used in software process assessment methods SPICE Process Assessment Model or SCAMPI. Based on the result of assessment of processes and the resources available for a multi-model |

| Activity ID | 1.1 |
|---|---|
| | process improvement project, processes are selected for improvement. |
| Inputs | Process level (theoretical/organizational/project),<br>Process, organizational, project goals,<br>Possible process strengths and weaknesses |
| Outputs | Selected processes for improvement or creation |
| Roles/ responsibili-ties | SEPG team |

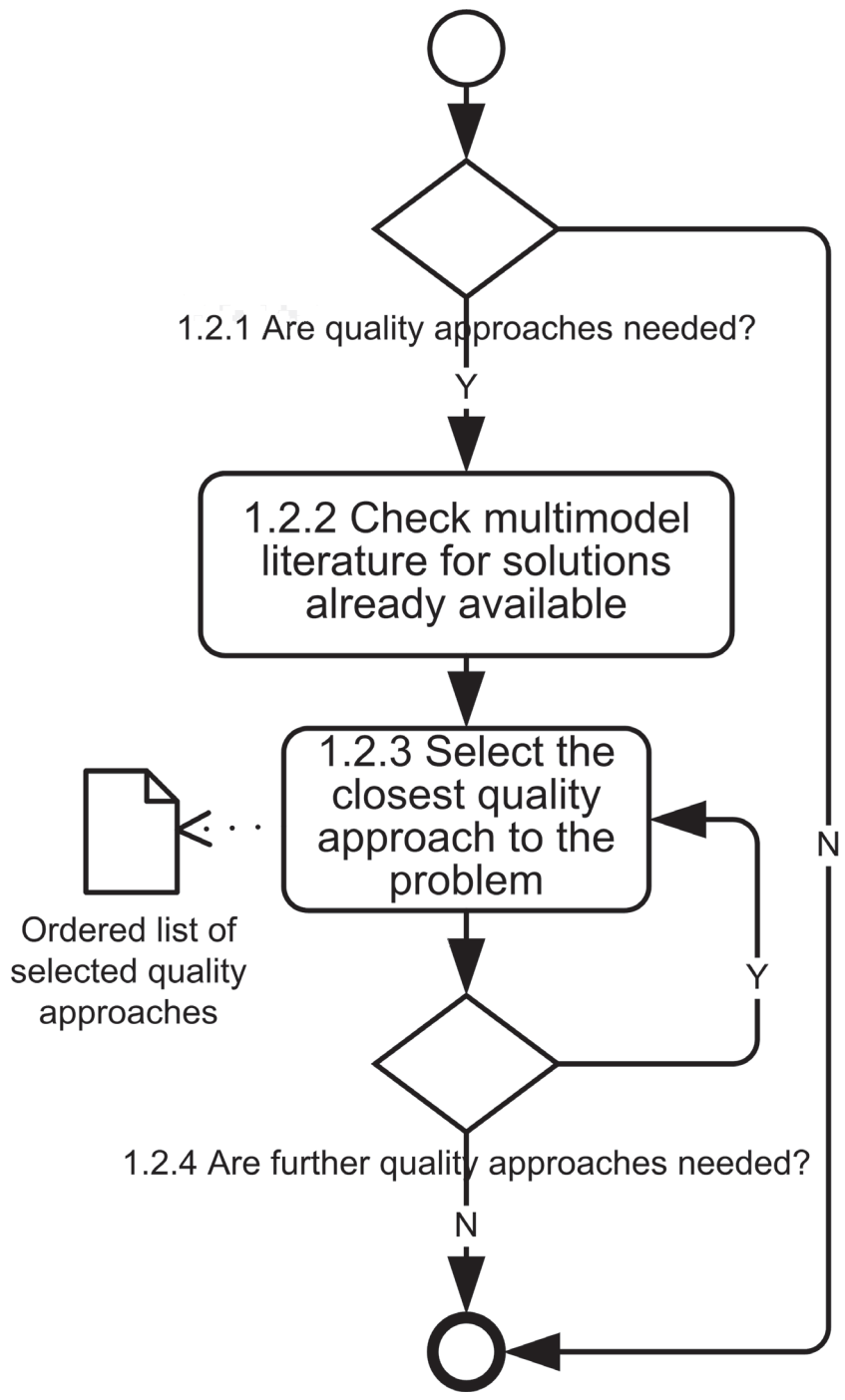

Figure 12 – Subprocess 1.2: Select quality approaches to be used

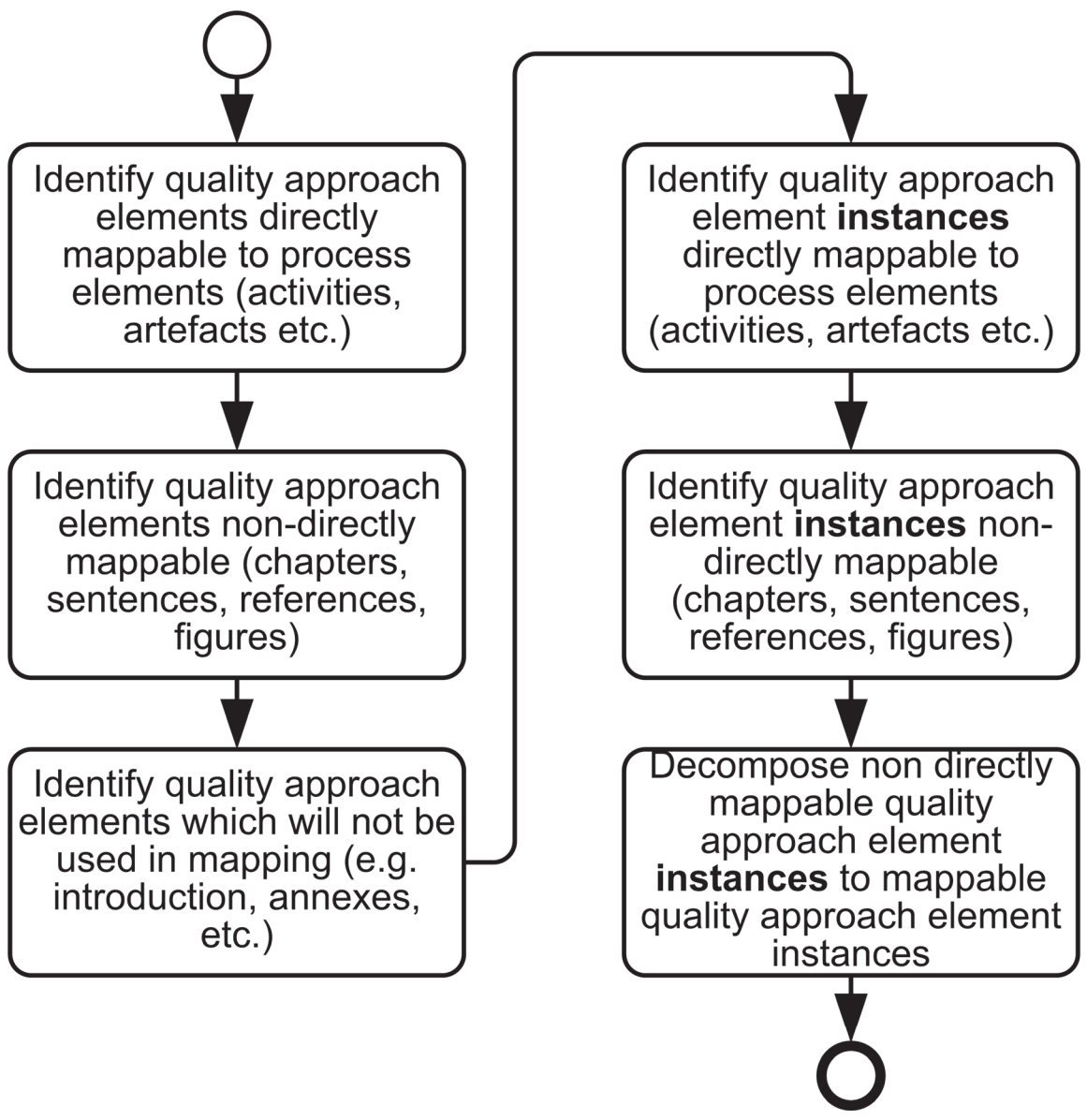

Figure 13 – Subprocess 2.3: Identify quality approach element instances

### 5.4.1.2. **Select quality approaches to be used**

| Process ID | 1.2 |
|---|---|
| Parent ID | 1 Selection (of processes, quality approaches and process representation) |
| Process name | Select quality approaches to be used |
| Process description | The purpose of this subprocess is to select quality approaches to be used. In order to select the quality approaches, the following activities should be performed:<br>- defining if any quality approach is needed,<br>- analysing the current multi-model literature for possible solutions already available,<br>- selecting quality approaches which have the process or process area which needs to be improved in their scope,<br>- identifying and handling process-related terms used in the quality approaches.<br>In selecting the quality approaches, classification frameworks can be used e.g. QMIM or others (Balla et al., 2001; Halvorsen & Conradi, 2001; M. C. Paulk, |

| Process ID | 1.2 |
|---|---|
| | 2008). Classifications will help to decide what kind of quality approaches are needed for the current situation.<br>Activities of this process can be seen in Figure 12. |
| **Entry criteria** | N/A |
| **Inputs** | - Quality approach directory/database/list,<br>- quality approach classification frameworks. |
| **Outputs** | Ordered list of selected quality approaches |
| **Exit criteria** | N/A |
| **Roles/ responsibilities** | SEPG team |

#### 5.4.1.3. **Select process (textual and graphical) representation format**

| Activity ID | 1.3 |
|---|---|
| **Parent ID** | 1 Selection (of processes, quality approaches and process representation) |
| **Activity name** | Select process (textual and graphical) representation format |
| **Activity description** | In order to represent the process select a textual format and a modeling language. A textual representation of a process can be e.g. the one proposed in chapter 5.3 and the process modeling language can be e.g. one of those analysed in chapter 6.5.3. The choice should be aligned with the organizational goals and requirements against such a representation and with possible other formats used at the organization.<br>If the selected process already has a representation (e.g. a textual description and/or a process model) then the representation format should be taken into account as a possible representation format. |
| **Inputs** | - Possible textual process representation formats,<br>- process modeling languages,<br>- organizational requirements. |
| **Outputs** | Process representation format for textual descriptions and process models |
| **Roles/ responsibilities** | SEPG team |

### 5.4.2. Analysis of quality approaches

| Process ID | 2 |
|---|---|
| **Parent ID** | 0 (PBU process) |
| **Name** | Analysis of quality approaches |
| **Process description** | The purpose of the „Analysis of quality approaches” subprocess is to:<br>- understand the purpose, terms, structure and other characteristics of quality approaches,<br>- identify quality approach elements and instances of these elements which could be mapped to process element. |
| **Entry criteria** | - The „Selection (of processes, quality approaches and process representation)” subprocess was performed |
| **Inputs** | - List of processes to be improved,<br>- list of quality approaches. |
| **Outputs** | - A quality approach selected,<br>- elements and elements instances of quality approaches identified. |
| **Exit criteria** | Characteristics of a selected quality approach has been understood and elements |

| Process ID | 2 |
|---|---|
| | were identified for mapping |

### 5.4.2.1. **Select a quality approach to analyse**

| Activity ID | 2.1 |
|---|---|
| **Parent ID** | 2 Analysis of quality approaches |
| **Activity name** | Select a quality approach to analyse |
| **Activity description** | From the ordered list of quality approaches to be used for refining/creating processes (identified in the 1.2 „Selection (of processes, quality approaches and process representation)" subprocess) select the first quality approach to be analysed. |
| **Inputs** | Ordered list of selected quality approaches |
| **Outputs** | Quality approach to be analyzed |
| **Roles/ responsibilities** | SEPG team |

### 5.4.2.2. **Analyse characteristics of quality approach**

| Activity ID | 2.2 |
|---|---|
| **Parent ID** | 2 Analysis of quality approaches |
| **Activity name** | Analyse characteristics of quality approach |
| **Activity description** | During this activity an analysis of main characteristics of quality approaches should be performed. At defining the quality approach set, classification frameworks were used, e.g. QMIM or others (Balla et al., 2001; Halvorsen & Conradi, 2001; M. C. Paulk, 2008). At this activity scope, terminology, granularity, structure and elements, content, size and complexity can be discovered. For doing this, result of chapter 4.4 can be used. |
| **Inputs** | - Selected quality approach,<br>- classification frameworks,<br>- elements and structure of quality approaches (4.4). |
| **Outputs** | Understanding characteristics of quality approach (e.g. scope, terminology, granularity, structure and elements, content, size and complexity) |
| **Roles/ responsibilities** | SEPG team |

### 5.4.2.3. **Identify quality approach element instances**

| Process ID | 2.3 |
|---|---|
| **Parent ID** | 2 Analysis of quality approaches |
| **Name** | Identify quality approach element instances |
| **Process description** | The purpose of this subprocess is to:<br>- identify quality approach elements directly mappable to process elements,<br>- identify quality approach elements non-directly mappable to process elements (these quality approach elements need to be processed manually)<br>- identify quality approach elements which will not be used in mapping (e.g. annexes, page numbers etc.),<br>- identify quality approach elements instances directly mappable to process elements (e.g. in case of CMMI, "specific practice" can be identified as a quality approach element, then instances of specific practice can be selected for improving/creating a process.), |

| Process ID | 2.3 |
|---|---|
| | - identify quality approach elements instances non-directly mappable to process elements,<br>- decompose non-directly mappable quality approach element instances to mappable quality approach element instances (e.g. in case of ISO 9001, chapters and sentences can be decomposed and elements of these can be identified as mappable element instances).<br>Activities of this process can be seen in Figure 13. |
| **Entry criteria** | SEPG team understands characteristics of the quality approach |
| **Inputs** | Quality approach |
| **Outputs** | - Directly and indirectly mappable quality approach elements,<br>- directly and indirectly mappable quality approach element instances,<br>- decomposition of indirectly mappable quality approach element instances. |
| **Exit criteria** | Quality approach element instances were identified |

## 5.4.3. Deriving process from quality approaches

| Process ID | 3 |
|---|---|
| **Parent ID** | 0 (PBU process) |
| **Name** | Deriving process from quality approaches |
| **Process description** | The purpose of this subprocess is to create or refine a selected process based on the quality approach element instances identified in the Analysis of quality approaches subprocess. In order to do this the following activities should be performed:<br>- mapping quality approach element instances to process element instances,<br>- selecting a process modeling language if it is not already in use,<br>- map quality approach element(s and their instances) to process modeling language element instances,<br>- create or refine the processes based on the quality approach element instances and their mapping to the process element instances. |
| **Entry criteria** | Quality approach element instances were identified |
| **Inputs** | - A selected process for improvement,<br>- quality approach element instances which are related to the process,<br>- process description and model (if it is already available). |
| **Outputs** | - Mapping of quality approach element instances to process element instances,<br>- process description and process model reflecting the refinements made based on the mapping to quality approach element instances. |
| **Exit criteria** | All the needed quality approaches were mapped to process elements and the process description and process model was created or modified accordingly |

### 5.4.3.1. Map quality approach element instances to process representation elements

| Activity ID | 3.1 |
|---|---|
| **Parent ID** | 3 Deriving process from quality approaches |
| **Activity name** | Map quality approach element instances to process representation elements |
| **Activity description** | In this activity the quality approach element instances identified in „Analysis of quality approaches” subprocess are mapped to process representation elements (to textual and process modeling language elements).<br>E.g. the CMMI-DEV v1.3 VER SP1.2 “Prepare for Peer reviews” quality approach element instance can be mapped to a subprocess element which can have a textual |

| Activity ID | 3.1 |
|---|---|
| | and a process modeling language representation as well. |
| Inputs | - Quality approach element instances,<br>- process modeling language,<br>- the textual process description format. |
| Outputs | A mapping of quality approach element instances to process representation elements |
| Roles/ responsibilities | SEPG team |

### 5.4.3.2. **Create/refine process**

| Activity ID | 3.2 |
|---|---|
| Parent ID | 3 Deriving process from quality approaches |
| Activity name | Create/refine process |
| Activity description | The purpose of this activity is to create a new or to refine an existing process model using the mapping of quality approach element instances to process representation elements. |
| Inputs | The mapping of quality approach element instances to process representation elements |
| Outputs | - Refined/created process model,<br>- refined/created textual representation. |
| Roles/ responsibilities | Process owner, SEPG team, Process stakeholders |

### 5.4.3.3. **Verification of the mapping**

| Activity ID | 3.3 |
|---|---|
| Parent ID | 3 Deriving process from quality approaches |
| Activity name | Verification of the mapping |
| Activity description | Within this activity after mapping a new quality approach to the Unified Process, and refining the Unified Process, a consistency check is performed. This consistency check ensures that previous mappings will be still consistent after the new mapping and process refinements. |
| Inputs | - The mapping of quality approach element instances to process representation elements,<br>- refinements of the selected process. |
| Outputs | Refined previous mappings (if needed) |
| Roles/ responsibilities | Process owner, SEPG team |

### 5.4.3.4. **Decision: Are further quality approaches needed?**

| Activity ID | 3.4 |
|---|---|
| Parent ID | 3 Deriving process from quality approaches |
| Activity name | Decision: Are further quality approaches needed? |
| Activity description | Based on the list of quality approaches and possible newly arisen needs it should be decided if further quality approaches are needed in refining the process.<br>Y: Start analysing the next quality approach<br>N: Continue to Validation |
| Inputs | List of quality approaches |
| **Activity ID** | 3.4 |
| **Outputs** | Y/N |
| **Roles/ responsibilities** | SEPG team |

## 5.4.4. Validation

| | |
|---|---|
| **Process ID** | 4 |
| **Parent ID** | 0 (PBU process) |
| **Name** | Validation |
| **Process description** | The purpose of the validation subprocess is to validate the process representation created/improved by using multiple quality approaches and to refine the process based on feedback obtained.<br>The activities of this process are:<br>- choose a validation technique,<br>- apply validation,<br>- analyse results and perform refinement. |
| **Entry criteria** | Process was created/refined using the PBU process |
| **Inputs** | The process created/refined using the PBU process |
| **Outputs** | Validated and refined process |
| **Exit criteria** | Stakeholders have understood the process and provided feedback. Refinements were made based on the feedback received. |

### 5.4.4.1. **Choose validation technique**

| | |
|---|---|
| **Activity ID** | 4.1 |
| **Parent ID** | 4 Validation |
| **Activity name** | Choose validation technique |
| **Activity description** | Choose the most appropriate validation technique which fits best with your company's goals. Validation can be achieved by various ways, e.g. expert peer reviews, a less formal walk-through to a formal inspection, interviews, using questionnaires or other techniques.<br>Formal validation of a process model makes it possible to check both the correctness and to establish that the representation of the process is clearly and unambiguously expressed. Validation techniques implemented in automatic process modeling tools and interactive simulation can also help to ensure the validity of the process model (e.g. the correct PML elements were used or the relations among them are correct). |
| **Inputs** | The unified process |
| **Outputs** | - Validation technique,<br>- stakeholders,<br>- validation plan. |
| **Roles/ responsibilities** | SEPG team, process stakeholders |

### 5.4.4.2. **Apply validation**

| | |
|---|---|
| **Activity ID** | 4.2 |
| **Parent ID** | 4 Validation |
| **Activity name** | Apply validation |
| **Activity ID** | 4.2 |
| **Activity description** | Within this activity the chosen validation technique(s) is performed. Feedback of stakeholders is collected. |
| **Inputs** | - Validation technique,<br>- stakeholders,<br>- validation plan. |
| **Outputs** | Feedback on the process |
| **Roles/ responsibilities** | SEPG team, process stakeholders |

#### 5.4.4.3. **Analyse results and perform refinement**

| | |
|---|---|
| **Activity ID** | 4.3 |
| **Parent ID** | 4 Validation |
| **Activity name** | Analyse results and perform refinement |
| **Activity description** | Within this activity the results obtained during validation are analysed and refinements of process are performed. |
| **Inputs** | Feedback results on the process |
| **Outputs** | - Results of analysis,<br>- refinements of process. |
| **Roles/ responsibilities** | Process owner, SEPG team |

#### 5.4.4.4. **Decision: Are there further processes to be improved?**

| | |
|---|---|
| **Activity ID** | 4.4 |
| **Parent ID** | 4 Validation |
| **Activity name** | Decision: Are there further processes to be improved? |
| **Activity description** | Y: in case if there is one or more further process to be improved then select the next process (activity 1.2)<br>N: end of the PBU process. |
| **Inputs** | List of processes to be improved |
| **Outputs** | Y/N |
| **Roles/ responsibilities** | SEPG team |

### 5.4.5. Roles and responsibilities

#### 5.4.5.1. **SEPG team leader**

| | |
|---|---|
| **Role ID** | R1 |
| **Role** | SEPG team leader |
| **Responsibilities** | The SEPG team leader is responsible for leading the SEPG team and main (multi-model) process improvement projects. |

#### 5.4.5.2. **SEPG team**

| | |
|---|---|
| **Role ID** | R2 |
| **Role** | SEPG team leader |
| **Responsibilities** | The SEPG team is responsible for performing the PBU process at the organization |

| Role ID | R2 |
|---|---|
| | including the selection of processes to be improved/created, selection of quality approaches, analysis of quality approaches, making process refinements in “Deriving process from quality approaches” and “Validation and refinement”subprocesses. |

#### 5.4.5.3. **Process stakeholder**

| Role ID | R3 |
|---|---|
| **Role** | Process stakeholder |
| **Responsibilities** | Process stakeholders are those people who perform the process at the organization. In case of any process improvement, their feedback should be collected, analysed and the process should be refined accordingly. It is also advisable to involve them into the “Deriving process from quality approaches” subprocess. |

#### 5.4.5.1. **Process owner**

| Role ID | R3 |
|---|---|
| **Role** | Process owner |
| **Responsibilities** | The process owner is the ultimate responsible for the process. The process owner has the authority and ability to make necessary changes on the process. |

### 5.4.6. Important data objects

#### 5.4.6.1. **Ordered list of selected quality approaches**

| Data object ID | A1 |
|---|---|
| **Data object name** | Ordered list of selected quality approaches |
| **Data object description** | Quality approaches which are selected for improving a selected process are placed in the ordered list of selected quality approaches. This list is ordered by the relevance of the quality approaches to a selected process, having the most relevant quality approach at the first place in the list. When a quality approach is analysed, the quality approach is taken out from the list. |

## 5.5. Limitations

Here we discuss limitations of each step we performed in chapter 5.

**Rationale of the subprocesses of a PBU process** – As (Polyvyanyy, 2012) states, a process can be described in many different ways. This means subprocesses can be organized and detailed in many different ways, however we tried to corroborate each subprocess of the proposed PBU process with a rationale.

**The representation format of PBU process** – Many different representation formats can be chosen for representing a process. We defined the representation format based on results of chapter 4 and the element set of BPMN. These provide a basis for representing the process both textually and graphically. In order to ensure the suitability of the representation format used, we will refine the process element list after performing a case study.

**A PBU process** – The main limitation of the PBU process at this stage is that it is theoretical, so it should be tested. This will be done in chapters 6-8 through a case study, recording all the deviations occurred and making refinements in chapter 10. We described the PBU process on high level. Another possible limitation is the level of detail on which the PBU process is described. In order to overcome this problem we included several subprocesses where it was necessary. Furthermore, the case study described in chapters 6-8 can also be used as a guide for performing a PBU process.

## 5.6. Conclusion

In this chapter we addressed the following research question:

*Q6 How can mapping of a set of quality approaches to a unified process model take place?*

In order to answer this question the research steps RS6 was identified in chapter 2: “RS6 Design a PBU process”.

Looking to the literature we have seen that the current multi-model initiatives such as quality approach harmonization, quality approach integration and respectively the quality approach mapping do not provide a solution for the multi-model problem and do not satisfy the multi-model criteria (see chapter 3).

In this chapter we identified the PBU framework which includes the following components: (1) **the PBU concept** –the concept of mapping quality approaches to process is called the concept of Process Based Unification or PBU concept; (2) **a PBU process** – the goal of the proposed PBU process is to provide practical guidance for the implementation of the PBU concept; (3) **the unified process** – the key PBU result is a single, unified process, to which quality approaches are mapped.

In chapter 5.3 we described the process representation format used in representing the PBU process (RS6-2), while in chapter 5.4 a possible PBU process was described (RS6-3).

In order to practically validate the PBU framework and the usage of the PBU process, it should be used in practice. In the followings (chapters 6-8) we present its application in a real world case study (RS7).

# 6. PREPARING THE CASE STUDY[4]

In order to validate the PBU framework in practice, we apply it in a real world case study. In chapters 6-8 the research step 7 (RS7) “Perform a case study on the PBU framework” is discussed.

This chapter presents the preparatory steps of the case study. 6.1 describes the approach of preparing for the case study. 6.2 describes the project in which the PBU framework was applied, 6.3 presents the reasons for selecting a process for improvement, 6.4 describes the selection of quality approaches to be used for creating a unified process and 6.5 describes the reasons of selecting a process representation format.

During the case study discussed through chapters 6-8, deviations from the original PBU process may happen and will be discussed at the end of each chapter. In 6.6 we address these deviations from the original PBU process. In 6.7 we present limitations and in 6.8 we summarise results of preparation steps.

## 6.1. Approach

Here we have the same logic of preparation as we had at the selection subprocess of the PBU process, namely: processes, quality approaches and process representation formats should be selected before any unification. Reasons are discussed in 5.2.1 and concrete steps are explained in 5.4.1. In this chapter we will perform selection in accordance with the PBU process and chapter 5. These steps depend on context, therefore prior to the selection of a process, quality approaches and process representation format, the context will be introduced.

The structure of this chapter follows the logical flow of the first subprocess of the PBU process: “1. Selection (of processes, quality approaches and process representation)” described in 5.4.1 completing with two additional steps:

[4] Preliminary results of this chapter have been presented at (Balla, Kelemen, & Bóka, 2009).

1. Selecting a case study context and presenting the reasons of choice – not included into the PBU process, needed for the case study (chapter 6.2),
2. Selecting a process to be improved / created – corresponds to activity 1.1 of the PBU process (chapter 6.3),
3. Selecting quality approaches to be used in unification – corresponds to activity 1.2 of PBU process (chapter 6.4),
4. Selecting a suitable process representation format – corresponds to activity 1.3 of PBU process (chapter 6.5),
5. Understanding deviation from the PBU process – provides input for research step 10 Identification of refinements on the PBU framework (chapter 6.6).

## 6.2. Selecting the case context

In this chapter we first describe guidelines for selecting a context for the case study (6.2.1), then we introduce the project which we select based on the guidelines (6.2.2), and finally we summarise our selection, analysing the project based on guidelines (6.2.3).

### 6.2.1. Guidelines for selecting the case context

In order to perform the case study, the usage of multiple quality approaches should be required in the project. This will generate the need for using a concept for the simultaneous usage of multiple quality approaches and the PBU framework can be used. Since we have a concrete concept with a well-defined process and activities, we need to have enough freedom to use the PBU concept. And finally, the most important thing, at the end of the case study, the study should be validated, it should be performed in a real project and results should be good enough to be accepted by participants / users.

Summarising our needs for the case context we can define the following *context guidelines*:

1. The case context should require the usage of multiple quality approaches.
2. The case should provide enough freedom to use the PBU framework.
3. The case should be a real industrial project.

In the next chapter we introduce a project which we think could meet the guidelines for the case context. Afterwards we review the project using the guidelines.

### 6.2.2. The Q-works project

**Polygon – the company that launched the Q-works project**

“Polygon Informatics Ltd. was established in 1990. Its main profile is the selling and popularizing of IBM technologies.” The “company is the largest Hungarian Premier Business Partner of IBM. It deals with server operation, the integration of servers, systems, networks and applications, and it has significant revenue from selling hardware as well.”

“Polygon, over the years, has established a stable and continuously expanding clientele – primarily in the Hungarian bank sector and involving companies and institutions that operate IBM machines.” Its “clientele is stabile and the regular assignments ordered by them engage our full capacity” (POLYGON Informatikai Kft., 2007).

**IBM Maximo – the software environment of the Q-works project**

The IBM Maximo software portfolio (as part of IBM Tivoli) provides a component based, interoperable, business process automation (BPA) environment. “The Maximo software portfolio is the leading asset and service management software in the marketplace. Based on key technologies and standards, this architecture leverages the latest Web concepts, standards and technologies, helping to ensure optimum compatibility with today’s Web-based infrastructures.” (IBM, 2010)

“Maximo software provides proactive BPA capabilities through the combination of Maximo Workflow and Maximo Escalations. These components help monitor events in Maximo software, including static data, and automate the processing of these events. Any Maximo data point, process or event can be monitored and managed by these components” (IBM, 2010). For more information on asset management and workflow implementation in IBM Maximo see (IBM, 2010, 2011a, 2011b).

**The Q-work project**

“Polygon Informatics Ltd, Hungary started the development of a special system for supporting software quality control procedures. This two year development project of a budget of 650.000 EUR received 400.000 EUR subsidy from the EU based on its outstandingly high professional evaluation ranking. The purpose of this joint project carried out together with the Budapest University of Technology and Economics was the development of a software workflow system easing the implementation of CMMI” (POLYGON Informatikai Kft., 2008).

“CMMI defines requirements against software processes through its process areas, goals and practices. Introducing this very detailed model covering all the particularities is extremely difficult, and yet today it is not supported by any IT solution (that is by any software program). Recognising this deficiency Polygon and the acknowledged domestic expert of this topic, the Department of Control Engineering and Information Technology of the Budapest University of Technology and Economics signed a co-operation agreement for the development of an application called Q-works” (POLYGON Informatikai Kft., 2008).

The purpose of Q-works was to facilitate the introduction of CMMI and to provide continuous support during compliance assessments. The Q-works project itself was the first phase of the research, covering a high level adaptation of CMMI process areas to a software built in IBM Maximo environment. During this phase the team examined how to develop and model processes which satisfy CMMI requirements in IBM Maximo environment.

### 6.2.3. Analysing the project based on context guidelines

We have chosen the Q-works project for performing the case study because it carries all the characteristics we needed, and meets the guidelines for the case context defined in 6.2.1:

Guideline 1: The case context should require the usage of multiple quality approaches.

- The main goal of the project is the implementation of a CMMI-based workflow system. CMMI defines what should be done, but concrete steps (how) are not explicitly provided. Due to this characteristic of CMMI, further quality approaches are needed in defining con-crete processes. As in Q-works we should use multiple quality approaches, therefore it is an excellent opportunity to use the PBU framework.

Guideline 2: The case should provide enough freedom to use the PBU framework.

- We were actively involved in the Q-works project as project members. The main task of our team was defining CMMI-conform processes. So we had the appropriate level of control and freedom to develop processes.

Guideline 3: The case should be a real industrial project.

- Q-works was a real project with a realistic goal supporting CMMI processes. The project was funded by EU and an industrial company. It was a requirement of the project and had to be presented both for the company and to the EU, that processes developed in the project were real.

We think all the guidelines were met by the Q-works project, therefore we selected it for performing the case study. In the next chapter we present the selection of a process for the case study.

# 6.3. Selecting a process

In this chapter we describe first guidelines for selecting a process for the case study (6.3.1), then we introduce the process which we selected based on the guidelines (6.3.2), and finally we summarise our selection, reviewing the process using the guidelines (6.3.3).

## 6.3.1. Guidelines for selecting a process

We cannot unify all the software processes and all the process based quality approaches at once. That may consume more resources than we have in the frame of a PhD project, therefore we have to make selections. Since we already selected the case context we have context constraints as well.

In order to make this selection systematically, we define selection guidelines, which contain context-dependent and context-independent parts.

In the Q-works project it was needed to implement CMMI-conform processes. This constraint reduced the number of processes: in CMMI-DEV there are 22 process areas, but we still cannot unify all the processes of 22 process areas, a selection still need to be made.

We cannot map all the CMMI components at once, we needed a starting process. Since we were applying the PBU framework for the first time, we needed a process simple enough to be performed quickly and complex enough to draw conclusions and refine the PBU framework if needed. Due to project needs it would be also useful if the process would be usable in multiple situations and reusable as part of other processes.

Summarising, we identified the following *process guidelines*:

1. The process should be part of CMMI (context-dependent),
2. The process should be usable in multiple situations (reusable, context dependent),
3. The process should be simple enough to be implemented quickly and be complex enough to provide basis for lessons learned in refining the PBU framework.

In the next chapter we introduce a process which we think could meet the guidelines then we review the project against the guidelines.

### 6.3.2. Introduction to peer reviews

After reviewing the 22 CMMI process areas, peer reviews (a specific goal in the Verification process area)seemed an appropriate process for the case study. Here we present a brief introduction to the peer review process.

Besides the fact that peer reviews are present in the software industry, they are more widely known in other fields and they have been a touchstone of scientific methods since the 20th century. In software industry peer reviews had a wider acceptance since 1976, when Fagan wrote his famous article on design and code inspections (Fagan, 1976).

Many formal (white-box and black-box) testing techniques are used at companies, such as boundary-value analysis, equivalence-class testing, decision-table based testing, structural testing and others. However, researchers have shown that inspections and reviews also have a significant effect on companies ROI (Rico, 2002, 2003, 2004) and a high percentage of saving in specification, design, code and test planning phases can be achieved (Harjumaa, Tervonen, & Huttunen, 2005) using them. Fagan shows that inspections detected 82% of the faults (Fagan, 1976). In measurements of D. F. Rico, inspections had the second highest ROI (3,272%) in comparison with SW-CMM (871%), CMMI (173%), ISO 9001 (229%), TSP (2,826%) and PSP (4,133%) (Rico, 2002, 2003). Further studies (Gilb & Graham, 1993) and software testing books (Graham et al., 2007; Hambling et al., 2007; Morgan, 2010) also show that by using peer reviews as a preliminary testing technique, a significant increase in productivity and product quality can be achieved.

Various benefits can be achieved from different stakeholder views. Karl E. Wiegers describes such benefits of peer reviews for developers, project managers, maintainers, quality assurance managers, requirements analysts and test engineers (K. E. Wiegers, 2002a).

Many standards and models consider peer reviews being important and focus on describing requirements and best practices for peer reviewing (such as SW-CMM (M. Paulk et al., 1995), CMMI (CMMI Product Team, 2010a, 2010b, 2010c), ISO 12207 (ISO/IEC/IEEE, 2008), IEEE 1028:2008 (IEEE, 2008b) or PSP (Humphrey, 1997)).

The International Software Testing Qualification Board (ISTQB, 2009) considers inspections / peer reviews a basic static testing technique which is required knowledge for the Certified Tester – Foundation Level exam (Graham et al., 2007; Hambling et al., 2007; Morgan, 2010).

### 6.3.3. Analysing the process based on process guidelines

We have chosen the peer review process for performing the case study because it carries all the characteristics we needed, and meets the guidelines for the process defined in chapter 6.3.1:

Guideline 1: The process should be part of CMMI (context-dependent):

- Peer reviews are present in CMMI-DEV as part of the Verification process area on specific goal and specific practice level. If specific practices are not performed, goals are not achieved and the process area is not implemented. Therefore, peer reviews are considered important part of CMMI-DEV and required for implementing the Verification process area.

Guideline 2: The process should be usable in multiple situations (reusable, context dependent):

- Peer reviews can be used for reviewing multiple work products during a project lifecycle, e.g. requirements review, project plan review and code review. Furthermore, peer reviews have multiple types such as: inspections, audits, technical reviews, walkthroughs and others. Due to this diversity of reviews it can be used in implementing different process areas of CMMI. E.g. code reviews during the Verification process, requirements review during the Requirements Development process, configuration audits in the Configuration Management process or product and process audit during the Process and Product Quality Assurance process.

Guideline 3: The process should be simple enough to be implemented quickly and be complex enough to provide basis for lessons learned in refining the PBU framework.

- The peer review process itself is not a long process, it takes a limited preparation, performing and analysis time. Therefore, it is simple enough to be implemented quickly. In the same time it has its complexity:
    - has a number of all the important process components e.g. subprocesses, activities, roles, artifacts etc.,
    - can be performed in many different ways,
    - it is included in multiple quality approaches in many different ways.

Besides meeting the guidelines, there were some additional reasons to choose the peer review process:

- Peer reviews are defined only on a high level in CMMI, therefore further sources should be used.
- Peer reviews are defined in multiple quality approaches, so the PBU framework can be used.
- Prior to the Q-works project we gained knowledge in the peer review process (Kelemen, 2008; Kelemen & Balla, 2009).

According to the reasons above, as a starting example of the Q-works project we selected the peer review process.

# 6.4. Selecting quality approaches

In this chapter first we describe guidelines for selecting quality approaches (6.4.1), then the search for quality approaches is shown (6.4.2) and finally we summarise our selection, reviewing the quality approaches found based on the guidelines for quality approaches (6.4.3).

## 6.4.1. Guidelines for selecting quality approaches

We faced a problem in describing the peer review process based only on CMMI-DEV: CMMI does not provide enough information for implementing the peer review process. It provides the requirements and best practices for peer reviews, but it does not provide a concrete pro-cess. It is often mentioned that CMMI describes what should be implemented however, it does not describe how. Therefore, we needed to look into other approaches which contain peer reviews and are detailed enough to define a peer review process. Certainly, some practical point of view would also be useful to consider. We can translate these needs to the following *quality approach guidelines*:

1. the quality approaches should include the selected process,
2. the set of quality approaches should provide enough information for developing the selected process,
3. the set quality approaches should provide practically usable guidance on implementing the selected process,
4. the quality approaches should be relevant and recognizable to the organization.

In the next chapter we describe the search for quality approaches.

### 6.4.2. Searching for quality approaches

According to the activity 1.2 of the PBU process (described in 5.4.1.2) we looked into the multi-model literature collected in chapter 3.2 and no peer-review related multi-model results were found, therefore we looked at quality approaches.

By working at SQI – Hungarian Software Quality Consulting Ltd. we gained access to the SQI Information Repository which includes documents related to software process improvement. This knowledge-base contains papers, presentations, templates, course materials, research results, guidelines, case studies, quality approaches, books, methodology descriptions, tutorials, reports and presentations. The content of SQI Information Repository is collected during years of practical and research work in the field of software process improvement and it is used at assessments, consultancy, research and development projects and training. In addition to this repository, SQI has access to SEIR(SEI Information Repository) which is one of the largest library over the world in the field of software process improvement (SQI, 2008).

We used these sources to collect quality approaches suitable for implementing the peer review process.

We searched our database and we found the following quality approaches:

1. ISO-IEC-90003:2004 Software engineering — Guidelines for the application of ISO 9001:2000 to computer software,
2. CMMI-DEV v1.2 (the case study was performed in 2010 before CMMI v1.3 was released),
3. CMMI-SVC v1.2,
4. Enterprise SPICE (ISO/IEC 15504) PAM & PRM,
5. FAA-iCMM v2.0,
6. IEEE 1028:2008,
7. ISO 12207,
8. ISO 9001:2000,
9. ISO/IEC 15939:2005,
10. ITIL v3,
11. Peer review process descriptions from Process Impact (Process Impact, 2010).

The discussion of the selected quality approaches is included in chapter 7.

### 6.4.3. Understanding the scope and analysing quality approaches based on quality approach guidelines

In order to understand how peer reviews appear in these quality approaches, first we looked for peer review definitions and types. 11 process oriented software quality approaches were analysed in which 30 definitions connected to reviews and audits were found of which 20 were different (see Appendix C). In 4 from 11 quality approaches analysed, no definitions related to reviews and audits were found.

Comparing all the definitions found would generate a 30x30 matrix. Narrowing the comparison to only the different definitions would result in a 20x20 matrix. Instead of building these matrices, we considered understanding the definitions and choosing a starting one.

After reading and understanding each definition found, we structured them as follows:

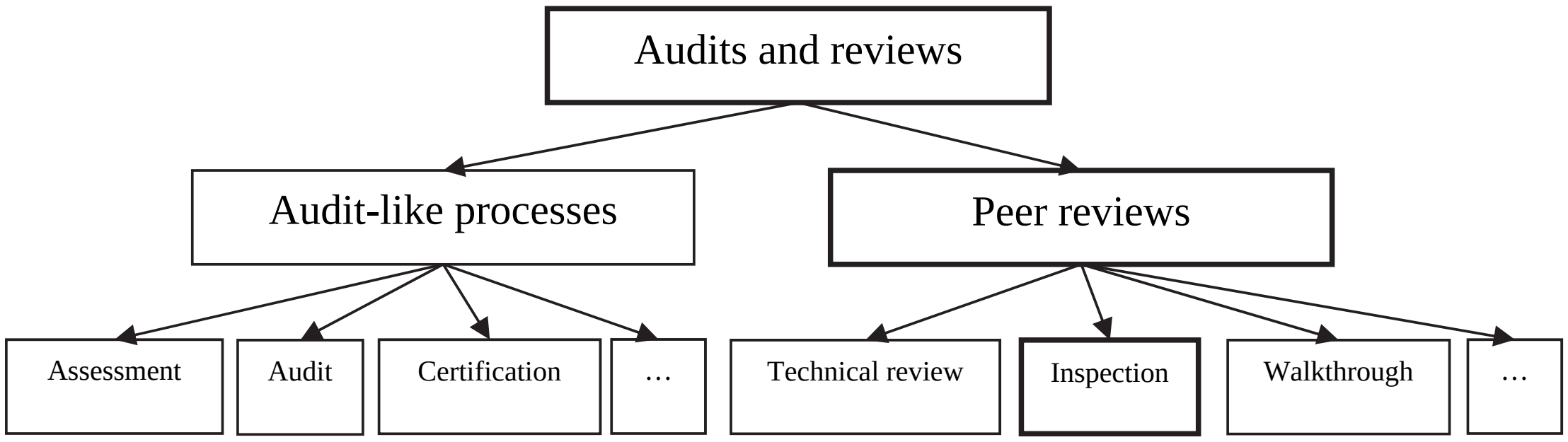

Figure 14 – Audit and review types found in quality approaches

Due to the fact that the main project goal was to implement CMMI processes, we selected the CMMI peer review definition as primary definition. Analysing the definitions and review-related content of quality approaches selected, we decided to implement first the *inspection* process, because it is the most formal type of peer review. Further, less formal review types can be implemented by simplifying the inspection process.

We narrowed our goal to the inspection process, excluding the audit-like processes and process definitions and we focused only on the inspection process from the many peer review process types.

Reaching the targeted type of review on the tree of terms in Figure 14 we identify the definitions of review, peer review and inspection.

**The *Review* definition: "***A process or meeting during which a software product, set of software products, or a software process is presented to project personnel, managers, users, customers, user representatives, auditors or other interested parties for examination, comment or approval." (IEEE 1028-2008)*

**The *Peer review* definition: "***The review of work products performed by peers during development of the work products to identify defects for removal." (CMMI-DEV v1.2)*

**The *Inspection* definition: "***A visual examination of a software product to detect and identify software anomalies, including errors and deviations from standards and specifications." (IEEE 1028-2008)*

No definitions were found in the following approaches:

1. Enterprise SPICE (ISO/IEC 15504) PAM & PRM (Process Dimension) – October 2009 – review phase (NO GLOSSARY accessible)
2. ISO-IEC-90003:2004 Software engineering — Guidelines for the application of ISO 9001:2000 to computer software
3. ISO/IEC 15939:2004 (no peer review related terms)
4. Based on the definitions found in the selected quality approaches we decided to exclude several quality approaches from the list. Table 18 summarizes the quality approaches and the reasons of exclusion.

Table 18 – Exclusion of quality approaches

| **Quality Approach** | **Reason of exclusion** |
|---|---|
| ISO-IEC-90003:2004 Software engineering — Guidelines for the application of ISO 9001:2000 to computer software | No detailed description is available. |
| Enterprise SPICE (ISO/IEC 15504) PAM & PRM | It is in development phase. No detailed description is available and it is mainly based on FAA-iCMMI v2.0. |
| FAA-iCMMI v2.0 | It includes similar descriptions and process elements to CMMI. Some elements such as timing the review might be useful, but most of other elements can easily be mapped to CMMI elements. |
| ISO 12207:2008 | It contains high level requirements for software audits, project management reviews and technical reviews. It recommends the usage of IEEE 1028, which was chosen. |
| ISO 9001:2000 | In contains general, high level requirements. |
| ISO/IEC 15939:2005 | It is not relevant. |
| ITIL v3 | It contains high level and mainly service oriented requirements. |
| CMMI-SVC 1.2 | It contains the same peer review requirements as CMMI-DEV. However, these requirements are on a higher level (specific practice level). In CMMI-DEV these are discussed on both specific goal and specific practice level. |

According to Table 18, we narrowed the list of quality approaches to be included to the followings:

1. CMMI-DEV v1.2,
2. IEEE 1028:2008,
3. Peer review process descriptions of Process Impact.

In the followings we review these quality approaches to see whether they carry all the characteristics we needed, and meets the quality approach guidelines defined in chapter 6.4.1.

Guideline 1: the quality approaches should include the selected process:

- The selected 3 approaches all include peer reviews. CMMI contains peer reviews on specific goal, specific practice levels, but also on informational component levels such as typical work products and subprocesses. The IEEE 1028 standard is dedicated specially to peer reviews and audits, while the scope of peer review process description of Process Impact is also the peer reviews.

Guideline 2: the set of quality approaches should provide enough information for developing the selected process:

- CMMI-DEV contains what should be performed during a peer review, while the other two approaches IEEE 1028 and Process impact's peer review process descriptions are well detailed and describe how various types of peer reviews should be performed.

Guideline 3: the set quality approaches should provide practically usable guidance on implementing the selected process:

- The Process Impact peer review process descriptions were developed by a process consultant company using practical experiences of software process improvement. The author of this document has also published several other documents on how to perform peer reviews in practice (K. Wiegers, 2002; K. E. Wiegers, 1998, 2002a, 2002b). Therefore, we consider this document provides the practical view to the peer reviews.

Guideline 4: the quality approaches should be relevant and recognizable to the organization:

- The goal of the Q-works project development of a special system for supporting software quality control procedures, especially to provide CMMI-based workflows. The project was also open to involve other approaches in order to make the developed workflows stronger. Therefore we consider the selected quality approaches relevant and recognizable to the organization.

Summarising, we will use the selected 3 quality approaches for creating a unified peer review process.

## 6.5. Selecting a process representation format

In this chapter we present the textual and graphical process representation we choose for representing the unified peer review process in the Q-works project.

### 6.5.1. Textual representation of quality element instances

After several discussions with the Q-works software development team we have chosen a generic table format for storing quality approach element instance texts. The format of this table is different from those introduced in 5.3. This is due that our goal with the new format was to make the information easily workable for Q-works software developers. Our guidelines for storing the textual information were the following:

- use a general format in which we can easily enter, store and modify textual information and relations among these,
- should be easy to be processed by IT experts.

The columns of the quality approach text table are the following:

- Parent – if the process is divided into multiple subprocesses, this column indicates the name and ID of the parent process of subprocess or activity
- Item ID – Since the process oriented software quality approach requirements are primarily task (or activity) oriented, this column indicates the ID of task also
- TO – Task Order: completed if the approach contains information regarding the order of task

- Inputs – task or process inputs (will be represented as artifacts)
- Outputs – task or process outputs (will be represented as artifacts)
- Entry criteria – task or process entry criteria
- Exit criteria – task of process exit criteria
- Roles

This table also served as an indicator: empty cells indicated that further information might be required from other sources (mainly from other quality approaches).

After filling the table with quality approach element instances, it can have two state types:

1. fully filled – a process could be (textually and graphically) described based on the table
2. partially filled – further information (e.g. quality approach element instances) might be required

Table 19 – Example of the quality approach element instance table

| Parent | Item ID | TO | Process/Activity description | Input | Output | Entry Criteria | Exit Criteria | Roles |
|---|---|---|---|---|---|---|---|---|
| [filled] | [filled] | [empty] | [filled] | [empty] | [filled] | [empty] | [empty] | [empty] |
| [filled] | [filled] | [empty] | [filled] | [filled] | [empty] | [filled] | [empty] | [empty] |
| [filled] | [filled] | [empty] | [filled] | [empty] | [filled] | [empty] | [empty] | [empty] |

Tables in Appendix L, Appendix M and Appendix N were developed using the structure above. (Columns were left out where elements were not found.)

Table 20 shows mapping quality approach table elements to process elements identified in chapter 4.

Table 20 – Matching quality approach table elements to process elements

| Process element (Table 17) | Quality approach storing table element (Table 19) |
|---|---|
| Process name | Item ID |
| Parent process | Parent process |
| Process description | Process/Activity description |
| Activity | Process/Activity description |
| Role, responsibility | Roles |
| Data object | Input and output |
| Element relations | Task order, Activity – input/output associations |
| Entry/exit criteria | Entry criteria, exit criteria |

**Explanation of mapping in Table 20**

While in Table 17 no element identification was included, in order to support maintainability of the high number of mapping we introduced the "item ID" element in Table 19 for unique element instance identification. In case of (sub)process, instead of identifying them by their name, they are identified with an unique ID. The mapping of process description and activity to "Process/Activity description" was done due to practical reasons: this way the resulting tables will contain one less column. In order to support maintainability of roles and to avoid duplications, responsibilities related to roles and the role-responsibility mappings are kept in a separate table (Table 48).

Table 19 represents data objects as inputs and outputs. This introduces additional information compared to Table 17: inputs and outputs denote the direction of the relation between the data object and the activity. Table 19 contains multiple element relations such as: task orders, activity-activity relations, activity-subprocess relations and input-output associations to activities.

Entry and exit criteria are included in Table 50 and Table 53. It is important to mention that the information in the tables were all mapped to process element instances and were also included into the tool used for process modeling (itp-commerce, 2012).

### 6.5.2. Graphical representation

Before starting to model any kind of process, we considered important studying the most significant modeling languages.

The practice indicates that graphical representation of processes helps understanding them, and can be a good support tool to understand the attached textual descriptions. A graphical representation beside that it is more easily and quickly interpretable than a multi-page text, condenses information and can be applied for representing unambiguously the junctions, cycles and decision points, enhances the clarity of information.

Due to the fact that sometimes it is really difficult to create unambiguous textual process descriptions, which are understandable for everyone, a graphical representation often is the best accessory of the process descriptor. A graphical representation helps solving these problems, by eliminating the potential textual inconsistencies and it is easier to process, verify, validate and maintain the process representation.

A process descriptor best practice is to prepare the graphical representation first, then to attach a text explanation to the elements of the graphical representation, primarily to processes, steps, activities and inputs/outputs. Obviously, the process descriptor decides the order of description and modeling, based on his/her situation and information available.

**Modeling methods and architectures**

Process modeling can be categorized from several points of views, e.g. according to modeling approach or modeling architecture.

According to the minimalist approach, only the inevitably important elements are necessary to model. In this case, the modeler focuses on a representation which can be digested quickly and easily. In the maximalist approach the goal is to create models which can be digested by computers. Models created in a maximalist way may imply a number of details which are more difficultly understandable and less clear for people.

From the architectural point of view, two principal directions can be observed: the top-down and bottom-up approach. The top-down method starts from the idea of the process, first describing it, writing down an ideal process and then performing it. As opposed to this, the bottom-up method tries to model existing, working processes.

In this research our goal was to explore existing processes and to model only their important elements; thus we followed a bottom-up, minimalist approach. The comparison made in this chapter is based on this approach as well.

**Workflow patterns**

Workflow (or control flow) patterns show the expressiveness of a graphical representation language. Workflow patterns are e.g. the sequence, different types of junctions or synchronization.

At the selection of a modeling language an important viewpoint may be the way in which the most frequently used workflow patterns are represented. In one work at the Technical University of Eindhoven, 21 workflow patterns were identified and assigned to different categories (White, 2004). Researchers examined how the selected patterns can be realized in BPMN and EPC. Another researcher, (Weske, 2007) on a similar manner to the former research, demonstrates how the workflow patterns can be represented in different process modeling languages. Having their results, one of our primary comparison aspects has to be the ability of the modeling language of expressing workflow patterns.

### 6.5.3. Selection of candidate process modeling languages

In this chapter we present the information gathered while we were looking for the characteristics and features of process modeling languages.

In software developer community UML (Unified Modeling Language) activity diagrams are applied in many cases for process descriptions, but we may face several other methods, as: Petri nets, EPC (Event Driven Process Chain), Workflow networks, YAWL (Yet Another Workflow Language), GPWL (Graph-Based Workflow Language), BPMN (Business Process Modeling Notation) and BPD (Business Process Diagram).

Besides these, less formalised solutions recommended by Enterprise Modeling and POEM (Process Oriented Enterprise Modeling) should be mentioned. Enterprise modeling is a situation-dependent modeling in enterprise environment, having the goal similar to BPMN: to relief the communication amongst the parties with different background. In order to support this goal, different type of diagrams can be used in different situations. Amongst the mentioned solutions BPMN, EPC and UML seems the most plausible for us, because:

- well-known workflow patterns can be easily used with them,
- they cover process elements identified,
- they have standardised and uniform components,
- they are suitable for human and machine processing equally,
- they are widespread,
- a wide-range software support is available for them.

A wide scientific and practical literature exists on process modeling methods, therefore only a brief overview will be presented here.

### 6.5.4. BPMN

BPMN (Business Process Modeling Notation) is a standardised graphical notation for representing business processes in workflow. BPMN was developed by BPMI (Business Process Management Initiative) and it is handled by OMG (Object Management Group) since the fusion of the two organizations in 2005. BPMN version 1.1 is included into this discussion. The goal of BPMI with BPMN was to create a standardized notation which is understandable for all relevant stakeholders. Such stakeholders might be business analysts – who create and refine the processes, developers – who are responsible for the realisation of certain processes or

managers – who manage and control the processes. Therefore, BPMN is a common graphical notation with the goal of eliminating the communication gaps amongst stakeholders with different background. It is important to emphasize that BPMN is built on flowchart technique; during its development several notation systems and methods were used, such as UML activity diagrams, UML EDOC Business Process, IDEF, ebXML BPSS, Activity-Decision Flow (ADF) diagram, RosettaNet, LOVeM and EPC (Wahl & Sindre, 2005).

### 6.5.4.1. **Elements of BPMN**

BPMN includes some well-defined basic elements, which can be ordered into the next categories:

- flow objects: event, activity and gateway;
- connecting objects: sequence flow, message flow, association;
- swimlanes: pool, lane;
- artifacts: data objects, group, annotation;

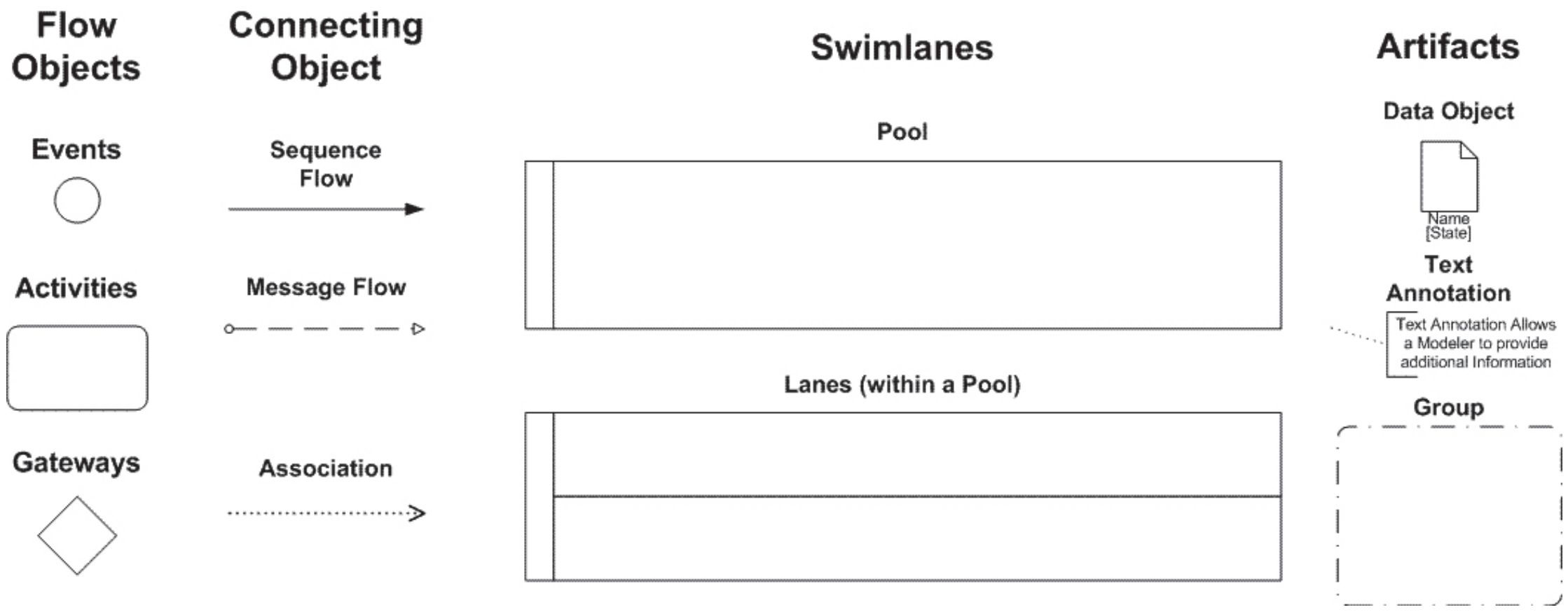

Figure 15 – Basic symbols of BPMN, source: (OMG, 2012)

### 6.5.4.2. **Diagram: BPD**

BPD-s (Business Process Diagram) can be created using BPMN notation. The following types of BPDs exist:

- high-level business,
- detailed business,
- cooperating with unknown external processes,
- collaboration BPD.

In a single model, different types of BPDs can be used. However, if too many types of BPDs are used simultaneously in one diagram, it can turn into one which can be understood quite difficultly. Therefore, it is advisable to use always only one type of BPD.

### 6.5.4.3. **Storage format: WS-BPEL**

The language WS-BPEL (Web Services Business Process Execution Language) or BPEL was created to describe executable business processes. The specification of BPMN includes a description about how the BPMN can be converted to BPEL codes, but unfortunately this mapping is informal and incomplete (White, 2005). As BPMN and BPEL are becoming widespread, several tools are trying to implement the conversion between these two. An open-source example tool is BPMN2BPEL.
In the course of the development of such tools, fundamental differences between BPMN and BPEL came into light. Because of these differences, in a number of cases BPDs cannot be converted to BPEL. The situation may become more complicated when the goal is to generate BPEL codes readable by humans, or round-trip engineering is used during the development (Gao, 2006).

#### 6.5.4.4. **Storage format: XPDL, software support**

XPDL (XML Process Definition Language) is a format standardised by the Workflow Management Coalition (WfMC). The standardized format provides the permeability amongst different workflow products, modelers and business process management tools. XPDL defines an XML scheme for describing the declarative part of business processes. In XPDL graphical and semantic information of the business processes can be stored. XPDL was designed so it can store all the information needed in BPDs, e.g. two dimensional coordinates, which show the position of BPMN elements in the diagram. It also contains information related to the execution. This is an important difference between BPEL and XPDL, because BPEL concentrates only on the executable aspects of BPDs. BPEL does not contain elements that would help in the graphical representation. For more details see the XPDL specification at (WfMC, 2012).

### 6.5.5. EPC

The EPC method was developed within the frame of project ARIS by prof. Wilhelm-August Scheer at Saarlandes University at the beginning of 90’s. It was firstly used at SAP, but nowadays a number of companies are using it for modeling, analysing and redesigning their business processes. EPC is a directed graph of events and functions. Implies diverse connecting elements, with which the alternative and parallel execution routes can be described. It uses logical operators such OR, AND and XOR. The intelligibility and simplicity are one of its largest strengths. Unfortunately its syntax and semantics are not well-defined (Aalst, 1999).

#### 6.5.5.1. **Elements of EPC**

EPC elements are the following: event, function, process interface, connectors (AND, OR, XOR), control flow arc, participant (e.g. organization unit), application, data (information, material, resource object), relation.

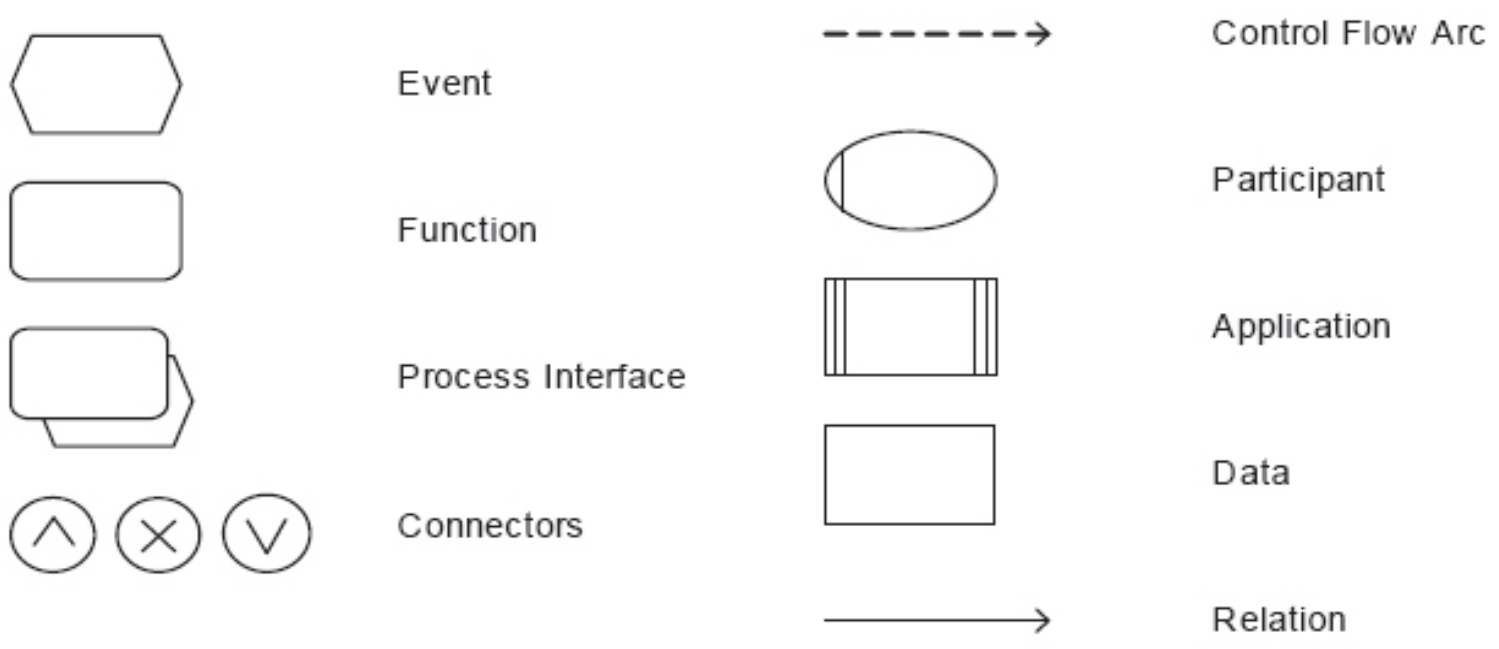

Figure 16 – Symbols of EPC diagram, source: (Mendling & Nüttgens, 2005)

6.5.5.2. **Storage format: EPML**

It is important to mention, that a standard, portable XML format was developed for storing EPCs, which is called EPML (EPC Markup Language). See (Mendling, 2012) about the format, tools supporting the format and connected scientific and practical articles. A brief description of the format can be found in technical report (Mendling & Nüttgens, 2005).

6.5.5.3. **Software support**

The best-known software (10), which support EPC are the following: SAP R/3 (SAP AG), ARIS (IDS Scheer), LiveModel/Analyst (Intellicorp Inc.), Visio (Visio Corp.), Visio (Microsoft), ADONIS (BOC Group), Semtalk (Semtation), Bonapart (Pikos), SmartDraw, EPC Tools (Paderborn Univ.).

## 6.5.6. UML activity diagrams

OMG defined 13 types of diagrams in UML 2.x, among which more are suitable for describing processes. Such types are e.g. sequence diagrams, activity diagrams, state-machine diagrams and communication diagrams (Altova, 2011). From among these, the activity diagram is suitable and frequently used for describing processes.

6.5.6.1. **Elements of UML activity diagrams**

The list of elements of UML activity diagrams can be seen on the figure below. It includes the following main elements: state, transition, flow, decision, swimlane, signal recepit, signal send, constraint and note.

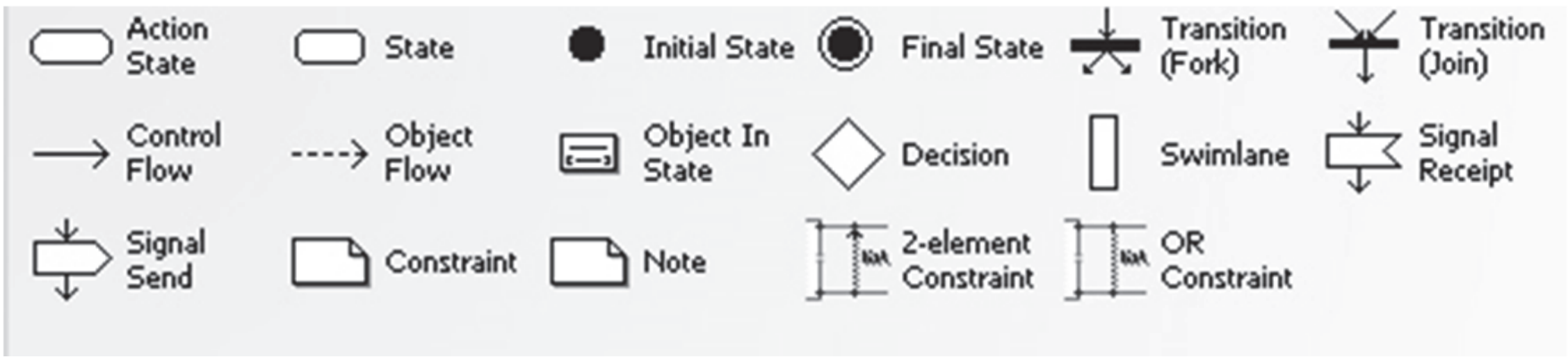

Figure 17 – Elements of UML activity diagrams, source: MS Visio

#### 6.5.6.2. **Storage format: XMI**

Standard XMI (ISO/IEC 19503:2005 Information technology - XML Metadata Interchange) (ISO/IEC, 2005) was developed by OMG for describing XML metadata. It can be used to describe any kind of metadata, which can be expressed in MOF (Meta-Object Facility). Most frequent applications of XMI are the storage of UML models, but it is used to store models of other languages as well. Unfortunately, the XMI-based portability between UML modeling tools is still not solved because of incompatible XMI formats.

#### 6.5.6.3. **Software support**

We could write much about software support of UML (activity diagrams), because it is a widespread modeling language on the software market. UML is supported by such software industry giants like IBM, Borland, Microsoft or Oracle. Besides them, several further commercial and freeware UML modeler products are available on the market. The list of tools supporting UML is expanding continuously; one list can be attained at (Objects by Design, 2011).

It is difficult to make a selection or comparison without using guidelines; therefore first we define the aspects which could be the most important at selecting a modeling language for representing the peer review process, and then we make the selection based on the guidelines defined.

### 6.5.7. Comparison guidelines

The primary guideline is the intelligibility, since models will be interpreted, used or modified by people coming from different environments. The intelligibility of a modeling language is a characteristic depending on an ability and background of the interpreter, thus it is difficultly measurable. Nevertheless some researchers address this topic, e.g. (Mendling, Reijers, & Cardoso, 2007) or (Becker, Rosemann, & von Uthmann, 2000).

The ability to represent workflow patterns shows the expressiveness of the language. The complexity of models increases with a less expressive language. The intelligibility decreases and the number of mistakes grows by the increase of the complexity of models.

Besides the expressiveness of the language, coverage of process elements identified in chapter 4 is required.

As we would like to create and modify our process models quickly and easily, another goal is to have a suitable software support for the chosen language.

The administrative organizations providing different services may use software which are independent from each other therefore, we consider important the selected approach not to be dependent on one single manufacturer. Besides freely choosing the software used for modeling, saving into a portable format may also be useful.

The more a technology or a modeling solution is used, the easier its introduction and acceptance. As a result, it is important to choose a solution as widely applied as possible.

Summarizing, we considered the following aspects:

- intelligibility,

- coverage of process elements,
- ability of expressing workflow patterns,
- software support,
- portable format,
- widespread in different areas.

These are the comparison guidelines for selecting the process modeling language.

### 6.5.8. Comparison

In the course of comparison we value the approaches based on the guidelines presented. We deal with the guidelines in order of their importance. The order of importance was decided by us taking into account project needs.

#### 6.5.8.1. **Intelligibility**

73 undergraduates and 12 process modeling experts were involved from the University of Vienna, Technical University of Eindhoven and the University of Madeira into a research performed by Mendling et al (Mendling, Reijers, et al., 2007). They asked participants to fill in questionnaires, then they analysed how the participants can interpret the 12 models presented. The research showed that understanding the models is primarily influenced by personal factors like the participants' previous theoretical and practical process modeling knowledge and experience. The other important factor in terms of the intelligibility is the model size, which can be measured with different metrics (Mendling, Neumann, & van der Aalst, 2007; Mendling, Reijers, et al., 2007). Additional possible influencing factors of the intelligibility could be the expertise of the interpreter in the modeling language, the components of modeling language, the layout and arrangement of figures. In terms of intelligibility (White, 2004) we did not see any difference between UML 2.0 activity diagrams and BPDs.

From among the three modeling languages, software designers and developers use UML. EPC is mainly spread in the business sphere, while BPMN is present in both areas. This is not by chance; the goal of BPMI with the BPMN was to create a standard notation which is intelligible for people with different backgrounds.

Therefore, we consider BPMN the most appropriate approach from the first point of view.

#### 6.5.8.2. **Coverage of process elements identified**

In order to model processes described using process elements identified in chapter 4, we need to select a language which can represent these process elements. In Table 21 we map process elements identified in Table 17 to the three preselected process modeling language elements. In order to understand how people perceive PML elements, besides the official specifications we looked at additional tutorial sites e.g. (Macke, 2011) or (Sparx Systems, 2011).

Table 21 shows that most of the process elements identified previously are present in BPMN, EPC and UML activity diagrams, therefore we consider them equally good for representing processes.

Table 21 – Coverage of process elements

| Process element | BPMN | EPC | UML Activity diagram |
|---|---|---|---|
| **Process name** | Process name | Function name | Activity name |
| **Process parent** | Parent-child relation (unlimited levels) | – | Activity-Action relation |
| **Process purpose** | In textual descriptions | In textual descriptions | – |
| **Activity** | Activity | Function | Activity |
| **Role, responsibility** | Lanes | Organisation unit | Partition or Swimlane |
| **Artifact** | Artifact | Document, Information object | Object |
| **Element relation** | Element relations | Element relations | Control flow, Object flow |
| **Entry/exit criteria** | – | – | Action constraints (pre and postconditions) |

#### 6.5.8.3. **Ability of expressing workflow patterns**

All three modeling languages are built on similar elements. Such elements are the fork-join, branch-merge, basic and extended activities, but these are defined using different notations. The equivalent of some elements do not exist in the other diagrams, which make the automatic conversion difficult. Wienberg describes such notation differences between EPC and UML activity diagrams (Wienberg, 2001). Making use of Aalst and other's results, White compared BPMN and UML activity diagrams based on 21 fundamental workflow patterns (White, 2004). As the result of the comparison, he came to the conclusion that both languages are equally appropriate for describing the selected workflow patterns. He remarks one single exception, that the meta-model of activity diagram does not have a suitable structure to describe a certain workflow pattern. White concludes that the two approaches are different views of a single meta-model (White, 2004). Since OMG handles both BPMN and UML activity diagrams, these might converge to each other in the future.

EPC was examined by Mendling et al, based on the 20 patterns of Aalst. Deficiencies were discovered and corrections were also proposed (Mendling, Neumann, & Nüttgens, 2005).

It is difficult to rank the three approaches from this point of view; all the three can represent the most of fundamental workflow patterns. However, a distinction is needed to be made: based on White's results we consider BPMN the most appropriate.

#### 6.5.8.4. **Software support**

Comparing their software support, it is visible that the most supported solution is the UML (activity diagram), BPMN comes after this, then EPC. The disadvantage of EPC's software support is connected to its origin; it was developed and used at a single company, while UML and BPMN are company-independent, handled by OMG. Therefore, we consider UML the most software-supported solution. In spite of the fact that most modeler tools support UML, it is possible to find suitable solutions for the other two languages too. E.g. all three types of diagrams can be created with MS Visio.

#### 6.5.8.5. **Portable format**

In terms of the portability EPC's strength is EPML defined by Mendling and Nüttgens (Mendling & Nüttgens, 2005), which is a standard and independent XML format. The question is that, how the different applications will support it. In the case of UML diagrams, several UML modeler tools support the XMI format (defined by OMG, currently still with incompatibility issues) or can export to some other XML formats. In the case of BPMN, in order to support portability the Workflow Management Coalition defined the XPDL format.
It is visible that all three formats are based on XML and freely available, but deficiencies exist in all three cases on the side of implementation. The opportunity is given; the application of the standardised format depends on the developers of modeling tools. Formats supporting BPMN and UML were specified by a consortium. Therefore, they may count on a bigger support than EPC.
From the portability point of view, XMI format was implemented by several manufacturers therefore, we consider the UML the most portable modeling language.

#### 6.5.8.6. **Widespread in different areas**

Due to the OMG's professional past, its standardized notations are accepted and adopted quickly by developer companies. We consider this a determining factor in the acceptance and adoption of UML and BPMN. EPC was introduced at a concrete company with no consortium behind it, which resulted a more modest spreading compared to the former two. The IT sector quasi exclusively uses UML for modeling. In the business modeling multiple notations are used, starting from the simple flowcharts towards the BPMN, EPCs and UML; therefore it is difficult to tell which one is the most widespread in this area. Because of UML is spread in both areas, we consider it the best solution for our goal.

### 6.5.9. Selecting a process modeling language

Based on the comparison made, we can state that all three notations have a number of benefits; in addition, in the most cases the differences between them are minimal. At the same time, taking into account the previously defined guidelines, BPMN seems the most appropriate solution from the intelligibility and expressiveness point of view. On the other hand, UML activity diagrams have the largest software support, best portable format and are widespread.
The next table summarises the result of comparison, listing the aspects according to their order of importance:

It is difficult to make the "best choice" among the presented modeling solutions without compromise. In spite of the fact that UML is adequate from more aspects than BPMN, *we choose BPMN as a final solution*, because:

- it is adequate from the first three most important aspects,

- a proper software support is available,
- very similar formats are available for all three languages in terms of portability,
- it is widespread in the business sphere; due to its intelligibility and OMG's support it's widespread can be growing quickly.

Table 22 – Comparison of modeling languages

| Guideline | BPMN | EPC | UML activity diagram |
|---|---|---|---|
| Intelligibility | + | | |
| Coverage of process elements | + | + | + |
| Ability of expressing workflow patterns | + | | |
| Software support | | | + |
| Portable format | | | + |
| Widespread | | | + |

We analysed the structure and content of different quality approaches and we distinguished several quality approach elements. Such elements of process-based quality approaches are e.g. inputs, outputs, typical work products, tasks or activities (see chapter 4.4). We mapped these quality approach elements to process elements in chapter 4.6. Based on the quality approach elements mapped to process elements in chapter 4 and the comparison made above, we think BPMN can be a good solution for representing graphically the processes developed by using elements of quality approaches (e.g. by making use of PBU process introduced in chapter 5).

## 6.6. Deviations from the PBU process

In chapter 6 we followed the first subprocess of the PBU process defined in chapter 5.4.1 however, two deviations from the original process occurred:

- selecting the context,
- searching and understanding the definitions of the process we would like to implement.

The first deviation is an additional step to the first subprocess of the PBU process and it occurred because we needed a context for the case study, it was included only to get acquainted to the case context. In practice the context is given and generates the need for multi-model solution, therefore we do not include it into the PBU process.

The second deviation is also an addition to the PBU process, and it occurred due to practical reasons: we needed a starting point, we needed to know what exactly we want to define. At this situation looking for the definitions seemed the best starting point. It involved 1. a search for definitions, 2. the analysis of definitions and 3. selection of what we need. We think this could be a useful guideline to the PBU process therefore we will discuss it further in chapter 10, "Lessons learned from the case study".

## 6.7. Limitations

In this chapter we discuss limitations of each step we performed in chapter 6.

**Selecting a context** – As any project, the Q-works project also had its constraints in terms of resources and requirements. One such constraint was that the primary goal of the project was the implementation of CMMI process areas, therefore it was necessary to include CMMI in the selection. We consider this a minor issue, since CMMI is one widely accepted quality approach in the field of software process improvement.

**Selecting a process to be improved / created** – at this step we needed to select any software process to be improved. The scope was narrowed by the context because we were required to define CMMI-conform processes.

We still had the option to select any process within the CMMI scope (related to at least one of the 22 CMMI process areas). We could select other processes e.g. the process of Project Planning, Measurement and Analysis or Requirements Management and many others, but a choice had to be made. This choice meets our guidelines. Of course if we would have chosen another process, this choice would have affected next steps e.g. the step of selecting quality approaches.

**Selecting quality approaches to be used in unification** – This selection was affected by the previous steps (selecting the context and the process). The context required that CMMI should be included in the set of selected quality approaches and the selected peer review process caused the inclusion of two peer-review related quality approaches.

Regarding the search: we tried to perform a comprehensive search for peer reviews within the quality approaches we had access to. However, there might be other quality approaches involving peer reviews which include different terminology, or are present in different databases we have not checked.

**Selecting a process representation format** – In the step of selecting a process representation format we selected a textual and graphical process representation. The selection of textual representation format was done in collaboration with the Q-works team and it was developed to be processed by software developers. This means that the representation format was not chosen for reading purposes, for that purposes other formats e.g. a format described in 5.3 could have been more appropriate. In selecting a graphical representation format the main guideline was the intelligibility for which BPMN seemed an excellent solution. Projects with other scope and goal may select other textual and graphical representation formats.

# 6.8. Conclusion

Chapters 6-8 are intended to answer Q7 “How can we provide proof of concept for the PBU framework” In order to answer this question a case study is performed. Discussions in this chapter were part of performing RS7 “Perform a case study on the PBU framework”. In chapter 6 we presented preparatory activities for the case study execution. After selecting the context of the case, all activities were performed according to the first subprocess “Selection of processes, quality approaches and process representation” of a PBU process (chapter 5.4.1), namely:

- 1.1 select processes to be improved/created (chapter 6.3),
- 1.2 select quality approaches to be used (chapter 6.4),
- 1.3 select process (textual and graphical) representation format (chapter 6.5).

In chapter 6.2 we introduced the Q-works project, the context of the case, in chapter 6.3 we selected the peer review process to be developed. In chapter 6.4 we selected 3 quality ap-proaches to be used for creating the unified peer review process, while in chapter 6.5 we selected a textual and graphical process representation format. In chapter 6.7 we discussed limitations of steps made in chapter 6.

In the next chapter the execution of the case study will be discussed aligned with the activities of second and third subprocess of the PBU process “Analysis of quality approaches” (introduced in chapter 5.4.1.3) and respectively “Deriving process from quality approaches” (introduced in chapter 5.4.3).

# 7. EXECUTING THE CASE STUDY[5]

In chapters 6-8 the research step 7 (RS7) “Perform a case study on the PBU framework” is discussed. In chapter 6 we presented the preparation of the case study, selecting the context, a process, quality approaches as well as a textual and a graphical process representation format. These steps were performed according to the first subprocess of the PBU process introduced in chapter 5.4.1. In this chapter we continue applying the PBU framework and we will perform activities of PBU subprocesses “Analysis of quality approaches” introduced in chapter 5.4.2 and “Deriving process from quality approaches” described in 5.4.3.

## 7.1. Approach

As a part of PBU process, peer review related quality approach elements and quality approach element instances need to be identified then used in creating a unified peer review process. The structure of this chapter follows the logical flow of the second and third subprocess of the PBU process.

RS7-2 “Executing the case study” can be broken down to the following operational steps:

1. Analysis of the three selected quality approaches (chapters 7.2, 7.4 and 7.6),
2. Deriving process from quality approach (chapters 7.3, 7.5 and 7.7),
3. Creating a unified peer review process (chapter 7.8),
4. Understanding deviations from the PBU process (chapter 7.9).

Limitations are discussed in 7.10 and conclusion in 7.11.

## 7.2. Analysis of quality approaches: first iteration

In this chapter we follow the activities of the first iteration of the “Analysis of quality approaches” PBU subprocesses.

[5] Results of this chapter were presented on the Q-Works workshop (Kelemen, 2010a, 2010b).

### 7.2.1. Select a quality approach to analyse: CMMI

This activity (Select a quality approach to analyse) is defined in chapter 5.4.2.1. Previously we selected 3 quality approaches which will be used in building a multi-model peer review process. These were: CMMI for Development v1.2, IEEE 1028-2008 and the peer review descriptions of Process Impact. As our primary goal in Q-works project was to implement a CMMI-DEV v1.2 conform processes, therefore CMMI is the first quality approach which will be analysed.

### 7.2.2. Analyse characteristics of quality approach

The purpose of the "Analysis of quality approaches" subprocess is to understand the scope, terminology, structure and other characteristics of quality approaches, as well as identifying quality approach elements and their instances which could be mapped to process elements.

**Scope**

"CMMI® (Capability Maturity Model® Integration) models are collections of best practices that help organizations to improve their processes". " Best practices in the model focus on activities for developing quality products and services to meet the needs of customers and end users" (CMMI Product Team, 2006).

**Terminology – Peer reviews**

CMMI-DEV v1.2 defines the peer reviews as follows: "*The review of work products performed by peers during development of the work products to identify defects for removal.*" We already reviewed definitions in the selection phase of the PBU process (see chapter 6.4.3)

**Quality approach structure and elements**

In chapter 4 we analysed the structure and elements of a number of quality approaches. Chapter 4.4.3 includes the discussion of CMMI v1.3. The Q-work project was launched in 2009, when v1.2 was the actual version of the model, thus the 1.2 version of CMMI was used. V1.3 contains similar elements to v1.2. Figure 18 represents the structure of CMMI v1.2. Comparing the two versions of CMMI, (see Figure 18 for the structure of v1.2 and Figure 4 for structure of v1.3) the only structural difference is that CMMI v1.2 uses the term "typical work product" while in CMMI v1.3 this term was replaced by "example work product". Both terms denote possible work products when implementing a CMMI practice. We think this is a minor difference between the two versions so we do not repeat the analysis and the description of the elements. These can be found in chapter 4.4.3.

In 4.4.3 we defined CMMI elements which can be mapped to process elements. From the peer review related part of CMMI (Verification Process Area, Specific Goal 2) the following quality approach elements can be used:

- specific goal,
- specific practice,
- subpractice,
- typical work products.

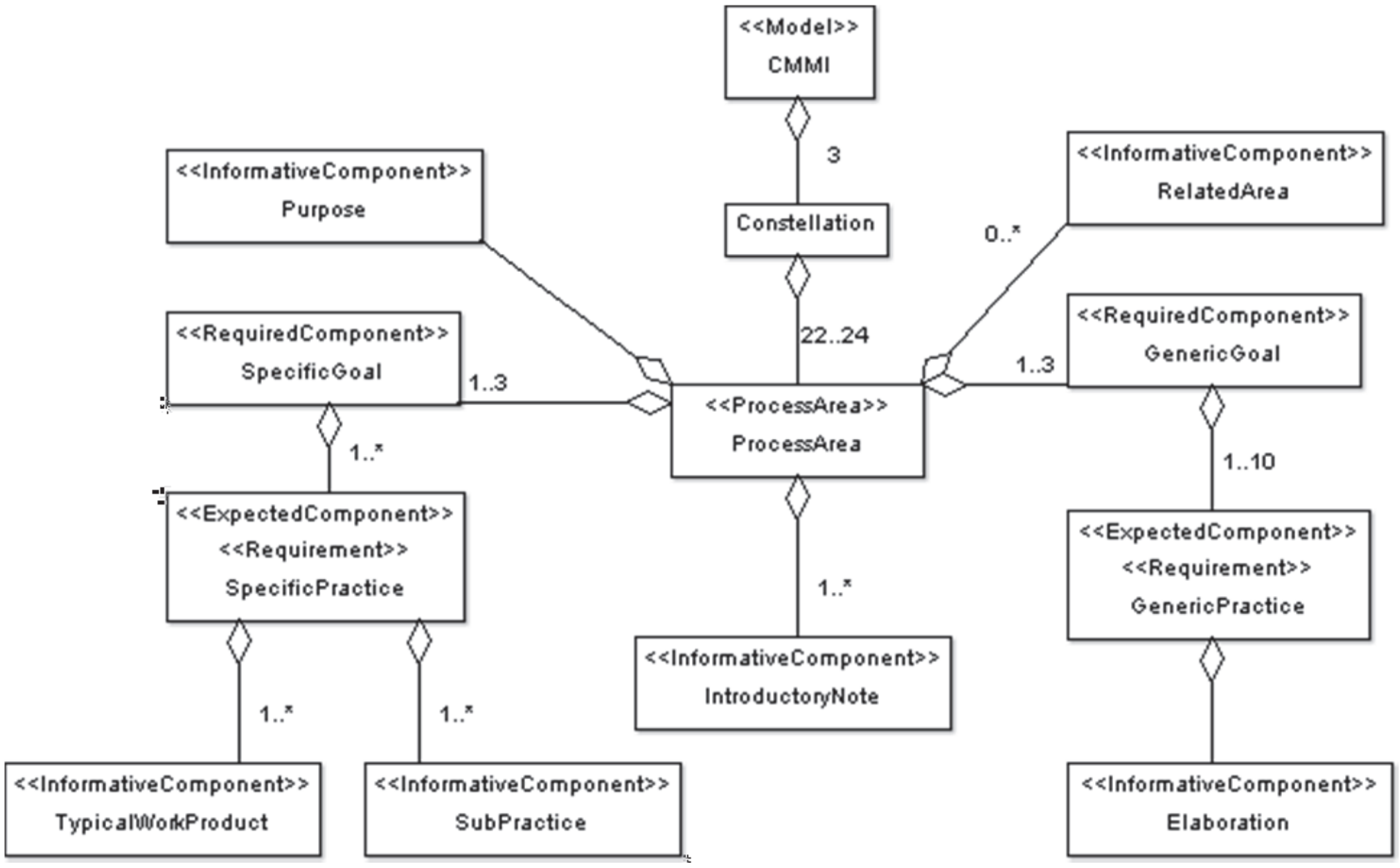

Figure 18 – The structure of CMMI-v1.2

From these quality approach elements, the implementation of specific practices is required by CMMI. Lower-level elements (see Figure 18) are only informational components and are available to help the implementation of practices. Due to the structure and goals of CMMI, more details are present on lower levels (typical work products and subpractices). In order to build a process we need more details. Therefore, we will analyse all the quality approach elements related to peer reviews on specific goal, specific practice, subpractice, and typical work product level.

**Granularity, size and complexity**

35 peer review-related quality approach element instances were identified in CMMI-DEV v1.2. These consist of 1 specific goal, 3 specific practices, 11 typical work products, 21 subpractices and a list of 4 roles. The specific goal, specific practices and some of the subpractices have descriptions which were identified as quality approach elements. These are not included in the tables because of their length. In total 100% of elements of Specific Goal 2 of Verification Process Area were identified, thus 100% of peer review-related requirements of CMMI-DEV v1.2 were covered. More details of handling the content (the identification of quality approach element instances) can be found in the next chapter (7.2.3).

## 7.2.3. Identify quality approach element instances – content

**Specific goals and specific practices**

In CMMI there are 4 requirements on performing the peer review process defined in Verification process area:

- VER SG 2: Perform Peer Reviews,
- VER SP 2.1: Prepare for Peer Reviews,
- VER SP 2.2: Conduct Peer Review,

- VER SP 2.2: Analyze Peer Review Data.

We will identify quality approach element instances from VER SG2 “Perform Peer reviews” chapter of CMMI-DEV v1.2 (pages 502-505).

**Subpractices and typical work products**

Due to the coarse granularity of CMMI, besides the instances of required specific goal, expected specific practices, informative components (subpractices and typical work products) were also identified.

Table 23 contains the subpractices related to peer reviews including their unique identifier. IDs are defined to make the identification of quality approach element instances easier. For example the ID “VER SP2.1 SUBP1” was given to the first subpractice of specific practice 2.1 in Verification process area. It can be observed that we used the CMMI acronyms for generating the IDs, this way it is easy to identify each CMMI element instance.

Table 23 – Subpractices related to peer reviews in CMMI-DEV v1.2

| Subpractice instance | ID |
|---|---|
| Determine what type of peer review will be conducted. | VER SP2.1 SUBP1 |
| Define requirements for collecting data during the peer review. | VER SP2.1 SUBP2 |
| Establish and maintain entry and exit criteria for the peer review. | VER SP2.1 SUBP3 |
| Establish and maintain criteria for requiring another peer review. | VER SP2.1 SUBP4 |
| Establish and maintain checklists to ensure that the work products are reviewed consistently. | VER SP2.1 SUBP5 |
| Develop a detailed peer review schedule, including the dates for peer review training and for when materials for peer reviews will be available. | VER SP2.1 SUBP6 |
| Ensure that the work product satisfies the peer review entry criteria prior to distribution. | VER SP2.1 SUBP7 |
| Distribute the work product to be reviewed and its related information to the participants early enough to enable participants to adequately prepare for the peer review. | VER SP2.1 SUBP8 |
| Assign roles for the peer review as appropriate. | VER SP2.1 SUBP9 |
| Prepare for the peer review by reviewing the work product prior to conducting the peer review. | VER SP2.1 SUBP10 |
| Perform the assigned roles in the peer review. | VER SP2.2 SUBP1 |
| Identify and document defects and other issues in the work product. | VER SP2.2 SUBP2 |
| Record the results of the peer review, including the action items. | VER SP2.2 SUBP3 |
| Collect peer review data. | VER SP2.2 SUBP4 |
| Identify action items and communicate the issues to relevant stakeholders. | VER SP2.2 SUBP5 |
| Conduct an additional peer review if the defined criteria indicate the need. | VER SP2.2 SUBP6 |
| Ensure that the exit criteria for the peer review are satisfied. | VER SP2.2 SUBP7 |
| Record data related to the preparation, conduct, and results of the peer reviews. | VER SP2.3 SUBP1 |
| Store the data for future reference and analysis. | VER SP2.3 SUBP2 |
| Protect the data to ensure that peer review data are not used inappropriately. | VER SP2.3 SUBP3 |
| Analyze the peer review data. | VER SP2.3 SUBP4 |

Table 24 contains the typical work products related to peer reviews and their IDs. IDs were given in a similar manner to the subpractice IDs, the only difference is that typical work prod-uct is denoted by the “TWP” acronym.

Table 24 – Peer review-related typical work products in CMMI-DEV v.12

| Typical work product instance | ID |
|---|---|
| Peer review schedule | VER SP 2.1 TWP1 |
| Peer review checklist | VER SP 2.1 TWP2 |
| Entry and exit criteria for work products | VER SP 2.1 TWP3 |
| Criteria for requiring another peer review | VER SP 2.1 TWP4 |
| Peer review training material | VER SP 2.1 TWP5 |
| Selected work products to be reviewed | VER SP 2.1 TWP6 |
| Peer review results | VER SP 2.2 TWP1 |
| Peer review issues | VER SP 2.2 TWP2 |
| Peer review data | VER SP 2.2 TWP3 |
| Peer review data | VER SP 2.3 TWP1 |
| Peer review action items | VER SP 2.3 TWP2 |

**Roles**

There are also roles listed in the text for which responsibilities are not defined, these roles are: Leader, Reader, Recorder and Author.

## 7.3. Deriving process from quality approaches: first iteration

Since we do not have a peer review process already available, in this chapter we create the process from element instances identified in the first quality approach (CMMI). For doing so, first we map quality approach element instances, and then we derive a process from the mapped quality approach element instances.

### 7.3.1. Map quality approach element instances to process representation elements

In this chapter we map each quality approach element to process elements and process modeling language elements. We mapped CMMI elements to process elements in chapter 4.4.3 which discusses the mapping of quality approach instances to process elements.

**Specific goals**

We start on the top level from the specific goal. There is only one specific goal related to peer reviews: VER SG2 – Perform Peer Reviews. Since this is on the top level, we can map this quality approach element instance to the process of peer reviews itself.

**Specific practices**

The next level in CMMI structure is the specific practice level. From the process point of view processes can have subprocesses. We know that we still have further elements in CMMI below the specific practice level, so we need to map specific practices to a process element which is between the process and activity level. It is logical to map specific practices to subprocesses. Subprocesses can be part of a process and can contain activities.

CMMI includes three peer review related specific practices. All of these can be mapped to subprocesses (prepare for peer reviews, conduct peer reviews and analyse peer review data).

**Subpractices and typical work products**

On the lowest level in the CMMI structure, two quality approach elements can be found: subpractice and typical work product. These are called informative components in CMMI, which means it is not mandatory to use these elements, but since the CMMI is not fine grained, we identify and map these elements as well.

Going through

Table 23, it can be concluded that all the element instances included into the table can be most appropriately mapped to activity process element, while quality approach element instances included in Table 24 can be mapped to the artifact process element.

Table 25 summarises the mapping of quality approach element instances to process elements. Since the next step is the mapping to PML elements and in chapter 6.5 we selected BPMN as a process modeling language; therefore we also map quality approach elements to the elements of BPMN.

The first column of Table 25 contains only quality approach element which are related to peer reviews. This means that several elements of CMMI were excluded such as related process areas, generic practices and generic goals among many others.

The second column of the Table 25 contains process elements identified in Table 16, while the third column contains BPMN element to which process elements can be mapped. In the third column only those BPMN elements are included to which a mapping is possible. Table 28 and Table 30 in the next two iterations will be developed in the same way as Table 25.

Figure 18, chapter 7.2.3 and Table 25 show that some elements are missing from CMMI which are needed to build a process, these are:

- elements mappable to entry criteria, exit criteria,
- elements mappable to responsibilities,
- element relations (e.g. the order of activities or role-activity relations).

In the next chapter we will create a process from quality approach elements mapped to process modeling language elements available.

Table 25 – Mapping peer review-related CMMI v1.2 elements to process and process modeling language elements

| **Quality approach element (based on chapter 7.2)** | **Process element (based on Table 16)** | **BPMN element (based on the element set of BPMN)** |
|---|---|---|
| Specific goal name | Process name | Process name |
| Specific goal description | Process description | Process description |
| Specific practice name | Process name | Subprocess name |
| Specific practice description | Process description | Subprocess description |
| Subpractice | Activity | Activity,<br>Gateway |
| Typical work product | Data object | Data object |
| Role | Role | Swim lane name |
| *Not present* | Entry/exit criteria | *Not present* |

| Quality approach element (based on chapter 7.2) | Process element (based on Table 16) | BPMN element (based on the element set of BPMN) |
|---|---|---|
| Specific goal name | Process name | Process name |
| *Not present* | Responsibility | Swim lane description |
| *Not present* | Element relations | Various relations |

### 7.3.2. Create/refine process

Since we do not have any starting process we need to create one. In order to understand the first approach on peer reviews we will create a process by using the element instances of the first quality approach (CMMI).

We create a CMMI-conform peer review process in a top-down way, similarly to the mapping process discussed in the previous chapter. If we create the process in a top-down way, first we shall use the specific goal which we mapped to the process itself. The process name could be identical to the goal name "Perform Peer Reviews", or could be anything else reflecting that the process is about peer reviews. For preserving simplicity, in this process we use the quality approach element instance names.

On the next level, 3 specific practices were identified and mapped to 3 subprocesses. If we would like to build a valid BPMN process from these three subprocesses we need to add further process elements such as process start, process end and process flow between subprocesses.

The order of specific practices is not provided in CMMI, therefore we create a logical order. The logical flow of any process having preparation, conduction and analysis subprocesses suggests us that it should be started with a preparation then conduction and analysis should come. According to this logic, Figure 19 shows a high level BPMN representation of the peer review process. Process starting, process ending and process flow elements (arrows) are additional information, which were not included in CMMI, but are needed to perform the process and keep the BPMN process valid.

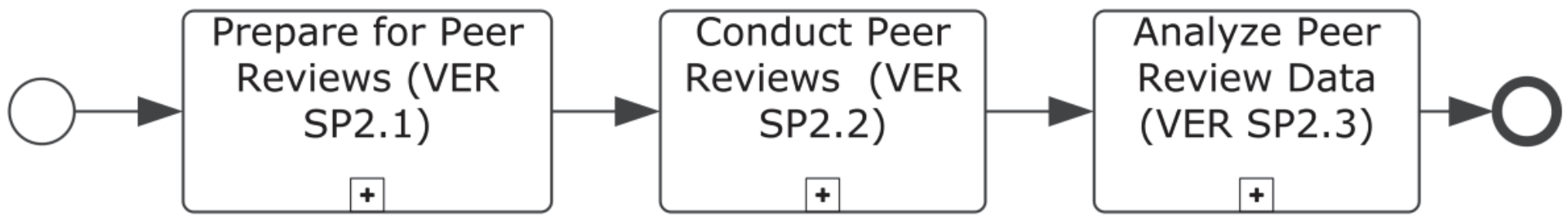

Figure 19 – Perform peer reviews

**Creating the process on subpractice level**

Each specific practice contains subpractices and typical work products which were mapped to activities and data objects in chapter 7.3.1.

At this level we face the very same problem we had at specific practice level: some of the elements are missing which are needed to build a process flow. One further problem is that we do not have information regarding which typical work product can be used as an input and which will serve as an output, and to which activities to connect them to. In fact all elements on this level are examples and all the element relations are missing. At this level, similarly to the previous one we tried to define element relations based on simple logic, knowing that these can be added in many different ways, but in creating a working process decisions should be taken.

Such decisions were:

- use all typical work products and subpractices,
- create a process flow from subpractices by connecting them,
- connect typical work products to subpractices.

All our decisions resulted in additional elements and additional information in the process model. Subpractice and typical work product level information are not needed at a compliance check because they are not required by CMMI, however, we need them to understand the process.

Looking at subpractices of SP 2.1 we have seen two types of subpractices: SUBP 3-5 begin with "Establish and maintain". Establishment and maintenance is usually repetitive, therefore we separated these subpractices from the flow and we indicated that these can be performed parallel with the simple activities (see Figure 47).

**Connecting data objects created from typical work products to activities**

In order to have a process we needed to connect data objects to activities. Here we also tried to be logical e.g. data object and activities were connected in case if similarities were found among them. For example we concluded that the data object "peer review entry criteria" should be connected to "establish and maintain entry and exit criteria for the peer review" activity. Similarly, the activity "ensure that the work product satisfies the peer review entry criteria prior to distribution" should be connected with "entry and exit criteria for work products".

We also used gateways in representing activities which outputs a decision influencing the process flow. One such gateway can be seen in Figure 48. Activity "Conduct an additional peer review if the defined criteria indicate the need" influences the process flow: if a new peer review is needed, the process should be restarted, otherwise the process will continue.

We created a table in which we maintain the identified elements and their relations. All the mapped elements are included in Table 49. A graphical process representation of this table is included into Appendix O.

### 7.3.3. Verification of the mapping

Since this is the first mapping of quality approach elements to the process no consistency check of the process is needed to comply with previous mappings.

### 7.3.4. Decision: Are further quality approaches needed?

Since many elements and element relations are missing from CMMI, the analysis of further quality approaches is needed.

## 7.4. Analysis of quality approaches: second iteration

In this chapter we follow the activities of the second iteration of the “Analysis of quality approaches” PBU subprocesses.

### 7.4.1. Select a quality approach to analyse

This selection activity is defined in chapter 5.4.1.3. As we have seen, CMMI does not provide enough information to build a decent peer review process. Thus, we need to analyse further quality approaches. In the selection subprocess we selected 3 quality approaches: CMMI, IEEE 1028 and the peer review descriptions of ProcessImpact.com. In this chapter we analyse the IEEE 1028 approach on peer reviews.

### 7.4.2. Analyse characteristics of quality approach: IEEE 1028

**Scope**

The scope of IEEE 1028 is to provide “minimum acceptable requirements for systematic software reviews, where “systematic” includes the following attributes: team participation, documented results of the review, documented procedures for conducting the review” (IEEE, 2008b).

“The purpose of this standard is to define systematic reviews and audits applicable to software acquisition, supply, development, operation, and maintenance. This standard describes how to carry out a review. Other standards or local management define the context within which a review is performed, and the use made of the results of the review. Software reviews can be used in support of the objectives of project management, system engineering (for example, functional allocation between hardware and software), verification and validation, configuration management, quality assurance, and auditing. Different types of reviews reflect differences in the goals of each review type. Systematic reviews are described by their defined procedures, scope, and objectives” (IEEE, 2008b).

**Terminology**

“The definitions, requirements, and procedures for the following five types of reviews are included” within IEEE 1028: Management reviews, Technical reviews, Inspections, Walkthroughs, Audits” (IEEE, 2008b). Table 26 includes the definitions of review and its 5 types discussed in IEEE 1028.

Table 26 – Peer review types and their definitions in IEEE 1028, source: (IEEE, 2008b)

| Peer review type | Definitions in (IEEE, 2008b) |
|---|---|
| Audit | “An independent examination of a software product, software process, or set of software processes performed by a third party to assess compliance with specifications, standards, contractual agreements, or other criteria.” |
| Inspection | “A visual examination of a software product to detect and identify software anomalies, including errors and deviations from standards and specifications.<br>NOTE – Inspections are peer examinations led by impartial facilitators who are trained in inspection techniques. Determination of remedial or investigative action for an anomaly is |

| Peer review type | Definitions in (IEEE, 2008b) |
|---|---|
| | a mandatory element of a software inspection, although the solution should not be determined in the inspection meeting.” |
| Management review | “A systematic evaluation of a software product or process performed by or on behalf of management that monitors progress, determines the status of plans and schedules, confirms requirements and their system allocation, or evaluates the effectiveness of management approaches used to achieve fitness for purpose.” |
| Review | “A process or meeting during which a software product, set of software products, or a software process is presented to project personnel, managers, users, customers, user representatives, auditors or other interested parties for examination, comment or approval.” |
| Technical review | “A systematic evaluation of a software product by a team of qualified personnel that examines the suitability of the software product for its intended use and identifies discrepancies from specifications and standards.<br>NOTE – Technical reviews may also provide recommendations of alternatives and examination of various alternatives.” |
| Walk-through | “A static analysis technique in which a designer or programmer leads members of the development team and other interested parties through a software product, and the participants ask questions and make comments about possible anomalies, violation of development standards, and other problems.” |

After reviewing definitions in Table 26 and peer review descriptions in the standard we have chosen to implement Inspections, because this is the most rigorous and most formal peer review type. We have chosen the most formal approach, because after this implementation it is easier to simplify a complex process and implement less formal types, than to implement a less formal review and then looking for missing parts to make it more formal.

**Quality approach structure and elements**

The structure and elements of IEEE 1028-2008 is discussed in chapter 4.4.5. Since the same version was used in the case study, we do not repeat here the previous discussion on the structure and elements. Summarizing, we identified the following elements: 1. Introduction to review, 2. Responsibilities, 3. Input, 4. Entry criteria, 5. Procedures, 6. Exit criteria, 7. Output. These elements were mapped to 1. Purpose, 2. Role and Responsibility, 3. Artifact, 4. Entry criteria, 5. Process and Activity, 6. Exit criteria and 7. Artifact.

**Granularity, size and complexity**

83 inspection-related quality approach element instances were identified in IEEE-1028. These consist of 1 “Introduction to inspections”, 5 roles and 5 role descriptions (responsibilities), 13 inputs, 15 outputs, a list of entry criteria, a list of exit criteria, 33 activities and 9 subprocesses. In total 100% of elements of the chapter 6. Inspections were identified. More details on handling the content of IEEE and the identification of quality approach element instances can be found in the next chapter (7.4.3).

## 7.4.3. Identify quality approach element instances – content

After understanding the 5 different review types discussed in IEEE-1028, we have chosen to implement the Inspection process. In this chapter we identify quality approach element instances related to Inspections found in IEEE-1028, discussing these by instances.

In this chapter we will identify quality approach element instances from chapter 6, “Inspections” of IEEE 1028 (pages 16-25).

**Introduction to review**

The introduction of inspection is included into the “Introduction to Inspections” chapter of the standard. This is general textual information, therefore we did not separate it further and it can be used in the textual introduction of the Peer review/Inspection process. For the single instance of this element see Appendix I.

**Roles and Responsibilities**

Roles and Responsibilities related to the Inspection process are included in the chapter 6.2 Responsibilities of IEEE 1028-2008. Here we identified subchapter names as roles and chapter text as responsibilities. We identified 5 roles: inspection leader, recorder, reader, author and inspector. These roles and their descriptions (responsibilities) are included into Table 48.

**Inputs and Outputs**

Inputs and outputs of the Inspection process are textually described in chapters 6.3 Input and 6.7 Output of the standard. Additionally, the level of conformance in case of inputs, outputs and procedures varies. “The word “shall” is used to express a requirement, “should,” to express a recommendation, and “may,” to express alternative or optional methods of satisfying a requirement” (IEEE, 2008b). Since these are described in chapter 1.4 “Conformance” of IEEE we call conformance keywords and conformance levels. Table 27 summarizes conformance levels and keywords of IEEE 1028.

Table 27 – Conformance keywords and related levels in IEEE 1028-2008

| Conformance keyword | Conformance level |
|---|---|
| shall | mandatory, requirement |
| should | recommendation |
| may | optional |

Inputs and outputs were identified taking into the account conformance levels and keywords. Table 42 contains the text of the Input and Output chapters of the standard, the input/output element instances identified based on the text and an ID for all inputs and outputs identified. IDs were defined to make the identification of the quality approach instance easy, for example the ID “IEEE 1028 In should a” contains the following information: “IEEE 1028”- the element instance is part of IEEE 1028. The word “In” – shows that it is an input, while in other cases “Out” is used for outputs. “should” is a conformance keyword, denoting the level of conformance required and finally “a” identifies the concrete input/output element instance. It is the same letter as it is used in the enumeration in the text of the standard.

**Entry and exit criteria**

Similarly to the quality approach elements discussed previously, entry and exit criteria are also textually described in IEEE 1028 in chapters 6.4 and 6.6. In BPMN – the process modeling language we choose for this project these elements are not graphically represented, therefore we do not see the need to process the instances of entry and exit criteria. All IEEE 1028 entry and exit criteria related to Inspections can be found in Table 43 without any transformation.

**Procedures, Data collection, Improvement**

IEEE 1028 contains elements which can be mapped to Activities and subprocesses as it was discussed previously. These elements are textually described and included in chapters 6.5 Procedures, 6.8 Data collection and 6.9 Improvement. We followed a similar identification of element instances included in these chapters to the input/output element instances. Results of the identification can be found in Table 44.

# 7.5. Deriving process from quality approaches: second iteration

Due to the high amount of quality approach elements identified, we need to model the Inspection process solely based on IEEE 1028. In this chapter we create a process from element instances identified in IEEE 1028-2008. For doing so, first we map quality approach element instances to process and process modeling language elements, and then we derive a process from the mapped quality approach element instances.

## 7.5.1. Map quality approach element instances to process representation elements

In this chapter we map each quality approach element instance we identified in the previous chapter to process elements and process modeling language elements. We mapped IEEE 1028-2008 elements to process elements in chapter 4.4.5; this chapter discusses the mapping of quality approach instances to process elements.

**Introduction to inspections**

The “Introduction to inspections” is a textual element of the standard. Therefore it was not sliced to further parts, but directly mapped to introduction of process description.

**Roles and responsibilities**

Role names were mapped to process roles. Process roles are represented as swim lanes in BPMN. Responsibilities are explicitly stated in role descriptions. Role descriptions were mapped to process responsibilities and BPMN swim lane descriptions.

**Inputs and outputs**

Inputs and outputs were mapped to artifacts. Artifacts are mapped to BPMN data objects.

**Entry, exit criteria**

Since there is no graphical BPMN element for entry and exit criteria, these quality approach elements were not sliced further, but were simply kept as part of textual description of the process.

**Procedures, Data collection, Improvement**

Since procedure names are on the highest level after the process level, procedure names were mapped to subprocesses. After procedure names, procedure descriptions were mapped to activity and gateway descriptions. In certain cases, procedures contain subchapters, for example chapter “6.5.6 Examination” includes 5 subchapters. These subchapters were mapped to activities. One subchapter of “6.5.6 Examination”, subchapter 6.5.6.3 Review software product and record anomalies” was further sliced and sentences were mapped to activities. This was done because sentences of this chapter contain multiple roles and we considered easier to understand if it is represented graphically. (This will be further discussed in the next chapter create/refine process.)

Chapters “Data collection” and “Improvement” were also mapped to subprocesses and their content were analysed, sliced and their parts were mapped to activities.

Table 28 – Mapping inspection-related IEEE-1028 elements to process and process modeling language elements

| **Quality approach element (based on chapter 7.4)** | **Process element (based on Table 16)** | **BPMN element (based on the element set of BPMN)** |
|---|---|---|
| Introduction to Inspections | Process description | Subprocess description |
| Procedure name | Process name | Subprocess name |
| Procedure description | Activity | Activity,<br>Gateway |
| Input, output | Data object | Data object |
| Role | Role | Swim lane name |
| Entry/exit criteria | Entry/exit criteria | *Not present* |
| Role description | Responsibility | Swim lane description |
| *Not present* | Element relations | Various relations |

### 7.5.2. Create/refine process

We have already created a peer review process in chapter 7.3 using CMMI element instances. If we follow the PBU process, now we should map and refine the process with IEEE 1028 element instances. However, after having 83 element instances identified in chapter 7.4, it seems quite difficult to map these elements to the CMMI-based peer review process without understanding the process flow of IEEE 1028 inspection process. A solution for understanding the IEEE 1028 process flow would be first to create an inspection process model solely using element instances of IEEE 1028.

Therefore, in this chapter we create the IEEE 1028 inspection process using only IEEE 1028 element instances, and later, in chapter 7.8 we will create the unified peer review process using the three process models developed using element instances of CMMI, IEEE 1028 and Process Impact.

Similarly to CMMI, in IEEE 1028 we still do not have all the information we would need for creating the process flow. Element relations are not explicitly defined in this standard. For example chapters are ordered; however the explicit order of subprocesses/activities is not included in IEEE 1028. We will use this implicit information to create the order of activities, for example we will create subprocesses in an order as subchapters follow each other unless there is no clear logical reason to reorder them. Since element relation information is missing, processes can be created in many different ways. However, similarly to the previously created CMMI-based peer review process, if we would like to build a process, decisions should be made and supplementary information should be added (e.g. the order of activities).

As we did in the case of CMMI we perform the process creation in a similar manner and we use a top-down approach.

**Introduction to inspections**

As we discussed previously the "Introduction to inspections" is included into textual process descriptions and not included into the graphical representation of the process.

**Entry, exit criteria**

Again, entry and exit criteria are not included into the graphical process representation because they cannot be represented in BPMN. These are included into the general, textual process descriptions.

**Inputs, outputs**

Inputs and outputs can be represented in BPMN as data object. These data objects can be connected to processes, subprocesses or activities and depending on the direction of connection they can serve as inputs or outputs. However, in IEEE 1028 inputs and outputs are provided in a process level and are not connected to other elements. Since our primary goal in this chapter is to understand the process flow, we do not connect inputs and outputs to process, subprocesses or activities. Therefore it is not needed to represent them graphically. We will represent these inputs and outputs later, in creating the unified process.

**Procedures, Data collection, Improvement**

Chapters "Procedures", "Data collection" and "Improvement" are probably the most important chapters of the Inspection process since these chapters describe the process itself. After reading these chapters it can be seen that the latter two chapters Data collection and Improvement can be clearly distinguished from the Procedures chapter.

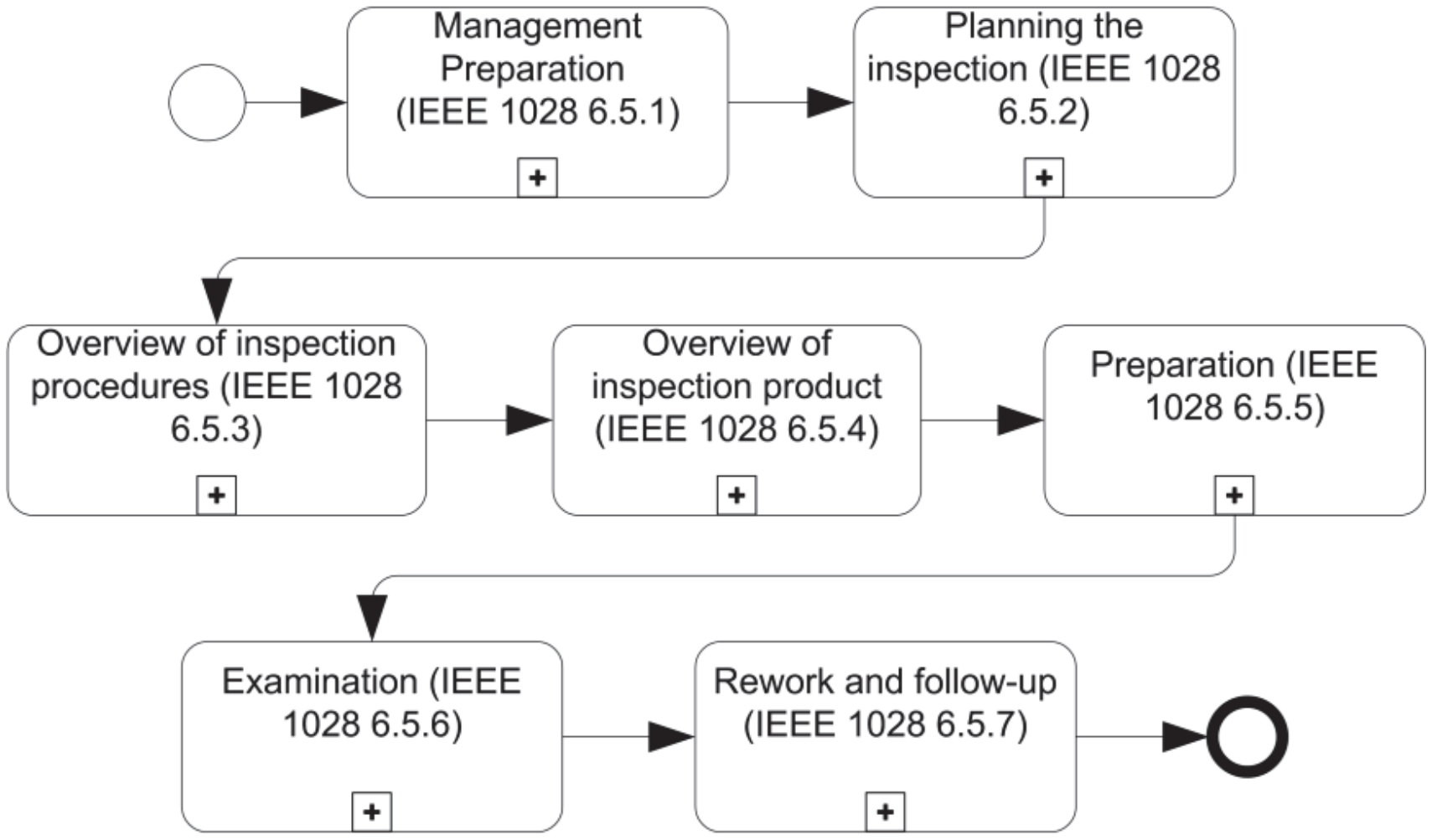

Figure 20 – Inspection process and its subprocesses

While the Procedures chapter describes the main Inspection process which should be performed at each Inspection, the other two chapters describe general process data collection and improvement guides. Data collection and improvement can be performed more independently from the Inspection procedures. For example in the Improvement chapter it is stated that "inspection data shall be analyzed regularly". The frequency of improvement activities are not explicitly stated, and could depend on organizational policies and resources. Thus it might be performed monthly, once in every six month or with other frequency. This suggests us that the improvement is not necessarily has the same process flow as the Procedures chapter.

Figure 20 shows the inspection process and its subprocesses developed from subchapter titles of the Procedures chapter. Graphical representation of chapters Data collection and Improvement are represented in Figure 57 and Figure 58 respectively.

As it can be seen in Figure 20, we created the main process flow as subchapters of the Procedures chapter follow each other. Process starting point, endpoint and process flow are added information on this figure.

**Subchapters of the "Procedures" chapter**

A sign (+) can be seen at each subprocess in Figure 20. This denotes that subprocesses (and thus subchapters of the Procedures chapter) are further decomposed and represented graphically.

**Typical decisions we made in creating the process**

As it is represented in Figure 21, sentences in the text were manually sliced.

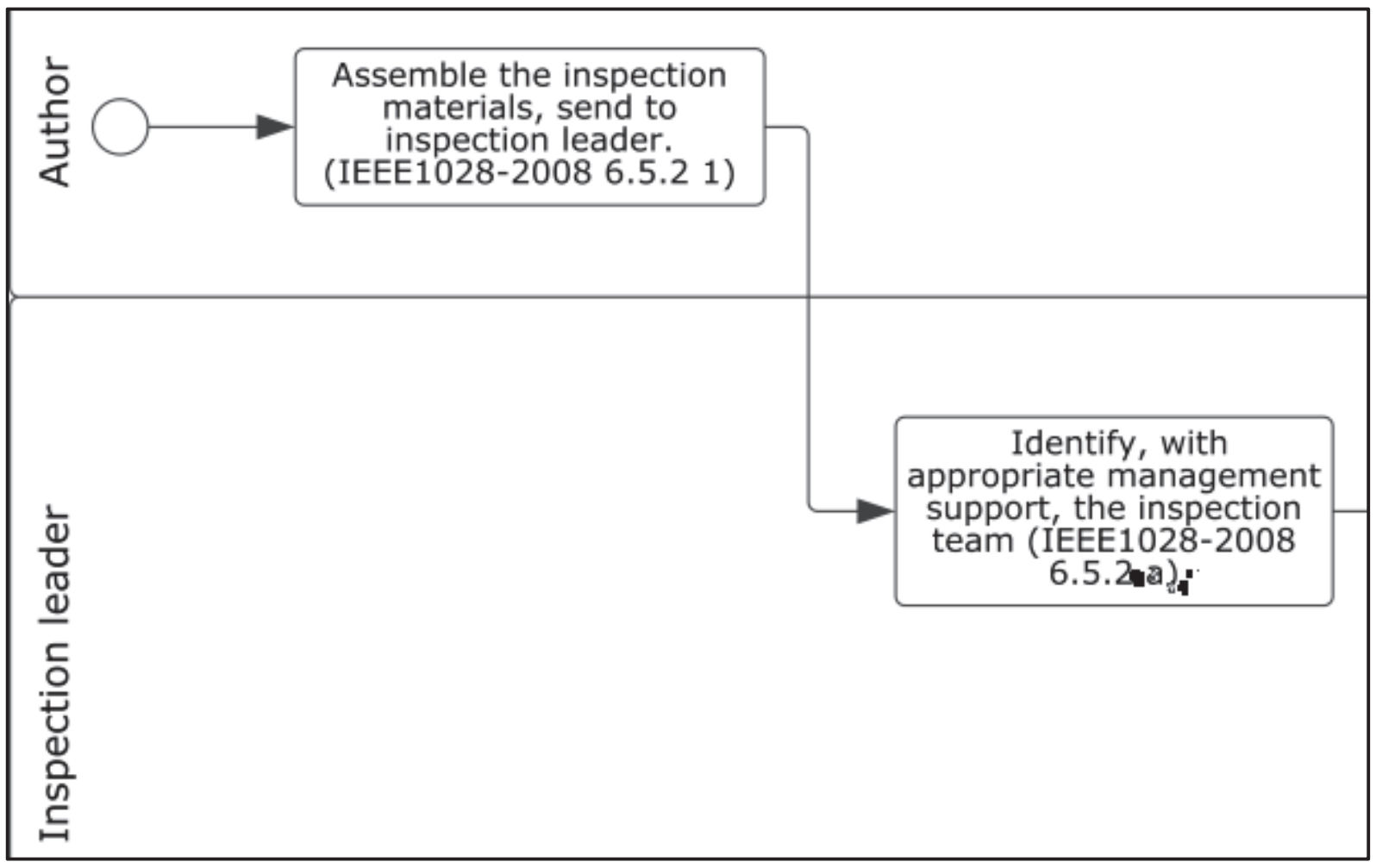

Figure 21 – Planning the inspection – part of the subprocess

For example the text "The author shall assemble the inspection materials for the inspection leader." has been processed and a swimlane for the Author was created according to the first part of the text, then the rest of it was placed as an activity of the swimlane („Assemble the inspection materials, send to inspection leader"). It is not explicitly stated in the text that it should be sent to the Inspection leader, but it is obvious that the Author will not keep the inspection material for him/herself. As the inspection leader is the responsible for the inspection, we added the following text to the activity: "send to inspection leader"). This way we clearly show what should be performed. Of course we could put the send or sharing to a separate activity, but we wanted to develop a process which can be easily perceived with not too many activities. Therefore we kept these two activities in one. The activity has its description which includes the original source text. This way changes in the activity can be checked if they correspond to the original source text.

A second example of processing text manually is the next swimlane. It was created from the following text: "The inspection leader shall be responsible for the following activities: a) Iden-tify, with appropriate management support, the inspection team”. This text starts with the role of inspection leader and continues with a shall-level requirement which can be mapped to an activity. Similarly to the previous activity, it is noted in parentheses that it is mapped to a quality approach element instance which has the id IEEE 1028-2008 6.5.2 a). As it was explained previously this means the quality approach element instances is part of IEEE 1028-2008 standard, was found in the chapter 6.5.2. In this case “a” means the place of the text in an enumeration. If no enumeration is included in the text, letters are used for denoting other parts of the text (e.g. sentences).

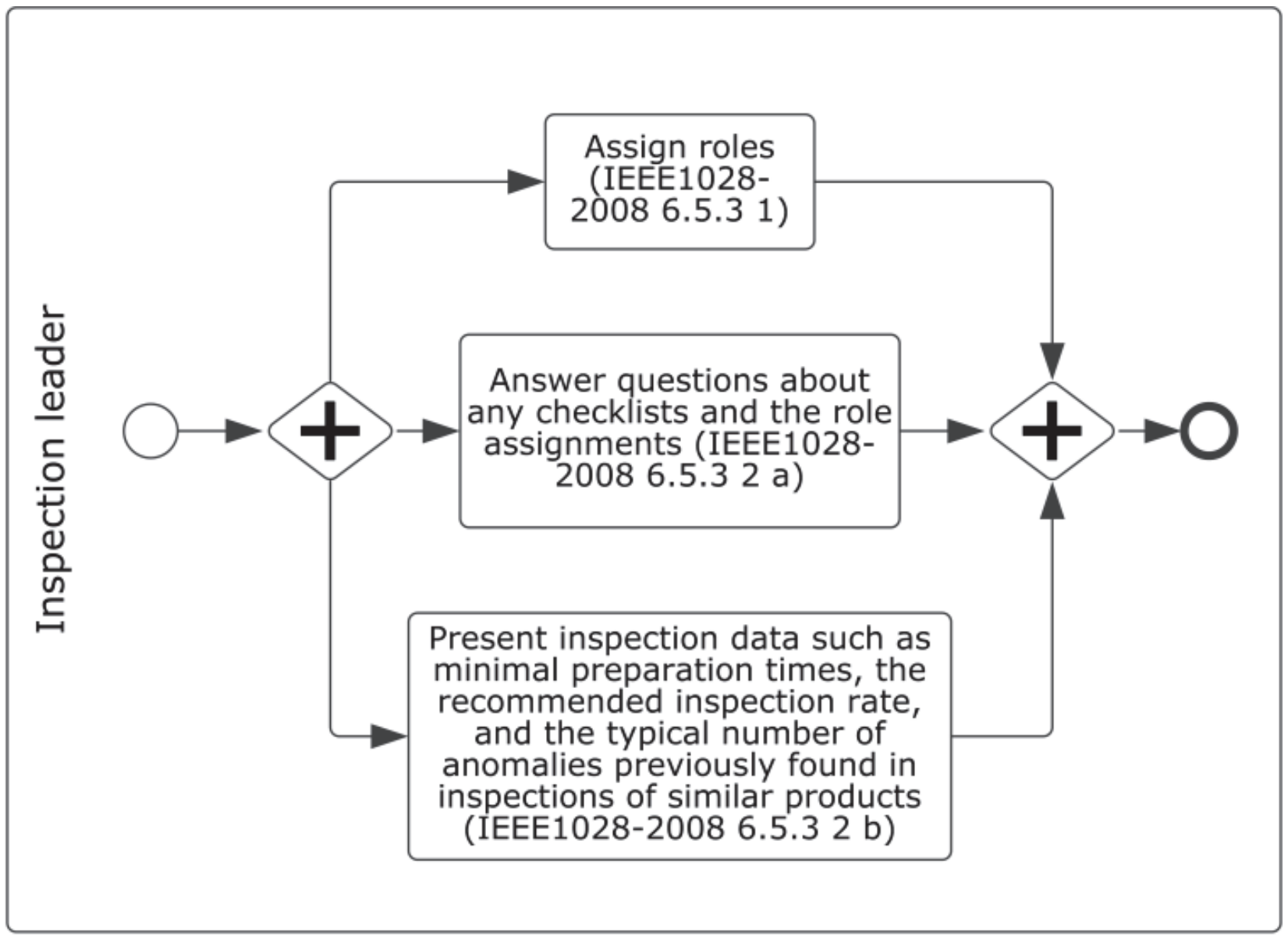

Figure 22 – Example of parallel activities in a process (with “and” gateway)

In certain cases where the order of activities could not be defined, we used parallelisation to describe the process flow. In this case all the activities should be performed, but the order of activities is not important. Figure 22 shows such an example process when the Inspection leader has to perform 3 different activities in any order.

As we summarised in Table 28, procedure descriptions were mapped to activities and gateways. We used gateways to express decisions which influence the process flow (A gateway outputs two or more process flows).

Table 29 and Figure 23 show an example of using gateways in the inspection process. As it can be seen in the first column of the table, an ID was given for the text: IEEE 1028-2008 6.5.5.3 – representing the third part of the chapter 6.5.5. The previous two sentences of chapter 6.5.5 were mapped to other activities. If we look at the text in the second column, we can see it starts with a role “The inspection leader should”, therefore these activities were placed to the Inspection leader swim lane (not shown on the figure). The rest of the text was sliced up and mapped to 3 activities and to one gateway.

Table 29 – Example of creating the process

| ID | Text in IEEE 1028 | Process element instance | Process element |
|---|---|---|---|
| IEEE1028-2008 6.5.5 3 | The inspection leader should classify anomalies as described in 6.8.1 to determine whether they warrant cancellation of the inspection meeting, and in order to plan efficient use of time in the inspection meeting. If the inspection leader determines that the extent or seriousness of the anomalies warrants, the inspection leader may cancel the inspection, requesting a later inspection when the software product meets the minimal entry criteria and is reasonably defect-free. The inspection leader should forward the anomalies to the author of the software product for disposition. | a Classify anomalies | Activity |
| | | b Cancel Inspection, request later inspection | Activity |
| | | b Extent or seriousness of the anomalies is high? | Gateway |
| | | c Forward the anomalies to the author of the software product for disposition. | Activity |

As it can be seen in the third column of the table and on the figure, the text was simplified in order to make it easier to follow. In order to sustain traceability the whole original text is also added to the documentation of the process (not shown on the figure).

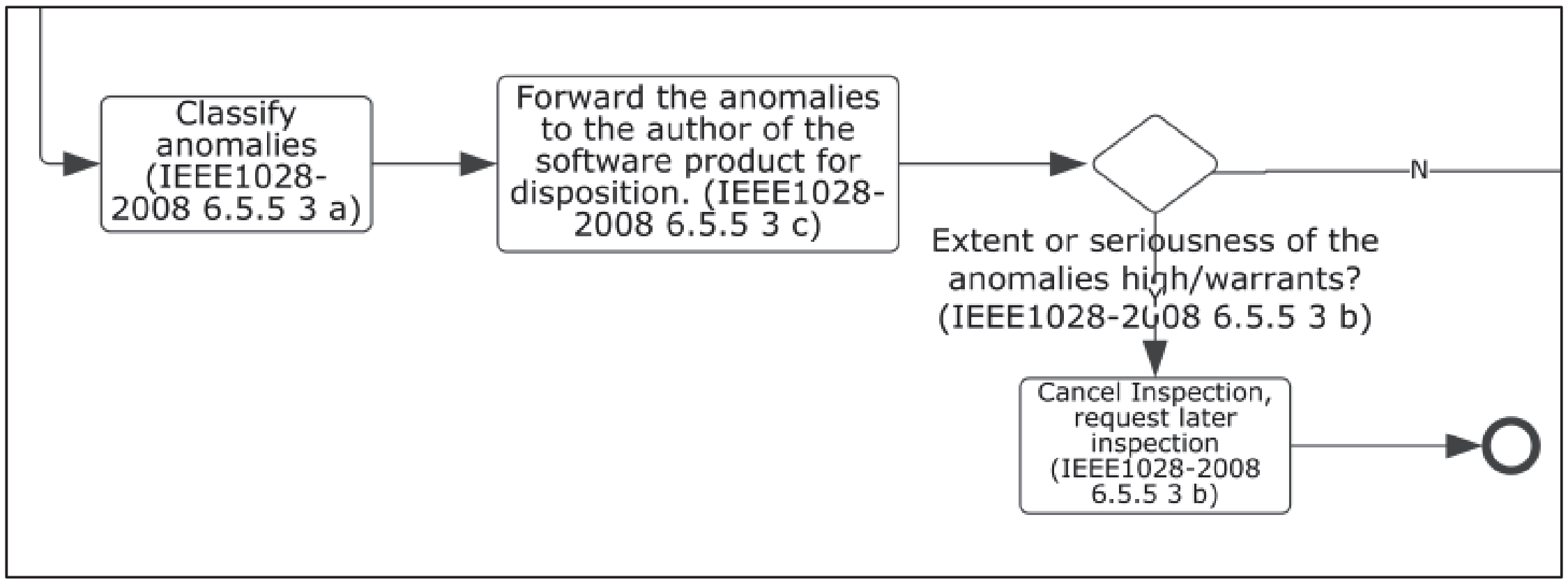

Figure 23 – “Or” gateway in an inspection process

Figure 24 shows another example of creating a process from quality approach elements identified. The only difference on this figure is that we used the ‘Group’ element of BPMN. The Group element serves a visual grouping of process element instances. We used this element to visualise a quality approach element instance (Review Software Product and record anomalies) which was sliced to 4 different, but consecutive process element instances (IEEE1028-2008 6.5.6.3 a-d). This was done for visualisation reasons and to enhance the intelligibility of the process model.

The whole IEEE 1028-2008 based Inspection process model is included in the appendix in Figure 50-Figure 58.

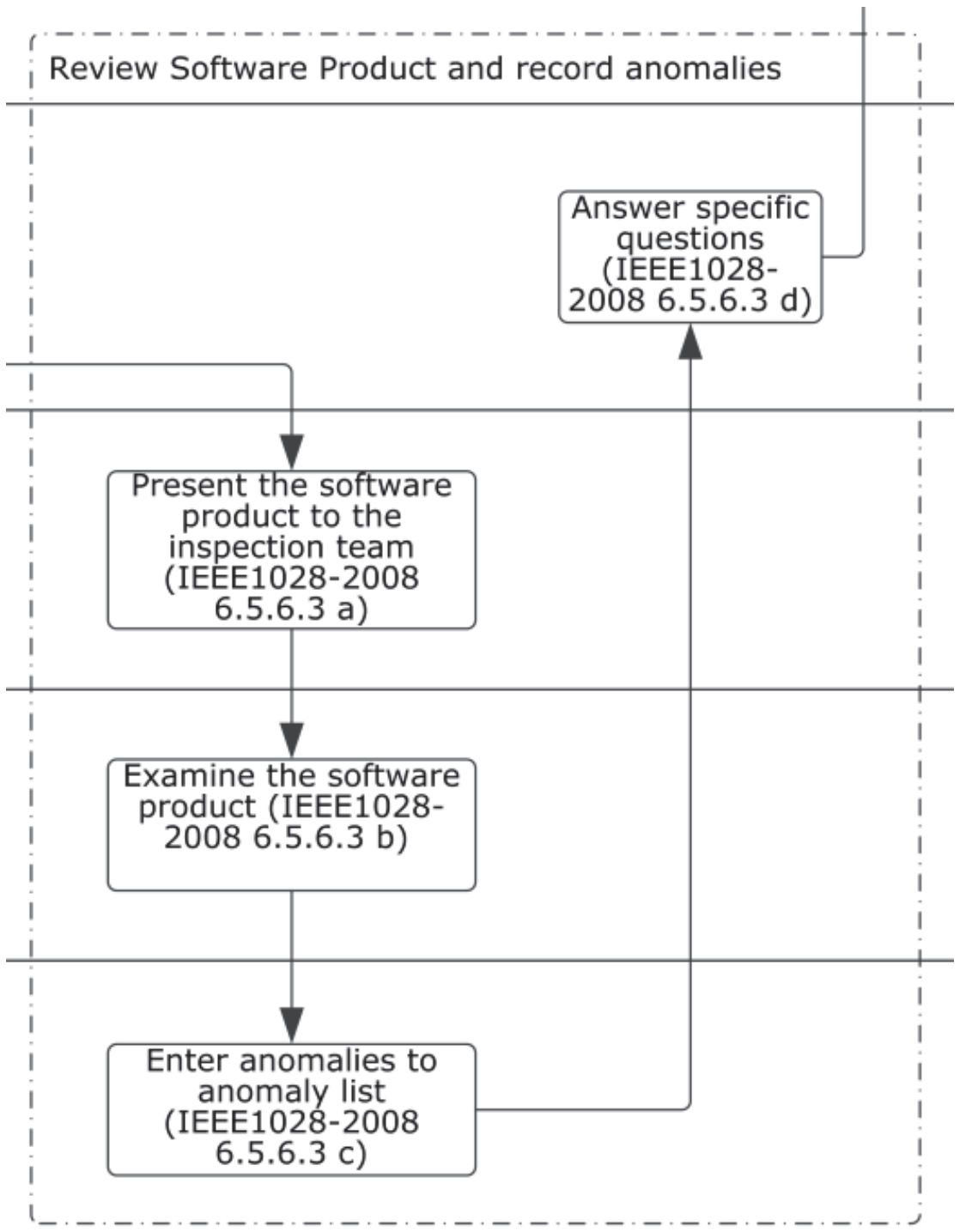

Figure 24 – Making easier to understand the process

### 7.5.3. Verification of the mapping

Since this mapping of quality approach elements to the process included no prior mapping, the consistency check is not needed.

### 7.5.4. Decision: Are further quality approaches needed?

Since many elements and element relations are missing from both CMMI and IEEE 1028-2008, the analysis of further quality approaches is needed.

## 7.6. Analysis of quality approaches: third iteration

In this chapter we follow the activities of the third iteration of the "Analysis of quality approaches" PBU subprocesses.

### 7.6.1. Select a quality approach to analyse

This selection activity is defined in chapter 5.4.1.3. As we have seen that CMMI and IEEE do not provide enough information on the relations among process elements (e.g. process flow, or which artifact is input to which activity) for building the peer review process we continue analysing quality approaches. In the selection subprocess we selected 3 quality approaches: CMMI, IEEE 1028 and the peer review descriptions of Process Impact. In this chapter we analyse the third selected quality approach: peer review descriptions of Process Impact.

### 7.6.2. Analyse characteristics of quality approach

**Scope**

"This document defines an overall peer review process. It includes procedures for conducting inspections and two types of informal peer review, a walkthrough and a passaround, as well as guidance for selecting the appropriate approach for each review" (Process Impact, 2010).

**Terminology – Peer reviews**

Process Impact defines the peer reviews as follows: "In a peer review, co-workers of a person who created a software work product examine that product to identify defects and correct shortcomings. A peer review:

- Verifies whether the work product correctly satisfies the specifications found in any predecessor work product, such as requirements or design documents
- Identifies any deviation from standards, including issues that may affect maintainability of the software
- Suggests improvement opportunities to the author
- Promotes the exchange of techniques and education of the participants." (Process Impact, 2010).

As we selected the Inspections as the most formal review type in 6.3, here we also select the Inspections for analysis.

**Quality approach structure and elements**

The structure and elements of Process Impact peer review description are discussed in chapter 4.4.6. Since the same version was used in the case study, we do not repeat here the previous discussion. Shortly summarizing we have identified the following quality approach elements: 1. Overview, 2. Work aid, 3. Risk assessment guidance, 4. Participants, 5. Procedure, 6. Role, 7. Responsibility, 8. Entry criteria, 9. Phase, 10. Task, 11. Sequence of tasks, 12. Task-role association, 13. Deliverable, 14. Exit criteria, 15. Verification, 16. Data item and 17. Metric. Some of these textual elements can be further sliced if needed or kept as one text.

**Granularity, Size and complexity**

The structure and granularity of this quality approach allowed us to analyse it even in more details: 160 inspection-related quality approach element instances were identified in Process Impact peer review description.

These consist of

- 1 overview chapter,
- 6 work aids,
- 1 risk assessment guidance,
- 1 description of suggested participants for work product types,
- 7 roles and 7 responsibility descriptions,
- a list of entry criteria,
- 6 phases,
- 32 tasks,
- 32 task order information,
- 32 task-role associations,
- 5 deliverables,

- a list of exit criteria,
- 12 data items which can be measured, 12 metrics which can be measured,
- 1 description on process measurement,
- 1 description of process maintenance,
- 1 description on how to record defects,
- 1 description of possible appraisals of inspected work products.

More details on handling the content of peer review descriptions of Process Impact and the identification of quality approach element instances can be found in the next chapter (7.6.3).

### 7.6.3. Identify quality approach element instances – content

After understanding the 3 different peer review types discussed in Process Impact's peer preview process description, we have chosen to implement the Inspection process.
In this chapter we will identify quality approach element instances from chapter 5 "Inspection procedure" of Process Impact peer review descriptions (pages 4-9) as it was planned at the beginning of the case study. In order to show that quality approach element instances can be identified also from other parts of this description we included all of these in Appendix J, but according to our goal, in this case study we will focus only on pages 4-9 of the quality approach. This is because quality approach element instances identified outside pages 4-9 are more measurement related and will be included into the measurement process, after performing this case study.

**Textual element instances**

Due to the capabilities of the PML chosen, we will not use the following elements in the graphical representation: overview, risk assessment guidance, description of suggested participants for work product types, data items metrics, description on process measurement, description of process maintenance, description on how to record defects and the description of possible appraisals of inspected work products. Despite that these elements can be used only in the textual descriptions we identified all instances of these which are included in Table 45. The table contains the element, the element instance text and the element instance ID for each quality approach element instance. The IDs were typically defined to recognize the element instance, for example Entry criteria has the ID "PI ent", where PI stands for Process Impact and "ent" is an abbreviation of entry criteria.

**Entry, exit criteria**

Since there is no graphical BPMN element for entry and exit criteria, these quality approach elements were not sliced further, but were simply kept as separate part of textual description of the process. These element instances are included in Table 45.

**Work aids and deliverables**

Work aids and deliverables of the Inspection process are textually described on pages 1 and 8 of the peer review process definition document. Table 47 contains the quality approach element (work aid or deliverable), the quality approach element instance and the quality approach element instance ID. All work aids and deliverables were given an ID similarly to the previous elements, for example: PI WA1 identifies the first work aid in the inspection process or PI D5 is the ID of the fifth deliverable.

**Roles and responsibilities**

Roles and Responsibilities of the Inspection process are described on pages 4-5 of peer review description of Process Impact. 7 roles and related responsibilities are described: author, moderator, reader, recorder, inspector, verifier and peer review coordinator. These roles and their descriptions (responsibilities) are included into Table 48.

**Phases, tasks, phase-task associations, task order and task-role associations**

The Peer review description of Process Impact includes several elements which can be used in building a process (flow), these are: phases, tasks, the associations of tasks to phases, the order of tasks and task-role associations. Quality approaches (CMMI and IEEE 1028) discussed previously did not include the order of tasks. Results of the identification can be found in Table 46. The table contains subprocess name, task id, task order, task description and the responsible for the task. As it can be seen only tasks have IDs in this table. Of course, the information in Table 46 could be represented in many different ways e.g. giving IDs to each single instance of subprocesses, task orders, responsibilities, associations of subprocess and tasks, tasks and responsibilities; then describing each element with element instances in separate tables, and creating tables of association; or a database structure. However, for simplicity reasons we gave IDs only for task instances. An ID also identifies the phase in which the task was found, e.g. the ID “pi11” identifies the first task of the first phase (there were less than 10 phases). Roles are already identified with a name, task order is only a number and associations are represented by the rows of the table, this way information loss was avoided.

## 7.7. Deriving process from quality approaches: third iteration

Similarly to the two previous quality approaches we will model a process based on the peer review process description of Process Impact. In this chapter we create a process from element instances identified in peer review process description of Process Impact. For doing so, first we map quality approach element instances to process and process modeling language elements and then we derive a process from the mapped quality approach element instances.

### 7.7.1. Map quality approach element instances to process representation elements

In this chapter we map each quality approach element instance we identified in the previous chapter to process elements and process modeling language elements. We mapped Process Impact peer review description elements to process elements in chapter 4.4.6; this chapter discusses the mapping of quality approach instances to process elements.

**Textual element instances**

Due to the capabilities of the PML chosen, we will not use the following elements in the graphical representation: overview, risk assessment guidance, description of suggested participants for work product types, the lists of entry and exit criteria, data items metrics, description on process measurement, description of process maintenance, description on how to record defects and the description of possible appraisals of inspected work products. Since there is no graphical BPMN element for these quality approach elements, they were not sliced further, but were simply kept as part of textual description of the process.

**Roles and responsibilities**

Roles were mapped to process roles. Process roles are represented as swim lanes in BPMN. Responsibilities were mapped to process responsibilities and BPMN swim lane descriptions.

**Work aids and deliverables**

Inputs and outputs were mapped to artifacts. Artifacts are mapped to BPMN data objects.

**Phases, tasks, phase-task associations, phase order, task order and task-role associations**

Since phases are on the highest level after the process level, these are mapped to subprocesses. Tasks are mapped to activities depending on the influence on the process flow. The task order is mapped to activity order and the task role associations are mapped to activity-swimlane associations.

Table 30 – Mapping inspection-related Process Impact elements to process and process modeling language elements

| **Quality approach element (based on chapter 7.6)** | **Process element (based on Table 16)** | **BPMN element (based on the element set of BPMN)** |
|---|---|---|
| Overview, risk assessment guidance, description of suggested participants for work product types, data items, metrics, description on process measurement, description of process maintenance, description on how to record defects, description of possible appraisals of inspected work products | Process description | Process description |
| Phase | Process name | Subprocess name |
| Task | Activity | Activity |
| Work aid, deliverable | Data object | Data object |
| Role | Role | Swim lane name |
| Entry/exit criteria | Entry/exit criteria text | *Not present* |
| Responsibility | Responsibility | Swim lane description |
| Task order | Element relation | Activity order |
| Phase order | Element relation | Subprocess order |
| Task-role association | Element relation | Activity- swim lane association |

## 7.7.2. Create/refine process

After having a high number of quality approach element instances identified in chapter 7.6.3, in order to understand the inspection process of Process Impact, similarly to the IEEE 1028 Inspection process created in chapter 7.5.2, first we create an inspection process model solely using element instances of Process Impact peer review process description. Later, in chapter 7.8 we will create the unified peer review process using the three process models developed using element instances of CMMI, IEEE 1028 and Process Impact.

Comparing to CMMI and IEEE 1028 we have more information in this quality approach. For example there is a task order which can be mapped to activity order or a phase order which can be mapped to subprocess order. We will map all the element instances which can be mapped to graphical representation. As we did in the case of CMMI and IEEE 1028, we perform the process creation in a similar manner and we use a top-down approach.

**Textual descriptions**

Overview, risk assessment guidance, description of suggested participants for work product types, data items, metrics, description on process measurement, description of process maintenance, description on how to record defects, description of possible appraisals of inspected work products, entry and exit criteria are the textual-only part of the process description.

As we discussed previously these elements are included into textual process descriptions, and not included into the graphical representation of the process. These are included into the general, textual process descriptions.

**Work aids and deliverables**

Work aids and deliverables can be represented in BPMN as data objects. These data objects can be connected to processes, subprocesses or activities and depending on the direction of connection they can serve as inputs or outputs. However, similarly to the inputs and outputs in IEEE 1028, work aids and deliverables of this quality approach are provided in a process level and are not connected to other elements. Since our primary goal in this chapter is to understand the process flow, we do not connect work aids and deliverables to process, subprocesses or activities. Therefore it is not needed to represent them graphically. We will represent data objects later, in creating the unified process.

**Phases, tasks, phase-task associations, phase order, task order and task-role associations**

Figure 25 shows the Process Impact's inspection process and its subprocesses. On this figure subprocesses are tailored from phases and subprocess order is tailored from phase order.

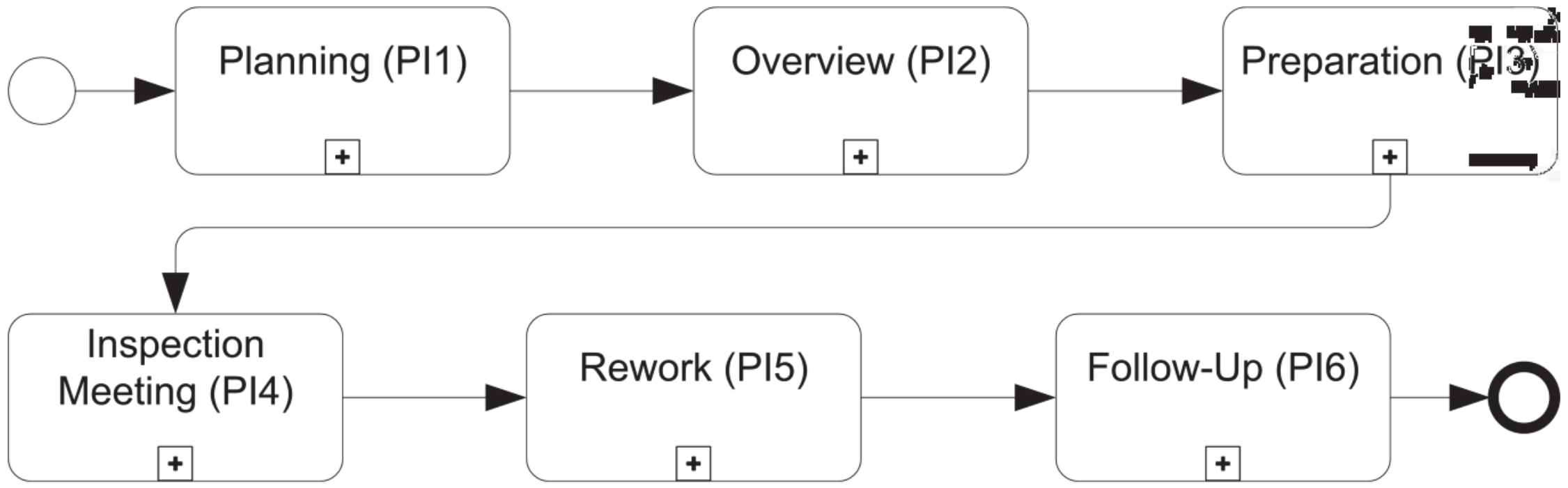

Figure 25 – The Inspection process in Process Impact's peer review description

Similarly to the subprocesses, activities were tailored from tasks, while the order of activities was tailored from the order of tasks. Figure 26 shows an example of using quality approach elements such as: role, role-task associations (mapped to role-activity association), task order (mapped to activity order). As it can be seen, information were added similarly to the case of previous process models created using element instances of other quality approaches. Such additional information are for example (sub)process start, (sub)process end. Processes also include IDs of subprocesses and activities in parentheses after subprocess/activity name.

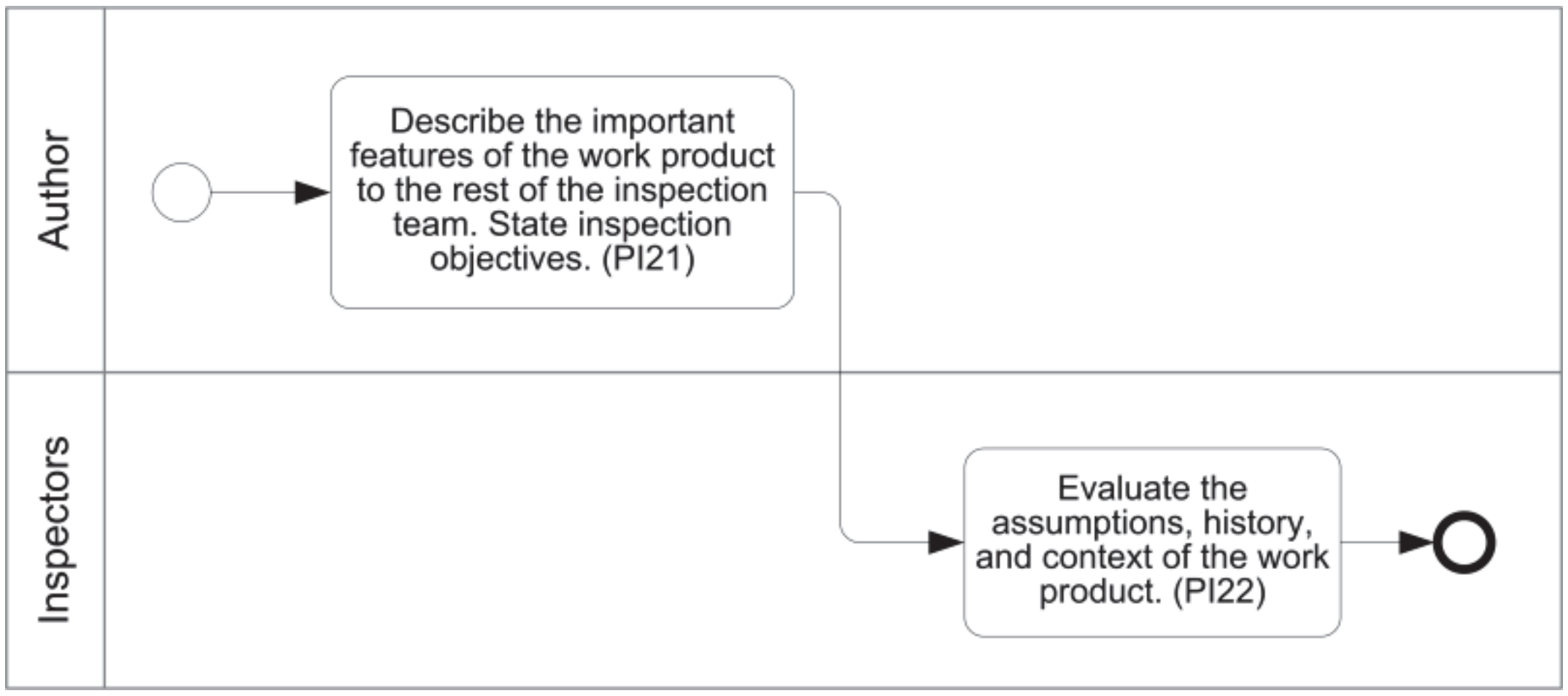

Figure 26 – Example of using quality approach element instances found in Process Impact's peer review description

All the process models were created in a similar manner; therefore we do not present there all the details of creating the process model. The subprocesses of the Inspection process created using element instances of Process Impact's peer review description are included in Figure 59 (Planning), Figure 60 (Overview), Figure 61 (Preparation), Figure 62 (Inspection Meeting), Figure 63 (Rework) and Figure 64 (Follow-up) in the appendix.

### 7.7.3. Verification of the mapping

Since the mapping of quality approach elements to the process did not include prior mapping, the consistency check is not needed.

### 7.7.4. Decision: Are further quality approaches needed?

We analysed 3 quality approaches and identified, mapped their element instances to process elements, we also created 3 different processes based on the element instances of the three quality approaches. Due to the amount of quality approach element instances identified, we consider that the three processes contain enough information to create the unified peer review process and no further quality approaches are needed at this point.

## 7.8. Creating the unified peer review process

In the previous chapters we created three separate process models using elements of three different quality approaches: CMMI, IEEE 1028 and the peer review descriptions of Process Impact.

In this chapter we show how we performed the unification (mapping and creating a unified process) focusing on the concrete examples from the case study, concluding what we can learn, refine and use further in multi-model environments.

We used the three process models in creating the final process model. We do not distinguish iterations in this chapter as we did previously. This is because we created the final process elements in a different order for each quality approach element. In chapter 7.8.1we present how we mapped different quality approach element instances using the process models.
(There is huge effort being spent on researching process models, one recent article deals with mapping process models (Romero, Dijkman, Grefen, & Weele, 2012). As process unification concepts evolve, results could be used in the in multi-model process improvement as part of process based unification.)

### 7.8.1. Mapping process elements

In this activity we created a unified peer review process resulting in 13 subprocesses (Table 54), 67 activities (Table 55), 10 gateways (Table 56), 19 data objects (Table 57), 20 entry and exit criteria (Table 58), 7 roles and responsibilities (Table 59).

**Subprocesses, Activities and Gateways**

We mapped subprocesses, activities and gateways using the three process models. Since IEEE 1028:2008-based process was the most detailed one, we took this as a base process model. Then the mapping activity had two iterations: first mapping to CMMI, then mapping to Process Impact peer review descriptions. Table 54, Table 55 and Table 56 contain the results of the mapping.

Final process elements were created on the basis of the strongest quality approach element instance. The ID of these quality approach elements instances are always shown on the figures in parentheses after the process element instance names. Mappings to others quality approach element instances are contained in the tables.

**Process flow**

We used IEEE 1028 as a base process in which no order of activities were found, however chapters and text are implicitly arranged in the order as an inspection process would be performed. We used this information in building the IEEE 1028 peer review process and we kept this information in the unified process.

During the mapping of subprocesses, activities and gateways of the second two processes first we followed the process flow of the base process model (IEEE 1028), then the process flow of the mapped process models (in the first iteration the CMMI then in the second iteration the Process Impact process model).

This way the mapping of new subprocesses, activities and gateways caused additions and modifications of subprocesses, activities and gateways in the base process. No deletion was needed. The process flow can be viewed in figures of the unified process, we did not described it textually.

**Roles and responsibilities**

Comparing the mapping of roles to mapping of activities, mapping of roles was easy. This is because in almost all the cases roles had the same name in the different quality approaches and quality approaches contained only up to 7 roles. Responsibilities were shortly described in IEEE 1028 and Process Impact, and no responsibility descriptions were included in CMMI, therefore we kept both texts of IEEE 1028 and Process Impact.

Roles and responsibilities are included in Table 59.

**Data objects**

Data objects were mapped in a different order. This is because the most concrete approach for data objects was the inspection description of Process Impact. This quality approach defines concrete templates for data objects. Therefore we mapped CMMI, then IEEE 1028 to Process Impact data objects. Data objects are included in Table 57.

**Data object - Activity relations**

Since there were no description of relations among activities and data objects in the quality approaches, data objects created previously were connected to activities simply based on logic. For example in Figure 27, the activity "Decide if further peer review is needed" is connected to "Criteria for requiring another peer review". These were connected because these are obviously related: if there is a criteria it should be required in such a decision and the direction should be set as an input.

**Entry and exit criteria**

We used entry and exit criteria textually, however we showed how those texts could be mapped, if one decides to use a modeling language which can represent entry/exit criteria. The mapping is included in Table 58.

**Textual elements**

A number of textual elements were also identified and discussed previously. Despite that these textual elements cannot be represented in a process model we consider all of these useful and they are included in the unified process description.

**Mapping types**

In chapter 4.2 we discussed the following types of mapping: 1:1, 1:N and N:M. In this case study we have encountered all these three types of mappings. Since we performed all the mapping in the same way, here we discuss only one example for each.

1:1 mapping: The gateway "Determine what type of peer review will be conducted." Figure 27 is a simple example of the result of a 1:1 mapping. In this case the quality approach element instance "VER SP2.1 SUBP1: Determine what type of peer review will be conducted." was mapped to the gateway "Determine what type of peer review will be conducted." Practically, the gateway was created using the quality approach element.

1:N mapping: The activity "VER SP2.2 SUBP6: Conduct an additional peer review if the defined criteria indicate the need" and the gateway "Is further peer review needed?" in Figure 27 are both mapped to the quality approach element instance "VER SP2.2 SUBP6: Conduct an additional peer review if the defined criteria indicate the need." This is a simple example of 1:N mapping. This mapping was created because data object cannot be associated with gateways; therefore we needed an activity before the gateway to which the data object can be connected.

N:M mapping: this type of mapping happened when in an iteration one quality approach element instance was mapped to more than one process element, then another quality approach element instance was mapped to the same process elements. This type of mapping is important because when the process model changes, the change can have effect on mapping. Since during the mapping no deletion happened and modifications were always verified, consistency was ensured in this case as well.

We also needed to implement CMMI generic practices in this project, therefore we mapped

GPs in our mapping as well. One N:M example of mapping is the mapping of two GPs to multiple process element instances: “VER GP 2.4 Assign Responsibility” and “VER.GP 2.7 Identify and Involve Relevant Stakeholders”. For the concrete mapping of GP 2.4 and GP 2.7 see Table 55 in the appendix.

### 7.8.2. Creating the process model

We created the final unified process model in two iterations. In fact these two iterations were relatively easy and were performed more quickly than identifying and extracting quality approach element instances and using them in creating the three separate processes. At this point we had a good understanding how these quality approaches defined various types of peer reviews, what are the key roles, subprocesses and data objects of each process and most importantly we understood the process flow.

When we created the three process models we thought we will use the process models only for understanding the process, but in fact these process models were a very good guide in the final mapping.

Peer reviews have various types. In this research we developed an inspection process, but we would like to allow the organization to perform other types of peer reviews as well. Therefore we created a top-level process which in current state contains only the Inspection part, but later can be extended with further review types. Figure 27 shows this top level process.

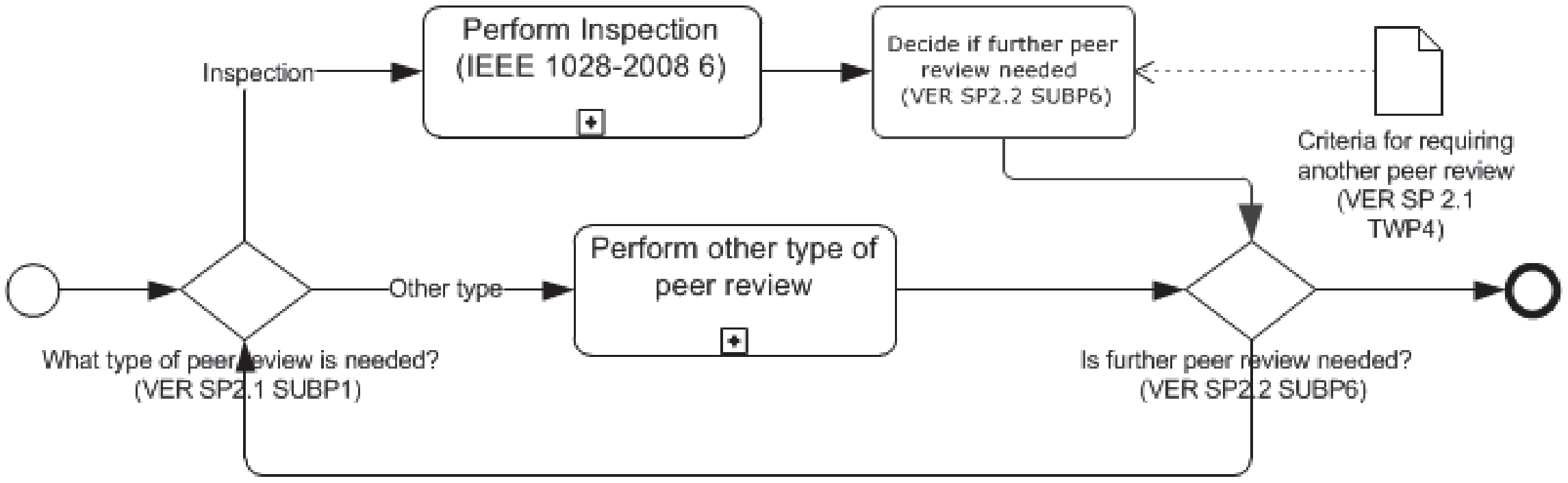

Figure 27 – A unified peer review process

As it can be seen in the text of quality approaches, management preparation is a separate subprocess: it can be performed in a different time with different roles, therefore we separated this process from peer review subprocesses. The situation is similar with data collection and improvement, which can be performed separated from a concrete peer review subprocess. However these are connected, because the improvement is based on peer review data and therefore needs the data collection. High level process model of these subprocesses can be seen in Figure 28 and Figure 29, while the more detailed process model representations can be seen in Figure 65, Figure 74 and Figure 75.

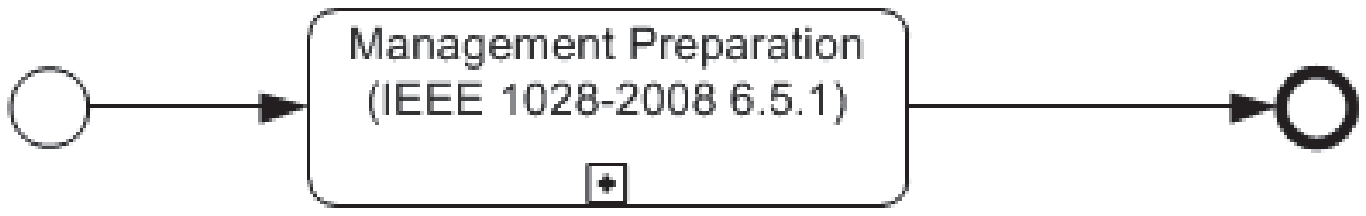

Figure 28 – Management Preparation

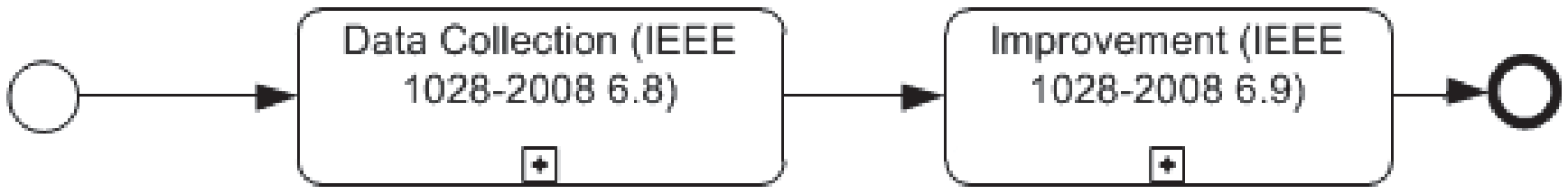

Figure 29 – Data collection and Improvement

Figure 30 contains a unified inspection process created using element instances of three quality approaches CMMI, IEEE 1028 and Process Impact's peer review description.
Looking at the figure three different sources can be observed: IEEE 1028 (6 element instance IDs), Process Impact (2 element instance IDs) and one added activity ("Check entry criteria").
The "Check entry criteria" activity was added because the "Check exit criteria" is present in the quality approaches. Since we have both entry and exit criteria it was logical to add a "check entry criteria" activity. Figure 30 contains two gateways originated from Process Impact's descriptions. These gateways were included because overview activities are optional.
For a figure on which all the three quality approaches are represented as primary process element instances see the Examination subprocess in Figure 70 in the appendix.
The subprocesses of the Inspection process are included in Figure 66 (Planning the inspection), Figure 67 (Overview of inspection product), Figure 68 (Overview of inspection procedures), Figure 69 (Preparation), Figure 70 (Examination), Figure 71 (Rework and follow-up), Figure 72 (Rework) and Figure 73 (Follow-up) in the appendix.

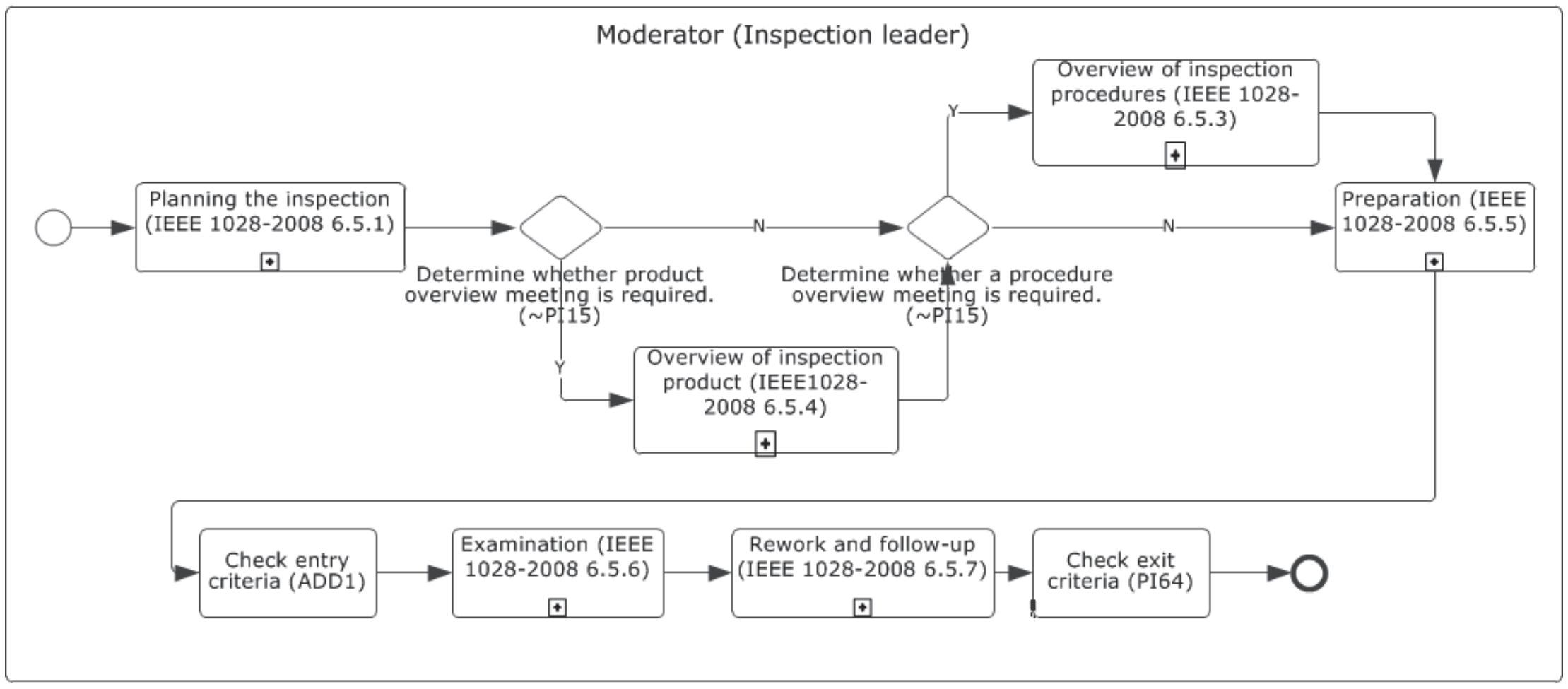

Figure 30 – A unified inspection process

### 7.8.3. Verification of the mapping

This activity is needed in each iteration starting from the second iteration. This was performed at every single modification on the process model. When a modification occurred, the consistency of previous mappings was checked.

In practice, this activity also included a rearranging of mappings. For example if there was an activity for which the primary source was IEEE 1028, then if a more precise or intelligible description was found in any of the other two quality approaches, those descriptions got higher priority and the process element instance got a new ID and all quality approach element instance mappings were kept in the documentation. One such example is the activity of “Describe the important features of the work product to the rest of the inspection team (PI21)”.

At the first iteration this activity looked like: “Present an overview of the software product to be inspected (IEEE1028-2008 6.5.4)” and had the following documentation: “IEEE1028-2008 6.5.4: Present an overview of the software product to be inspected. This overview should be used to introduce the inspectors to the software product. The overview may be attended by other project personnel who could profit from the presentation.”

At the second iteration, when the IEEE 1028 based process model was mapped to CMMI, nothing happened with this activity because no similar process element instances were found in the CMMI based process.

At the third iteration we found a similar activity in Process Impact process and we found it easier to understand, therefore the activity was changed to “Describe the important features of the work product to the rest of the inspection team (PI21)” and both the process model (Figure 67) and the documentation (Table 55) was updated accordingly.

We had 1:1, 1:N and N:M mappings. In these cases verification of the mapping included the verification of all elements in the mapping.

## 7.9. Deviations from the PBU process

In this chapter we followed the PBU process defined in chapter 5.4.2 (Analysis of quality approaches) and 5.4.3 (Deriving process from quality approaches), however three deviations from the original process occurred:

- Creating separate process models,
- Using separated process models in creating the unified process,
- Verification of the mapping.

The first deviation is an additional step to the third subprocess of the PBU process, and we included it because we needed to understand what is a possible process flow behind the requirements of different quality approaches. This way we understood the 3 different peer review processes.

The second deviation from the process is that instead of pure mapping of quality approach element instances to the process we performed the mapping using the three processes designed based on element instances of quality approaches.

This reduced the number of possible mappings, because

- we did not map different process elements (e.g. data objects to activities),
- in case of subprocesses, activities and gateways: we used the process flow in creating the mapping. This means we did not try to map activities of totally different subprocesses. For example in the first iteration, activities of CMMI perform peer reviews were not mapped to activities of management preparation because after modeling the three separate processes we understood that these are totally different parts of the process.

The third deviation was caused by the first deviation so that no further verification was needed.

We think experiences obtained in performing the case study could serve as a useful guideline to the PBU process, therefore we will discuss it further in chapter 10, “Lessons learned from the case study”.

## 7.10. Limitations

In this chapter we discuss limitations of each step we performed in chapter 7.

**Analysis of the three selected quality approaches** (chapters 7.2, 7.4 and 7.6) – During the analysis of quality approaches we mapped some of the quality approach element instances to textual process elements. This was done for simplicity reasons; however there might be situations when these textual elements (e.g. the measurement of the process) should be modeled.

**Deriving process from quality approach** (chapters 7.3, 7.5 and 7.7) – we used quality approach element instances identified in the previous step to build three different peer review processes. We tried to create a logical process and in many cases we added elements (like process flow, or activity - data object relations). It is important to mention that the peer review processes we created are the results of many consecutive decisions, for which we highlighted the most important examples, but same processes could be modeled in different ways and levels, therefore we should state that the process models we created are possible process models, but not the ultimate or definitive process models.

**Creating a unified peer review process** (chapter 7.8) – the most important threat to validity could be that we started building the peer review process from scratch. In many situations organizations do not have a process implemented, however there are cases when a process is already implemented in the organization. In this latter case, a process in use might have effect on the unified process to be developed, however we do not see change effects on the PBU process.

**Understanding deviations from the PBU process** (chapter 7.9) – Deviations from the PBU process were discussed and related guidelines, process refinements will be included in chapter 10, “Lessons learned from the case study”.

## 7.11. Conclusion

Chapters 6-8 are intended to answer Q7 “How can we provide proof of concept for the PBU framework” In order to answer this question a case study is performed. Discussions in this chapter were part of performing RS7-2 “Executing the case study”.

We performed RS7-2 in following two subprocesses of the PBU process. Both of these PBU subprocesses (“Analysis of quality approaches” and “Deriving process from quality approaches”) were performed in three iterations.

In chapters 7.2, 7.4 and 7.6 we analysed elements and element instances of three quality approaches: CMMI, IEEE 1028 and Process Impact. In chapters 7.3, 7.5 and 7.7 we derived separated peer review and inspection processes using element instances of quality approaches mentioned. In chapter 7.8 we created a unified peer review process based on the quality approach element instances and the three process models. In chapter 7.9 we discussed deviations from the PBU process. In chapter 7.10 we discussed limitations of steps made in chapter 7.
In the next chapter the validation of the case study will be discussed aligned with the activities of fourth subprocess of the PBU process "Validation" introduced in chapter 5.4.4.

# 8. VALIDATING THE RESULTS OF THE CASE STUDY[6]

In chapters 6-8 the research step 7 (RS7) "Perform a case study on the PBU framework" is discussed. In chapter 6 we presented the preparation of the case study (RS7-1): selecting the context, a process, quality approaches to be used and a textual and graphical process representation format.

In chapter 7 we presented the execution of the case study (RS7-2) performing the analysis and process creation from the element instances of quality approaches in three iterations, then a unified peer review process was developed using element instances of three quality approaches. These chapters were described according to the first three subprocess of the PBU process presented in chapter 5.4. In this chapter we describe the validation of the case study results (RS7-3) according to the last PBU subprocess ("Validation") described in chapter 5.4.4.

## 8.1. Approach

Here we continue with the PBU process and perform activities of the last subprocess "Validation" described in chapter 5.4.4. As a validation of the case study our goal is to investigate if the Q-works project accepts the result of the PBU process: the unified peer review process.

In order to validate results of the case study, we will perform 3 operational steps, which correspond to the activities defined in chapter 5.4.4:

- in chapter 8.2 we describe choosing the validation technique,
- in chapter 8.3 we describe how the validation technique was applied,

[6] Preliminary results of this chapter were presented on the Q-Works workshop (Kelemen, 2010a, 2010b).

- in chapter 8.4 we discuss analysing validation results, possible refinements and application of the validated unified peer review process,
- as part of the PBU process, a decision about further processes need to be made, this will be briefly addressed in chapter 8.5.

After performing Validation deviations from the PBU process in chapter 8.6 and in chapter 8.7 RS3-i3 “Refinement on the process elements” will be discussed.

# 8.2. Choosing validation technique

Our objective in validation of the case study was to get the product accepted by users. For a validation which focuses on the acceptance of a process, different methods can be used such as interviews, structured questionnaires with open or closed questions, brainstorming and walkthroughs among many others. Research methods usable for validation are described in multiple sources, e.g.: (Given, 2008; Kojo Arhinful, Mohan Das, Prawitasari Hadiyono, Heggenhougen, & Higginbotham, 2000; Robinson, 2006; Saunders, Lewis, & Thornhill, 2007; Yin, 2009).

Three major circumstances influenced our selection of validation technique, *first*: the Q-works project consisted of two separate teams: one of process experts and one of the implementation team. This meant that the two teams worked basically in separate environments and had monthly meetings. The *second* important parameter was that we did not have previous experience in applying the PBU framework. Therefore, a validation of preliminary results was desired. *Additionally*, the PBU concept was new and we wanted to validate its results both by process improvement experts and users.

According to the circumstances, we choose to validate PBU results using walkthrough technique in an iterative way. The walkthrough technique did not require any additional effort from the Q-work project team and could be iteratively applied on project meetings by process experts and users.

In Table 31 we included validation techniques and reasons of selecting a validation technique for the unified peer review process.

Table 31 – Choosing a validation technique

| **Validation technique** | **Yes / No** | **Reason** |
|---|---|---|
| Interview | No | Due to the monthly project meetings in the Q-works project it was more efficient to show the unified peer review process than organizing interviews. |
| Questionnaire | No | Due to the monthly project meetings in the Q-works project it was more efficient to show the unified peer review process than to create questionnaires and ask users about a process which can be presented during the project meetings. |
| Brainstorming | No | Since our approach was to develop a process using 3 different quality approaches, users did not need to provide input for the unified process. Therefore no brainstorming was needed during the development. In this case, their acceptance was needed rather than a brainstorming about the process. |
| Walkthrough | Yes | A walkthrough seemed perfect choice of validation because:<br>- monthly project meetings could be used for walkthroughs without requiring additional effort/time from the Q-works project team; |

| Validation technique | Yes / No | Reason |
|---|---|---|
| | | - monthly project meetings give the opportunity to walkthrough intermediate work products allowing us to get feedback on preliminary results;<br>- allowed us to quickly involve both process experts and users. The Q-works project involved both software process improvement experts and users. It consisted of an integrated team from two main parties:<br>  - from Polygon Ltd (project management, administration, Maximo experts and developers),<br>  - and BME - SQI process improvement and CMMI experts. |

In applying the walkthrough technique we defined the following main iterations:

1. an iteration of walkthroughs on the preliminary results,
2. an iteration of walkthroughs on the final result.

The three separate peer review processes were considered the preliminary results, while the unified peer review process was considered the final result.

For both iterations first we needed process experts and then users, therefore results were first presented to process experts and later to users.

According to the above mentioned, the following steps were defined for validation:

1. walkthrough on the preliminary PBU results with the process experts,
2. walkthrough on the preliminary PBU results with the whole Q-works team,
3. walkthrough on the final PBU results with the process experts,
4. walkthrough on the final PBU results with the whole Q-works team.

A validation has value if validators have real interest in the validated product; therefore here we describe the interest of the two groups of validators:

The BME-SQI process improvement consultants were the process experts, while the Q-works implementation team were the users.

1. **Process experts**: Since the unified peer review process was the first process we implemented in the Q-works project, process experts wanted to use lessons learned from creating the peer review process later during the development of further processes. All process experts were involved in creating processes in the Q-works project. They started their work after the development of the unified peer review process, using multiple quality approaches and the process modeling language chosen for representing the unified peer review process. Therefore they were interested to see:
    - how quality approach elements would be used in creating processes,
    - how processes can be represented in the process modeling tool,
    - where and how to use the unified peer review process,
    - was the unified peer review process mature enough to present it to users group.

Experience of process experts: 4 process experts were involved to the process team from BME and SQI having (20+6+6+6 years of both academic and practical experience in process assessment and process improvement). One of the process experts was a certified SCAMPI Lead Appraiser for CMMI for Development and CMMI for Services. All process experts have participated in various process improvement projects and internal and formal CMM/CMMI as-sessments at several international companies.

2. **Users:** As the goal of the Q-works project was to implement CMMI-conform processes, the Q-works implementation team was interested to understand:
    - how such a peer review process can be implemented in Maximo environment,
    - were there all information they needed is included in the process.

Experience of users: people involved in the Q-works project team from Polygon Ltd. cover a broad scale of experience starting from general manager of the company, a project manager with 10+ years of project management experience, developers with 10+ years of experience and young managers and organizers with various levels of experience.

According to the above we consider that both experts and users were interested and possessed enough experience to judge if they accept the unified peer review process after each walkthrough.

In the next chapter we present how Validation steps defined in this chapter were applied in the Q-works project.

## 8.3. Apply validation

We selected the Q-works project for the case study based on context criteria (see chapter 6.2). Here we present the project progress from validation point of view, focusing on steps defined in 8.2. Additional information on the Q-works project progress can be found in Appendix T.

1. **Walkthrough on the preliminary PBU results with the process experts**

The three iterations presented in chapters 7.2-7.7 were performed and three separate processes were developed based on CMMI, IEEE 1028 and Process Impact peer review descriptions (see Appendix O-Appendix Q for process models developed).

At the end of this work, results were presented to the process team in a half-day walkthrough procedure as a first validation step. At the end of this walkthrough, results were considered usable by the process team. Consensus was reached and a decision was taken: preliminary results can be presented to the whole team.

2. **Walkthrough on the preliminary PBU results with the whole Q-works team**

A week after the first walkthrough the three processes were presented to the whole team on a project meeting. The walkthrough included a general introduction to peer reviews, peer review types, the PBU concept, decisions made, quality approach elements and element instances identified, graphical representation of peer review processes built using element instances of three quality approaches.

Project participants were asked to share their thoughts on the possible implementation barriers of the peer reviews in IBM Maximo environment. At the end of this project meeting a consensus was reached that a unified peer review process should be developed using the PBU concept.

3. **Walkthrough on the final PBU result with the process experts**

After a consensus was reached with the whole process team to apply the PBU framework and create the unified peer review process, the unified peer review process was developed using the three separate peer review processes. The development of the unified peer review process is described in chapter 7.8. At the end, similarly to the first iteration of the validation, the unified peer review process was presented to the process team. The process team considered the unified peer review process mature enough to be accepted and presented it to the whole project team.

4. **Walkthrough on the final PBU result with the whole Q-works team**

A week after the second walkthrough the unified peer review process was presented to the whole team on a project meeting. The walkthrough included the mapping (e.g. mapped roles, activities etc.), graphical representation of unified peer review process built using element instances of three quality approaches.

Project participants were asked for the second time to share their thoughts on the possible implementation barriers of the peer reviews in IBM Maximo environment. At the end of this project meeting a consensus was reached that the unified peer review process will be used in the Q-works project. All related materials were shared with all team members and the development work has started. Results of the software development work are discussed in chapter 8.4.

5. **Additional step: using lessons learned in the development of further processes in Q-works project**

After the unified peer review process was accepted by the whole project team and the implementation in Maximo environment has begun, further CMMI processes were selected for implementation in a multi-model basis. In the first phase these were:

- Measurement and Analysis,
- Requirements related process areas: Requirements Management and Requirements Development,
- Project related process areas: Project Planning, Project Monitoring and Control and Integrated Project Management.

Due to scope, size and confidentiality reasons the discussion of the development of these process areas are not included into this thesis.

## 8.4. Analyse results and perform refinement

Theoretical refinement of the unified peer review process was not needed in the Q-works project. However, due to practical reasons and project needs two major simplifications were made at the beginning of the development. These simplifications were made due to restrictions of Maximo environment and the Q-works system later was updated to reflect the original whole, unified peer review process.

**1st simplification:**

Cause: any company using Q-works system relying on IBM Maximo environment should buy licences for all process participants, therefore a decision was made to minimize and tailor roles to project needs.

Simplification: In the first version of the Q-works the following three roles were used: peer review leader, author and inspector.

Update: After a couple of months of development, developers found the way how they can use multiple roles within a single license and then the Q-works implementation was updated with further process roles (such as reader or recorder) reflecting the original unified peer review process.

**2nd simplification:**

Cause: Developers were not experienced in IBM Maximo environment.

Simplification: as a first step, developers implemented only the high level unified peer review process in Q-works. This was done as a trial with the goal to exploit how the IBM Maximo environment works.

Update: After several development revisions they included the whole peer review process. They also included the mapping documentation in the activity descriptions of the Q-works activities.

At the end of the iterative development process the unified peer review process was finalized in IBM Maximo environment as a part of the Q-works system. Since the implementation of the unified peer review process was performed by Polygon and was not part of this research and also due to confidentiality reasons the discussion of implementation is not included into this thesis.

However, to show that the implementation was indeed performed, Figure 31 shows IBM Maximo job plans of the unified peer review process in the Q-works system (due to localization settings some of the texts are in Hungarian). Subprocesses (e.g. Planning the inspection, Overview of inspection product etc.) are included as job plans and are highlighted with grey background. Further screenshots of the implementation of the unified peer review process in Q-works project can be found in the appendix (see Figure 77-Figure 81).

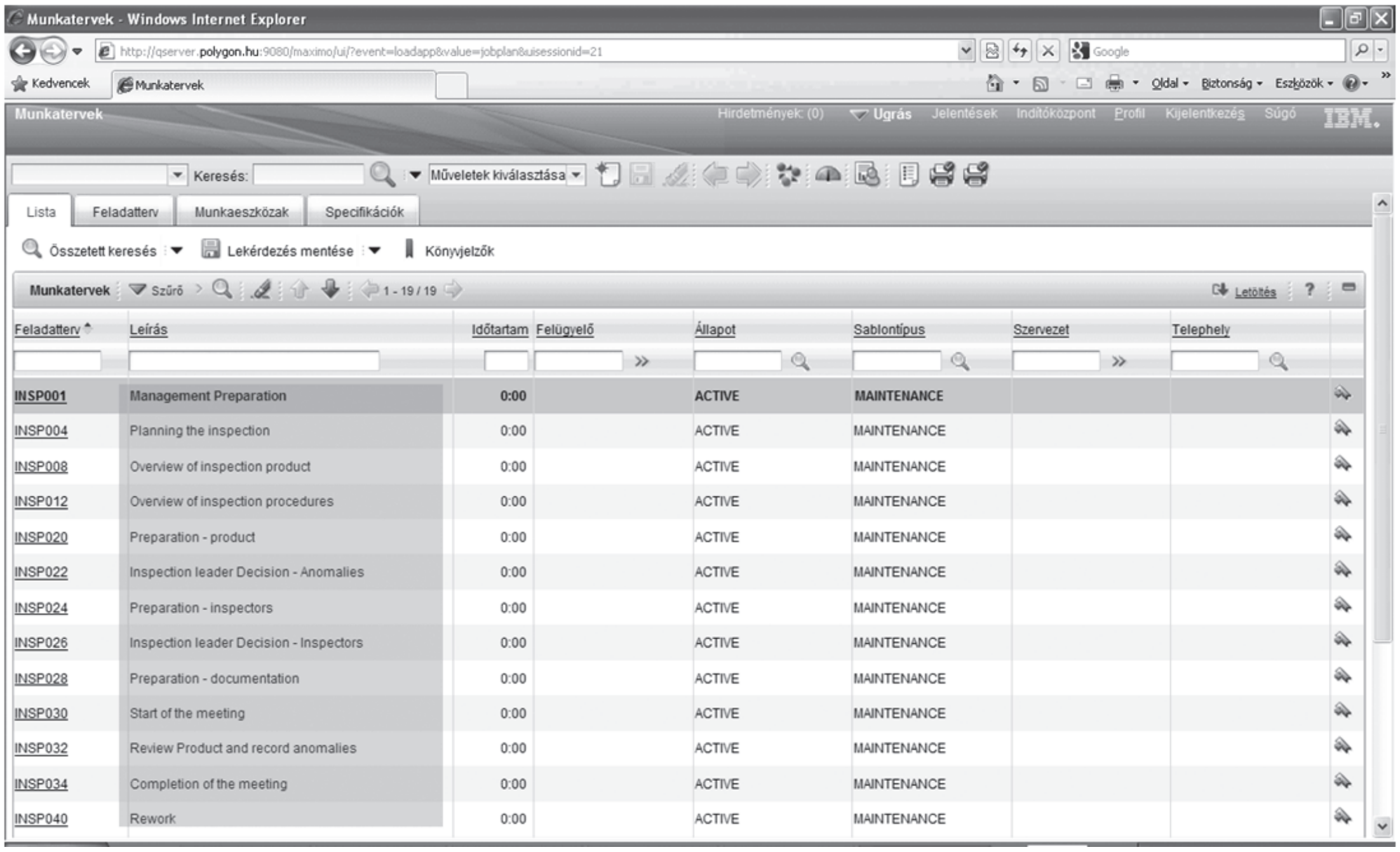

Figure 31 – Job plans for the unified peer review process (INSP001 – INSP054)

Additional evidence that the Polygon Ltd. found the unified peer review process usable is that they organized an open workshop at the headquarters of IBM Hungary on 19 May, 2010. Top-ics of the workshop included introduction to Q-works project, presenting the PBU concept, the unified peer review process, presenting the IBM Maximo environment and representation of the unified peer review process in IBM Maximo environment. After the finalization of this project there were further collaborations between the two groups and it is planned that implementation of CMMI process areas will continue using the PBU concept.

## 8.5. Decision: Are there further processes to be improved?

The purpose of this case study was to develop a unified peer review process, therefore no further processes are needed. In the frame of Q-works project further processes (compliant to CMMI-DEV v1.2 ML3 process areas) were developed based on multiple quality approaches. These are not in the focus of this case study.

## 8.6. Deviations from the PBU process

In the Validation subprocesses no deviations from the PBU process occurred, the PBU process was performed according to the Validation subprocess described in chapter 5.4.4.

## 8.7. Refinements on the process elements

Since the amount of element instances which were not mapped to process elements is minimal and the unified peer review process was accepted, it is not needed to refine process elements.

## 8.8. Limitations

Here we discuss limitations of steps performed in this chapter.
**Validation** – A limitation of validation is that the developed unified process was used as an input for the Q-works system and it was not used in practice by the users. In order to overcome this problem, the validation included iterations of walkthroughs with both software developers and process experts.
**Refinements on the process element list** – It was not needed to refine the process elements after performing the case study, therefore we do not see limitations of refinements (RS3-i3).

## 8.9. Conclusion

In chapters 6-8 the research step 7 (RS7) “Perform a case study on the PBU framework” was discussed. Our objective in this chapter was to get the unified peer review process accepted by users (RS7-3) and to make refinements on the process elements according to RS3-i3 “Refinements on the process elements” (if needed).
In this chapter we presented the Validation of the case study, which is based on the fourth subprocess of the PBU process also called “Validation”, introduced in chapter 5.4.4. In this practical validation we had several options to practically validate the unified peer review process. Being participant of the Q-work project we have chosen the informal walk-through validation technique, however there might be situations when a more formal validation technique could be needed. Despite that we used an informal validation technique, project participants and Polygon Ltd. accepted the PBU results in three different iterations. The walkthrough was successful, users accepted the unified peer review process and they used it as an input for the Q-works system implementation. Moreover, internal process experts agreed on the unified peer review process before each external validation, therefore we consider the PBU results valid.

Chapters 6-8 intended to answer research question Q7 “How can we provide a proof of concept for the PBU framework?”. This question was answered by applying the PBU framework in a case study. The case study showed that such a mapping is feasible using the PBU framework.

Regarding Q3, the suitable set of process elements was not changed at this stage (RS3-i3).

# 9. VALIDITY AND RELIABILITY OF THE RESEARCH

The hypothesis of this thesis (formulated in 2.1) was that mapping quality approaches to a unified process can provide a multi-model solution.

In order to determine if a mapping of quality approach elements to process elements is possible we analysed the structure and elements of various quality approaches in chapter 4. We concluded that the mapping of quality approach elements and process elements is possible for the quality approaches analysed. Findings of chapters 3 and 4 led us to develop a multi-model solution which we call PBU framework (mapping quality approaches to processes). A central component of the PBU framework is the PBU process which describes the process of mapping quality approaches to a process; this is described in chapter 5. In chapters 6-8 we presented a case study reflecting how the PBU process was used in practice by developing a unified peer review process which conforms to three different quality approaches: CMMI, IEEE 1028 and Process Impact's peer review descriptions. The case study discussed in chapters 6-8 shows an example of a mapping of quality approach element instances to process element instances and it provides a proof of concept that such a mapping is indeed practically feasible. Now discussion of validity and reliability of the research is also needed to assess its quality.

## 9.1. Approach

The trueness of the hypothesis can be assessed by answering two research questions:

1. Is the PBU framework adhering to MSPI criteria? (Q8)
2. Which conclusions to validity and reliability of the PBU framework can be drawn from research results? (Q9)

These research questions can be translated to two research steps:

1. Verify the PBU framework against MSPI criteria (RS8)

2. Investigate validity and reliability of the PBU framework (RS9)

**Ad Q8/RS8:** In chapter 3 we identified MSPI criteria which we will apply to the PBU framework as follows:

1. Traceability of PBU results,
2. Adaptability and expandability of PBU results,
3. Completeness of PBU results,
4. Appraisal support,
5. Repeatability of the PBU process.

Handling differences of quality approaches were understood more deeply during the case study and thus we discuss each of these separately:

6. Handling the differences in terminology,
7. Handling the differences in granularity,
8. Handling the differences in structure and elements,
9. Handling the differences in content,
10. Handling the differences in size and complexity.

At designing the PBU framework and during the case study, details of MSPI criteria were deeper understood, but were not discussed until this point. Therefore, before assessing if MSPI criteria were satisfied, a discussion of MSPI criteria is needed reflecting the three phases of understanding.

After understanding MSPI criteria, a discussion is needed how it was satisfied in the research: both in the PBU framework and the case study.

Since only a single case study was performed to show the feasibility of PBU concept, after discussing how the PBU framework and the case study satisfy the MSPI criteria, it is also needed to show how the PBU framework can satisfy MSPI criteria in other settings.

According to the above mentioned, in order to assess if the PBU framework adheres to MSPI criteria we identified the following operational steps:

a) **Problem definition** – short reflection of the problem based on the MSPI criteria identified in RS1-2. In RS1 we identify problems based on literature, here we include possible problem refinements based on experiences of the case study.
b) **Satisfying the MSPI criteria in the PBU framework** – discussion if the PBU framework satisfies the MSPI criteria or with which modifications are required to achieve.
c) **Satisfying the MSPI criteria in the case study** – discussion if the current application of the PBU framework satisfies the MSPI criteria or which modifications are required.
d) **Conclusion –** discussion of applying the PBU framework in other settings.

We will go through each criterion performing each operational step defined above. We will always focus on presenting information on how the PBU framework and its application in the case study satisfy the MSPI criteria. In chapters 9.2-9.11 we discuss MSPI criteria and we perform steps a-d.

**Ad Q9/RS9:** We also need to show that mapping of quality approaches to the unified process is possible in a valid and reliable way.

The case study discussed in chapters 6-8 provides a proof of concept that such a mapping is indeed practically feasible. Additionally, discussion on validity and reliability of the research is needed in order to assess its quality.

*Choosing a validation approach*

Ensuring validity and reliability in quantitative research is well defined in the literature (Given, 2008; Golafshani, 2003; Hoepfl, 1997; Kaplan & Maxwell, 2005; Shenton, 2004; Trochim, 2006). However, according to (Gibbert & Ruigrok, 2010; Golafshani, 2003; Shenton, 2004) ensuring validity and reliability in qualitative research is not obvious, and several researchers argue about what and how should be ensured in assessing the quality of a less standardized qualitative research.

In a literature discussion Golafshani shows that the concept of reliability could be irrelevant or misleading in qualitative research. She acknowledges that terms such as “Credibility, Neutrality or Conformability, Consistency or Dependability and Applicability or Applicability and Transferability” could be important in qualitative research (Golafshani, 2003).

Shenton considers Guba’s criteria for qualitative research: “credibility (in preference to internal validity), transferability (in preference to external validity/generalizability), dependability (in preference to reliability) and confirmability (in preference to objectivity)” (Shenton, 2004).

Based on an extensive review of case study publications, Gibbert and Ruigrok consider internal, construct and external validity and reliability as main criteria in qualitative research. They point out that case studies often neglect internal and construct validity and have emphasis on external validity. Validity types are not independent, prerequisites of external validity are internal and construct validity. They suggest prioritization of internal and construct validity over external validity (Gibbert & Ruigrok, 2010).

In this thesis we discussed a case study which we consider more qualitative than quantitative research, therefore validity and reliability of the research are needed to be discussed in terms of qualitative research.

As validity and reliability are not universal concepts in qualitative research (Amis & Silk, 2007; Gibbert & Ruigrok, 2010; Golafshani, 2003) we need to:

1. understand concepts of validity and reliability in the context of our research,
2. discuss validity and reliability according to their meaning in our research.

We will base our discussion of validity and reliability on (Fisher, 2010; Gibbert & Ruigrok, 2010; Yin, 2009), which are:

- ecological validity,
- construct validity,
- internal validity,
- external validity,
- reliability.

Chapters 9.2-9.11 focus on the discussion of the MSPI criteria; chapters 9.12.1-9.12.5 focus on discussing validity and reliability of the research.

# 9.2. Repeatability of the PBU process

## 9.2.1. Understanding issues of repeatability

**First iteration: at literature review (chapter 3.3 and 3.5)**

During the literature review we found that several multi-model initiatives describe only the multi-model result and questions may arise such as "What process/methodology was followed?", "Is the process repeatable?" or "Is the solution itself documented?".

As source quality approaches are often described textually, harmonization, mapping or integration are frequently done in a subjective manner. In order to ensure the repeatability and transparency, a description of the process by which a multi-model result is achieved should be available (Yoo et al., 2006).

**Second iteration: before designing a PBU process (chapter 3.6)**

As a second iteration we argued about how the PBU framework would satisfy the repeatability criterion:

"Many of multi-model solutions (quality approach mappings, integrations, harmonisations) are not documented (Kelemen et al., 2011), therefore it is arguable how the multi-model result was achieved. In order to satisfy the repeatability criterion, the PBU framework will include a clearly documented and repeatable process. The PBU process will be described using a process modeling language (PML), defining its process activities, inputs, outputs (PBU results) and guidelines for applying the PBU process."

**Third iteration: after performing the case study**

After defining and performing the PBU process we had a more clear view on issues related to repeatability. As we mentioned previously, the repeatability can be supported by a documentation of the process and the results. In the case study we encountered decision points multiple times. These included decisions regarding the process description format, identification of quality approach element instances, mapping and terminology.

During process based unification we built a process model; according to Polyvyanyy "One can fairly adopt the ideas of Donald E. Knuth to conclude that process modeling is both a science and an art. Process modeling does have an aesthetic sense. Similar to composing an opera or writing a novel, process modeling is carried out by humans who undergo creative practices when engineering a process model. Therefore, the very same process can be modeled in a myriad number of ways." (Polyvyanyy, 2012).

Decisions made in the modeling and mapping process influence the PBU results (unified process and mappings); therefore documentation on how results were achieved (what decisions were made) should also be established in order to ensure the repeatability of the PBU process. (In the case study these decisions were documented and explained through chapters 6-8.)

Summarizing, after the third iteration three major things were found to be documented in order to achieve a sufficient level of repeatability:

1. the process (what should be performed in order to achieve the multi-model results),
2. results achieved (multi-model results),
3. how results were achieved, what decisions were taken.

### 9.2.2. Ensuring repeatability

**Repeatability in the PBU process**

In order to support the repeatability, we provided a clearly documented PBU process in chapter 5.4, this satisfies the first documentation criterion – the documentation of the process.

The PBU process results in the unified process and mappings to unified process. If these are documented, then the second documentation criterion – documentation of results is satisfied.

The third documentation criterion, documenting the decisions are not included in the PBU process, but we include a guideline for documenting decisions in chapter 10.6, lessons learned.

**Repeatability in the case study**

*Documentation criterion 1: document the process* – The case study was performed according to the PBU process. Since we used the PBU process as the process of developing the results, and the PBU process is documented in chapter 5.4, we satisfied the first documentation criterion in the case study.

Problems encountered in performing the PBU process: at the end of each case study chapter we discussed what deviations from the PBU process occurred and how we handled these deviations. The documentation of the deviations can be found in chapters 6.6, 7.9 and 8.5. Based on case study experiences on the deviations from the PBU process we refined the PBU process in chapter 10.

*Documentation criterion 2: document results* – Documentation of the results is included in the appendix: quality approach element instances, mapping of quality approaches to unified process, separate and unified process models. With this we satisfied the second documentation criteria. We did not encounter any problems in documenting the results, however the documentation of results with a more accurate tool which stores the quality approaches, unified process and the mapping between these could support the maintainability of the resulting documentation. For supporting the maintainability of the PBU results we propose a database in chapter 11.3.

*Documentation criterion 3: document how results were achieved* – In chapters 6-8 we described how differences between quality approaches were handled, how the identification of quality approach elements and quality approach element instances were performed, and examples of decisions made were shown (e.g. how various mappings were done, how process models were developed, what terminology issues were encountered and how these were solved). All the case study documentation in chapters 6-8 are documentations which help the repeatability of the PBU process and reflect decisions made, and therefore satisfy the third documentation criterion.

### 9.2.3. Conclusion – ensuring repeatability in other settings

In order to ensure repeatability we defined three documentation criteria in chapter 9.2, here we discuss how these can be ensured when the PBU framework is applied in other settings.

*Documentation criterion 1: document the process* – the PBU process is documented on a generic level. For building a unified process from other quality approaches, same PBU subpro-cesses and their activities should be performed. Therefore we assume satisfying this criterion is generalized across quality approaches and processes.

*Documentation criterion 2: document results* – one output of the PBU process is a mapping between quality approach element instances to the unified process element instances. Both the mapping and the resulting unified process should be documented in case of any selected quality approaches and processes, therefore we assume this criterion is also generalizable across quality approaches and processes.

*Documentation criterion 3: document how results were achieved* – decisions in case of other processes should also be documented. Especially those decisions which are finer grained than the PBU process. The PBU process defined in chapter 5 covers two levels of granularity. Finer grained in this case means that it might happen that as a part of PBU subprocesses or activities, further decisions should be taken, which are not described in the PBU process as they are too specific or detailed. For example selecting quality approaches and processes are on the same granularity as the PBU process: these activities are included in the first subprocess and the documentation of these decisions will be reflected by the results (e.g. in the mappings and the unified process).

Those decisions which are on a finer granularity level and not included into the PBU process still need to be documented in order to ensure repeatability (e.g. decisions to include a mutually exclusive process flow or exclude possible contradictions in the content of different quality approaches). If these are documented (similarly to the case study documentation in chapters 6-8) then the PBU process can be repeated.

During the case study we did not encounter problems caused by quality approaches in documenting how results were achieved. However, in case of other quality approaches there might be (e.g. copyright) issues which do not allow the identification and usage of quality approach element instances in a mapping documentation.

Summarizing, the repeatability of the PBU process is ensured:

- by the PBU process documentation (for documentation criterion 1),
- by PBU results (for documentation criterion 2),
- by documenting decisions (for documentation criterion 2).

The guideline in chapter 10.6 requires the documentation of decisions and choices. Acting according to this guideline will result in a documentation on how results are achieved (documentation criterion 3).

A consequence of these three types of documentation could be a decreased productivity in the process development phase (during the PBU process), which later can have positive effects. For instance anyone can understand how results were achieved, results can be reused, errors and mistakes made during the development of the unified process can be traced and corrected. However there might be situations in which the conformity to quality approaches is not required, in those cases a decision on the level of documentation can be taken to increase productivity in the process development phase. This should be done knowing that this may cause the results will not satisfy the MSPI criteria and will be difficult e.g. to ensure repeatability, traceability and appraisal support.

## 9.3. Traceability of PBU results

### 9.3.1. Understanding the issue of traceability

**First iteration: at literature review (chapter 3.3 and 3.5)**

Multiple sources of the literature have shown that in order to achieve traceability, a mapping of quality approaches to multi-model result is needed (Ghose et al., 2007; Salviano, 2009b; Siviy et al., 2008a). Besides providing traceability, the mapping can support appraisals, changeability and adaptability. At the end of literature review we concluded that the result of a multi-model solution should clearly be traceable back to source quality approaches and this can be achieved by mapping.

**Second iteration: before designing a PBU process (chapter 3.6)**

As a second iteration we argued about how the PBU framework would satisfy the traceability criterion: in a process based unification both processes and quality approaches will be decomposed, and clear relationships (mapping) between quality approach element instances and process element instances will be identified. These relationships will describe traceability.

**Third iteration: after performing the case study**

After applying the PBU framework we had a more clear view on traceability. In the case study we performed a high number of mappings. These include mappings of quality approach element instances to various process element instances e.g. activity, role or data object instances.

The notions of traceability and mapping have not changed after the case study, but became more concretized as follows: traceability means that relation of multi-model results to quality approaches is identified and documented. The multi-model result can contain additional information for which the source is not a quality approach. For this information traceability is not ensured and required.

Summarizing the issue and possible solution after the three iterations:

*The issue of traceability*: relating multi-model results to source quality approaches.

*A possible solution*: identifying and documenting a mapping between the multi-model results and source quality approaches, which is part of the PBU process and therefore the PBU framework supports mapping.

### 9.3.2. Supporting traceability

**Traceability in the PBU process**

In the PBU process both processes and quality approaches are decomposed and mapping among process element instances and quality approach element instances are required to be identified. This is described in chapter 5.4.3. The mapping is done on instance level and not on the type (or class) level. This is because there are cases where type level mappings cannot be performed and mappings should be done on instance level., e.g. in case of shall statements in ISO 9001-structured standards quality approach elements which always can be mapped to the same process elements cannot be identified. As a solution for this problem, shall statements can be decomposed and processed manually and mapped to various process element instances such as instances of activities, roles and data objects; this requires instance level mapping. In an opposite case, typical work products of CMMI always can be mapped to data objects, and this means mapping of typical work products could be done on type level. Due to the varying granularity of any process, typical work products can be mapped to data objects or data items contained in data objects; this implies an instance level mapping.

The PBU process contains high level activities for mapping and does not go into details of how to map certain quality approach element instances to process element instances, however it suggests a mapping on instance level.

Quality approach element instances identified are mapped to the unified process; and instance level mapping describes the relations between quality approaches and the unified process thus provides traceability.

**Traceability in the case study**

Chapter 7 discusses how the mapping of quality approach instances to process element instances was performed. Multiple examples show how quality approach element instances were mapped. One example screenshot of mapping multiple quality approach element instances to process element instances can be seen in Figure 32. The figure shows how traceability is achieved with the tool used: mapping is included in the documentation of the process on process element instance level. This shows how any process element instance of the unified process model can be related to quality approach element instances.

Table 54, Table 55, Table 56, Table 57, Table 58 and Table 59 in the appendix contain the mapping of the unified peer review process discussed in the case study to the element instances of three different quality approaches, providing traceability of the unified process to the (source) quality approaches.

### 9.3.3. Conclusion – ensuring traceability in other settings

In the PBU framework traceability is ensured by a mapping of process element instances to quality approach element instances. In chapter 4 and 1 we identified process elements to which quality approach elements can be mapped.

Using the process representation format we proposed in chapter 4 and BPMN for graphical representation, a mapping can be performed for other processes similarly to the case study: identifying quality approach elements, element instances and mapping them to process element instances. Since the quality approach element instances identified are process-oriented a mapping will be always possible in case of any other processes and quality approaches.

In case of quality approaches we analysed in chapter 4 the mapping can be done even in case of quality approaches “weak” structures and semantics (see e.g. ISO 9001-like standards). In textually described standards manual mapping should be performed, texts should be understood and mapped. For mapping a weakly structured quality approach see e.g. how ISO 9001 was decomposed and mapped to CMMI in (Mutafelija & Stromberg, 2009b). Even in this case mapping and thus traceability can be ensured.

# 9.4. Adaptability and expandability of PBU results

## 9.4.1. Understanding issues of adaptability and expandability

**First iteration: at literature review (chapter 3.3 and 3.5)**

During the literature review we found that when a new quality approach is released it may have change effects on the multi-model results and a company's processes. This could also happen when a new version of already used quality approach is released. In order to emphasize the importance of the changes occurring in quality approaches we distinguished two criteria related to changes: adaptability (in case that a new version of an existing approach appears) and expandability (in case that a new quality approach appears).

**Second iteration: before designing a PBU process (chapter 3.6)**

As a second iteration we argued about how the PBU framework can ensure changeability.

Adaptability: In order to ensure adaptability, changes in the new versions of quality approaches will be discovered and reflected in quality approach element mappings. In case of a new version of a quality approach 3 things can happen regarding quality approach elements: deletion, modification and appearance of new quality approach elements. The mapping of quality approach element instances to process element instance will be stored and this will handle adaptability issues related to these three kinds of changes.

Expandability: The expandability of the PBU results will be ensured by an iterative process, so that in each iteration, a new quality approach can be added to the PBU result. When a new quality approach is mapped to the unified process then a new iteration should be performed. If such an iterative process can be defined, then the expandability of the PBU result can be ensured.

**Third iteration: after performing the case study**

We had a clear view on adaptability and expandability through the literature review and before performing the case study. We did not encounter adaptability nor expandability issues in the case study, therefore no refinements on these terms and their definitions were made. Summarizing we can define the problem and possible solutions as follows:

*The issue of adaptability:* adherence to a new version of a quality approach.

*A possible solution for ensuring adaptability:* In order to ensure adaptability, changes in the new versions of quality approaches should be identified and mapping should be updated accordingly. In case a new version of a quality approach appears, 3 things can happen regarding the quality approach element instances: *deletion*, *modification* and appearance of *new* quality approach element instances. Same can happen at quality approach element level.

*The issue of expandability:* adherence to a new quality approach.

*A possible solution for ensuring expandability:* The application of an iterative PBU process and inclusion of a new quality approach in each iteration.

## 9.4.2. Supporting adaptability and expandability

**Adaptability in the PBU process:**

Mapping of quality approach element instances to process element instances is required in the PBU process. In order to support adaptability in the PBU framework, mapping and the unified process should be updated based on the changes in the new version of the quality approach. In chapter 11.3 we provide a database in which multiple versions of quality approach elements/element instances and their relations can be stored. A tool built upon the proposed database can support adaptability of PBU results.

**Adaptability in the case study:**

In the case study we used a different version (v1.2) of the CMMI model than we analysed in chapter 4 (v1.3). There were small differences between the two versions e.g.: textual clarifications in the new version such as renaming the quality approach element “typical work product” to “example work product”. Changes in new versions of the quality approaches can be mainly handled if the organization which releases the new version provides information on the changes made. In this case one possible solution could be the mapping of the new version to the old version. This mapping could be stored in a database. In chapter 11.3 we provide a database structure for storing the quality approach element instances and their mapping to process elements. The database also includes a solution for storing the mapping between (different versions of) quality approach element instances.

**Expandability in the PBU process:**

The expandability of the PBU results is ensured by an iterative process, so that in each iteration a new quality approach can be added to the PBU result (unified process). When a new quality approach needs to be mapped to the unified process then a new iteration should be performed. Such an iterative process is defined in chapter 5.4, and expandability of the PBU result is ensured.

**Expandability in the case study:**

Problems encountered when quality approach element instances were needed to be mapped. We found that mapping is difficult without having an overview of the process behind the quality approach, therefore a separate process model for each quality approach was developed which helped to understand the process in the quality approach. After creating the process model it was easier to map quality approach element instances to the unified process.

The iterative process performed in the case study resulted in a unified peer review process. Iterations of the case study are described in chapters 7.2-7.7, and the development of the unified peer review process is included in chapter 7.8. These chapters include discussion on how the quality approach element instances were identified, how separate process models were built and how these were mapped to the unified process.

Building separate process models was not part of the first version of the PBU process but it was added as a guideline based on case study experiences and discussed in chapter 10.

### 9.4.3. Conclusion – ensuring adaptability and expandability in other settings

In a mapping adaptability and expandability issues can occur, e.g. if we do not use tools for storing the mapping information it could be difficult to adapt or expand the current mapping.

*Adaptability* in the PBU framework is supported by a database for storing the PBU results. Since the database can store various quality approach elements and their instances, it will support adaptability in case of other quality approaches and processes.

*Adapting* PBU results (the unified process and mapping) to new versions of quality approaches is not automated and is performed manually for both adapting the mapping and the unified process. A further support of adaptability can be a database which allows storing relations between versions of quality approach element instances (see a data model of such a database in 11.3). Adaptability could be even further enhanced and supported if quality approaches would have a semantic structure with clearly identified process oriented quality approach elements and changes in new versions would be available in a semantically described format (e.g. in an XML). If that would be provided, organizations could use a tool which updates old versions to the new quality approach element instances in a mapping. Then a manual verification would be needed on the unified process to verify the implementation of necessary changes on the unified process caused by the changes in mapping.

*Expandability* is supported by an iterative process in which further quality approaches can be added. Same iterative process will be applied in case of other quality approaches and processes. In each iteration quality approach element instances are identified and mapped to process element instances. Since this is done on iterative basis it can be repeated to a theoretically infinite number of quality approaches. Based on case study experiences, only a limited number of quality approaches will probably needed in practice.

Expanding to other processes is not a problem since processes always have the same elements and mapping of quality approach element instances is possible to instances of the same process elements.

# 9.5. Completeness of PBU results

## 9.5.1. Understanding issues of completeness

**First iteration: at literature review (chapter 3.3 and 3.5)**

The literature review showed that a multi-model solution should provide a clear indication on the coverage of source quality approaches. Furthermore, the completeness of multi-model results should be sufficient for users.

**Second iteration: before designing a PBU process (chapter 3.6)**

As a second iteration we argued about how the PBU framework can ensure completeness.

A well-documented mapping between the PBU results and source quality approaches will provide the level of coverage (e.g. what parts of source quality approaches are covered by the PBU result). In order to ensure sufficient level of details of PBU result, the resulting process is represented in a Process Modeling Language (PML). PMLs (e.g. BPM or EPM) have the abilities to provide appropriate completeness as processes can be decomposed to unlimited levels and each (sub)process can be described in a chosen/needed level of granularity.

**Third iteration: after performing the case study**

As we have discussed previously, a multi-model result should indicate the coverage of source quality approaches. Furthermore, the information provided by PBU result should be sufficient for users. These are two different views on completeness of results and we can give their definitions as follows:

*The issue of coverage:* providing information on what parts of source quality approaches are included in the multi-model result.

*A possible solution for ensuring coverage:* after performing the case study, information can be collected on which quality approach element instances were mapped to the unified process. This information provides a quantitative view on what parts of the quality approach requirements were implemented in the unified process.

*The issue of sufficiency:* understanding how sufficient is the information gained from quality approaches.

*A possible solution for ensuring sufficiency:* sufficiency of information can be ensured first by selecting the proper quality approaches. Second, results can be checked both by experts and users when the PBU framework is applied then refinements can be made based on their feedback. Based on results and experiences from applying the PBU framework, the sufficiency could be readdressed and eventually improved in a learning cycle.

Furthermore, a PML which allows theoretically unlimited process levels can be selected, providing the basis of detailing most important parts.

### 9.5.2. Ensuring completeness

**Ensuring completeness of the PBU results in the PBU process:**

*Coverage* of quality approaches can be assessed by checking the mapping of quality approach element instances to the unified process. This mapping shows what element instances were mapped and also provides implicit information what was excluded (what was not mapped). The coverage of the quality approaches is verified at the activity ”Verification of the mapping” discussed in chapter 5.4.3.3.

For verifying the *sufficiency of the information* the PBU process provides three process elements which are aimed to ensure sufficiency of the unified process:

- At the first subprocess quality approaches are selected based on their scope (chapter 5.4.1)
- At the third subprocess process experts decide after each iteration if further quality approaches are needed (activity “Decision: Are further quality approaches needed?” described in chapter 5.4.3.4)
- In the Validation subprocess user are asked if they can accept the resulting unified process and can decide if the information is sufficient for them, then the unified process can be refined based on the user feedback. (see chapter 5.4.4)

**Completeness of the PBU results in the case study:**

**Coverage of quality approach (requirements) in the case study**

In order to indicate the level of completeness we counted the number of quality approach element instances identified at the end of each iteration in the case study. Coverage was discussed in chapters 7.2.2, 7.4.2 and 7.6.2.

Additionally, a well-documented mapping between the PBU results and source quality approaches were developed. Mapping tables are included in Table 54, Table 55, Table 56, Table 57, Table 58 and Table 59 in the appendix, thus coverage can be reassessed.

Our research focused on identifying quality approach elements and element instances which can be mapped to process elements and respectively process element instances, therefore the number of quality approach element instances itself does not reflect the completeness of the PBU results. Some information was not included into the mapping such as page numbers, introductions, remarks and notes, text formatting etc.

In order to review these exclusions, content related to the peer review/inspection process in the three quality approaches was verified at the end of the case study.

Since a single process was selected in the case study, the inputs of the completeness verification focused on these parts of the quality approaches which describe the selected process (peer reviews). The following parts of quality approaches were verified for completeness:

- VER SG2 “Perform Peer reviews” chapter of CMMI-DEV v1.2 (pages 502-505),
- chapter 6, “Inspections” of IEEE 1028 (pages 16-25),
- and chapter 5 “Inspection procedure”, of Process Impact peer review descriptions (pages 4-9).

These verified parts were identical to those used in identification of quality approach element instances in the case study.

The output of the completeness verification of the mapping was a 20+ pages document, containing source texts and verification of their coverage. Since most of the content of this document coincides to the information contained in the mapping tables Table 54, Table 55, Table 56, Table 57, Table 58 and Table 59 in the appendix, here we present only the text excluded from the mapping. Table 32 includes texts excluded from the mapping of quality approaches.

Table 32 – Texts excluded from the mapping of quality approaches to unified process.

| **Quality approach** | **Text** |
|---|---|
| IEEE 1028-2008 | “Additional reference material may be made available by the individuals responsible for the software product when requested by the inspection leader.” |
| IEEE 1028-2008 | “Although this standard sets minimum requirements for the content of the documented evidence, it is left to local procedures to prescribe additional content, format requirements, and media.” |
| Process Impact’s peer review description | “Participants<br>The roles and responsibilities shown below pertain to the inspection process. All participants are inspectors, in addition to any specialized role they might have. At least three participants, including the author, are required for an inspection. If only three people participate in an inspection, the moderator shall also serve as recorder or reader. The author may not serve as reader, moderator, or recorder.” |

As it can be seen in Table 32, excluded texts are general guidelines regarding the application of the rest of the texts, which were mapped. Of course, a decision of inclusion/exclusion should be made at each relevant quality approach element instance. Depending on organizational procedures, these information can be useful and can be used e.g. in an introductory text of a process description or at the description of the roles (e.g. 3$^{rd}$ row of the table).

During the case study, excluded texts were discussed on meetings with the Q-works project members and consensus was reached in each case. Participants of the Q-works project agreed that they do not want to use excluded text, furthermore a number of case specific decisions were made in using the unified peer review process, which are discussed in chapter 8. See Table 60 for the meetings related to the development of the unified peer review process and the progress of the case study.

Regardless of minor textual exclusions, 100% of the requirements were mapped: from CMMI all peer review related goals and practices were mapped. Furthermore, generic goals and practices were also mapped due to the project scope. From IEEE 1028 all “shall”, “should” and “may” statements were discussed and mapped. Process Impact peer review description has no requirements, it is more a procedure than a standard for implementing the peer review process, but all process-related quality approach element instances were mapped from this approach as well.

**Sufficiency of information in the case study**

In order to ensure sufficient level of details of PBU result (the unified peer review process), the resulting process was represented in a BPMN. BPMN has the ability to provide appropriate level of completeness of processes. The unified peer review process was represented in multiple process levels on BPMN diagrams. Additionally, the tool we used for modeling the unified peer review process (Process Modeler for Microsoft Visio) allowed us to include mappings to each process element instance in its documentation.

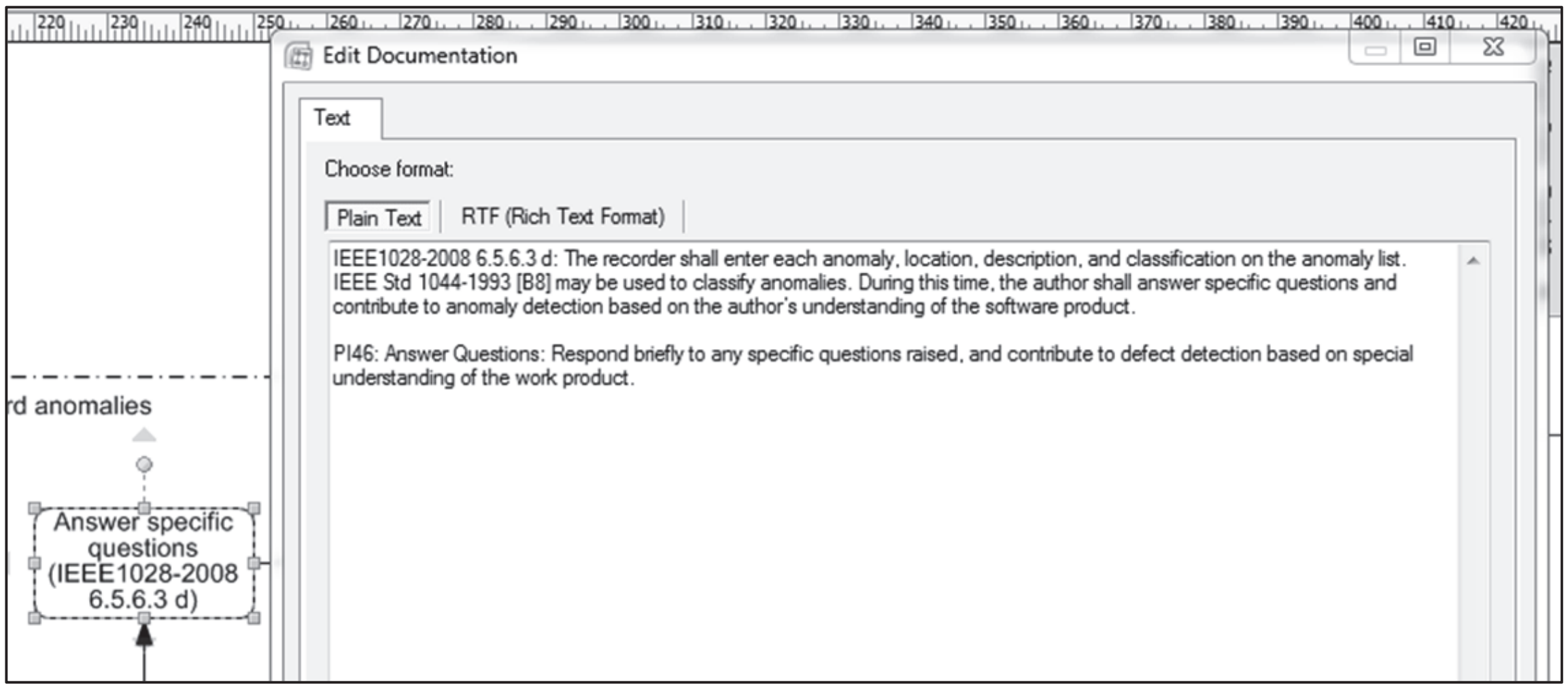

Figure 32 – A process element instance and the mapped quality approach element instances in the process element instance documentation

In order to balance sufficiency of information and align the level of completeness with user needs, the unified process was discussed and validated by users. This (practical validation) is described in chapter 8.

We showed that all relevant information in the quality approach can be included into the process.

Regarding the validity of the PBU framework, further questions may arise, e.g. Can all information in the unified process be mapped to source quality approaches? In other words: is all the information in the unified process relevant to the quality approaches? If not all the information in the unified process can be mapped to the quality approaches we can have two cases:

- the information is useless (and was included by a mistake or with reason),
- the information is useful (and was included by a mistake or with reason).

For the first case if something useless was included it can be excluded during the continuous improvement of the processes at the organization. For the second case, if the information is useful it could mean quality approaches can be improved using practical feedback. This feedback could be sent to the quality approach releasing organizations.

### 9.5.3. Conclusion – ensuring completeness in other settings

Completeness can be discussed in terms of coverage of quality approaches and sufficiency of information for users.

*Coverage* of quality approaches can be assessed easily by counting and summarizing quality approach element instances mapped and not mapped. More important is the mapping of requirements because this is in the focus of appraisals.

In a process based unification quality approach element instances are identified and mapped to process element instances. As we have seen in the case study there could be texts/parts of quality approaches which will not be mapped. These elements will not be covered by the unified process. Despite that some texts were left out from the mapping, the case study showed that all the requirements were mapped to the unified process. The coverage of requirements was 100%, even if some texts were not included into the mapping.

Even in the case of weakly structured quality approaches such as ISO 9001, the needed percentage of coverage can be ensured (see e.g. (Mutafelija & Stromberg, 2009b) for a mapping of ISO 9001 requirements to CMMI v1.2 requirements).

*Sufficiency of information:* in the PBU process one of the activities is the selection of quality approaches. At this activity main characteristics of quality approaches are analysed and a selection is made based on their scope. This scope should be proportional to the scope of the unified process. At this stage a set of quality approaches are selected which should provide sufficient level of information. Later, at identifying and mapping quality approach element instances to the unified process a further selection is made: only those element instances will be mapped which are relevant to the process. Decisions are iteratively made if further approaches are needed. These decision points and user feedback ensure sufficiency of information in case of any quality approaches and processes. Furthermore a PML (e.g. BPMN) can also support sufficiency of information by allowing theoretically unlimited process levels.

Summarizing we conclude completeness of PBU result can be ensured in case of other quality approaches and processes.

## 9.6. Appraisal support

### 9.6.1. Understanding issues of supporting appraisals

**First iteration: at literature review (chapter 3.3 and 3.5)**

The literature review showed that mapping can provide traceability and traceability can support appraisals. Multi-model appraisal support means the conformity of processes can be appraised against multiple quality approaches. This can happen at once or separately in time. In order to ensure the conformance to source quality approaches, results of a multi-model solution should be appraisable based on multiple (source) quality approaches.

**Second iteration: before designing a PBU process (chapter 3.6)**

As a second iteration we argued about how the PBU framework can support appraisals:

Using a well-documented mapping, the relations between the unified process and source quality approaches will be ensured. Furthermore we will propose a database schema describing the relations (mapping) between quality approach element instances and process element instances, and this will also support the appraisals as it will make easier to follow the links between the unified process and source quality approaches.

**Third iteration: after performing the case study**

*The issue of appraisal support:* results of a multi-model solution should be appraisable.

*A possible solution for appraisal support:* in applying the PBU framework a mapping of the unified process to multiple quality approaches is developed. This mapping supports appraisals because element instances of the unified process can be related to the quality approach requirements.

## 9.6.2. Supporting appraisals

**Appraisal support in the PBU process:**

Since the PBU process requires a mapping of quality approach element instances to process element instances, processes created using the PBU process are appraisable based on the mapping.

Appraisals can be supported by tools which include the needed information at an appraisal, e.g. the mapping between the process and quality approaches.

In the frame of this thesis we do not have enough resources to develop a tool, but as an additional support for appraisals, we include a data model describing the mapping between quality approach elements and process elements in chapter 11.3, which can enhance the appraisal support and makes easier to store the mapping between the unified process and source quality approaches.

**Appraisal support in the case study:**

In the case study formal multi-model appraisal was not needed, the only quality approach for which the requirements were needed to be supported was CMMI. As we discussed in chapter 8 the main goal of the project in which we developed the unified peer review process was to implement CMMI-DEV v1.2 process areas. Since CMMI was the less detailed quality approach regarding the peer review it was not a problem to prove the unified process we developed conforms to CMMI. A SCAMPI LA was included into the process expert team and she agreed that the process satisfies the CMMI requirements against peer reviews.

In chapter 7 we showed the level of coverage of source quality approaches, and we have showed that 100% of the peer review/inspection requirements of the quality approaches are mapped to the unified process, furthermore the mapping between the unified process and source quality approaches are included in tables Table 54, Table 55, Table 56, Table 57, Table 58 and Table 59 in the appendix.

### 9.6.3. Conclusion – ensuring appraisal support in other settings

In the PBU framework appraisal support and traceability is ensured by a mapping between the unified process and quality approaches.

As we have shown, 100% of the quality approach requirements can be covered by a unified process developed using the PBU framework in case of CMMI, IEEE 1028 and Process Impact peer review descriptions. In case of other quality approaches and processes this is also possible when a mapping of quality approach requirements to the unified process can be developed. When discussing traceability of multi-model results in chapter 9.3 we argued that this is possible even in the case of less structured quality approaches. The unified process can always have an attached documentation which can include the mapping (see Figure 32).

A further appraisal support could be a *tool for multi-model appraisals* for which we provide a basis as a database for storing quality approach element instances, process element instances and the mapping of these two. Of course such a tool would need multiple functions which are out of scope of this thesis, but the database can serve as a basis for further research and tool development.

Another appraisal support could be a *process for multi-model appraisals* which guides the process experts in performing appraisals based on multiple quality approaches. Since the appraisal process often differs at different quality approaches (see e.g. differences between SCAMPI for CMMI and ISO 15504 part 2 and 3) a unified appraisal process would probably require elements of multiple appraisal/assessment methods. The development of such a unified appraisal process is not in the focus of this thesis, however if these appraisal methods are process oriented, the PBU framework could also be used.

## 9.7. Handling differences in terminology of quality approaches

### 9.7.1. Understanding issues of terminology

**First iteration: at literature review (chapter 3.3)**

During literature review we found that one of the main problems in multi-model SPI is the difference in terminology of the quality approaches and the difficulty of recognizing the similarities between them. This was the initial finding about possible issues associated with terminology.

**Second iteration: before designing a PBU process (chapter 3.6)**

As a second iteration we argued about how the PBU framework would handle the differences in terminology.

With the PBU framework, we map quality approach element instances to process element instances. All quality approaches processed are process-based, therefore mapping of terms should also be possible. Take the peer review process as an example: terms such as “Inspection checklist” in IEEE 1028 and “peer review checklist” in CMMI-DEV v1.3 (VER SP 2.1 TWP 2) would be mapped to a single quality approach element instance using one single term for such a checklist. Decisions in mapping the terminology have to be taken and documented.

**Third iteration: after performing the case study**

After defining and performing the case study we had a more clarified concept about issues related to terminology, which we discuss here.

***Issue:*** according to our experiences, homonyms (one of a group of words that share the same spelling and the same pronunciation but have different meanings) and synonyms (different expressions with same meaning) exist in any language; furthermore there is usually no consistent language across or even within quality approaches.

For example CMMI includes a Product Integration process area, while in agile environment the area which can be mapped to CMMI Product Integration is called Continuous Integration (CMMI Product Team, 2010a). Tools are often called continuous integration tools which can be used to satisfy CMMI Product Integration practices (see e.g. the tools Hudson, Jenkins or CruiseControl). In case of project management there is another terminological difference between tools used in practice and (theoretical) quality approaches: specific practices of "Project Monitoring and Control" process area included in CMMI can be satisfied by "project tracking" tools.

***Possible issue resolution:*** By handling the terminology we mean that synonyms and homonyms in quality approaches should be identified, analysed, and a resolution should be given as follows:

- If synonyms are identified synonymic terms should be mapped to one single term – this way clarifying that different terms have the same meaning (this implies the mapper has to decide what terms to use).
- If homonyms are identified a definition of the term should be given – this way clarifying that terms with the same spelling or pronunciation mean one single thing (and again making choices with regard to the terms being selected and the meaning they will be assigned).

## 9.7.2. Handling the differences in terminology

**Handling the differences in terminology in the PBU process**

In order to highlight the terminology issue in the PBU process, the analysis of the terms is required

- to be performed at high level in the first (Selection) subprocess (see chapter 5.4.1),
- and in second (Analysis) subprocess (see chapter 5.4.2.2).

In the first (Selection) subprocess the analysis of the terms is required to set the scope of the unified process.

In the second (Analysis) subprocess, terminology of quality approaches is analysed to understand what exactly needed to be done within the scope.

**Handling the terminology in the case study**

In the case study our selected process was the peer review process. Accordingly, various types of peer reviews and their definitions for these peer review types were reviewed and used in selecting the quality approaches for creating the peer review process.

A problem we faced in analysing terms related to the peer reviews was that while IEEE 1028 and Process Impact defines explicitly the different peer review types, in CMMI only a list of peer review types is included. A further problem we faced was that peer review types are dif-ferent in each quality approach. This required a number of decisions regarding terminology. Table 33 summarizes the peer review types of the 3 quality approaches.

Table 33 – Peer review types in quality approaches

| Quality approach | Peer review types |
|---|---|
| CMMI | **inspection**, structured walkthroughs, active reviews (not defined, listed as examples) |
| IEEE 1028-2008 | **inspection**, walkthrough, passaround |
| Process Impact | management review, technical review, **inspection**, walk-through, audit |

In the case study we selected the inspections as it is the most formal and well described peer review process, and also because less formal peer review types can be tailored from inspections. We also selected definitions for the review from IEEE 1028, peer review from CMMI and inspection from IEEE 1028 (see chapter 6).

Terms connected to data objects, roles and activities and other process elements were also handled. This required a lower level analysis of the terminology and was mainly performed in the mapping subprocess (see chapter 7).

An example of identification and handling synonyms in case of roles: the role of moderator in Process Impact descriptions has similar responsibilities to inspection leader in IEEE 1028, therefore these were mapped. As a result of the mapping, the moderator role has assigned responsibilities of both inspection leader and moderator in the unified process.

An example of identification and handling synonyms in case of data objects: the input “Inspection checklist” in IEEE 1028 and the typical work product “peer review checklist” in CMMI-DEV v1.3 (VER SP 2.1 TWP 2) and the Process Impact’s work aid “inspection moderator’s checklist” were mapped to the data object “inspection moderator’s checklist”.

We did not encounter homonyms in the case study.

The consequence of the decisions was that in case of synonyms only one term was used instead of multiple terms. If anyone was looking for other quality approach terms than those we used in the unified process, it was not a problem to find them because mappings are always attached to the unified process and the texts of the mappings are searchable. Therefore handling terminology did not have any drawbacks in the case study.

### 9.7.3. Conclusion – handling terminology in other settings

Based on our understanding handling terminology consists of handling synonyms and homonyms.

The identification of synonyms worked in the case study at multiple quality approach elements (e.g. input, work aid and typical work product). Synonyms can be present in other quality approaches and processes; a solution for handling synonyms at other quality approach element instances is identification and mapping synonymic quality approach element instances.

We did not encounter homonyms in the case study, but the handling (identification and definition) of homonyms should be applicable in case of other quality approaches and processes since this issue is not quality approach nor process specific. It is more a linguistic problem which could be present in any case when the mapping of texts from different sources is needed.

We consider that the handling of synonyms and homonyms can be generalized only on a high level, because these need manual interaction and intuition and are rather language specific problem than a process or quality approach specific. Since it is a linguistic problem, to define solution is not in the scope of this thesis. However, the PBU process and its application shows subprocesses where homonyms and synonyms can be identified. These are: selection, analysis and mapping. The issue can be solved on a high level as follows:

- synonyms when encountered should be identified and mapped,
- homonyms when encountered should be identified and defined.

The consequence of mapping synonyms is using only one instead of multiple terms. If anyone is looking for other quality approach terms than those used in the unified process, those should be findable. If unused synonyms of terms can be found in the mapping there is no consequence of using one term instead of synonyms, this depends on the tool used.

The identification and definition of homonyms results in clarification of the term and its meaning in the context, which help users to get a clear view of the process. We do not see any negative consequences of handling homonyms.

# 9.8. Handling differences in granularity of quality approaches

## 9.8.1. Understanding issues of granularity

**First iteration: at literature review (chapter 3.3 and 3.5)**

In the literature review we found that granularity is an issue when comparing or mapping quality approaches (Yoo et al., 2006). In order to create a mapping between quality approaches the granularity to be adopted should be “established” (Thiry et al., 2010). Furthermore, “misunderstanding the granularity of quality approaches can erode the benefits of SPI efforts” (A. Ferreira & Machado, 2009; Siviy et al., 2008a).

**Second iteration: before designing a PBU process (chapter 3.6)**

As a second iteration we argued about how the PBU framework would handle the granularity:

Quality approaches have their descriptions in different levels of granularity (see e.g. the discussion of the peer review process in the introduction). Fortunately, processes carry a similar characteristic; they also can be decomposed, and described at various levels of granularity. This coincidence in characteristics of quality approaches and processes will allow handling granularity differences in quality approaches. We will find a process element instance on a proper, corresponding granularity level for each quality approach element instance which will be mapped. This way granularity differences in quality approaches will be hidden by the single process model.

**Third iteration: after performing the case study**

As we iteratively discussed in earlier chapters, quality approaches are described in different levels of granularity. Therefore, a multi-model solution should recognize and handle differ-ences in granularity.

*Issues of differences in granularity of quality approaches:* quality approaches are described with different granularity, which can cause problems in mapping them to the multi-model result. The following questions can arise related to the granularity of quality approaches:

- How a mapping can be developed if a quality approach is more detailed (granular) than the multi-model result?
- How a mapping can be developed if a quality approach is less detailed (granular) than the multi-model result?

*A possible solution for handling granularity differences:*

By identifying quality approach element instances of different quality approaches and mapping them to a single process, granularity differences of quality approaches can be handled, because PMLs (e.g. BPMN) allow unlimited level of decomposition of processes. From granularity point of view quality approach element instances on the highest granularity level can be mapped to process element instances and lower level quality approach element instances then can be mapped to lower level process element instances. In case no lower level process element instance exists, it can be created as this is allowed by PMLs. This way the unified process will hide granularity differences and the unified process is described on multiple granularity levels and a solution is given for both cases.

## 9.8.2. Handling the differences in granularity

**Handling granularity in the PBU process**

In the PBU framework granularity is handled at the identification and mapping of quality approach element instances to the unified process.

Literature shows that “granularity in process modeling depends on the model (modeling) purpose” (Kalpic & Bernus, 2002), other sources discuss decomposition and multi-granularity of process models, see e.g. (Holschke, Rake, & Levina, 2009; Ma, Zhou, Zhu, & Wang, 2009; Zhu, Sun, Huang, & Liu, 2010). All these publications show that decomposition of processes is possible and needed. In Table 34 we discuss how we decompose process elements identified in chapter 4.

Table 34 – Process elements and their possible decomposition

| Process Element | Possible Process Element Decomposition |
|---|---|
| **(Sub)Process** | Processes can be decomposed to subprocesses. This is also allowed by PMLs (e.g. by BPMN or EPC). The decomposition processes to subprocesses allows to include finer grained instances of any other process elements including (sub)process, process purpose, activity, role, responsibility, data object, element relation, entry/exit criteria, because in any process level same elements can be used. This characteristic of processes allows decomposing and describing processes at any level. |
| **(Sub)Process name** | The new, lower process level will have different name, therefore the process name is not needed to be decomposed, rather it can reflect that it is a part of a process. |
| **Process parent** | Allows process decomposition and denotes to which process is the current process level belongs to. |

| Process Element | Possible Process Element Decomposition |
|---|---|
| **(Sub)Process** | Processes can be decomposed to subprocesses. This is also allowed by PMLs (e.g. by BPMN or EPC). The decomposition processes to subprocesses allows to include finer grained instances of any other process elements including (sub)process, process purpose, activity, role, responsibility, data object, element relation, entry/exit criteria, because in any process level same elements can be used. This characteristic of processes allows decomposing and describing processes at any level. |
| **Process purpose** | In case of process decomposition the new subprocess can have a purpose. This is textual information which can be edited freely (preferably based on and consistent with the parent process). |
| **Activity** | Activity is atomic, however if it needs to be further detailed, it can be replaced by a sub(process). Replacing activity with a subprocess allows activity decomposition. In practice PML tools allow activity to subprocess replacement – often called conversion, see e.g. (itp-commerce, 2012). |
| **Role, responsibility** | Roles can be present at any process level, so that in case of process decomposition finer grained roles can be added to the subprocesses of the decomposed process. Furthermore, a role can have assigned responsibilities. One option for decomposition of a role can be adding a new role and splitting responsibilities of the original role between the original and the new role. Then same responsibilities will be performed by two different roles. In this case one role is decomposed to (replaced by) two roles. |
| **Data object** | Data objects can be present at any process level, so that in case of process decomposition finer grained data objects can be added to the subprocesses of the decomposed process. This allows the decomposition of the original data objects to finer grained data objects. Furthermore, a data object can have data items. One option for decomposition of a data object, (similarly to the role decomposition) can be adding a new data object and splitting data items of the original data object between the original and the new data object. Then same data items will be present in two different data objects, so that one data object is decomposed to (replaced by) two data objects. |
| **Element relation** | An element relation is present between two elements (e.g. activity-activity, activity-data object) and it is atomic. E.g. if we have activity 1 and activity 2 as two related activities. If activity 1 and/or activity 2 are decomposed, the relation between them not needed to be decomposed. Same is true for relations between activities and data objects. Thus process decomposition can be done without decomposing element relations. Of course at any decomposition of related elements a review of their relation can be performed and relation can be modified if needed. |
| **Entry/exit criteria** | Entry/exit criteria can be present at any process level. Furthermore, entry/exit criteria consist of a list of criteria; decomposition can also be performed based on splitting the list of criteria, similarly to role and data object decomposition. |

Since process element instances can be divided to unlimited levels, granularity is handled in in the PBU framework through process element instance decomposition at mapping of quality approach element instances to process element instances.

**Handling granularity in the case study**

The identification of quality approach element instances in multiple granularity levels were performed in three iterations in the case study in chapters 7.2, 7.4 and 7.6. These included a process, multiple subprocesses and activities with related further process elements such as roles, data objects, etc. This way quality approach element instances were mapped to the unified process on multiple granularity levels, reflecting how decomposition was performed in multiple cases.

In order to understand the selected process in quality approaches we created three separate process models before mapping quality approach element instances to the unified process. When we achieved the three separate process models we always took the most detailed process model elements for the basis of the unified process. For example we mapped activities to the most detailed IEEE 1028:2008 process model, the mapping of data objects were performed in different order starting with Process Impact's data objects as these were the most detailed. The mapping order of quality approach element instances identified to a unified process is described in chapter 7.8.1. Based on case study experiences guidelines related to mapping and granularity are defined in chapter 10.6.

### 9.8.3. Conclusion – handling granularity in other settings

Handling differences in granularity of quality approaches can be generalized to other quality approaches and processes because both quality approach instances and processes element instances can be decomposed to unlimited level of details. Since the mapping is always done from one direction, an 1:N mapping of quality approach element instances to process element instances is always possible. The unified process will hide granularity differences of source quality approaches. Main process elements (activity, subprocess and process) can always be decomposed and described in finer grains. As we have shown in Table 34 any further process element can be described at multiple granularity levels. As a consequence, processes having elements identified in chapter 4 can also be described at multiple granularity levels. Due to the granularity characteristic of the process elements and PMLs, multi-granularity can be ensured for other processes.

The granularity level of quality approach element instances only depends on the granularity of the (textual and graphical) quality approach descriptions. In case if a finer grained quality approach element instance is identified than the current process level, a new, corresponding finer process level can be included into the unified process. In this case prior mappings can be revisited and decided if previously mapped quality approach element instances need to be mapped to lower process element instances.

# 9.9. Handling differences in structure and elements of quality approaches

## 9.9.1. Understanding issues of difference in structure and elements

**First iteration: at literature review (chapter 3.3 and 3.5)**

In the literature review we found that quality approaches differ in structure and this increases the complexity of mapping. Several new quality approaches follow the structure and elements of previous ones e.g. SAFE+ extension of CMMI, or the three constellations of the CMMI v1.3 (for development, services and acquisition). There is a tendency to develop quality approaches reusing structures and elements of previous models, especially in the case of capability/maturity models, see e.g. (von Wangenheim et al., 2010) for a review of capability/maturity models or (Helgesson, Höst, & Weyns, 2011) for a review of methods for evaluation of maturity models.

**Second iteration: before designing a PBU process (chapter 3.6)**

"Knowing the structure and elements of quality approaches will help to determine which elements can and which cannot be mapped to process elements. If (the majority of) quality approach elements can be mapped to process elements, the PBU concept can work; as processes could be build/enhanced using quality approach elements of multiple quality approaches.

In order to handle the differences in structure and elements of quality approaches, their elements will be mapped to process elements of a single process model. This process model will have a single, unified structure, having only process model elements, this way differences in structure and elements of quality approaches will be hidden."

**Third iteration: after performing the case study**

As we have shown in chapter 4 quality approaches have different structure and elements. The literature showed that structural differences can make difficult the simultaneous usage of multiple quality approaches.

Structural recognitions (e.g. recognizing that CMMI typical work products can be mapped to data objects and not to activities) can accelerate and ease the usage of quality approaches. The more structured the quality approach is, the easier it is to use. Similar recognitions can help in case of other quality approaches.

*Issues of differences in structure and elements of quality approaches:*

- less structured quality approaches are difficult to map,
- differences in structure and element of quality approaches make their mapping a complex task.

*A possible solution for handling differences in structure and elements of quality approaches:*

Structural and elemental differences should be recognized and handled by a multi-model solution. End users should be able to use the multi-model result without focusing on differences in structure and elements. As we mentioned previously this can be solved by introducing a unified process (having a unified structure and unified process elements) to which quality approaches are mapped and their structure is hidden by the unified process. Due to the mapping and hidden structure users do not need to know and handle differently structured quality ap-proaches. They can simply use a process having a unified and consistent structure.

### 9.9.2. Handling the differences in structure and elements

**Handling structure and elements in the PBU process:**

In chapter 4 we identified elements of 7 quality approaches and discussed which elements and how they can be mapped to process elements. This analysis led us to create the PBU process which requires the identification of quality approach element instances and their mapping to process element instances.

After the mapping is done to the unified process, the structure and elements of quality approaches are hidden from users, they only use the unified peer review process and they do not need to handle various structures and elements. They only need to be aware of the process elements represented textually and/or graphically. The more quality approaches are mapped to the unified process the more structures and elements need to be understood and mapped by experts, while from the user point of view the structures and elements remain hidden. This means a simplification for the users in using multiple quality approaches and can ease and accelerate their work. The identification of quality approach elements and quality approach element instances in the PBU process is described in chapter 5.4.2.3.

**Handling structure and elements in the case study:**

In the case study we used the quality approach structures and elements identified in chapter 4 as a basis for identifying quality approach element instances. Due to project settings we needed to process CMMI v1.2, meanwhile finalizing this thesis, a new version of CMMI was released, therefore we discussed structure and elements of the new version in chapter 4, while in the case study we discussed the prior version (v1.2) of CMMI (see chapter 7.2.2).

Understanding the structure and elements of three different quality approaches helped us to categorize and identify various quality approach element instances which were used later in creating the unified peer review process.

The identification of quality approach element instances were performed in three iterations in the case study in chapters 7.2, 7.4 and 7.6. We did not encounter problems of identifying and mapping quality approach element instances to process element instances.

### 9.9.3. Conclusion – handling structure and elements of other quality approaches

Structure and elements of 7 quality approaches were discussed in chapter 4. These discussions helped us in identifying quality approach element instances of that 7 quality approaches. Many other quality approaches are also well structured e.g. SPICE, Enterprise Spice or TMMi and their elements can be clearly identified.  There are quality approaches for which quality approach elements can only be identified as chapters, subchapters sentences, shall statements (e.g. ISO 9001-like standards). The usage of these latter, less structured approaches is more difficult since text should be processed manually. We have shown that quality approach elements in the more structured cases can be identified and used in building a unified process. We did not include less structured quality approaches in the case study, but we created a mapping of MSZ EN ISO 9001:2000 requirements to CMMI-DEV v1.2 in (Kelemen & Balla, 2008). In this mapping we used sentences and chapters of the less structured ISO 9001 to map them to practices and goals of the more structured CMMI and we have shown that identification of elements and element instances of a less structured quality approach is also possible. As element instances of less structured quality approaches can be used in a mapping, their mapping to the unified process should also be possible.

If identification and mapping of quality approach element instances is performed in applying the PBU framework then handing structure and elements can be generalized for other processes and quality approaches.

# 9.10. Handling differences in content of quality approaches

## 9.10.1. Understanding issues of differences in content

**First iteration: at literature review (chapter 3.3 and 3.5)**

In literature review we found that quality approaches have different content. In certain cases content of different quality approaches are harmonized in order to reduce the dependence on domain expertise.

**Second iteration: before designing a PBU process (chapter 3.6)**

As a second iteration of understanding issues of different content we argued on how the PBU framework would handle content of quality approaches:

"The PBU framework will provide a unified process to which the content of quality approaches will be mapped. Content will be handled by using structural and elemental recognitions in quality approaches. For example content present in quality approach elements will be mapped to proportional process elements (e.g. in the most cases, content of CMMI subpractices or ISO 9001 shall statements will be mapped to process activities)".

**Third iteration: after performing the case study**

After performing the case study we had a more clear view on the issues and possible solutions of handling different content of different quality approaches:

We have observed that differences can occur because of overlapping or contradiction between the content of the quality approaches.

*Issues of differences in content:*

- overlapping between quality approach element instances (not 100% mapping),
- contradictions between quality approach element instances (contradictory statements –e.g. during an activity something opposite/else should be performed according to another quality approaches).

*Possible solutions for the issues caused by different content:*

- Overlapping: decomposition of quality approach element instances should be done until overlapping intersections are identified. Intersections of quality approach element instances will be mapped to one process element instance of the unified process and the remaining parts of the quality approach element instances will be mapped independently to the element instances of the unified process.

- Contradictions: a decision can be made to include both versions using a decision point (gateway in BPMN) in the process model or to include only one of the contradictory content. This is up to the developer of the unified process. In both cases decisions should be documented in order to ensure repeatability of the PBU process. However, if anything is left out, its implementation and assessment will not be possible.

## 9.10.2. Handling the differences in content

**Handling content in the PBU process**

Regardless of the reasons of the difference, differences in content can be handled by mapping quality approach element instance to process element instances. In case of contradictions there can be multiple solutions e.g. a decision point (represented as a gateway in BPMN) can be included where process experts describe the multiple cases of the processes or a decision can be made for the favour of one option from the contradicting ones. In case of contradictions in roles and responsibilities a decision should be made on which to include/exclude or split, similarly with data objects and entry/exit criteria. Exclusion could be useful in those situations when only one instance of the contradicting quality approach element instances can be present and splitting when contradictions can be present e.g. in different process flows/roles/data objects entry/exit criteria for different (sub)processes.

In case of overlapping: Decomposing quality approach element instances to the level on which the overlapping is present will cause a 1:1 mapping of the intersecting elements.

Parts which are not overlapped will be mapped to different process element instances. Such a decomposition and mapping can be done for overlapping activities and (sub)processes. For data objects, since these can include any number of data items, overlapping can be handled on a data item level. For roles, overlapping can be handled in responsibility level as roles can have unlimited number of responsibilities. For entry and exit criteria overlapping can be handled on criterion level as entry and exit criteria can contain unlimited list of criterion. In all cases the intersection of overlapping quality approach element instances will be mapped to one process element instance. The remaining parts of quality approach element instances will be mapped independently to other process element instances.

In the PBU process quality approach elements (described in chapter 5.4.2.3), their instances are identified and used in building process(es) in chapter 5.4.3.1.

**Handling content in the case study**

We did not encounter contradictions in the three quality approaches. Overlapping in the case study were handled according to the above mentioned. Every process element instance to which more than one quality approach element instance is mapped shows an overlapping between quality approaches.

Since we use the mapping of quality approach element instances to process element instances, mapping also means a summarization of the content. Process elements can be represented textually and graphically. If we take “activity” as an example of a process element we can see it has a name and a description which contains the mapping. See Figure 32 for an example of such an activity. In all cases where a mapping was performed minimum one quality approach element instance is mapped to a process element instance. In all cases of mapping (such as 1:1, 1:N, N:M) a description with multiple mappings was developed. The description is usually read when the activity name is not clear enough. This means if the activity is clear enough and we have multiple mappings, users do not need to read various parts of different quality approaches but they simply go through a process. If they do not understand something, then they can read the documentation of the process element instance which contains the mapping. This kind of representation of the content of quality approaches saves time and energy for the users and also gives a good overview where and how the content of different quality approaches interrelate.

In the case study we identified quality approach element instances in three iterations, we mapped these elements to separate processes and then we mapped processes to a unified process. See the content of overlapped quality approach element instances in Table 54, Table 55, Table 56, Table 57, Table 58 and Table 59 in the appendix.

### 9.10.3. Conclusion – handling differences in content of other quality approaches

Differences in content of quality approaches exist, which can include overlapping and contradictions.

**Handling overlapping:** process element instances can always be decomposed to finer grains until a level where overlapping between quality approach element instances is reached. Then quality approach element instances can be mapped to the new elements of the unified process. Due to the unlimited levels in a process decomposition and mapping can be performed in case of any quality approaches and processes.

**Handling contradictions:** contradictory statements can be included into the unified process by defining two separate process flows, which can be introduced by a gateway or one of the contradictory statements can be included and the rest excluded. Handling contradictions can be performed in case of any quality approaches and processes. In order to ensure repeatability each of these decisions should be documented. It is important to mention that exclusion implies the impossibility of later implementation and assessment of excluded elements, therefore this option should be used carefully.

Summarizing, different content of quality approaches can always be handled by decomposition, exclusion and mutually exclusive solutions regardless of the selected process and quality approaches.

## 9.11. Handling differences in size and complexity of quality approaches

### 9.11.1. Understanding issues of size and complexity

**First iteration: at literature review (chapter 3.3 and 3.5)**

In the literature review the main finding was that size and complexity can influence the selection of quality approaches in a multi-model software process improvement.

**Second iteration: before designing a PBU process (chapter 3.6)**

Before designing the PBU process we argued about how the PBU framework would handle size and complexity differences of quality approaches:

"Complexity can be measured by the coupledness (the number of cross-references inside a quality approach). For example in case of CMMI, it can be observed that there are process areas which are highly interconnected, and there are more separated ones (Balla & Kelemen, 2011; Kelemen, 2011b). These different kinds of process areas should be handled differently. In the PBU framework cross-references among quality approaches will be discovered and referred elements will be checked for further mapping. "

The unified process will hide size and complexity differences among quality approaches as there will be (only one) size and complexity, which is the size and complexity of the unified process. The unified process will also keep the references to source quality approaches, this is discussed in chapter 9.3.

**Third iteration: after performing the case study**

*Issues of size and complexity of quality approaches:*

- Size of quality approaches: In case when a too detailed quality approach is selected available resources may not be sufficient to perform the multi-model SPI.
- Complexity of quality approaches: element instances of a quality approach may be linked with/rely on other element instances of the same quality approach. The more relations are present, the more the quality approach is coupled. Relations and coupledness mean that satisfying one requirement of a quality approach could require the satisfaction of another requirement of the same quality approach.

*A possible solution regarding size and complexity of quality approaches:*

Handling complexity: Quality approaches can contain elements which are related. Relations between quality approach element instances should be identified and taken into consideration when a multi-model result is developed.

Handling size: when selecting quality approach quality approaches should be selected according to their size and complexity. This selection should be made according to the resources available.

Both size and complexity can be measured. Size can be measured e.g. by the number of quality approach elements and element instances. Complexity can be measured e.g. by the number of cross-references inside a quality approach.

### 9.11.2. Handling the differences in size and complexity

**Handling size and complexity in the PBU process**

Size – if the quality approaches are too long compared to the resources available but contain relevant information for the selected process, specific parts can be selected. This will reduce the number of quality approach element instances needed to be handled.

In the PBU process when selecting the quality approaches size is considered and decision has to be taken by experts: which quality approaches and which parts of quality approaches are needed in implementing the selected process.

Complexity – quality approach elements and relations of quality approach elements were discussed in chapter 4.4. This identification of quality approach elements and their relations (structures of quality approaches) led us to define the PBU concept and the PBU process. When creating the mapping, element relations are taken into account and mapping of quality approach element instances to process element instances is done in that way that relations discovered are used.

**Handling size and complexity in the case study**

Size – In the case study size of quality approaches were considered and specific parts were selected from each quality approach. This way we reduced the size of quality approach parts used to less than 50 pages. At the end of each iteration number of element instances identified in that iteration were counted and summarized.

Complexity – In building the unified peer review process we identified relations among different quality approach elements (e.g. between work products and subpractices; inputs and tasks etc.). In case of CMMI subpractices are related to practices, practices are related to goals. In CMMI the specific practice “Conduct Peer Reviews” cannot be performed without performing the specific practice “Prepare for Peer Reviews”. Similar relations among elements were identified in case of IEEE 1028 and Process Impact peer review descriptions.

Additionally, in chapter 11 we will briefly show further interrelations among the element instances of a quality approach (CMMI v1.3). Concepts presented in chapter 11 can be used in developing a unified process.

### 9.11.3. Conclusion – handling size and complexity in other settings

In the three quality approaches we analysed in the case study: CMMI-DEV v1.2 has about 500 pages, IEEE 1028 has 52 pages and the Process Impact’s peer review process definitions are described in 14 pages. Logically, a selection should be made and parts from quality approaches should be used, this reduces both the size and complexity which need to be handled.

Size – If structure and elements of quality approaches are understood then such a selection can always be made. In less structured quality approaches such as ISO 9001 chapters and sentences can always be selected and this way size can be reduced (handled).

Regarding the complexity some of the approaches have clear element relations (especially the more structured ones), others have less clear, mostly textually described element relations. In the first case relations and references can be quickly identified and used. In the latter case texts and possible relations and references between element instances needed to be processed manually.

The case study showed that the PBU process handles differences in size and complexity of quality approaches by a unified process.

## 9.12. Validity and reliability

As Figure 33 illustrates, validity and reliability are generic research criteria which should be discussed in any research. MSPI criteria are specific criteria which should be discussed in developing an MSPI solution (in this case the PBU framework). Generic research criteria overlap with more specific MSPI criteria. These overlapping occur at construct and external validity, and will be discussed in chapter 9.12.2 and chapter 9.12.4 respectively.

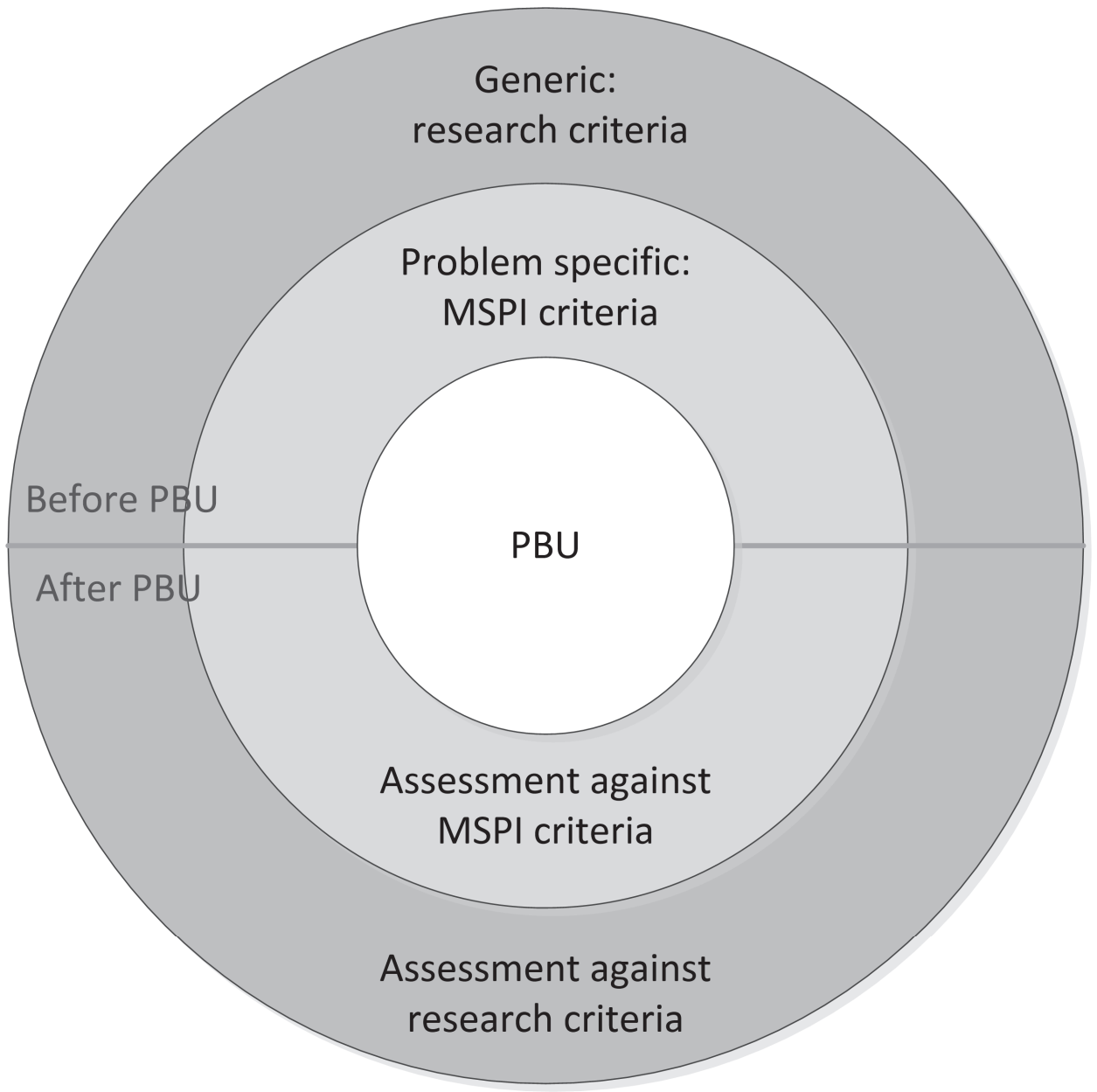

Figure 33 – Onion diagram of the needed assessments
against research criteria and MSPI criteria

In this chapter ecological, internal, construct and external validity as well as reliability will be discussed in two steps:

1. understanding the research criteria,
2. discussing research criteria for the PBU framework.

### 9.12.1. Ecological validity

**Understanding ecological validity**

"Research is often done in ways that are not natural, for example psychological experiments, filling in questionnaires, taking part in role plays and simulations. Such methods raise the question of ecological validity." (Fisher, 2010). The main question of Fisher regarding ecological validity is: Are these results valid in the "messy complexity of real life?" According to (Fisher, 2010) ecological validity the environment of the case study should be discussed, and should be pointed out how 'natural' the case study was.

Other sources assume that case studies are performed in a real life environment and they do not include ecological validity in their research criteria (Gibbert & Ruigrok, 2010; Yin, 2009).

**Ecological validity of the research**

According to (Fisher, 2010) we include ecological validity in our discussion, in order to make it clear what the environment of the case study was.

The environment of the case study on performed on the PBU framework was the Q-works project introduced in chapters 6.2.2. The goal of the Q-work project was to develop a system which supports the application of CMMI process areas.

Environment is important in Q-works project because the project team consisted of two groups: process experts and software developers, so as the project achieved new phases the environment changed from process experts to joint environment.

Table 35 includes the environments related to PBU subprocesses and teams involved.

Table 35 – Case study environments

| Environment | Team | PBU subprocesses |
|---|---|---|
| Unified Process Development | Process experts | 1 Selection,<br>2 Analysis of quality approaches,<br>3 Deriving process from quality approaches. |
| Feedback | Process experts and software developers | 4 Validation |

**1. Environment of developing the unified process**

Subprocesses 1-3 of the PBU process was performed in process expert's environment. This means no involvement from other parties was needed, however project settings influenced the selection of quality approaches. The goal of Q-works was developing a system conforming to CMMI, so that we included CMMI in the selected quality approaches. This might seems unnatural, however, requirement of CMMI conformance is widespread at IT organizations, as thousands of companies are CMMI certified and thousands of further companies base their processes on CMMI without official certification. This means we could select CMMI for a completely different project as well (e.g. when it is not required). Since CMMI is not an unexpected quality approach, and quality approaches did not influence the environment we consider the selection subprocess ecologically valid.

Subprocesses 2 and 3 of the PBU process were also performed in process expert's environment. This means no involvement from other parties was needed because the unified process was developed based on quality approaches.

Therefore we consider the environment of "Analysis of quality approaches" and "Deriving process from quality approaches" as these were performed in their natural environment. Furthermore, real-life quality approaches were used; real-life problems were faced and solved during the development of the unified peer review process.

Q-works specific project settings had no effect on ecological validity of these three subprocesses. Decisions such as how to identify quality approach element instances or how to create the mapping of quality approaches to the unified process were made by process experts.

**2. Feedback environment**

In the previous environment the unified peer review process was developed. However, the PBU process is not complete without validation, which is the fourth (last) subprocess of the PBU process.

Project settings can influence the selection and the acceptance of the unified process and as we pointed out, the selection was done with the goal of developing a system conforming to CMMI. This is quite natural, as thousands of organizations are CMMI certified. Project settings imply that different acceptance results might be achieved in case of a more direct application of the peer review process (e.g. in a project in which users apply the unified peer re-view process in their own work). Q-works project members did not use the case study results for reviewing their work products, instead of this they used it as an input for software development.

However we can state that the acceptance process had several iterations involving both process experts and software developers. Developers needed to understand the process and to use it for development. A SACAMPI Lead Appraiser was involved in the process expert team who reviewed and accepted the resulting unified peer review process. (See chapter 8 for discussion of the acceptance procedure of the case study results).

The result of the case study is a unified peer review process and it was used within the frame of Q-works project.

According to the above mentioned we consider the whole case study ecologically valid.

## 9.12.2. Construct validity

**Understanding construct validity in qualitative research**

“The construct validity of a procedure refers to the extent to which a study investigates what it claims to investigate, that is, to the extent to which a procedure leads to an accurate observation of reality”(Gibbert & Ruigrok, 2010). According to Yin construct validity is the identification of “correct operational measures for the concepts being studied” (Yin, 2009). One main challenge for case study researchers to develop an objective and well-considered set of actions which will be used instead of subjective judgements (Gibbert & Ruigrok, 2010; Yin, 2009). In contrast with the positivist literature above, interpretivists such as (Silverman, 2006) reject construct validity from their set of research criteria. However we think construct validity can be translated to our settings and therefore we should discuss it.

Concrete strategies provided by positivists for ensuring construct validity are (1) multiple sources of evidence (also called triangulation), (2) thick descriptions (also called chain of evidences) and (3) review of the draft case study report by key informants(Creswell & Miller, 2000; Gibbert & Ruigrok, 2010; Yin, 2009). Triangulation ensures looking at the same phenomenon from different angles and can strengthen research results using different sources of data. Thick descriptions provide a chain of evidences and logical, detailed descriptions of the case study. A review by subjects of the case study can assess if they agree with the investigator’s conclusions and interpretations (Yin, 2009).

**Construct validity of the research**

In our case study we used two strategies from the three mentioned above.

1. Thick descriptions and chain of evidences

Objective set of actions were developed before conducting the case study and were included into the PBU process. Well defined set of actions and reasons for choosing this set of actions were discussed in chapter 5 prior performing the case study. After defining the set of actions in chapter 5, a thick and rich description was developed in which we discussed the case study through chapters 6-8, including further evidences in the appendices such as mappings and resulting process models and the unified process model. Chapter 5 provides set of actions on how the case study should be performed. Chapters 6-8 provide a chain of evidence how the case study was performed and what decisions were taken. Furthermore, at the end of each chapter deviations from the PBU process are discussed.

2. Review by case study participants

We did not research human behaviour or opinions about a topic, but rather we developed a unified process by making use of the PBU framework. Project participants were not subject of the case study, however they reviewed results in the following steps:

1. walkthrough on the preliminary PBU results with the process experts,
2. walkthrough on the preliminary PBU results with the whole Q-works team,
3. walkthrough on the final PBU results with the process experts,
4. walkthrough on the final PBU results with the whole Q-works team.

A summary of the PBU process and how PBU results were achieved was also presented to participants. Both experts and software developers accepted the unified peer review process. This part of the case study was considered a validation and it is discussed in chapter Summarizing, we used two tactics from Yin's recommendations and thus we ensured construct validity.

### 9.12.3. Internal validity

**Understanding internal validity**

"Internal validity is also called ''logical validity'' and refers to the presence of causal relationships between variables and results" (Gibbert & Ruigrok, 2010). "Internal validity [is] (for explanatory or casual studies only and not for descriptive or exploratory studies): seeking to establish a causal relationship, whereby certain conditions are believed to lead to other conditions, as distinguished from spurious relationships" (Yin, 2009). "Internal validity is concerned with whether the evidence presented justifies the claims cause and effect" (Fisher, 2010).

**Internal validity of the research**

This research is more descriptive and exploratory than explanatory or casual. We did not investigate if certain conditions led to other conditions, thus according to Yin's definition internal validity is not relevant in our research.

If we look at the logic of steps (subprocesses and activities) of the case study, we discussed each why and how were defined and organized. In this sense we consider our research internal validity sufficient. At the end of each case study subprocess we discussed deviations and refinements are discussed in chapter 10.

### 9.12.4. External validity

**Understanding external validity**

We used the PBU framework in a case study, therefore discussing external validity is extremely important.

External validity is "generalisations or interpretations that a researcher has proved in a particular context apply equally well to other populations of other contexts" (Fisher, 2010). According to Yin, "external validity [is about]: defining the domain to which a study's findings can be generalized" (Yin, 2009). "External validity is grounded in the intuitive belief that theories must be shown to account for phenomena not only in the setting in which they are studied but also in other settings" (Gibbert & Ruigrok, 2010). External validity is also called transferability or generalizability (Lincoln & Guba, 1985).

**External validity of the research**

Our hypothesis in this research was: if mapping of quality approach elements to processes is possible, the MSPI criteria can be satisfied.

In case of external validity we need to answer both parts of the hypothesis, namely:

1. Is mapping possible in broader scale?
2. Can MSPI criteria be satisfied in broader scale?

Since the basic concept of PBU is the mapping of quality approaches to a unified process, broader scale means applying the PBU framework in case of other quality approaches or other processes.

Table 36 summarizes where the case study can be positioned, furthermore what generalizability issues need to be discussed.

Table 36 – Generalizability of results on Process and Quality Approach

| | **Process included (Peer reviews/Inspections)** | **Process excluded (Software processes)** |
|---|---|---|
| **Quality approaches included (CMMI, IEEE 1028, Process Impact)** | 1. Case study | 2. Need to be discussed |
| **Quality approaches excluded** | 3. Need to be discussed | Can be derived from 2. AND 3. |

In order to answer the first question we should discuss:

1. Mapping other quality approaches to a unified process
2. Developing other processes using the PBU framework

**Mapping other quality approaches to a unified process**

In order to determine if the mapping of quality approach elements to process elements is possible besides the identification of process elements, an analysis of quality approach elements was performed.

More than 300 quality approaches were identified by Moore (Moore, 1999). Covering all quality approaches was infeasible within the frame of this research. However, in order to overcome this problem we analysed 7 different quality approaches which represent a high variety of quality approach characteristics. These included quality approaches released by important bodies of standardization such as ISO, IEEE, IEC standards, models from SEI CMU, a governmental quality approach, and also a non-standardized quality approach. These approaches were chosen so that they differ at least in the scope, content, terminology, granularity, structure, size and complexity (the characteristics current multi-model initiatives have problems to handle). Coarse (e.g. CMMI) and finer grained (e.g. IEEE 1028), well (e.g. CMMI) and less structured (e.g. ISO 9001) quality approaches were included into this analysis. We pointed out in all cases that identifying quality approach elements (see chapter 4.4) and their mapping to processes (see chapter 4.6) is possible, even in case of less structured quality approaches. As we covered a high variety of quality approaches and we only experienced solvable problems in case of less structured quality approaches (e.g. manual analysis and mapping of sentences) we can conclude that identification and mapping of other quality approaches should also be possible.

**Developing other processes using the PBU framework**

Since other processes are also described in the quality approaches (e.g. CMMI for Development v1.3 distinguishes 22 process areas). Certainly some processes are explained in detail in quality approaches (e.g. peer reviews in IEEE 1028) and so that there is source to use, some others do not appear in the same quality approach (e.g. only peer reviews are included in IEEE 1028). Thus, for implementing other processes, other quality approaches may be chosen and analysed. As we concluded that identification and mapping of other quality approaches should also be possible, we do not see barriers of developing other processes based on quality approaches.

**Satisfying MSPI criteria in other settings**

In order to answer the second question we should discuss satisfying MSPI criteria in case of other quality approaches and other processes. As we mentioned MSPI criteria and general research criteria overlap. This overlapping happens in case of external validity: in chapters 9.2.3-9.11.3 we already discussed generalizability of the PBU framework for each MSPI criterion pointing out how the PBU framework can satisfy MSPI criteria in case of other quality approaches and other processes where applicable. We concluded at the end of discussing MSPI criteria that it can be satisfied in case of other quality approaches and other processes, thus the external validity of satisfying MSPI criteria is ensured.

### 9.12.5. Reliability

**Understanding reliability**

“Reliability refers to the absence of random error, enabling subsequent researchers to arrive at the same insights if they conducted the study along the same steps again” (Gibbert & Ruigrok, 2010).

Yin defines reliability similarly: “Reliability: [is] demonstrating that the operations of a study – such as the data collection procedures – can be repeated, with the same results.” … “The objective is to be sure that, if a later investigator followed the same procedures as described by an earlier investigator and conducted the same case study all over again, the later investigator should arrive at the same finding and conclusions” (Yin, 2009).

One prerequisite for reliability is the documentation of the case study (Yin, 2009). In order to ensure reliability Yin identifies two tactics: (1) usage of a case study protocol “to deal with the documentation problem” and (2) the development of a case study database. He also adds: “a good guideline for doing case studies is … to conduct the research so that an auditor could in principle repeat the procedures and arrive at the same results” (Yin, 2009).

**Reliability of the research**

According to the literature sources, reliability of this research means that same PBU result can be achieved when the PBU process is repeated by another researcher and this can be supported by documentation.

The PBU process can be performed and a resulting unified process can be developed in many different ways, taking several decisions. Subjective decisions influence the repeatability of the process. In case of decisions, explanation of what happened is required. This is because other people may take other decisions with acceptable results in different cases. If decisions are documented a comparison of results is possible and differences can be explained. Together this should provide sufficient repeatability. Repeatability will improve over time on the basis of practical experiences (of others) and best practices.

In order to ensure repeatability we defined 3 documentation criteria: (1) documentation of the process (the PBU process in chapter 5) (2) documentation of the PBU results (unified process and mapping) and (3) documentation of the decisions taken.

The achievement of these three documentation criteria and repeatability of the PBU process were discussed in detail in chapter 9.2. Additionally, ensuring repeatability in other settings than the case study was discussed in chapter 9.2.3.

The first documentation criterion corresponds to the Yin's case study protocol tactic. For the second tactic we used software tools for collecting quality approach element instances, and developing the mapping. We primarily used MS excel for quality approach element instances and later the ITP Commerce's free BPMN modeler for developing process models.

Reliability is ensured in the PBU framework through a three level documentation (first criterion corresponding to Yin's first tactic) and by the usage of software tools with which various data files were generated which can be comprehended as a case study database (corresponding to Yin's second tactic).

## 9.13. Limitations

Here we summarize limitations of steps performed in this chapter.

**Verification of the PBU framework against MSPI criteria** – The verification of the PBU framework against MSPI criteria can be strengthened by further application of the PBU framework (e.g. based on results of using it in other case studies).

**Validity and reliability** – According to (Gibbert & Ruigrok, 2010; Golafshani, 2003; Shenton, 2004) there is no definitive agreement on what and how should be ensured in assessing the quality of a qualitative research. In order to overcome this problem we used multiple literature sources and techniques in discussing validity and reliability of the research.

## 9.14. Conclusion

In this chapter we assessed the quality of our research from two overlapping perspectives: (1) MSPI criteria and (2) general research criteria.

We developed a deeper understanding what MSPI criteria mean and how can be satisfied. We assessed first the PBU framework and the case study based on MSPI criteria (RS8) and we found that the PBU framework satisfies MSPI criteria; however certain enhancements (such as documenting decisions) still can be made. This will be discussed in chapter 10.

For the second, research criteria (RS9) we went through the PBU framework from ecological, construct, internal and external validity as well from reliability perspective, sometimes refer-ring back to the discussions for MSPI where overlapping between the two criteria happened. Based on our discussions we can conclude that the research satisfies the general research criteria, however if certain modification would be made on the source quality approaches, these could be satisfied even easier (e.g. adding structure and semantics for the less structured approaches).

In the next chapter the lessons learned in this case study will be presented.

# 10. LESSONS LEARNED FROM THE CASE STUDY

In chapter 6-8 we discussed the case study including its preparation, execution and validation. The case study and these chapters were described according to the PBU process presented in chapter 5. In this chapter we present lessons learned during the case study which will be used as a basis for refinement of the PBU process.

## 10.1. Approach

The goal of this chapter is to answer research questions Q10 "*What did we do differently in applying the PBU framework as compared to its original design?*" and Q11 "*What omissions can we identify when comparing the results of applying the PBU framework to the MSPI criteria*?"

In order to answer Q10 and Q11 we perform RS10 "Modify the PBU framework based on RS7 (based on concrete experiences of the case study)" and RS11 "Modify the PBU framework based on RS8".

Lessons learned will be discussed based on:

a. deviations from the PBU process identified at the end of case study chapters 6-8,
b. requirements identified during the analysis of adherence to MSPI criteria in chapter 9.

Lessons learned and refinements will be discussed in chapters 10.2-10.6.

In order to show a coherent picture of what the lessons learned are and what refinement these lessons cause, each discussion will include (1) the source of the lesson learned which can be (a) deviation from PBU process during the case study or (b) a MSPI criteria requirement, then the (2) lesson learned and (3) the resulting refinement on the PBU framework.

Figure 34 represents the refined PBU process, refined activities having grey background.

The following lessons learned will be discussed:

1. Searching and understanding the definitions of the process to be implemented (based on a deviation),
2. Creating separate process models (based on a deviation),
3. Using the separated process models in creating the unified process (based on a deviation),
4. Verification of the mapping (based on a deviation),
5. Documenting decisions (based on a MSPI criteria requirement).

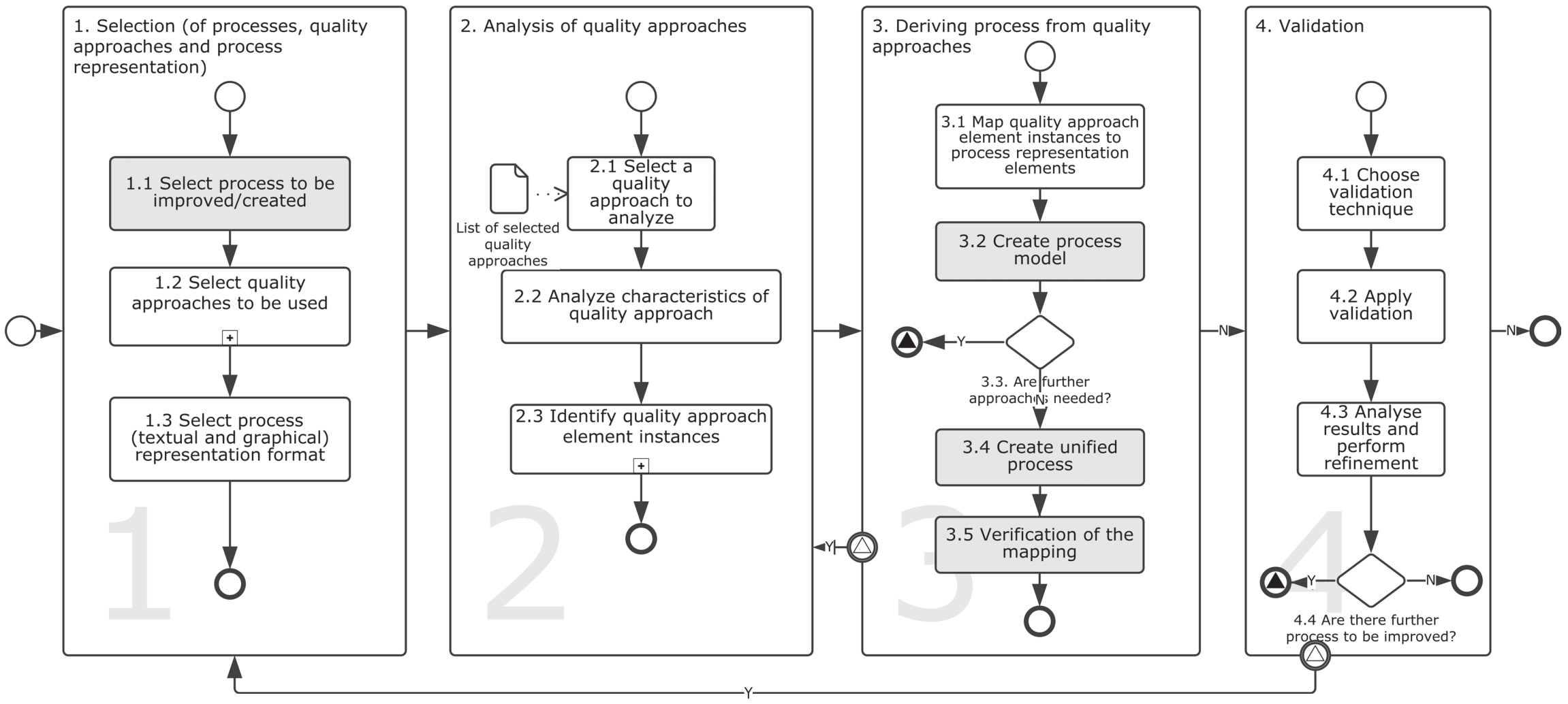

Figure 34 – A refined PBU process (refined activities are grey)

## 10.2. Searching and understanding the definitions of the process to be implemented

**Deviation from PBU process**

In chapter 6 we followed the PBU process defined in chapter 5.4.1, however two deviations from the original process occurred:

1. selecting the context,
2. searching and understanding the definitions of the process.

The first deviation is an additional step to the first subprocess of the PBU process. It occurred because we needed to carefully select the context for performing a case study research. In practice context is given, therefore it is not needed to include it into the PBU process.

The second deviation is an addition to the PBU process and it occurred due to practical reasons: we needed to know what process exactly we want to develop. At this situation looking for the definitions was a good starting point.

The activity of defining a starting point involved:

1. a search for definitions,
2. the analysis of definitions and
3. selecting the definitions needed (definitions of review, peer review, inspection).

**Lesson learned**

Process implementation without understanding basic notions is not possible, so most im-portant terms should be defined at the beginning as these will provide the basis for the whole process development. In the case study the analysis of terms helped clarifying the scope of unification and this can be useful in performing process based unification in case of other quality approaches or processes. An analysis of the important terms could be useful in understanding what processes are really needed and how they will be implemented.

**Refinement on PBU process**

As a part of the activity “Select processes to be improved/created” of the “Selection (of processes, quality approaches and process representation)” subprocess (1) a search for the most important terms, (2) their analysis and (3) selection of basic definitions related to the process are needed.

# 10.3. Creating separate process models

**Deviation from PBU process**

Chapter 7 describes the application of two PBU subprocesses in the case study: Analysis of quality approaches and Deriving process from quality approaches. In these two subprocesses three deviations from the original PBU process occurred:

1. creating separate process models,
2. using the separate process models in creating the unified process,
3. verification of the mapping.

Here we discuss the first deviation and we will continue in chapters 10.4 and 10.5 with the remaining two.

The first deviation is an additional activity to the third subprocess of the PBU process. Remember that in the first version of the PBU process the identification of quality approach element instances (subprocess 2.3) was followed by the mapping of quality approach element instances to process (activity 3.1). In the case study we needed to create separate process models because we needed to understand what is a possible process flow behind the requirements of each different quality approach. This way we developed 3 different peer review process models before the mapping.

The process model provided us an overview about how a process can look like if it is developed based on a single quality approach. In the three separate process models the most important process elements were the activities and subprocesses and their relations – these provided the overview of the process. In most cases the quality approaches did not contain all the information (e.g. the process flow, order of activities), therefore we completed the process with element relations which seemed logical (as it is discussed in chapter 7).

**Lesson learned**

Separate process models can help understanding a possible process flow behind each individual quality approach. Understanding the process flow behind quality approaches helps developing a proper mapping between quality approaches and the unified process, namely: which quality approach element instances to which process element instances to map.

**Refinement**

The most important part in this deviation was to understand the process flow and how requirements can be mapped to a process (what process these requirements reflect), therefore we include a new activity in the PBU process "Create process model". Table 37 contains the activity description in the same format as other activities were described in chapter 5.4. Changes in the numbering of activities and updates of the PBU process are reflected in Figure 34.

Table 37 – Create process model

| | |
|---|---|
| **Activity ID** | 3.2 |
| **Parent ID** | 3 Deriving process from quality approaches |
| **Activity name** | Create process model |
| **Activity description** | Create separate process model using element instances of the quality approach analysed. Later this can be used in building the unified process. The result of this activity is an intermediate product which helps understanding what process is behind the quality approach. |
| **Inputs** | Quality approach element instances |
| **Outputs** | A process model for the selected process based on each individual quality approach |
| **Roles/ responsibilities** | SEPG team |

## 10.4. Use separate process models in creating the unified process

**Deviation from PBU process**

As we discussed in chapter 10.2, creating separate process models (the first deviation in chapter 7) was an additional activity to the third subprocess of the PBU process. We included this because we needed to understand what is a possible process flow behind quality approaches.

The second deviation in chapter 7 was "Use separate process models in creating the unified process" – instead of pure mapping of quality approach element instances to the process. We performed the mapping using the three processes defined.

This activity reduced the number of possible mappings: in case of subprocesses, activities and gateways we used the process flow in creating the mapping and we did not try to map activities of totally different subprocesses. For example in the first iteration, activities of CMMI 'perform peer reviews' were not mapped to activities of 'management preparation'. This is because after modeling the three separate processes we understood that these were totally different parts of the process and we did not need to try to map them.

After each process model was built, the mapping was created – which was easier, because we already understood what the quality approaches require and we used the process models during the mapping.

**Lesson learned**

Understanding and using separate process models can reduce the number of possible mappings between quality approach element instances and process element instances.

**Refinement**

The "Create/refine process" activity of the first version of the PBU process is split into activities "Create process model" and "Create unified process" (see Figure 34). After creating par-ticular process models using quality approaches separately, we can use them in creating the unified process model within the activity “Create unified process”. Table 38 includes the resulting “Create unified process” activity.

Table 38 – Create unified process

| | |
|---|---|
| **Activity ID** | 3.4 |
| **Parent ID** | 3 Deriving process from quality approaches |
| **Activity name** | Create unified process |
| **Activity description** | The role of this activity is to create a new process model using separate process models. Creation of unified process model involves the mapping to source quality approaches. For doing this, quality approach element instances and separate process models can be used. |
| **Inputs** | - The mapping of quality approach element instances to process representation elements,<br>- separate process models. |
| **Outputs** | - Unified process model,<br>- the mapping of quality approach element instances to unified process model. |
| **Roles/ responsibilities** | Process owner, SEPG team, Process stakeholders |

# 10.5. Verification of the mapping

**Deviation from PBU process**

The third deviation occurred in chapter 7: we performed the activity “verification of the mapping” after iteration in subprocesses 3 and 4. This was caused by the first deviation (create separate process models). In performing the PBU process, separate process models were created and no mapping and verification of mapping was needed inside the iteration. The creation of the unified process model was performed outside (after) the three iterations of creating separate process models (incorporating the mapping). Thus, the verification of mapping was needed to be placed after creating the unified process model.

**Lesson learned**

The activity “Verification of the mapping” should be performed once after creating the unified process model.

**Refinement**

The place of activity “Verification of the mapping” in the PBU process was changed. It is now performed once after the iterations of subprocesses 3 and 4. See Figure 34 for the place of activity 3.5 “Verification of the mapping” in the process flow.

# 10.6. Document decisions

**MSPI criteria requirement**

In chapter 9 the case study was discussed against MSPI criteria. In order to ensure repeatability (chapter 9.2) three criteria on documentation were defined: (1) documenting the process, (2) documenting the result and (3) documenting how results were achieved (documenting subjective decisions).

**Lesson learned**

As the PBU framework should fully satisfy MSPI criteria, all the three documentation criteria should be satisfied.

**Refinement**

The PBU process is documented in chapter 5 and refinements based on case study in chapter 10. The PBU result is the unified process, which is also automatically documented if the PBU process is performed.

In order to ensure the repeatability of the PBU process decisions should also be documented (the decision itself and its reason). Such decisions can encounter at various activities of the PBU process: e.g. at selecting the quality approaches, making decisions on the terms used or deciding the granularity of the unified process. So that the refinement at this point is adding the third documentation criterion to the PBU process: in case if repeatability is required, decisions should be documented.

## 10.7. Limitations

Here we summarize limitations of steps performed in this chapter.

**Refinements made on the PBU framework based on case study experiences** – One single case study was performed in this research. Testing the PBU framework in further case studies could strengthen current refinements and may reveal new ones.

**Refinements made on the PBU framework based on the verification of the PBU framework against MSPI criteria** – During this step we performed one single modification of the PBU framework by adding the documentation of decisions criterion. One limitation of this criterion is that it may slow down the PBU process. Therefore this addition should mainly be used where repeatability is required.

## 10.8. Conclusion

The goal of this chapter was to discuss lessons learned and refinements of the PBU framework based on experiences of the case study (RS10) and based on the verification of the PBU framework against MSPI criteria (RS11). In this chapter we discussed four deviations from the PBU process (Q10) and an MSPI criterion requirement (Q11). For all these we discussed lessons learned and refinements on the PBU framework.

Further testing and case study experiences from the community can help making the PBU process an even more reliable tool of multi-model process improvement.

The next chapter includes reflections to the research which include (1) guidelines for applying the PBU framework, (2) a data model for supporting maintainability of the PBU results and (3) strategies for analysing quality approaches.

# 11. REFLECTIONS TO THE RESEARCH

## 11.1. Approach

The objective of this chapter is to present a number of issues that arose during the research and are relevant, but outside of the scope of the researched thesis. These include (1) guidelines for those wishing to use the PBU process in their work, (2) a data model for further supporting maintainability of the PBU results and (3) preliminary results of analysing a quality approach: the CMMI model.

Guidelines include practical experiences of applying the PBU framework and they can also support its application. Compared to the lessons learned presented in chapter 10, these guidelines are more practical observations than research results and have no refinement effect on the PBU framework. We present guidelines in chapter 11.2.

In chapter 9 we identified options for enhancing maintainability of PBU results. In order to give a reflection on how maintainability can be enhanced, we will discuss the structure of a data model which can support maintainability. The data model will be presented in chapter 11.3.

As a further reflection to the research, we present preliminary results of applying two qualitative strategies for analysing quality approaches: complexity analysis and text mining applied on quality approaches. These will be included in chapter 11.4.

## 11.2. Guidelines for applying the PBU framework

In this chapter we summarize experiences in using the PBU framework which did not have change effects on the PBU framework itself but can facilitate its practical usage.

### 11.2.1. If certification is not required use those elements of quality approaches that you need most

We have experienced that especially finer grained quality approaches have element instances which are probably not needed at every organization. Quality approaches are usually developed by committees, reflecting knowledge, research results and practical experiences of its members. Specific content in the quality approaches may not fit in the expectations of the organisation due to the differences in culture, knowledge, process view, rigor etc. between the organization and the (authors of the) quality approach.

Some quality approaches are more while others are less rigorous. An organizational culture can also be more formal or informal and processes can be developed in a variety of levels. In case certification is not required, a good option could be to use only the most useful element instances of quality approaches and focus more on practical acceptance and usage of the process rather than the conformity and traceability to quality approaches. In this case a complete mapping, traceability and appraisal support will not be fully provided.

### 11.2.2. Make the “Analysis” subprocess independent (if possible)

According to our experience, analysis of quality approaches is a time consuming (sub)process and can be performed independently from any company. Thus, a further refinement could be to make the second subprocess independent and create an open repository of quality approaches from which anyone can use the approaches and their element instances needed. Furthermore, if standardisation organizations could provide their standards in a more semantic format (e.g. denoting basic process elements) it would also help in identifying quality approach element instances. Taking these changes would accelerate the whole PBU process.

### 11.2.3. Order the mapping: start from the finer grained elements

In (Madison, 2008) bottom-up, top-down, vertical and horizontal process mappings are distinguished and these primarily refer to building processes from green field (mapping organizational processes to process model). In such situation the process is described based on experiences and not based on various sources such as quality approaches.

Since mapping often requires decomposition to achieve alignment between notions from different quality approaches having different granularity, starting at the most detailed level gives a good foundation. Do the following: first map the finer grained elements to the process and then continue with the less detailed, less elaborated ones.

In the case study we used a different order of mapping for activities than at mapping data objects. For activities we started with IEEE 1028 – since we considered it the most detailed (finer grained) quality approach, then we continued with Process Impact and finally with CMMI. In case of data objects Process Impact provided templates and thus it was easier to map Process impact’s quality approach elements to data objects then continue with IEEE and CMMI. This approach facilitates that even the lowest level quality approach element instances will be mapped to the unified process.

### 11.2.4. Identify quality approach elements according to their importance

In focusing on the process flow, activities, subprocesses and transitions between them are the most important elements. This is also typical in various process representation formats such as automata machines or Petri nets (Booth, 1967; Murata, 1989). Of course there are other representations of processes (see e.g. a variety of UML diagrams or data flow diagrams), but these usually do not have the primary focus on the process flow. If activities and subprocesses and transitions are identified, a process flow can be quickly drawn, then other elements can be added.

In building processes using element instances of process oriented quality approaches, important elements are mappable to process flow, e.g. activities/subprocesses and activity/subprocess relations. In case if the goal is to understand the process flow, element instances should be identified and mapped first, then other element instances can be added if required.

### 11.2.5. Use a tool

According to our experience, maintaining a high number of quality approach element instances, process element instances and their mapping could be a difficult task which requires time and effort in case the maintenance is done in simple tables. A possible solution for enhancing maintainability is using a tool which can ease the adaptability, expandability, traceability and the appraisal support of the PBU results.

In the case study a BPMN process modeling tool (itp-commerce, 2012) was used in which mappings of process element instances to quality approach element instances were stored as texts. See Figure 32 for the mapping stored in the tool. This tool helped us to ensure traceability to quality approaches; however it did not support adaptability and expandability. This is because the tool is only able to store mappings as textual properties of process element instances but it cannot store/maintain links to the source quality approach element instances.

In chapter 11.3 a data model for enhancing maintainability of the PBU results will be presented. This could serve as a general basis for a tool and also for understanding what the interrelations between main entities of the PBU result are.

## 11.3. A data model for enhancing maintainability of PBU results

If we analyse tables in the appendix we can see that most important data to be stored are:

- quality approaches, their elements and element instances,
- process representation, process elements and process element instances,
- mapping between the quality approaches and the processes.

**Reasons for a data model**

1. Entities and their relations can be stored in tables as we did in the case study or in a more structured and easier accessible way e.g. in XML or a database. More structured and easier accessible storing of data is important in any system when the amount of data and their relationships is high and multiple views of the same data are needed. In our case such multi-ple views could be: quality approaches, unified processes, mappings of processes to quality approaches etc. A good solution could be to store all the entities in a database on which a tool can be built.

2. A data model also can make clear what interrelations are among important entities. This helps us understanding how core entities can be defined and what the relations between them are.

We use the term *maintainability* to cover the following set of MSPI criteria: adaptability, expandability, traceability and appraisal support.

As we mentioned in 11.2.5, a software tool would be a good option to enhance maintainability. In the limits of this thesis we are not going to develop a software tool, but we provide a data model, which can be used as a basis for developing a tool. This data model is not an ultimate solution for the maintainability, it is rather a general discussion on how such a database can be built for enhancing maintainability.

In chapter 9 we have seen that MSPI criteria can be satisfied, however improvements can be done to better support adaptability, expandability, traceability and the appraisal support. A usual activity during appraisals is the conformance check of processes against quality approach requirements. In order to provide this kind of appraisal support, traceability is needed. In this chapter we present a data model which can be used to enhance maintainability of PBU results.

In developing a data model we distinguish the following two steps:

1. identifying requirements of a data model based on MSPI criteria (what should be done),
2. developing an entity-relationship diagram based on requirements (how should be done).

In chapter 11.3.1 we present the requirements for a data model, while in chapter 11.3.2 the data model itself is presented.

### 11.3.1. Identifying requirements for a data model

In this chapter we identify requirements for a data model. Requirements are identified based on the maintainability issues of the PBU results.

Table 39 – Requirements of a data model

<table>
<tr><th>MSPI criteria</th><th>Implementation requirements for a data model</th></tr>
<tr><td>Adaptability</td><td rowspan="2">Required entities for storing:<br>1. Element instances of different versions of quality approaches,<br>2. Process instances,<br>3. Mapping of quality approach element instances to process element instances.</td></tr>
<tr><td>Expandability</td></tr>
<tr><td>Traceability</td><td rowspan="2">4. Required entities for storing the mapping of quality approach element instances to process element instances (same as requirement 3)</td></tr>
<tr><td>Appraisal support</td></tr>
</table>

Table 39 contains the MSPI criteria and requirements against the data model which were formulated by us based on discussion in chapter 9 and our data modeling experience.

According to Table 39 and experience of the case study the following main entities and their related entities should be represented in the data model.

For requirement 1 identified in Table 39:

1. quality approach elements (type level),
2. quality approach element instance (instance level).

For requirement 2 identified in Table 39:

3. textual and graphical process model elements (type level),
4. textual and graphical process model element instances (instance level).

For requirement 3-4 identified in Table 39:

5. mapping quality approach element instances to process element instances.

The five main entities identified above and their related entities are discussed in chapter 11.3.2.

### 11.3.2. A data model

In this chapter we present elements of a data model which were developed to reflect requirements identified in the previous chapter. The entity-relationship diagram was developed by using the free ER modeler tool 'MySQL Workbench'. We used MySQL data types, but these can easily be substituted in case of using other databases. The emphasis is more on entities and relations than on data types and attributes. However some example attributes and data types are shown in figures to illustrate the purpose of entities.

Figure 82 and Figure 83 in the appendix show an overview of the five major parts (quality approach element, quality approach element instance, mapping, process instance and process language) of the ER model.

Figure 35 shows an overview of main entities and their relations:

- the entity "QualityApproachElement" stores the element types (e.g. in case of CMMI: goals, practices among others) of a quality approach,
- the entity "QualityApproachElementInstance" stores the quality approach element instances (e.g. in case of CMMI instances of goals, practices among others),
- the entity "QualityApproach_Process_Mapping" stores the relation among quality approach element instances and process element instances, this is the main entity which provides traceability and appraisal support through mapping,
- the entity "ProcessElementInstance" stores the element instances of the processes,
- the entity "ProcessElement" stores the process element types and their textual and possible graphical representations – this latter is needed when a process is also represented in a process modeling language.

In chapter 4 we mapped quality approach elements to process elements, however, as it can be observed in Figure 35 we did not include a relation between the entities "QualityApproachElement" and "ProcessElement". As we have shown in the case study, all mappings are performed on instance level between "QualityApproachElementInstance" and "ProcesElementInstance". Reasons of this:

- Granularity of processes is not generalized thus the same quality approach element instance can be mapped to process, subprocess or activity depending on granularity level of the process (description).
- Weakly structured textual information (e.g. shall statements) could be mapped to multiple process elements, e.g. to activities but also to roles-responsibilities. One such statement could be: “The inspection leader shall invite inspection participants”. In this case the statement can be mapped to at least one role as a responsibility and to an activity.
- As we have stated in chapter 4.6 a mapping should always be performed on instance level. This is because it is not trivial which quality approach element instances should be mapped to which process element instances. (This could be supported by automatic recognition of similarities between the text of quality approach element instances and process element instances, but this still cannot provide the type level mapping, because automatic mapping of texts should also be confirmed by users.)

If mapping could be performed on type level, a M:N relation and entity between the entities “QualityApproachElement” and “ProcessElement” could be included.

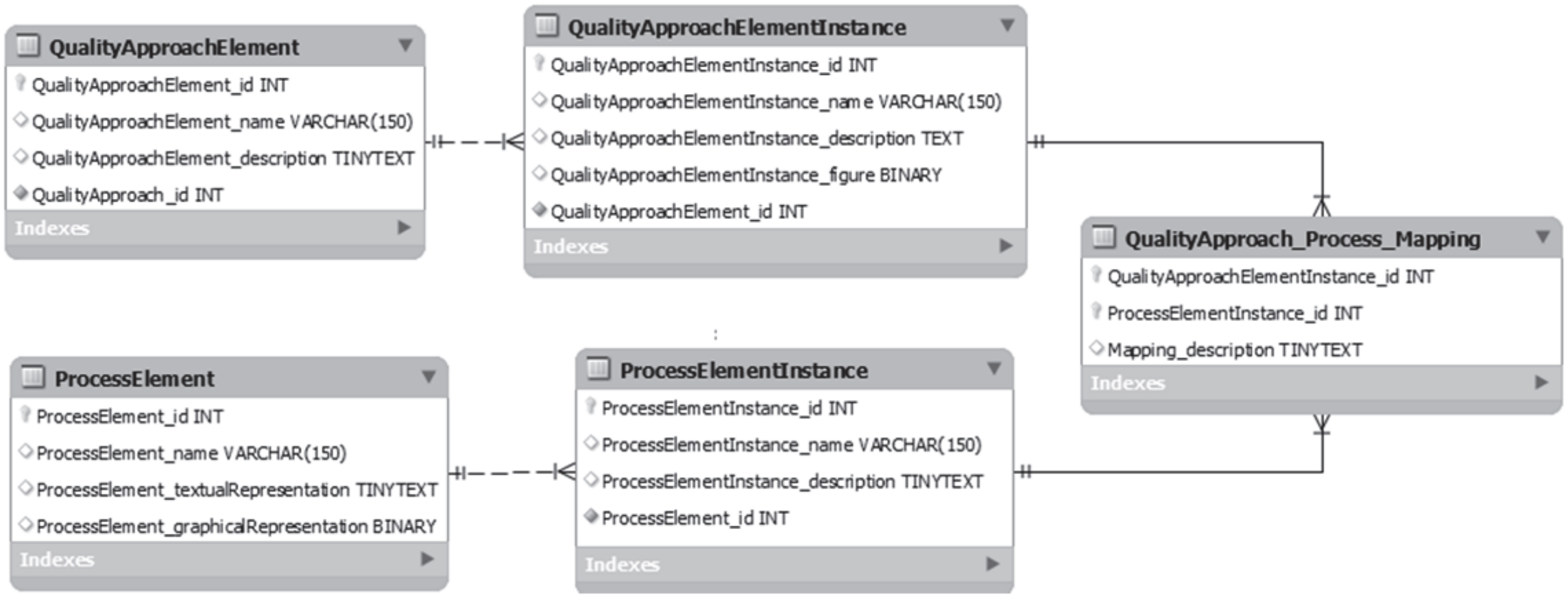

Figure 35 – Main entities of a data model

Figure 36 shows a view of the “ProcessElement” entity and its related entities. The “ProcessElement” entity describes how textual and graphical attributes of a process element can be stored. For simplicity reasons, attributes related to the graphical representation were not included into the table such as size or position. The “ProcessElement” entity stores the possible process elements of a process modeling language such as activity, role, inputs-outputs, etc. The description of the process modeling language is included into the “ProcessModelingLanguage” entity. It can happen that different process modeling languages can have the very same element (e.g. the “role” element is present in multiple languages, therefore there is an additional entity called “ProcessModelingLanguageElement” in which the association of process elements and process modeling languages can be stored.

Process elements can be connected to each other, but not all kind of connections are allowed in a process modeling language, therefore the table “AllowedProcessElementRelation” stores the allowed relations among process elements. The types of allowed relations between two process elements are stored in the "AllowedProcessElemenetRelationType" entity.

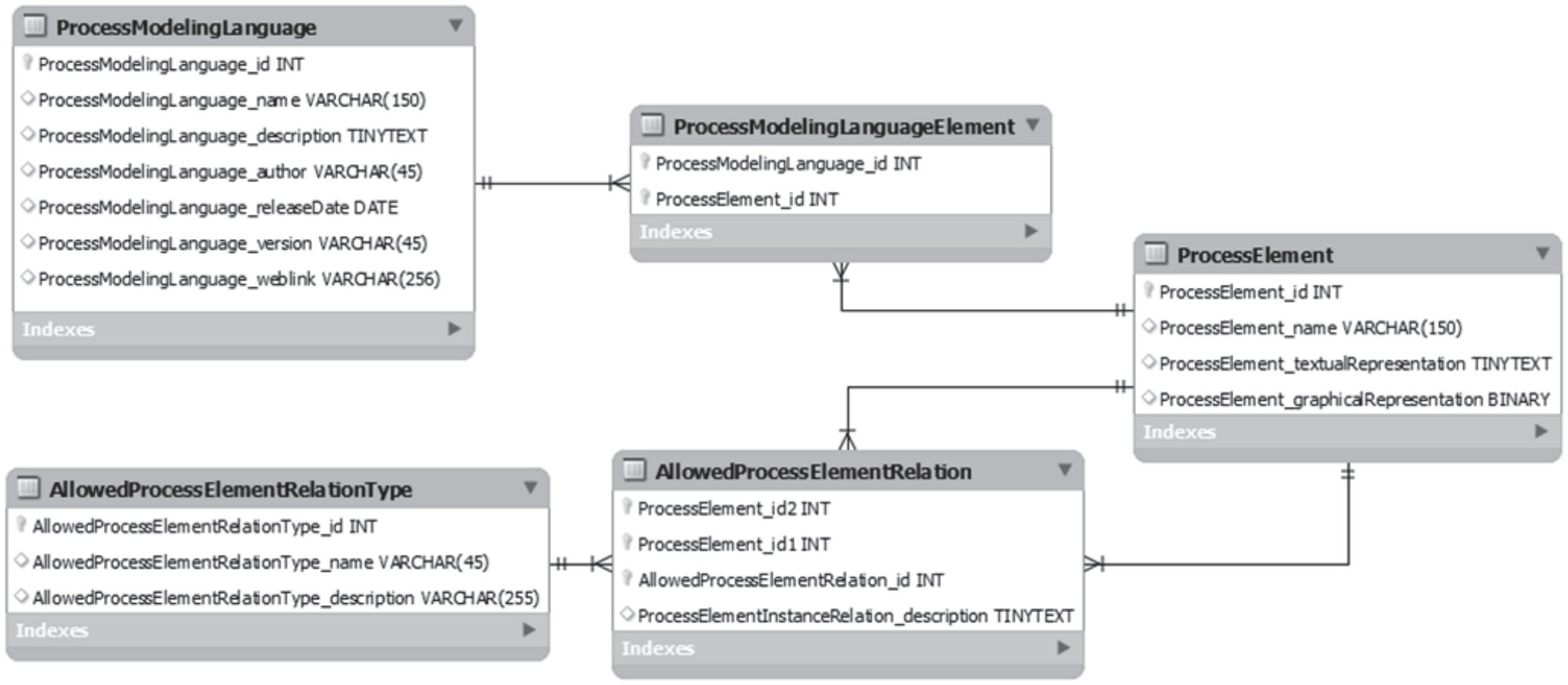

Figure 36 – The ProcessElement and its related entities

Figure 37 presents the entities "ProcessElementInstance", "ProcessElementIntanceRelation" and "ProcessElementIntanceRelationType". The entity "ProcessElementInstance" is an instance of a "ProcessElement" shown in Figure 36. Examples of peer review process element instances in BPMN could be: instances of subprocesses (e.g. prepare for peer review), roles (e.g. author or reviewer) or instances of any other process elements. The relation between two process element instances is stored in the "ProcessElementInstanceRelation" entity. Such relations among many others can be for example:

- parent-children relation in case of process – subprocess,
- process flow relation (which subprocess follows the other subprocess),
- activity – role (which role is needed to perform a certain activity),
- new version of, e.g. in case if a tool built on this database allows storing multiple versions of processes, relations between versions can be stored in the proposed data model.
- tailored from – if an organization describes processes on multiple levels (e.g. on theoretical, organizational and project levels).

Since we do not know the full extent of the possible relation types, relation types are stored in "ProcessElementInstanceRelationType" – and this structure allows the addition of new types over the time.

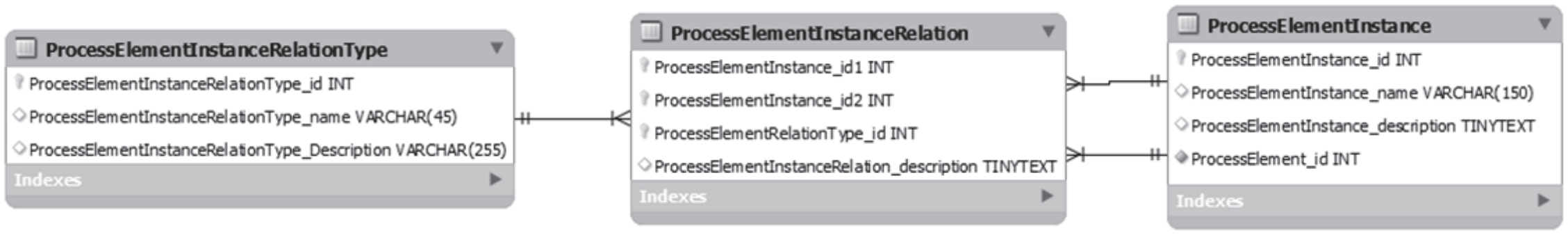

Figure 37 – The ProcessElementInstance and its related entities

Figure 38 includes entities "QualityapproachElement", "QualityApproach", "QualityApproachAttributeList", "QualityApproachAttributeValue" and "QualityApproachAttirbute".

The "QualityApproach" entity includes the attributes of a quality approach; it can have ele-ments which are stored in the entity “QualityapproachElement”.

A further requirement against such a data model could be an enhanced support of the selection of quality approaches. In the literature review included in chapter 3 we discussed initiatives which help selecting from quality approaches based on their attributes. Since these attributes are quite flexible and not ultimately set and probably will be developed further in the future, we store these attributes using three entities: “QualityApproachAttributeList”, “QualityApproachAttributeValue” and “QualityApproachAttirbute”. Possible quality approach attributes could be: origin, popularity, assessment approach, goal, scope among many others. As these attributes can have various types of values, values are stored in a separate entity, “QualityApproachAttributeValue”. We also allow any number of attributes for quality approaches, this is represented by the entity “QualityApproachAttributeList”.

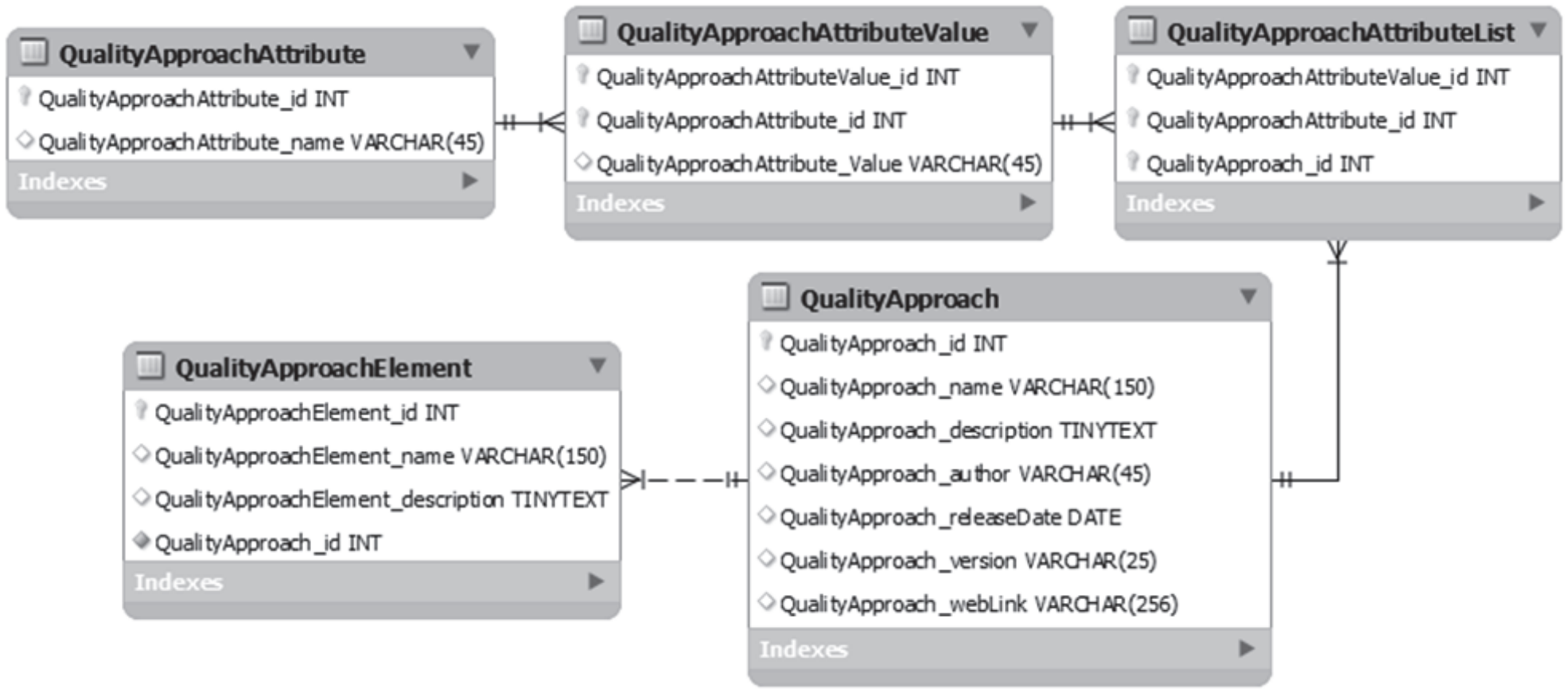

Figure 38 – The QualityApproach and its related entities

Figure 39 shows the “QualityApproachElementInstance” and its related entities “QualityApproachElementInstanceRelation” and “QualityApproachElementInstanceRelationType”. These three entities works similarly to those described in Figure 36.

Similarly to process element instance relation types, possible quality approach element instance relation types can be (but not restricted to): part of / contains, refers to, requires or version of.

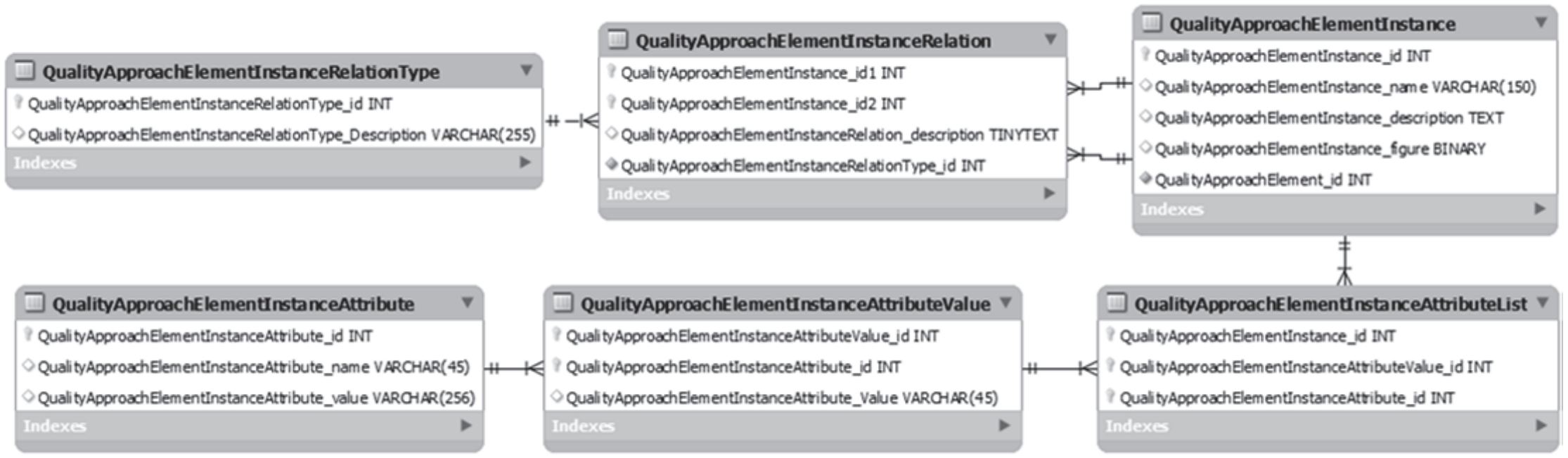

Figure 39 – The QualityApproachElementInstance and its related entities

Quality approach element instances can have attributes which are related only to certain instances e.g. the level of requirement in IEEE and ISO standards shall, should may. These attributes are stored in entities “QualityApproachElementInstanceAttribute”, “QualityAp-proachElementInstanceAttributeValue" and "QualityApproachElementInstanceAttributeList".

## 11.4. Strategies for quantitative analysis of quality approaches[7]

In this chapter we discuss preliminary results of applying two qualitative strategies for analysing quality approaches:

- complexity analysis of quality approaches,
- text mining applied on quality approaches.

We call *complexity analysis* the analysis of the cross-references of quality approaches and discovery of isolated and coupled element instances. In chapter 11.4.1 we present complexity analysis of CMMI v1.3. In chapter 11.4.2 we present suggestions related to the text mining used for understanding the focus of quality approaches. Both strategies could be interesting because they do not need too much manual or qualitative investigation and could provide a quick overview on quality approaches. Quantitative analysis could help in understanding focus, decisions in implementation and training as well as in tool development and appraisals.

### 11.4.1. Complexity analysis

We analysed structure and elements of 7 different process based quality approaches in chapter 4 and some other versions in (Kelemen et al., 2008a), which can provide a basis for further analysis of quality approaches. However, deeper investigation and analysis (e.g. complexity of) quality approaches were not provided and are not common in the literature, even though these could contribute to better understanding and processing of quality approaches. Preliminary complexity analyses were presented in (Balla & Kelemen, 2011; Chen, Staples, & Bannerman, 2008; A. L Ferreira et al., 2010; Andre L. Ferreira et al., 2010; Kelemen, 2011b; Monteiro, Machado, Kazman, & Henriques, 2010). These investigations of quality approaches are mainly focusing on cross-references and interrelations inside quality approaches (among their elements or element instances). In this chapter we present preliminary results of investigating cross-references in CMMI. These were briefly presented in (Balla & Kelemen, 2011; Kelemen, 2011b).

**Understanding cross-references in CMMI**

CMMI version 1.3 has three constellations: CMMI for Development, CMMI for Services and CMMI for Acquisition.

In chapter 4 we presented the structure of CMMI v1.3. Figure 40 presents a summary of CMMI versions 1.1-1.3, constellations and their elements, showing that constellations have about 400-500 pages each, and version 1.3 has 1440 pages in total. Standards such as SPICE, Enterprise SPICE, TMMi among others have similar size. The amount of information in these standards makes it difficult to understand and apply them; therefore in this chapter we show preliminary results of analysing CMMI as an example of possible complexity analysis of quality approaches.

[7] Results of this chapter were presented on SEPG Europe 2011 (Balla & Kelemen, 2011) and on the International ODF conference, Budapest (Kelemen, 2011b).

| Measure | CMMI for Development | | | | CMMI for Acquisition | | CMMI for Services | |
|---|---|---|---|---|---|---|---|---|
| | V1.1 Staged | V1.1 Cont | V1.2 | V1.3 | V1.2 | V1.3 | V1.2 | V1.3 |
| Pages | 715 | 710 | 560 | 482 | 428 | 438 | 531 | 520 |
| Process Areas | 25 | 25 | 22 | 22 | 22 | 22 | 24 | 24 |
| Generic Goals | 2 | 5 | 5 | 3 | 5 | 3 | 5 | 3 |
| Generic Practices | 12 | 17 | 17 | 13 | 17 | 13 | 17 | 13 |
| Specific Goals | 55 | 55 | 50 | 49 | 46 | 47 | 52 | 53 |
| Specific Practices | 185 | 189 | 173 | 167 | 161 | 163 | 182 | 181 |

Figure 40 – CMMI versions and constellations and their elements.
Source: (Forrester & Wemyss, 2011)

#### 11.4.1.1. **Searching for cross-references in CMMI**

In order to understand relations among CMMI element instances we conducted a systematic search within CMMI documents.

**Search space**

In this search we focused on CMMI v1.3 and we analyzed all three constellations. As the goal was to understand relations among element (instance)s of CMMI, the search was narrowed to the chapter 2 of all three CMMI documents – these contain process areas and their related elements (e.g. specific and generic practices). We filtered out all the elements which were irrelevant for our search. These were introductions, appendices, remarks, notes, design elements etc.

**Search terms**

First we conducted a manual analysis on CMMI element instances such as process areas, specific and generic practices. Based on this manual analysis it was easy to discover cross-references, as they intentionally appeared in the following patterns:

- "Refer to the" … "process area …"
- "Refer to the" … "specific practice in" … "process area…"

We have not found other reference patterns in CMMI, therefore we searched for these two types of references. Some irrelevant search results were filtered out when searching for patterns such as "the organization's set of standard processes can ***refer to the*** standard processes established at the organization level".

**Search results**

The search resulted in 1016 cross-references in total, and 992 after applying filters. Out of these 311 were found in CMMI-DEV, 388 in CMMI-SVC and 293 in CMMI-ACQ.
As CMMI-DEV is the most common constellation in the following we show some results gained from this constellation, excluding CMMI-ACQ and CMMI-SVC from the remaining discussion.

#### 11.4.1.2. **Complexity analysis: understanding cross-references in CMMI**

In CMMI-DEV, cross-references were found in various element instances e.g. in instances of introductory notes, related process areas, specific practices and generic practices. Besides that references can be found on different levels, these are pointing to different element instances such as instances of process areas, specific goals and specific practices.
Figure 41 shows generic practices and process areas referenced in CMMI. For better visibility, references found below process area level (e.g. in introductory notes, specific goals and specific practices) are represented on process area level. The figure shows that the least referred element instances are the generic practices and the PPQA (Process and Product Quality Assurance) process area, while most referred process areas are IPM (Integrated Project Management), TS (Technical Solution), QPM (Quantitative Project Management), PI (Product Integration) and OPM (Organizational Performance Management).

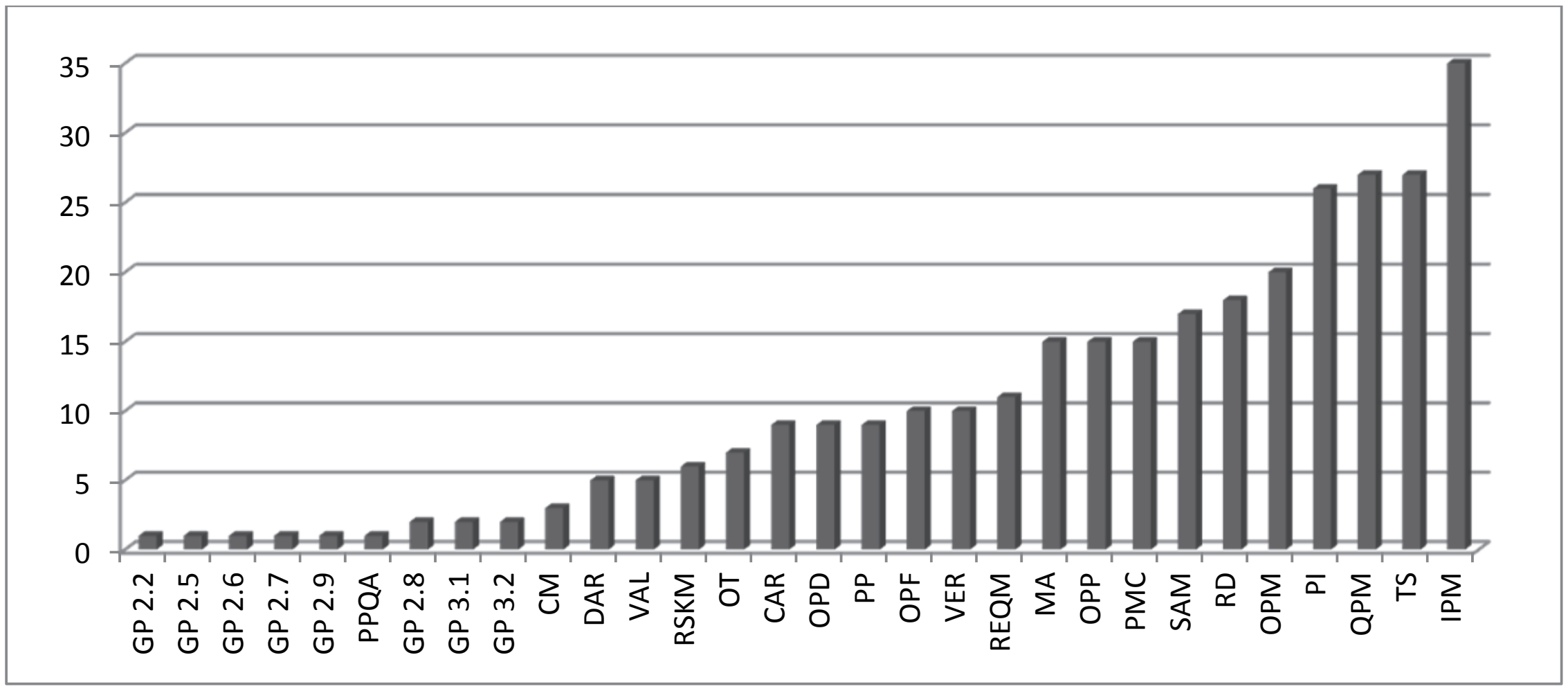

Figure 41 – References to process areas and generic practices in CMMI-DEV v1.3

Less referenced elements could be more independent and thus easier to understand and implement, e.g. at teaching CMMI process areas PPQA could be useful to start with. At the same time understanding and implementing highly referred process areas might be crucial, but also more difficult as several other process areas rely on these. Better understanding of highly referred process areas could be especially useful in MSPI: e.g. implementing IPM using multiple quality approaches might be challenging, because other process areas will need to be taken into consideration when developing the IPM process.
Figure 42 shows process areas to which other CMMI element instances refer to. The figure shows that the process areas containing a small number of references are: PPQA, CAR (Caus-al Analysis and Resolution) and OT (Organizational Training), and the processes areas containing the most of the references are: MA (Measurement and Analysis), PP (Project Planning) and PMC (Project Monitoring and Control). Process areas which contain a few references are probably more independent from other areas. For instance in Process and Product Quality Assurance it is important that the process should not depend on other processes and to be as independent as possible, the Organizational Training is also usually independent from other processes.

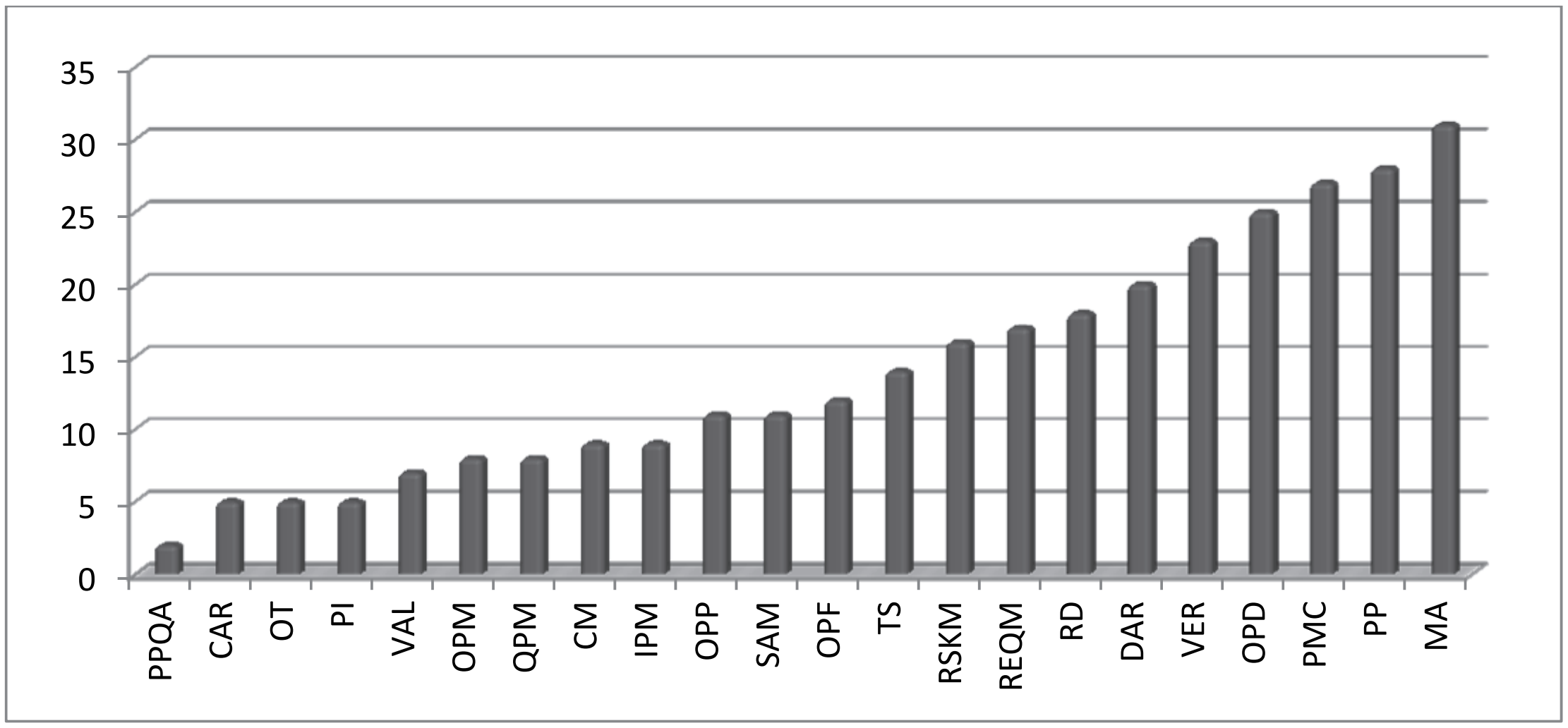

Figure 42 – References from CMMI-DEV v1.3 process areas

The Measurement and Analysis, Project Planning and Project Monitoring and Control process areas contain many references to other CMMI element instances, therefore these could depend on many other process areas. Measurements, project planning and project monitoring and control activities are integral parts of maturity level 2 project and upwards. The implementation of these process areas may depend on the operation of a number of other process areas, therefore it deserves special attention to be paid on their appropriate implementation.

Figure 41 and Figure 42 showed process areas and generic practices which are the most and the least referenced and those which contain the most and the least references. Figure 43 goes further and shows coupledness of elements. On the horizontal (X) axis those process areas and generic practices are represented which contain references, while on the vertical (Y) axis the referred elements are represented. The colour of the dots on the figure varies proportionally to the number of references between 1 and 10. The warmer the colour the more references are present from X to Y. The figure shows clearly that PMC has the most references to another process area: the PP. This seems logical as monitoring and control should rely on planning. For the opposite direction there is only one reference (from PP to PMC).

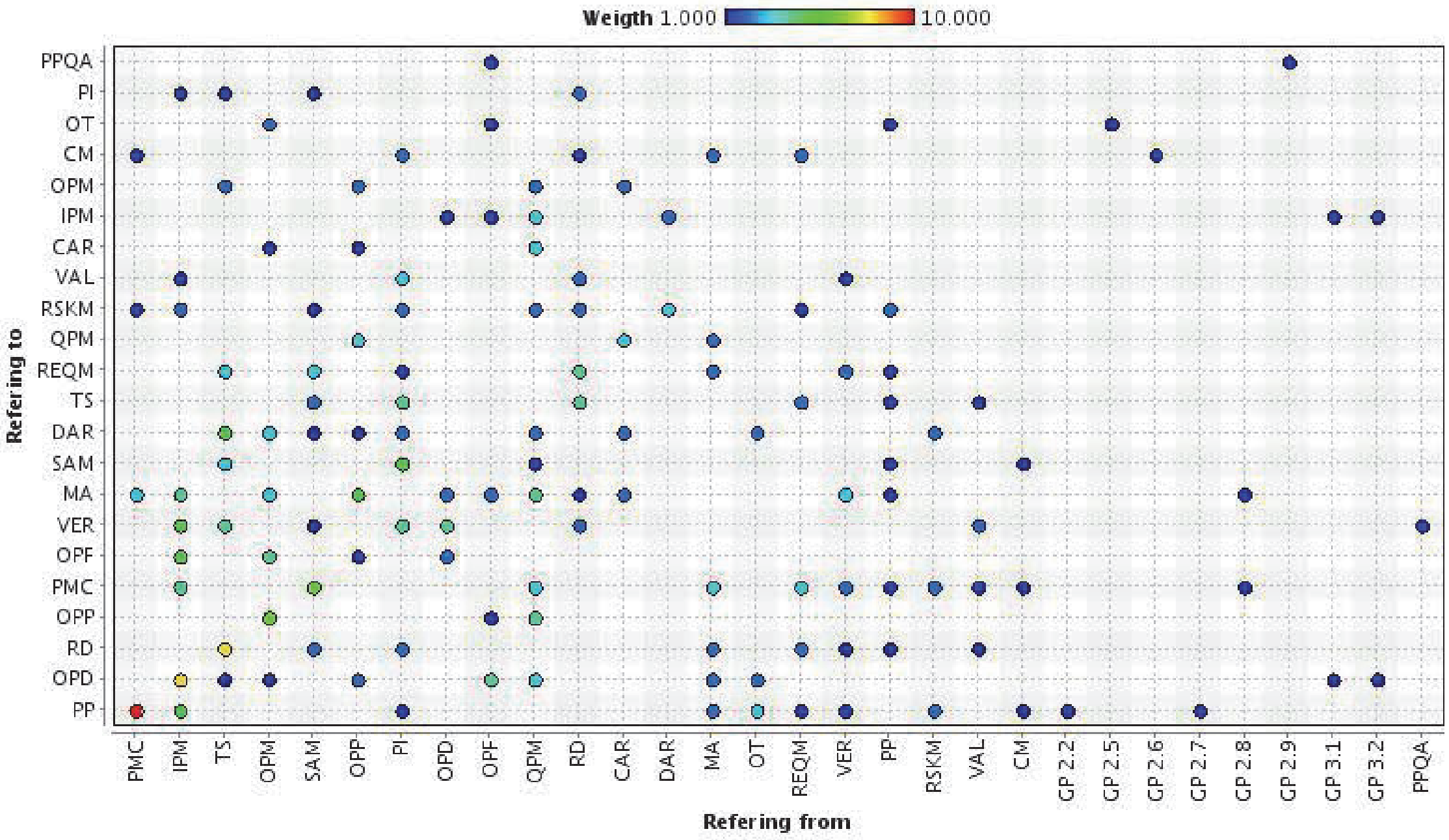

Figure 43 – References among CMMI-DEV v1.3 process areas and generic practices

Summarizing cross-references in Figure 43, we can see the followings: the most references are between TS (Technical Solution) and RD (Requirements Development) (8,4). The number of TS-RD references was not expected, and would be an interesting subject of further investigation. Further highly coupled elements are PP-PMC, IPM-OPD (8,1), OPM-OPD (6,2) and QPM-OPD (4,3). Similarly to the previous two figures, Figure 43 also shows the isolated element instances (e.g. PPQA).

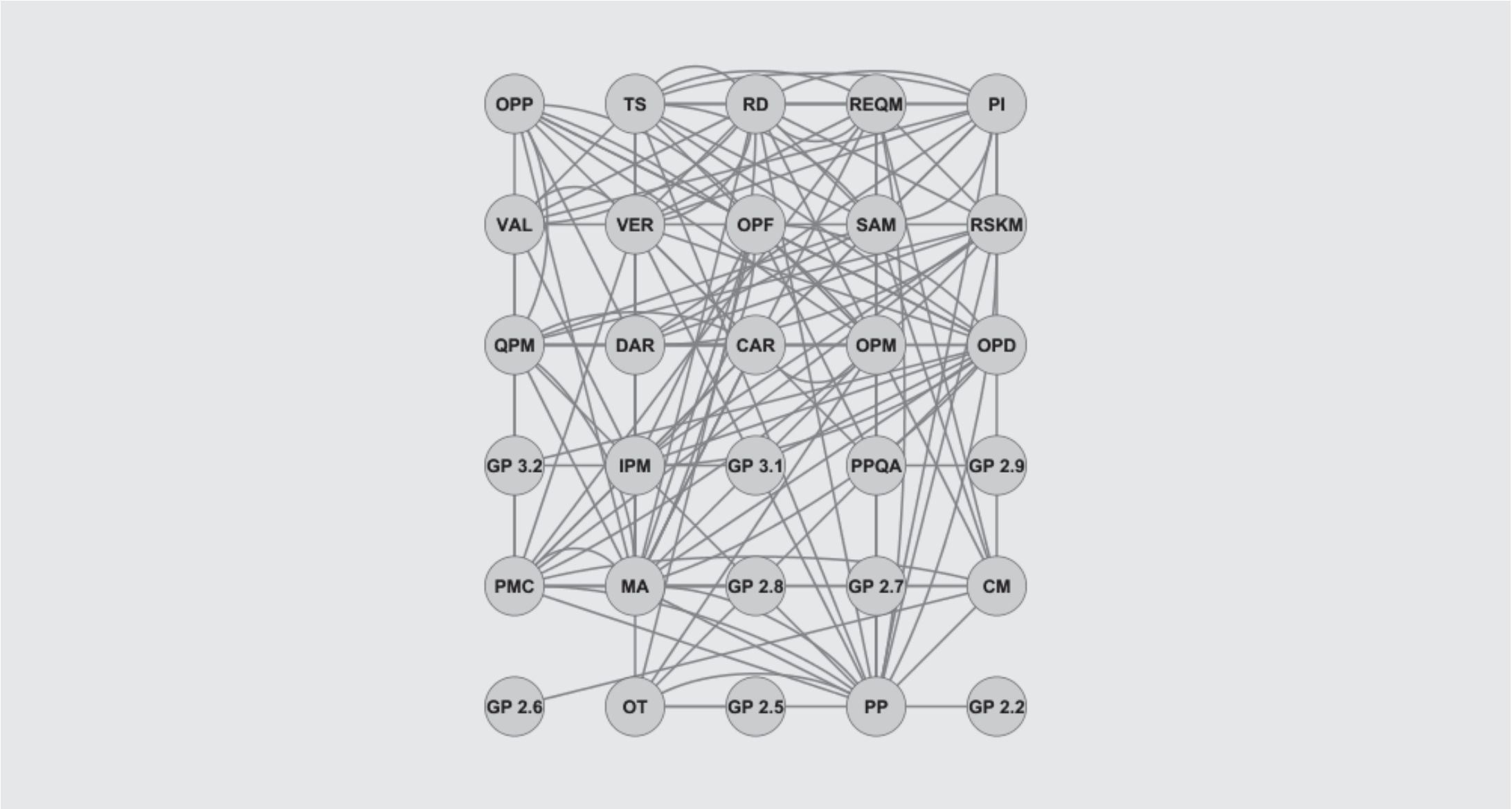

Figure 44 – A graphical representation of references among process areas and generic practices in CMMI-DEV v1.3

It could be worthwhile to implement highly coupled process areas together and independent ones more independently. We also notice that only 16.6% of elements have more than 6 cross-references (counting both directions) with one another element, which shows a closely Pareto distribution of cross-references.

Figure 44, Figure 84 and Figure 85 present various views on CMMI-DEV v1.3 cross-references.

#### 11.4.1.3. **Preliminary results**

Cross-references are present in quality approaches. In this chapter we showed preliminary results of analysing cross-references in CMMI. Based on these references highly coupled and independent element instances can be distinguished and these might need different implementation. A further systematic analysis could help in understanding the complexity and interrelations within a quality approach. Understanding cross-references could serve to a better implementation, training, (e.g. simplified) appraisals or better tool development. Similar references might be present in other quality approaches; therefore both CMMI and other approaches would need further complexity-analyses.

### 11.4.2. Text mining applied on quality approaches

#### 11.4.2.1. **Steps of a basic text mining**

As we discussed in chapter 1, there are several solutions for choosing quality approaches. Unfortunately these are not often updated to include new (and new versions of) quality approaches, therefore evolvement of easily usable quantitative techniques could fill an important gap in understanding focus of quality approaches by providing a quick overview on the quality approaches to be used. The application of quantitative tools could serve a good starting point especially in applying long-described quality approaches such as CMMI, ITIL, SPICE or Enterprise SPICE. In this chapter we present preliminary results of applying a simple text mining technique on CMMI. This technique is general and can be applied to any document and thus to other quality approaches.

In order to get an overview about the most frequent words in CMMI we defined and performed the following steps:

1. Selecting relevant document to be analysed
2. Analysing relevant documents
    a) Removing useless document parts
    b) Tokenization
    c) Filtering stopwords
    d) Transforming canonization
    e) Truncation
3. Understanding results

For complexity analysis we used a free text mining tool Rapid Miner (Rapid - I, 2012). Figure 45 shows a generic process for analysing documents with a text mining tool.

#### 11.4.2.1. **Understanding focus of quality approaches with text mining**

1. Selecting and filtering relevant document parts to be analysed

For analysing CMMI we selected all constellations (CMMI-DEV, CMMI-SVC, and CMMI-ACQ) of version 1.3.

2. Analysing relevant documents

2a Removing useless document parts – useless parts of CMMI were removed (e.g. figures, formatting)

2b Tokenization – Tokenization can be performed in two ways: (1) by single words and (2) by N-grams (expressions including multiple words). In this quick search we primarily focused on single word-count analysis.

2c Filtering stopwords

We found that several words are common in documents and could mislead the result of text mining. These are called stopwords in text mining. Such words were: page (appeared on each page of each document), “this”, “that” among many others. For filtering stopwords we used two dictionaries:

- A generic English dictionary included in the tool (Rapid - I, 2012),
- An additional self-defined dictionary for filtering further common, but irrelevant words. We considered irrelevant those common words which have no connection with the topic (e.g. this, that, etc.)

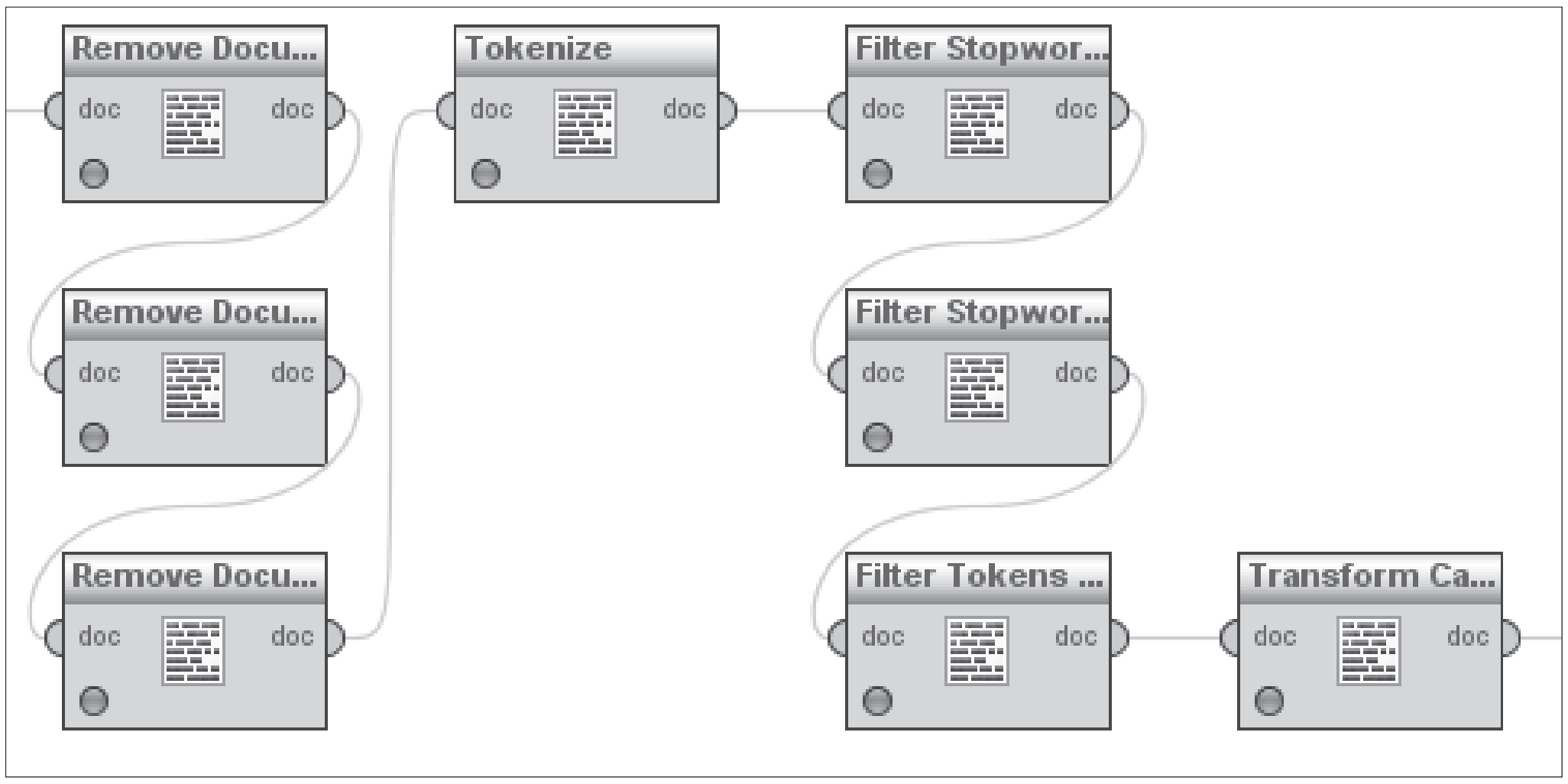

Figure 45 – Steps of filtering most frequent words in a document

2d) Transforming canonization – in order to avoid different counting of upper and lower case words we transformed all words to lower case.

2e) Truncation – several words and expressions are present in documents in various forms (e.g. work, working), therefore in order to achieve a clear view, a truncation of words can be performed. For truncation we used two well-known algorithms Porter and Snowball, which gave similar results.

#### 11.4.2.2. **Preliminary results**

Table 61 in the appendix includes the list of 30 most frequently occurring words and trunks (after applying Snowball on the wordlist).

Based on Table 61 most frequent words (concepts) in CMMI are:

- Process (8946 words, 10853 trunks),
- Product (2338 words, 4370 trunks),
- Work (3706 words, 3751 trunks),
- Project (3170 words, 3556 trunks),
- Service (2934 words, 4219 trunks).

Results suggest that CMMI is not a clearly specific quality approach but it is a more widely applicable one. It is also interesting that CMMI is highly process oriented quality approach (which is clearly stated in the model). Going through the first 30 words and trunks we can see that the focus might also be on organization, management, performance, suppliers, training, risks, planning and measurement.

It is important to mention that preliminary results presented here show rather a feasibility of such a quick analysis than a decent and tested result, thus further investigation is needed.

Preliminary results show that text mining tools could be used in practice in understanding focus and selecting quality approaches.

## 11.5. Limitations

Here we summarize limitations of reflections presented in this chapter.

**Guidelines for using the PBU process** – One single case study was performed in this research. Testing the PBU framework in further case studies could strengthen current guidelines and may reveal new ones.

**A data model for enhancing the maintainability of PBU results** – The limitation of the data model presented in this chapter is that it is general and covers only most important aspects of maintainability of PBU results. Therefore when building a software the data model should be tailored for the context.

**Limitations of quantitative analysis of quality approaches** – The three constellations of CMMI contain core process areas which appear in all the three documents. These were duplicated and duplications were not filtered. A later sentence duplication analysis of CMMI showed hundreds of sentence duplications across CMMI constellations – these duplications should be filtered and counted only once. CMMI contains expressions which contain process, project and work (e.g. process area, work product, project planning, project monitoring and control) – these expressions (or N-grams) influence search results and must be taken into account.

Limitations of this initial quick search show that the text mining process should be chosen carefully and should be developed systematically. Both results and limitations motivate further research of applying text mining on quality approaches.

## 11.6. Conclusion

The goal of this chapter was to present a number of issues that arose during the research and are relevant, but outside of the scope of the researched thesis. These included:

- guidelines for performing the PBU process (in chapter 11.2),
- a data model to support maintainability of PBU results (in chapter 11.3),
- strategies for quantitative analysis of quality approaches (in chapter 11.4).

Five guidelines based on practical experiences of applying the PBU framework were presented in chapter 11.2 which can support its application: (1) if certification is not required use those elements of quality approaches that you need most, (2) make the “Analysis” subprocess independent, (3) start the mapping from the finer grained elements, (4) identify quality approach elements according to their importance, and (5) use a tool. Comparing to the lessons learned presented in chapter 10 guidelines are more practical observations than research results and have no refinement effect on the PBU framework.

The data model presented in chapter 11.3 supports the maintainability of PBU results. The data model itself does not give full maintainability support, however a basic discussion is given on what data should such a tool handle. In chapter 11.3.1 general requirements were identified for a data model for supporting maintainability of PBU results; in chapter 11.3.2 main entities of such a data model were discussed.

Limited number of literature deals with the quantitative analysis of quality approaches, despite that this can be useful when understanding complexity and scope of the quality approaches. Data mining tools and basic algorithms are already available for text mining. In chapter 11.4.1 we presented preliminary results of analysing CMMI cross-references and in chapter 11.4.2 a text mining technique for understanding the scope of CMMI. The CMMI was chosen as a widely known and accepted quality approach, however further versions of same techniques may be applicable to other quality approaches.

# 12. CONCLUSION AND FURTHER WORK

The goal of this thesis was to understand the problem of *multi-model software process improvement* and providing a *Process Based Unification* for simultaneous usage of multiple software quality approaches. We call Process Based Unification (PBU) when multiple quality approaches are mapped to a single process. This unification of multiple quality approaches can be used in several situations e.g. in those discussed in chapter 1.3. The PBU framework can provide a solution for the multi-model problem by unifying structure, elements, content and terminology to a single process. It can enhance the usage of multiple process oriented quality approaches and thus can accelerate multi-model software process improvement.
In this chapter we present contributions of this research (12.1), limitations of research results (12.2) and recommendations for further work (12.3).

## 12.1. Contributions

Many organizations struggle with the application of multiple software quality approaches. The problem is in particular caused by the amount and variety of software quality approaches that are available and have been introduced over the years. In the introduction we showed that classifications and taxonomies for software quality approaches only support organizations to some extent, i.e. to characterize, compare and choose from approaches. However, no acceptable solution is offered for the simultaneous usage of multiple software quality approaches in organizations. The main goal of this thesis was to develop a solution for simultaneous usage of multiple quality approaches. In this chapter we summarize major contributions of this research having the discussions in line with the research questions identified in chapter 2.
In this research the basic hypothesis was that mapping quality approaches to a unified process can provide a multi-model solution.

On the basis of this we identified the following research objective: investigation whether the PBU concept can lead to an acceptable solution for the simultaneous usage of multiple quality approaches, by designing a PBU process.

Our hypothesis was formulated as the following research question: *Does the PBU framework provide a solution of sufficient quality for current problems of simultaneous usage of multiple quality approaches?*

The main research question was broken down to 5 high level questions (H1-H5):

H1 How can we recognize a solution of sufficient quality for current problems of simultaneous usage of multiple quality approaches?

H2 How can we design a PBU process that allows mapping of quality approaches to a unified process?

H3 How can we provide a proof of concept for the PBU framework?

H4 How can we validate the PBU framework?

H5 What improvements can be made on the proposed PBU framework?

In order to answer high level questions, these were broken down to 11 operational research questions (Q1-Q11):

H1/Q1: What criteria should a multi-model solution satisfy?

H1/Q2: Do current MSPI initiatives satisfy MSPI criteria?

H2/Q3: What is a suitable set of process elements to base the unified process on?

H2/Q4: What are the elements of quality approaches?

H2/Q5: Which characteristics of (a number of) current quality approaches further and/or hinder mapping of these approaches to a process?

H2/Q6 How can mapping of a set of quality approaches to a unified process take place?

H3/Q7 How can we provide proof of concept for the PBU framework?

H4/Q8 Is the PBU framework adhering to MSPI criteria?

H4/Q9 Which conclusions to validity and reliability of the PBU framework can be drawn from the research results?

H5/Q10 What did we do differently in applying the PBU framework as compared to its original design?

H5/Q11 What omissions can we identify when comparing the results of applying the PBU framework to the MSPI criteria?

Answering operational research questions (Q1-Q11) above also provide answer to high level questions (H1-H5), giving full answer to the main question and proof for the hypothesis. Chapters 12.1.1-12.1.7 are structured to discuss answers for questions Q1-Q11 and to provide full answer to the main question.

### 12.1.1. MSPI criteria (Q1-Q2)

In order to provide an MSPI solution first we needed to know what makes an MSPI solution 'good'. We also needed a basis (e.g. criteria) on which multi-model solutions can be assessed and compared. In order to develop such criteria we needed to understand problems and initiatives of MSPI, strengths and weaknesses of current initiatives.

Regarding the problems of MSPI, we discovered problems caused by differences in quality approaches which are the inputs of MSPI solutions, problems of documentation and repeatability of the MSPI process and problems connected to the quality of multi-model results.

Regarding the current initiatives for solving the problem of using multiple quality approaches, we derived three categories. These are: quality approach harmonization, quality approach integration, and quality approach mapping. Based on an analysis of the strengths and weaknesses of current initiatives in these three categories, we derived criteria for multi-model software process improvement solutions. These are: criteria for handling the differences of inputs of MSPI solutions (in terminology, structure, granularity, content, size and, complexity), criteria for documentation and repeatability of the MSPI process and finally criteria related to the quality of multi-model results or outputs (such as adaptability, expandability, completeness, traceability and appraisal support).

As a result of a systematic review we found that no initiative satisfies all MSPI criteria, therefore a multi-model initiative satisfying all MSPI criteria would fill an important gap.

The MSPI criteria were used in assessing current MSPI initiatives and the PBU framework. It can also serve as a basis for developing or assessing the quality of further MSPI solutions.

### 12.1.2. Options and limitations of mapping quality approaches to processes (Q3-Q5)

A main target of this research was to investigate the feasibility of mapping quality approaches to processes.

In order to investigate if such a mapping is indeed possible we needed to answer the following questions: (1) what can be considered common process elements, (2) what can be considered quality approaches elements and (3) how these two sets can be mapped.

First, in order to understand what common process elements are to which quality approach elements can be mapped, we reviewed the literature. Based on literature review we delivered a set of common process elements. Second, in order to understand what elements quality approaches have, we discussed structure and elements of 7 well-chosen quality approaches.

Third, in order to understand if the elements of quality approaches can be mapped to process elements we discussed how such mappings could be possible.

Since the identification of process elements is a crucial point of this research, we performed this step in three iterations: (1) based on literature, (2) after identifying quality approach elements and (3) after performing a case study on PBU framework.

The set of common process elements can be used when building processes and developing mappings between quality approaches and processes.

The structure and elements of quality approaches can be used for understanding quality approaches and developing a mapping to processes. Besides that the analysis of structure and elements of quality approaches helps reusing them in other settings, due to the variety and diversity of quality approaches reviewed, it also provides a basis for the way other quality approaches might be analysed and mapped.

### 12.1.3. A Process Based Unification (Q6)

With the MSPI criteria and investigating options and limitations of mapping quality approaches to processes, a basis for developing an MSPI solution was achieved. The MSPI solution developed in this research is called Process Based Unification (abbreviated as PBU) which is a framework for developing unified processes. Unification in this case means mapping of various elements of different quality approaches to a single process.

In order to investigate if such a mapping is indeed practically feasible we developed the PBU framework, which consist of the following elements:

**The PBU concept** – The concept of mapping quality approaches to process is called the concept of Process Based Unification or PBU concept.

**The unified process** – The key PBU result is a single, unified process, to which quality approaches are mapped. This resulting process is called the unified process. In order to ensure that the unified process conforms to multiple quality approaches, quality approaches are decomposed and their element instances are mapped to process element instances. A unified process is both usable in practice (e.g. at a software company) as a process and also supports the simultaneous usage of multiple quality approaches.

**A PBU process** – In order to guide the practical implementation of the PBU concept, we provide a process. This process is called PBU process, which relies on the PBU concept. The goal of the proposed PBU process is to provide practical guidance for the implementation of the PBU concept. It includes a set of subprocesses and activities which can be used in such unification.

### 12.1.4. Case study on the PBU framework (Q8)

After designing the PBU framework, we used it in a case study for unifying relevant parts of 3 different quality approaches (IEEE 1028:2008, CMMI-DEV v1.2 and Peer review descriptions of Process Impact) to a single unified peer review process.

A case study was conducted successfully which provides a proof of concept for the PBU framework.

Besides that the case study showed that the PBU framework is indeed practically feasible, it had further contributions, such as:

- Comparison and selection guidelines for process representation for the case study (discussed in chapter 6) – these guidelines and the result of the comparison can be used in other cases where textual and graphical representations of processes are needed.
- Discussion of problems encountered in applying the PBU framework and possible solutions (included in chapters 6-8) – these results can be used in understanding what mapping and process development problems can be faced and how these can be solved.
- A unified peer review process which primarily was a contribution to the Q-works project (environment of the case study) but also can provide a basis how peer reviews and other processes can be tailored from multiple quality approaches.

### 12.1.5. Adhering to MSPI criteria (Q8)

Problems faced in designing the PBU framework and in the case study helped us to understand the MSPI criteria in more detail, this in fact resulted a more elaborated MSPI criteria. After applying the PBU framework in a case study, we assessed it based on the refined MSPI criteria. This assessment showed that using the PBU framework in the case study satisfies MSPI criteria. The criteria of repeatability of the PBU process can be satisfied if decisions are documented as were in the case study. Furthermore, we discussed that the PBU framework would satisfy the MSPI criteria in other settings (in case of developing other processes or using other quality approaches).

### 12.1.6. Validity and reliability of the research (Q9)

Research should be performed in a valid and reliable way. We discussed how various research methodologists understand validity and reliability of research. We discussed how ecological, construct, internal, external validity as well as reliability of this research were achieved according to multiple viewpoints and tactics suggested by methodologists.
We also pointed out that validity and reliability criteria of research overlap MSPI criteria:

1. external validity overlaps applying MSPI criteria in other settings,
2. MSPI criteria/repeatability of PBU process overlaps reliability.

External validity and reliability were discussed taking these overlapping into MSPI criteria. We showed that we satisfied validity and reliability criteria as follows:

1. the PBU framework was applied in its natural environment (ecological validity),
2. the PBU process was defined as objective and well-considered set of activities and were used instead of subjective judgements (construct validity),
3. the logic of activities were discussed as well as deviations from the process (internal validity),
4. the PBU framework can be applied in case of other processes and other quality approaches (external validity),
5. the same PBU result can be achieved when the PBU process is repeated by other researchers (reliability).

### 12.1.7. Refinements on the PBU framework (Q10-Q11)

The first version of the PBU framework was a theoretical design. In the case study we performed the PBU process with deviations, recoding deviations after performing each subprocess. Four deviations from the PBU process (Q10) and one MSPI criteria requirement: documenting decisions (Q11) were used in refining the PBU framework. For all these, we discussed what lessons were learned and what refinements were made on the PBU framework.

## 12.2. Limitations of the contributions

Here we give a brief summary on the limitations of contributions.
**MSPI criteria**

The MSPI criteria were developed based on a systematic literature review and were further discussed based on case study results. Given the method used, the most important criteria have been identified. The criteria can be and probably will be extended / refined as current problems tend to be solved or new problems and initiatives arise.

**Options and limitations of mapping quality approaches to processes**

In identifying a set of process elements several literature sources as well as quality approaches were taken into account.

The three iterations of identifying common process elements provide a good starting point: process elements identified were sufficient for using them in the Q-works project, were sufficient for representing all quality approach requirements needed and were sufficient for process modeling purposes in BPMN. This list might be further developed or refined depending on the context and requirements of the project in which it will be applied.

The analysis of structure of quality approach elements provided a basis for further investigation of other quality approaches and could be used for developing good structures and semantics for newer versions of quality approaches.

A limitation of this research is that we analysed 7 quality approaches from hundreds available. However, the most important types were discussed including the structured CMMI and the unstructured ISO 9001 approach. We showed for each of them that approaches can be mapped to process, so that it is fairly likely that other approaches are also mappable to processes.

**A Process Based Unification**

The developed PBU framework provides a multi-model solution, however satisfying certain MSPI criteria can be further enhanced. Such MSPI criteria are e.g. adaptability, expandability and appraisal support. These could be further supported by a software tool developed for the PBU framework.

**Case study on the PBU framework**

The PBU framework was used in a case study in which one single process was developed based on three quality approaches. The external validity of this case could be further enhanced by testing the PBU framework in other cases for developing other unified processes and using other quality approaches. However after discussing external validity and satisfying MSPI criteria in other situations, we do not see real reason to suspect that performing further case studies would lead to radically different conclusions.

**Adhering to MSPI criteria**

Despite that the explicit requirement of documenting decisions for ensuring repeatability was not part of the first version of the PBU framework, decisions were documented in the case study and MSPI criteria was satisfied. It was also discussed how MSPI criteria could be satisfied in other settings. However, decisions could be documented even more rigorously and systematically: by not just discussing most important examples, but documenting extensively each single decision. This requirement was added as a refinement for the PBU framework, therefore we do not see limitations of satisfying MSPI criteria.

## 12.3. Further work

Work can be continued based on results presented in this thesis in several directions. In this chapter we discuss these possible further directions.

### 12.3.1. Developments related to MSPI criteria

As it is discussed in (C. Pardo, Pino, García, Piattini Velthius, et al., 2011) MSPI is trending, improving and new MSPI solutions are expected to appear. With the MSPI criteria a basis is provided for developing further and improved MSPI solutions as well as a basis for comparing these solutions. The MSPI criteria themselves could also be further developed.
MSPI criteria have already been used as a basis for identifying criteria for process architecture in (Pesantes, Lemus, Mitre, & Mejıa, 2012) but it could also be used and developed further by other MSPI research projects which have been recently started in the field e.g. (Buglione, Hauck, von Wangenheim, & McCaffery, 2012; A. Ferreira & Machado, 2009; C. Pardo, Pino, García, Piattini, & Baldassarre, 2010; Peldzius & Ragaisis, 2012; SEI, 2008).

### 12.3.2. Development of software tools

Software tools are being developed for supporting MSPI initiatives, these are e.g. QMIM Quality Organizer (Bóka et al., 2006; Kelemen, Balla, & Bóka, 2007) for supporting the QMIM framework, HProcessTool (C. Pardo, Pino, García, Romero, et al., 2011) for Pardo's recently published homogenization process, the one presented in (Ekert, 2009), MethodPark's Stages (MethodPark, 2012) or a CMMI-Scrum supporting tool called SPIALS (Homchuen-chom, Piyabunditkul, Lichter, & Anwar, 2011). As multi-model initiatives are trending, a trending of their supporting applications is expected and tools which today seem to be more theoretical or not yet widely spread will be used more in practice.

### 12.3.3. Further support and testing of the refined PBU framework

In chapter 10 we included guidelines and refinements for the PBU framework as well as a generic database structure for storing quality approach elements, process elements and their mapping. The refined PBU framework should be tested in practice as well as a supporting tool using concepts discussed in this thesis can be developed.

### 12.3.4. Quality approach improvement

In this thesis we discussed several problems caused by diversity in characteristics of quality approaches (e.g. regarding structure, elements, terminology, content, size and complexity). Improvement of quality approaches could result in their easier usage. Based on the quality approaches analysed, we conclude that emphasis should be put on researching options and limitations of quality approach improvement such as:

- Structuring unstructured quality approaches – could ease understanding and mapping quality approaches to processes.

- Semantic representation of quality approach element instances – could help in automating of mapping element instances to process element (instance)s, could accelerate process development.
- Indicating the source of information of quality approach elements – so that users can trust in them. Source types could include practical experiences, research result, measurements etc.
- Arguments and justification on each requirement – inclusions and removal of requirements with no sound reasoning in the quality approach could cause users to start questioning the validity and quality of a quality approach. Such disappointments could be prevented and well-grounded quality approaches can be developed with reasoning and answering the why question for every single requirement in a quality approach.

# APPENDICES

## Appendix A. List of sources included into literature search

1. Apithanataveepa, A. (2008, March 18). *Maximizing your SPI efforts through Multi-model Harmonization*. Retrieved from http://www.swpark.or.th/spi@ease/register/SPI_Multiframework_by%20Tuvnord%20&%20Gosoft_03122008.pdf
2. Automotive Special Interest Group (SIG). (2008a). Automotive SPICE© Process Reference Model (PRM). Retrieved from http://www.automotivespice.com/automotiveSIG_PRM_v44.pdf
3. Automotive Special Interest Group (SIG). (2008b). Automotive SPICE Process Assessment Model (PAM). Retrieved from http://www.automotivespice.com/automotiveSIG_PAM_v24.pdf
4. Baldassarre, M., Caivano, D., Pino, F., Piattini, M., & Visaggio, G. (2010). A Strategy for Painless Harmonization of Quality Standards: A Real Case. In *Product-Focused Software Process Improvement*, Lecture Notes in Computer Science (Vol. 6156, pp. 395-408). Springer Berlin / Heidelberg. Retrieved from http://dx.doi.org/10.1007/978-3-642-13792-1_30
5. Baldassarre, M. T., Piattini, M., Pino, F. J., & Visaggio, G. (2009). Comparing ISO/IEC 12207 and CMMI-DEV: Towards a mapping of ISO/IEC 15504-7. In *2009 ICSE Workshop on Software Quality* (pp. 59-64). Presented at the 2009 ICSE Workshop on Software Quality (WOSQ), Vancouver, BC, Canada. doi:10.1109/WOSQ.2009.5071558
6. Balla, K. (2004a). *The complex quality world : developing quality management systems for software companies*. Eindhoven: Technische Universiteit Eindhoven.
7. Balla, K. (2004b). SYNERGIC USE OF SOFTWARE QUALITY MODELS. *Production Systems and Information Engineering*, *Volume 2*, 73-89.
8. Balla, K., Bemelmans, T., Kusters, R., & Trienekens, J. (2001). Quality through Managed Improvement and Measurement (QMIM): Towards a Phased Development and Implementation of a Quality Management System for a Software Company. *Software Quality Journal*, *9*(3), 177-193.
9. Bendell, T. (2005). Structuring business process improvement methodologies. *Total Quality Management and Business Excellence*, *16*(8-9), 969-978. doi:10.1080/14783360500163110
10. Bueno, P., Crespo, A., & Jino, M. (2006). Analysis of an Artifact Oriented Test Process Model and of Testing Aspects of CMMI. In *Product-Focused Software Process Improvement*. Retrieved from http://dx.doi.org/10.1007/11767718_23

11. Calvo-Manzano, J. A., Cuevas, G., Gómez, G., Mejia, J., Muñoz, M., & San Feliu, T. (2010). Methodology for process improvement through basic components and focusing on the resistance to change. *Journal of Software Maintenance and Evolution: Research and Practice*, n/a-n/a. doi:10.1002/smr.505
12. CMMI Product Team, C. S. (2010a, November). CMMI® for Acquisition, Version 1.3. CMU SEI. Retrieved from http://www.sei.cmu.edu/reports/10tr032.pdf
13. CMMI Product Team, C. S. (2010b, November). CMMI® for Services, Version 1.3. CMU SEI. Retrieved from http://www.sei.cmu.edu/reports/10tr034.pdf
14. CMMI Product Team, C. S. (2010c, November). CMMI® for Development, Version 1.3. CMU SEI. Retrieved from http://www.sei.cmu.edu/reports/10tr033.pdf
15. Diaz, J., Garbajosa, J., & Calvo-Manzano, J. A. (2009). Mapping CMMI Level 2 to Scrum Practices: An Experience Report. In R. V. O'Connor, N. Baddoo, J. Cuadrago Gallego, R. Rejas Muslera, K. Smolander, & R. Messnarz (Eds.), *Software Process Improvement*, Communications in Computer and Information Science (Vol. 42, pp. 93-104). Springer Berlin Heidelberg. Retrieved from http://dx.doi.org/10.1007/978-3-642-04133-4_8
16. DMO. (2007, March). +SAFE, V1.2: A Safety Extension to CMMI-DEV, V1.2. SEI CMU. Retrieved from http://www.sei.cmu.edu/reports/07tn006.pdf
17. Ehsan, N., Perwaiz, A., Arif, J., Mirza, E., & Ishaque, A. (2010). CMMI / SPICE based process improvement. In *2010 IEEE International Conference on Management of Innovation & Technology* (pp. 859-862). Presented at the Technology (ICMIT 2010), Singapore, Singapore. doi:10.1109/ICMIT.2010.5492803
18. FAA. (2001, September). Integrated Capability Maturity Model® (FAA-iCMM®) version 2.0. The Federal Aviation Administration. Retrieved from http://www.faa.gov/about/office_org/headquarters_offices/aio/library/media/FAA-iCMMv2.pdf
19. Ferchichi, A., Bigand, M., & Lef`ebvre, H. (2008). An Ontology for Quality Standards Integration in Software Collaborative Projects. In *Proceedings of MDISIS 2008*. Presented at the MDISIS 2008.
20. Ferreira, A. L., Machado, R. J., & Paulk, M. C. (2010). Size and Complexity Attributes for Multi-model Improvement Framework Taxonomy. In *Software Engineering and Advanced Applications (SEAA), 2010 36th EUROMICRO Conference on* (pp. 306 -309). doi:10.1109/SEAA.2010.54
21. Ferreira, A., & Machado, R. (2009). Software Process Improvement in Multi-model Environments. In *2009 Fourth International Conference on Software Engineering Advances* (pp. 512-517). Presented at the 2009 Fourth International Conference on Software Engineering Advances (ICSEA), Porto, Portugal. doi:10.1109/ICSEA.2009.80
22. Ferreira, A. L., Machado, R. J., & Paulk, M. C. (2010). Quantitative Analysis of Best Practices Models in the Software Domain. Presented at the 2010 Asia Pacific Software Engineering Conference. Retrieved from http://74.125.155.132/scholar?q=cache:xpz3NnEDcvoJ:scholar.google.com/+%22multi-model%22+%22software+process+improvement%22&hl=hu&num=100&as_sdt=2000
23. Fong-hao, L., & Shu-Hsien, L. (2003). Building A Knowledge Base of IEEE/EIA 12207 And CMMI With Ontology. Retrieved from http://protege.stanford.edu/conference/2003/Fong-hao_Liu_Shu-Hsien_Lin_Building_A_knowledge_Base_of_IEEE_EAI_12207_and_CMMI_with_Ontology.pdf
24. Fonseca, J. A., & de Almeida Júnior, J. R. (2005). CMMI RAMS Extension Based on CENELEC Railway Standard. In *Computer Safety, Reliability, and Security*. Retrieved from http://dx.doi.org/10.1007/11563228_1
25. Fuggetta, A. (2000). Software process: A Roadmap. In *Proceedings of the conference on The future of Software engineering - ICSE '00* (pp. 25-34). Presented at the the conference, Limerick, Ireland. doi:10.1145/336512.336521
26. Giaglis, G. M. (2001). A Taxonomy of Business Process Modeling and Information Systems Modeling Techniques. *International Journal of Flexible Manufacturing Systems*, *13*(2), 209-228. doi:10.1023/A:1011139719773
27. Guzmán, J., Mitre, H., Amescua, A., & Velasco, M. (2010). Integration of strategic management, process improvement and quantitative measurement for managing the competitiveness of software engineering organizations. *Software Quality Journal*, *18*, 341-359.

28. Habib, M., Ahmed, S., Rehmat, A., Khan, M. J., & Shamail, S. (2008). Blending Six Sigma and CMMI - an approach to accelerate process improvement in SMEs. In *2008 IEEE International Multitopic Conference* (pp. 386-391). Presented at the 2008 IEEE International Multitopic Conference (INMIC), Karachi, Pakistan. doi:10.1109/INMIC.2008.4777768
29. Halvorsen, C. P., & Conradi, R. (1999). A Taxonomy of SPI Frameworks. In *in Proc. 24th NASA Software Engineering Workshop* (pp. 1–2).
30. Halvorsen, C. P., & Conradi, R. (2001). A Taxonomy to Compare SPI Frameworks. *Lecture notes in computer science*, *2077*, 217–235.
31. Henderson-Sellers, B. (2006). SPI-A Role for Method Engineering.
32. Heston, K. M., & Phifer, W. (2010). The multiple quality models paradox: how much 'best practice' is just enough? *Journal of Software Maintenance and Evolution: Research and Practice*. doi:10.1002/spip.434
33. Hwang, S., & Yeom, H. (2009). Analysis of Relationship among ISO/IEC 15504, CMMI and K-model. In *2009 10th ACIS International Conference on Software Engineering, Artificial Intelligences, Networking and Parallel/Distributed Computing* (pp. 306-309). Presented at the 2009 10th ACIS International Conference on Software Engineering, Artificial Intelligences, Networking and Parallel/Distributed Computing (SNPD), Daegu, Korea. doi:10.1109/SNPD.2009.87
34. Ibrahim, L. (2010). Enterprise SPICE (ISO/IEC 15504) Process Assessment Model (Process Dimension). The SPICE User Group. Retrieved from http://enterprisespice.com/EnterpriseSPICEsources.pdf
35. ISCN. (2010). Capability Adviser functions list. Retrieved March 11, 2010, from http://www.iscn.com/capadv/functions.html
36. ISO/IEC. (2004). ISO/IEC 15504-4:2004 Information technology -- Process assessment -- Part 4: Guidance on use for process improvement and process capability determination (SPICE).
37. ISO/IEC/IEEE. (2008). ISO/IEC/IEEE 12207:2008 Systems and software engineering — Software life cycle processes.
38. Kelemen, Z. D. (2008). Structure of Process-Based Quality Approaches - Elements of a research developing a common meta-model for proces-based quality approaches and methods. In*Proceedings of EuroSPI 2008 Doctoral Symposium*. Presented at the European Systems & Software Process Improvement and Innovation, Dublin, Ireland.
39. Kelemen, Z. D., Balla, K., & Bóka, G. (2007). Quality Organizer: a support tool in using multiple quality approaches. In *Proceedings of 8th International Carpathian Control Conference (ICCC 2007)*. Presented at the 8th International Carpathian Control Conference (ICCC 2007), Strbske Pleso, Slovakia.
40. Kelemen, Z. D., Kusters, R., Trienekens, J., & Balla, K. (2009). A Process Based Unification of Process-Oriented Software Quality Approaches. In *2009 Fourth IEEE International Conference on Global Software Engineering* (pp. 285-288). Presented at the 2009 Fourth IEEE International Conference on Global Software Engineering (ICGSE), Limerick, Ireland: IEEE Computer Society. doi:10.1109/ICGSE.2009.39
41. Kelemen, Z. (2008). Towards supporting simultaneous use of process-based quality approaches. In *Proceedings of 9th international Carpathian Control Conference: ICCC 2008* (pp. 291-295). Presented at the 9th international Carpathian Control Conference: ICCC 2008, Sinaia, Romania: University of Craiova.
42. Lemus, S. M., Pino, F. J., & Velthius, M. P. (2010). Towards a model for information technology governance applicable to the banking sector. In *Information Systems and Technologies (CISTI), 2010 5th Iberian Conference on* (pp. 1 -6).
43. Marcal, A. S. C., Soares, F. S. F., & Belchior, A. D. (2007). Mapping CMMI Project Management Process Areas to SCRUM Practices. In *SEW '07: Proceedings of the 31st IEEE Software Engineering Workshop* (pp. 13–22). Washington, DC, USA: IEEE Computer Society. doi:http://dx.doi.org/10.1109/SEW.2007.64
44. Matulevičius, R., & Heymans, P. (2007). Comparing Goal Modelling Languages: An Experiment. In *Requirements Engineering: Foundation for Software Quality*. Presented at the 13th InternationalWorking Conference, REFSQ 2007, Trondheim, Norway. Retrieved from http://dx.doi.org/10.1007/978-3-540-73031-6_2
45. Moore, J. (1999). An integrated collection of software engineering standards. *IEEE Software*, *16*(6), 51-57. doi:10.1109/52.805473

46. Mutafelija, B., & Stromberg, H. (2009, July). Mapping ISO standards to CMMI v1.2. *SEIR - Software Engineering Information Repository*. Retrieved October 7, 2010, from https://seir.sei.cmu.edu/seir/domains/CMMi/General/BSCW2009-ISO-to-CMMI/frmset.BSCW2009-ISO-to-CMMI.asp?DOMAIN=CMMi&SECTION=General&SUBSECTION=Documents&SUBSUBSECTION=Mappings&POSITION_ON_SEIR_PAGE=266&OUTLINE=Mappings
47. Pardo, C. J., Pino, F. J., García, F., & Piattini, M. (2009). Homogenization of Models to Support Multi-model Processes in Improvement Environments. In *ICSOFT (1)* (pp. 151-156).
48. Paulk, M. C. (1993). Comparing ISO 9001 and the Capability Maturity Model for Software. *Software Quality Journal*, *2*(4), 245-256. doi:10.1007/BF00403767
49. Paulk, M. C. (2008). A Taxonomy for Improvement Frameworks. Presented at the World Congress for Software Quality. Retrieved from World Congress for Software Quality
50. Pino, F. J., Baldassarre, M. T., Piattini, M., & Visaggio, G. (2009). Harmonizing maturity levels from CMMI-DEV and ISO/IEC 15504. *Journal of Software Maintenance and Evolution: Research and Practice*, n/a-n/a. doi:10.1002/spip.437
51. Pino, F. J., Baldassarre, M. T., Piattini, M., Visaggio, G., & Caivano, D. (2010). Mapping Software Acquisition Practices from ISO 12207 and CMMI. *Communications in Computer and Information Science, Evaluation of Novel Approaches to Software Engineering*, *69*(II.), 234-247. doi:10.1007/978-3-642-14819-4_17
52. Ragaisis, S., Peldzius, S., & Simenas, J. (2010). Mapping CMMI-DEV maturity levels to ISO/IEC 15504 capability profiles. In *Proceedings of the 9th WSEAS international conference on Telecommunications and informatics*, TELE-INFO'10 (pp. 13–18). Stevens Point, Wisconsin, USA: World Scientific and Engineering Academy and Society (WSEAS). Retrieved from http://portal.acm.org/citation.cfm?id=1844648.1844649
53. Rout, T. P., & Tuffley, A. (2007). Harmonizing ISO/IEC 15504 and CMMI. *Software Process: Improvement and Practice*, *12*(4), 361-371. doi:10.1002/spip.329
54. Rout, T. P., Tuffley, A., & Cahill, B. (2001). CAPABILITY MATURITY MODEL INTEGRATION MAPPING TO ISO/IEC 15504-2:1998. Software Quality Institute, Griffith University. Retrieved from http://www.sqi.gu.edu.au/cmmi/report/docs/MappingReport.pdf
55. Rubio, D. M., Andriano, N., Ruiz de Mendarozqueta, & Bartó, C. (2008). An integrated improvement framework for sharing assessment lessons learned. In *Proceedings del XIV Congreso Argentino de Ciencias de la Computación*.
56. Salviano, C. F. (2009a). PRO2PI Methodology - A Methodology on "Process Capability Profile to drive Process Improvement" - version 3.0. Divisão de Melhoria de Processo de Software Centro de Tecnologia da Informação Renato Acher, Campinas, SP, Brasil. Retrieved from http://pro2pi.wdfiles.com/local--files/publicacoes-sobre-a-metodologia/Salviano2009TR_PRO2PIv3p0.pdf
57. Salviano, C. F. (2009b). PRO2PI website. Retrieved March 15, 2010, from http://pro2pi.wikidot.com/
58. Salviano, C. F. (2009c). A Multi-model Process Improvement Methodology Driven by Capability Profiles. In *2009 33rd Annual IEEE International Computer Software and Applications Conference* (pp. 636-637). Presented at the 2009 33rd Annual IEEE International Computer Software and Applications Conference (COMPSAC), Seattle, Washington, USA. doi:10.1109/COMPSAC.2009.94
59. Salviano, C. F., Martinez, M. R. M., Zoucas, A., & Thiry, M. (2010). Practices and Techniques for Engineering Process Capability Models. *CLEI ELECTRONIC JOURNAL*, *13*(1). Retrieved from http://www.clei.cl/cleiej/papers/v13i1p3.pdf
60. Salviano, C. F., Martinez, M. R. M., Banhesse, E. L., Enelize, A., Zoucas, A., & Thiry, M. (2010). A Method for Tridimensional Process Assessment Using Modelling Theory. In *2010 Seventh International Conference on the Quality of Information and Communications Technology* (pp. 430-435). Presented at the 2010 Seventh International Conference on the Quality of Information and Communications Technology (QUATIC), Porto, Portugal. doi:10.1109/QUATIC.2010.95
61. Salviano, C. F., Zoucas, R., Silva, J. V. L., Alves, A. M., Wangenheim, C. G. V., & Thiry, M. (2009). A Method Framework for Engineering Process Capability Models. In *EuroSPI 2009 Conference Proceedings*. Presented at the EuroSPI 2009, Alcala, Spain.

62. Salviano, C. F., & Figueiredo, A. M. C. M. (2008). Unified Basic Concepts for Process Capability Models. In *Proceedings of The Twentieth International Conference on Software Engineering and Knowledge Engineering (SEKE'08)* (pp. 173-178). Presented at the The Twentieth International Conference on Software Engineering and Knowledge Engineering (SEKE'08), San Francisco, CA, USA.
63. SEI, C. (2008). Process Improvement in Multi-model Environments (PrIME) website. Retrieved March 11, 2010, from http://sei.cmu.edu/process/research/prime.cfm
64. Siviy, J., Kirwan, P., Marino, L., & Morley, J. (2008a, March). Process Architecture in a Multi-model Environment. Carnegie Mellon University. Retrieved from http://www.sei.cmu.edu/library/assets/multi-modelSeries_wp4_processArch_052008_v1.pdf
65. Siviy, J., Kirwan, P., Marino, L., & Morley, J. (2008b, March). Improvement Technology Classification and Composition in Multi-model Environments. Carnegie Mellon University. Retrieved from http://www.sei.cmu.edu/library/assets/3.pdf
66. Siviy, J., Kirwan, P., Marino, L., & Morley, J. (2008c, March). The Value of Harmonizing Multiple Improvement Technologies: A Process Improvement Professional’s View. Carnegie Mellon University. Retrieved from http://www.sei.cmu.edu/library/assets/multi-modelSeries_wp1_harmonizationValue_052008_v1.pdf
67. Siviy, J., Kirwan, P., Marino, L., & Morley, J. (2008d, March). Maximizing your Process Improvement ROI through Harmonization. Carnegie Mellon University. Retrieved from http://www.sei.cmu.edu/library/assets/multi-modelExecutive_wp_harmonizationROI_032008_v1.pdf
68. Siviy, J., Kirwan, P., Marino, L., & Morley, J. (2008e, May). Implementation Challenges in a Multi-model Environment. Carnegie Mellon University. Retrieved from http://www.sei.cmu.edu/library/assets/4.pdf
69. Siviy, J., Kirwan, P., Marino, L., & Morley, J. (2008f, May). Strategic Technology Selection and Classification in Multi-model Environments. Carnegie Mellon University. Retrieved from http://www.sei.cmu.edu/library/assets/2.pdf
70. Siviy, J. M., Penn, M. L., & Stoddard, R. (2007). *Achieving Success via Multi-Model Process Improvement*. SEPG 2007. Retrieved from http://www.sei.cmu.edu/library/abstracts/presentations/multi-model.cfm
71. SPICE User Group. (2007). Enterprise SPICE - An enterprise integrated standards-based model. Retrieved March 9, 2010, from http://www.enterprisespice.com/
72. Thiry, M., Zoucas, A., & Tristão, L. (2010). Mapping Process Capability Models to Support Integrated Software Process Assessments. *CLEI ELECTRONIC JOURNAL*, *13*.
73. TMMi Foundation. (2009). The Test Maturity Model Integration (TMMi) v2.0. Retrieved from http://www.tmmifoundation.org/downloads/tmmi/TMMi%20Framework.pdf
74. Wang, Y., King, G., Dorling, A., & Wickberg, H. (1999). A unified framework for the software engineering process system standards and models. In *Proceedings 4th IEEE International Software Engineering Standards Symposium and Forum (ISESS'99). 'Best Software Practices for the Internet Age'* (pp. 132-141). Presented at the IEEE Computer Society 4th International Symposium and Forum on Software Engineering Standards, Curitiba, Brazil. doi:10.1109/SESS.1999.766587
75. Wang, Y., & Bryant, A. (2002). Process-Based Software Engineering: Building the Infrastructures. *Annals of Software Engineering*, *14*(1), 9-37.
76. von Wangenheim, C. G., Silva, D. A. D., Buglione, L., Scheidt, R., & Prikladnicki, R. (2010). Best practice fusion of CMMI-DEV v1.2 (PP, PMC, SAM) and PMBOK 2008. *Information and Software Technology*, *52*(7), 749-757. doi:10.1016/j.infsof.2010.03.008
77. Yoo, C., Yoon, J., Lee, B., Lee, C., Lee, J., Hyun, S., & Wu, C. (2006). A unified model for the implementation of both ISO 9001:2000 and CMMI by ISO-certified organizations. *Journal of Systems and Software*, *79*(7), 954-961. doi:10.1016/j.jss.2005.06.042
78. Yoo, C., Yoon, J., Lee, B., Chongwon Lee, Jinyoung Lee, Seunghun Hyun, & Chisu Wu. (2004). An Integrated Model of ISO 9001:2000 and CMMI for ISO Registered Organizations. In *11th Asia-Pacific Software Engineering Conference* (pp. 150-157). Presented at the 11th Asia-Pacific Software Engineering Conference, Busan, Korea. doi:10.1109/APSEC.2004.30

# Appendix B. List of articles included into basic text mining analysis

1. Baldassarre, M., Caivano, D., Pino, F., Piattini, M., & Visaggio, G. (2010). A Strategy for Painless Harmonization of Quality Standards: A Real Case. In *Product-Focused Software Process Improvement*, Lecture Notes in Computer Science (Vol. 6156, pp. 395-408). Springer Berlin / Heidelberg. Retrieved from http://dx.doi.org/10.1007/978-3-642-13792-1_30
2. Baldassarre, M. T., Piattini, M., Pino, F. J., & Visaggio, G. (2009). Comparing ISO/IEC 12207 and CMMI-DEV: Towards a mapping of ISO/IEC 15504-7. In *2009 ICSE Workshop on Software Quality* (pp. 59-64). Presented at the 2009 ICSE Workshop on Software Quality (WOSQ), Vancouver, BC, Canada. doi:10.1109/WOSQ.2009.5071558
3. Balla, K. (2004b). SYNERGIC USE OF SOFTWARE QUALITY MODELS. *Production Systems and Information Engineering*, *Volume 2*, 73-89.
4. Balla, K., Bemelmans, T., Kusters, R., & Trienekens, J. (2001). Quality through Managed Improvement and Measurement (QMIM): Towards a Phased Development and Implementation of a Quality Management System for a Software Company. *Software Quality Journal*, *9*(3), 177-193.
5. Bendell, T. (2005). Structuring business process improvement methodologies. *Total Quality Management and Business Excellence*, *16*(8-9), 969-978. doi:10.1080/14783360500163110
6. Bueno, P., Crespo, A., & Jino, M. (2006). Analysis of an Artifact Oriented Test Process Model and of Testing Aspects of CMMI. In *Product-Focused Software Process Improvement*. Retrieved from http://dx.doi.org/10.1007/11767718_23
7. Calvo-Manzano, J. A., Cuevas, G., Gómez, G., Mejia, J., Muñoz, M., & San Feliu, T. (2010). Methodology for process improvement through basic components and focusing on the resistance to change. *Journal of Software Maintenance and Evolution: Research and Practice*, n/a-n/a. doi:10.1002/smr.505
8. Diaz, J., Garbajosa, J., & Calvo-Manzano, J. A. (2009). Mapping CMMI Level 2 to Scrum Practices: An Experience Report. In R. V. O'Connor, N. Baddoo, J. Cuadrago Gallego, R. Rejas Muslera, K. Smolander, & R. Messnarz (Eds.), *Software Process Improvement*, Communications in Computer and Information Science (Vol. 42, pp. 93-104). Springer Berlin Heidelberg. Retrieved from http://dx.doi.org/10.1007/978-3-642-04133-4_8
9. Ehsan, N., Perwaiz, A., Arif, J., Mirza, E., & Ishaque, A. (2010). CMMI / SPICE based process improvement. In *2010 IEEE International Conference on Management of Innovation & Technology* (pp. 859-862). Presented at the Technology (ICMIT 2010), Singapore, Singapore. doi:10.1109/ICMIT.2010.5492803
10. Ferchichi, A., Bigand, M., & Lef`ebvre, H. (2008). An Ontology for Quality Standards Integration in Software Collaborative Projects. In *Proceedings of MDISIS 2008*. Presented at the MDISIS 2008.
11. Ferreira, A. L., Machado, R. J., & Paulk, M. C. (2010). Size and Complexity Attributes for Multi-model Improvement Framework Taxonomy. In *Software Engineering and Advanced Applications (SEAA), 2010 36th EUROMICRO Conference on* (pp. 306 -309). doi:10.1109/SEAA.2010.54
12. Ferreira, A., & Machado, R. (2009). Software Process Improvement in Multi-model Environments. In *2009 Fourth International Conference on Software Engineering Advances* (pp. 512-517). Presented at the 2009 Fourth International Conference on Software Engineering Advances (ICSEA), Porto, Portugal. doi:10.1109/ICSEA.2009.80
13. Ferreira, A. L., Machado, R. J., & Paulk, M. C. (2010). Quantitative Analysis of Best Practices Models in the Software Domain. Presented at the 2010 Asia Pacific Software Engineering Conference. Retrieved from http://74.125.155.132/scholar?q=cache:xpz3NnEDcvoJ:scholar.google.com/+%22multi-model%22+%22software+process+improvement%22&hl=hu&num=100&as_sdt=2000
14. Fong-hao, L., & Shu-Hsien, L. (2003). Building A Knowledge Base of IEEE/EIA 12207 And CMMI With Ontology. Retrieved from http://protege.stanford.edu/conference/2003/Fong-hao_Liu_Shu-Hsien_Lin_Building_A_knowledge_Base_of_IEEE_EAI_12207_and_CMMI_with_Ontology.pdf

15. Fonseca, J. A., & de Almeida Júnior, J. R. (2005). CMMI RAMS Extension Based on CENELEC Railway Standard. In *Computer Safety, Reliability, and Security*. Retrieved from http://dx.doi.org/10.1007/11563228_1
16. Fuggetta, A. (2000). Software process: A Roadmap. In *Proceedings of the conference on The future of Software engineering - ICSE '00* (pp. 25-34). Presented at the the conference, Limerick, Ireland. doi:10.1145/336512.336521
17. Giaglis, G. M. (2001). A Taxonomy of Business Process Modeling and Information Systems Modeling Techniques. *International Journal of Flexible Manufacturing Systems*, *13*(2), 209-228. doi:10.1023/A:1011139719773
18. Guzmán, J., Mitre, H., Amescua, A., & Velasco, M. (2010). Integration of strategic management, process improvement and quantitative measurement for managing the competitiveness of software engineering organizations. *Software Quality Journal*, *18*, 341-359.
19. Habib, M., Ahmed, S., Rehmat, A., Khan, M. J., & Shamail, S. (2008). Blending Six Sigma and CMMI - an approach to accelerate process improvement in SMEs. In *2008 IEEE International Multitopic Conference* (pp. 386-391). Presented at the 2008 IEEE International Multitopic Conference (INMIC), Karachi, Pakistan. doi:10.1109/INMIC.2008.4777768
20. Halvorsen, C. P., & Conradi, R. (1999). A Taxonomy of SPI Frameworks. In *in Proc. 24th NASA Software Engineering Workshop* (pp. 1–2).
21. Halvorsen, C. P., & Conradi, R. (2001). A Taxonomy to Compare SPI Frameworks. *Lecture notes in computer science*, *2077*, 217–235.
22. Henderson-Sellers, B. (2006). SPI-A Role for Method Engineering.
23. Heston, K. M., & Phifer, W. (2010). The multiple quality models paradox: how much 'best practice' is just enough? *Journal of Software Maintenance and Evolution: Research and Practice*. doi:10.1002/spip.434
24. Hwang, S., & Yeom, H. (2009). Analysis of Relationship among ISO/IEC 15504, CMMI and K-model. In *2009 10th ACIS International Conference on Software Engineering, Artificial Intelligences, Networking and Parallel/Distributed Computing* (pp. 306-309). Presented at the 2009 10th ACIS International Conference on Software Engineering, Artificial Intelligences, Networking and Parallel/Distributed Computing (SNPD), Daegu, Korea. doi:10.1109/SNPD.2009.87
25. Kelemen, Z. D. (2008). Structure of Process-Based Quality Approaches - Elements of a research developing a common meta-model for proces-based quality approaches and methods. In*Proceedings of EuroSPI 2008 Doctoral Symposium*. Presented at the European Systems & Software Process Improvement and Innovation, Dublin, Ireland.
26. Kelemen, Z. D., Balla, K., & Bóka, G. (2007). Quality Organizer: a support tool in using multiple quality approaches. In *Proceedings of 8th International Carpathian Control Conference (ICCC 2007)*. Presented at the 8th International Carpathian Control Conference (ICCC 2007), Strbske Pleso, Slovakia.
27. Kelemen, Z. D., Kusters, R., Trienekens, J., & Balla, K. (2009). A Process Based Unification of Process-Oriented Software Quality Approaches. In *2009 Fourth IEEE International Conference on Global Software Engineering* (pp. 285-288). Presented at the 2009 Fourth IEEE International Conference on Global Software Engineering (ICGSE), Limerick, Ireland: IEEE Computer Society. doi:10.1109/ICGSE.2009.39
28. Kelemen, Z. (2008). Towards supporting simultaneous use of process-based quality approaches. In *Proceedings of 9th international Carpathian Control Conference: ICCC 2008* (pp. 291-295). Presented at the 9th international Carpathian Control Conference: ICCC 2008, Sinaia, Romania: University of Craiova.
29. Lemus, S. M., Pino, F. J., & Velthius, M. P. (2010). Towards a model for information technology governance applicable to the banking sector. In *Information Systems and Technologies (CISTI), 2010 5th Iberian Conference on* (pp. 1 -6).
30. Marcal, A. S. C., Soares, F. S. F., & Belchior, A. D. (2007). Mapping CMMI Project Management Process Areas to SCRUM Practices. In *SEW '07: Proceedings of the 31st IEEE Software Engineering Workshop* (pp. 13–22). Washington, DC, USA: IEEE Computer Society. doi:http://dx.doi.org/10.1109/SEW.2007.64
31. Matulevičius, R., & Heymans, P. (2007). Comparing Goal Modelling Languages: An Experiment. In *Requirements Engineering: Foundation for Software Quality*. Presented at the 13th InternationalWorking

Conference, REFSQ 2007, Trondheim, Norway. Retrieved from http://dx.doi.org/10.1007/978-3-540-73031-6_2

32. Moore, J. (1999). An integrated collection of software engineering standards. *IEEE Software*, *16*(6), 51-57. doi:10.1109/52.805473
33. Pardo, C. J., Pino, F. J., García, F., & Piattini, M. (2009). Homogenization of Models to Support Multi-model Processes in Improvement Environments. In *ICSOFT (1)* (pp. 151-156).
34. Paulk, M. C. (1993). Comparing ISO 9001 and the Capability Maturity Model for Software. *Software Quality Journal*, *2*(4), 245-256. doi:10.1007/BF00403767
35. Paulk, M. C. (2008). A Taxonomy for Improvement Frameworks. Presented at the World Congress for Software Quality. Retrieved from World Congress for Software Quality
36. Pino, F. J., Baldassarre, M. T., Piattini, M., & Visaggio, G. (2009). Harmonizing maturity levels from CMMI-DEV and ISO/IEC 15504. *Journal of Software Maintenance and Evolution: Research and Practice*, n/a-n/a. doi:10.1002/spip.437
37. Pino, F. J., Baldassarre, M. T., Piattini, M., Visaggio, G., & Caivano, D. (2010). Mapping Software Acquisition Practices from ISO 12207 and CMMI. *Communications in Computer and Information Science, Evaluation of Novel Approaches to Software Engineering*, *69*(II.), 234-247. doi:10.1007/978-3-642-14819-4_17
38. Ragaisis, S., Peldzius, S., & Simenas, J. (2010). Mapping CMMI-DEV maturity levels to ISO/IEC 15504 capability profiles. In *Proceedings of the 9th WSEAS international conference on Telecommunications and informatics*, TELE-INFO'10 (pp. 13–18). Stevens Point, Wisconsin, USA: World Scientific and Engineering Academy and Society (WSEAS). Retrieved from http://portal.acm.org/citation.cfm?id=1844648.1844649
39. Rout, T. P., & Tuffley, A. (2007). Harmonizing ISO/IEC 15504 and CMMI. *Software Process: Improvement and Practice*, *12*(4), 361-371. doi:10.1002/spip.329
40. Rout, T. P., Tuffley, A., & Cahill, B. (2001). CAPABILITY MATURITY MODEL INTEGRATION MAPPING TO ISO/IEC 15504-2:1998. Software Quality Institute, Griffith University. Retrieved from http://www.sqi.gu.edu.au/cmmi/report/docs/MappingReport.pdf
41. Rubio, D. M., Andriano, N., Ruiz de Mendarozqueta, & Bartó, C. (2008). An integrated improvement framework for sharing assessment lessons learned. In *Proceedings del XIV Congreso Argentino de Ciencias de la Computación*.
42. Salviano, C. F. (2009c). A Multi-model Process Improvement Methodology Driven by Capability Profiles. In *2009 33rd Annual IEEE International Computer Software and Applications Conference* (pp. 636-637). Presented at the 2009 33rd Annual IEEE International Computer Software and Applications Conference (COMPSAC), Seattle, Washington, USA. doi:10.1109/COMPSAC.2009.94
43. Salviano, C. F., Martinez, M. R. M., Zoucas, A., & Thiry, M. (2010). Practices and Techniques for Engineering Process Capability Models. *CLEI ELECTRONIC JOURNAL*, *13*(1). Retrieved from http://www.clei.cl/cleiej/papers/v13i1p3.pdf
44. Salviano, C. F., Martinez, M. R. M., Banhesse, E. L., Enelize, A., Zoucas, A., & Thiry, M. (2010). A Method for Tridimensional Process Assessment Using Modelling Theory. In *2010 Seventh International Conference on the Quality of Information and Communications Technology* (pp. 430-435). Presented at the 2010 Seventh International Conference on the Quality of Information and Communications Technology (QUATIC), Porto, Portugal. doi:10.1109/QUATIC.2010.95
45. Salviano, C. F., Zoucas, R., Silva, J. V. L., Alves, A. M., Wangenheim, C. G. V., & Thiry, M. (2009). A Method Framework for Engineering Process Capability Models. In *EuroSPI 2009 Conference Proceedings*. Presented at the EuroSPI 2009, Alcala, Spain.
46. Salviano, C. F., & Figueiredo, A. M. C. M. (2008). Unified Basic Concepts for Process Capability Models. In *Proceedings of The Twentieth International Conference on Software Engineering and Knowledge Engineering (SEKE'08)* (pp. 173-178). Presented at the The Twentieth International Conference on Software Engineering and Knowledge Engineering (SEKE'08), San Francisco, CA, USA.
47. Siviy, J., Kirwan, P., Marino, L., & Morley, J. (2008a, March). Process Architecture in a Multi-model Environment. Carnegie Mellon University. Retrieved from http://www.sei.cmu.edu/library/assets/multimodelSeries_wp4_processArch_052008_v1.pdf

48. Siviy, J., Kirwan, P., Marino, L., & Morley, J. (2008b, March). Improvement Technology Classification and Composition in Multi-model Environments. Carnegie Mellon University. Retrieved from http://www.sei.cmu.edu/library/assets/3.pdf
49. Siviy, J., Kirwan, P., Marino, L., & Morley, J. (2008c, March). The Value of Harmonizing Multiple Improvement Technologies: A Process Improvement Professional's View. Carnegie Mellon University. Retrieved from http://www.sei.cmu.edu/library/assets/multi-modelSeries_wp1_harmonizationValue_052008_v1.pdf
50. Siviy, J., Kirwan, P., Marino, L., & Morley, J. (2008d, March). Maximizing your Process Improvement ROI through Harmonization. Carnegie Mellon University. Retrieved from http://www.sei.cmu.edu/library/assets/multi-modelExecutive_wp_harmonizationROI_032008_v1.pdf
51. Siviy, J., Kirwan, P., Marino, L., & Morley, J. (2008e, May). Implementation Challenges in a Multi-model Environment. Carnegie Mellon University. Retrieved from http://www.sei.cmu.edu/library/assets/4.pdf
52. Siviy, J., Kirwan, P., Marino, L., & Morley, J. (2008f, May). Strategic Technology Selection and Classification in Multi-model Environments. Carnegie Mellon University. Retrieved from http://www.sei.cmu.edu/library/assets/2.pdf
53. Thiry, M., Zoucas, A., & Tristão, L. (2010). Mapping Process Capability Models to Support Integrated Software Process Assessments. *CLEI ELECTRONIC JOURNAL*, *13*.
54. Wang, Y., King, G., Dorling, A., & Wickberg, H. (1999). A unified framework for the software engineering process system standards and models. In *Proceedings 4th IEEE International Software Engineering Standards Symposium and Forum (ISESS'99). 'Best Software Practices for the Internet Age'* (pp. 132-141). Presented at the IEEE Computer Society 4th International Symposium and Forum on Software Engineering Standards, Curitiba, Brazil. doi:10.1109/SESS.1999.766587
55. Wang, Y., & Bryant, A. (2002). Process-Based Software Engineering: Building the Infrastructures. *Annals of Software Engineering*, *14*(1), 9-37.
56. von Wangenheim, C. G., Silva, D. A. D., Buglione, L., Scheidt, R., & Prikladnicki, R. (2010). Best practice fusion of CMMI-DEV v1.2 (PP, PMC, SAM) and PMBOK 2008. *Information and Software Technology*, *52*(7), 749-757. doi:10.1016/j.infsof.2010.03.008
57. Yoo, C., Yoon, J., Lee, B., Lee, C., Lee, J., Hyun, S., & Wu, C. (2006). A unified model for the implementation of both ISO 9001:2000 and CMMI by ISO-certified organizations. *Journal of Systems and Software*, *79*(7), 954-961. doi:10.1016/j.jss.2005.06.042
58. Yoo, C., Yoon, J., Lee, B., Chongwon Lee, Jinyoung Lee, Seunghun Hyun, & Chisu Wu. (2004). An Integrated Model of ISO 9001:2000 and CMMI for ISO Registered Organizations. In *11th Asia-Pacific Software Engineering Conference* (pp. 150-157). Presented at the 11th Asia-Pacific Software Engineering Conference, Busan, Korea. doi:10.1109/APSEC.2004.30

# Appendix C. The structure of ISO/IEC/IEEE 12207-2008

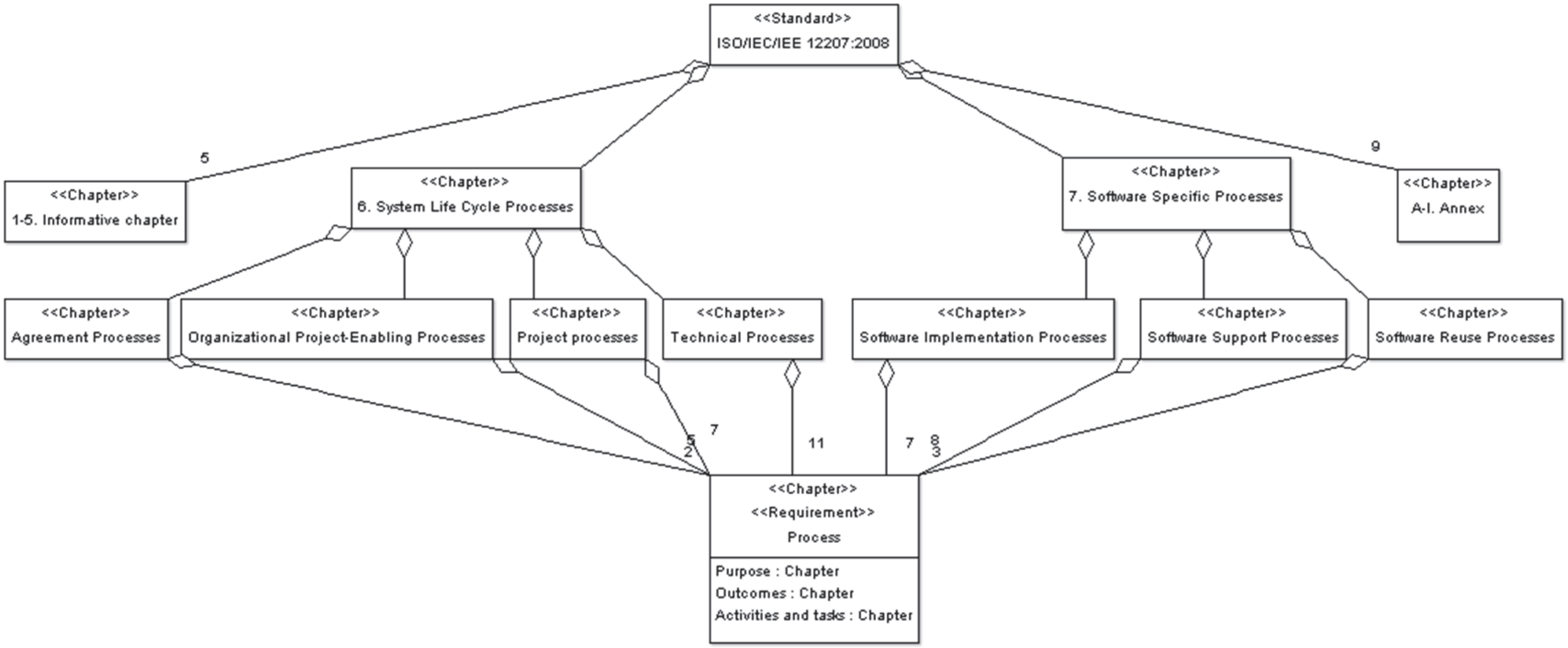

Figure 46 – The structure of ISO/IEC/IEEE 12207-2008

# Appendix D. Peer review definitions in process oriented software quality approaches

Table 40 – Peer review-related definitions in process oriented software quality approaches

| Term | Source | Source in the doc | Definition |
|---|---|---|---|
| | 90003:2004 | | No review related definitions |
| | Enterprise SPICE (ISO/IEC 15504) | | based on FAA-iCMMI |
| | ISO/IEC 15939:2005 | | No review related definitions |
| | ISO-IEC-90003:2004 | | No review related definitions |
| appraisal | CMMI v1.2 | | In the CMMI Product Suite, an examination of one or more processes by a trained team of professionals using an appraisal reference model as the basis for determining, at a minimum, strengths and weaknesses. (See also “assessment” and “capability evaluation.”) |
| appraisal | FAA-iCMMI v2.0 | | (See “process appraisal”) |
| assessment | CMMI v1.2 | | In the CMMI Product Suite, an appraisal that an organization does internally for the purposes of process improvement. The word assessment is also used in the CMMI Product Suite in an everyday English sense (e.g., risk assessment). (See also “appraisal” and “capability evaluation.”) |
| assessment | ITIL v3 | | Inspection and analysis to check whether a Standard or set of Guidelines is being followed, that Records are accurate, or that Efficiency and Effectiveness targets are being met. See also Audit. Certification |
| audit | CMMI v1.2 | | In CMMI process improvement work, an objective examination of a work product or set of work products against specific criteria (e.g., requirements). |
| audit | FAA-iCMMI v2.0 | iCMM, ISO 9000, 12207, 731, CMMI | An independent examination of products, services and processes to determine the extent to which audit criteria (e.g. policies, procedures, requirements, specifications, standards, contractual agreements, or other criteria) are fulfilled |
| audit | IEEE 1028:2008 | | An independent examination of a software product, software process, or set of software processes performed by a third party to assess compliance with specifications, standards, contractual agreements, or other criteria. |
| audit | ISO 12207 | | independent assessment of software products and processes conducted by an authorized person in order to assess compliance with requirements |

| Term | Source | Source in the doc | Definition |
| --- | --- | --- | --- |
| audit | ITIL v3 | | Formal inspection and verification to check whether a Standard or set of Guidelines is being followed, that Records are accurate, or that Efficiency and Effectiveness targets are being met. An Audit may be carried out by internal or external groups. See also Certification, Assessment. |
| capability evaluation (DEV) | CMMI v1.2 | | An appraisal by a trained team of professionals used as a discriminator to select suppliers, to monitor suppliers against the contract, or to determine and enforce incentives. Evaluations are used to gain insight into the process capability of a supplier organization and are intended to help decision makers make better acquisition decisions, improve subcontractor performance, and provide insight to a purchasing organization. (See also “appraisal” and “assessment.”) |
| certification | ITIL v3 | | Issuing a certificate to confirm Compliance to a Standard. Certification includes a formal Audit by an independent and Accredited body. The term Certification is also used to mean awarding a certificate to verify that a person has achieved a qualification. |
| configuration audit | CMMI v1.2 | | An audit conducted to verify that a configuration item, or a collection of configuration items that make up a baseline, conforms to a specified standard or requirement. (See also “audit,” “configuration item,” “functional configuration audit,” and “physical configuration audit.”) |
| design review | CMMI v1.2 | | A formal, documented, comprehensive, and systematic examination of a design to evaluate the design requirements and the capability of the design to meet these requirements, and to identify problems and propose solutions. |
| evaluation | FAA-iCMMI v2.0 | | The term “evaluation” is applicable to products, product elements, work products, and services. Evaluation of hardware, software, documentation, operations, and support services is accomplished through formal reviews, peer reviews, audits, tests, and demonstrations. The term “test” is often used informally to refer to certain evaluations although “test” (technically) is only one form of evaluation. The “requirements” that are the basis for verification are typically documented in product requirement specifications and sub-product design specifications. The “needs” that are the basis for validation may be documented in operational concept documents. Since the range of possible evaluation activities can be quite large and potentially expensive, cost-benefit analyses may be warranted to determine a scope of evaluation activities that provides a reasonable balance of defect removal and verification effectiveness versus cost. |
| evaluation | FAA-iCMMI v2.0 | 12207 | A systematic determination of the extent to which an entity meets its specified criteria. |
| formal evaluation process | CMMI v1.2 | | A structured approach to evaluating alternative solutions against established criteria to determine a recommended solution to address an issue. |
| functional configuration audit | CMMI v1.2 | | An audit conducted to verify that the development of a configuration item has been completed satisfactorily, that the item has achieved the performance and functional characteristics specified in the functional or allo- |

| Term | Source | Source in the doc | Definition |
|---|---|---|---|
| | | | cated configuration identification, and that its operational and support documents are complete and satisfactory. (See also “configuration audit,” “configuration management,” and “physical configuration audit.”) |
| inspection | IEEE 1028:2008 | | A visual examination of a software product to detect and identify software anomalies, including errors and deviations from standards and specifications. |
| inspection | IEEE 1028:2008 | | NOTE—Inspections are peer examinations led by impartial facilitators who are trained in inspection techniques. Determination of remedial or investigative action for an anomaly is a mandatory element of a software inspection, |
| inspection | ITIL v3 | | no definition for inspections |
| internal audit | ISO 9001:2000 | | The organization shall conduct internal audits at planned intervals to determine whether the quality management system<br>a) conforms to the planned arrangements (see 7.1), to the requirements of this International Standard and to the quality management system requirements established by the organization, and<br>b) is effectively implemented and maintained.<br>An audit programme shall be planned, taking into consideration the status and importance of the processes and areas to be audited, as well as the results of previous audits. The audit criteria, scope, frequency and methods shall be defined. Selection of auditors and conduct of audits shall ensure objectivity and impartiality of the audit process.<br>Auditors shall not audit their own work.<br>The responsibilities and requirements for planning and conducting audits, and for reporting results and maintaining records (see 4.2.4) shall be defined in a documented procedure.<br>The management responsible for the area being audited shall ensure that actions are taken without undue delay to eliminate detected nonconformities and their causes. Follow-up activities shall include the verification of the actions taken and the reporting of verification results (see 8.5.2).<br>NOTE See ISO 10011-1, ISO 10011-2 and ISO 10011-3 for guidance. |
| management review | IEEE 1028:2008 | | A systematic evaluation of a software product or process performed by or on behalf of management that monitors progress, determines the status of plans and schedules, confirms requirements and their system allocation, or evaluates the effectiveness of management approaches used to achieve fitness for purpose. |
| management review | ISO 9001:2000 | | Top management shall review the organization's quality management system, at planned intervals, to ensure its continuing suitability, adequacy and effectiveness. This review shall include assessing opportunities for |

| Term | Source | Source in the doc | Definition |
|---|---|---|---|
| | | | improvement and the need for changes to the quality management system, including the quality policy and quality objectives. |
| peer review | CMMI v1.2 | | The review of work products performed by peers during development of the work products to identify defects for removal. The term peer review is used in the CMMI Product Suite instead of the term work product inspection. (See also "work product.") |
| peer review | CMMI v1.2 | | SVC addition (however the term work product inspection is not defined in the glossary): The term "peer review" is used in the CMMI Product Suite instead of the term "work product inspection." |
| peer review | processimpact.com | | In a peer review, co-workers of a person who created a software work product examine that product to identify defects and correct shortcomings. A peer review:<br>• Verifies whether the work product correctly satisfies the specifications found in any predecessor work product, such as requirements or design documents<br>• Identifies any deviation from standards, including issues that may affect maintainability of the software<br>• Suggests improvement opportunities to the author<br>• Promotes the exchange of techniques and education of the participants. |
| physical configuration audit | CMMI v1.2 | | An audit conducted to verify that a configuration item, as built, conforms to the technical documentation that defines and describes it. (See also, "configuration audit," "configuration management," and "functional configuration audit.") |
| post-implementation review | ITIL v3 | | A Review that takes place after a Change or a Project has been implemented. A PIR determines if the Change or Project was successful, and identifies opportunities for improvement. |
| process appraisal | FAA-iCMMI v2.0 | iCMM,15504 | Comparison of processes being practiced to a reference model or standard; a disciplined evaluation of an organization's processes against a reference model |
| review | IEEE 1028:2008 | | A process or meeting during which a software product, set of software products, or a software process is presented to project personnel, managers, users, customers, user representatives, auditors or other interested parties for examination, comment or approval. |
| review | ITIL v3 | | An evaluation of a Change, Problem, Process, Project, etc. Reviews are typically carried out at predefined points in the Lifecycle, and especially after Closure. The purpose of a Review is to ensure that all Deliverables have been provided, and to identify opportunities for improvement. See also Post-Implementation Review. |

| Term | Source | Source in the doc | Definition |
|---|---|---|---|
| technical review | IEEE 1028:2008 | | A systematic evaluation of a software product by a team of qualified personnel that examines the suitability of the software product for its intended use and identifies discrepancies from specifications and standards. |
| technical review | IEEE 1028:2008 | | NOTE—Technical reviews may also provide recommendations of alternatives and examination of various alternatives. |
| walk-through | IEEE 1028:2008 | | A static analysis technique in which a designer or programmer leads members of the development team and other interested parties through a software product, and the participants ask questions and make comments about possible anomalies, violation of development standards, and other problems. |

# Appendix E. IEEE 12207 requirements for software audits and peer reviews

**7.2.6 Software Review Process**

**7.2.6.1 Purpose**

The purpose of the Software Review Process is to maintain a common understanding with the stakeholders of the progress against the objectives of the agreement and what should be done to help ensure development of a product that satisfies the stakeholders. Software reviews are at both project management and technical levels and are held throughout the life of the project.

**7.2.6.2 Outcomes**

As a result of successful implementation of the Software Review Process:
a) management and technical reviews are held based on the needs of the project;
b) the status and products of an activity of a process are evaluated through review activities;
c) review results are made known to all affected parties;
d) action items resulting from reviews are tracked to closure; and
e) risks and problems are identified and recorded.

**7.2.6.3 Activities and tasks**

The project shall implement the following activities in accordance with applicable organization policies and procedures with respect to the Software Review Process.

**7.2.6.3.1 Process implementation.** This activity consists of the following tasks:
**7.2.6.3.1.1** Periodic reviews shall be held at predetermined milestones as specified in the project plan(s). Stakeholders should determine the need for any ad hoc reviews in which agreeing parties may participate.
**7.2.6.3.1.2** All resources that are required to conduct the reviews shall be provided. These resources include personnel, location, facilities, hardware, software, and tools.
**7.2.6.3.1.3** The parties that participate in a review should agree on the following items at each review: meeting agenda, software products (results of an activity) and problems to be reviewed; scope and procedures; and entry and exit criteria for the review.
**7.2.6.3.1.4** Problems detected during the reviews shall be recorded and entered into the Software Problem Resolution Process (subclause 7.2.8) as required.
**7.2.6.3.1.5** The review results shall be documented and distributed. This communication includes adequacy of review (for example, approval, disapproval, or contingent approval) of the review results.
**7.2.6.3.1.6** Participating parties shall agree on the outcome of the review and any action item responsibilities and closure criteria.

**7.2.6.3.2 Project Management Reviews.** This activity consists of the following task:
**7.2.6.3.2.1** Project status shall be evaluated relative to the applicable project plans, schedules, standards, and guidelines. The outcome of the review should be considered by appropriate management and should provide for the following:
a) Making activities progress according to plan, based on an evaluation of the activity or software product status.
b) Maintaining global control of the project through adequate allocation of resources.
c) Changing project direction or determining the need for alternate planning.
d) Evaluating and managing the risk issues that may jeopardize the success of the project.

**7.2.6.3.3 Technical Reviews.** This activity consists of the following task:

**7.2.6.3.3.1** Technical reviews shall be held to evaluate the software products or services under consideration and provide evidence that:

a) They are complete.

b) They comply with their standards and specifications.

c) Changes to them are properly implemented and affect only those areas identified by the Configuration Management Process (subclause 7.2.2).

d) They are adhering to applicable schedules.

e) They are ready for the next planned activity.

f) The development, operation, or maintenance is being conducted according to the plans, schedules, standards, and guidelines of the project.

**7.2.7 Software Audit Process**

**7.2.7.1 Purpose**

The purpose of the Software Audit Process is to independently determine compliance of selected products and processes with the requirements, plans and agreement, as appropriate.

**7.2.7.2 Outcomes**

As a result of successful implementation of the Software Audit Process:

a) an audit strategy is developed and implemented;

b) compliance of selected software work products and/or services or processes with requirements, plans and agreement is determined according to the audit strategy;

c) audits are conducted by an appropriate independent party; and

d) problems detected during an audit are identified and communicated to those responsible for corrective action, and resolution.

**7.2.7.3 Activities and tasks**

The project shall implement the following activities in accordance with applicable organization policies and procedures with respect to the Software Audit Process.

**7.2.7.3.1 Process implementation.** This activity consists of the following tasks:

**7.2.7.3.1.1** Audits shall be held at predetermined milestones as specified in the project plan(s).

**7.2.7.3.1.2** Auditing personnel shall not have any direct responsibility for the software products and activities they audit.

**7.2.7.3.1.3** All resources required to conduct the audits shall be agreed by the parties. These resources include support personnel, location, facilities, hardware, software, and tools.

**7.2.7.3.1.4** The parties should agree on the following items at each audit: agenda; software products (and results of an activity) to be reviewed; audit scope and procedures; and entry and exit criteria for the audit.

**7.2.7.3.1.5** Problems detected during the audits shall be recorded and entered into the Software Problem Resolution Process (subclause 7.2.8) as required.

**7.2.7.3.1.6** After completing an audit, the audit results shall be documented and provided to the audited party. The audited party shall acknowledge to the auditing party any problems found in the audit and related problem resolutions planned.

**7.2.7.3.1.7** The parties shall agree on the outcome of the audit and any action item responsibilities and closure criteria.

**7.2.7.3.2 Software audit.** This activity consists of the following task:

**7.2.7.3.2.1** Software audits shall be conducted to ensure that:

a) As coded, software products (such as a software item) reflect the design documentation.

b) The acceptance review and testing requirements prescribed by the documentation are adequate for the ac-

ceptance of the software products.

c) Test data comply with the specification.

d) Software products were successfully tested and meet their specifications.

e) Test reports are correct and discrepancies between actual and expected results have been resolved.

f) User documentation complies with standards as specified.

g) Activities have been conducted according to applicable requirements, plans, and contract.

h) The costs and schedules adhere to the established plans.

# Appendix F. Peer Review requirements of FAA-iCMM v2.0

**All the Evaluation process area is related to Peer reviews.**

**Eg. BP 08.04 Evaluate incremental work products**

Evaluate incremental work products and services.

*Description*

Verify and validate incremental work products. Incremental evaluation of products, product elements, documentation and services provides for relatively efficient removal of defects and deficiencies. Peer reviews (or an equally effective alternative method of incremental work product review) should be performed at appropriate phases of work product development. Incremental evaluations include reviews of in-progress requirement, design, and element documentation and specifications, formal and informal reviews and audits, evaluation of product assemblies, and product and service elements under development. Solution components are evaluated based on design specifications and interface requirements. Standards or criteria should be established and used to specify the evaluation data to be collected and documented. Evaluation of incremental work products and services with or by the customer/user and in the intended operational environment (incremental validation) is performed at appropriate phases of incremental product and service development. The goal of evaluating developed products incrementally is to ensure that problems are found, and defects and deficiencies eliminated, as early in the development process as possible, saving the considerable cost of fault isolation, problem mitigation, and rework associated with resolving problems in a complex, integrated system.

*Typical Work Products*

- peer review results, for requirement, design, and code work products (see Notes below for detailed peer review work products)
- system requirement review minutes
- design review minutes
- test reports on component tests
- software unit test reports
- monthly reports of service performance evaluations
- incremental capability work product verification results
- incremental capability work product validation results

*Notes*

Practice runs of service elements (e.g., a customer service department, product trouble report processing function/group) can be used to evaluate the design of service solution definitions under development. These may require simulated inputs, and their evaluation procedures and analysis of results should be as rigorous as any tangible/actual product test and evaluation.

An efficient strategy should be considered for selecting the level of subsystem assembly for incremental test and evaluation. Incremental validation should be performed on partial assemblies of products and on elements (e.g., components, subsystems) when their development maturity allows operation by users (operators/maintainers) in the intended operational environment or in an appropriate representation of the operational environment. User evaluation of a software user interface is an example of incremental validation of an element. User evaluation of integrated, but incomplete products (e.g., limited functionality and features) is an example validation of a partially completed product.

The use of the peer review method has proven to be an effective means of early defect removal. Reviews by “peers” (as opposed to reviews by supervisors or managers of work product producers) are typically effective in identifying defects, since they are conducted in a nonthreatening, collegial environment. Peer reviews should be performed on key work products including requirement, design, and test documentation, implementation activities, and specific planning work products (e.g., software development plan, risk management plan, or test plan). Incremental evaluation of requirement, design, and test work products should be accompanied by established evaluation criteria.

Peer reviews, or equally effective incremental work product reviews, should be performed well in advance of completion of work products to allow for the efficient detection and removal of defects and deficiencies, as well as on near-final work products. Peer reviews should be supported by appropriate procedures, participant roles, stakeholder lists, training, and tools. The product and service requirements, resulting from the practices of the Requirements process area (PA 02), should be verified by peer review or other effective methods to insure that the requirements meet established quality criteria (e.g., unambiguous, complete, traceable, feasible, and verifiable). They should also be validated to assure that the operational need is accurately represented in the requirements.

Peer review work products include:

- checklists of review criteria
- actions resulting from the review
- identification of the work product
- size of the work product
- size and composition of the review team
- preparation time per reviewer
- length of the review meeting
- types and number of defects found and fixed
- rework effort
- defect description
- defect category
- severity of the defect
- units containing the defect
- units affected by the defect
- activity where the defect was introduced

The following are typical peer review activities:

- Identify peers who will be the reviewers. They may include subject matter experts, stakeholders, etc.
- Ensure that the peer review leader and other participants are aware of their roles.
- Distribute review materials to reviewers in advance so they can adequately prepare for the peer review.
- Specify and enforce readiness and completion criteria for peer reviews.
- Use checklists to identify criteria for the review of the work products in a consistent manner
- Track actions identified in peer reviews until they are resolved.
- Use successful completion of peer reviews, including the rework to address the items identified in the peer reviews, as a completion criterion for the associated task.
- After peer reviews are conducted, collect, record, and analyze data about the reviews.

*Additional Practice Guidance*

- Include product evaluation issues (e.g., unanticipated or unintended functions or behavior) as an integral part of all formal, system-level design reviews.
- Evaluate designs prior to implementation through analysis, modeling, prototypes or simulations to gain confidence in the design functionality and robustness.
- Test new and unproven designs (i.e., highest risk) at the lowest assembly level to verify their compliance with established requirements early in the development life cycle.
- Review the incremental verification results vis-à-vis requirements with key stakeholders on an on-going basis.
- Verify system, subsystem, and work products against requirements established in an earlier phase.
- Evaluate initial, new, and changed requirements against established quality criteria, such as feasibility, verifiability, traceability, etc.
- Evaluate design specifications against established criteria.
- Evaluate design specifications for proper sequencing of events, inputs, outputs, interfaces, logic flow, allocation of timing and sizing budgets, and error definition, isolation, and recovery.
- Evaluate design specifications against safety, security, and other critical requirements.
- Validate the requirements through customer/user interaction to gain confidence that a product or service implemented in accordance with the requirements would meet the operational need.

# Appendix G. ISO 9001:2000 review requirements

Table 41 – ISO 9001:2000 review requirements

| Item id | Requirement (Task) |
|---|---|
| ISO 9001:2000 7.1 c | The organization shall plan and develop the processes needed for product realization. Planning of product realization shall be consistent with the requirements of the other processes of the quality management system (see 4.1).<br>In planning product realization, the organization shall determine the following, as appropriate:<br>c) required verification, validation, monitoring, inspection and test activities specific to the product and the criteria for product acceptance; |
| ISO 9001:2000 7.4.3 c | The organization shall establish and implement the inspection or other activities necessary for ensuring that purchased product meets specified purchase requirements. |
| ISO 9001:2000 7.3.5 | **7.3.5 Design and development verification**<br>Verification shall be performed in accordance with planned arrangements (see 7.3.1) to ensure that the design and development outputs have met the design and development input requirements. Records of the results of the verification and any necessary actions shall be maintained (see 4.2.4). |
| ISO 9001:2000 7.3.4 | **7.3.4 Design and development review**<br>At suitable stages, systematic reviews of design and development shall be performed in accordance with planned arrangements (see 7.3.1)<br>a) to evaluate the ability of the results of design and development to meet requirements, and b) to identify any problems and propose necessary actions.<br>Participants in such reviews shall include representatives of functions concerned with the design and development stage(s) being reviewed. Records of the results of the reviews and any necessary actions shall be maintained (see 4.2.4). |
| ISO 9001:2000 7.2.2 | **7.2.2 Review of requirements related to the product**<br>The organization shall review the requirements related to the product. This review shall be conducted prior to the organization's commitment to supply a product to the customer (e.g. submission of tenders, acceptance of contracts or orders, acceptance of changes to contracts or orders) and shall ensure that<br>a) product requirements are defined,<br>b) contract or order requirements differing from those previously expressed are resolved, and<br>c) the organization has the ability to meet the defined requirements.<br>Records of the results of the review and actions arising from the review shall be maintained (see 4.2.4).<br>Where the customer provides no documented statement of requirement, the customer requirements shall be confirmed by the organization before acceptance. |

| Item id | Requirement (Task) |
| --- | --- |
| | Where product requirements are changed, the organization shall ensure that relevant documents are amended and that relevant personnel are made aware of the changed requirements. |

# Appendix H. Quality approach element instances in CMMI

As CMMI is a well-structured quality approach we do not include here a table of quality approach element instances. We include quality approach element instances directly into the process element table in Appendix L.

# Appendix I. Quality approach element instances in IEEE 1028

**Introduction to Inspections**

The purpose of an inspection is to detect and identify software product anomalies. An inspection is a systematic peer examination that does one or more of the following:

a) Verifies that the software product satisfies its specifications
b) Verifies that the software product exhibits specified quality attributes
c) Verifies that the software product conforms to applicable regulations, standards, guidelines, plans, specifications, and procedures
d) Identifies deviations from provisions of item a), item b), and c)
e) Collects software engineering data (for example, anomaly and effort data)
f) Provides the collected software engineering data that may be used to improve the inspection process itself and its supporting documentation (for example, checklists)
g) Requests or grants waivers for violation of standards where the adjudication of the type and extent of violations are assigned to the inspection jurisdiction
h) Uses the data as input to project management decisions as appropriate (e.g., to make trade-offs between additional inspections versus additional testing)

Inspections consist of two to six participants (including the author). An inspection is led by an impartial trained facilitator who is trained in inspection techniques. Determination of remedial or investigative action for an anomaly is a mandatory element of a software inspection, although the resolution should not occur in the inspection meeting. Collection of data for the purpose of analysis and improvement of software engineering procedures is a mandatory element of software inspections.

Examples of software products subject to inspections include, but are not limited to, the following:

— Software requirements specification
— Software design description

- Source code
- Software test documentation
- Software user documentation
- Maintenance manual
- System build procedures
- Installation procedures
- Release notes
- Software models
- Specifications
- Software development process descriptions
- Policies, strategies, and plans
- Marketing and publicity documents
- Software architectural descriptions

Table 42 – Input/output element instances and their IDs identified based on the text in IEEE

| Text | Input/Output element instance | ID |
|---|---|---|
| Input to the inspection shall include the following:<br>a) A statement of objectives for the inspection<br>b) The software product(s) to be inspected<br>c) Documented inspection procedure<br>d) Inspection reporting forms<br>e) Anomalies or issues list<br>f) Source documents such as specifications and software product inputs that serve as documents that have been used by the author as inputs to development the software product<br>Input to the inspection may also include the following:<br>g) Inspection checklists<br>h) Quality criteria for requiring a reinspection<br>i) Predecessor software product that has previously been inspected, approved, or established as a baseline | Inspection statement of objectives | IEEE 1028 In should a |
| | Software product to be inspected | IEEE 1028 In should b |
| | Inspection procedure | IEEE 1028 In should c |
| | Inspection reporting forms | IEEE 1028 In should d |
| | Issue list | IEEE 1028 In should e |
| | Source products of the software product to be inspected | IEEE 1028 In should f |
| | Inspection checklist | IEEE 1028 In should g |
| | Quality criteria for requiring a reinspection | IEEE 1028 In may h |
| | Predecessor software product that has previously been inspected, approved, or established as a baseline | IEEE 1028 In may i |

| Text | Input/Output element instance | ID |
|---|---|---|
| j) Any regulations, standards, guidelines, plans, specifications, and procedures against which the software product is to be inspected<br>k) Hardware, instrumentation, or other software product specifications<br>l) Performance data<br>m) Anomaly categories (see IEEE Std 1044-1993 [B8]) | Regulations, standards, guidelines, plans, specifications, and procedures against which the software product is to be inspected | IEEE 1028 In may j |
| | Product specification | IEEE 1028 In may k |
| | Performance data | IEEE 1028 In may l |
| | Anomaly categories | IEEE 1028 In may m |
| The output of the inspection shall be documented evidence that identifies the following:<br>a) The project that created the software product under inspection<br>b) The inspection team members<br>c) The inspection meeting duration<br>d) The software product inspected<br>e) The size of the materials inspected (for example, the number of text pages)<br>f) Specific inputs to the inspection<br>g) Inspection objectives and whether they were met<br>h) The anomaly list, containing each anomaly location, description, and classification<br>i) The disposition of the software product<br>j) Any waivers granted or waivers requested<br>k) Individual and total preparation time of the inspection team<br>l) The total rework time<br>The inspection output should include the following:<br>m) The inspection anomaly summary listing the number of anomalies identified by each anomaly category<br>n) An estimate of the rework effort and rework completion date, if the rework effort is expected to be significant<br>The inspection output may include the following:<br>o) An estimate of the savings by fixing items found in inspection, compared to their cost to fix if identified later<br>Although this standard sets minimum requirements for the content of the documented evi- | Project identifier | IEEE 1028 Out shall a |
| | Inspection team members (name and/or id) | IEEE 1028 Out shall b |
| | Inspection meeting duration | IEEE 1028 Out shall c |
| | Software product to be inspected | IEEE 1028 Out shall d |
| | Size of materials inspected | IEEE 1028 Out shall e |
| | Specific inputs to the inspection | IEEE 1028 Out shall f |
| | Inspection objectives and whether were met | IEEE 1028 Out shall g |
| | Anomaly list | IEEE 1028 Out shall h |
| | Software product disposition | IEEE 1028 Out shall i |
| | Waivers granted or requested | IEEE 1028 Out shall j |
| | Preparation times | IEEE 1028 Out shall k |
| | Rework time | IEEE 1028 Out shall l |
| | Inspection anomaly summary | IEEE 1028 Out should m |
| | Estimate of rework effort | IEEE 1028 Out should n |
| | Estimate of savings | IEEE 1028 Out may o |

| Text | Input/Output element instance | ID |
|---|---|---|
| dence, it is left to local procedures to prescribe additional content, format requirements, and media. | | |

Table 43 – IEEE 1028 inspection entry and exit criteria element instances

| Entry criteria | Exit criteria |
|---|---|
| 6.4.1 Authorization<br>Inspections shall be planned and documented in the appropriate project planning documents (for example, the project plan, the software quality assurance plan, or the software verification and validation plan).<br>Additional inspections may be conducted during acquisition, supply, development, operation, and maintenance of the software product at the request of project management, quality management, or the author, according to local procedures.<br><br>6.4.2 Preconditions<br>An inspection shall be conducted only when the relevant inspection inputs are available.<br><br>6.4.3 Minimum entry criteria<br>An inspection shall not be conducted until all of the following events have occurred, unless there is a documented rationale, accepted by management, for exception from these provisions:<br>a) The inspection leader determines that the software product to be inspected is complete and conforms to project standards for format.<br>b) Automated error-detecting tools (such as spell-checkers and compilers) have been used to identify and eliminate errors prior to the inspection.<br>c) Prior milestones upon which the software product depends are satisfied as identified in the appropriate planning documents.<br>d) Required supporting documentation is available.<br>e) For a reinspection, all items noted on the anomaly list that affect the software product under inspection are resolved. | An inspection shall be considered complete when the activities listed in 6.5 have been accomplished, and the output described in 6.7 exists. |

Table 44 – IEEE 1028 element instances related to activities and their IDs

| Subprocess | ID | Activity description | In process |
|---|---|---|---|

| Subprocess | ID | Activity description | In process |
|---|---|---|---|
| Management preparation (GP 2.10) | IEEE1028-2008 6.5.1 | Management shall ensure that the inspection is performed as required by applicable standards and procedures and by requirements mandated by law, contract, or other policy. | |
| | IEEE1028-2008 6.5.1 a | Plan time and resources required for inspection, including support functions, as required in IEEE Std 1058-1998 [B9] or other appropriate standards | |
| | IEEE1028-2008 6.5.1 b | Provide funding, infrastructure, and facilities required to plan, define, execute, and manage the inspection | |
| | IEEE1028-2008 6.5.1 c | Provide training and orientation on inspection procedures applicable to a given project | |
| | IEEE1028-2008 6.5.1 d | Ensure that inspection team members possess appropriate levels of expertise and knowledge sufficient to comprehend the software product under inspection | |
| | IEEE1028-2008 6.5.1 e | Ensure that inspections are planned, and that planned inspections are conducted | |
| | IEEE1028-2008 6.5.1 f | Act on inspection team recommendations in a timely manner | |
| Planning the inspection | IEEE1028-2008 6.5.2 1 | The author shall assemble the inspection materials for the inspection leader. Inspection materials include the software product to be inspected, standards and documents that have been used to develop the software product, etc. | Assemble the inspection materials, send to inspection leader. |
| | IEEE1028-2008 6.5.2 a | Identify, with appropriate management support, the inspection team | |
| | IEEE1028-2008 6.5.2 b | Assign specific responsibilities to the inspection team members | |
| | IEEE1028-2008 6.5.2 c | Schedule the meeting date and time, select the meeting place, and notify the inspection team | |
| | IEEE1028-2008 6.5.2 d | Distribute inspection materials to participants, and allow adequate time for their preparation | |
| | IEEE1028-2008 6.5.2 e | Set a timetable for distribution of inspection material and for the return of comments and forwarding of comments to the author for disposition | |
| | IEEE1028- | Specify the scope of the inspection, including the priority of sections of the docu- | |

| Subprocess | ID | Activity description | In process |
|---|---|---|---|
| | 2008 6.5.2 f | ments to be inspected | |
| | IEEE1028-2008 6.5.2 g | Establish the anticipated inspection rate for preparation and meeting | |
| Overview of inspection procedures | IEEE1028-2008 6.5.3 1 | Roles shall be assigned by the inspection leader. | Assign roles/Insp leader |
| | IEEE1028-2008 6.5.3 2 | The inspection leader shall answer questions about any checklists and the role assignments and should present inspection data such as minimal preparation times, the recommended inspection rate, and the typical number of anomalies previously found in inspections of similar products. | /a: Answer questions about any checklists and the role assignments<br>/b: Present inspection data such as minimal preparation times, the recommended inspection rate, and the typical number of anomalies previously found in inspections of similar products |
| Overview of inspection product | IEEE1028-2008 6.5.4 | The author should present an overview of the software product to be inspected. This overview should be used to introduce the inspectors to the software product. The overview may be attended by other project personnel who could profit from the presentation. | |
| Preparation (VER SP 2.1 SUBP 10) | IEEE1028-2008 6.5.5 1 | Each inspection team member shall examine the software product and other inputs prior to the review meeting. | Examine the software product and other inputs prior to the review meeting. |
| | IEEE1028-2008 6.5.5 2 | Anomalies detected during this examination shall be documented and sent to the inspection leader. | |
| | IEEE1028-2008 6.5.5 3 | The inspection leader should classify anomalies as described in 6.8.1 to determine whether they warrant cancellation of the inspection meeting, and in order to plan efficient use of time in the inspection meeting. If the inspection leader determines that the extent or seriousness of the anomalies warrants, the inspection leader may cancel the inspection, requesting a later inspection when the software product meets the minimal entry criteria and is reasonably defect-free. The inspection leader should forward the anomalies to the author of the software product for disposition. | /a Classify anomalies<br>/b1 Cancel Inspection, request later inspection<br>/b2 Extent or seriousness of the anomalies is high?<br>/c Forward the anomalies to the author of the software product for disposition. |
| | IEEE1028- | The inspection leader or reader shall specify a suitable order in which the software | Specify a suitable order in which the software |

| Subprocess | ID | Activity description | In process |
|---|---|---|---|
| | 2008 6.5.5 4 | product will be inspected (such as sequential, hierarchical, data flow, control flow, bottom up, or top down). | product will be inspected |
| | IEEE1028-2008 6.5.5 5 | The reader(s) shall prepare sufficiently to be able to present the software product at the inspection meeting. | |
| | IEEE1028-2008 6.5.5 6 | The inspection leader shall verify that inspectors are prepared for the inspection. The inspection leader shall reschedule the meeting if the inspectors are not adequately prepared. The inspection leader should gather individual preparation times and record the total in the inspection documentation. | /a Are inspectors prepared for the inspection? |
| | | | /b Reschedule meeting |
| | | | /c Gather individual preparation times and record the total in the inspection documentation. |
| Examination | IEEE1028-2008 6.5.6.1 (title) | Introduce meeting | |
| | IEEE1028-2008 6.5.6.2(title) | Review general items | |
| | IEEE1028-2008 6.5.6.3 (title) | Review software product and record anomalies | /a Present the software product to the inspection team |
| | | | /b Examine the software product |
| | | | /c Enter anomalies to anomaly list |
| | | | /d Answer specific questions |
| | IEEE1028-2008 6.5.6.4 (title) | Review the anomaly list | |
| | IEEE1028-2008 6.5.6.5 (title) | Make exit decision | |
| Rework/Follow-up | IEEE1028-2008 6.5.7 | The inspection leader shall verify that the action items assigned in the meeting are closed. | Take corrective actions based on action items (added) |
| | | | Verify that the action items assigned in the meet- |

| Subprocess | ID | Activity description | In process |
|---|---|---|---|
| | | | ing are closed. |
| Data collection | IEEE1028-2008 6.8 | Inspections shall provide data for the analysis of the quality of the software product, the effectiveness of the acquisition, supply, development, operation and maintenance processes, and the effectiveness and the efficiency of the inspection itself. In order to maintain the effectiveness of inspections, data from the author and inspectors shall not be used to evaluate the performance of individuals. To enable these analyses, anomalies that are identified at an inspection meeting shall be classified in accordance with 6.8.1, 6.8.2, and 6.8.3.<br>Inspection data shall contain the identification of the software product, the date and time of the inspection, the inspection team, the preparation and inspection times, the volume of the materials inspected, and the disposition of the inspected software product. The capture of this information shall be used to optimize local guidance for inspections.<br>The management of inspection data requires a capability to enter, store, access, update, summarize, and report classified anomalies. The frequency and types of the inspection analysis reports, and their distribution, are left to local standards and procedures. | Collect inspection data |
| Improvement | IEEE1028-2008 6.9 | Inspection data shall be analyzed regularly in order to improve the inspection process itself, and should be used to improve the activities used to produce software products.<br>Frequently occurring anomalies shall be included in the inspection checklists or role assignments.<br>The checklists themselves shall also be inspected regularly for superfluous or misleading questions.<br>Consistently granted or requested waivers shall be analyzed to determine if the standards need to be changed.<br>The preparation times, meeting times, and number of participants shall be analyzed to determine connections between preparation (checking) rate, meeting rate, and number and severity of anomalies found. | /a Inspection data shall be analyzed regularly in order to improve the inspection process itself, and should be used to improve the activities used to produce software products. |
| | | | /b Frequently occurring anomalies shall be included in the inspection checklists or role assignments. |
| | | | /c The checklists themselves shall also be inspected regularly for superfluous or misleading questions. |
| | | | /d Consistently granted or requested waivers shall be analyzed to determine if the standards need to be changed. |

| Subprocess | ID | Activity description | In process |
|---|---|---|---|
| | | | /e The preparation times, meeting times, and number of participants shall be analyzed to determine connections between preparation (checking) rate, meeting rate, and number and severity of anomalies found. |
| | | Benefits (savings) achieved should be assessed regularly, and the inspection process should be continually adapted to achieve greater effectiveness at maximum efficiency | Benefits (savings) achieved should be assessed regularly, and the inspection process should be continually adapted to achieve greater effectiveness at maximum efficiency |

# Appendix J. Quality approach element instances of Process Impact

Table 45 – Element instances in Process Impact Peer review descriptions which will be used only textually

| Element | Element Instance Description | Element inst. ID |
|---|---|---|
| Overview | In a peer review, co-workers of a person who created a software work product examine that product to identify defects and correct shortcomings. A peer review:<br>- Verifies whether the work product correctly satisfies the specifications found in any predecessor work product, such as requirements or design documents<br>- Identifies any deviation from standards, including issues that may affect maintainability of the software<br>- Suggests improvement opportunities to the author<br>- Promotes the exchange of techniques and education of the participants.<br>All interim and final development work products are candidates for review, including:<br>- Requirements specifications<br>- User interface specifications and designs<br>- Architecture, high-level design, and detailed designs and models<br>- Source code<br>- Test plans, designs, cases, and procedures | PI ov |

| Element | Element Instance Description | Element inst. ID |
|---|---|---|
| | - Software development plans, including project management plan, configuration management plan, and quality assurance plan<br>This document defines an overall peer review process. It includes procedures for conducting inspections and two types of informal peer review, a walkthrough and a passaround, as well as guidance for selecting the appropriate approach for each review. | |
| Risk assessment guidance | To judge which software components (or portions of components) to review and what type of review method to use, consider the following risk criteria:<br>• Components that use new technology, techniques, or tools<br>• Key architectural components<br>• Complex logic or algorithms that are difficult to understand but must be accurate and optimized<br>• Mission-, security-, or safety-critical components with dangerous failure modes<br>• Components having many exception conditions or failure modes<br>• Exception handling code that cannot easily be tested<br>• Components that are intended to be reused<br>• Components that will serve as models or templates for other components<br>• Components that affect multiple portions of the product<br>• Complex user interfaces<br>• Components created by less experienced developers<br>• Code modules having high complexity<br>• Modules having a history of many defects or changes<br><br>Work products that fit in one or more of these categories are considered high risk. A product is considered low risk if an undetected error will not significant affect the project's ability to meet its schedule, quality, cost, and feature objectives. Use inspections for high-risk work products, or the high-risk portions of large products, and for major work products that are about to be baselined. Less formal reviews are acceptable for other work products. | PI RA |
| Participants | Table 1 suggests project roles who might review different work products. Not all of these perspectives need to be represented. In general, a work product should be reviewed by:<br><br>• The author of any predecessor document or specification<br>• Someone who must base their subsequent work on the work product<br>• Peers of the author<br>• Anyone responsible for a component to which the work product interfaces<br><br>Attendance by anyone with supervisory authority over the author is by invitation of the author only. | PI Part |

| Element | Element Instance Description | Element inst. ID |
|---|---|---|
| | **Table 1. Review Participants for Different Types of Work Products.** | |
| Entry criteria | o The author selected an inspection approach for the product being reviewed.<br>o All necessary supporting documentation is available<br>o The author has stated his objectives for this inspection.<br>o Reviewers are trained in the peer review process.<br>o Documents to be inspected are identified with a version number. All pages are numbered and line numbers are displayed. The documents have been spell-checked.<br>o Source code to be inspected is identified with a version number. Listings have line numbers and page numbers. Code compiles with no errors or warning messages using the project's standard compiler switches. Errors found using code analyzer tools have been corrected. | PI Ent |

**Table 1. Review Participants for Different Types of Work Products.**

| ork Product Type | Suggested Reviewers |
|---|---|
| Architecture or High-Level Design | architect, requirements analyst, designer, project manager, integration test engineer |
| Detail Design | designer, architect, programmer, integration test engineer |
| Process Documentation | process improvement group leader, process improvement working group members, management-level process owner, practitioner representatives who will use the process |
| Project Plans | project manager, program manager, business sponsor, marketing or sales representative, technical lead, quality assurance manager |
| Requirements Specification | requirements analyst, project manager, architect, designer, system test engineer, quality assurance manager, user or marketing representative, documentation writer, subject matter expert, technical support representative |
| Source Code | programmer, designer, unit test engineer, maintainer, requirements analyst, coding standards expert |
| System Technical Documentation | author, project manager, maintainer, programmer |
| Test Documentation | test engineer, programmer (unit testing) or architect (integration testing) or requirements analyst (system testing), quality assurance representative |
| User Interface Design | user interface designer, requirements analyst, user, application domain expert, usability or human factors expert, system test engineer |
| User Manual | documentation writer, requirements analyst, user or marketing representative, system test engineer, maintainer, designer, instructional designer, trainer, technical support representative |

<table>
<tr><th>Element</th><th>Element Instance Description</th><th>Element inst. ID</th></tr>
<tr><td></td><td>o For a re-inspection, all issues from the previous inspection were resolved.<br>Any additional entry criteria defined for the specific type of work product are also satisfied.</td><td></td></tr>
<tr><td>Information to Record for Each Defect Found</td><td><table>
<tr><th>Column</th><th>Description</th></tr>
<tr><td>Origin</td><td>development phase in which the defect was introduced</td></tr>
<tr><td>Type</td><td>• Missing (something needs to be there but is not)<br>• Wrong (something is erroneous or conflicts with something else)<br>• Extra (something unnecessary is present)<br>• Usability<br>• Performance<br>• Non-defect issue (question, point of style, suggestion, clarification needed)</td></tr>
<tr><td>Severity</td><td>• Major (could cause product failure or cost significantly more to correct in the future)<br>• Minor (non-fatal error, cosmetic problem, annoyance, or a workaround is available)</td></tr>
<tr><td>Location</td><td>page and line or section number where the defect is located</td></tr>
<tr><td>Description</td><td>concise description of the issue or possible defect</td></tr>
</table></td><td>PI Def</td></tr>
<tr><td>Possible Appraisals of Inspected Work Products</td><td><table>
<tr><th>Appraisal</th><th>Meaning</th></tr>
<tr><td>Accepted As Is</td><td>Modifications may be required in the work product, but verification of the modification is not necessary.</td></tr>
<tr><td>Accept Conditionally</td><td>Defects must be corrected, and the changes must be verified by the individual named on the Inspection Summary Report.</td></tr>
<tr><td>Re-inspect Following Rework</td><td>A substantial portion of the product must be modified, or there are many changes to make. A second inspection is required after the author has completed rework.</td></tr>
<tr><td>Inspection Not Completed</td><td>A significant fraction of the planned material was not inspected, or the inspection was terminated for some reason.</td></tr>
</table></td><td>PI App</td></tr>
<tr><td>Exit criteria</td><td>o All of the author’s inspection objectives are satisfied.<br>o Issues raised during the inspection are tracked to closure.<br>o All major defects are corrected.<br>o Uncorrected defects are logged in the project’s defect tracking system.<br>o The modified work product is checked into the project’s configuration management system.</td><td>PI Ex</td></tr>
</table>

<table>
<tr><th>Element</th><th>Element Instance Description</th><th>Element inst. ID</th></tr>
<tr><td></td><td>o If changes were required in earlier project deliverables, those deliverables have been correctly modified, checked into the project's configuration management system, and any necessary regression tests were passed.<br>o Moderator has collected and recorded the inspection data.<br>Moderator has delivered the completed Inspection Summary Report and defect counts to the peer review coordinator.</td><td></td></tr>
<tr><td>Data Item (12)</td><td>Table 4. Data Items Collected From Each Inspection.
<table>
<tr><th>Data Item</th><th>Definition</th></tr>
<tr><td>Effort.Planning</td><td>total labor hours spent by the moderator and author in planning, scheduling meetings, assembling, duplicating, and distributing materials, and any other related tasks</td></tr>
<tr><td>Effort.Overview</td><td>total labor hours spent by the participants in an overview meeting, if one was held</td></tr>
<tr><td>Effort.Preparation</td><td>total labor hours spent by the inspectors and author preparing for the inspection</td></tr>
<tr><td>Effort.Rework</td><td>total labor hours the author spent correcting defects in the initial deliverable and making other improvements; include verification time from the follow-up stage</td></tr>
<tr><td>Time.Meeting</td><td>duration of the inspection meeting in hours</td></tr>
<tr><td>Defects.Found.Major, Defects.Found.Minor</td><td>total number of major and minor defects found by the inspection team; do not include non-defect issues raised, such as questions, requests for clarification, points of style, or items from the Typo Lists</td></tr>
<tr><td>Defects.Corrected.Major, Defects.Corrected.Minor</td><td>total number of major and minor defects corrected during rework</td></tr>
<tr><td>Size.Planned, Size.Actual</td><td>total physical lines of code (not including comments and blank lines) or number of document pages that were planned for inspection and that were actually inspected</td></tr>
<tr><td>Number.of.Inspectors</td><td>number of active participants in the inspection meeting</td></tr>
<tr><td>Inspection.Appraisal</td><td>inspection team's decision about disposition of the inspected work product (accepted as is, accepted conditionally, re-inspect following rework)</td></tr>
</table></td><td>PI DI</td></tr>
<tr><td>Metrics (12)</td><td><table>
<tr><th>Metric</th><th>How Calculated</th></tr>
<tr><td>Defect.Density</td><td>Defects.Found.Total / Size.Actual</td></tr>
<tr><td>Defects.Found.Total</td><td>Defects.Found.Major + Defects.Found.Minor</td></tr>
<tr><td>Defects.Corrected.Total</td><td>Defects. Corrected.Major + Defects. Corrected.Minor</td></tr>
<tr><td>Effort.Inspection</td><td>Effort.Planning + Effort.Overview + Effort.Preparation + Effort.Meeting + Effort.Rework</td></tr>
</table></td><td>PI Met</td></tr>
</table>

| Element | Element Instance Description | Element inst. ID |
|---|---|---|
| | Effort.Meeting: Number.of.Inspectors * Time.Meeting<br>Effort.per.Defect: Effort.Inspection / Defects.Found.Total<br>Effort.per.Unit.Size: Effort.Inspection / Size.Actual<br>Percent.Inspected: 100 * Size.Actual / Size.Planned<br>Percent.Majors: 100 * Defects.Found.Major / Defects.Found.Total<br>Rate.Inspection: Size.Actual / Time.Meeting<br>Rate.Preparation: Size.Planned / (Effort.Preparation / Number.of.Inspectors)<br>Rework.per.Defect: Effort.Rework / Defects.Corrected.Total | |
| Description on process measurement | The moderator shall collect the data items in Table 4 from each inspection. These data items are used to calculate the process metrics in Table 5 and to monitor and improve the inspection process. The moderator shall record the data items in the appropriate spaces on the Inspection Summary Report and Issue Log and report them to the organization's peer review coordinator. The peer review coordinator shall maintain these metrics in a repository and produce periodic reports of summary data for practitioners and managers.<br>*<add section to describe tools and procedures for peer review coordinator to enter inspection data into the repository and generate reports>* | PI Measurement |
| Description of process maintenance | Submit suggestions for improvements to be made in this peer review process to *<organization>*'s peer review process owner. | PI Maintenance |

Table 46 – Tasks in Process Impact process descriptions

| Phase | Task id | TO | Task | Responsible |
|---|---|---|---|---|
| **Planning** | pi11 | 1 | Give moderator the work product to be inspected and supporting documents, such as specifications, predecessor documents, or pertinent test documentation. | Moderator |
| | pi12 | 2 | Determine whether work product satisfies inspection entry criteria. | Author |
| | pi13 | 3 | Based on the size and complexity of the work product, determine how many inspection meetings will be required. | Moderator and Author |
| | pi14 | 4 | Select inspectors and assign roles to individuals. Gain agreement from the other inspectors to participate. | Moderator and Author |

| Phase | Task id | TO | Task | Responsible |
|---|---|---|---|---|
| | pi15 | 5 | Determine whether an overview meeting is required. | Author |
| | pi16 | 6 | Schedule the inspection, and possibly overview, meetings and distribute a meeting notice. | Moderator |
| | pi17 | 7 | Distribute the inspection package to the participants at least 3 working days prior to the inspection meeting. | Moderator or Author |
| **Overview** | pi21 | 1 | Describe the important features of the work product to the rest of the inspection team. State inspection objectives. | Author |
| | pi22 | 2 | Evaluate the assumptions, history, and context of the work product. | Inspectors |
| **Preparation** | pi31 | 1 | Ask individual inspectors to prepare with specific objectives in mind, such as: checking the consistency of cross-references; checking for interface errors; checking traceability to, and consistency with, predecessor specifications; or checking conformance to standards. | Moderator and Author |
| | pi32 | 2 | Examine the work product, to understand it, find defects, and raise questions about it. Use the appropriate defect checklist to focus attention on defects commonly found in the type of product being inspected. Use other analysis methods to look for defects as appropriate. | Inspectors |
| | pi33 | 3 | Log minor defects found, such as typographical errors or style inconsistencies, on the Typo List. Deliver this to the author at or prior to the inspection meeting. | Inspectors |
| **Inspection Meeting** | pi41 | 1 | **Open the Meeting:** Introduce the participants (if necessary) and state their roles, state the purpose of the inspection, and direct inspectors to focus their efforts toward finding defects, not solutions. Remind participants to address their comments to the work product under review, not to the author. | Moderator |
| | pi42 | 2 | **Establish Preparedness:** Ask each inspector for his preparation time and record the times on the Inspection Summary Report. If preparation is insufficient, reschedule the meeting. | Moderator |
| | pi43 | 3 | **Present Work Product:** Describe portions of the work product to the inspection team. | Reader |
| | pi44 | 4 | **Raise Defects and Issues:** Point out concerns, potential defects, questions, or improvement opportunities after the reader presents each section. | Inspectors |
| | pi45 | 5 | **Record Issues:** Capture the information in Table 2 on the Issue Log for each issue raised. State aloud what was recorded to make sure it was recorded accurately | Recorder |
| | pi46 | 6 | **Answer Questions:** Respond briefly to any specific questions raised, and contribute to defect detection based on special understanding of the work product. | Author |
| | pi47 | 7 | **Make Product Appraisal:** After all meetings scheduled for a given inspection are complete, decide on the work product appraisal, selecting from the options in Table 3. If the inspectors disagree, assign the most conservative appraisal offered by any of the inspectors. | Inspectors |

| Phase | Task id | TO | Task | Responsible |
|---|---|---|---|---|
| | pi48 | 8 | **Sign Inspection Summary Report:** All participants sign the Inspection Summary Report to indicate their agreement with the inspection outcome. | Inspectors |
| | pi49 | 9 | **Collect Inspection Feedback.** Ask the inspectors to evaluate the inspection and suggest improvements, using the Inspection Lessons Learned Questionnaire. | Moderator |
| **Rework** | pi51 | 1 | Correct defects and typos found, resolve issues raised, and modify work product accordingly. Mark issues list to indicate action taken. | Author |
| | pi52 | 2 | Correct any other project documents based on defects identified in the inspected work product. | Author |
| | pi53 | 3 | Record any uncorrected defects in the project's defect tracking system. | Author |
| | pi54 | 4 | If rework verification is not needed, report the number of major and minor defects found and corrected and the actual rework effort to the moderator. | Author |
| | pi55 | 5 | Record the actual rework effort on the Inspection Summary Report. | Moderator |
| **Follow-Up** | pi61 | 1 | Confirm that the author has addressed every item on the Issue Log. Determine whether the author made appropriate decisions as to which defects not to correct and which improvement suggestions not to implement. | Verifier |
| | pi62 | 2 | Examine the modified work product to judge whether the rework has been performed correctly. Report any findings to the author, so rework can be declared complete, incorrect rework can be redone, or items that were not originally pursued can be addressed. | Verifier |
| | pi63 | 3 | Report the number of major and minor defects found and corrected and the actual rework effort to the moderator. | Author |
| | pi64 | 4 | Check whether the exit criteria for the inspection and for the peer review process have been satisfied. If so, the inspection is complete. | Moderator |
| | pi65 | 5 | Check the baselined work product into the project's configuration management system. | Author |
| | pi66 | 6 | Deliver Inspection Summary Report and counts of defects found and defects corrected to peer review coordinator. | Author |

Table 47 – Peer review-related work aids and deliverables in Process Impact's peer review descriptions

| Quality approach element | Quality approach element instance | Element ID |
|---|---|---|
| Work aid | Inspection Summary Report | PI WA1 |
| Work aid | Issue Log | PI WA2 |

| Quality approach element | Quality approach element instance | Element ID |
|---|---|---|
| Work aid | Typo List | PI WA3 |
| Work aid | Inspection Moderator’s Checklist | PI WA4 |
| Work aid | Inspection Lessons Learned Questionnaire | PI WA5 |
| Work aid | Review checklists for several types of software work products | PI WA6 |
| Deliverable | Baselined work product | PI D1 |
| Deliverable | Completed Inspection Summary Report | PI D2 |
| Deliverable | Completed Issue Log | PI D3 |
| Deliverable | Completed Typo Lists | PI D4 |
| Deliverable | Counts of defects found and defects corrected | PI D5 |

# Appendix K. Peer review role definitions in quality approaches

Table 48 – Roles related to peer reviews and inspections in quality approaches

| Name | Source | Description |
|---|---|---|
| Author | processimpact.com | • Creator or maintainer of the work product to be inspected. Initiates the inspection process by asking the peer review coordinator to assign a moderator.<br>• States his objectives for the inspection.<br>• Delivers work product and its specification or predecessor document to moderator.<br>• Works with moderator to select inspectors and assign roles.<br>• Addresses items on the Issue Log and Typo Lists.<br>• Reports rework time and defect counts to moderator. |

| Name | Source | Description |
|---|---|---|
| Moderator | processimpact.com | • Plans, schedules, and leads the inspection events.<br>• Works with author to select inspectors and assign roles.<br>• Assembles inspection package and delivers it to inspectors at least 3 days prior to the inspection meeting.<br>• Determines whether preparation is sufficient to hold the meeting. If not, reschedules the meeting.<br>• Facilitates inspection meeting. Corrects any inappropriate behavior. Solicits input from inspectors as reader presents each section of the work product. Records any action items or side issues that arise during the inspection.<br>• Leads inspection team in determining the work product appraisal.<br>• Serves as verifier or delegates this responsibility to someone else.<br>• Delivers completed Inspection Summary Report to the organization’s peer review coordinator. |
| Reader | processimpact.com | • Presents portions of the work product to the inspection team to elicit comments, issues, or questions from inspectors. |
| Recorder | processimpact.com | • Records and classifies issues raised during inspection meeting. |
| Inspector | processimpact.com | • Examines work product prior to the inspection meeting to find defects and prepare for contributing to the meeting.<br>• Records preparation time.<br>• Participates during the meeting to identify defects, raise issues, and suggest improvements. |
| Verifier | processimpact.com | Performs follow-up to determine whether rework has been performed appropriately and correctly. |
| Peer review coordinator | processimpact.com | • Custodian of the organization’s inspection metrics database.<br>• Maintains records of inspections conducted and data from the Inspection Summary Report for each inspection.<br>• Generates reports on inspection data for management, process improvement team, and peer review process owner. |
| Inspection leader | IEEE 1028-2008 | The inspection leader shall be responsible for planning and organizational tasks pertaining to the inspection, shall determine the parts/components of the software product and source documents to be inspected during the meeting (in conjunction with the author), shall be responsible for planning and preparation as described in 6.5.2 and 6.5.4, shall ensure that the inspection is conducted in an orderly manner and meets its objectives, shall ensure that the inspection data is collected, and shall issue the inspection output as described in 6.7. |
| Recorder | IEEE 1028-2008 | The recorder shall document anomalies, action items, decisions, waivers, and recommendations made by the inspection team. The recorder should record inspection data required for process analysis. The inspection leader may be the recorder. |

| Name | Source | Description |
|---|---|---|
| Reader | IEEE 1028-2008 | The reader shall lead the inspection team through the software product in a comprehensive and logical fashion, interpreting sections of the work (for example, generally paraphrasing groups of 1 to 3 lines), and highlighting important aspects. The software product may be divided into logical sections and assigned to different readers to lessen required preparation time. |
| Author | IEEE 1028-2008 | The author shall be responsible for the software product meeting its inspection entry criteria, for contributing to the inspection based on special understanding of the software product, and for performing any rework required to make the software product meet its inspection exit criteria. |
| Inspector | IEEE 1028-2008 | Inspectors shall identify and describe anomalies in the software product. Inspectors shall be chosen based on their expertise and should be chosen to represent different viewpoints at the meeting (for example, sponsor, end user, requirements, design, code, safety, test, independent test, project management, quality management, and hardware engineering). Only those viewpoints pertinent to the inspection of the product should be present.<br>Some inspectors should be assigned specific topics to ensure effective coverage. For example, one inspector may focus on conformance with a specific standard or standards, another on syntax or accuracy of figures, and another for overall coherence. These viewpoints should be assigned by the inspection leader when planning the inspection, as provided in item b) of 6.5.2. |

# Appendix L. Process elements created based on CMMI element instances

Table 49 – CMMI process element instances and their relations

| Parent | Item id | Process/Activity | Input | Output |
|---|---|---|---|---|
| Prepare for Peer Reviews (VER SP2.1) | VER SP2.1 SUBP1 | Determine what type of peer review will be conducted. | Review Types, review type descriptions | Decision: review type |
| Prepare for Peer Reviews (VER SP2.1) | VER SP2.1 SUBP2 | Define requirements for collecting data during the peer review. | | Peer review data collection require-ments |
| Prepare for Peer Reviews (VER SP2.1) | VER SP2.1 SUBP3 | Establish and maintain entry and exit criteria for the peer review. | | Entry and exit criteria for work products (VER SP2.1 TWP3) |

| Parent | Item id | Process/Activity | Input | Output |
|---|---|---|---|---|
| Prepare for Peer Reviews (VER SP2.1) | VER SP2.1 SUBP4 | Establish and maintain criteria for requiring another peer review. | | Criteria for requiring another peer review (VER SP 2.1 TWP4) |
| Prepare for Peer Reviews (VER SP2.1) | VER SP2.1 SUBP5 | Establish and maintain checklists to ensure that the work products are reviewed consistently. | | Peer review checklist (VER SP2.1 TWP2) |
| Prepare for Peer Reviews (VER SP2.1) | VER SP2.1 SUBP6 | Develop a detailed peer review schedule, including the dates for peer review training and for when materials for peer reviews will be available. | | Peer Review Schedule (VER SP 2.1 TWP1), Peer review training material (VER SP2.1 TWP5) |
| Prepare for Peer Reviews (VER SP2.1) | VER SP2.1 SUBP7 | Ensure that the work product satisfies the peer review entry criteria prior to distribution. | Peer review entry criteria (VER SP 2.1 TWP 3a) | |
| Prepare for Peer Reviews (VER SP2.1) | VER SP2.1 SUBP8 | Distribute the work product to be reviewed and its related information to the participants early enough to enable participants to adequately prepare for the peer review. | Selected work products to be reviewed (VER SP2.1 TWP6) | |
| Prepare for Peer Reviews (VER SP2.1) | VER SP2.1 SUBP9 | Assign roles for the peer review as appropriate. | role types | assigned roles |
| Prepare for Peer Reviews (VER SP2.1) | VER SP2.1 SUBP10 | Prepare for the peer review by reviewing the work product prior to conducting the peer review. | work product to be reviewed | |
| Conduct Peer Reviews (VER SP2.2) | VER SP2.2 SUBP1 | Perform the assigned roles in the peer review. | assigned roles, work product to be reviewed | |
| Conduct Peer Reviews (VER SP2.2) | VER SP2.2 SUBP2 | Identify and document defects and other issues in the work product. | work product to be reviewed | Peer review issues (VER SP2.2 TWP2) |

| Parent | Item id | Process/Activity | Input | Output |
|---|---|---|---|---|
| Conduct Peer Reviews (VER SP2.2) | VER SP2.2 SUBP3 | Record the results of the peer review, including the action items. | | Peer review results (VER SP2.2 TWP1) |
| Conduct Peer Reviews (VER SP2.2) | VER SP2.2 SUBP4 | Collect peer review data. | | Peer review data (VER SP2.2 TWP3, VER SP2.3 TWP2) |
| Conduct Peer Reviews (VER SP2.2) | VER SP2.2 SUBP5 | Identify action items and communicate the issues to relevant stakeholders. | | Peer review action items (VER SP2.3 TWP2) |
| Conduct Peer Reviews (VER SP2.2) | VER SP2.2 SUBP6 | Conduct an additional peer review if the defined criteria indicate the need. | | DECISION: additional review needed |
| Conduct Peer Reviews (VER SP2.2) | VER SP2.2 SUBP7 | Ensure that the exit criteria for the peer review are satisfied. | Peer review exit criteria (VER SP 2.1 TWP 3b) | |
| Analyze Peer Review Data (VER SP2.3) | VER SP2.3 SUBP1 | Record data related to the preparation, conduct, and results of the peer reviews. | | Peer review data (VER SP2.2 TWP3, VER SP2.3 TWP2) |
| Analyze Peer Review Data (VER SP2.3) | VER SP2.3 SUBP2 | Store the data for future reference and analysis. | Peer review data (VER SP2.2 TWP3, VER SP2.3 TWP2) | |
| Analyze Peer Review Data (VER SP2.3) | VER SP2.3 SUBP3 | Protect the data to ensure that peer review data are not used inappropriately. | Peer review data (VER SP2.2 TWP3, VER SP2.3 TWP2) | |
| Analyze Peer Review Data (VER SP2.3) | VER SP2.3 SUBP4 | Analyze the peer review data. | Peer review data (VER SP2.2 TWP3, VER SP2.3 TWP2) | Peer review data analysis |

# Appendix M. Process elements created based on IEEE 1028 element instances

Table 50 – Inspection process inputs, outputs, entry and exit criteria in IEEE 1028

| Input | Output | Entry criteria | Exit criteria |
|---|---|---|---|
| Input to the inspection shall include the following:<br>a) A statement of objectives for the inspection<br>b) The software product(s) to be inspected<br>c) Documented inspection procedure<br>d) Inspection reporting forms<br>e) Anomalies or issues list<br>f) Source documents such as specifications and software product inputs that serve as documents that have been used by the author as inputs to development the software product<br>Input to the inspection may also include the following:<br>g) Inspection checklists<br>h) Quality criteria for requiring a reinspection<br>i) Predecessor software product that has previously been inspected, approved, or established as a baseline<br>j) Any regulations, standards, guidelines, plans, specifications, and procedures against which the software product is to be inspected<br>k) Hardware, instrumentation, or other | The output of the inspection shall be documented evidence that identifies the following:<br>a) The project that created the software product under inspection<br>b) The inspection team members<br>c) The inspection meeting duration<br>d) The software product inspected<br>e) The size of the materials inspected (for example, the number of text pages)<br>f) Specific inputs to the inspection<br>g) Inspection objectives and whether they were met<br>h) The anomaly list, containing each anomaly location, description, and classification<br>i) The disposition of the software product<br>j) Any waivers granted or waivers requested<br>k) Individual and total preparation time of the inspection team<br>l) The total rework time<br>The inspection output should include the following:<br>m) The inspection anomaly summary listing the number of anomalies identified by | 6.4.1 Authorization<br>Inspections shall be planned and documented in the appropriate project planning documents (for example, the project plan, the software quality assurance plan, or the software verification and validation plan).<br>Additional inspections may be conducted during acquisition, supply, development, operation, and maintenance of the software product at the request of project management, quality management, or the author, according to local procedures.<br><br>6.4.2 Preconditions<br>An inspection shall be conducted only when the relevant inspection inputs are available.<br><br>6.4.3 Minimum entry criteria<br>An inspection shall not be conducted until all of the following events have occurred, unless there is a documented rationale, accepted by management, for exception from these provisions:<br>a) The inspection leader determines that the software product to be inspected is complete and conforms to project standards for format.<br>b) Automated error-detecting tools (such as spell-checkers and compilers) have been used to identify | An inspection shall be considered complete when the activities listed in 6.5 have been accomplished, and the output described in 6.7 exists. |

| Input | Output | Entry criteria | Exit criteria |
| --- | --- | --- | --- |
| software product specifications<br>l) Performance data<br>m) Anomaly categories (see IEEE Std 1044-1993 [B8]) | each anomaly category<br>n) An estimate of the rework effort and rework completion date, if the rework effort is expected to be significant<br>The inspection output may include the following:<br>o) An estimate of the savings by fixing items found in inspection, compared to their cost to fix if identified later<br>Although this standard sets minimum requirements for the content of the documented evidence, it is left to local procedures to prescribe additional content, format requirements, and media. | and eliminate errors prior to the inspection.<br>c) Prior milestones upon which the software product depends are satisfied as identified in the appropriate planning documents.<br>d) Required supporting documentation is available.<br>e) For a reinspection, all items noted on the anomaly list that affect the software product under inspection are resolved. | |

Table 51 – IEEE 1028 process element instances and their relations

| Parent | Item id | Process/Activity description | Roles |
| --- | --- | --- | --- |
| Inspection | | | |
| Management preparation | IEEE1028-2008 6.5.1 | Management shall ensure that the inspection is performed as required by applicable standards and procedures and by requirements mandated by law, contract, or other policy. | |
| | IEEE1028-2008 6.5.1 a | Plan time and resources required for inspection, including support functions, as required in IEEE Std 1058-1998 [B9] or other appropriate standards | |
| | IEEE1028-2008 6.5.1 b | Provide funding, infrastructure, and facilities required to plan, define, execute, and manage the inspection | |
| | IEEE1028-2008 6.5.1 c | Provide training and orientation on inspection procedures applicable to a given project | |
| | IEEE1028-2008 6.5.1 d | Ensure that inspection team members possess appropriate levels of expertise and knowledge sufficient to comprehend the software product under inspection | |

Process elements created based on IEEE 1028 element instances

| Parent | Item id | Process/Activity description | Roles |
|---|---|---|---|
| | IEEE1028-2008 6.5.1 e | Ensure that inspections are planned, and that planned inspections are conducted | |
| | IEEE1028-2008 6.5.1 f | Act on inspection team recommendations in a timely manner | |
| Planning the inspection | IEEE1028-2008 6.5.2 1 | The author shall assemble the inspection materials for the inspection leader. Inspection materials include the software product to be inspected, standards and documents that have been used to develop the software product, etc. | Author |
| | IEEE1028-2008 6.5.2 a | Identify, with appropriate management support, the inspection team | Inspection leader |
| | IEEE1028-2008 6.5.2 b | Assign specific responsibilities to the inspection team members | Inspection leader |
| | IEEE1028-2008 6.5.2 c | Schedule the meeting date and time, select the meeting place, and notify the inspection team | Inspection leader |
| | IEEE1028-2008 6.5.2 d | Distribute inspection materials to participants, and allow adequate time for their preparation | Inspection leader |
| | IEEE1028-2008 6.5.2 e | Set a timetable for distribution of inspection material and for the return of comments and forwarding of comments to the author for disposition | Inspection leader |
| | IEEE1028-2008 6.5.2 f | Specify the scope of the inspection, including the priority of sections of the documents to be inspected | Inspection leader |
| | IEEE1028-2008 6.5.2 g | Establish the anticipated inspection rate for preparation and meeting | Inspection leader |
| Overview of inspection procedures | IEEE1028-2008 6.5.3 1 | Roles shall be assigned by the inspection leader. | Inspection leader |
| | IEEE1028-2008 6.5.3 2 | The inspection leader shall answer questions about any checklists and the role assignments and should present inspection data such as minimal preparation times, the recommended inspection rate, and the typical number of anomalies previously found in inspections of similar products. | Inspection leader |
| Overview of inspection product | IEEE1028-2008 6.5.4 | The author should present an overview of the software product to be inspected. This overview should be used to introduce the inspectors to the software product. The overview may be attended by other project personnel who could profit from the presentation. | Author |

| Parent | Item id | Process/Activity description | Roles |
| --- | --- | --- | --- |
| Preparation | IEEE1028-2008 6.5.5 1 | Each inspection team member shall examine the software product and other inputs prior to the review meeting. | Each inspection team member |
| | IEEE1028-2008 6.5.5 2 | Anomalies detected during this examination shall be documented and sent to the inspection leader. | Each inspection team member |
| | IEEE1028-2008 6.5.5 3 | The inspection leader should classify anomalies as described in 6.8.1 to determine whether they warrant cancellation of the inspection meeting, and in order to plan efficient use of time in the inspection meeting. If the inspection leader determines that the extent or seriousness of the anomalies warrants, the inspection leader may cancel the inspection, requesting a later inspection when the software product meets the minimal entry criteria and is reasonably defect-free. The inspection leader should forward the anomalies to the author of the software product for disposition. | Inspection leader |
| | IEEE1028-2008 6.5.5 4 | The inspection leader or reader shall specify a suitable order in which the software product will be inspected (such as sequential, hierarchical, data flow, control flow, bottom up, or top down). | Inspection leader or reader |
| | IEEE1028-2008 6.5.5 5 | The reader(s) shall prepare sufficiently to be able to present the software product at the inspection meeting. | Reader |
| | IEEE1028-2008 6.5.5 6 | The inspection leader shall verify that inspectors are prepared for the inspection. The inspection leader shall reschedule the meeting if the inspectors are not adequately prepared. The inspection leader should gather individual preparation times and record the total in the inspection documentation. | Inspection leader |
| Examination | IEEE1028-2008 6.5.6.1 (title) | Introduce meeting | Inspection leader |
| | IEEE1028-2008 6.5.6.2(title) | Review general items | |
| | IEEE1028-2008 6.5.6.3 (title) | Review software product and record anomalies | |
| | IEEE1028-2008 6.5.6.4 (title) | Review the anomaly list | |
| | IEEE1028-2008 6.5.6.5 (title) | Make exit decision | Inspection team |
| Rework/Follow-up | IEEE1028-2008 6.5.7 | The inspection leader shall verify that the action items assigned in the meeting are closed. | Inspection leader |

Process elements created based on IEEE 1028 element instances

# Appendix N. Process elements created based on Process Impact element instances

Table 52 – Process Impact process element instances and their relations

| Parent | Item id | TO | Process/activity description | Input | Output | Roles |
|---|---|---|---|---|---|---|
| Planning | pi11 | 1 | Give moderator the work product to be inspected and supporting documents, such as specifications, predecessor documents, or pertinent test documentation. | work product, supporting documents | | Moderator |
| | pi12 | 2 | Determine whether work product satisfies inspection entry criteria. | work product, supporting documents, inspection entry criteria | decision Y/N (work product satisfies the input criteria) | Author |
| | pi13 | 3 | Based on the size and complexity of the work product, determine how many inspection meetings will be required. | work product size, complexity | number of inspection meetings | Moderator and Author |
| | pi14 | 4 | Select inspectors and assign roles to individuals. Gain agreement from the other inspectors to participate. | possible inspection participants | inspection participants | Moderator and Author |
| | pi15 | 5 | Determine whether an overview meeting is required. | | decision Y/N (overview meting) | Author |
| | pi16 | 6 | Schedule the inspection, and possibly overview, meetings and distribute a meeting notice. | | inspection schedule, meeting notice [distributed] | Moderator |
| | pi17 | 7 | Distribute the inspection package to the participants at least 3 working days prior to the inspection meeting. | inspection package | inspection package [distirbuted] | Moderator or Author |
| Overview | pi21 | 1 | Describe the important features of the work product to the rest of the inspection team. State inspection objectives. | | description of important features of the work product, inspection objectives | Author |
| | pi22 | 2 | Evaluate the assumptions, history, and context of the work product. | | | Inspectors |
| Preparation | pi31 | 1 | Ask individual inspectors to prepare with specific objectives in mind, such as: checking the consistency of cross-references; checking for interface errors; checking traceability to, and consistency with, predecessor specifications; or checking conformance to standards. | | | Moderator and Author |

| Parent | Item id | TO | Process/activity description | Input | Output | Roles |
|---|---|---|---|---|---|---|
| | pi32 | 2 | Examine the work product, to understand it, find defects, and raise questions about it. Use the appropriate defect checklist to focus attention on defects commonly found in the type of product being inspected. Use other analysis methods to look for defects as appropriate. | defect checklist | | Inspectors |
| | pi33 | 3 | Log minor defects found, such as typographical errors or style inconsistencies, on the Typo List. Deliver this to the author at or prior to the inspection meeting. | typo list template | minor defects found, typos [delivered to author] | Inspectors |
| Inspection Meeting | pi41 | 1 | **Open the Meeting:** Introduce the participants (if necessary) and state their roles, state the purpose of the inspection, and direct inspectors to focus their efforts toward finding defects, not solutions. Remind participants to address their comments to the work product under review, not to the author. | | | Moderator |
| | pi42 | 2 | **Establish Preparedness:** Ask each inspector for his preparation time and record the times on the Inspection Summary Report. If preparation is insufficient, reschedule the meeting. | | preparation times [inspection summary report] | Moderator |
| | pi43 | 3 | **Present Work Product:** Describe portions of the work product to the inspection team. | | | Reader |
| | pi44 | 4 | **Raise Defects and Issues:** Point out concerns, potential defects, questions, or improvement opportunities after the reader presents each section. | | | Inspectors |
| | pi45 | 5 | **Record Issues:** Capture the information in Table 2 on the Issue Log for each issue raised. State aloud what was recorded to make sure it was recorded accurately | Issue log | Issue log | Recorder |
| | pi46 | 6 | **Answer Questions:** Respond briefly to any specific questions raised, and contribute to defect detection based on special understanding of the work product. | | | Author |
| | pi47 | 7 | **Make Product Appraisal:** After all meetings scheduled for a given inspection are complete, decide on the work product appraisal, selecting from the options in Table 3. If the inspectors disagree, assign the most conservative appraisal offered by any of the inspectors. | | decision about wp acceptance [in inspection summary report] | Inspectors |

Process elements created based on Process Impact element instances

| Parent | Item id | TO | Process/activity description | Input | Output | Roles |
|---|---|---|---|---|---|---|
| | pi48 | 8 | **Sign Inspection Summary Report:** All participants sign the Inspection Summary Report to indicate their agreement with the inspection outcome. | | sign [in inspection summary report] | Inspectors |
| | pi49 | 9 | **Collect Inspection Feedback.** Ask the inspectors to evaluate the inspection and suggest improvements, using the Inspection Lessons Learned Questionnaire. | Inspection Lessons Learned Questionnaire | Inspection Lessons Learned Questionnaire [filled] | Moderator |
| Rework | pi51 | 1 | Correct defects and typos found, resolve issues raised, and modify work product accordingly. Mark issues list to indicate action taken. | | work product [corrected] | Author |
| | pi52 | 2 | Correct any other project documents based on defects identified in the inspected work product. | | project documents [corrected] | Author |
| | pi53 | 3 | Record any uncorrected defects in the project's defect tracking system. | | list of uncorrected defects [recorded in defect tracking system] | Author |
| | pi54 | 4 | If rework verification is not needed, report the number of major and minor defects found and corrected and the actual rework effort to the moderator. | | major and minor defects found, actual rework effort, + decision! | Author |
| | pi55 | 5 | Record the actual rework effort on the Inspection Summary Report. | Inspection Summary Report template | rework effort [in Inspection Summary Report] | Moderator |
| Follow-Up | pi61 | 1 | Confirm that the author has addressed every item on the Issue Log. Determine whether the author made appropriate decisions as to which defects not to correct and which improvement suggestions not to implement. | Issue log | [eg. In follow-Up report] | Verifier |
| | pi62 | 2 | Examine the modified work product to judge whether the rework has been performed correctly. Report any findings to the author, so rework can be declared complete, incorrect rework can be redone, or items that were not originally pursued can be addressed. | modified wp | findings [eg in follow-ip report] | Verifier -> Author |
| | pi63 | 3 | Report the number of major and minor defects found and corrected and the actual rework effort to the moderator. | | major and minor defects found, actual rework effort [report] | Author -> Moderator |

| Parent | Item id | TO | Process/activity description | Input | Output | Roles |
| --- | --- | --- | --- | --- | --- | --- |
| | pi64 | 4 | Check whether the exit criteria for the inspection and for the peer review process have been satisfied. If so, the inspection is complete. | exit criteria for inspection and peer review process | decision: inspection complete or not | Moderator |
| | pi65 | 5 | Check the baselined work product into the project's configuration management system. | | | Author |
| | pi66 | 6 | Deliver Inspection Summary Report and counts of defects found and defects corrected to peer review coordinator. | | Inspection summary report | Author -> Peer review coordinator |

Table 53 – Entry and exit criteria in Process Impact's Peer review Descriptions

| Entry/exit criteria | Item ID | Criteria description |
| --- | --- | --- |
| Entry criteria | PI EnC1 | The author selected an inspection approach for the product being reviewed. |
| Entry criteria | PI EnC3 | The author has stated his objectives for this inspection. |
| Entry criteria | PI EnC4 | Reviewers are trained in the peer review process. |
| Entry criteria | PI EnC5 | Documents to be inspected are identified with a version number. All pages are numbered and line numbers are displayed. The documents have been spell-checked. |
| Entry criteria | PI EnC6 | Source code to be inspected is identified with a version number. Listings have line numbers and page numbers. Code compiles with no errors or warning messages using the project's standard compiler switches. Errors found using code analyzer tools have been corrected. |
| Exit criteria | PI ExC2 | Issues raised during the inspection are tracked to closure. |
| Exit criteria | PI ExC3 | All major defects are corrected. |
| Exit criteria | PI ExC4 | Uncorrected defects are logged in the project's defect tracking system. |
| Exit criteria | PI ExC5 | The modified work product is checked into the project's configuration management system. |
| Exit criteria | PI ExC6 | If changes were required in earlier project deliverables, those deliverables have been correctly modified, checked into the project's configuration management system, and any necessary regression tests were passed. |
| Exit criteria | PI ExC7 | Moderator has collected and recorded the inspection data. |
| Exit criteria | PI ExC8 | Moderator has delivered the completed Inspection Summary Report and defect counts to the peer review coordinator. |

# Appendix O.A peer review process in BPMN based on CMMI

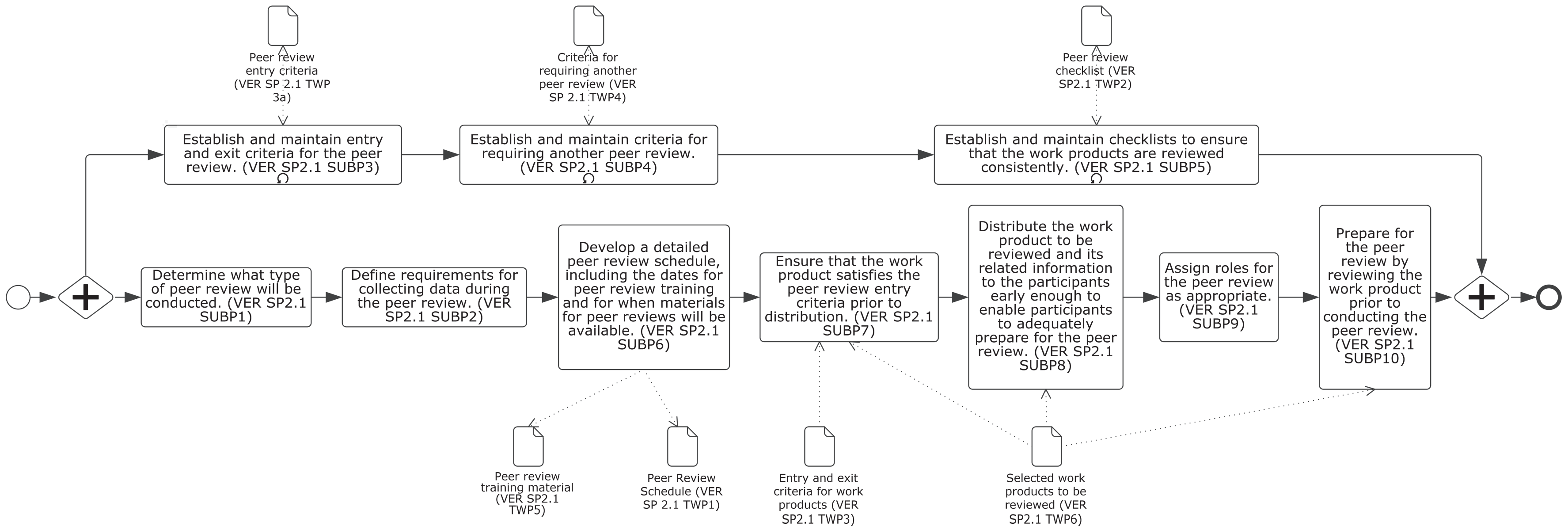

Figure 47 – Prepare for peer review

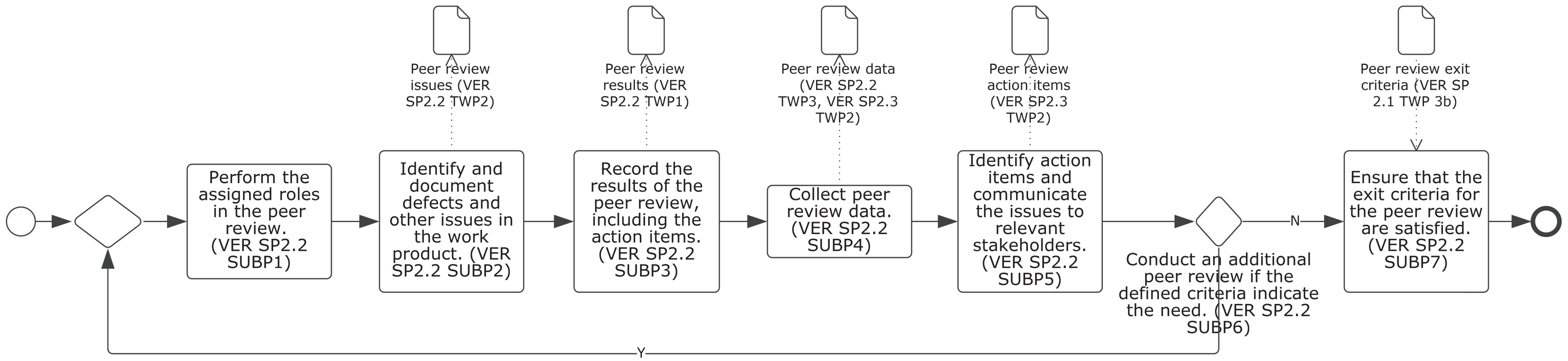

Figure 48 – Conduct peer reviews

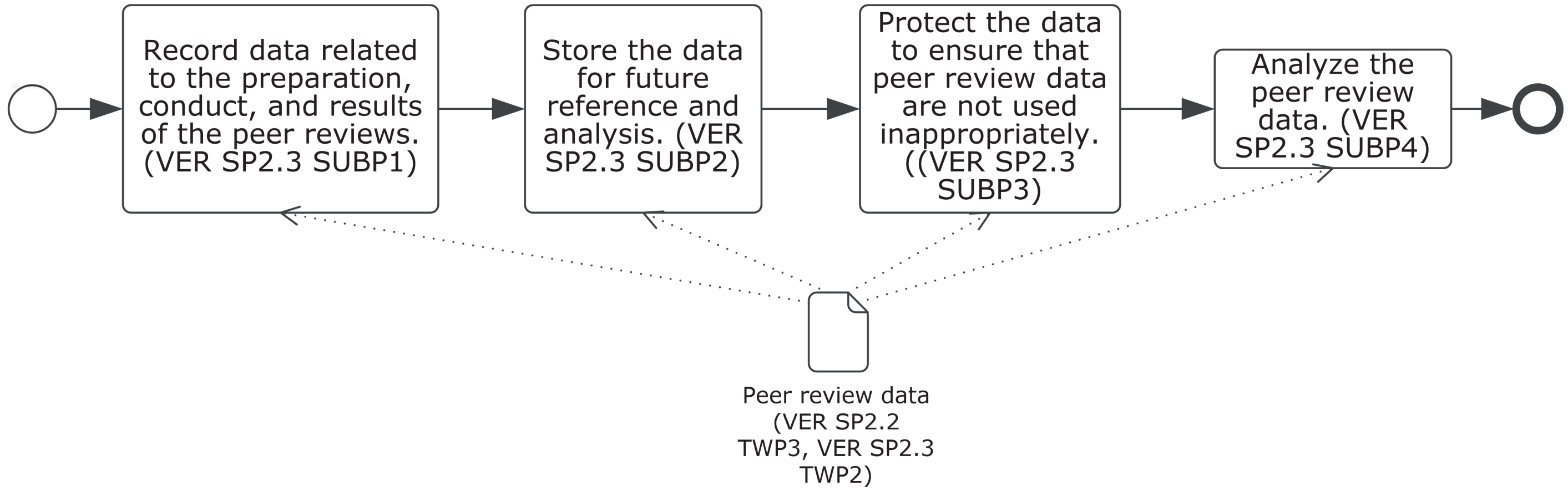

Figure 49 – Analyze peer review data

# Appendix P. An inspection process in BPMN based on IEEE 1028-2008

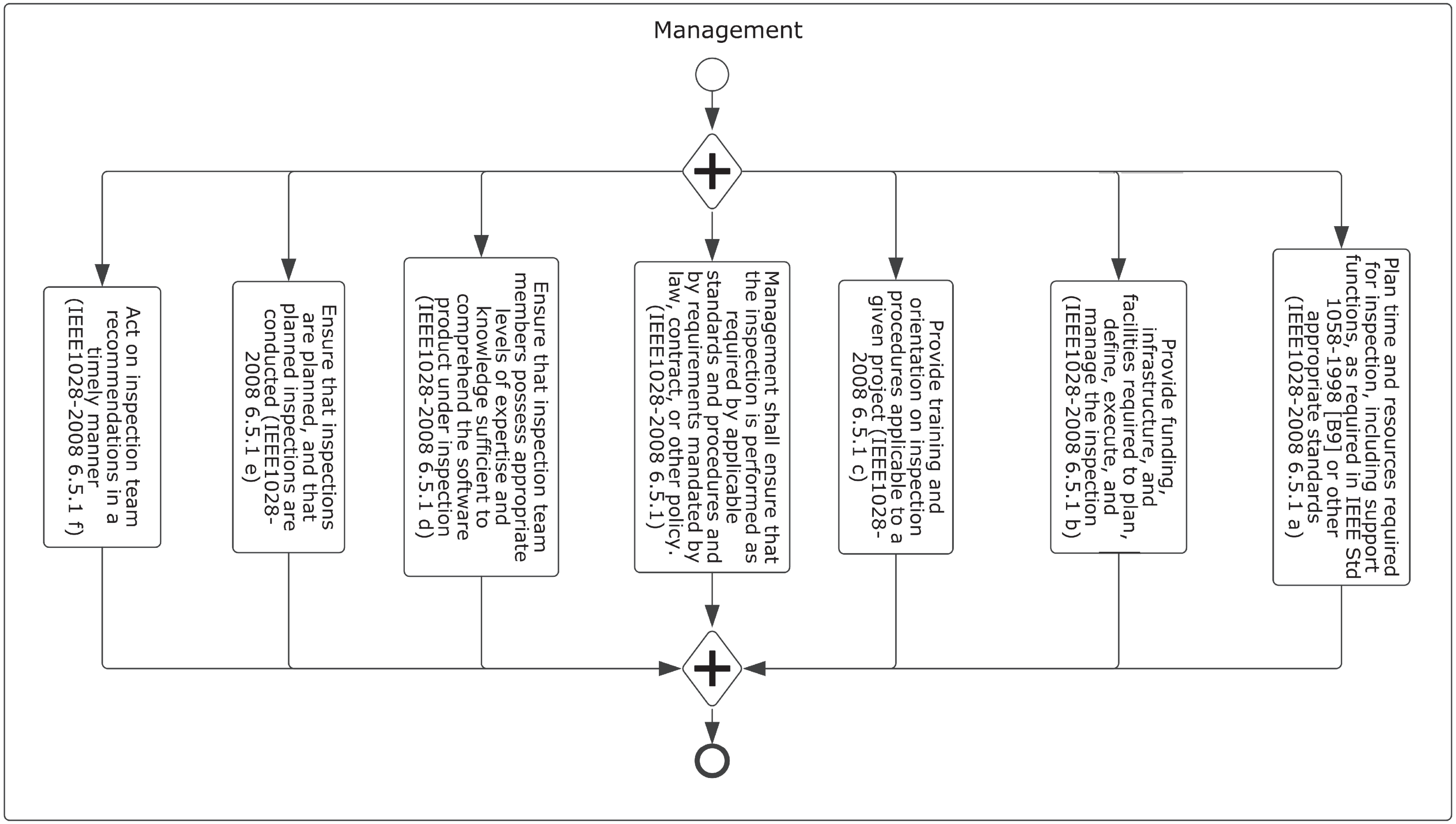

Figure 50 – Management preparation

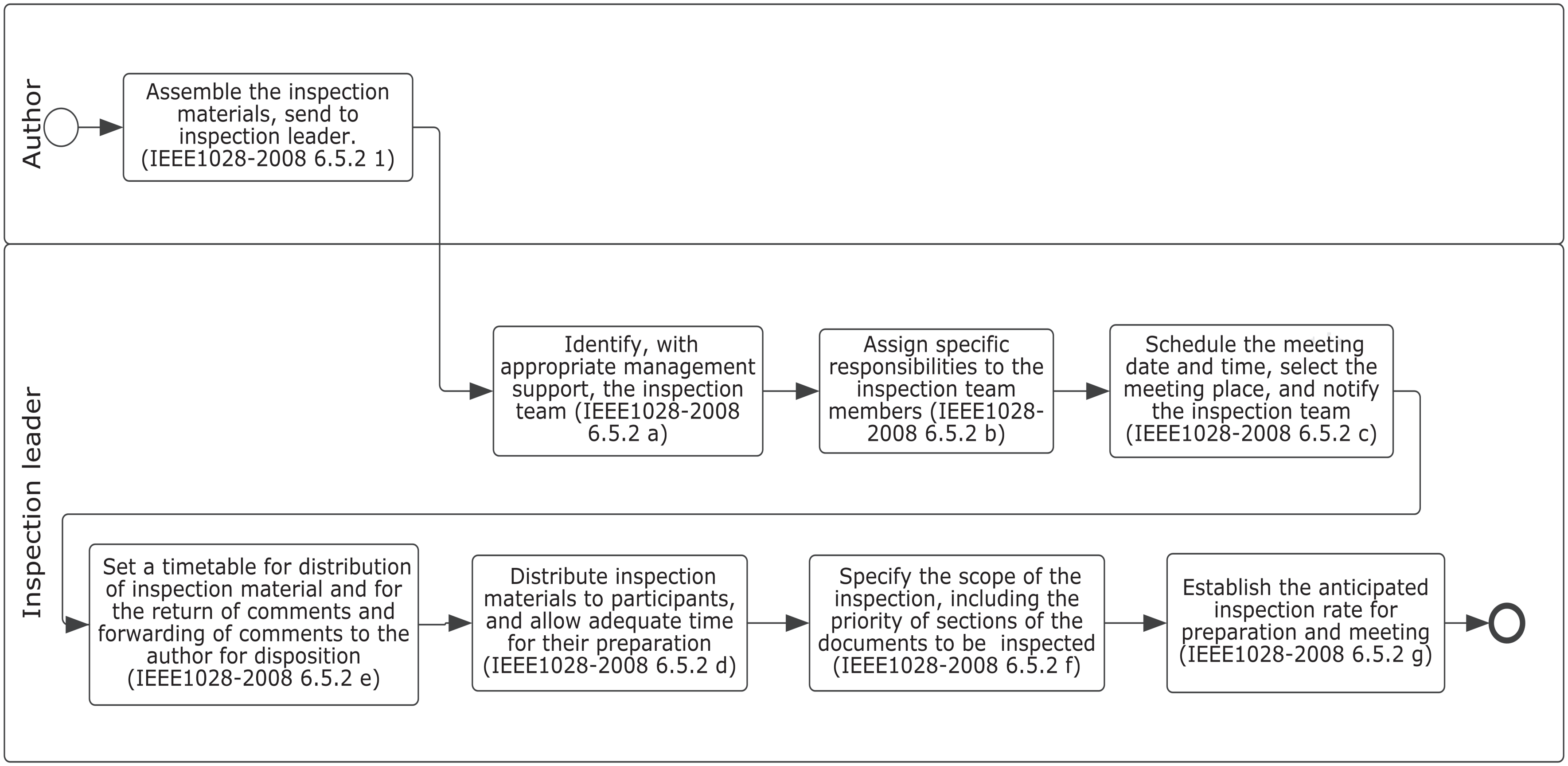

Figure 51 – Planning the inspection

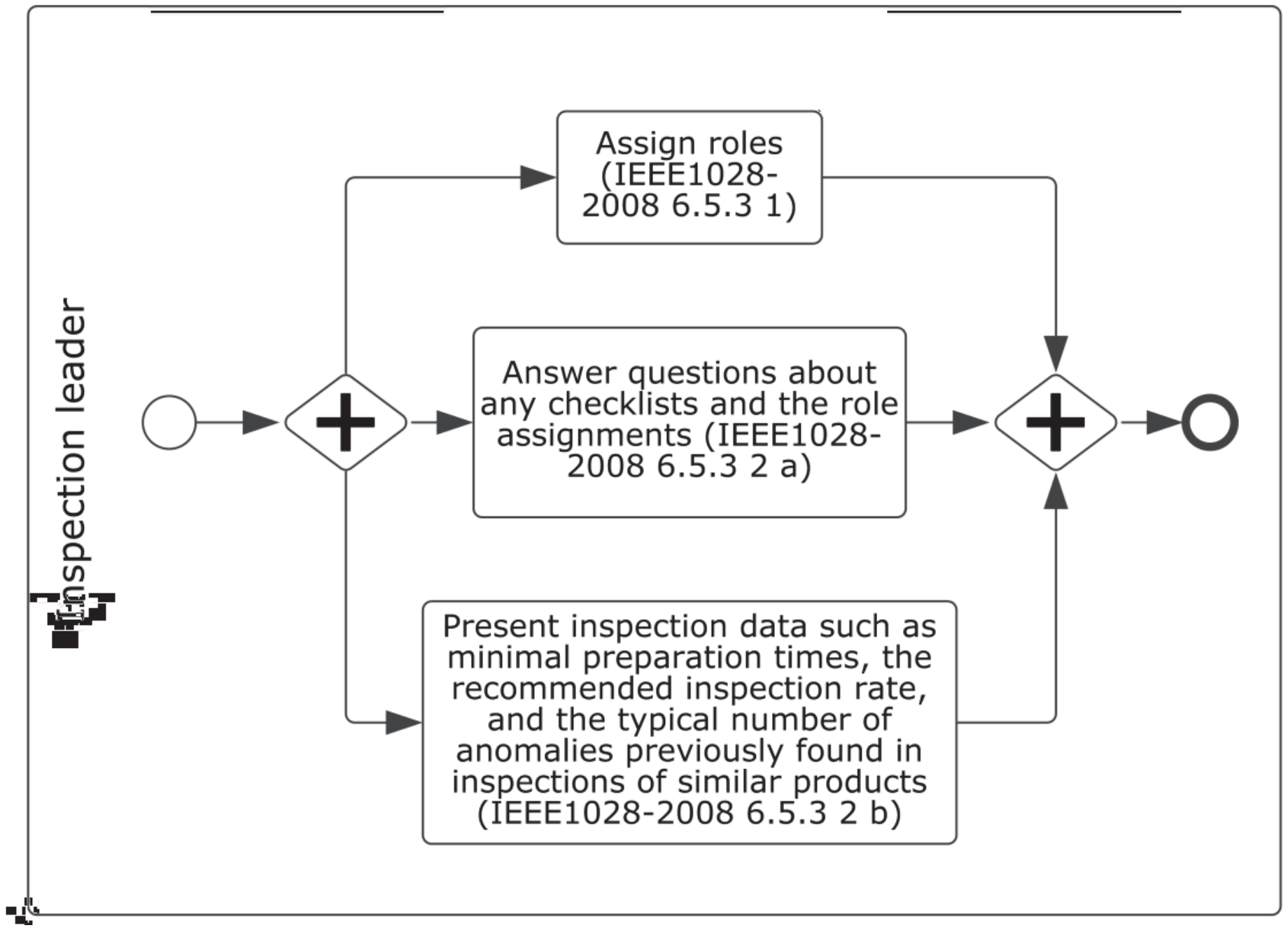

Figure 52 – Overview of the inspection procedures

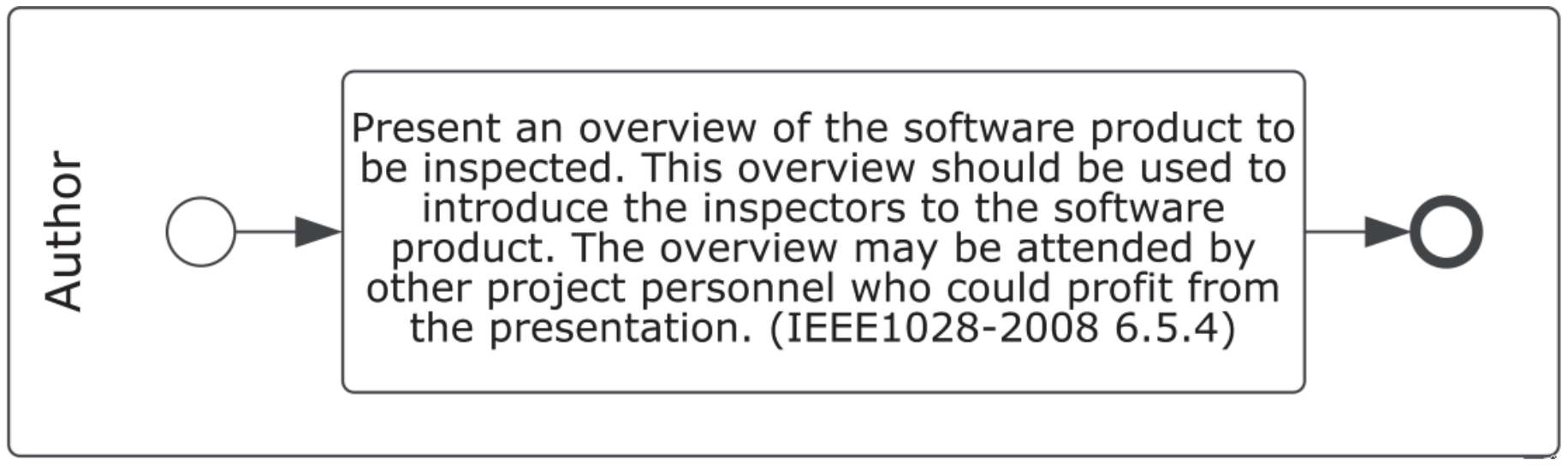

Figure 53 – Overview of the inspection product

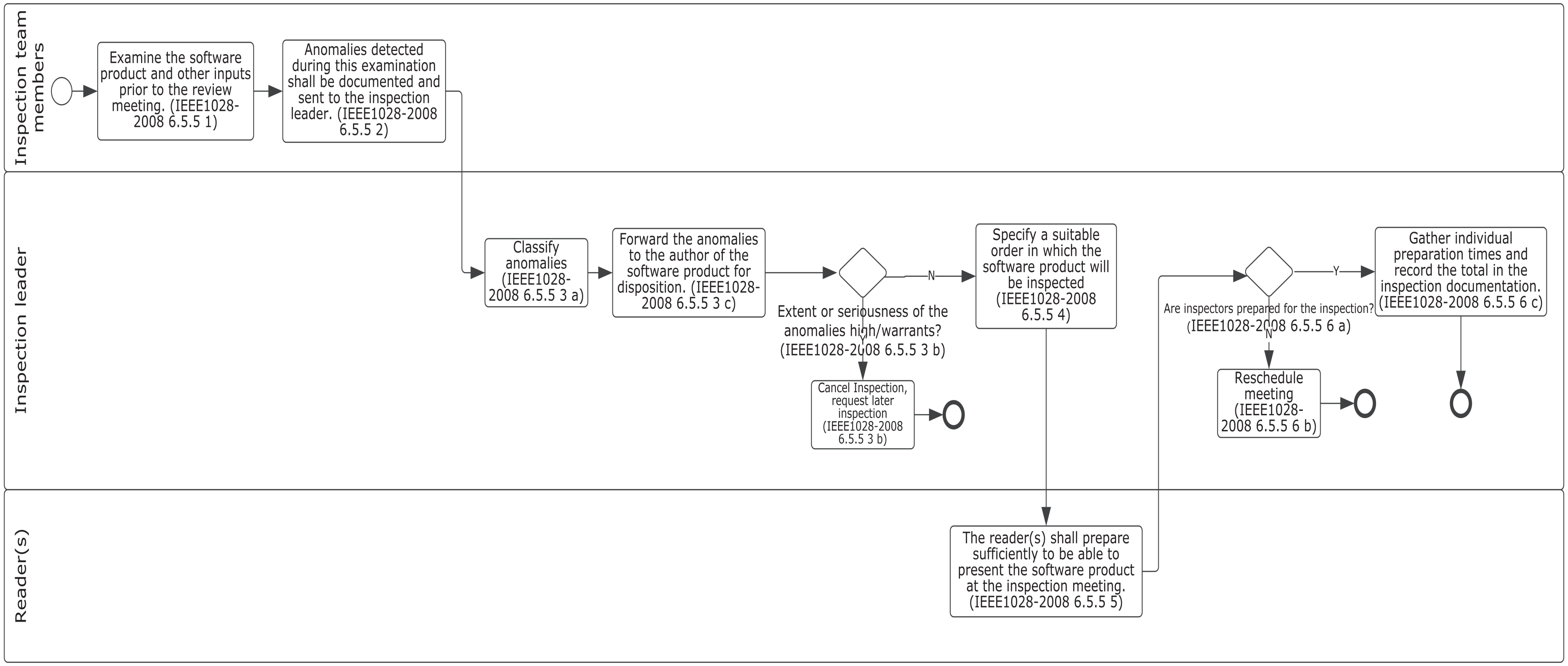

Figure 54 – Preparation

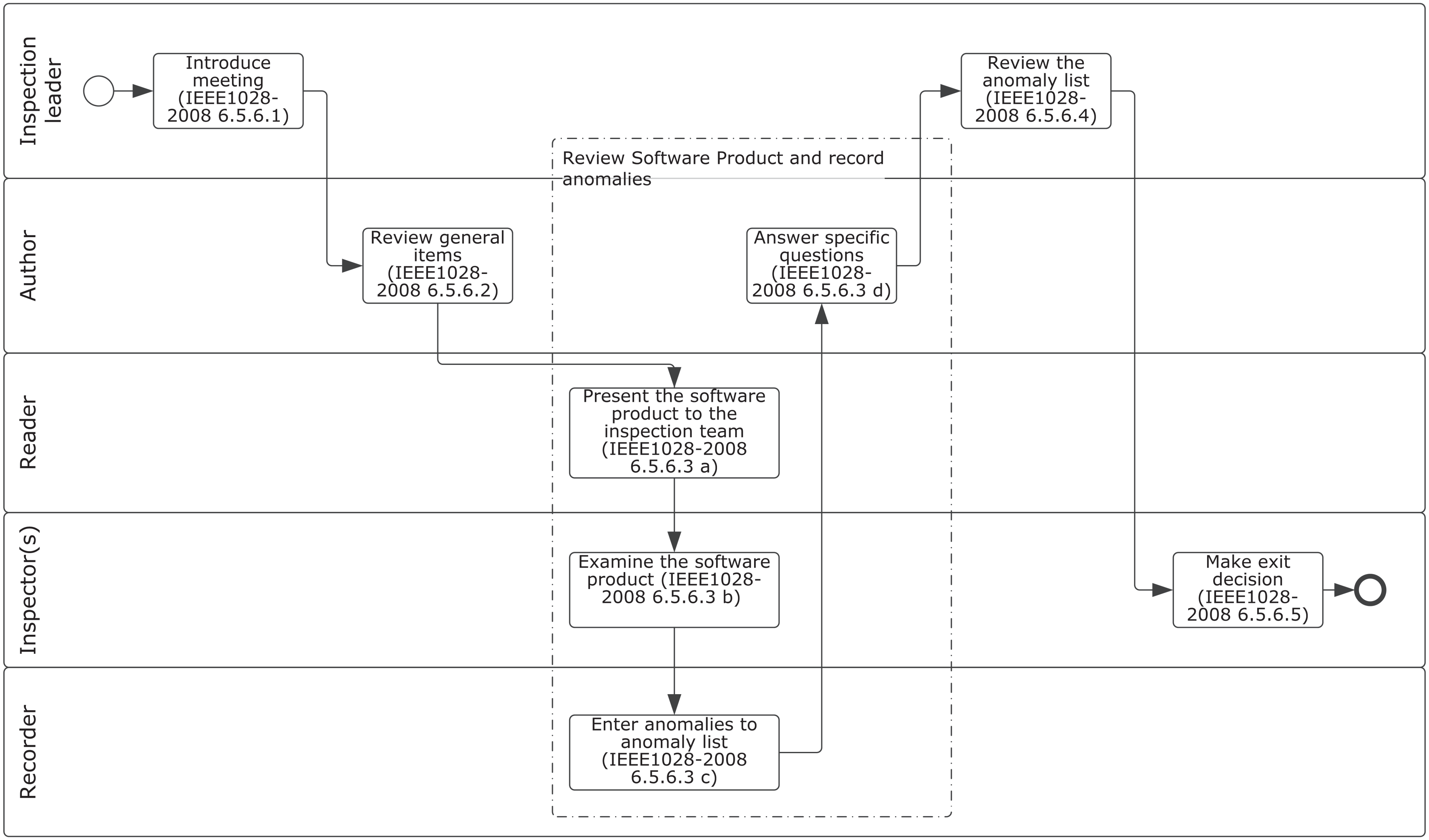

Figure 55 – Examination

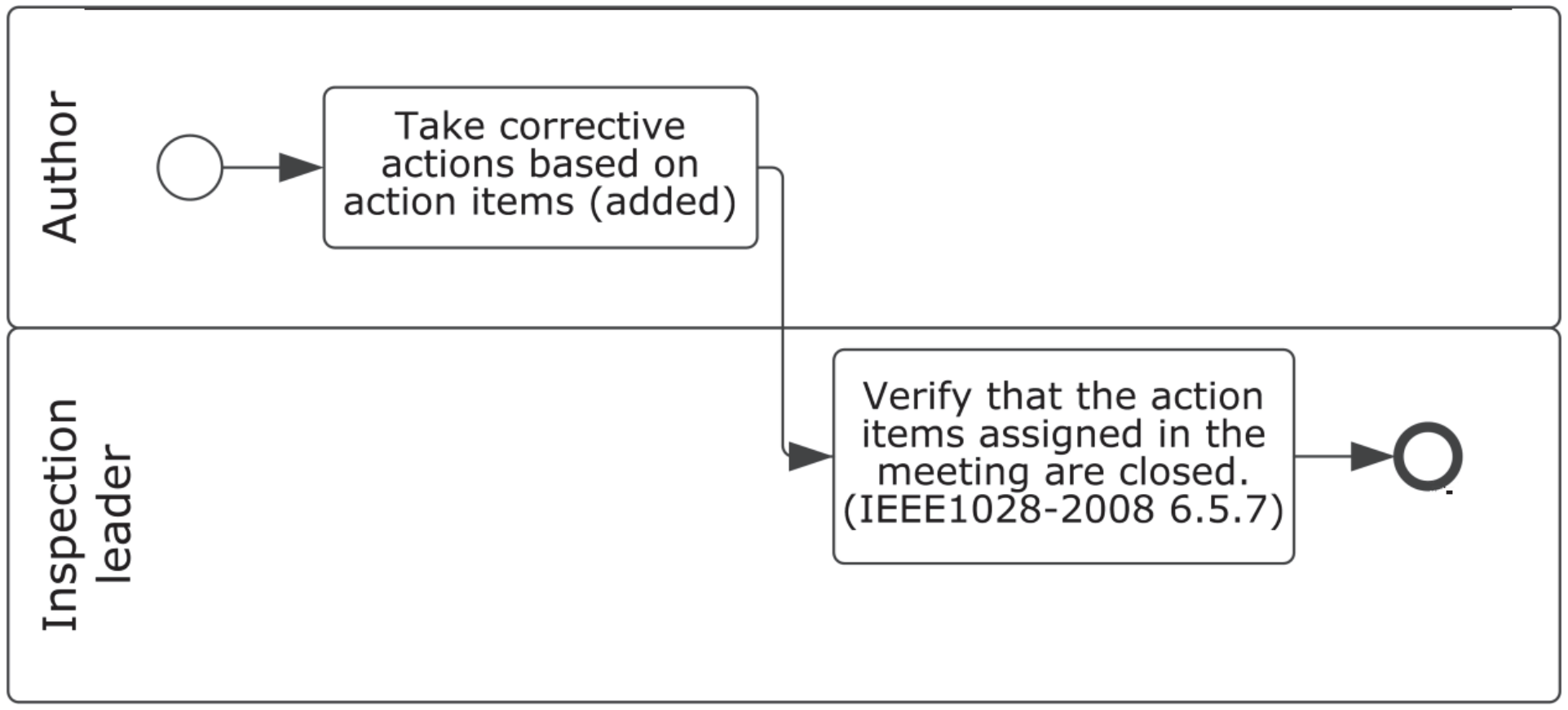

Figure 56 – Rework/follow-up

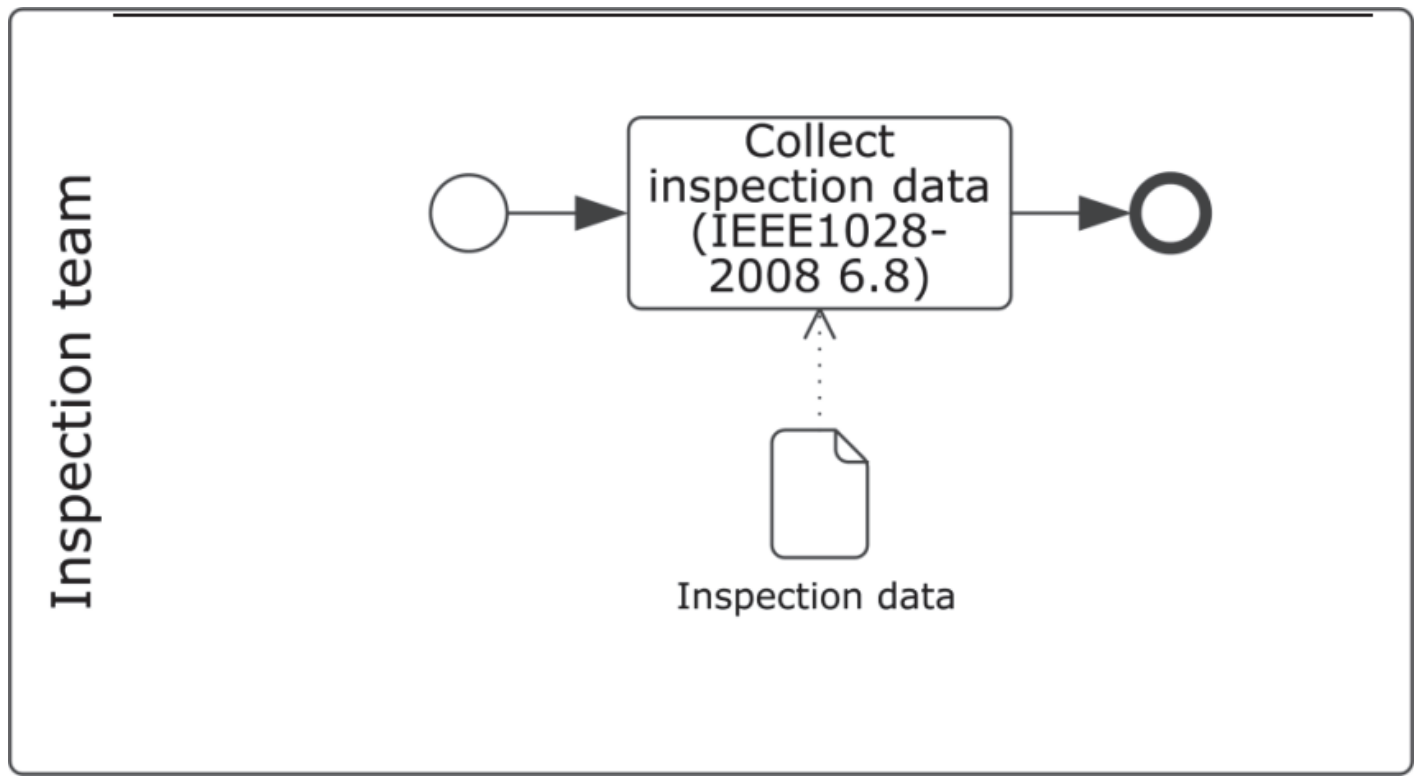

Figure 57 – Data collection

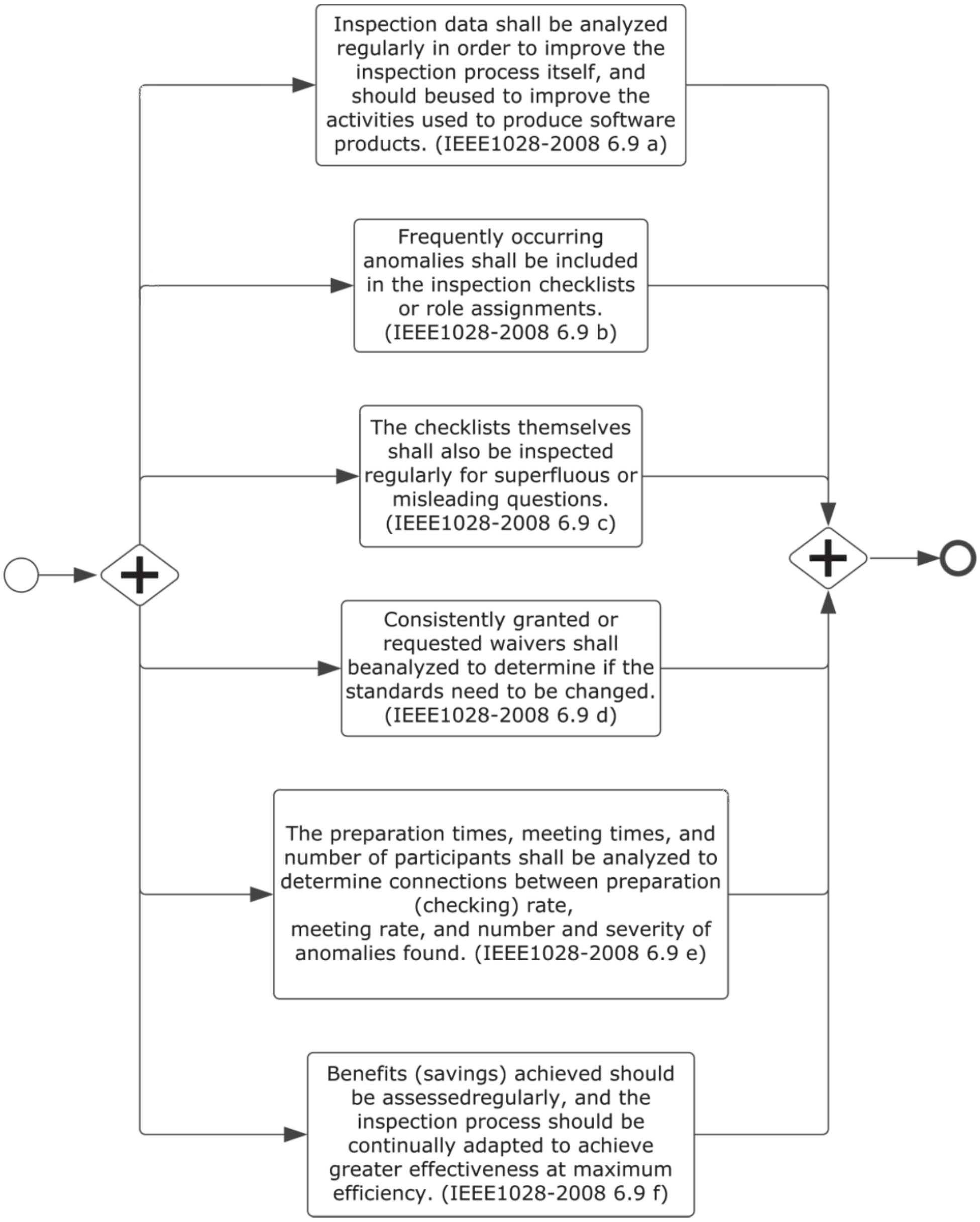

Figure 58 – Improvement

# Appendix Q. A peer review process in BPMN based on Process Impact

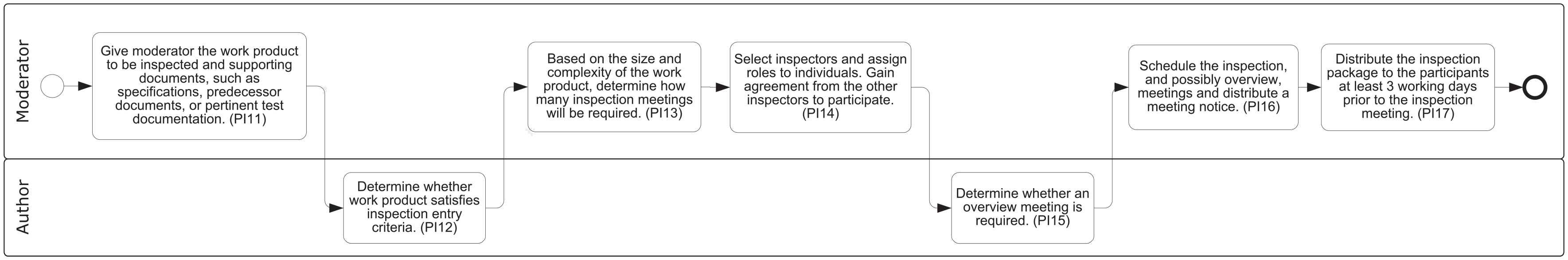

Figure 59 – Planning

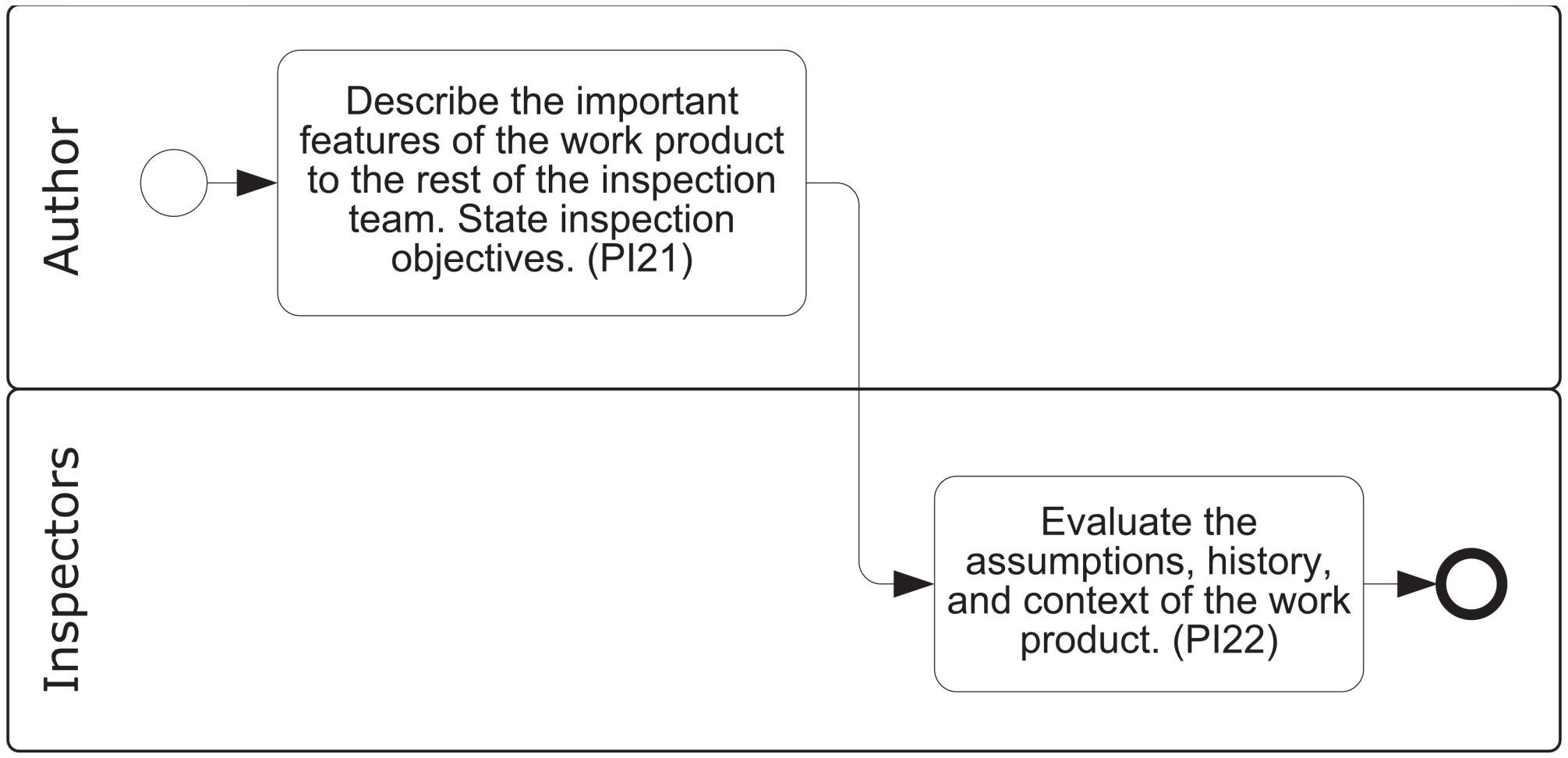

Figure 60 – Overview

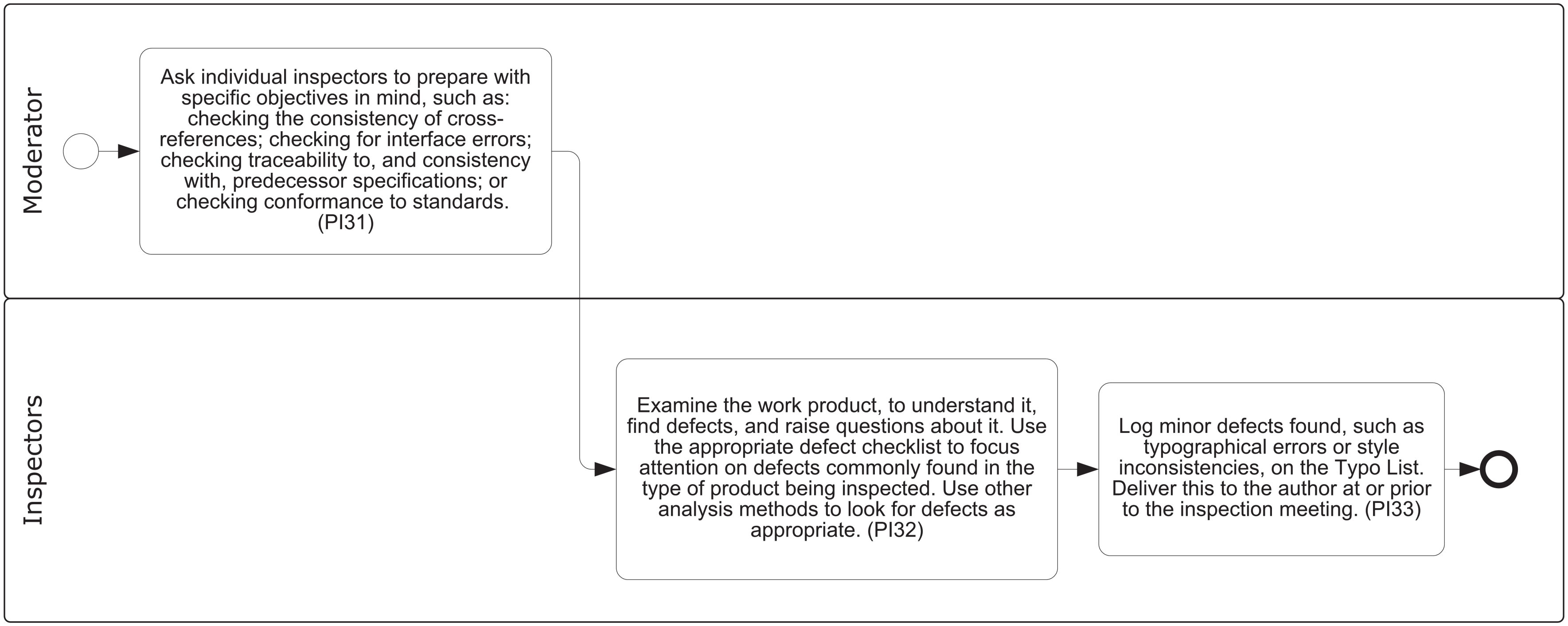

Figure 61 – Preparation

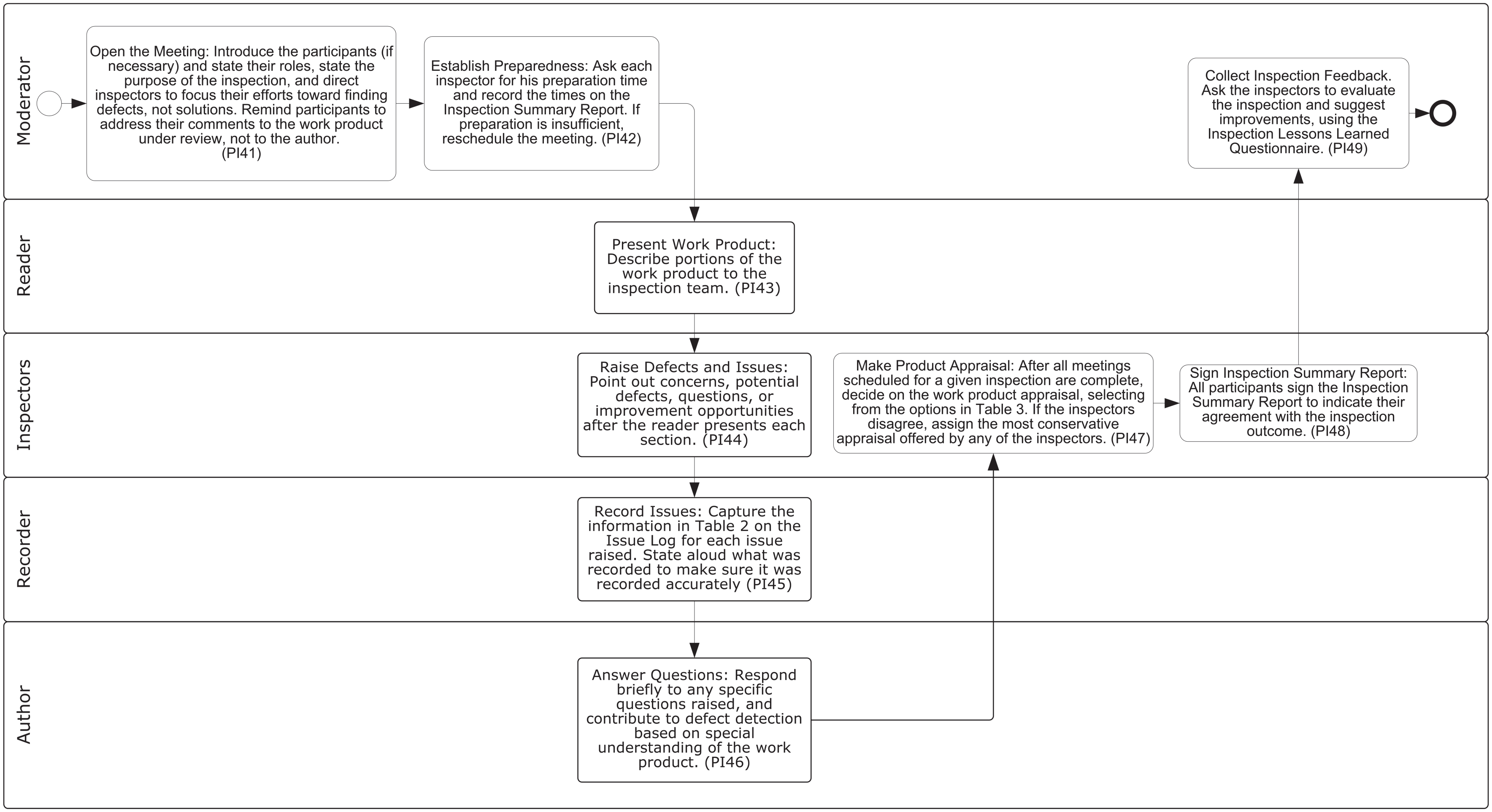

Figure 62 – Inspection meeting

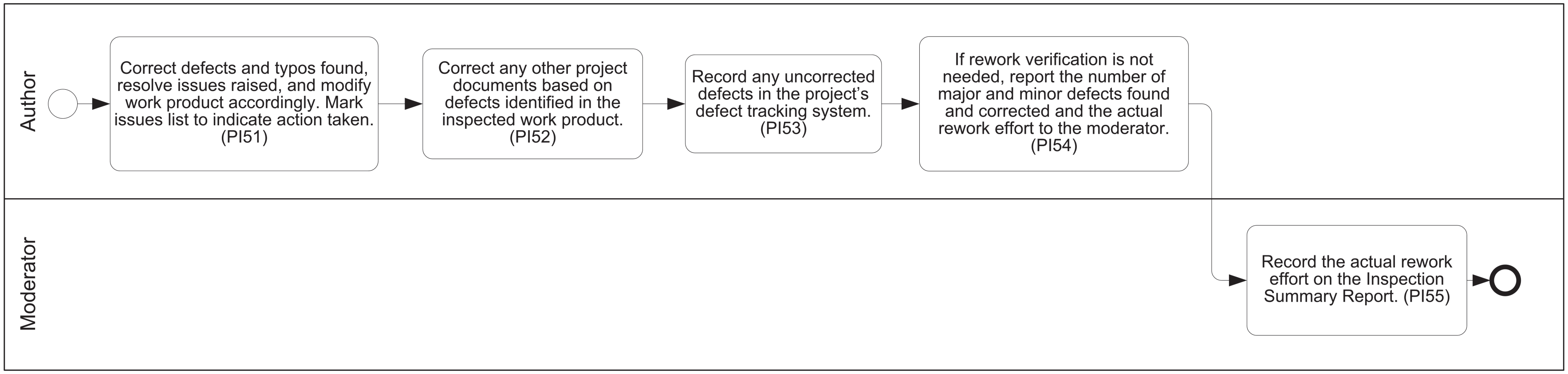

Figure 63 – Rework

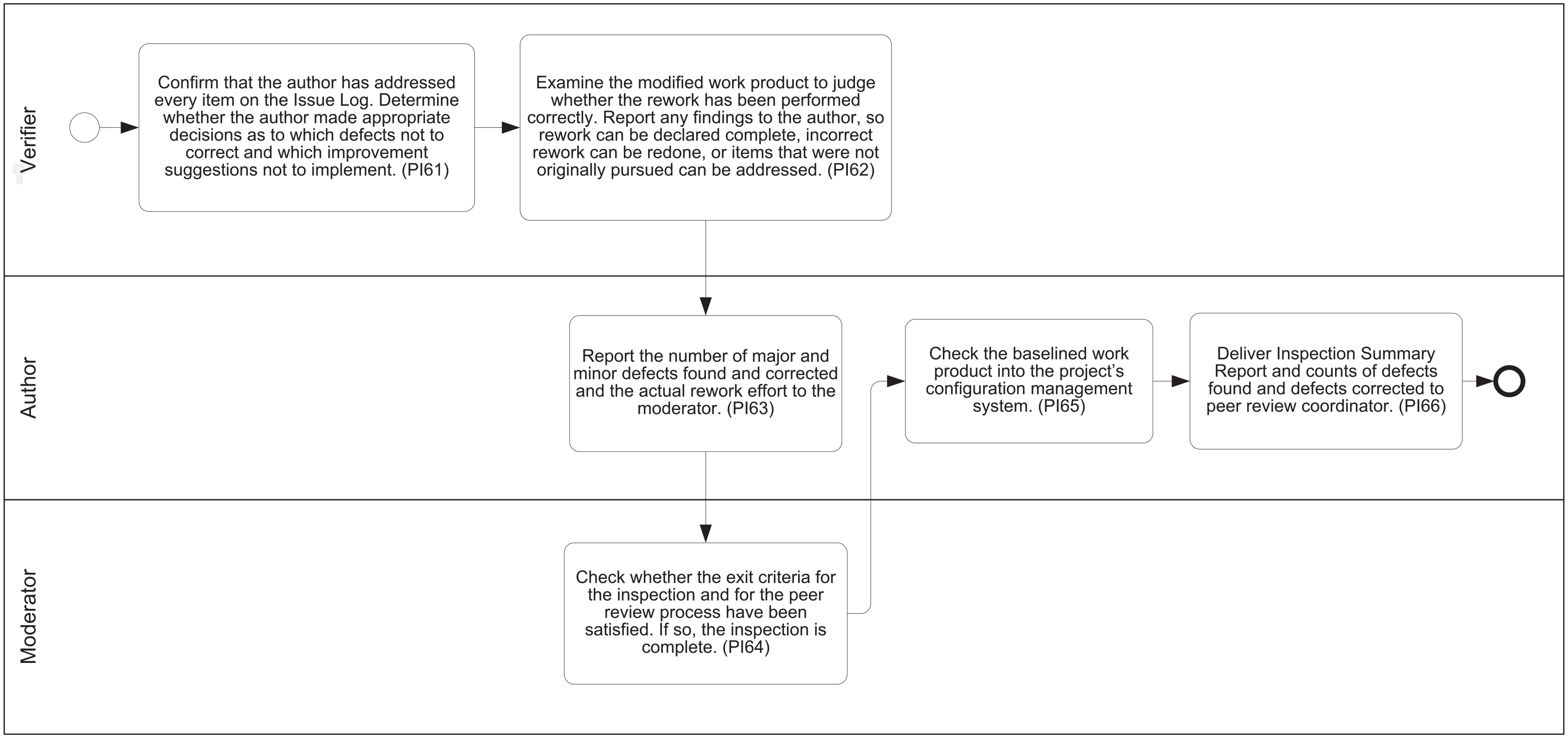

Figure 64 – Follow-up

# Appendix R. Mapping of the unified process to source quality approaches

Table 54 – Subprocesses of a unified peer review process

| Subprocess name | Documentation |
|---|---|
| Verify that the action items assigned in the meeting are closed (IEEE 1028-2008 6.5.7) | IEEE1028-2008 6.5.7: The inspection leader shall verify that the action items assigned in the meeting are closed.<br><br>PI6: the Follow-Up subprocess |
| Management Preparation (IEEE 1028-2008 6.5.1) | IEEE 1028-2008 6.5.1: Management Preparation |
| Improvement (IEEE 1028-2008 6.9) | IEEE 1028-2008 6.9: Improvement |
| Data Collection (IEEE 1028-2008 6.8) | IEEE 1028-2008 6.8: Data Collection |
| Overview of inspection product (IEEE1028-2008 6.5.4) | IEEE1028-2008 6.5.4: The author should present an overview of the software product to be inspected. This overview should be used to introduce the inspectors to the software product. The overview may be attended by other project personnel who could profit from the presentation.<br><br>PI21: Describe the important features of the work product to the rest of the inspection team. State inspection objectives.<br><br>PI22: Evaluate the assumptions, history, and context of the work product. |
| Overview of inspection procedures (IEEE 1028-2008 6.5.3) | IEEE 1028-2008 6.5.3: Overview of inspection procedures |
| Rework and follow-up (IEEE 1028-2008 6.5.7) | IEEE 1028-2008 6.5.7: Rework and follow-up |
| Take corrective actions based on action items (added, PI5) | PI5 subprocess<br><br>PI51: Correct defects and typos found, resolve issues raised, and modify work product accordingly. Mark issues list to indicate action taken.<br>PI52: Correct any other project documents based on defects identified in the inspected work product. |
| Perform other type of peer review | |
| Perform Inspection (IEEE 1028-2008 6) | IEEE 1028-2008 6. Inspections |
| Examination (IEEE 1028-2008 6.5.6) | IEEE 1028-2008 6.5.6: Examination |

| Subprocess name | Documentation |
|---|---|
| Planning the inspection (IEEE 1028-2008 6.5.1) | IEEE 1028-2008 6.5.2: Planning the inspection |
| Preparation (IEEE 1028-2008 6.5.5) | IEEE 1028-2008 6.5.5 Preparation |

Table 55 – Activities and their mapping to quality approach element instances in a unified peer review process

| Activity (task) name | Documentation |
|---|---|
| **Data collection <Process Level>** | |
| Collect inspection data<br>(IEEE 6.8) | IEEE 6.8. Data Collection:<br>Inspections shall provide data for the analysis of the quality of the software product, the effectiveness of the acquisition, supply, development, operation and maintenance processes, and the effectiveness and the efficiency of the inspection itself. In order to maintain the effectiveness of inspections, data from the author and inspectors shall not be used to evaluate the performance of individuals. To enable these analyses, anomalies that are identified at an inspection meeting shall be classified in accordance with 6.8.1, 6.8.2, and 6.8.3.<br>Inspection data shall contain the identification of the software product, the date and time of the inspection, the inspection team, the preparation and inspection times, the volume of the materials inspected, and the disposition of the inspected software product. The capture of this information shall be used to optimize local guidance for inspections.<br>The management of inspection data requires a capability to enter, store, access, update, summarize, and report classified anomalies.<br>The frequency and types of the inspection analysis reports, and their distribution, are left to local standards and procedures.<br><br>CMMI-DEV v1.2:<br>VER SP2.2 SUBP4 Collect peer review data.<br>VER SP2.3 SUBP1 Record data related to the preparation, conduct, and results of the peer reviews.<br>VER SP2.3 SUBP3 Protect the data to ensure that peer review data are not used inappropriately.<br>Typical data are product name, product size, composition of the peer review team, type of peer review, preparation time per reviewer, length of the review meeting, number of defects found, type and origin of defect, and so on. Additional information on the work product being peer reviewed may be collected, such as size, development stage, operating modes examined, and requirements being evaluated.<br>VER SP3.2 SUBP2 Store the data for future reference and analysis. |

| Activity (task) name | Documentation |
| --- | --- |
| | Refer to the Measurement and Analysis (MA) (CMMI-DEV) process area for more information about data collection.<br>VER.GP 2.6 Manage Configurations<br>Place designated work products of the Verification (VER) (CMMI-DEV) process under appropriate levels of control.<br>Elaboration:<br>Examples of work products placed under control include the following:<br>* Verification procedures and criteria<br>* Peer review training material<br>* Peer review data<br>* Verification reports |
| **Examination <Process Level>** | |
| Sign Inspection Summary Report (PI48) | PI48: Sign Inspection Summary Report: All participants sign the Inspection Summary Report to indicate their agreement with the inspection outcome. |
| Collect Inspection Feedback. (PI49) | PI49: Collect Inspection Feedback. Ask the inspectors to evaluate the inspection and suggest improvements, using the Inspection Lessons Learned Questionnaire. |
| Record the results of the peer review, including the action items.<br>(VER SP2.2 SUBP3) | VER SP2.2 SUBP3: Record the results of the peer review, including the action items. |
| Establish Preparedness (PI42) | PI42: Establish Preparedness: Ask each inspector for his preparation time and record the times on the Inspection Summary Report. If preparation is insufficient, reschedule the meeting. |
| Make exit decision (IEEE1028-2008 6.5.6.5) | IEEE1028-2008 6.5.6.5: Make exit decision<br>The purpose of the exit decision is to bring an unambiguous closure to the inspection meeting. The exit decision shall determine if the software product meets the inspection exit and quality criteria. As part of this decision, any appropriate rework and verification shall be prescribed. Specifically, the inspection team shall identify the software product disposition as one of the following:<br>Accept with no verification or with rework verification. The software product is accepted as is or with only minor rework (for example, that would require no further verification).<br>Accept with rework verification. The software product is to be accepted after the inspection leader or a designated member of the inspection team (other than the author) verifies rework.<br>Reinspect. The software product cannot be accepted. Once anomalies have been resolved a reinspection should be scheduled to verify rework. At a minimum, a reinspection shall examine the |

| Activity (task) name | Documentation |
|---|---|
| | software product areas changed to resolve anomalies identified in the last inspection, as well as side effects of those changes.<br><br>VER SP 2.2 SUBP 7: Ensure that the exit criteria for the peer review are satisfied.<br><br>PI47: Make Product Appraisal: After all meetings scheduled for a given inspection are complete, decide on the work product appraisal, selecting from the options in Table 3. If the inspectors disagree, assign the most conservative appraisal offered by any of the inspectors. |
| Review the anomaly list (IEEE1028-2008 6.5.6.4) | IEEE1028-2008 6.5.6.4: Review the anomaly list<br>At the end of the inspection meeting, the inspection leader shall have the anomaly list reviewed with the team to ensure its completeness and accuracy. The inspection leader shall allow time to discuss every anomaly when disagreement occurred. The inspection leader shall not allow the discussion to focus on resolving the anomaly but on clarifying what constitutes the anomaly. If a disagreement as to the existence or severity of an anomaly cannot be quickly resolved during the meeting, that disagreement shall be documented in the anomaly report. |
| Introduce meeting (IEEE 6.5.6.1) | IEEE 6.5.6.1 Introduce meeting<br>The inspection leader shall introduce the participants and describe their roles. The inspection leader shall state the purpose of the inspection and should remind the inspectors to focus their efforts toward anomaly detection, not resolution. The inspection leader shall remind the inspectors to direct their remarks to the recorder and to comment only on the software product, not its author. Inspectors may pose questions to the author regarding the software product. The inspection leader shall resolve any special procedural questions raised by the inspectors. Extensive discussion about issues should be postponed to the end of the meeting or to a separate meeting.<br><br>PI41: Open the Meeting: Introduce the participants (if necessary) and state their roles, state the purpose of the inspection, and direct inspectors to focus their efforts toward finding defects, not solutions. Remind participants to address their comments to the work product under review, not to the author. |
| Review general items (IEEE1028-2008 6.5.6.2) | IEEE1028-2008 6.5.6.2 Review general items<br>Anomalies referring to the software product in general (and thus not attributable to a specific instance or location) shall be presented to the inspectors and recorded. |
| Examine the software product (IEEE1028-2008 6.5.6.3 b) | IEEE1028-2008 6.5.6.3 b: The inspection team shall examine the software product objectively and thoroughly, and the inspection leader shall focus this part of the meeting on creating the anomaly list. |

| Activity (task) name | Documentation |
|---|---|
| | PI44: Raise Defects and Issues: Point out concerns, potential defects, questions, or improvement opportunities after the reader presents each section. |
| Present the software product to the inspection team (IEEE1028-2008 6.5.6.3 a) | IEEE1028-2008 6.5.6.3 a: The reader shall present the software product to the inspection team.<br><br>PI43: Present Work Product: Describe portions of the work product to the inspection team. |
| Answer specific questions (IEEE1028-2008 6.5.6.3 d) | IEEE1028-2008 6.5.6.3 d: The recorder shall enter each anomaly, location, description, and classification on the anomaly list. IEEE Std 1044-1993 [B8] may be used to classify anomalies. During this time, the author shall answer specific questions and contribute to anomaly detection based on the author’s understanding of the software product.<br><br>PI46: Answer Questions: Respond briefly to any specific questions raised, and contribute to defect detection based on special understanding of the work product. |
| Enter anomalies to anomaly list (IEEE1028-2008 6.5.6.3 c, e) | IEEE1028-2008 6.5.6.3 c: The recorder shall enter each anomaly, location, description, and classification on the anomaly list. IEEE Std 1044-1993 [B8] may be used to classify anomalies.<br><br>IEEE1028-2008 6.5.6.3 e: If there is disagreement about an anomaly, the potential anomaly shall be logged and marked for resolution at the end of the meeting.<br><br>PI45: Record Issues: Capture the information in Table 2 on the Issue Log for each issue raised. State aloud what was recorded to make sure it was recorded accurately. |
| **Follow-up <Process Level>** | |
| Examine the modified work product to judge whether the rework has been performed correctly. Report any findings to the author, so rework can be declared complete, incorrect rework can be redone, or items that were not originally pursued can be addressed. (PI62) | PI62: Examine the modified work product to judge whether the rework has been performed correctly. Report any findings to the author, so rework can be declared complete, incorrect rework can be redone, or items that were not originally pursued can be addressed. |
| Report the number of major and | PI63: Report the number of major and minor defects found and corrected and the actual rework effort to the moderator. |

| Activity (task) name | Documentation |
| --- | --- |
| minor defects found and corrected and the actual rework effort to the moderator. (PI63) | |
| Confirm that the author has addressed every item on the Issue Log. Determine whether the author made appropriate decisions as to which defects not to correct and which improvement suggestions not to implement. (PI61) | PI61: Confirm that the author has addressed every item on the Issue Log. Determine whether the author made appropriate decisions as to which defects not to correct and which improvement suggestions not to implement. |
| Check the baselined work product into the project's configuration management system. (PI65) | PI65: Check the baselined work product into the project's configuration management system. |
| Deliver Inspection Summary Report and counts of defects found and defects corrected to peer review coordinator. (PI66) | PI66: Deliver Inspection Summary Report and counts of defects found and defects corrected to peer review coordinator. |
| **Improvement <Process Level>** | |
| Act on inspection team recommendations in a timely manner (IEEE1028-2008 6.5.1 f) | IEEE1028-2008 6.5.1 f Act on inspection team recommendations in a timely manner<br>(Moved from management preparation subprocess) |
| Inspect checklists (IEEE1028-2008 6.9 c) | IEEE1028-2008 6.9 c: The checklists themselves shall also be inspected regularly for superfluous or misleading questions.<br><br>CMMI-DEV v1.2<br>VER SP2.1 SUBP5. Establish and maintain checklists to ensure that the work products are reviewed consistently.<br>Examples of items addressed by the checklists include the following:<br>* Rules of construction<br>* Design guidelines<br>* Completeness<br>* Correctness |

| Activity (task) name | Documentation |
|---|---|
| | * Maintainability<br>* Common defect types<br><br>The checklists are modified as necessary to address the specific type of work product and peer review. The peers of the checklist developers and potential users review the checklists. |
| Include frequently occurring anomalies in the inspection checklists (IEEE1028-2008 6.9 b) | IEEE1028-2008 6.9 b: Frequently occurring anomalies shall be included in the inspection checklists or role assignments. |
| Analyze inspection data (IEEE1028-2008 6.9 a) | IEEE1028-2008 6.9 a: Inspection data shall be analyzed regularly in order to improve the inspection process itself, and should be used to improve the activities used to produce software products.<br><br>SP2.3 SUBP4: Analyze the peer review data.<br>Examples of peer review data that can be analyzed include the following:<br>* Phase defect was injected<br>* Preparation time or rate versus expected time or rate<br>* Number of defects versus number expected<br>* Types of defects detected<br>* Causes of defects<br>* Defect resolution impact |
| Analyze consistently granted or requested waivers (IEEE1028-2008 6.9 d) | IEEE1028-2008 6.9 d: Consistently granted or requested waivers shall be analyzed to determine if the standards need to be changed. |
| Assess benefits achieved and improve inspection process (IEEE1028-2008 6.9 f) | IEEE1028-2008 6.9 f: Benefits (savings) achieved should be assessed regularly, and the inspection process should be continually adapted to achieve greater effectiveness at maximum efficiency. |
| Analyze preparation times, meeting times, and number of participants (IEEE1028-2008 6.9 e) | IEEE1028-2008 6.9 e: The preparation times, meeting times, and number of participants shall be analyzed to determine connections between preparation (checking) rate, meeting rate, and number and severity of anomalies found. |
| Establish and maintain criteria for requiring another peer review. (VER | VER SP2.1 SUBP4: Establish and maintain criteria for requiring another peer review. |

| Activity (task) name | Documentation |
|---|---|
| SP2.1 SUBP4) | |
| Establish and maintain entry and exit criteria for the peer review. (VER SP2.1 SUBP3) | VER SP2.1 SUBP3: Establish and maintain entry and exit criteria for the peer review. |
| Define requirements for collecting data during the peer review. (VER SP2.1 SUBP2) | VER SP2.1 SUBP2: Define requirements for collecting data during the peer review. |
| **Inspection <Process Level>** | |
| Check entry criteria (ADD1) | PI IEEE Main description PI<br>IEEE 1028-2008 6.4.1 Auth a Inspections shall be planned and documented in the appropriate project planning documents (for example, the project plan, the software quality assurance plan, or the software verification and validation plan).<br>IEEE 1028-2008 6.4.2 Precond An inspection shall be conducted only when the relevant inspection inputs are available.<br>IEEE 1028-2008 6.4.3 min a The inspection leader determines that the software product to be inspected is complete and conforms to project standards for format.<br>IEEE 1028-2008 6.4.3 min c Prior milestones upon which the software product depends are satisfied as identified in the appropriate planning documents.<br>PI EnC2 IEEE 1028-2008 6.4.3 min d Required supporting documentation is available. All necessary supporting documentation is available<br>PI EnC7 IEEE 1028-2008 6.4.3 min e For a reinspection, all items noted on the anomaly list that affect the software product under inspection are resolved. For a re-inspection, all issues from the previous inspection were resolved.<br><br>PI EnC1 The author selected an inspection approach for the product being reviewed.<br>PI EnC3 The author has stated his objectives for this inspection.<br>PI EnC4 Reviewers are trained in the peer review process.<br>PI EnC5 IEEE 1028-2008 6.4.3 min b Documents to be inspected are identified with a version number. All pages are numbered and line numbers are displayed. The documents have been spell-checked. Automated error-detecting tools (such as spell-checkers and compilers) have been used to identify and eliminate errors prior to the inspection.<br>PI EnC6 IEEE 1028-2008 6.4.3 min b Source code to be inspected is identified with a version number. Listings have line numbers and page numbers. Code compiles with no errors or warning messages using the project’s standard compiler switches. Errors found using code analyzer tools have been corrected. Automated error-detecting tools (such as spell-checkers and compilers) have been used to |

| Activity (task) name | Documentation |
|---|---|
| | identify and eliminate errors prior to the inspection. |
| Check exit criteria (PI64) | PI64: Check whether the exit criteria for the inspection and for the peer review process have been satisfied. If so, the inspection is complete.<br><br>PI IEEE Main Source Main description<br>IEEE 1028-2008 6.6 IEEE 1028-2008 An inspection shall be considered complete when the activities listed in 6.5 have been accomplished, and the output described in 6.7 exists.<br>PI ExC2 PI Issues raised during the inspection are tracked to closure.<br>PI ExC3 PI All major defects are corrected.<br>PI ExC4 PI Uncorrected defects are logged in the project's defect tracking system.<br>PI ExC5 PI The modified work product is checked into the project's configuration management system.<br>PI ExC6 PI If changes were required in earlier project deliverables, those deliverables have been correctly modified, checked into the project's configuration management system, and any necessary regression tests were passed.<br>PI ExC7 PI Moderator has collected and recorded the inspection data.<br>PI ExC8 PI Moderator has delivered the completed Inspection Summary Report and defect counts to the peer review coordinator. |
| **Management Preparation <Process Level>** | |
| Provide funding, infrastructure, and facilities required to plan, define, execute, and manage the inspection (IEEE1028-2008 6.5.1 b) | IEEE1028-2008 6.5.1 b: Provide funding, infrastructure, and facilities required to plan, define, execute, and manage the inspection<br><br>CMMI-DEV v1.2:<br>GP 2.3 Provide Resources |
| Plan time and resources required for inspection, including support functions (IEEE1028-2008 6.5.1 a) | IEEE1028-2008 6.5.1 a: Plan time and resources required for inspection, including support functions, as required in IEEE Std 1058-1998 [B9] or other appropriate standards<br><br>CMMI-DEV v1.2:<br>GP 2.2 Plan the Process |
| Ensure that inspections are planned, and that planned inspections are | IEEE1028-2008 6.5.1 e: Ensure that inspections are planned, and that planned inspections are conducted |

| Activity (task) name | Documentation |
|---|---|
| conducted (IEEE1028-2008 6.5.1 e) | CMMI-DEV v1.2<br>GP 2.2 Plan the Process<br><br>GP 2.8 Monitor and Control the Process |
| Ensure that inspection team members possess appropriate levels of expertise and knowledge sufficient to comprehend the software product under inspection (IEEE1028-2008 6.5.1 d) | IEEE 1028-2008:<br>6.5.1 d: Ensure that inspection team members possess appropriate levels of expertise and knowledge sufficient to comprehend the software product under inspection<br><br>CMMI-DEV v1.2:<br>VER SP 2.1 SUBP6: Develop a detailed peer review schedule, including the dates for peer review training and for when materials for peer reviews will be available.<br><br>GP2.5 Train People |
| Management shall ensure that the inspection is performed as required by applicable standards and procedures and by requirements mandated by law, contract, or other policy. (IEEE1028-2008 6.5.1) | IEEE1028-2008 6.5.1: Management shall ensure that the inspection is performed as required by applicable standards and procedures and by requirements mandated by law, contract, or other policy.<br><br>~VER SP2.1 SUBP3: Establish and maintain entry and exit criteria for the peer review.<br><br>GG1: Achieve Specific Goals GP 1.1 Perform Specific Practices |
| Provide training and orientation on inspection procedures applicable to a given project (IEEE1028-2008 6.5.1 c) | IEEE1028-2008 6.5.1 c: Provide training and orientation on inspection procedures applicable to a given project<br><br>CMMI-DEV v1.2:<br>VER SP 2.1 SUBP6: Develop a detailed peer review schedule, including the dates for peer review training and for when materials for peer reviews will be available.<br><br>GP2.5 Train People |
| **Overview of inspection procedures <Process Level>** | |
| Present inspection data (IEEE1028-2008 6.5.3 2 b) | IEEE1028-2008 6.5.3 2 b:The inspection leader shall answer questions about any checklists and the role assignments and should present inspection data such as minimal preparation times, the recommended inspection rate, and the typical number of anomalies previ- |

| Activity (task) name | Documentation |
|---|---|
| | ously found in inspections of similar products. |
| Answer questions about any check-lists and the role assignments (IEEE1028-2008 6.5.3 2 a) | IEEE1028-2008 6.5.3 2: The inspection leader shall answer questions about any checklists and the role assignments and should present inspection data such as minimal preparation times, the recommended inspection rate, and the typical number of anomalies previously found in inspections of similar products. |
| Assign roles (IEEE1028-2008 6.5.3 1) | IEEE1028-2008 6.5.3 1: Roles shall be assigned by the inspection leader.<br><br>VER GP 2.4 Assign Responsibility<br>VER.GP 2.7 Identify and Involve Relevant Stakeholders<br><br>PI14: Select inspectors and assign roles to individuals. Gain agreement from the other inspectors to participate. |
| **Overview of inspection product <Process Level>** | |
| Evaluate the assumptions, history, and context of the work product. (PI22) | PI22: Evaluate the assumptions, history, and context of the work product. |
| Describe the important features of the work product to the rest of the inspection team. (PI21) | PI21: Describe the important features of the work product to the rest of the inspection team. State inspection objectives.<br><br>IEEE1028-2008 6.5.4: Present an overview of the software product to be inspected. This overview should be used to introduce the inspectors to the software product. The overview may be attended by other project personnel who could profit from the presentation. |
| **Peer Review Process <Process Level>** | |
| Decide if further peer review needed (VER SP2.2 SUBP6) | VER SP2.2 SUBP6: Conduct an additional peer review if the defined criteria indicate the need. |
| **Planning the inspection <Process Level>** | |
| Schedule the meeting date and time, select the meeting place, and notify the inspection team (IEEE1028-2008 6.5.2 c) | IEEE1028-2008 6.5.2 c: Schedule the meeting date and time, select the meeting place, and notify the inspection team<br><br>VER SP 2.1 SUBP6: Develop a detailed peer review schedule, including the dates for peer review training and for when materials for peer reviews will be available. |

| Activity (task) name | Documentation |
| --- | --- |
| | PI16: Schedule the inspection, and possibly overview, meetings and distribute a meeting notice. |
| Assemble the inspection materials, send to inspection leader.<br>(IEEE1028-2008 6.5.2 1) | IEEE1028-2008 6.5.2 1: The author shall assemble the inspection materials for the inspection leader. Inspection materials include the software product to be inspected, standards and documents that have been used to develop the software product, etc.<br><br>PI11 Give moderator the work product to be inspected and supporting documents, such as specifications, predecessor documents, or pertinent test documentation. |
| Assign specific responsibilities to the inspection team members<br>(IEEE1028-2008 6.5.2 b) | IEEE1028-2008 6.5.2 b: Assign specific responsibilities to the inspection team members<br><br>CMMI DEV v1.2:<br>GP 2.4: Assign Responsibility<br>VER SP2.1 SUBP9: Assign roles for the peer review as appropriate.<br><br>PI14: Select inspectors and assign roles to individuals. Gain agreement from the other inspectors to participate. |
| Distribute inspection materials to participants, and allow adequate time for their preparation (IEEE1028-2008 6.5.2 d) | IEEE1028-2008 6.5.2 d: Distribute inspection materials to participants, and allow adequate time for their preparation<br><br>VER SP2.1 SUBP8: Distribute the work product to be reviewed and its related information to the participants early enough to enable participants to adequately prepare for the peer review.<br><br>PI17: Distribute the inspection package to the participants at least 3 working days prior to the inspection meeting. |
| Set a timetable for distribution of inspection material and for the return of comments and forwarding of comments to the author for disposition (IEEE1028-2008 6.5.2 e) | IEEE1028-2008 6.5.2 e: Set a timetable for distribution of inspection material and for the return of comments and forwarding of comments to the author for disposition<br><br>~VER SP 2.1 SUBP6: Develop a detailed peer review schedule, including the dates for peer review training and for when materials for peer reviews will be available.<br><br>~PI16: Schedule the inspection, and possibly overview, meetings and distribute a meeting notice. |
| Establish the anticipated inspection rate for preparation and meeting<br>(IEEE1028-2008 6.5.2 g) | IEEE1028-2008 6.5.2 g: Establish the anticipated inspection rate for preparation and meeting |

| Activity (task) name | Documentation |
|---|---|
| Identify, with appropriate management support, the inspection team (IEEE1028-2008 6.5.2 a) | IEEE1028-2008 6.5.2 a: Identify, with appropriate management support, the inspection team<br><br>GP 2.7: Identify and Involve Relevant Stakeholders<br><br>PI41: Select inspectors and assign roles to individuals. Gain agreement from the other inspectors to participate. |
| Specify the scope of the inspection, including the priority of sections of the documents to be inspected (IEEE1028-2008 6.5.2 f) | IEEE1028-2008 6.5.2 f: Specify the scope of the inspection, including the priority of sections of the documents to be inspected<br><br>>~ VER SP2.1 SUBP2: Define requirements for collecting data during the peer review. |
| Based on the size and complexity of the work product, determine how many inspection meetings will be required. (PI13) | PI13: Based on the size and complexity of the work product, determine how many inspection meetings will be required.<br><br>VER SP2.2 SUBP6: Conduct an additional peer review if the defined criteria indicate the need. |
| Ensure that the work product satisfies the peer review entry criteria prior to distribution. (VER SP2.1 SUBP7) | VER SP2.1 SUBP7: Ensure that the work product satisfies the peer review entry criteria prior to distribution.<br><br>PI12: Determine whether work product satisfies inspection entry criteria. |
| **Preparation <Process Level>** | |
| Ask individual inspectors to prepare (PI31) | PI31: Ask individual inspectors to prepare with specific objectives in mind, such as: checking the consistency of cross-references; checking for interface errors; checking traceability to, and consistency with, predecessor specifications; or checking conformance to standards. |
| Reschedule meeting (IEEE1028-2008 6.5.5 6 b) | IEEE1028-2008 6.5.5 6 b: The inspection leader shall verify that inspectors are prepared for the inspection. The inspection leader shall reschedule the meeting if the inspectors are not adequately prepared. |
| Classify anomalies (IEEE1028-2008 6.5.5 3 a) | IEEE1028-2008 6.5.5 3 a: The inspection leader should classify anomalies as described in 6.8.1 to determine whether they warrant cancellation of the inspection meeting, and in order to plan efficient use of time in the inspection meeting. |
| The reader(s) shall prepare sufficiently to be able to present the software product at the inspection meeting. (IEEE1028-2008 6.5.5 5) | IEEE1028-2008 6.5.5 5: The reader(s) shall prepare sufficiently to be able to present the software product at the inspection meeting. |

| Activity (task) name | Documentation |
|---|---|
| Specify a suitable order in which the software product will be inspected (IEEE1028-2008 6.5.5 4) | IEEE1028-2008 6.5.5 4: The inspection leader or reader shall specify a suitable order in which the software product will be inspected (such as sequential, hierarchical, data flow, control flow, bottom up, or top down). |
| Gather individual preparation times and record the total in the inspection documentation. (IEEE1028-2008 6.5.5 6 c) | IEEE1028-2008 6.5.5 6 c: The inspection leader should gather individual preparation times and record the total in the inspection documentation. |
| Anomalies detected during this examination shall be documented and sent to the inspection leader. (IEEE1028-2008 6.5.5 2) | IEEE1028-2008 6.5.5 2: Anomalies detected during this examination shall be documented and sent to the inspection leader.<br><br>PI33: Log minor defects found, such as typographical errors or style inconsistencies, on the Typo List. (Deliver this to the author at or prior to the inspection meeting.) |
| Examine the software product and other inputs prior to the review meeting. (IEEE1028-2008 6.5.5 1) | IEEE1028-2008 6.5.5 1: Each inspection team member shall examine the software product and other inputs prior to the review meeting.<br><br>PI32: Examine the work product, to understand it, find defects, and raise questions about it. Use the appropriate defect checklist to focus attention on defects commonly found in the type of product being inspected. Use other analysis methods to look for defects as appropriate. |
| Forward the anomalies to the author of the software product for disposition. (IEEE1028-2008 6.5.5 3 c) | IEEE1028-2008 6.5.5 3 c: The inspection leader should forward the anomalies to the author of the software product for disposition.<br><br>PI33: (Log minor defects found, such as typographical errors or style inconsistencies, on the Typo List.) Deliver this to the author at or prior to the inspection meeting. |
| Cancel Inspection, request later inspection (IEEE1028-2008 6.5.5 3 b) | IEEE1028-2008 6.5.5 3 b: If the inspection leader determines that the extent or seriousness of the anomalies warrants, the inspection leader may cancel the inspection, requesting a later inspection when the software product meets the minimal entry criteria and is reasonably defect-free. |
| **Rework <Process Level>** | |
| Correct defects and typos found, resolve issues raised, and modify work product accordingly. Mark issues list to indicate action taken. | PI51: Correct defects and typos found, resolve issues raised, and modify work product accordingly. Mark issues list to indicate action taken. |

| Activity (task) name | Documentation |
|---|---|
| (PI51) | |
| Correct any other project documents based on defects identified in the inspected work product. (PI52) | PI52: Correct any other project documents based on defects identified in the inspected work product. |
| If rework verification is not needed, report the number of major and minor defects found and corrected and the actual rework effort to the moderator. (PI54) | PI54: If rework verification is not needed, report the number of major and minor defects found and corrected and the actual rework effort to the moderator. |
| Record the actual rework effort (PI55) | PI55: Record the actual rework effort on the Inspection Summary Report. |
| Record any uncorrected defects (PI53) | PI53: Record any uncorrected defects in the project's defect tracking system. |
| **Rework and follow-up <Process Level>** | |

Table 56 – Gateways and their documentation

| Gateway name | Documentation |
|---|---|
| **Inspection <Process Level>** | |
| Determine whether product overview meeting is required. (~PI15) | PI15: Determine whether an overview meeting is required. |
| Determine whether a procedure overview meeting is required. (~PI15) | PI15: Determine whether an overview meeting is required. |
| **Peer Review Process <Process Level>** | |
| Is further peer review needed? (VER SP2.2 SUBP6) | VER SP2.2 SUBP6: Conduct an additional peer review if the defined criteria indicate the need. |
| What type of peer review is needed? (VER SP2.1 SUBP1) | VER SP2.1 SUBP1: Determine what type of peer review will be conducted. |
| **Preparation <Process Level>** | |

| Gateway name | Documentation |
| --- | --- |
| Are inspectors prepared for the inspection? (IEEE1028-2008 6.5.5 6 a) | IEEE1028-2008 6.5.5 6 a: The inspection leader shall verify that inspectors are prepared for the inspection. |
| Extent or seriousness of the anomalies is high? (IEEE1028-2008 6.5.5 3 b) | IEEE1028-2008 6.5.5 3 b: If the inspection leader determines that the extent or seriousness of the anomalies warrants, the inspection leader may cancel the inspection, requesting a later inspection when the software product meets the minimal entry criteria and is reasonably defect-free. |

Table 57 – Data objects

| Name | CMMI reference | IEEE 1028 reference | PI reference | main source/ID |
| --- | --- | --- | --- | --- |
| Inspection Summary Report | VER SP 2.2 TWP1, VER SP 2.2 TWP3 | IEEE 1028 In should d,<br>IEEE 1028 Out shall a,<br>IEEE 1028 Out shall b,<br>IEEE 1028 Out shall c,<br>IEEE 1028 Out shall e,<br>IEEE 1028 Out shall k,<br>IEEE 1028 Out shall l,<br>IEEE 1028 Out should n (added),<br>IEEE 1028 Out may o (added),<br>IEEE 1028 In should a (added) | PI WA1, PI D2 | PI WA1 |
| Issue Log | VER SP 2.2 TWP2, VER SP 2.2 TWP3,<br>VER SP 2.3 TWP1 | IEEE 1028 In should d,<br>IEEE 1028 In should e,<br>IEEE 1028 Out shall h,<br>IEEE 1028 Out should m | PI WA2, PI D3, PI D5 | PI WA2 |
| Typo List | VER SP 2.2 TWP3 | IEEE 1028 In should d | PI WA3, PI D4 | PI WA3 |
| Inspection Moderator’s Checklist | | | PI WA4 | PI WA4 |
| Inspection Lessons Learned Questionnaire | | IEEE 1028 Out shall g | PI WA5 | PI WA5 |
| Review checklist | VER SP 2.1 TWP2 | IEEE 1028 In should g | PI WA6 | PI WA6 |
| Work product to be inspected | VER SP 2.1 TWP6 | IEEE 1028 In should b,<br>IEEE 1028 Out shall d | PI D1 | VER SP 2.1 TWP6 |

| Name | CMMI reference | IEEE 1028 reference | PI reference | main source/ID |
|---|---|---|---|---|
| Peer review schedule | VER SP 2.1 TWP1 | | | VER SP 2.1 TWP1 |
| Entry and exit criteria for work products | VER SP 2.1 TWP3 | | | VER SP 2.1 TWP3 |
| Criteria for requiring another peer review | VER SP 2.1 TWP4 | IEEE 1028 In may h | | VER SP 2.1 TWP4 |
| Peer review training material | VER SP 2.1 TWP5 | | | VER SP 2.1 TWP5 |
| Peer review action items | VER SP 2.3 TWP2 | | | VER SP 2.3 TWP2 |
| Inspection procedure | | IEEE 1028 In should c | | IEEE 1028 In should c |
| Source products of the software product to be inspected | | IEEE 1028 In should f | | IEEE 1028 In should f |
| Predecessor software product | | IEEE 1028 In may i | | IEEE 1028 In may i |
| Regulations, standards, guidelines, plans, specifications, and procedures against which the software product is to be inspected | | IEEE 1028 In may j | | IEEE 1028 In may j |
| Product specification | | IEEE 1028 In may k | | IEEE 1028 In may k |
| Performance data | | IEEE 1028 In may l | | IEEE 1028 In may l |
| Waivers granted or requested | | IEEE 1028 Out shall j | | IEEE 1028 Out shall j |

Table 58 – Entry/exit criteria

| En/Ex Crit. | PI | IEEE | Main Source | Main description | Secondary description |
|---|---|---|---|---|---|
| Entry criteria | | IEEE 1028-2008 6.4.1 Auth a | IEEE 1028-2008 | Inspections shall be planned and documented in the appropriate project planning documents (for example, the project plan, the software quality assurance plan, or the software verification and validation plan). | |
| Entry criteria | | IEEE 1028-2008 6.4.2 Precond | IEEE 1028-2008 | An inspection shall be conducted only when the relevant inspection inputs are available. | |
| Entry criteria | | IEEE 1028-2008 6.4.3 min a | IEEE 1028-2008 | The inspection leader determines that the software product to be inspected is complete and conforms to project standards for format. | |
| Entry criteria | | IEEE 1028-2008 6.4.3 min c | IEEE 1028-2008 | Prior milestones upon which the software product depends are satisfied as identified in the appropriate planning documents. | |
| Entry criteria | PI EnC2 | IEEE 1028-2008 | IEEE 1028- | Required supporting documentation is available. | All necessary supporting documenta- |

| En/Ex Crit. | PI | IEEE | Main Source | Main description | Secondary description |
|---|---|---|---|---|---|
| | | 6.4.3 min d | 2008 | | tion is available |
| Entry criteria | PI EnC7 | IEEE 1028-2008 6.4.3 min e | IEEE 1028-2008 | For a reinspection, all items noted on the anomaly list that affect the software product under inspection are resolved. | For a re-inspection, all issues from the previous inspection were resolved. |
| Entry criteria | PI EnC1 | | PI | The author selected an inspection approach for the product being reviewed. | |
| Entry criteria | PI EnC3 | | PI | The author has stated his objectives for this inspection. | |
| Entry criteria | PI EnC4 | | PI | Reviewers are trained in the peer review process. | |
| Entry criteria | PI EnC5 | IEEE 1028-2008 6.4.3 min b | PI | Documents to be inspected are identified with a version number. All pages are numbered and line numbers are displayed. The documents have been spell-checked. | Automated error-detecting tools (such as spell-checkers and compilers) have been used to identify and eliminate errors prior to the inspection. |
| Entry criteria | PI EnC6 | IEEE 1028-2008 6.4.3 min b | PI | Source code to be inspected is identified with a version number. Listings have line numbers and page numbers. Code compiles with no errors or warning messages using the project's standard compiler switches. Errors found using code analyzer tools have been corrected. | Automated error-detecting tools (such as spell-checkers and compilers) have been used to identify and eliminate errors prior to the inspection. |
| Exit criteria | | IEEE 1028-2008 6.6 | IEEE 1028-2008 | An inspection shall be considered complete when the activities listed in 6.5 have been accomplished, and the output described in 6.7 exists. | |
| Exit criteria | PI ExC2 | | PI | Issues raised during the inspection are tracked to closure. | |
| Exit criteria | PI ExC3 | | PI | All major defects are corrected. | |
| Exit criteria | PI ExC4 | | PI | Uncorrected defects are logged in the project's defect tracking system. | |
| Exit criteria | PI ExC5 | | PI | The modified work product is checked into the project's configuration management system. | |
| Exit criteria | PI ExC6 | | PI | If changes were required in earlier project deliverables, those deliverables have been correctly modified, checked into the project's configuration management system, and any necessary regression tests were passed. | |
| Exit criteria | PI ExC7 | | PI | Moderator has collected and recorded the inspection data. | |
| Exit criteria | PI ExC8 | | PI | Moderator has delivered the completed Inspection Summary Report and defect counts to the peer review coordinator. | |

Table 59 – Roles

| Name | PI description (responsibility) | IEEE description |
|---|---|---|
| author | • Creator or maintainer of the work product to be inspected. Initiates the inspection process by asking the peer review coordinator to assign a moderator.<br>• States his objectives for the inspection.<br>• Delivers work product and its specification or predecessor document to moderator.<br>• Works with moderator to select inspectors and assign roles.<br>• Addresses items on the Issue Log and Typo Lists.<br>• Reports rework time and defect counts to moderator. | The author shall be responsible for the software product meeting its inspection entry criteria, for contributing to the inspection based on special understanding of the software product, and for performing any rework required to make the software product meet its inspection exit criteria. |
| moderator (IEEE inspection leader) | • Plans, schedules, and leads the inspection events.<br>• Works with author to select inspectors and assign roles.<br>• Assembles inspection package and delivers it to inspectors at least 3 days prior to the inspection meeting.<br>• Determines whether preparation is sufficient to hold the meeting. If not, reschedules the meeting.<br>• Facilitates inspection meeting. Corrects any inappropriate behavior. Solicits input from inspectors as reader presents each section of the work product. Records any action items or side issues that arise during the inspection.<br>• Leads inspection team in determining the work product appraisal.<br>• Serves as verifier or delegates this responsibility to someone else.<br>• Delivers completed Inspection Summary Report to the organization's peer review coordinator. | The inspection leader shall be responsible for planning and organizational tasks pertaining to the<br>inspection, shall determine the parts/components of the software product and source documents to be inspected during the meeting (in conjunction with the author), shall be responsible for planning and preparation as described in 6.5.2 and 6.5.4, shall ensure that the inspection is conducted in an orderly manner and meets its objectives, shall ensure that the inspection data is collected, and shall issue the inspection output as described in 6.7. |
| reader | • Presents portions of the work product to the inspection team to elicit comments, issues, or questions from inspectors. | The reader shall lead the inspection team through the software product in a comprehensive and logical fashion, interpreting sections of the work (for example, generally paraphrasing groups of 1 to 3 lines), and highlighting important aspects. The software product may be divided into logical sections and assigned to different readers to lessen required preparation time. |

| Name | PI description (responsibility) | IEEE description |
|---|---|---|
| recorder | • Records and classifies issues raised during inspection meeting. | The recorder shall document anomalies, action items, decisions, waivers, and recommendations made by the inspection team. The recorder should record inspection data required for process analysis. The inspection leader may be the recorder. |
| inspector | • Examines work product prior to the inspection meeting to find defects and prepare for contributing to the meeting.<br>• Records preparation time.<br>• Participates during the meeting to identify defects, raise issues, and suggest improvements. | Inspectors shall identify and describe anomalies in the software product. Inspectors shall be chosen based on their expertise and should be chosen to represent different viewpoints at the meeting (for example, sponsor, end user, requirements, design, code, safety, test, independent test, project management, quality management, and hardware engineering). Only those viewpoints pertinent to the inspection of the product should be present.<br>Some inspectors should be assigned specific topics to ensure effective coverage. For example, one inspector may focus on conformance with a specific standard or standards, another on syntax or accuracy of figures, and another for overall coherence. These viewpoints should be assigned by the inspection leader when planning the inspection, as provided in item b) of 6.5.2. |
| verifier | Performs follow-up to determine whether rework has been performed appropriately and correctly. | |
| peer review coordinator | • Custodian of the organization's inspection metrics database.<br>• Maintains records of inspections conducted and data from the Inspection Summary Report for each inspection.<br>• Generates reports on inspection data for management, process improvement team, and peer review process owner. | |

# Appendix S. The unified peer review process model

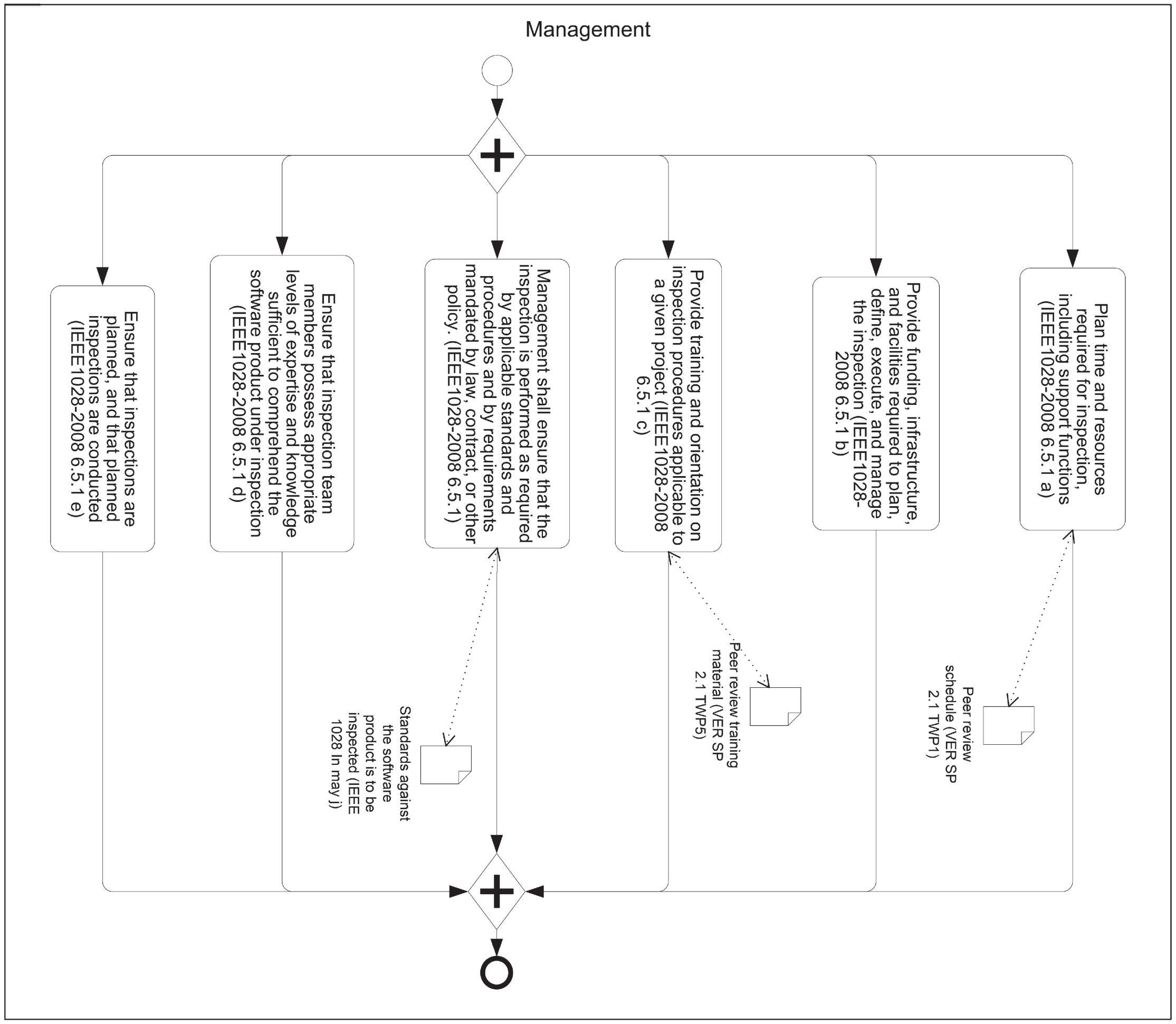

Figure 65 – Management preparation

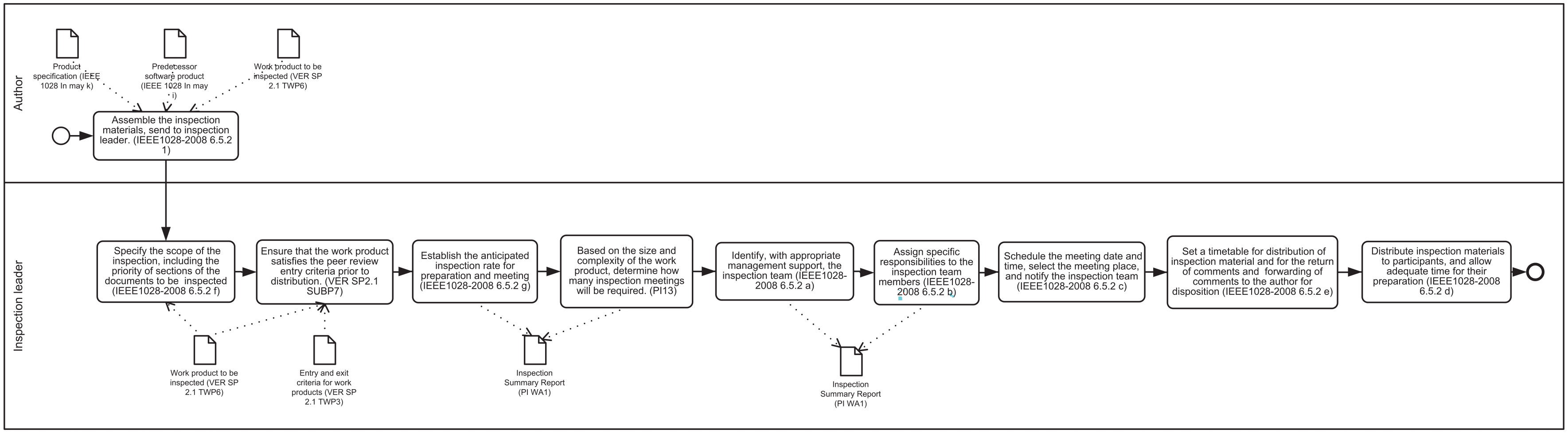

Figure 66 – Planning the inspection

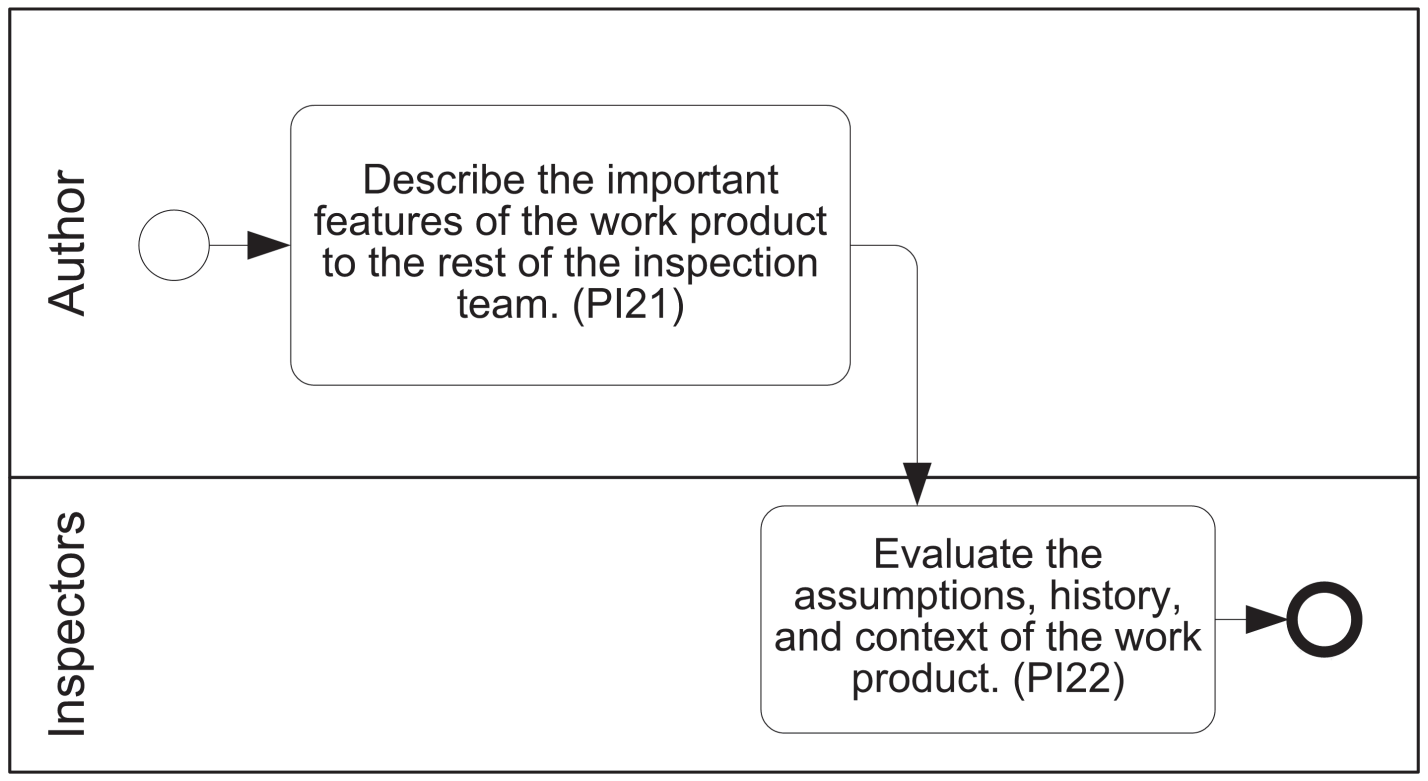

Figure 67 – Overview of inspection product

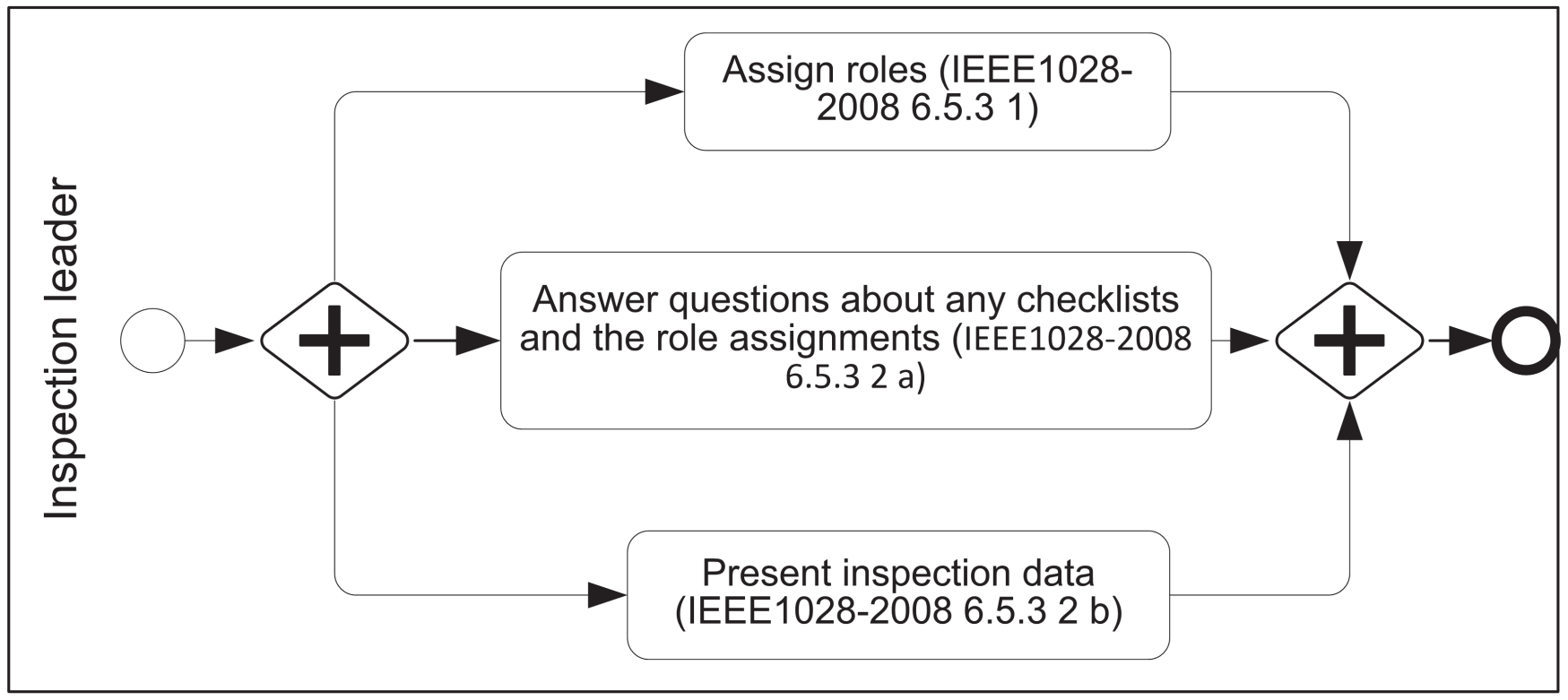

Figure 68 – Overview of inspection procedures

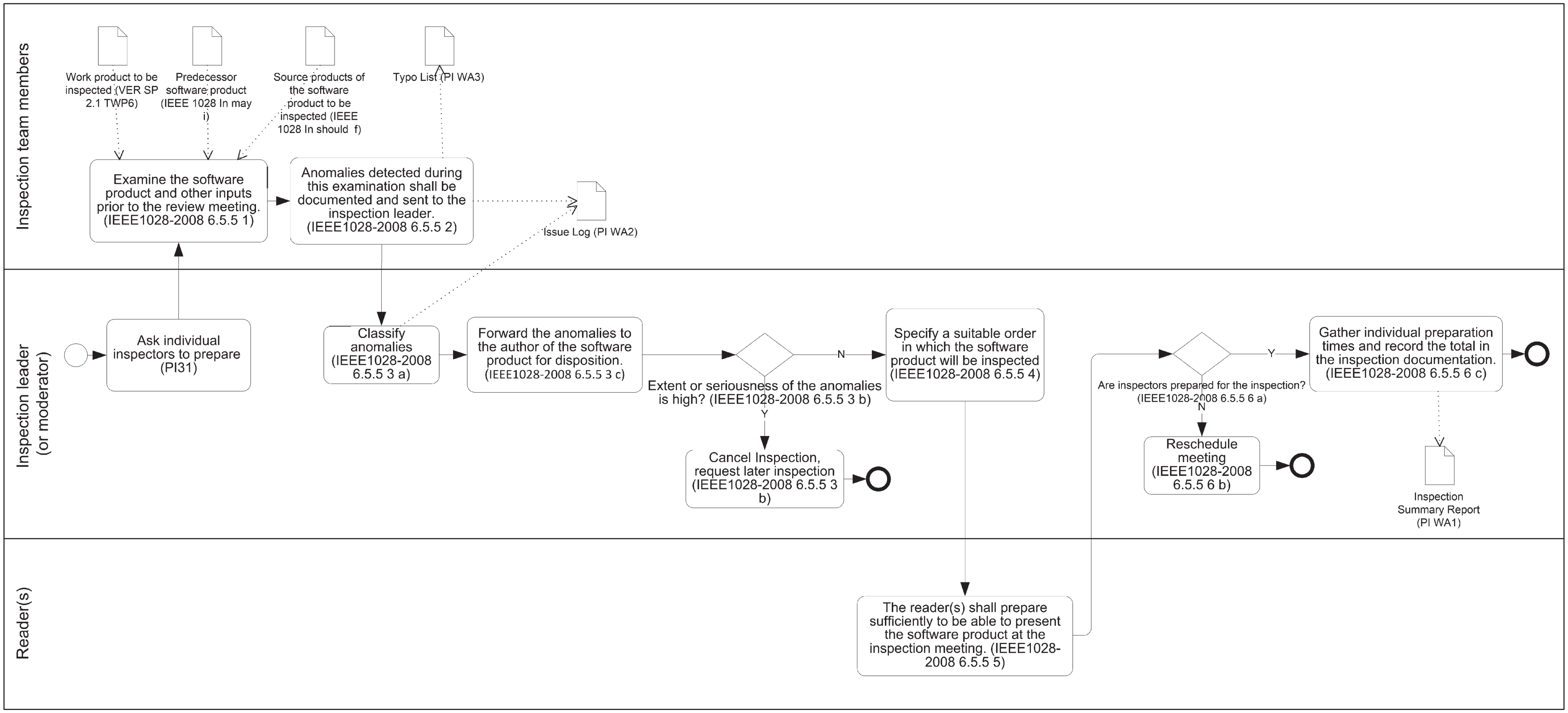

Figure 69 – Preparation

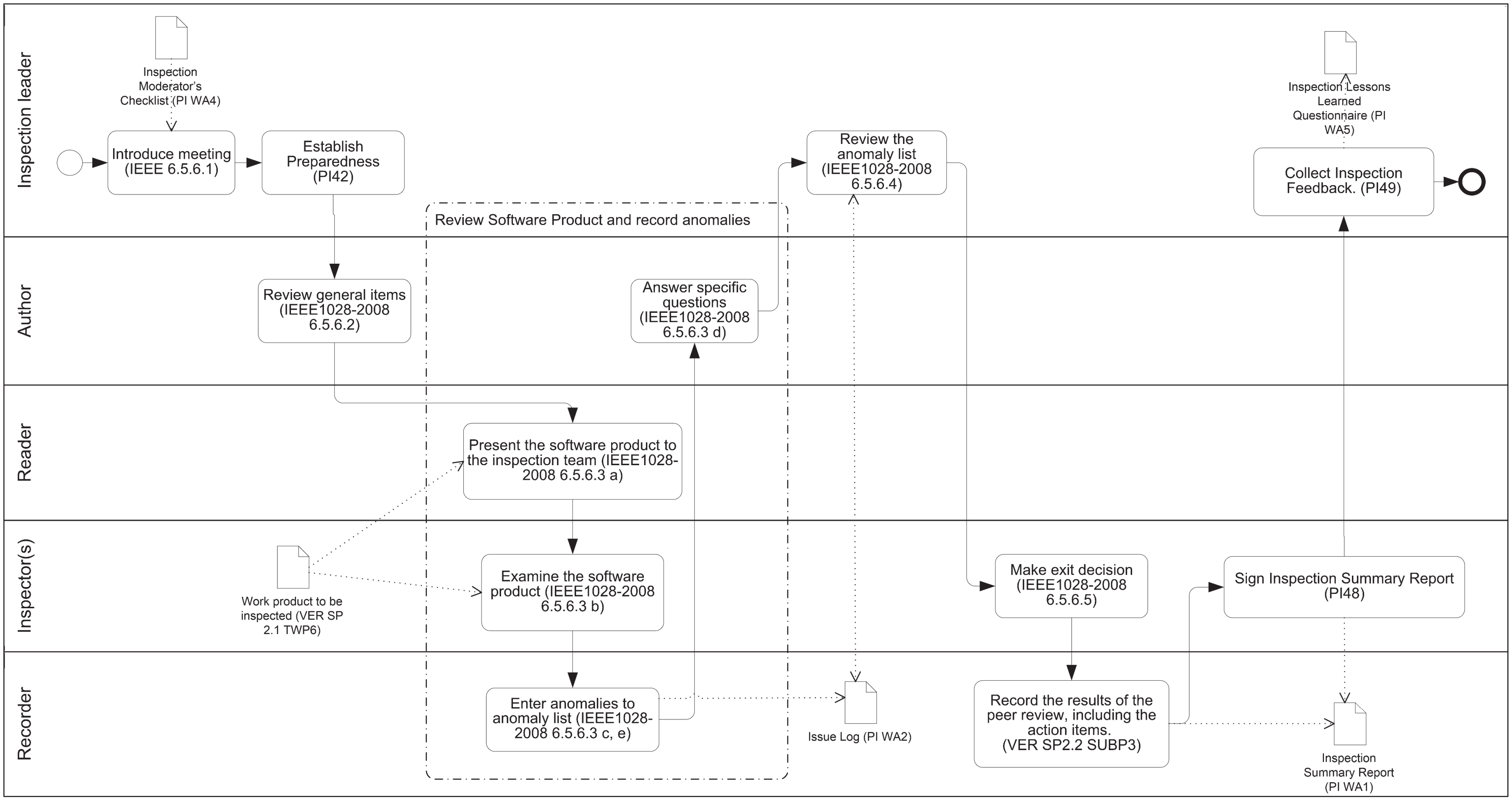

Figure 70 – Examination

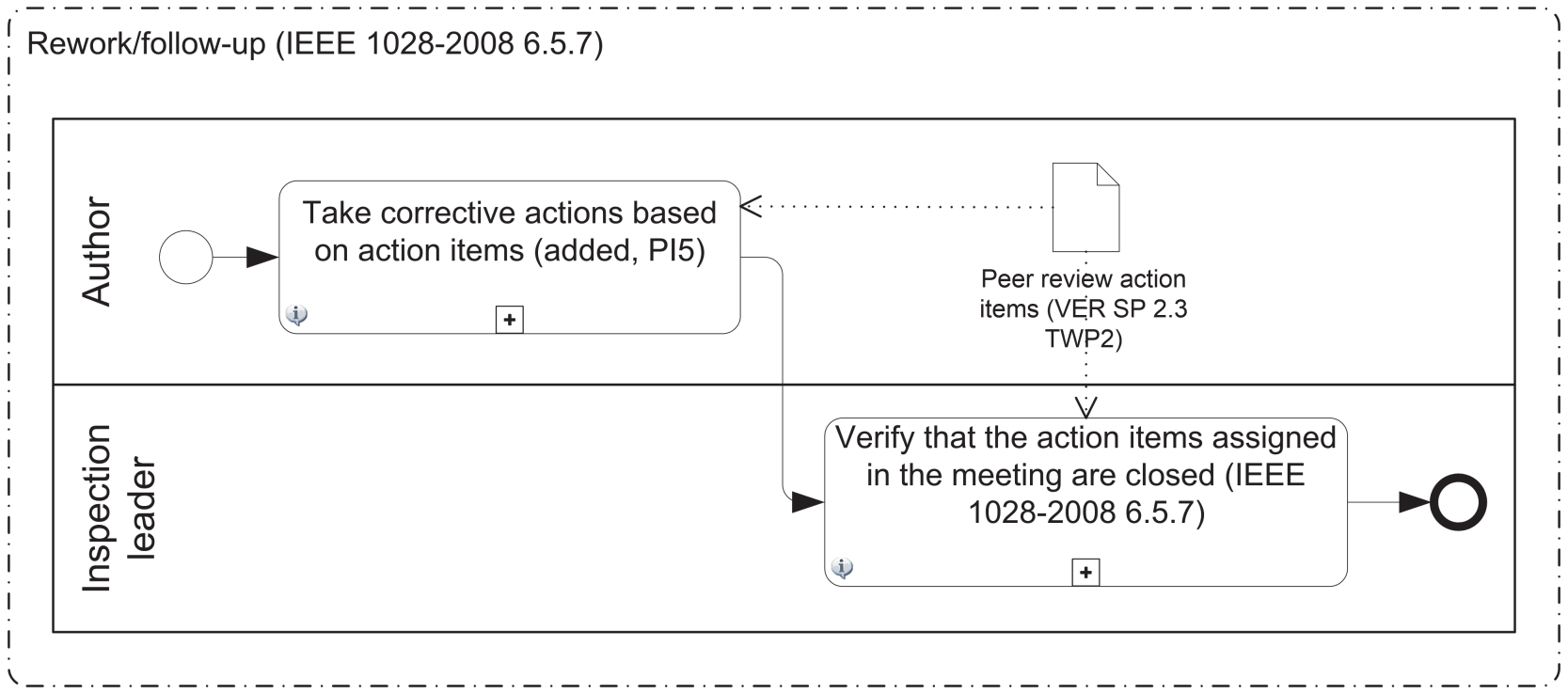

Figure 71 – Rework and follow-up

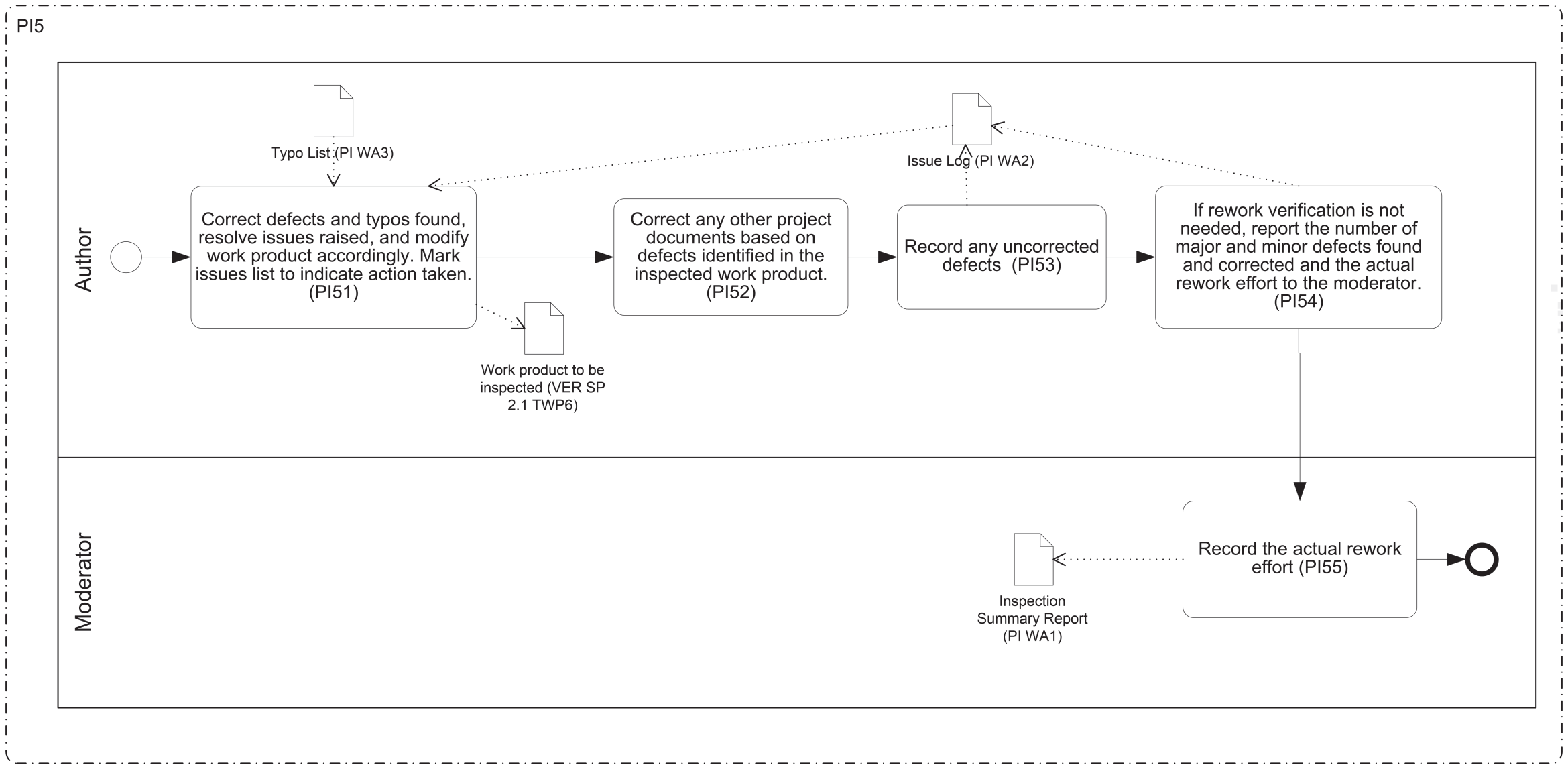

Figure 72 – Rework

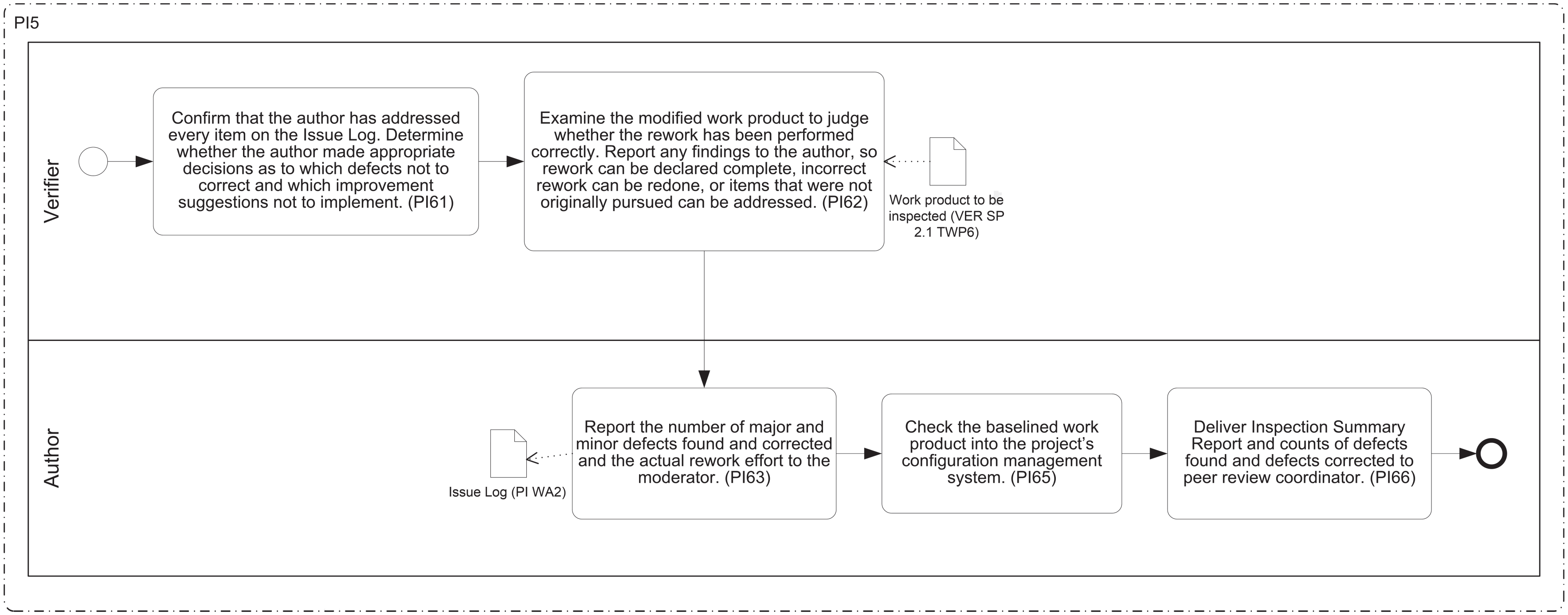

Figure 73 – Follow-up

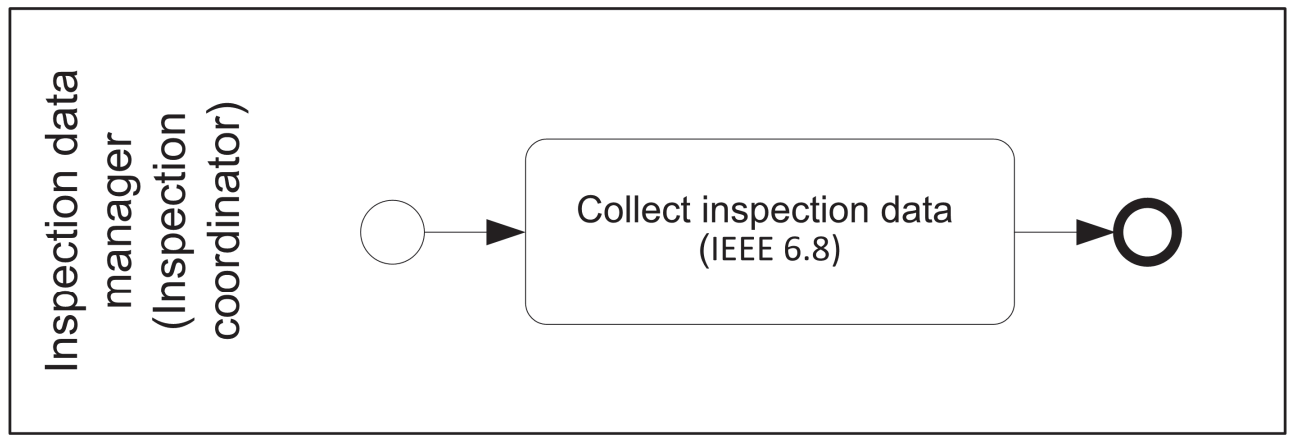

Figure 74 – Data collection

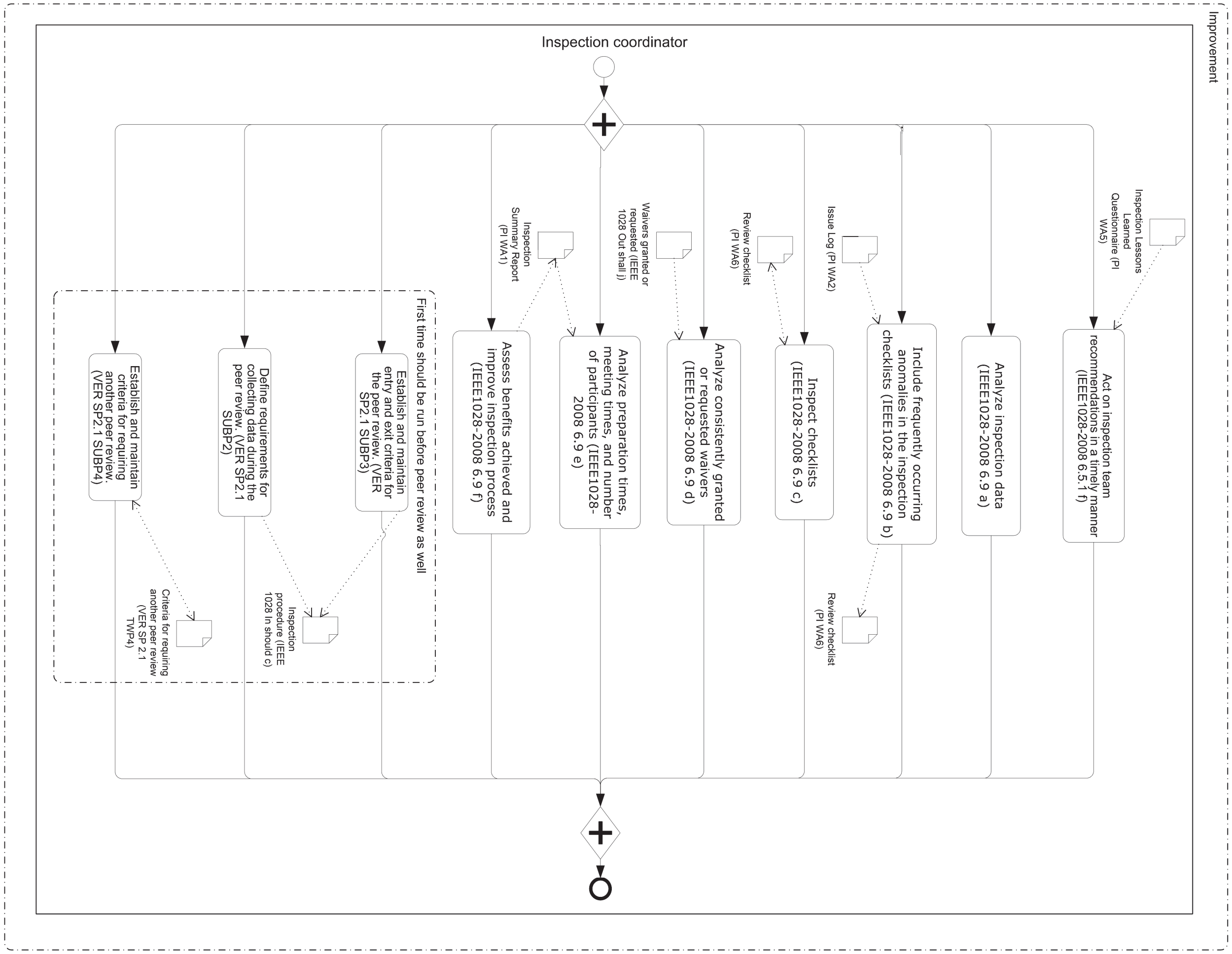

Figure 75 – Improvement

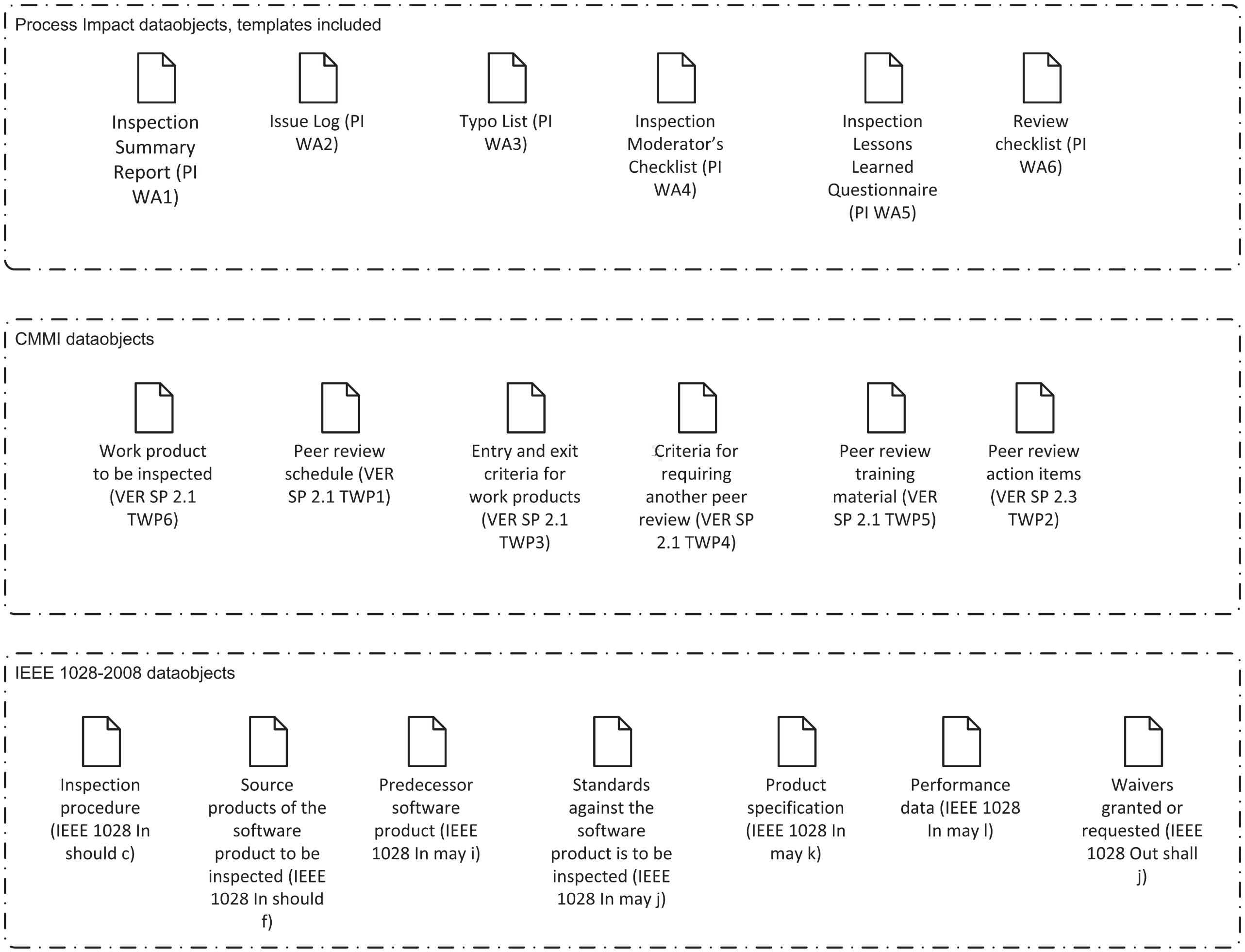

Figure 76 – Data objects used in the inspection process

# Appendix T. Meetings related to the development of the unified peer review process

In this appendix we present how the two teams in the Q-works project worked together; describing project in its timeline.
In order to keep the project on track integrated project team meetings with participants from all parties on a monthly basis between 2009.09.14 and 2011.04.28.

After more than one year passed after the project kick-off meeting, the following multi-model peer review related results were achieved:

- The PBU process was applied to develop the unified peer review process
- A multi-model peer review process model was finished (after several versions and revisions)
- The unified peer review process was converted to Maximo environment
- Refinements and simplifications were made on the process model based on the project needs.

Table 60 summarizes the joint Q-works project meetings, meeting topics and participants, reflecting the evolution of the unified peer review process. We call joint meetings when both parties were present from Polygon Ltd. and BME-SQI.

Table 60 – Peer review related joint meetings in the Q-works project

| **Activity type** | **Date** | **Peer-review related topic** | **Description** |
|---|---|---|---|
| Training on technical IT processes | 2009.08.xx (before the project started) | Training on peer reviews for Polygon, Szeged site | A training was provided for Polygon including a practical part on a non-formally developed multi-model peer review process. |
| Project and Technical Meeting | 2009.11.11 | Selecting a prototype process: peer reviews | Decision: agreement on the peer review process and on multi-model approach. |
| Project and Technical Meeting | 2009.12.17 | Peer review process and multi-model (the PBU framework and unified process) presentation | Presentation of 3 separated peer review processes and the PBU concept. |
| Project and technical meeting | 2010.02.11 | Presentation of the *first version* (activities) of the multi-model peer review process | Tailored process from the 3 process model elements. |
| Technical-only meeting | 2010.02.25 | Presentation and discussion on the second version of the multi-model process, excluded texts | The second version of peer review process was completed with data objects and roles. |
| Project and | 2010.03.30 | Presentation about including | Almost all types of multi-model peer |

| Activity type | Date | Peer-review related topic | Description |
|---|---|---|---|
| technical meeting | | the multi-model Peer review process into Maximo Environment (Polygon technical staff)<br>Further topics:<br>Final agreement on using BPMN for describing processes.<br>Organizing a workshop, accepting workshop programme | review process elements (process, subprocess, activity and data objects) were included into IBM Maximo environment, except roles. |
| Technical-only meeting | 2010.04.15 | Presentation and discussion on peer review process documents (templates, number of documents) and roles and modifications made on the *third version* of the multi-model process. | Peer review process was divided into three separate subprocesses due to their separate timeline.<br>A discussion was requested by Polygon on process roles and documentation. |
| Technical-only meeting | 2010.05.27 | Selecting important roles for Q-Works project (peer review leader, author, inspector). Selecting important sub-processes (cutting out the management preparation and improvement/data collection parts)<br>Decision about Maximo and Portal functions and user access. | |

# Appendix U. The unified peer review process in IBM Maximo environment

Figures of this chapter represent the creation of work plans for the unified peer review process. The source of the figures in this chapter is the Q-works test plan (POLYGON Informatikai Kft., 2010).

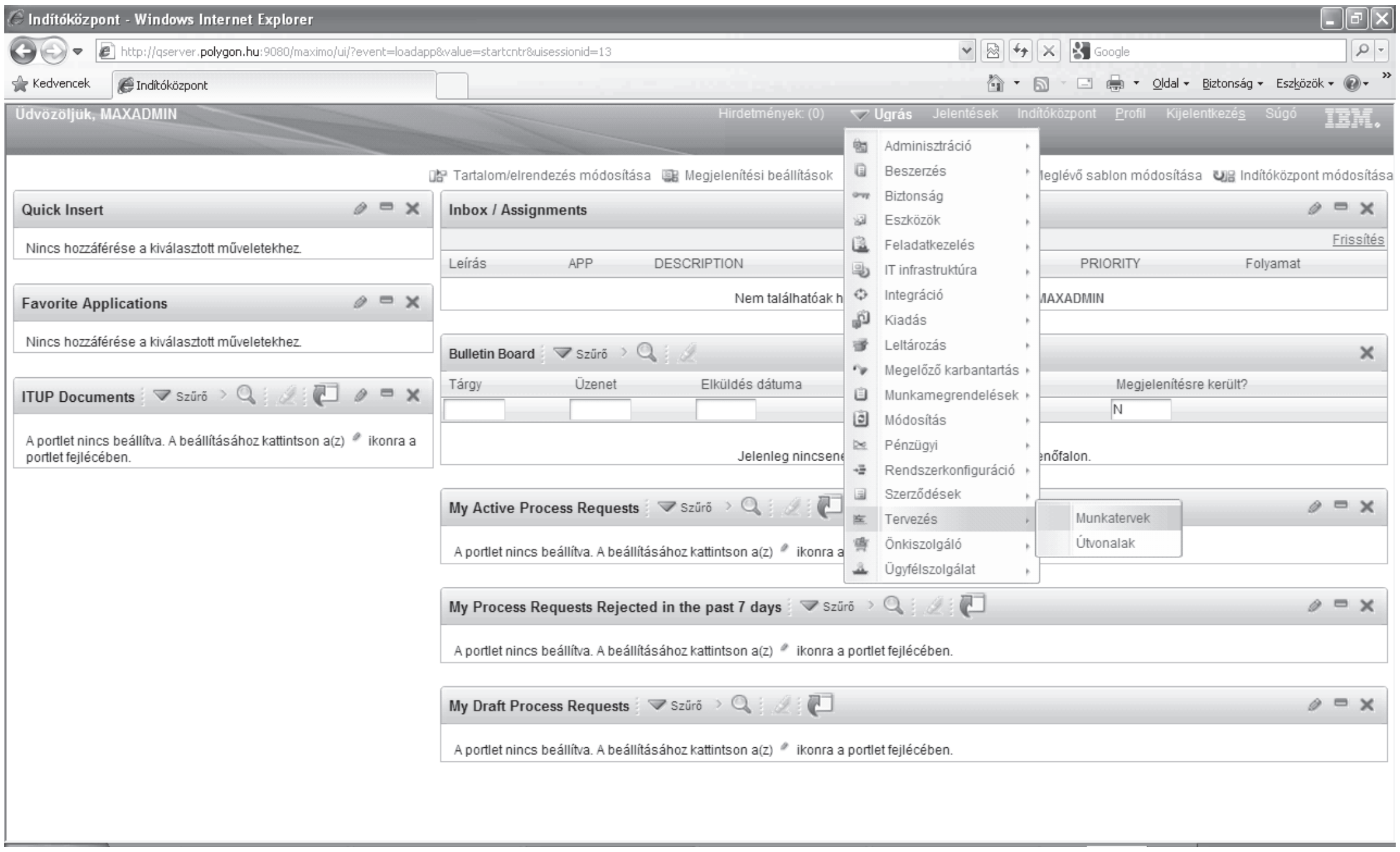

Figure 77 – Accessing job plans in Maximo (Go to > Planning > Job Plans)

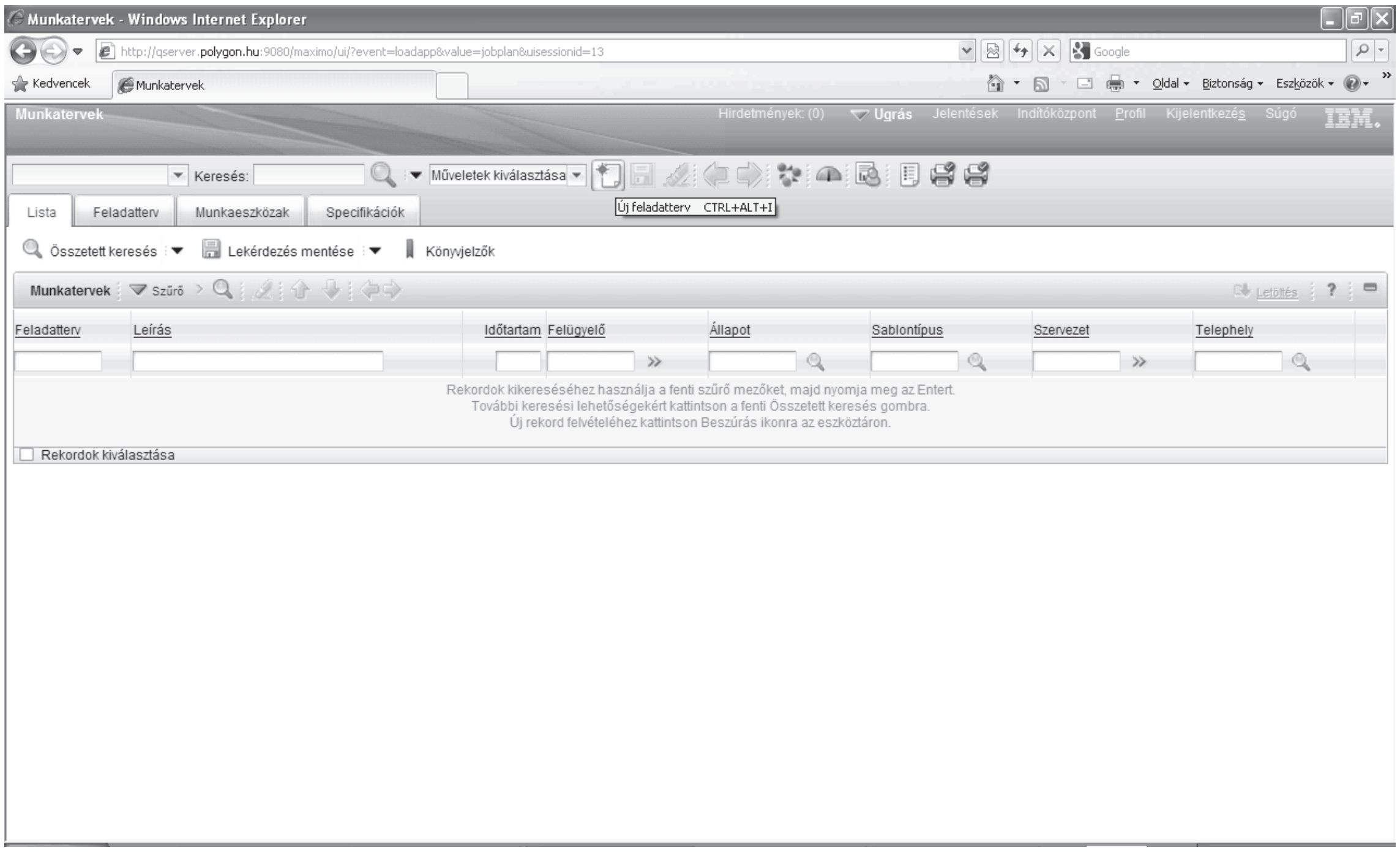

Figure 78 – Creating a new record by clicking on the “Új feladatterv” (new task plan) button

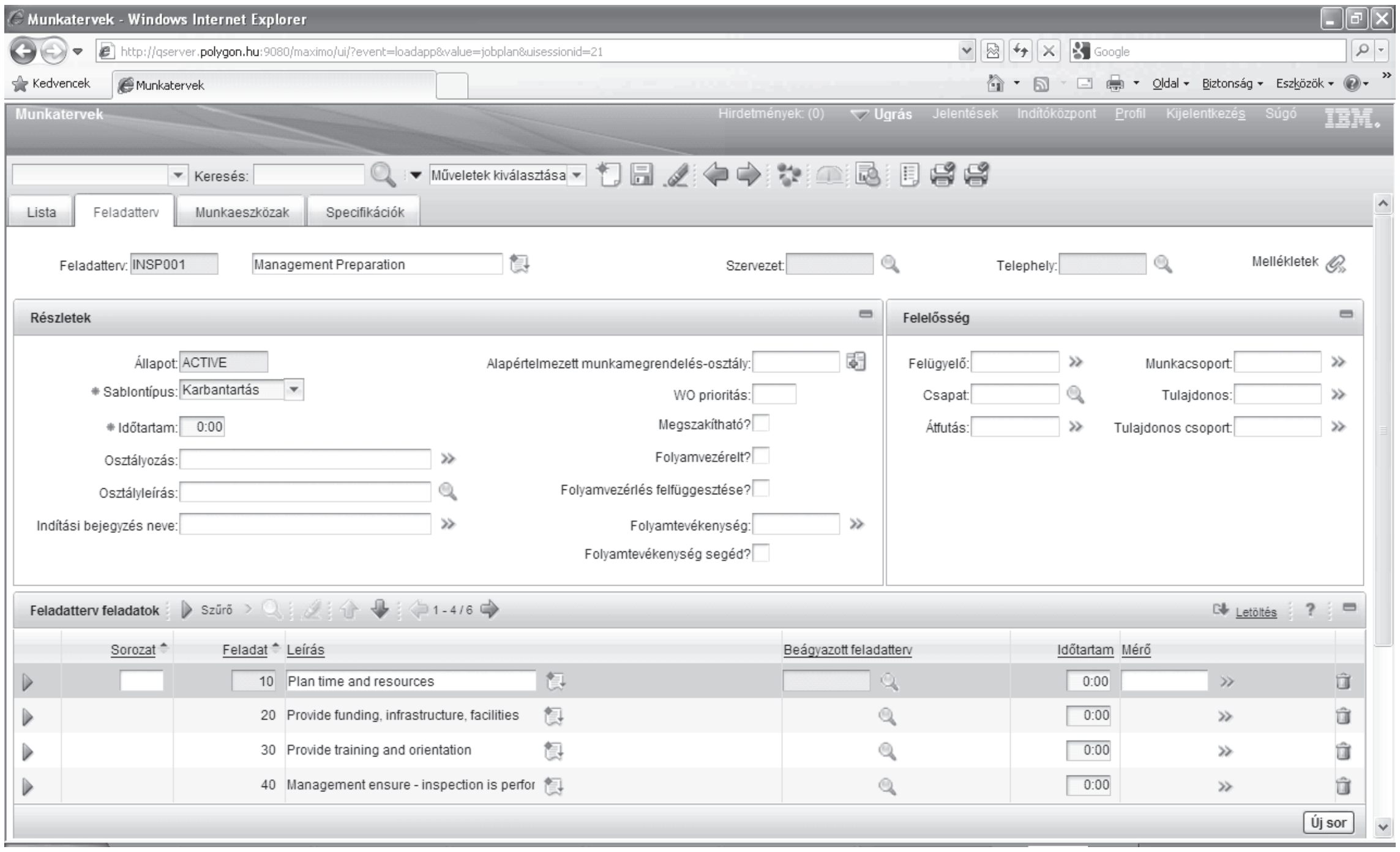

Figure 79 – Creating the job plan INSP001

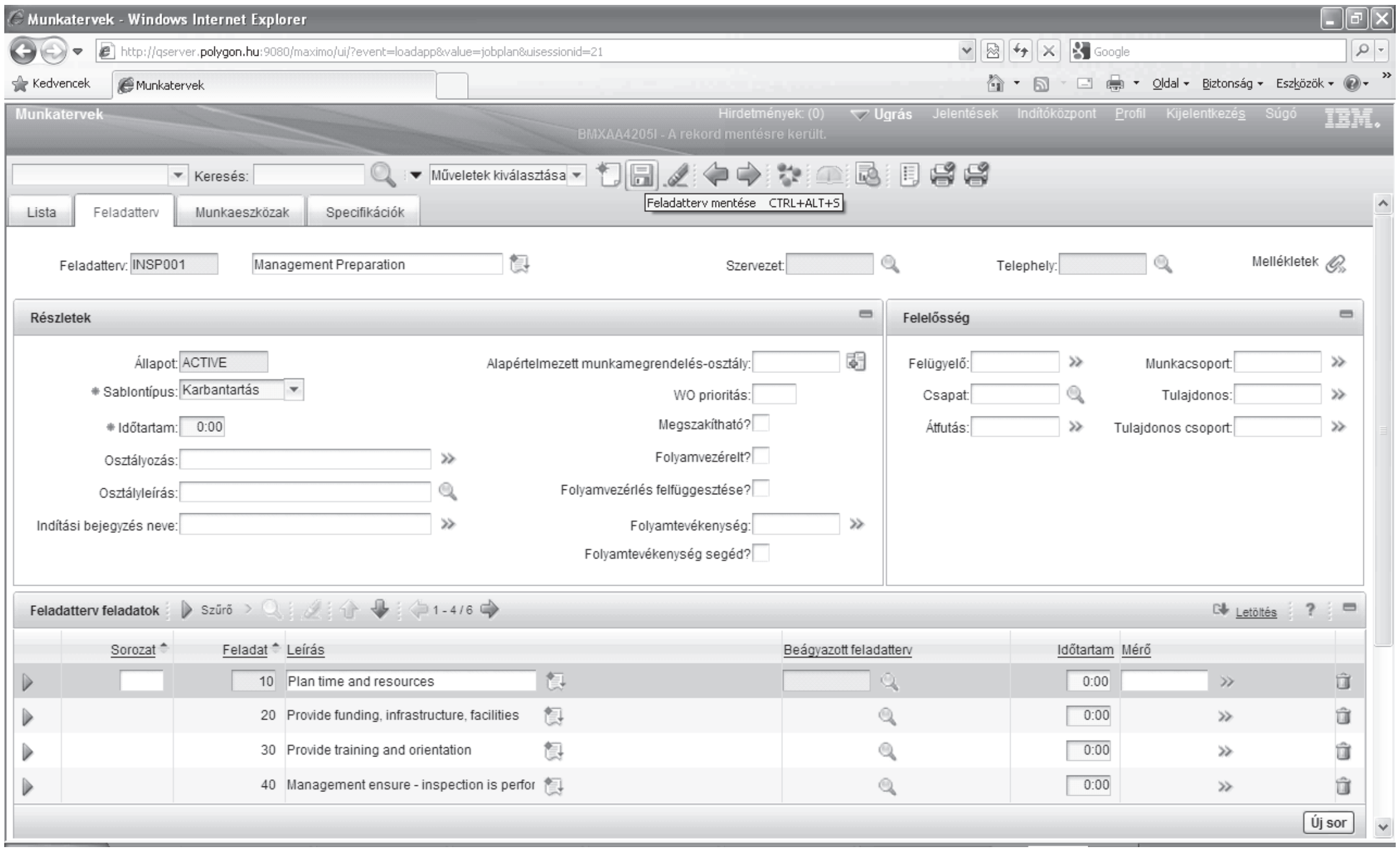

Figure 80 – Activating and saving the job plan

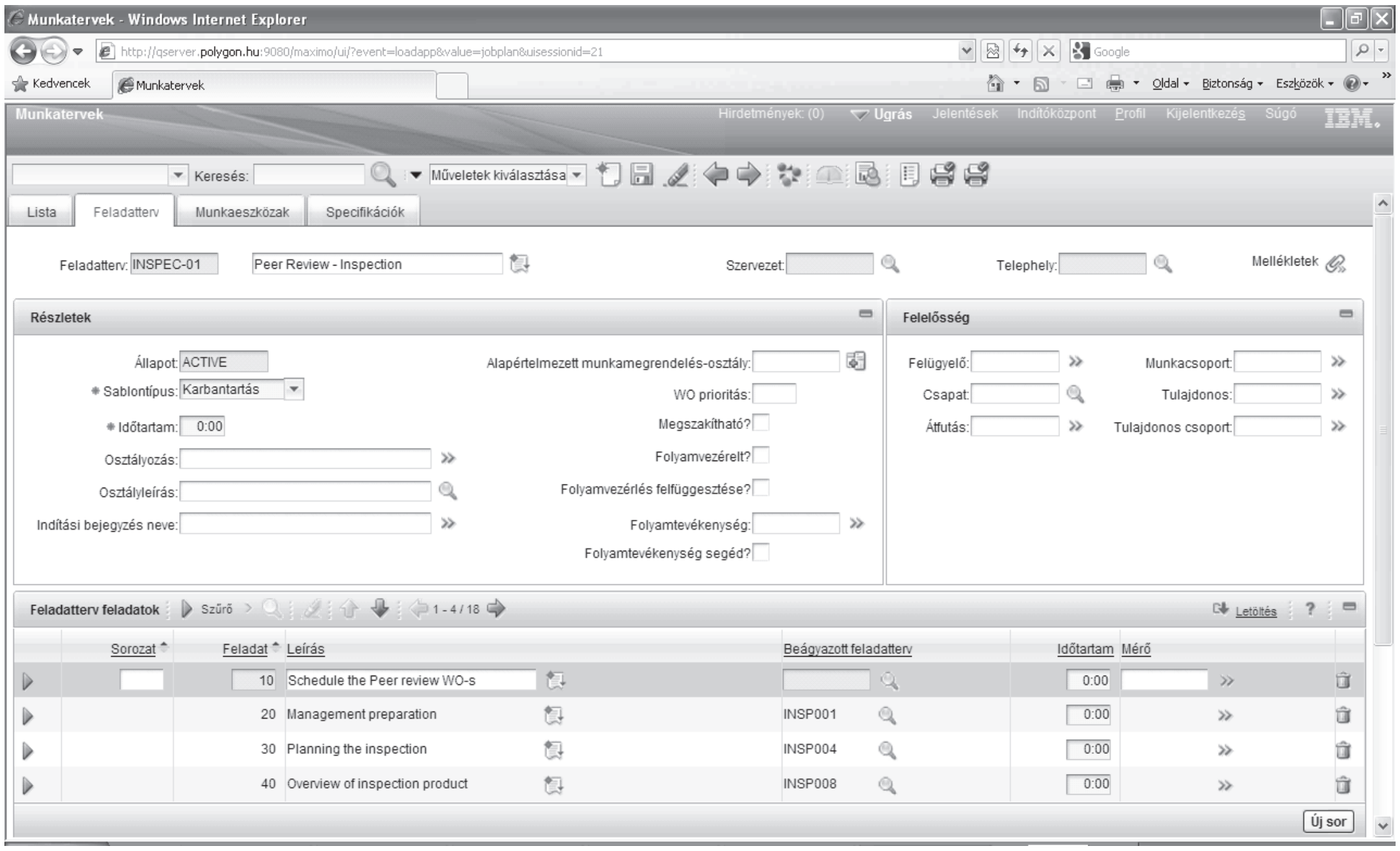

Figure 81 – Using nested job plans: INSPEC-01

# Appendix V. A data model for enhancing maintainability of PBU results

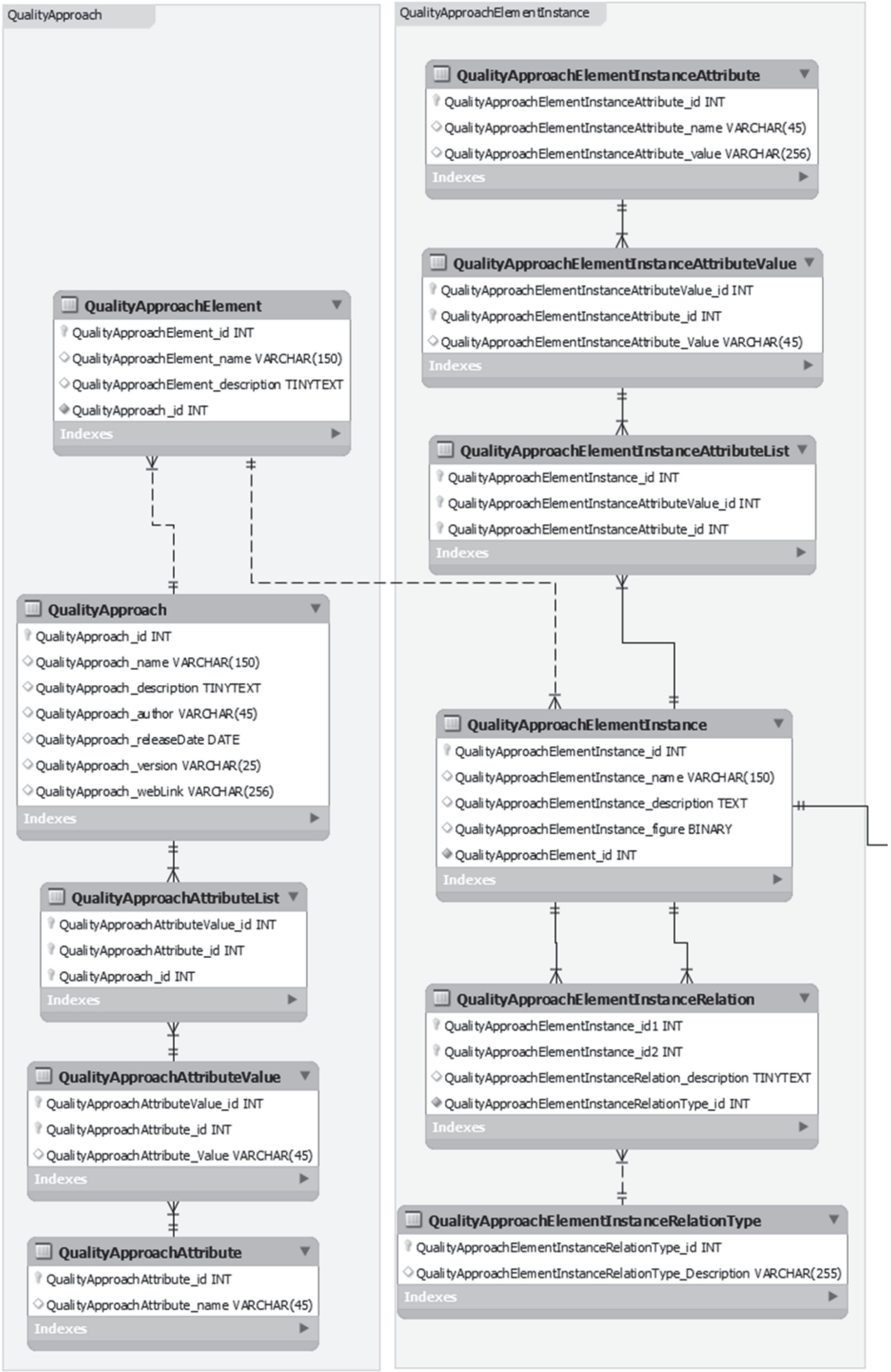

Figure 82 – An overview of the data model (part 1)

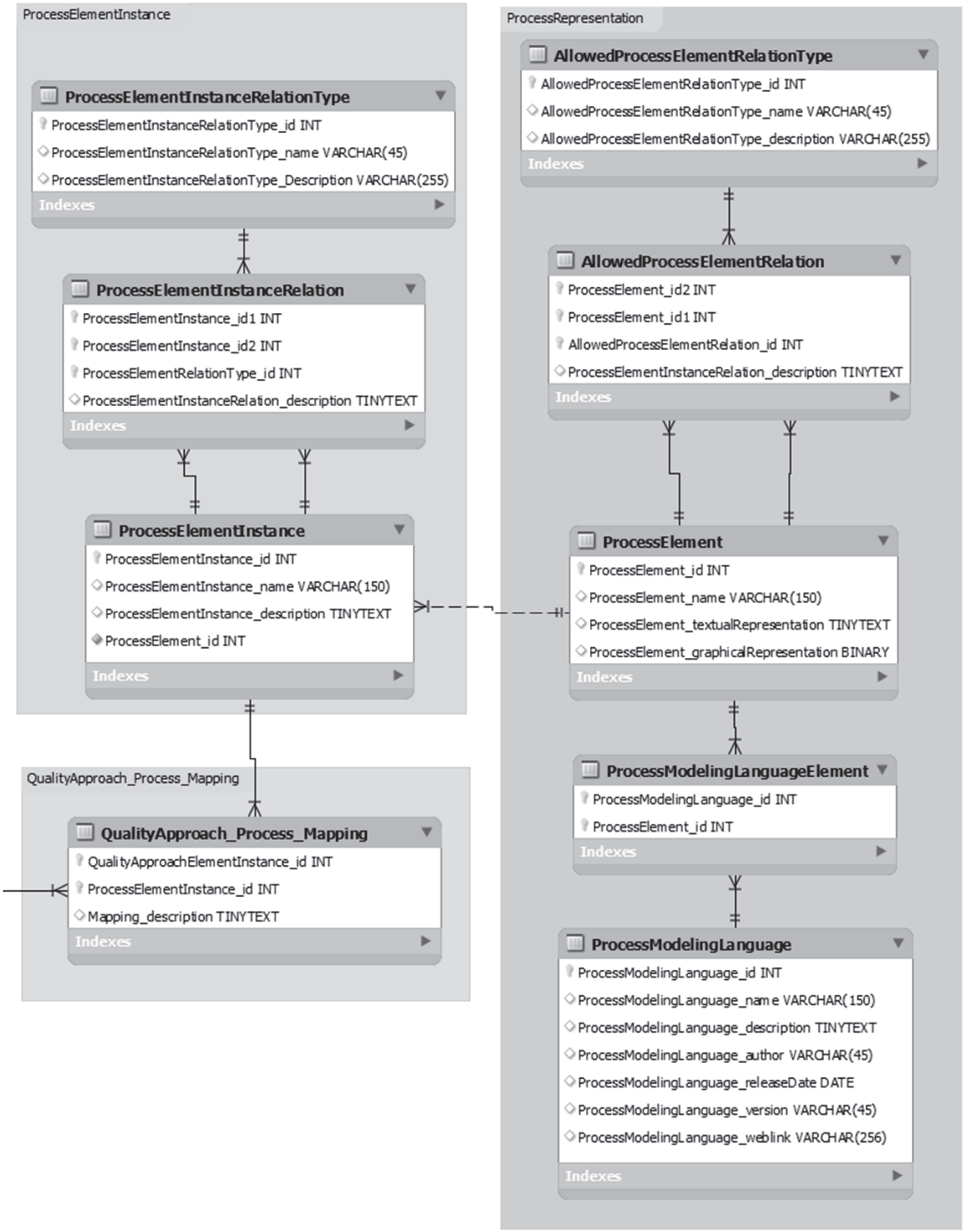

Figure 83 – An overview of the data model (part 2)

# Appendix W. Graphical representation of CMMI-DEV cross-references

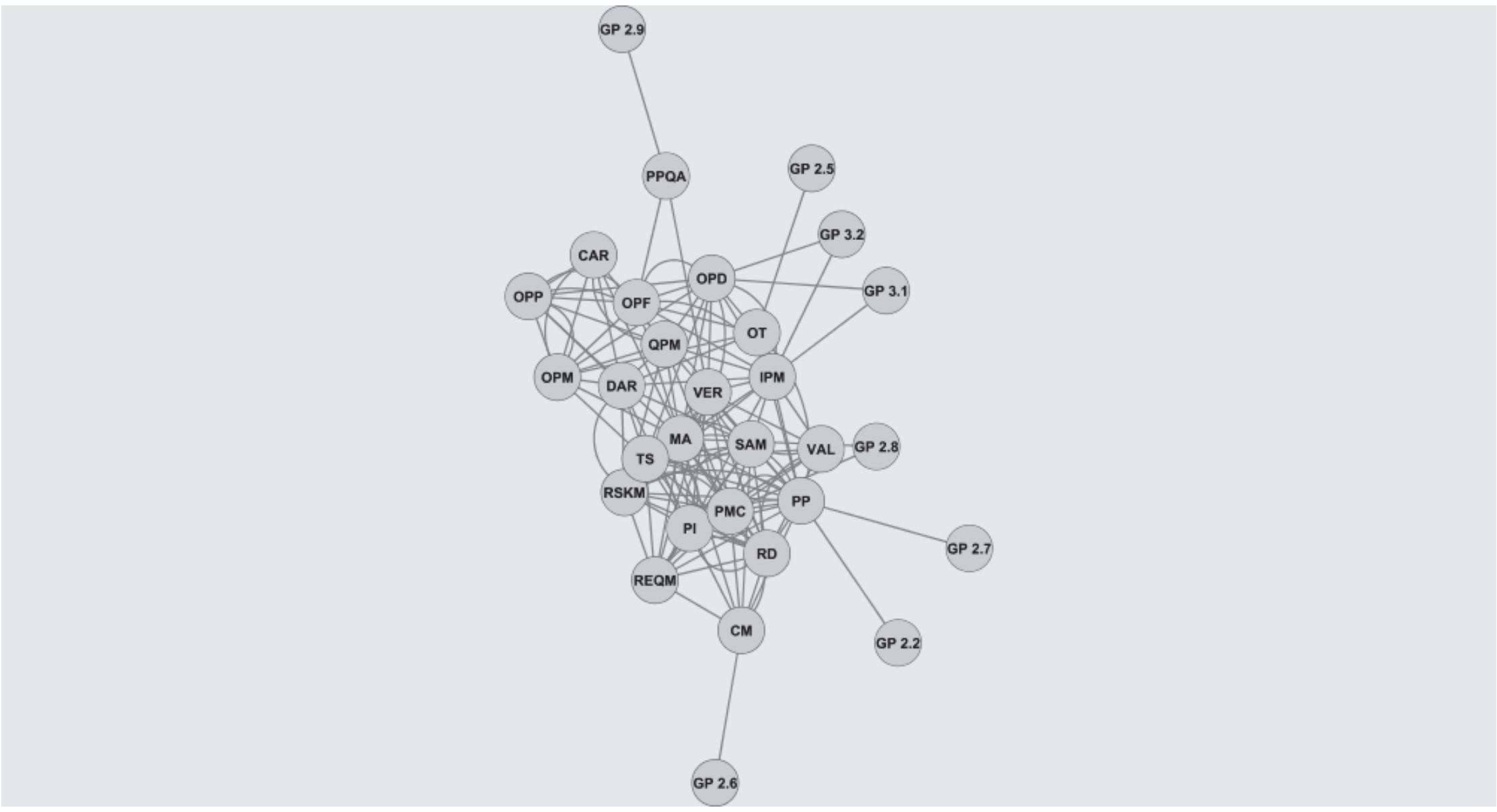

Figure 84 – A view of CMMI-DEV cross references in an unordered network

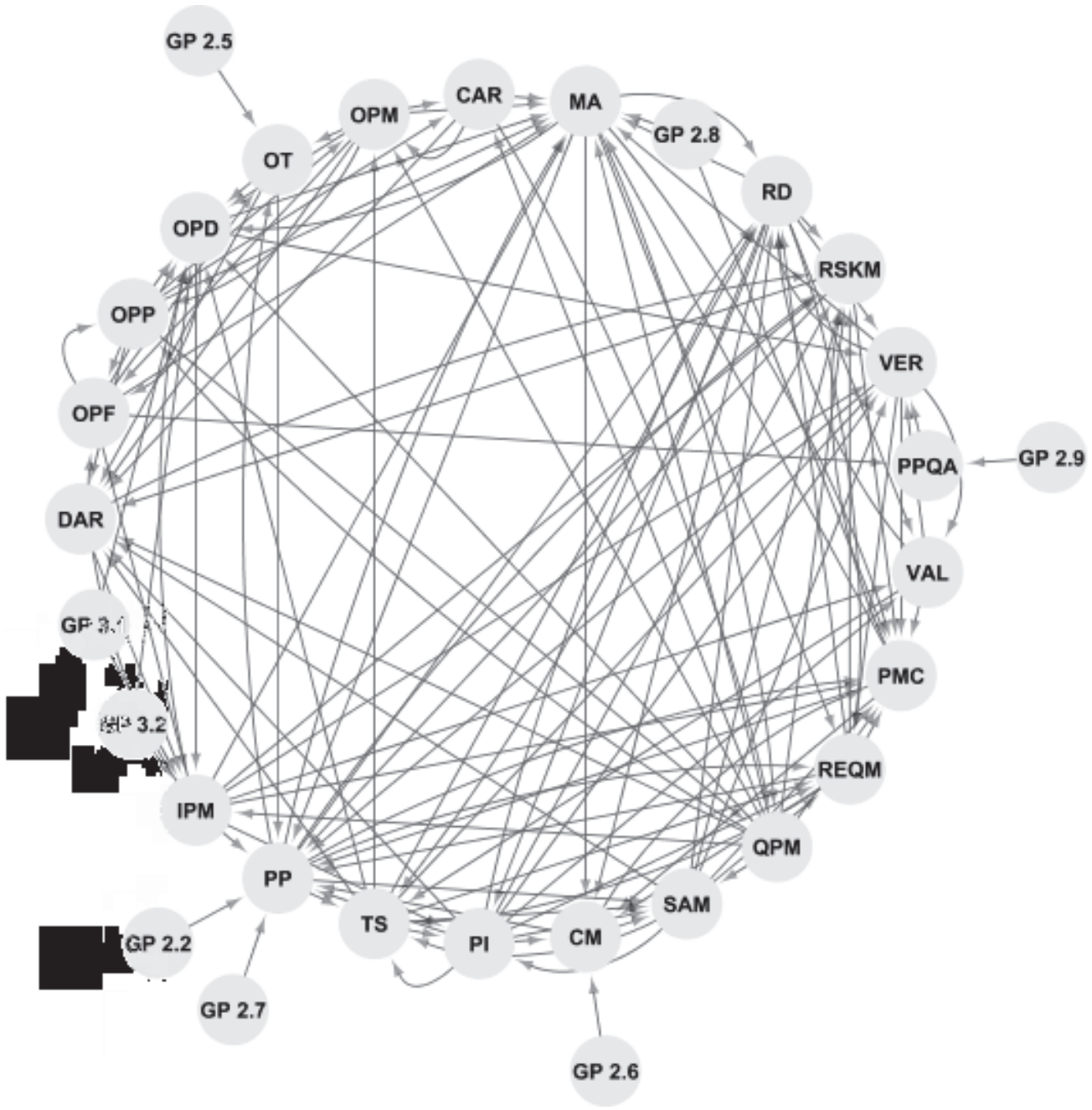

Figure 85 – A CMMI-DEV cross-references in a directed graph

# Appendix X. Most frequent words in CMMI v1.3

Table 61 – list of 30 most frequent words and trunks in CMMI v1.3

| | Tokenized wordlist | | | Truncated (Snowball) wordlist | | |
|---|---|---|---|---|---|---|
| # | Word | No of documents | Word count | Word | No of documents | Word count |
| 1 | process | 3 | 8946 | process | 3 | 10853 |
| 2 | work | 3 | 3706 | product | 3 | 4370 |
| 3 | project | 3 | 3170 | servic | 3 | 4219 |
| 4 | service | 3 | 2934 | work | 3 | 3751 |
| 5 | cmmi | 3 | 2682 | project | 3 | 3556 |
| 6 | management | 3 | 2532 | perform | 3 | 3501 |
| 7 | performance | 3 | 2437 | manag | 3 | 3459 |
| 8 | requirements | 3 | 2406 | requir | 3 | 3022 |
| 9 | product | 3 | 2338 | plan | 3 | 2988 |
| 10 | organization | 3 | 2194 | area | 3 | 2930 |
| 11 | area | 3 | 2044 | cmmi | 3 | 2682 |
| 12 | products | 3 | 1903 | organ | 3 | 2546 |
| 13 | processes | 3 | 1879 | includ | 3 | 2319 |
| 14 | organizational | 3 | 1641 | measur | 3 | 2124 |
| 15 | information | 3 | 1589 | risk | 3 | 2089 |
| 16 | version | 3 | 1577 | develop | 3 | 2017 |
| 17 | objectives | 3 | 1545 | establish | 3 | 1969 |
| 18 | include | 3 | 1538 | improv | 3 | 1924 |
| 19 | analysis | 3 | 1366 | exampl | 3 | 1863 |
| 20 | supplier | 3 | 1359 | object | 3 | 1798 |
| 21 | data | 3 | 1298 | inform | 3 | 1769 |
| 22 | services | 3 | 1285 | supplier | 3 | 1714 |
| 23 | training | 3 | 1274 | organiz | 3 | 1650 |
| 24 | development | 3 | 1262 | level | 3 | 1638 |
| 25 | quality | 3 | 1261 | identifi | 3 | 1636 |
| 26 | risk | 3 | 1225 | use | 3 | 1603 |
| 27 | plan | 3 | 1215 | version | 3 | 1594 |
| 28 | activities | 3 | 1203 | select | 3 | 1567 |
| 29 | level | 3 | 1113 | practic | 3 | 1549 |
| 30 | system | 3 | 1110 | model | 3 | 1446 |

# TERMS AND DEFINITIONS

Table 62 contains terms and definitions used in this research.

Table 62 – Terms and definitions

| Term | Definition |
|---|---|
| Standard | When we use the term *standard* we refer to the materials officially standardized and published by standardization organizations. Such international standardization organizations are e.g. the International Organization for Standardization (ISO, 2010), Institute of Electrical and Electronics Engineers (IEEE, 2010) or International Electrotechnical Commission (IEC, 2010). |
| Model | We use the term *model* when we refer to a published material which calls itself a model, e.g. CMM, CMMI or TMM. The models or standards are not necessarily standardised, but some of them are (e.g. SPICE is standardised as ISO 15504). |
| Method | *A method is a systematic procedure, technique or mode of inquiry, employed by or proper to a particular discipline; or: a body of skills or techniques.* (Blokdijk & Blokdijk, 1987) |
| Methodology | *Methodology means the science of method: a treatise or dissertation on method.* (Blokdijk & Blokdijk, 1987) |
| Technique | *A technique is a procedure or body of technical actions.* (Blokdijk & Blokdijk, 1987) |
| Notation | *A notation is a system of characters, symbols or abbreviated expressions used to express technical facts of quantities.* (Blokdijk & Blokdijk, 1987) |
| Improvement technology | *SEI deifintion:* "...we use the terms *improvement technologies*, technologies, or models somewhat interchangeably as shorthand when we are referring in general to the long list of reference models, standards, best practices, regulatory policies, and other types of practice-based improvement technologies that an organization may use simultaneously." (Siviy et al., 2008b). |
| Multi-model problem | In this research we call the *multi-model problem* the problem of the simultaneous usage of multiple quality approaches. |

| Term | Definition |
|---|---|
| Multi-model initiative | We call *multi-model initiative* all the initiatives which are aimed at solving the multi-model problem. |
| Multi-approach process improvement | See multi-model process improvement. |
| Multi-model Process Improvement | *Multi-approach process improvement* or *multi-model process improvement* mean process improvement based on multiple software quality approaches. According to our terminology, the first term would be more logical, but we use the latter one because it is getting emphasized in the field of process improvement (Apithanataveepa, 2008; A. L Ferreira et al., 2010; Malzahn, 2008; C. J. Pardo et al., 2009; Salviano, 2009a; Siviy et al., 2008e). These two terms have the same meaning. |
| Multi-model solution | A multi-model initiative which solves the multi-model problem is called a *multi-model solution*. |
| PBU concept | The concept of mapping quality approaches to a process is called the concept of Process Based Unification (PBU) or PBU concept. |
| PBU framework | The PBU framework is a multi-model solution which consist of: the PBU concept, PBU process and the PBU result (unified process + mapping) |
| PBU result | PBU result is the result of the PBU process and consist of a unified process and the mapping of quality approaches to the unified process |
| Unified process | The key PBU result is a single, unified process, to which quality approaches are mapped. This resulting process is called the unified process. |
| Process Based Unification | For the Process Based Unification (abbreviated as PBU): see the definition of the PBU concept. |
| Approach | In order to emphasize that each standard (e.g. ISO 9001 or ISO 12207), method, and (improvement) technology framework (e.g. CMMI, SPICE) is a specific approach to software quality, we call each of them an *approach.* |
| Quality approach | An approach which is connected to quality is called *quality approach.* |
| Software quality approach | A quality approach which can be used in software industry is called *software quality approach*. |
| Process-oriented quality approach | A (software) quality approach which mainly focuses on a process or more processes is called *process-oriented* (software) quality approach (abbreviated as po(s)qa) |
| Quality approach element | We call *quality approach element (type or class)* – an identified element of a quality approach (e.g. activity, artifact, role, chapter or requirement). These are types which can be mapped to process elements. E.g. specific goal and specific practice are elements of CMMI. We use the terms “quality approach element” and “quality approach element class” as synonyms. |
| Quality approach element instance | *Quality approach element instance* is the instantiation of a quality approach element. E.g. “prepare for review” can be an instance of “activity” element or a “project plan” can be an instance of the “artifact” element. |

| Term | Definition |
|---|---|
| Quality approach harmonization | Quality approach harmonization is the process of releasing a modified quality approach or the extension of an existing quality approach in accordance (or in a harmonized way) with one or more quality approach(es). In this sense harmonization results in a modified approach or addition that carries several characteristics of one or more quality approach(es) with which it is harmonized. Such common characteristics can be e.g. the common terminology, the common structure, or the common way of process descriptions. |
| Quality approach integration | An integrated quality approach is a quality approach which has been established on the basis of multiple quality approaches. Quality approach integration is the process of developing an integrated quality approach. A clear difference between this category and the previous harmonization category is that the source quality approaches are not left standalone, but that they are put together into a single, integrated quality approach carrying new shared characteristics, often replacing the source approaches. |
| Quality approach mapping | Quality approach mapping focuses on the identification of the requirements of two different quality approaches. Subsequently the identified requirements of the first approach are mapped to the requirements of the second approach. |

# ACRONYMS

Table 63 – Common acronyms used in this thesis

| Acronym | Definition |
|---|---|
| ADF | Activity-Decision Flow |
| ARC | Appraisal Requirements for CMMI |
| BPA | Business Process Automation |
| BPD | Business Process Diagram |
| BPEL | Business Process Execution Language |
| BPM | Business Process Modeling |
| BPMN | Business Process Modeling Notation |
| BPSS | Business Process Specification Schema |
| CMM | Capability Maturity Model |
| CMMI | Capability Maturity Model Integration |
| CMMI-ACQ | CMMI for Acquisition |
| CMMI-DEV | CMMI for Development |
| CMMI-SVC | CMMI for Services |
| ebXML | Electronic Business using eXtensible Markup Language |
| EDOC | Enterprise Distributed Object Computing |
| EPC | Event Driven Process Chain |
| EPML | EPC Markup Lanuage |
| GG | Generic Goal |
| GP | Generic Practice |
| iCMM | Integrated Capability Maturity Model |
| IEC | International Electrotechnical Commission |
| IEEE | Institute of Electrical and Electronics Engineers |
| ISO | International Organization for Standardization |
| IDEF | Integration Definition |
| LOVeM | Line of Visibility Enterprise Modeling |
| MSPI | Multi-model Software Process Improvement |

| Acronym | Definition |
|---|---|
| OMG | Object Management Group |
| PA | Process Area |
| PAM | Process Assessment Model |
| PBU | Process Based Unification |
| PML | Process Modeling Language |
| PRM | Process Reference Model |
| ROI | Return on Investment |
| QMIM | Quality through Managed Improvement and Measurement |
| SCAMPI | Standard CMMI Appraisal Method for Process Improvement |
| SEI | Software Engineering Institute |
| SG | Specific Goal |
| SP | Specific Practice |
| SPI | Software Process Improvement |
| SPICE | Software Process Improvement and Capability dEtermination |
| TMM | Test Maturity Model |
| TMMi | Test Maturity Model Integration |
| UML | Unified Modeling Language |
| UMM | UN/CEFACT Modelling Methodology |
| WfMV | Workflow Management Coalition |
| WS-BPEL | Web Services Business Process Execution Language |
| XMI | XML Metadata Interchange |
| XML | eXtensible Markup Language |
| XPDL | XML Process Definition Language |
| YAWL | Yet Another Workflow Language |
| QMS | Quality Management System |

# REFERENCES

Aalst, W. (1999). Formalization and verification of event-driven process chains. Information and Software Technology, 41, 639–650. doi:10.1016/S0950-5849(99)00016-6

Altova. (2011). UML Application Flow Diagrams. Retrieved December 12, 2011, from http://altova.com/umodel/application-flow-diagrams.html

Amis, J. M., & Silk, M. L. (2007). The Philosophy and Politics of Quality in Qualitative Organizational Research. Organizational Research Methods, 11(3), 456–480. doi:10.1177/1094428107300341

Apithanataveepa, A. (2008, March 18). Maximizing your SPI efforts through Multimodel Harmonization. Retrieved from http://www.swpark.or.th/spi@ease/register/SPI_Multiframework_by%20Tuvnord%20&%20Gosoft_03122008.pdf

Automotive Special Interest Group (SIG). (2008). Automotive SPICE© Process Reference Model (PRM). Retrieved from http://www.automotivespice.com/automotiveSIG_PRM_v44.pdf

Baldassarre, M., Caivano, D., Pino, F., Piattini, M., & Visaggio, G. (2010). A Strategy for Painless Harmonization of Quality Standards: A Real Case. Product-Focused Software Process Improvement, Lecture Notes in Computer Science (Vol. 6156, pp. 395–408). Springer Berlin / Heidelberg. Retrieved from http://dx.doi.org/10.1007/978-3-642-13792-1_30

Baldassarre, M. T., Piattini, M., Pino, F. J., & Visaggio, G. (2009). Comparing ISO/IEC 12207 and CMMI-DEV: Towards a mapping of ISO/IEC 15504-7. 2009 ICSE Workshop on Software Quality (pp. 59–64). Presented at the 2009 ICSE Workshop on Software Quality (WOSQ), Vancouver, BC, Canada. doi:10.1109/WOSQ.2009.5071558

Balla, K. (2001). The complex quality world : developing quality management systems for software companies. Eindhoven: Technische Universiteit Eindhoven.

Balla, K. (2004). SYNERGIC USE OF SOFTWARE QUALITY MODELS. Production Systems and Information Engineering, Volume 2, 73–89.

Balla, K., Bemelmans, T., Kusters, R., & Trienekens, J. (2001). Quality through Managed Improvement and Measurement (QMIM): Towards a Phased Development and Implementation of a Quality Management System for a Software Company. Software Quality Journal, 9(3), 177–193.

Balla, K., & Kelemen, Z. D. (2011, June 7). Important Concepts In CMMI and What is Difficult to Understand. Presented at the SEPG Europe 2011, Dublin. Retrieved from http://mycite.omikk.bme.hu/doc/105012.rar

Becker, J., Rosemann, M., & von Uthmann, C. (2000). Guidelines of Business Process Modeling. Retrieved from http://dx.doi.org/10.1007/3-540-45594-9_3

Bekkers, W., Weerd, I., Spruit, M., & Brinkkemper, S. (2010). A Framework for Process Improvement in Software Product Management. In A. Riel, R. O'Connor, S. Tichkiewitch, & R. Messnarz (Eds.), Systems, Software and Services Process Improvement (Vol. 99, pp. 1–12). Berlin, Heidelberg: Springer Berlin Heidelberg. Retrieved from http://www.springerlink.com/index/10.1007/978-3-642-15666-3_1

Bhuta, J., Boehm, B., & Meyers, S. (2006). Process Elements: Components of Software Process Architectures. In M. Li, B. Boehm, & L. J. Osterweil (Eds.), Unifying the Software Process Spectrum (Vol. 3840, pp. 332–346). Berlin, Heidelberg: Springer Berlin Heidelberg. Retrieved from http://www.springerlink.com/index/10.1007/11608035_28

Blank, S. (2006). The four steps to the epiphany : successful strategies for products that win. [Foster City, Calif.]: Cafepress.com.

Blokdijk, A., & Blokdijk, P. (1987). Planning and Desing of Information Systems.

Bóka, G., Balla, K., Kusters, R., & Trienekens, J. (2006). Towards Tool Support for Situation-Dependent Software Process Improvement. EuroSPI 2006 Industrial Proceedings (pp. 9.1–9.11.). Presented at the EuroSPI 2006, Joensuu, Finland.

Booth, T. L. (1967). Sequential machines and automata theory. Retrieved from http://platon.serbi.ula.ve/librum/librum_ula/ver.php?ndoc=220007

Buglione, L., Hauck, J. C. R., von Wangenheim, C. G., & McCaffery, F. (2012). Hybriding CMMI and Requirement Engineering Maturity & Capability Models - Applying the LEGO Approach for Improving Estimates. In S. Hammoudi, M. van Sinderen, & J. Cordeiro (Eds.), ICSOFT (pp. 55–61). SciTePress. Retrieved from http://dblp.uni-trier.de/db/conf/icsoft/icsoft2012.html#BuglioneHWM12

Center for History and New Media. (2009). Zotero. Retrieved September 21, 2010, from http://www.zotero.org/

Chen, X., Staples, M., & Bannerman, P. (2008). Analysis of Dependencies between Specific Practices in CMMI Maturity Level 2. In R. V. O'Connor, N. Baddoo, K. Smolander, & R. Messnarz (Eds.), Software Process Improvement (Vol. 16, pp. 94–105). Berlin, Heidelberg:

Springer Berlin Heidelberg. Retrieved from http://www.springerlink.com/index/10.1007/978-3-540-85936-9

Chow, T., & Cao, D.-B. (2008). A survey study of critical success factors in agile software projects. Journal of Systems and Software, 81(6), 961–971. doi:10.1016/j.jss.2007.08.020

CMMI Product Team, C. S. (2006, August). CMMI® for Development, Version 1.2. CMU SEI. Retrieved from http://www.sei.cmu.edu/reports/06tr008.pdf

CMMI Product Team, C. S. (2010a, November). CMMI® for Development, Version 1.3. CMU SEI. Retrieved from http://www.sei.cmu.edu/reports/10tr033.pdf

CMMI Product Team, C. S. (2010b, November). CMMI® for Services, Version 1.3. CMU SEI. Retrieved from http://www.sei.cmu.edu/reports/10tr034.pdf

CMMI Product Team, C. S. (2010c, November). CMMI® for Acquisition, Version 1.3. CMU SEI. Retrieved from http://www.sei.cmu.edu/reports/10tr032.pdf

Cooper, B., Vlaskovits, P., & Blank, S. G. (2010). The entrepreneur's guide to customer development : a "cheat sheet" to The four steps to the epiphany. [S.l.]: B. Cooper and P. Vlaskovitz.

Creswell, J. W., & Miller, D. L. (2000). Determining Validity in Qualitative Inquiry. Theory Into Practice, 39(3), 124–130. doi:10.1207/s15430421tip3903_2

Cugola, G., & Ghezzi, C. (1998). Software Processes: a Retrospective and a Path to the Future. Software Process: Improvement and Practice, 4(3), 101–123.

Curtis, B., Hefley, W., & Miller, S. (2001). People Capability Maturity Model®, The: Guidelines for Improving the Workforce. Boston: Addison-Wesley.

Curtis, B., Kellner, M. I., & Over, J. (1992). Process modeling. Communications of the ACM, 35(9), 75–90. doi:10.1145/130994.130998

Diaz, J., Garbajosa, J., & Calvo-Manzano, J. A. (2009). Mapping CMMI Level 2 to Scrum Practices: An Experience Report. In R. V. O'Connor, N. Baddoo, J. Cuadrago Gallego, R. Rejas Muslera, K. Smolander, & R. Messnarz (Eds.), Software Process Improvement, Communications in Computer and Information Science (Vol. 42, pp. 93–104). Springer Berlin Heidelberg. Retrieved from http://dx.doi.org/10.1007/978-3-642-04133-4_8

DMO. (2007, March). +SAFE, V1.2: A Safety Extension to CMMI-DEV, V1.2. SEI CMU. Retrieved from http://www.sei.cmu.edu/reports/07tn006.pdf

DTIC. (2008, November 18). Web page of 8th Annual CMMI Technology Conference & User Group "Investigation, Measures and Lessons Learned about the Relationship between CMMI® Process Capability and Project or Program Performance." Retrieved March 24, 2010, from http://www.dtic.mil/ndia/2008cmmi/2008cmmi.html

Dybå, T., & Dingsøyr, T. (2008). Empirical studies of agile software development: A systematic review. Information and Software Technology, 50(9-10), 833–859. doi:10.1016/j.infsof.2008.01.006

Ekert, D. (2009). Development of a Concept for Integrating various Quality Standards. EuroSPI 2009 Proceedings. Presented at the EuroSPI 2009, Alcala, Spain.

FAA. (2001, September). Integrated Capability Maturity Model® (FAA-iCMM®) version 2.0. The Federal Aviation Administration. Retrieved from http://www.faa.gov/about/office_org/headquarters_offices/aio/library/media/FAA-iCMMv2.pdf

Fagan, M. E. (1976). Design and Code Inspections to Reduce Errors in Program development. IBM Systems Journal, Vol. 15(No. 3), pp.182–211.

Fagan, M. E. (1986). Advances in Software Inspections. IEEE Transactions on Software Engineering, Vol. SE-12(No. 7), pp 744–751.

Fenton, N. E., & Pfleeger, S. L. (1997). Software metrics : a rigorous and practical approach. Boston: PWS Pub.

Ferreira, A. L, Machado, R. J., & Paulk, M. C. (2010). Size and Complexity Attributes for Multimodel Improvement Framework Taxonomy. Software Engineering and Advanced Applications (SEAA), 2010 36th EUROMICRO Conference on (pp. 306 –309). doi:10.1109/SEAA.2010.54

Ferreira, A., & Machado, R. (2009). Software Process Improvement in Multimodel Environments. 2009 Fourth International Conference on Software Engineering Advances (pp. 512–517). Presented at the 2009 Fourth International Conference on Software Engineering Advances (ICSEA), Porto, Portugal. doi:10.1109/ICSEA.2009.80

Ferreira, Andre L., Machado, R. J., & Paulk, M. C. (2010). Quantitative Analysis of Best Practices Models in the Software Domain. Presented at the 2010 Asia Pacific Software Engineering Conference. Retrieved from http://74.125.155.132/scholar?q=cache:xpz3NnEDcvoJ:scholar.google.com/+%22multimodel%22+%22software+process+improvement%22&hl=hu&num=100&as_sdt=2000

Fisher, C. M. (2010). Researching and writing a dissertation : the essential guidebook for business students. Harlow: Financial Times Prentice Hall.

Fonseca, J. A., & de Almeida Júnior, J. R. (2005). CMMI RAMS Extension Based on CENELEC Railway Standard. Computer Safety, Reliability, and Security. Retrieved from http://dx.doi.org/10.1007/11563228_1

Forrester, E., & Wemyss, G. (2011). CMMI® Version 1.3 and Beyond. Dublin, Ireland.

Fuggetta, A. (2000). Software process: A Roadmap. Proceedings of the conference on The future of Software engineering - ICSE ’00 (pp. 25–34). Presented at the the conference, Limerick, Ireland. doi:10.1145/336512.336521

Gao, Y. (2006). BPMN-BPEL transformation and round trip engineering. eClarus Software. Retrieved from http://eclarus.com/resources/BPMN_BPEL_Mapping.pdf

Garcia, S., Graettinger, C., & Kost, K. (2005). Proceedings of the First International Research Workshop for Process Improvement in Small Settings.

García-Mireles, G. A., Ángeles Moraga, M., & García, F. (2012). Development of maturity models: a systematic literature review (pp. 279–283). IET. doi:10.1049/ic.2012.0036

García-Mireles, G. A., Moraga, M. Á., García, F., & Piattini, M. (2012). Towards the Harmonization of Process and Product Oriented Software Quality Approaches. In D. Winkler, R. V. O'Connor, & R. Messnarz (Eds.), Systems, Software and Services Process Improvement (Vol. 301, pp. 133–144). Berlin, Heidelberg: Springer Berlin Heidelberg. Retrieved from http://www.springerlink.com/index/10.1007/978-3-642-31199-4_12

Ghose, A., Koliadis, G., & Chueng, A. (2007). Process Discovery from Model and Text Artefacts. 2007 IEEE Congress on Services (Services 2007) (pp. 167–174). Presented at the 2007 IEEE Congress on Services (Services 2007), Salt Lake City, UT, USA. doi:10.1109/SERVICES.2007.52

Gibbert, M., & Ruigrok, W. (2010). The What'' and How'' of Case Study Rigor: Three Strategies Based on Published Work. Organizational Research Methods, 13(4), 710–737. doi:10.1177/1094428109351319

Gilb, T., & Graham, D. (1993). Software inspection. Wokingham  England ;;Reading  Mass.: Addison-Wesley.

Given, L. M. (2008). The SAGE Encyclopedia of Qualitative Research Methods (Two Volume Set) (1st ed., Vol. 1, 2). Sage Publications, Inc.

Golafshani, N. (2003). Understanding Reliability and Validity in Qualitative Research. The Qualitative Report, 8(4), 597–607.

Graham, D., Van Veenendaal, E., Evans, I., & Black, R. (2007). Foundations of software testing : ISTQB certification. Australia: Intl Thomson Business Pr.

Gray, E. M., & Smith, W. L. (1998). On the limitations of software process assessment and the recognition of a required re-orientation for global process improvement. Software Quality Journal, 7(1), 21–34. doi:10.1023/B:SQJO.0000042057.89615.60

Griffith University. (2007). Appraisal Assistant-A CMMI appraisal tool/ISO 15504 assessment tool. Retrieved March 11, 2010, from http://www.sqi.gu.edu.au/AppraisalAssistant/about.html

Habra, N., Alexandre, S., Desharnais, J.-M., Laporte, C. Y., & Renault, A. (2008). Initiating software process improvement in very small enterprises. Information and Software Technology, 50(7-8), 763–771. doi:10.1016/j.infsof.2007.08.004

Halvorsen, C. P., & Conradi, R. (1999). A Taxonomy of SPI Frameworks. in Proc. 24th NASA Software Engineering Workshop (pp. 1–2).

Halvorsen, C. P., & Conradi, R. (2001). A Taxonomy to Compare SPI Frameworks. Lecture notes in computer science, 2077, 217–235.

Hambling, B., Morgan, P., Samaroo, A., Thompson, G., & Williams, P. (2007). Software testing : an ISEB foundation. Swindon  U.K.: British Computer Society.

Hammer, M. (2002). Process management and the future of Six Sigma. MIT Sloan Management Review, 43(2), 26–32.

Harjumaa, L., Tervonen, I., & Huttunen, A. (2005). Peer Reviews in Real Life - Motivators and Demotivators. Fifth International Conference on Quality Software (QSIC'05) (pp. 29–36). Presented at the Fifth International Conference on Quality Software (QSIC'05), Melbourne, Australia. doi:10.1109/QSIC.2005.48

Helgesson, Y. Y. L., Höst, M., & Weyns, K. (2011). A review of methods for evaluation of maturity models for process improvement. Journal of Software Maintenance and Evolution: Research and Practice, n/a–n/a. doi:10.1002/smr.560

Heston, K. M., & Phifer, W. (2010). The multiple quality models paradox: how much "best practice" is just enough? Journal of Software Maintenance and Evolution: Research and Practice. doi:10.1002/spip.434

Hoepfl, M. C. (1997). Choosing Qualitative Research: A Primer for Technology Education Researchers. Journal of Technology Education, 9(1).

Holschke, O., Rake, J., & Levina, O. (2009). Granularity as a Cognitive Factor in the Effectiveness of Business Process Model Reuse. In U. Dayal, J. Eder, J. Koehler, & H. A. Reijers (Eds.), Business Process Management (Vol. 5701, pp. 245–260). Berlin, Heidelberg: Springer Berlin Heidelberg. Retrieved from http://www.springerlink.com/index/10.1007/978-3-642-03848-8_17

Homchuenchom, D., Piyabunditkul, C., Lichter, H., & Anwar, T. (2011). SPIALS: A lightweight Software Process Improvement self-assessment tool (pp. 195–199). IEEE. doi:10.1109/MySEC.2011.6140668

Humphrey, W. (1997). Introduction to the Personal Software Process. Reading  Mass.: Addison-Wesley Pub.

IBM. (2010, March). IBM Maximo technology for business and IT agility (whitepaper no: TIW10356-USEN-01). Retrieved from http://public.dhe.ibm.com/common/ssi/ecm/en/tiw10356usen/TIW10356USEN.PDF

IBM. (2011a). IBM Maximo Asset Management Version 7 Release 5: Administering Maximo Asset Management. Retrieved from http://publib.boulder.ibm.com/infocenter/tivihelp/v49r1/topic/com.ibm.mbs.doc/pdfs/pdf_mbs_sysadmin.pdf

IBM. (2011b). IBM Maximo Asset Management Version 7 Release 5: Workflow Implementation Guide. Retrieved from http://publib.boulder.ibm.com/infocenter/tivihelp/v49r1/topic/com.ibm.mbs.doc/pdfs/pdf_mbs_workflow.pdf

Ibrahim, L. (2010). Enterprise SPICE (ISO/IEC 15504) Process Assessment Model  (Process Dimension). The SPICE User Group. Retrieved from http://enterprisespice.com/EnterpriseSPICEsources.pdf

IEC. (2010). Webpage of IEC - International Electrotechnical Commission. IEC - International Electrotechnical Commission > INTERNATIONAL STANDARDS AND CONFORMITY ASSESSMENT. Retrieved March 15, 2010, from http://www.iec.ch/

IEEE. (2010). Webpage of Institute of Electrical and Electronics Engineers. IEEE - the world's leading professional association for the advancement of technology. Retrieved March 15, 2010, from http://www.ieee.org/portal/site

IEEE, C. S. (2008a). IEEE 829-2008 IEEE Standard for Software and System Test Documentation.

IEEE, C. S. (2008b). IEEE 1028-2008 IEEE Standard for Software Reviews.

ISCN. (2010). Capability Adviser functions list. Retrieved March 11, 2010, from http://www.iscn.com/capadv/functions.html

ISO. (2008). ISO 9001:2008: Quality Management Systems. Requirements.

ISO. (2010). ISO - International Organization for Standardization Webpage. Retrieved March 15, 2010, from http://www.iso.org/iso/home.html

ISO/IEC. (2001). ISO/IEC 9126-1:2001 Software engineering -- Product quality -- Part 1: Quality model. Retrieved from http://www.iso.org/iso/catalogue_detail.htm?csnumber=22749

ISO/IEC. (2003a). ISO/IEC TR 9126-2:2003 Software engineering -- Product quality -- Part 2: External metrics. Retrieved from http://www.iso.org/iso/iso_catalogue/catalogue_tc/catalogue_detail.htm?csnumber=22750

ISO/IEC. (2003b). ISO/IEC TR 9126-3:2003 Software engineering -- Product quality -- Part 3: Internal metrics. Retrieved from http://www.iso.org/iso/iso_catalogue/catalogue_tc/catalogue_detail.htm?csnumber=22891

ISO/IEC. (2004a). ISO/IEC TR 9126-4:2004 Software engineering -- Product quality -- Part 4: Quality in use metrics. Retrieved from http://www.iso.org/iso/iso_catalogue/catalogue_tc/catalogue_detail.htm?csnumber=39752

ISO/IEC. (2004b). ISO/IEC 15504-4:2004 Information technology -- Process assessment -- Part 4: Guidance on use for process improvement and process capability determination (SPICE).

ISO/IEC. (2004c). ISO/IEC 90003:2004 Software Engineering – Guidelines for the application of ISO9001:2000 to computer software.

ISO/IEC/IEEE. (2008). ISO/IEC/IEEE 12207:2008 Systems and software engineering — Software life cycle processes.

ISTQB. (2009). International Software Testing Qualifications Board Homepage. Retrieved April 26, 2010, from http://www.istqb.org/

itp-commerce. (2012). Itp-commerce BPMN Process Modeler. Retrieved July 6, 2012, from http://www.itp-commerce.com/en/process-modeler/

Kalpic, B., & Bernus, P. (2002). Business process modelling in industry—the powerful tool in enterprise management. Computers in Industry, 47(3), 299–318. doi:10.1016/S0166-3615(01)00151-8

Kaplan, B., & Maxwell, J. A. (2005). Qualitative research methods for evaluating computer information systems. Evaluating the Organizational Impact of Health Care Information Systems. Retrieved from http://www.libreriafarmaceutica.com/cover_note/books/4/8/5/9780387245584/9780387245584-c1.pdf

Kasser, J. (2005). Introducing the role of process architecting. INCOSE 2005: 15th Annual International Symposium: Systems Engineering: Bridging Industry, Government and Academia: Proceedings (pp. p. 1–12 1). Presented at the INCOSE 2005: 15th Annual International Symposium: Systems Engineering: Bridging Industry , Government and Academia, Rochester, New York. Retrieved from http://therightrequirement.com/pubs/2005/Introducing%20the%20Role%20of%20Process%20Architecting%202p.pdf

Kelemen, Z. D. (2008, June 5). Verification and Validation in CMMI (in Hungarian). Presented at the Software Testing Workshop of IIR (Institute of International Research), Budapest.

Kelemen, Z. D. (2009). Simultaneous use_ of process-oriented software quality approaches. SEPG Europe 2009 Prague. Retrieved from http://mycite.omikk.bme.hu/doc/72215.ppsx

Kelemen, Z. D. (2010a, May 11). Implementing CMMI in IBM Maximo Environment. Presented at the Benchmarking Conference of Hungarian Software Innovation Pole Cluster, Szeged.

Kelemen, Z. D. (2010b, May 19). Developing a Multimodel Peer Review Process in Q-Works project. Presented at the Q-Works International Workshop on Automated workflow system for the quality control of software developments, Budapest.

Kelemen, Z. D. (2011a). Additional materials for the article Identifying Criteria for Multimodel Software Process Improvement Solution – Based on a Review of Current Problems and Initiatives. Retrieved February 14, 2011, from http://sqi.hu/documents/multimodel/literature/

Kelemen, Z. D. (2011b, June 28). What is important in CMMI and what are the interrelations among its elements? Presented at the International ODF Symposium, Budapest. Retrieved from http://www.multiracio.com/doc/szomin08_kelemendaniel.pdf

Kelemen, Z. D., & Balla, K. (2008). A CMMI-DEV v1.2 és az ISO 9001:2000 kapcsolata. Magyar Minőség, 17(2), 27–40.

Kelemen, Z. D., & Balla, K. (2009). Experiences of teaching code reviews for IT students. EuroSPI 2009 Conference Proceedings. Presented at the European Systems & Software Process Improvement and Innovation, University of Alcala, Spain. Retrieved from http://2009.eurospi.net/

Kelemen, Z. D., Balla, K., & Bóka, G. (2007). Quality Organizer: a support tool in using multiple quality approaches. Proceedings of 8th International Carpathian Control Conference (ICCC 2007). Presented at the 8th International Carpathian Control Conference (ICCC 2007), Strbske Pleso, Slovakia.

Kelemen, Z. D., Balla, K., Trienekens, J., & Kusters, R. (2008a). Structure of Process-Based Quality Approaches - Elements of a research developing a common meta-model for proces-based quality approaches and methods. Proceedings of EuroSPI 2008 Doctoral Symposium. Presented at the European Systems & Software Process Improvement and Innovation, Dublin, Ireland.

Kelemen, Z. D., Balla, K., Trienekens, J., & Kusters, R. (2008b). Towards supporting simultaneous use of process-based quality approaches. Presented at the International Carpathian Control Conference ICCC' 2008, Sinaia, Romania.

Kelemen, Z. D., Kusters, R., & Trienekens, J. (2011). Identifying Criteria for Multimodel Software Process Improvement Solutions - Based on a Review of Current Problems and Initiatives. Journal of Software Maintenance and Evolution: Research and Practice, incorporating Software Process: Improvement and Practice. doi:10.1002/smr.549

Kelemen, Z. D., Kusters, R., Trienekens, J., & Balla, K. (2009). A Process Based Unification of Process-Oriented Software Quality Approaches. 2009 Fourth IEEE International Conference on Global Software Engineering (pp. 285–288). Presented at the 2009 Fourth IEEE International Conference on Global Software Engineering (ICGSE), Limerick, Ireland: IEEE Computer Society. doi:10.1109/ICGSE.2009.39

Kellner, M. I., Madachy, R. J., & Raffo, D. M. (1999). Software process simulation modeling: Why? What? How? Journal of Systems and Software, 46(2-3), 91–105. doi:10.1016/S0164-1212(99)00003-5

Kitchenham, B. (2004). Procedures for Performing Systematic Reviews. Keele University. Retrieved from http://citeseerx.ist.psu.edu/viewdoc/download?doi=10.1.1.122.3308&rep=rep1&type=pdf

Kojo Arhinful, D., Mohan Das, A., Prawitasari Hadiyono, J., Heggenhougen, K., & Higginbotham, N. (2000). HOW TO USE APPLIED QUALITATIVE METHODS TO DESIGN DRUG USE INTERVENTIONS. Bostin University. Retrieved from http://dcc2.bumc.bu.edu/prdu/how_to_use_applied_qualitative_m.htm

Korsaa, M., Biro, M., Messnarz, R., Johansen, J., Vohwinkel, D., Nevalainen, R., & Schweigert, T. (2010). The SPI manifesto and the ECQA SPI manager certification scheme. Journal of Software Maintenance and Evolution: Research and Practice, n/a–n/a. doi:10.1002/smr.502

Kugler, M. (2008, November 4). Web page of Central and Eastern Europe SPI 2008 conference / Highlights and Summary. Retrieved March 24, 2010, from http://www.cee-spi.com/1708.html

Lanubile, F., Ebert, C., Prikladnicki, R., & Vizcaino, A. (2010). Collaboration Tools for Global Software Engineering. IEEE Software, 27, 52–55. doi:10.1109/MS.2010.39

Lincoln, Y. S., & Guba, E. G. (1985). Naturalistic inquiry. Beverly Hills, Calif.: Sage Publications.

Linders, B. (2011, March 4). Agile Process Improvement. Retrieved May 2, 2012, from http://www.agilejournal.com/articles/columns/column-articles/5847-agile-process-improvement

M. De Oliveira, J., De Oliveira, K., & Belchior, A. (2006). Measurement Process: A Mapping Among CMMI-SW, ISO/IEC 15939, IEEE Std 1061, Six Sigma and PSM. 2006 International Conference on Service Systems and Service Management (pp. 810–815). Presented at the 2006 International Conference on Service Systems and Service Management, Troyes, France. doi:10.1109/ICSSSM.2006.320566

Ma, Q., Zhou, N., Zhu, Y., & Wang, H. (2009). Evaluating Service Identification with Design Metrics on Business Process Decomposition (pp. 160–167). IEEE. doi:10.1109/SCC.2009.44

Macke, S. (2011). Event-driven process chain. Retrieved December 12, 2011, from http://dia-installer.de/shapes/edpc/

Madison, D. (2008). Process mapping, process improvement, and process management : a practical guide to enhancing work and information flow. Chico, Calif.: Paton Press.

Malzahn, D. (2008, June 12). Ontology based multi-model process improvement for business organizations. Presented at the SEPG Europe, Munich. Retrieved from http://www.accel-gmbh.de/LinkClick.aspx?fileticket=XYsCFXgvcnA%3D&tabid=81&mid=399

Mendling, J. (2012). EPML homepage. Retrieved May 29, 2012, from http://www.mendling.com/EPML/

Mendling, J., Neumann, G., & Nüttgens, M. (2005). Towards Workflow Pattern Support of Event-Driven Process Chains (EPC. Proc. of the 2nd Workshop XML4BPM 2005 (pp. 23–38). doi:10.1.1.59.1149

Mendling, J., Neumann, G., & van der Aalst, W. (2007). Understanding the Occurrence of Errors in Process Models Based on Metrics. On the Move to Meaningful Internet Systems 2007: CoopIS, DOA, ODBASE, GADA, and IS. Retrieved from http://dx.doi.org/10.1007/978-3-540-76848-7_9

Mendling, J., & Nüttgens, M. (2005). EPC Markup Language (EPML) - An XML-Based Interchange Format for Event-Driven Process Chains (EPC). Vienna University of Economics and Business Administration. Retrieved from http://www.wiso.uni-hamburg.de/fileadmin/wiso_fs_wi/Team/Mitarbeiter/Prof._Dr._Markus_Nuettgens/Publikationen/ISEB.pdf

Mendling, J., Reijers, H., & Cardoso, J. (2007). What Makes Process Models Understandable? Business Process Management. Retrieved from http://dx.doi.org/10.1007/978-3-540-75183-0_4

Merriam-Webster Inc. (1996). Dictionary and Thesaurus - Merriam-Webster Online. Merriam-Webster Online Dictionary and Thesaurus. Retrieved April 26, 2010, from http://www.merriam-webster.com/

MethodPark. (2012). Effectively Managing Process Compliance Systems Engineering in the Face of Multiple Models, Standards and Best Practices. Retrieved from http://www.methodpark.com/fileadmin/downloads/product/Factsheets/Whitepaper_EffectivelyManagingCompliance.pdf

Mirna, M., Jezreel, M., Giner, A., A., C.-M. J., & Tom´s, S. F. (2011). Advantages of Using a Multi-model Environment in Software Process Improvement (pp. 397–402). IEEE. doi:10.1109/CERMA.2011.85

Monteiro, P., Machado, R. J., Kazman, R., & Henriques, C. (2010). Dependency Analysis between CMMI Process Areas. In M. Ali Babar, M. Vierimaa, & M. Oivo (Eds.), Product-Focused Software Process Improvement (Vol. 6156, pp. 263–275). Berlin, Heidelberg: Springer Berlin Heidelberg. Retrieved from http://www.springerlink.com/index/10.1007/978-3-642-13792-1_21

Moore, J. W. (1999). An integrated collection of software engineering standards. IEEE Software, 16(6), 51–57. doi:10.1109/52.805473

Morgan, P. (2010). Software testing an ISTQB-ISEB foundation guide. Swindon, U.K. :: British Informatics Society,.

Murata, T. (1989). Petri nets: Properties, analysis and applications. Proceedings of the IEEE, 77(4), 541–580. doi:10.1109/5.24143

Mutafelija, B., & Stromberg, H. (2003). Systematic process improvement using ISO 9001:2000 and CMMI. Boston: Artech House.

Mutafelija, B., & Stromberg, H. (2009a). Process improvement with CMMI v1.2 and ISO standards. Boca Raton: CRC Press.

Mutafelija, B., & Stromberg, H. (2009b, July). Mapping ISO standards to CMMI v1.2. SEIR - Software Engineering Information Repository. Retrieved October 7, 2010, from https://seir.sei.cmu.edu/seir/domains/CMMi/General/BSCW2009-ISO-to-CMMI/frmset.BSCW2009-ISO-to-CMMI.asp?DOMAIN=CMMi&SECTION=General&SUBSECTION=Documents&SUBSUBSECTION=Mappings&POSITION_ON_SEIR_PAGE=266&OUTLINE=Mappings

NDIA. (2009, November 16). Web page of 9th ANNUAL TECHNOLOGY CONFERENCE “Investigation, Measures, and Lessons Learned About the Relationship Between CMMI Process Capability and Project or Program Performance.” Retrieved March 24, 2010, from http://www.dtic.mil/ndia/2009CMMI/2009CMMI.html

Objects by Design. (2011). UML Modeling Tools. Retrieved December 12, 2011, from http://www.objectsbydesign.com/tools/umltools_byCompany.html

OMG. (2012). BPMN Information Home. Retrieved May 29, 2012, from http://www.bpmn.org/

Osterweil, L. J. (1997). Software processes are software too, revisited: an invited talk on the most influential paper of ICSE 9, 540–548. doi:http://dx.doi.org/http://doi.acm.org/10.1145/253228.253440

Osterweil, L. J. (2007). What we learn from the study of ubiquitous processes. Software Process: Improvement and Practice, 12(5), 399–414. doi:10.1002/spip.331

Pardo, C. J., Pino, F. J., García, F., & Piattini, M. (2009). Homogenization of Models to Support Multi-model Processes in Improvement Environments. ICSOFT (1) (pp. 151–156).

Pardo, C., Pino, F., García, F., Romero, F. R., Piattini, M., & Baldassarre, M. T. (2011). HProcessTOOL: A Support Tool in the Harmonization of Multiple Reference Models. In B. Murgante, O. Gervasi, A. Iglesias, D. Taniar, & B. O. Apduhan (Eds.), Computational Science and Its Applications - ICCSA 2011 (Vol. 6786, pp. 370–382). Berlin, Heidelberg: Springer Berlin Heidelberg. Retrieved from http://www.springerlink.com/index/10.1007/978-3-642-21934-4_30

Pardo, C., Pino, F. J., García, F., Piattini, M., & Baldassarre, M. T. (2010). A process for driving the harmonization of models. Proceedings of the 11th International Conference on Product Focused Software - PROFES '10 (pp. 51–54). Presented at the the 11th International Conference, Limerick, Ireland. doi:10.1145/1961258.1961271

Pardo, C., Pino, F. J., García, F., Piattini, M., & Baldassarre, M. T. (2011). An ontology for the harmonization of multiple standards and models. Computer Standards & Interfaces. doi:10.1016/j.csi.2011.05.005

Pardo, C., Pino, F. J., García, F., Piattini Velthius, M., & Baldassarre, M. T. (2011). Trends in Harmonization of Multiple Reference Models. In L. A. Maciaszek & P. Loucopoulos (Eds.), Evaluation of Novel Approaches to Software Engineering (Vol. 230, pp. 61–73). Berlin, Heidelberg: Springer Berlin Heidelberg. Retrieved from http://www.springerlink.com/index/10.1007/978-3-642-23391-3_5

Paulk, M. C. (2008). A Taxonomy for Improvement Frameworks. Presented at the World Congress for Software Quality.

Paulk, M., Weber, C. V., Curtis, B., & Chrissis, M. B. (1995). The capability maturity model : guidelines for improving the software process. Reading Mass.: Addison-Wesley Pub. Co.

Peldzius, S., & Ragaisis, S. (2012). Framework for Usage of Multiple Software Process Models. In A. Mas, A. Mesquida, T. Rout, R. V. O'Connor, & A. Dorling (Eds.), Software Process Improvement and Capability Determination (Vol. 290, pp. 210–221). Berlin, Heidelberg: Springer Berlin Heidelberg. Retrieved from http://www.springerlink.com/index/10.1007/978-3-642-30439-2_19

Pesantes, M., Lemus, C., Mitre, H. A., & Mejıa, J. (2012). Identifying Criteria for Designing a Process Architecture in a Multimodel Environment. Presented at the ICSSP 2012, Zürich, Switzerland. Retrieved from https://files.ifi.uzh.ch/icse2012/proceedings/icse12-online/icssp12/p83-pesantes.pdf

Pino, F. J., García, F., & Piattini, M. (2007). Software process improvement in small and medium software enterprises: a systematic review. Software Quality Journal, 16(2), 237–261. doi:10.1007/s11219-007-9038-z

POLYGON Informatikai Kft. (2007). Company Information. Retrieved October 16, 2012, from http://www.polygon.hu/wps/wcm/connect/polygon2en/Polygon+Informatikai+Kft./Company/

POLYGON Informatikai Kft. (2008). POLYGON Informatikai Kft. - QWorks information page. Retrieved March 25, 2010, from http://www.polygon.hu/wps/wcm/connect/polygon2/Polygon+Informatikai+Kft./Palyazatok/EU+palyazatok/GOP+2008/

POLYGON Informatikai Kft. (2010). 6.3. Tesztelési terv készítése, teszt adatok felvitele.

Polyvyanyy, A. (2012). Structuring Process Models. University of Potsdam, Postdam. Retrieved from http://opus.kobv.de/ubp/volltexte/2012/5902/pdf/polyvyanyy_diss.pdf

Portillo-Rodriguez, J., Vizcaino, A., Ebert, C., & Piattini, M. (2010). Tools to Support Global Software Development Processes: A Survey (pp. 13–22). IEEE. doi:10.1109/ICGSE.2010.12

Process Impact. (2010, November 24). Peer Review Process Description v1.0 draft 1. Retrieved from http://processimpact.com/pr_goodies.shtml

Rahman, A. A., Sahibuddin, S., & Ibrahim, S. (2011). A unified framework for software engineering process improvement — A taxonomy comparative analysis (pp. 153–158). IEEE. doi:10.1109/MySEC.2011.6140661

Rapid - I. (2012). RapidMiner. Retrieved May 29, 2012, from http://rapid-i.com/content/view/181/190/

Richardson, I. (2006). SPI Models: What Characteristics Are Required for Small Software Development Companies? In J. Kontio & R. Conradi (Eds.), Software Quality — ECSQ 2002 (Vol. 2349, pp. 100–113). Berlin, Heidelberg: Springer Berlin Heidelberg. Retrieved from http://www.springerlink.com/index/10.1007/3-540-47984-8_14

Rico, D. F. (2002). How to estimate ROI for inspections, PSP, TSP, SW-CMM, ISO 9000, and CMMI. The DoD Softwaretech News.

Rico, D. F. (2003). Practical Metrics and Models for Return on Investment. Retrieved from http://davidfrico.com/rico03apdf.htm

Rico, D. F. (2004). ROI of software process improvement : metrics for project managers and software engineers. Boca Raton  Fla.: J. Ross Pub.

Ries, E. (2011). The lean startup. New York: Crown Business.

Robinson, W. C. (2006, November). IS 540: Research methods: Questionnaires. Retrieved April 9, 2010, from http://web.utk.edu/~wrobinso/540_lec_qaire.html

Romero, H., Dijkman, R., Grefen, P., & Weele, A. (2012). Harmonization of Business Process Models. In F. Daniel, K. Barkaoui, & S. Dustdar (Eds.), Business Process Management Workshops (Vol. 99, pp. 13–24). Berlin, Heidelberg: Springer Berlin Heidelberg. Retrieved from http://www.springerlink.com/index/10.1007/978-3-642-28108-2_2

Rout, T. P., & Tuffley, A. (2007). Harmonizing ISO/IEC 15504 and CMMI. Software Process: Improvement and Practice, 12(4), 361–371. doi:10.1002/spip.329

Rout, T. P., Tuffley, A., & Cahill, B. (2001). CAPABILITY MATURITY MODEL INTEGRATION MAPPING TO ISO/IEC 15504-2:1998. Software Quality Institute, Griffith University. Retrieved from http://www.sqi.gu.edu.au/cmmi/report/docs/MappingReport.pdf

Salviano, C. F. (2009a). A Multi-model Process Improvement Methodology Driven by Capability Profiles. 2009 33rd Annual IEEE International Computer Software and Applications Conference (pp. 636–637). Presented at the 2009 33rd Annual IEEE International Computer Software and Applications Conference (COMPSAC), Seattle, Washington, USA. doi:10.1109/COMPSAC.2009.94

Salviano, C. F. (2009b). PRO2PI Methodology - A Methodology on "Process Capability Profile to drive Process Improvement" - version 3.0. Divisão de Melhoria de Processo de Software Centro de Tecnologia da Informação Renato Acher, Campinas, SP, Brasil. Retrieved from http://pro2pi.wdfiles.com/local--files/publicacoes-sobre-a-metodologia/Salviano2009TR_PRO2PIv3p0.pdf

Saunders, M., Lewis, P., & Thornhill, A. (2007). Research methods for business students (4th ed.). Harlow England ;;New York: Financial Times/Prentice Hall.

SEI. (2010). SEIR - Software Engineering Information Repository. Retrieved March 11, 2010, from https://seir.sei.cmu.edu/seir/

SEI, C. (2008). Process Improvement in Multimodel Environments (PrIME) website. Retrieved March 11, 2010, from http://sei.cmu.edu/process/research/prime.cfm

Sheard, S. A. (2001). Evolution of the frameworks quagmire. Computer, 34(7), 96–98. doi:10.1109/2.933516

Shenton, A. K. (2004). Strategies for ensuring trustworthiness in qualitative research projects. Education for information, 22(2), 63–76.

Silverman, D. (2006). Interpreting qualitative data: Methods for analyzing talk, text, and interaction. Sage Publications Ltd.

Siviy, J., Kirwan, P., Marino, L., & Morley, J. (2008a, March). Improvement Technology Classification and Composition in Multimodel Environments. Carnegie Mellon University. Retrieved from http://www.sei.cmu.edu/library/assets/3.pdf

Siviy, J., Kirwan, P., Marino, L., & Morley, J. (2008b, March). Maximizing your Process Improvement ROI through Harmonization. Carnegie Mellon University. Retrieved from http://www.sei.cmu.edu/library/assets/multimodelExecutive_wp_harmonizationROI_032008_v1.pdf

Siviy, J., Kirwan, P., Marino, L., & Morley, J. (2008c, March). The Value of Harmonizing Multiple Improvement Technologies: A Process Improvement Professional's View. Carnegie Mellon University. Retrieved from http://www.sei.cmu.edu/library/assets/multimodelSeries_wp1_harmonizationValue_052008_v1.pdf

Siviy, J., Kirwan, P., Marino, L., & Morley, J. (2008d, March). Process Architecture in a Multimodel Environment. Carnegie Mellon University. Retrieved from http://www.sei.cmu.edu/library/assets/multimodelSeries_wp4_processArch_052008_v1.pdf

Siviy, J., Kirwan, P., Marino, L., & Morley, J. (2008e, May). Implementation Challenges in a Multimodel Environment. Carnegie Mellon University. Retrieved from http://www.sei.cmu.edu/library/assets/4.pdf

Siviy, J., Kirwan, P., Marino, L., & Morley, J. (2008f, May). Strategic Technology Selection and Classification in Multimodel Environments. Carnegie Mellon University. Retrieved from http://www.sei.cmu.edu/library/assets/2.pdf

Siviy, J., Penn, M. L., & Stoddard, R. W. (2007). CMMI and Six Sigma : partners in process improvement. Upper Saddle River NJ: Addison-Wesley.

Solingen, R. van. (2000). Product focused Software Process Improvement : SPI in the embedded software domain. (M. Aldham-Breary, Trans.). [Eindhoven]: Eindhoven University of Technology, Faculty of Technology Management. Retrieved from http://alexandria.tue.nl/extra2/200000702.pdf

Sparx Systems. (2011). UML 2 Tutorial - Activity Diagram. Retrieved December 12, 2011, from http://www.sparxsystems.com/resources/uml2_tutorial/uml2_activitydiagram.html

SQI. (2008). SQI Information Repository. Retrieved from http://sqi.hu/tools/information-repository/

Thiry, M., Zoucas, A., & Tristão, L. (2010). Mapping Process Capability Models to Support Integrated Software Process Assessments. CLEI ELECTRONIC JOURNAL, 13.

TMMi Foundation. (2009). The Test Maturity Model Integration (TMMi) v2.0. Retrieved from http://www.tmmifoundation.org/downloads/tmmi/TMMi%20Framework.pdf

Trochim, W. M. K. (2006, October 20). Reliability & Validity. Retrieved May 6, 2012, from http://www.socialresearchmethods.net/kb/relandval.php

TSO. (2007). ITIL Lifecycle Publication Suite 5volset. London, Great Britain.: The Stationery Office/Tso.

Tsui, F. F., & Karam, O. (2007). Essentials of software engineering. Sudbury, Mass.: Jones and Bartlett Publishers.

Tuffley, A., Rout, T. P., Stone-Tolcher, M., & Gray, I. (2004). ISO 9001:2000 and the Capability Maturity Model Integration. QUALCON 2004 Conference Proceedings. Presented at the QUALCON 2004, Adelaide.

von Wangenheim, C. G., Hauck, J. C. R., Salviano, C. F., & von Wangenheim, A. (2010). Systematic Literature Review of Software Process Capability/Maturity Models. Proceedings of International Conference on Software Process. Improvement And Capability dEtermination (SPICE). Presented at the International Conference on Software Process. Improvement And Capability dEtermination (SPICE), Pisa, Italy. Retrieved from http://www.inf.ufsc.br/~gresse/download/SPICE2010_Systematic_Literature_vf.pdf

Wahl, T., & Sindre, G. (2005). An Analytical Evaluation of BPMN Using a Semiotic Quality Framework. In T. A. Halpin, K. Siau, & J. Krogstie (Eds.), Proceedings of the Workshop on Evaluating Modeling Methods for Systems Analysis and Design (EMMSAD'05), held in conjunctiun with the 17th Conference on Advanced Information Systems (CAiSE'05), Porto, Portugal, EU (pp. 533–544). FEUP, Porto, Portugal, EU.

Weidlich, M., Dijkman, R., & Mendling, J. (2010). The ICoP Framework: Identification of Correspondences between Process Models (Vol. 6051, pp. 483–489). doi:10.1007/978-3-642-13094-6_37

Weske, M. (2007). Business process management : concepts, languages, architectures. Berlin ;;New York: Springer.

WfMC. (2012). XML Process Definition Language. Retrieved May 28, 2012, from http://www.wfmc.org/standards/on XPDL.htm

White, S. A. (2004, March). Process Modelin Notations and Workflow Patterns. BPTrends. Retrieved from http://www.omg.org/bp-corner/bp-files/Process_Modeling_Notations.pdf

White, S. A. (2005, March). Using BPMN to Model a BPEL Process. BPTrends. Retrieved from http://bptrends.com/publicationfiles/03-05%20WP%20Mapping%20BPMN%20to%20BPEL-%20White.pdf

Wibas. (2009). wibas GmbH – CITIL = CMMI+ITIL. Retrieved June 29, 2010, from http://www.wibas.com/publications/itil_and_citil/citil__cmmiitil/index_en.html

Wiegers, K. (2002). Peer reviews in software : a practical guide. Boston MA: Addison-Wesley.

Wiegers, K. E. (1998). Seven Deadly Sins of Software Reviews. Software Development. Retrieved from www.processimpact.com

Wiegers, K. E. (2002a). Humanizing Peer Reviews. Retrieved March 8, 2010, from http://www.processimpact.com/articles/humanizing_reviews.html

Wiegers, K. E. (2002b). Seven Truths about Peer Reviews. Cutter IT Journal. Retrieved from http://www.processimpact.com

Wienberg, A. (2001). A Comparison of Event-driven Process Chains and UML Activity Diagram for Denoting Business Processes. Technische Universität Hamburg-Harburg Arbeitsbereich Softwaresysteme. Retrieved from http://www.sts.tu-harburg.de/pw-and-m-theses/2001/Ferd01.pdf

Wortmann, H., & Kusters, R. (2007). Enterprise Modelling and Enterprise Information Systems.

Yin, R. (2009). Case study research : design and methods (4th ed.). Los Angeles Calif.: Sage Publications.

Yoo, C., Yoon, J., Lee, B., Lee, C., Lee, J., Hyun, S., & Wu, C. (2006). A unified model for the implementation of both ISO 9001:2000 and CMMI by ISO-certified organizations. Journal of Systems and Software, 79(7), 954–961. doi:10.1016/j.jss.2005.06.042

Zhu, J., Sun, H., Huang, Z., & Liu, X. (2010). MBPR: A Business Process Repository Supporting Multi-Granularity Process Model Retrieval (pp. 172–178). Presented at the SERVICE COMPUTATION 2010 : The Second International Conferences on Advanced Service Computing, Lisbon, Portugal: IARIA. Retrieved from http://www.thinkmind.org/index.php?view=article&articleid=service_computation_2010_8_10_20057

# SUMMARY

## PROCESS BASED UNIFICATION FOR MULTI-MODEL SOFTWARE PROCESS IMPROVEMENT

Many different quality approaches are available in the software industry. Some of the approaches, such as ISO 9001 are not software specific, i.e. they define general requirements for an organization and they can be used at any company. Others, such as Automotive SPICE have been derived from a software specific approach, and can be used for improving specific (in this case automotive) processes. Some are created to improve development processes (e.g. CMMI for Development), others focus on services (e.g. CMMI for Services), and again others are related to particular processes such as software testing (e.g. TMMi) or resource management (e.g. People CMM).
A number of differences among quality approaches exist and there can be various situations in which the usage of multiple approaches is required, e.g. to strengthen a particular process with multiple quality approaches or to reach certification of the compliance to a number of standards. First of all it has to be decided which approaches have potential for the organization. In many cases one approach does not contain enough information for process implementation. Consequently, the organization may need to use several approaches and the decision has to be made how the chosen approaches can be used simultaneously. This area is called Multi-model Software Process Improvement (MSPI). The simultaneous usage of multiple quality approaches is called the multi-model problem.

In this dissertation we propose a solution for the multi-model problem which we call the *Process Based Unification (PBU) framework*. The PBU framework consists of the PBU concept, a PBU process and the PBU result. We call PBU concept the mapping of quality approaches to a unified process. The PBU concept is operationalized by a PBU process. The PBU result includes the resulting unified process and the mapping of quality approaches to the unified process.

Accordingly, we addressed the following research question:
*Does the PBU framework provide a solution of sufficient quality for current problems of simultaneous usage of multiple quality approaches?*
In order to recognize a solution of sufficient quality we identified criteria for multi-model solutions based on current problems and initiatives. This is called MSPI criteria.
In order to determine if the mapping of quality approaches to a process is possible we analysed elements of quality approaches, elements of processes and the mapping of quality approach elements to process elements. Findings of the analysis lead us to design a multi-model solution, the PBU framework.
In order to show the feasibility of the results, we performed a case study reflecting how the PBU framework can be used in practice for developing a unified (peer review) process which conforms to three different quality approaches. Finally we assessed the PBU framework against MSPI criteria. With the case study and the assessment of the PBU framework we answered the research question and provided a proof of concept of the research.

# SAMENVATTING

## PROCES GEBASEERDE UNIFICATIE VOOR HET VERBETEREN VAN MULTI-MODEL SOFTWARE PROCESSEN

Er zijn vele kwaliteitsbenaderingen beschikbaar in de sofware industrie. Sommige hiervan, zoals ISO 9001 zijn niet software specifiek; ze definiëren algemene voorwaarden voor een organisatie en kunnen worden gebruikt door elk bedrijf. Andere kwaliteitsbenaderingen, zoals Automotive SPICE zijn gebaseerd op een software specifieke benadering, en kunnen worden gebruikt om specifieke processen (in dit geval automotieve processen) te verbeteren. Sommige benaderingen zijn gecreëerd om ontwikkelingsprocessen te verbeteren (bijv. CMMI for Development), andere focussen op services (bijv. CMMI for Services), en nog weer andere zijn gerelateerd met bepaalde processen zoals het testen van software (bijv. TMMi) of resource management (bijv. People CMM).
Er zijn meerdere verschillen tussen bovenstaande kwaliteitsbenaderingen en in bepaalde situaties is het gebruik van meerdere benaderingen nodig, bijvoorbeeld om een bepaald proces te versterken met meerdere kwaliteitsbenaderingen of om certificatie te bereiken voor de naleving van bepaalde standaards. Ten eerste moet er beslist worden welke benaderingen mogelijk zijn binnen de organisatie. In veel gevallen bevat één benadering niet genoeg informatie om het proces te implementeren. Daarom zal de organisatie meerdere benaderingen moeten gebruiken er zal beslist moeten worden hoe de gekozen benaderingen tegelijkertijd gebruikt kunnen worden. Dit gebied wordt Multi-model Software Process Improvement (MSPI) genoemd. Het tegelijkertijd gebruiken van meerder kwaliteitsbenaderingen wordt het multi-model probleem genoemd.

In dit proefschrift wordt een oplossing voor het multi-model probleem voorgelegd, dat we het Process Based Unification (PBU) framework noemen. Dit framework bestaat uit het PBU concept, het PBU proces en het PBU resultaat. Het PBU concept bestaat uit het in kaart brengen van kwaliteitsbenaderingen tot een geünificeerd proces. Het PBU concept wordt verwezenlijkt door het PBU proces. Het PBU resultaat bestaat uit het resulterende geünificeerde proces en het in kaart brengen van de kwaliteitsbenaderingen die leiden tot het geünificeerde proces.

Derhalve wordt de volgende onderzoeksvraag behandeld:
*Is het PBU framework een oplossing van voldoende kwaliteit voor de huidige problemen bij het gelijktijdig gebruik van meerdere kwaliteitsbenaderingen?*
Om een oplossing van voldoende kwaliteit te herkennen zijn criteria gedefinieerd voor een multi-model oplossing die gebaseerd zijn op de huidige problemen en initiatieven. Deze worden MSPI criteria genoemd. Om te onderzoeken of het in kaart brengen van de kwaliteitsbenadering van een proces mogelijk is hebben we elementen van kwaliteitsbenaderingen, elementen van processen en het in kaart brengen van de kwaliteitsbenadering elementen naar de proces elementen geanalyseerd. De resultaten van deze analyse hebben geleid tot het opstellen van de multi-model oplossing, het PBU framework.
Om de haalbaarheid van de resultaten te laten zien, hebben we een case study gedaan die reflecteert hoe het PBU framework in de praktijk gebruikt kan worden voor het ontwikkelen van een geünificeerd (peer review) proces dat in overeenstemming is met de drie verschillende kwaliteitsbenaderingen. Ten slotte is het PBU framework geëvalueerd ten opzichte van de MSPI criteria. Met de case study en de evaluatie van het PBU framework wordt de onderzoeksvraag beantwoordt en wordt een bewijs gegeven voor het concept van het onderzoek.

# CURRICULUM VITAE

Zádor Dániel Kelemen was born on 5 September 1980 in Székelyudvarhely (Odorheiu Secuiesc), Romania. He obtained his Bachelor's degree from the Gheorghe Asachi Technical University of Iași, Romania in 2002 and his Master's degree from Budapest University of Technology and Economics in Budapest, Hungary in 2006.

Between 2006 and 2009 he was a PhD student at Budapest University of Technology with specialization for intelligent systems, cooperating with Eindhoven University of Technology.

Since 2005 he is a Software Process Improvement consultant at SQI – Hungarian Software Quality Consulting Institute Ltd. Besides consultancy work, he participates in process and service assessments, in formal CMMI-based "SCAMPI A" appraisals and different R&D projects in the field of Software Quality, Software Testing and Software Process Improvement.

Since 2006 he participates in teaching Software Testing at Budapest University of Technology and Economics. He is also a trainer in the fields of Software Quality and Software Testing at SQI – Hungarian Software Quality Consulting Institute Ltd.

From 2007 he started a PhD project at Eindhoven University of Technology of which the results are presented in this dissertation.

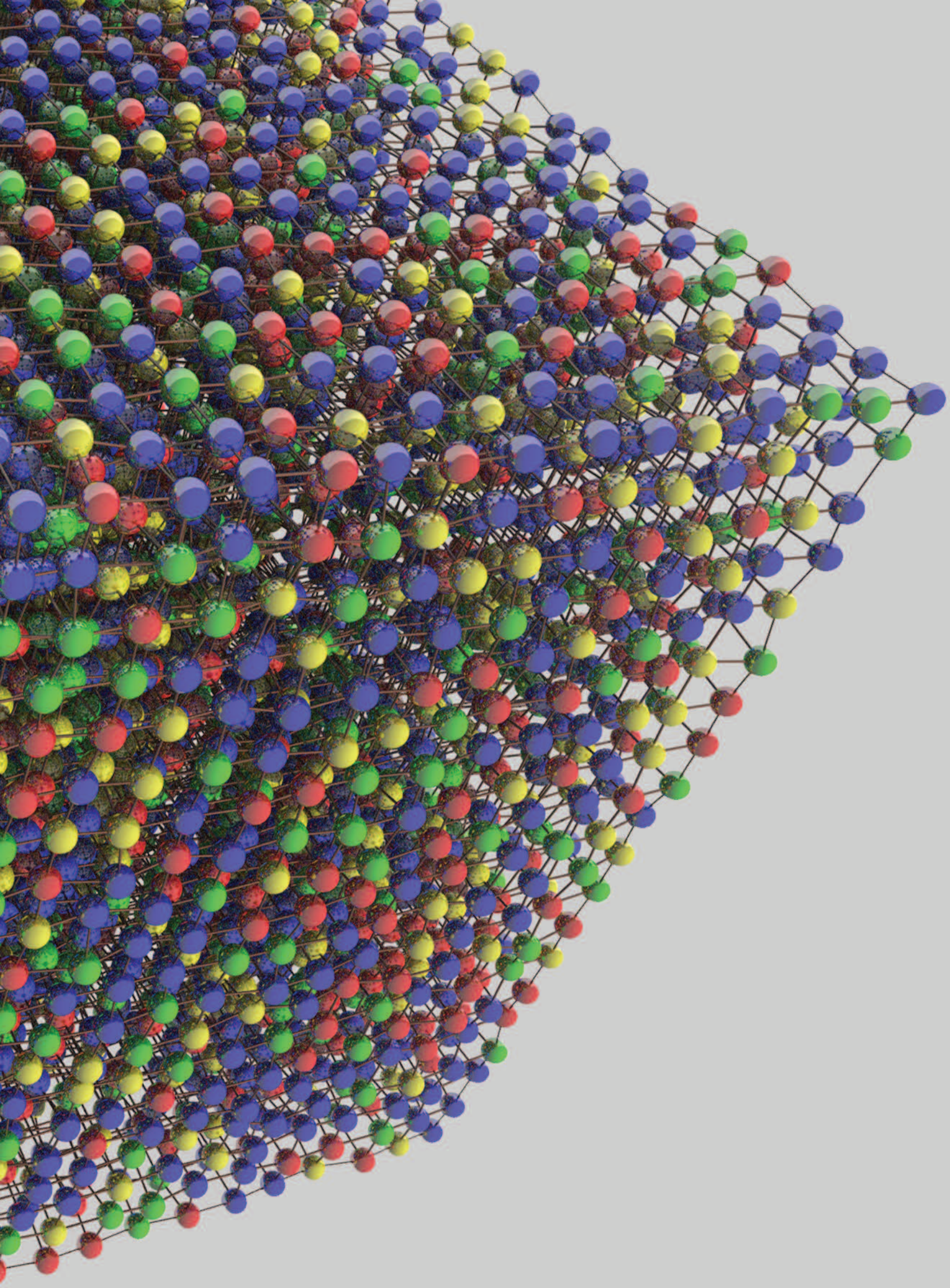

**Eindhoven University of Technology**
**Department of Industrial Engineering & Innovation Management**